\documentclass[aps,prx,10pt,twocolumn,superscriptaddress,notitlepage,floatfix]{revtex4-2}

\usepackage[utf8]{inputenc}
\usepackage{graphicx}
\graphicspath{{../}}

\pdfoutput=1 

\usepackage[ruled,vlined]{algorithm2e}

\usepackage{dsfont}
\usepackage{amsmath}
\usepackage{amssymb,bm}
\usepackage{amsthm}
\usepackage{mathtools}
\usepackage{url}
\usepackage{microtype}
\usepackage{array, makecell}
\usepackage{boldline}
\usepackage{enumitem}

\usepackage{color,dsfont}
\usepackage{ragged2e}
\usepackage[colorlinks=true, hyperindex, breaklinks, linkcolor=blue, urlcolor=blue, citecolor=blue]{hyperref} 
\usepackage[normalem]{ulem}
\usepackage[capitalise]{cleveref}
\usepackage{mathrsfs}
\usepackage[caption=false]{subfig}
\usepackage{float}
\usepackage{verbatim}
\usepackage{latexsym}
\usepackage{setspace}
\usepackage{amsfonts}
\usepackage{stmaryrd}
\usepackage{xcolor}
\usepackage{enumitem}
\usepackage{lipsum}
\usepackage{multirow}
\AtBeginDocument{\tabcolsep=6pt}
\usepackage{booktabs}

\definecolor{AWSdarkblue}{RGB}{26,35,47}
\definecolor{AWScharcoal}{RGB}{17,22,33}
\definecolor{AWSorange}{RGB}{252,134,8}
\definecolor{AWSlightblue}{RGB}{19,142,188}
\definecolor{AWSblue}{RGB}{19,142,188}
\definecolor{AWSlightgreen}{RGB}{88,162,40}
\definecolor{AWSdarkgreen}{RGB}{28,122,2}
\definecolor{AWSred}{RGB}{252,62,54}

\newcommand{\ket}[1]{|#1\rangle}  
\newcommand{\STOP}{\texttt{STOP}}
\newcommand{\tHubbard}{t}

\newtheorem{claim}{Claim} 
\newtheorem{definition}{Definition}
\newtheorem{lemma}{Lemma}

\newcommand{\aop}{\hat{a}}
\newcommand{\aopd}{\hat{a}^\dagger}
\newcommand{\bop}{\hat{b}}
\newcommand{\bopd}{\hat{b}^\dagger}
\newcommand{\EJ}{E_J}
\newcommand{\epd}{\epsilon_d}

\newcommand{\hc}{\text{h.c.}}

\newcommand{\kappab}{\kappa_b}
\newcommand{\omegaa}{\omega_a}
\newcommand{\omegab}{\omega_b}
\newcommand{\omegad}{\omega_d}
\newcommand{\omegap}{\omega_p}

\newcommand{\phiop}{\hat{\phi}}
\newcommand{\phizpa}{\varphi_a}
\newcommand{\phizpb}{\varphi_b}
\def\id{{\mathchoice {\rm 1\mskip-4mu l} {\rm 1\mskip-4mu l} {\rm 1\mskip-4.5mu l} {\rm 1\mskip-5mu l}}}
\newcommand{\cnot}{\text{CNOT}\xspace}
\newcommand{\ccz}{\text{CCZ}\xspace}
\newcommand{\tof}{\text{TOF}\xspace}
\usepackage{textcomp}
\usepackage{gensymb}

\usepackage{array}
\usepackage{multirow}
\newcolumntype{x}[1]{>{\centering\arraybackslash\hspace{0pt}}p{#1}}
\definecolor{amethyst}{rgb}{0.6, 0.4, 0.8}

\newcommand{\kb}[2]{\ensuremath{\vert #1 \rangle\! \langle #2 \vert}}
\newcommand{\BUTOFF}{\texttt{BUTOF}\xspace} 
\newcommand{\TDTOFF}{\texttt{TDTOF}\xspace} 

\newcommand{\REGone}{\texttt{REGIME 1}\xspace}
\newcommand{\REGtwo}{\texttt{REGIME 2}\xspace}
\newcommand{\REGthree}{\texttt{REGIME 3}\xspace} 
\newcommand{\jp}[1]{\textcolor{black}{#1}}  
\newcommand{\etc}[1]{\textcolor{black}{#1}}  
\newcommand{\hp}[1]{\textcolor{black}{#1}} 
\newcommand{\ji}[1]{\textcolor{black}{#1}} 
\newcommand{\CC}[1]{\textcolor{black} {#1}} 
\newcommand{\kn}[1]{\textcolor{black}{#1}} 
\newcommand{\ch}[1]{\textcolor{black}{#1}} 

\begin{document}

\title{Building a fault-tolerant quantum computer using concatenated cat codes}

\author{Christopher Chamberland}
\affiliation{AWS Center for Quantum Computing, Pasadena, CA 91125, USA}
\affiliation{IQIM, California Institute of Technology, Pasadena, CA 91125, USA}

\author{Kyungjoo Noh}
\affiliation{AWS Center for Quantum Computing, Pasadena, CA 91125, USA}

\author{Patricio Arrangoiz-Arriola}
\thanks{These authors contributed equally}
\affiliation{AWS Center for Quantum Computing, Pasadena, CA 91125, USA}

\author{Earl~T.~Campbell}
\thanks{These authors contributed equally}
\affiliation{AWS Center for Quantum Computing, Pasadena, CA 91125, USA}

\author{Connor T. Hann}
\thanks{These authors contributed equally}

\affiliation{AWS Center for Quantum Computing, Pasadena, CA 91125, USA}
\affiliation{Department of Physics, Yale University, New Haven, CT 06511, USA}

\author{Joseph Iverson}
\thanks{These authors contributed equally}

\affiliation{AWS Center for Quantum Computing, Pasadena, CA 91125, USA}

\author{Harald Putterman}
\thanks{These authors contributed equally}

\affiliation{AWS Center for Quantum Computing, Pasadena, CA 91125, USA}

\author{Thomas C. Bohdanowicz}
\affiliation{AWS Center for Quantum Computing, Pasadena, CA 91125, USA}
\affiliation{IQIM, California Institute of Technology, Pasadena, CA 91125, USA}

\author{Steven T. Flammia}
\affiliation{AWS Center for Quantum Computing, Pasadena, CA 91125, USA}

\author{Andrew Keller}
\affiliation{AWS Center for Quantum Computing, Pasadena, CA 91125, USA}

\author{Gil Refael}
\affiliation{AWS Center for Quantum Computing, Pasadena, CA 91125, USA}
\affiliation{IQIM, California Institute of Technology, Pasadena, CA 91125, USA}
\author{John Preskill}
\affiliation{AWS Center for Quantum Computing, Pasadena, CA 91125, USA}
\affiliation{IQIM, California Institute of Technology, Pasadena, CA 91125, USA}

\author{Liang Jiang}
\affiliation{AWS Center for Quantum Computing, Pasadena, CA 91125, USA}
\affiliation{Pritzker School of Molecular Engineering, The University of Chicago, Illinois 60637, USA}

\author{Amir H. Safavi-Naeini}
\affiliation{AWS Center for Quantum Computing, Pasadena, CA 91125, USA}
\affiliation{Department of Applied Physics and Ginzton Laboratory, Stanford University, Stanford, CA 94305, USA}

\author{Oskar Painter}
\affiliation{AWS Center for Quantum Computing, Pasadena, CA 91125, USA}
\affiliation{IQIM, California Institute of Technology, Pasadena, CA 91125, USA}

\author{Fernando G.S.L. Brand\~ao}
\affiliation{AWS Center for Quantum Computing, Pasadena, CA 91125, USA}
\affiliation{IQIM, California Institute of Technology, Pasadena, CA 91125, USA}

\begin{abstract}

We present a comprehensive architectural analysis for a \etc{proposed} fault-tolerant quantum computer based on cat codes concatenated with outer quantum error-correcting codes.  For the physical hardware, we propose a system of acoustic resonators coupled to superconducting circuits with a two-dimensional layout. Using estimated physical parameters for the hardware, we perform a detailed error analysis of measurements and gates, including CNOT and Toffoli gates.  Having built a realistic noise model, we numerically simulate quantum error correction when the outer code is either a repetition code or a thin rectangular surface code. Our next step toward universal fault-tolerant quantum computation is a protocol for fault-tolerant Toffoli magic state preparation that significantly improves upon the fidelity of physical Toffoli gates at very low qubit cost. To achieve even lower overheads, we devise a new magic-state distillation protocol for Toffoli states. Combining these results together, we obtain realistic full-resource estimates of the physical error rates and overheads needed to run useful fault-tolerant quantum algorithms. We find that with around 1,000 superconducting circuit components, one could construct a fault-tolerant quantum computer that can run circuits which are \etc{currently} intractable for classical computers. Hardware with 18,000 superconducting circuit components, in turn, could simulate the Hubbard model in a regime beyond the reach of classical computing.

\end{abstract}

\maketitle

\tableofcontents

\section{Introduction}

Building a fault-tolerant quantum computer is one of the great scientific and engineering challenges of the 21st century. 
A successful quantum computing architecture must meet many conflicting demands: it must have an error correction threshold that is achievable by hardware on a large scale, a convenient physical layout, and implement arbitrary quantum algorithms with low resource overhead requirements. 
All proposed quantum architectures require tradeoffs among these objectives. 
For example, the most popular proposed architecture, the surface code~\cite{Bravyi98}, has a convenient two-dimensional physical layout and relatively high threshold error rates, but the overhead for running useful algorithms remains daunting~\cite{Ogorman17,motta2018low,gidney2019factor,Campbell_2019,kivlichan2020improved,chamberland2021universal}, even after years of optimization.

\etc{Recent work has shown that qubits with highly \textit{biased} noise are a promising route to fault tolerance~\cite{Tuckett18,Tuckett19,Tuckett20,BonillaAtaides20}, at least when gates that preserve the noise bias can be easily implemented in the architecture~\cite{Aliferis08,Puri19,Guillaud2019,GM2020}. 
One possible route to realizing such qubits is via a two-component cat code~\cite{Mirrahimi2014,puri2017,Cohen2017}, a bosonic qubit encoded in an oscillator mode \cite{Albert2018,Joshi2020,Cai2020}, subjected to engineered two-photon dissipation~\cite{Mirrahimi2014,Leghtas2015,Touzard2018} or an engineered Kerr nonlinearity~\cite{Goto2016_bifurcation,Goto2016_universal,puri2017,Puri2019,Grimm2020}. The engineered interaction heavily suppresses population transfer between the two constituent coherent states of the cat qubit, causing an effective noise bias towards phase-flip errors on the cat qubits~\cite{Mirrahimi2014,puri2017}. Experiments suggest that it is possible to engineer highly biased noise with this approach~\cite{Lescanne2020}. Furthermore, bias-preserving \cnot and Toffoli (\tof) gates can be performed for these cat codes~\cite{Guillaud2019,Puri19}. }

\etc{The performance of dissipative cat qubits is influenced by three key parameters. The average number of excitations in each cat is $|\alpha|^2$, which determines the level of noise bias as bit flips are exponentially suppressed with $|\alpha|^2$. The rate of phase-flip errors is determined by the competition between two processes: $\kappa_1$ is the single-excitation loss rate (per time) that is the main cause of phase errors; and $\kappa_2$ is the engineered two-excitation dissipation rate (per time) stabilizing the cat-code subspace and suppressing errors. The ratio of these processes $\kappa_1/\kappa_2$ is a dimensionless quantity primarily determining the phase-flip error rate. Calculating accurate predictions for $\kappa_1/\kappa_2$ is crucial for estimating the performance of cat qubit architectures.} 

\etc{Concatenating the (inner) cat code with another (outer) quantum error-correcting code can reduce qubit requirements by tailoring the outer code to suppress the dominant phase-flip errors. We call these coding schemes \textit{concatenated cat codes}. While this idea has been explored previously for the case where the outer code is a repetition code~\cite{Guillaud2019,GM2020}, these proposals are completely reliant on increasing $|\alpha|^2$ to suppress bit-flips errors. As $|\alpha|^2$ increases, phase errors become more frequent and other physical mechanisms start to become important, so a fully scalable architecture must allow for some bit-flip protection from the outer code.  Furthermore, these previous proposals~\cite{Guillaud2019,GM2020} did not study the rate of bit-flips processes during CNOT gates and did not propose a 2D layout capable of implementing fault-tolerant logic.  The lack of such an analysis has left open several urgent questions, such as how a 2D architecture with dissipative cats concatenated with the surface code would perform in practice, and how parameters at the hardware level (such as $\kappa_1/\kappa_2$) relate to the needs of the larger architecture.}  
 
In this paper, we give a full-stack analysis of a fault-tolerant quantum architecture based on \etc{dissipative} cat codes concatenated with outer quantum error-correcting codes.
We propose a blueprint for a possible practical implementation based on hybrid electro-acoustic systems consisting of acoustic resonators coupled to superconducting circuits. These systems are a promising platform for realizing concatenated cat codes due to their small footprint \cite{Arrangoiz-Arriola2016}, potential for ultra-high coherence times \cite{MacCabe2020}, and easy integration with superconducting circuits for control and read-out \cite{CircuitQED,Arrangoiz-Arriola2019}. 

We give a comprehensive error analysis of this approach that provides a detailed picture of the physically achievable hardware parameters (including $\kappa_1$, $\kappa_2$ and $|\alpha|^2$) and error rates for gates and measurements based on estimated parameters for coupling strengths and phonon loss and dephasing rates. 
Using the obtained values of the hardware parameters, we then explicitly analyze quantum error correction when the outer code is either a repetition code or a thin rectangular surface code~\cite{dennis2002topological,fowler2012surface}. We then show how to build a fault-tolerant quantum computer in our architecture, combining lattice surgery and magic state distillation for Toffoli states. Finally, we provide a resource overhead estimate as a function of physical error rates required to fault-tolerantly run quantum algorithms.

\etc{Our analysis can be broadly classified into three categories: 1) a hardware proposal; 2) a physical-layer analysis of gate and measurements errors; and 3) a logical-level analysis of memory and computation failure rates.
More specifically, in \cref{sec:HardwareImplementation} we describe our hardware proposal for using phononic bandgap resonators and superconducting circuits to store and process quantum information at the physical level.  This section 
provides a range for what hardware parameters are feasible.  Then in \cref{sec:GatesMeas} we give a complete analysis of gate and measurement errors for phononic qubits using realistic noise parameters that we expect from the hardware proposal.  In Sections~\ref{sec:LogicalMemory}, \ref{sec:BottomUp}, and \ref{Sec:TopDown} we give a gate-level analysis of universal fault-tolerant quantum computation that looks at logical error rates across a physically relevant parameter regime.}

\begin{table*}
    \begin{center}
        \begin{tabular}{c c c c c c c c c}
            \toprule 
            & $\kappa_1/\kappa_2$ & $g_2/2\pi$ & $\kappa_b/2\pi$ & $\kappa_2/2\pi$    & $\kappa_{1, \text{i}} = T_{1, \text{i}}^{-1}$ 
            &  $|\alpha|^2$ & $p_{\mathrm{CNOT}}$ & Capabilities \\
            \midrule
            \REGone & $10^{-3}$ & $2 \, \text{MHz}$ & $57 \, \text{MHz}$ &  $280 \, \text{kHz}$    & $(570 \, \mu\text{s})^{-1}$ 
            & $8$ & 3.6 \% & repetition code QEC \\
            \midrule
            \REGtwo & $10^{-4}$ & $2 \, \text{MHz}$ &  $57 \, \text{MHz}$ & $280 \, \text{kHz}$    & $(5.7 \, \text{ms})^{-1}$ 
            & $8$ & 1.2 \% & surface code QEC  \\
            \midrule
            \REGthree & $10^{-5}$ & $2 \, \text{MHz}$ & $57 \, \text{MHz}$ & $280 \, \text{kHz}$   & $(57 \, \text{ms})^{-1}$ 
            & $8$ & 0.3 \% & useful quantum algorithms \\
            \bottomrule
        \end{tabular}
    \end{center}
    \caption{
    \etc{The three regimes studied in this work. The dimensionless loss $\kappa_1/\kappa_2$ is a key figure of merit of the cat code, as the $Z$-type error rates of the bias-preserving CNOT and Toffoli gates scale as $\sqrt{\kappa_1/\kappa_2}$. Therefore as we move from \REGone to \REGthree, the overall performance of the system improves, but the requirements imposed on the storage loss $\kappa_1$ become progressively more challenging. These requirements are shown in the table, in the more intuitive form of an energy decay time $T_{1,\text{i}} = 1/\kappa_{1, \text{i}}$. 
    We also show the required values of the nonlinear interaction strength $g_2$ and the buffer decay rate $\kappa_b$, from which we calculate $\kappa_2 = 4|g_2|^2/\kappa_b$ (see \Cref{sec:HardwareImplementation} for definitions of $g_2$ and $\kappa_b$, and a derivation of the expression for $\kappa_2$). 
    } } 
    \label{tab:regimes}
\end{table*}

\subsection{Overview of main results}

\etc{We frame our main results in terms of regimes for the hardware parameters, which we denote \REGone, \REGtwo and \REGthree, and which we summarize in \cref{tab:regimes}. All regimes assume the same number of excitations per cat ($|\alpha|^2=8$) but each regime corresponds to a different order of magnitude 
in the crucial $\kappa_1/\kappa_2$ parameter. In \REGone, the physical cat-qubit CNOT gate fails with probability $3.6*10^{-2}$, which is well above the threshold for the surface code error correction.  However, concatenating cat and repetition codes, \REGone is just below the repetition code phase-flip threshold and so is a suitable regime to demonstrate quantum error correction.  In \REGtwo, it is possible to demonstrate small proof-of-principle algorithms with the cat and repetition code.  However, while bit-flip errors are rare with $|\alpha|^2=8$, without additional bit-flip protection a quantum computer would decohere before it is able to demonstrate a useful algorithm. In \REGtwo, CNOT gates fail with probability $1.2*10^{-2}$, which is above the usually reported surface code error correction threshold with depolarizing noise. Nevertheless, due to noise bias and other aspects of the noise processes, \REGtwo would allow a demonstration of fully scalable quantum error correction and computation. However, the surface code overhead remains high in \REGtwo.  In \REGthree, CNOT gates fail with probability $3*10^{-3}$ and we estimate the resource overhead costs for the task of estimating the ground state energy density of the Hubbard model.  For this algorithm, we find $6\times$ fewer qubits are needed than for hardware assuming an unbiased, depolarizing noise model with CNOT gate infidelities of $10^{-3}$ as considered in Ref.~\cite{kivlichan2020improved}.   We now summarize in more detail how these conclusions were reached and the technical innovations needed along the way.}

In \cref{sec:HardwareImplementation} we describe our hardware proposal for using phononic-crystal-defect resonators (PCDRs), of the type reported in Ref.~\cite{Arrangoiz-Arriola2019}, as the storage elements. These are periodically patterned suspended nanostructures that support localized acoustic resonances in the gigahertz range. 
They are fabricated from a piezoelectric material such as $\text{LiNbO}_3$, which allows us to couple these resonances to superconducting circuits with nearly the same strength as ordinary electromagnetic cavities.  Following a recent demonstration~\cite{Lescanne2020}, we propose implementing the two-phonon dissipation by engineering an interaction through which the storage mode exchanges excitations with an ancillary ``buffer'' mode \emph{in pairs}. 
This buffer is strongly coupled to a bath, so these excitations decay rapidly. We compute $\kappa_2$ for a bath consisting of a multi-pole bandpass filter connected to a semi-infinite transmission line, or waveguide. The filter allows us to control the density of states of the bath, causing it to vanish at all frequencies except those within the filter passband (in this work this is modelled by connecting a dissipative circuit with an appropriate admittance function $Y(\omega)$ to the buffer resonator). The filter is useful not only to protect the storage mode from radiative decay, but also plays a crucial role in suppressing correlated phase-flip errors while stabilizing multiple storage modes simultaneously with the same buffer mode.

Given the stringent requirement for $\kappa_{1}/\kappa_{2}$, it is ideal to maximize $\kappa_{2}$. 
\ch{
In our architecture, however, $\kappa_2$ is ultimately limited by crosstalk. Indeed, we find that stronger engineered dissipation can simultaneously lead to increased crosstalk, such that there comes a point where further increasing $\kappa_2$ is no longer beneficial. Specifically, by quantifying crosstalk error rates and calculating their impacts on logical lifetimes, we find that the optimal value is $\kappa_2/2\pi \approx 280$kHz at $|\alpha|^2 = 8$.  }\etc{This constraint on $\kappa_2$ is discussed further in this section under the \cref{sec:GatesMeas} summary, and in more detail in \cref{subsection:Surface code logical failure rates in the presence of crosstalk errors}.} In turn, this imposes the requirement (shown in \cref{tab:regimes}) that the intrinsic relaxation time of the storage modes be at least $T_{1,i} \approx 57 \, \text{ms}$ to reach \REGthree, where it is possible to perform useful quantum algorithms. At present, piezoelectric PCDRs made of $\text{LiNbO}_3$ can only reach $T_{1,i} \approx 1 \, \mu s$~\cite{Wollack2020}.

The engineered dissipation needed to stabilize each cat code is provided by coupling each phononic resonator to nonlinear circuit elements. 
Specifically, we follow the approach of Ref.~\cite{Lescanne2020}, where the nonlinearity is provided by a circuit element variant of a superconducting quantum interference device (SQUID) called an asymmetrically-threaded SQUID (ATS). 
While Ref.~\cite{Lescanne2020} demonstrated an ATS can be used to stabilize a single mode into a cat code, our hardware layout necessitates that each ATS couple to and stabilize multiple resonators simultaneously. 
We present a simple scheme for this multiplexed stabilization, and provide a detailed analysis of the crosstalk that arises from coupling multiple modes to the same ATS. 
Moreover, we show that by employing a bandpass filter and carefully optimizing the phonon-mode frequencies, we are able to largely suppress the dominant sources of crosstalk in our system, though some residual crosstalk remains and we return to discuss this later.

In \cref{sec:GatesMeas}, we then analyze the errors in our gates and measurements. 
To do this, we introduce a method that we call the shifted Fock basis method.
This method allows us to efficiently perform a perturbative analysis of the dominant $Z$ error rates of the cat-qubit gates and improve the efficiency of numerical simulation of large cat qubits compared to the usual Fock basis method. 
The shifted Fock basis method allows us to compute the $Z$ error rates of various cat-qubit gates using a small Hilbert space dimension that is independent of the average excitation number $|\alpha|^{2}$ of the cat qubit. 

Using this method, we go on to show that the optimal $Z$ error rates (per gate) of the cat qubit gates at the optimal gate time scale as $\sqrt{\kappa_{1}/\kappa_{2}}$.
The optimal $Z$ error rates of the \cnot and \tof gates are in fact independent of the size of the cat qubit, whereas those of $Z$ and CZ rotations decrease linearly in $1/|\alpha|$.  We also study the effects of bosonic dephasing and thermal excitations on various cat-qubit gates.  Provided these additional effects are small, they do not disturb the noise bias or our main conclusion.    

\hp{We then develop and analyze schemes for readout in both the $X$ and $Z$ bases, enabling fast and hardware-efficient stabilizer measurements.  For $X$-basis readout we propose to use an additional dedicated mode in each unit cell of our architecture.  By exchanging the ancilla information with this mode and performing repeated quantum non-demolition (QND) parity measurements in parallel with the gates of the subsequent error correction cycle, we can suppress the infidelity mechanisms associated with the transmon while having minimal impact on the syndrome measurement cycle time.  We present a fast and high-fidelity $Z$-basis readout using the storage and buffer modes; the resulting error probability decays exponentially as a function of $|\alpha|^2$.  In the $Z$-basis readout scheme, excitation's are swapped to the buffer mode where they leak to the transmission line and are detected via a homodyne measurement. }

With a clear understanding of gate and measurement error rates, we proceed in \cref{sec:LogicalMemory} to analyze the logical failure rates for a quantum memory based on concatenating the cat code with one of two codes: a repetition code and a thin rectangular surface code. 
We compute logical $Z$ failure rates for both the repetition code and the surface code. 
In the case of the surface code, we compute explicit leading-order failure rates for logical $X$ errors as a function of the $Z$-distance of the code. 
Our thresholds are computed using a full circuit-level simulation and a minimum-weight perfect matching (MWPM) decoder.  These main simulation results, which inform the conclusions of \cref{tab:regimes}, neglected any crosstalk errors. While filters can suppress a wide class of crosstalk errors, there are still residual crosstalk errors that cannot be eliminated by the filters.  To investigate this, we perform additional simulations using the detailed information about the residual crosstalk errors from the hardware analysis (\cref{sec:multimode_stabilization}) and address these errors by adding extra edges in the matching graphs of the surface code decoder. These extra edges are constructed such that they can detect unique syndrome patterns created by the residual crosstalk errors.  \ch{We find that the performance of the surface code is largely unchanged in the presence of crosstalk, provided that the strength of the engineered coupling between the storage and buffer modes is less than a few MHz (this informs our choice of $g_2/2\pi = 2$MHz in \cref{tab:regimes}).
Were it not for crosstalk, however, the architecture could tolerate stronger engineered couplings and engineered dissipation, which would ease demands on the storage mode coherence. Crosstalk is thus ultimately a limiting factor for our architecture, so we also describe several future research directions that would allow us to further mitigate its effects in future designs.}

Using the thin surface code, we consider lattice surgery as a means of performing logical Clifford operations in \Cref{Sec:LatticeSurgery}.  By extending our full circuit-level simulation to model timelike errors during lattice surgery, we obtain logical error probabilities for Clifford operations.

To fault-tolerantly simulate universal quantum computation~\cite{campbell2017roads} with Toffoli gates, we introduce in \cref{sec:BottomUp} a new protocol to fault-tolerantly prepare \tof magic states encoded in the repetition code. 
Due to the fault-tolerant properties of our protocol, all gates required in our circuits can be implemented at the physical level. 
Hence we refer to such an approach as a bottom-up approach for preparing \tof magic states.
The main insight is that a \tof state can be prepared by measuring a single Clifford observable, which can be achieved using a sequence of physical \cnot and \tof gates.
To ensure fault-tolerance, this Clifford measurement has to be repeated a fixed number of times, but due to suppressed bit-flip noise the state does not significantly decohere during this measurement process.
Using the full circuit-level noise model of \cref{sec:GatesMeas} and assuming $\kappa_1/\kappa_2 = 10^{-5}$, we show that \tof magic states can be prepared with total logical $Z$ failure rates as low as $6 * 10^{-6}$, which is several orders of magnitude lower than what could be achieved using non-fault-tolerant methods to prepare \tof states. 
Furthermore, the noise on the prepared \tof state is dominated by one specific Pauli error, which is a feature we can further exploit.

In \cref{Sec:TopDown}, we show how \tof magic states probabilistically prepared using our bottom-up approach can be injected in a new magic state distillation scheme. 
This protocol distills 2 higher-fidelity \tof states from 8 lower-fidelity \tof states with high success probability. 
For generic noise, the protocol achieves quadratic error reduction. 
In the relevant case where a single Pauli error dominates, we can achieve cubic error reduction. 
The protocol is compiled down to architecture-level lattice surgery operations performed at the encoded level using repetition and surface codes. 
Hence we refer to such an approach as being top-down. Our top-down approach allows us to distill \tof magic states with low enough logical error rates for use in quantum algorithms of practical interest. 
Further, we note that given the low error rates achieved using our bottom-up approach, only one round of distillation is required in our top-down approach to prepare \tof states with the desired logical error rates. 

Finally, in \cref{sec:Overhead} we analyze the overhead required for running quantum algorithms in our architecture, \jp{based on our estimated gate error rates for \REGthree. We consider running circuits on 100 qubits with up to 1,000 Toffoli gates, which are comfortably beyond the reach of classical simulability using the best currently known simulation algorithms.
For circuits of this size, and for our estimated gate error rates, bit flip errors are sufficiently rare that it suffices to concatenate the cat code with a repetition code. We find that a device with 1,000-2,000 ATS's could execute the circuit reliably.}
This number of hardware components is compatible with next generation cryogenic dilution refrigerators, \jp{indicating that our proposal holds promise} for early implementations of fault-tolerant quantum computation. 

\jp{For known applications of quantum computing with potential commercial value, substantially larger circuits are needed. Again assuming \REGthree parameters, we find that for these larger circuits the cat code should be concatenated with a thin surface code which protects against bit flips as well as phase errors, and the overhead cost is correspondingly higher. As a representative application, we consider }
the task of estimating the ground state energy density of the Hubbard model. \jp{A quantum computer with about 100 logical qubits executing about 1 million Toffoli gates could perform this task in
a parameter regime that is very challenging for classical computers \jp{running the best currently known classical algorithms}. For this purpose we estimate that our architecture could be implemented using 18,000 ATS components and that the quantum algorithm could be executed in 32-89 minutes depending on the physical parameters of the Hubbard model.} 

Notably, for this problem the magic-state factory uses at most $9.5\%$ of the total resources and is never a bottleneck on algorithm execution time.  This low factory overhead is due to a combination of factors. Firstly, the bottom-up procedure gives initial \tof states with a cost which is not much more than a physical \tof but with orders of magnitude lower error rates. Secondly, at the required $\tof$ error rate it suffices to implement one round of the top-down protocol using a mixture of repetition codes and surface codes, which dramatically reduces the factory footprint.  In contrast, the best performing T state factories (in architectures without biased noise) rely completely on surface codes and either require multiple rounds of distillation to achieve the same error suppression~\cite{Bravyi12,Meier13,campbell2018magic} or only produce 1 T state at a time so that 8 rounds are needed to realize 2 \tof gates~\cite{BraKit05}.

\section{Hardware implementation and stabilization schemes}
\label{sec:HardwareImplementation}

\etc{In our proposal, the lowest-level protection from errors occurs directly at the hardware level and is based on the idea of autonomous quantum error correction (QEC)~\cite{Paz1998}, where rather than correcting errors at the ``software level'', one instead engineers a system whose unitary evolution and dissipation is sufficient to protect the encoded information from Markovian errors. One can think of this process as the continuous analog of the standard, discrete QEC cycle consisting of syndrome measurements and correcting unitaries. The value of autonomous QEC is that it eliminates the need for active measurements and classical feedback.}

Historically, proposals for the implementation of autonomous QEC have been formulated in the language of coherent feedback control~\cite{Ahn2002} or reservoir engineering~\cite{Sarovar2005,kerckhoff2010designing}, where the evolution is described via a stochastic master equation or a Lindblad master equation, respectively. Here we specifically adopt a bosonic autonomous QEC technique that more neatly fits into the latter category. It was first introduced by Mirrahimi \emph{et al.} in 2014~\cite{Mirrahimi2014} and demonstrated for individual qubits in recent experiments~\cite{Leghtas2015, Touzard2018, Lescanne2020}. We summarize the most relevant pieces here for convenience.

\subsection{Overview of cat codes and driven-dissipative stabilization} \label{subsec:cat_codes_overview}

The basic idea is to encode a qubit in a \etc{two-dimensional subspace $S=\text{span}\{|-\alpha\rangle, |+\alpha\rangle\}$} of a harmonic oscillator, spanned by the two quasi-orthogonal coherent states $|\pm \alpha\rangle$~\cite{Cochrane1999, Jeong2002}. The qubit states can be defined in the $X$ basis as the following two-component Schr{\" o}dinger cat states:
\begin{equation}
    |\pm\rangle = \mathcal{N_\pm}(|\alpha\rangle \pm |-\alpha\rangle).
\end{equation}
These states are eigenstates of the parity operator $\hat{P} = \exp{(i\pi \aopd \aop)}$ with eigenvalues $\pm 1$, and $\mathcal{N}_\pm = 1/\sqrt{2(1 \pm e^{-|2\alpha|^2})}$. The codewords of this code are
\begin{align}
    |0\rangle &= |+\alpha\rangle + \mathcal{O}(e^{-2|\alpha|^2})|-\alpha\rangle \\
    |1\rangle &= |-\alpha\rangle + \mathcal{O}(e^{-2|\alpha|^2})|+\alpha\rangle.
\end{align}
Note that $|0 \rangle \approx |+\alpha\rangle$ and $|1 \rangle \approx |-\alpha\rangle$ is a very good approximation for $|\alpha|^2 \gg 1$, as will typically be assumed throughout this paper. \etc{The notation $|0 \rangle $ and $|1 \rangle$ is reserved for these cat qubit computational states throughout, and to avoid ambiguity we use $\vert \hat{n}=0 \rangle$ and $\vert \hat{n}=1 \rangle$ for the vacuum and single phonon (or photon in some alternative architectures) Fock states.}

\begin{figure}[t!]
\includegraphics[width=\linewidth]{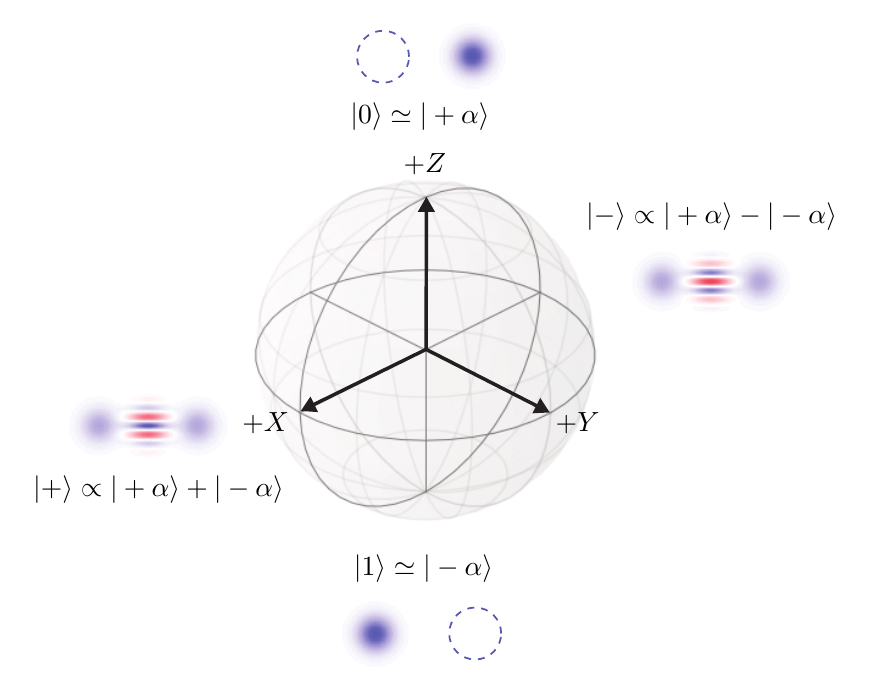}
\caption{\label{fig:bloch_sphere} Bloch sphere of the cat qubit. 
The codewords $|0\rangle$, $|1\rangle$ and the $|\pm\rangle$ states are indicated on the $Z$ and $X$ axes, respectively, along with their Wigner function representations (shown for $\alpha = 2$).}
\end{figure}

The usual error channels that affect real oscillators, such as energy relaxation and dephasing, will eventually corrupt the information encoded in this manner. To protect against these common errors, one can engineer an artificial coupling to a bath such that the oscillator only emits and absorbs excitations to and from this bath \emph{in pairs}. Such dynamics can be modeled by a Lindblad master equation of the form
\begin{equation}
\label{eq:two-phonon_ME_1}
\frac{d\hat{\rho}(t)}{dt} = \kappa_{2}\mathcal{D}[\hat{a}^{2}-\alpha^{2}]\hat{\rho}(t) + \kappa_{1}\mathcal{D}[\hat{a}]\hat{\rho}(t) + \kappa_{\phi}\mathcal{D}[\hat{a}^{\dagger}\hat{a}]\hat{\rho}(t)
\end{equation}
where $\mathcal{D}[\hat{L}]\hat{\rho} := \hat{L}\hat{\rho} \hat{L}^\dagger - \frac{1}{2}(\hat{L}^\dagger \hat{L} \hat{\rho} + \hat{\rho} \hat{L}^\dagger \hat{L})$, $\kappa_1$ is the usual single-phonon (or photon) dissipation rate, $\kappa_\phi$ is the pure dephasing rate, and $\kappa_2$ is a two-phonon (or two-photon) dissipation rate. In the case where $\kappa_1 = \kappa_\phi = 0$, any linear combination of the codewords $|0\rangle, \, |1\rangle$ is a steady state of \cref{eq:two-phonon_ME_1}. This is straightforward to see, as any state for which $\aop^2|\psi\rangle = \alpha^2|\psi\rangle$ is stationary under this master equation, and this includes both the even- and odd-parity cats. \etc{Furthermore, outside this subspace, there are no further steady states of the Lindblad master equation. Therefore, any initial state will eventually evolve to a mixture of states within this subspace.} We refer to the rate at which this decay happens as the confinement rate, $\kappa_\text{conf}$; using the displaced Fock basis (see \cref{appendix:Shifted Fock Basis}) one can show that $\kappa_\text{conf} = 4|\alpha|^2 \kappa_2$. For finite $\kappa_1, \kappa_\phi$, this description of the dynamics no longer holds true exactly. In particular, the stationary solutions of \cref{eq:two-phonon_ME_1} are no longer pure states. However, if $\kappa_\text{conf} > \kappa_\text{err}$, where $\kappa_\text{err}$ is the effective error rate,  then the codewords are still metastable states. The threshold $\kappa_\text{err}$ depends on the error channel in question: $\kappa_\text{err} = \kappa_1$ for phonon (or photon) loss, and $\kappa_\text{err} = \kappa_\phi$ for dephasing~\cite{Lieu2020}. 

The key feature of this code is that, above the threshold $\kappa_\text{conf} > \kappa_\text{err}$, the bit-flip rate (or rate of $X$-type errors) $\Gamma_{0 \leftrightarrow 1}$ decays exponentially with the ``code distance'' $|\alpha|^2$ as
\begin{equation}
    \Gamma_{0 \leftrightarrow 1} \sim |\alpha|^2 e^{-c|\alpha|^2} \kappa_\text{err},    
\end{equation}
where $2 \leq c \leq 4$ for phonon (or photon) loss~\cite{GM2020} and $c = 2$ for dephasing~\cite{Mirrahimi2014}. On the other hand, the phase-flip rate (or rate of $Z$-type errors) $\Gamma_{+ \leftrightarrow -}$ increases linearly as
\begin{equation}
    \Gamma_{+ \leftrightarrow -} \sim |\alpha|^2\kappa_\text{err}.
\end{equation}
For sufficiently small values of the dimensionless loss parameter $\kappa_\text{err}/\kappa_2$, and sufficiently large $|\alpha|^2$, this translates to a large noise bias, i.e. a large discrepancy between the $X$ and $Z$ error rates. As alluded to earlier, this bias is a key feature of our proposal and will be exploited when designing the outer error-correcting codes.

The driven-dissipative dynamics of \cref{eq:two-phonon_ME_1} can be physically realized by using a cleverly designed nonlinear element to couple the storage mode $\aop$ to an engineered environment, or reservoir. 
Following Refs.~\cite{Leghtas2015, Lescanne2020}, the idea is to generate a nonlinear interaction of the form $g_2^* \hat{a}^{2}\bopd + \text{h.c.}$ between the storage mode and an ancillary mode $\bop$, which here we refer to as the ``buffer mode'' in keeping with existing terminology. 
The buffer mode is in turn strongly coupled to a bath --- it is designed to have a large energy relaxation rate $\kappab$ so that it rapidly and irreversibly emits the photons it contains into the environment. 
If $\kappab \gg g_2$, the $\bop$ mode is in the vacuum state $|\bopd \bop = 0\rangle$ most of the time, and its excited states can be adiabatically eliminated from the Hamiltonian~\cite{Reiter2012, Leghtas2015}. In this picture, there exists an effective Markovian description of the $\aop$ mode dynamics where the $\bop$ mode \etc{is considered as part of the environment} and where the emission of excitations via $g_2^* \hat{a}^{2} \bopd$ can be accurately modeled as a dissipative process acting on the $\aop$ mode alone. To stimulate the absorption process $g_2 \hat{a}^{\dagger 2} \bop$ a linear drive $\epd^* \bop e^{-i\omegad t} + \hc$ on the buffer mode is added to supply the required energy. With this drive tuned on resonance with the buffer ($\omega_d = \omega_b$), the evolution of the combined system is described by
\begin{multline}
\label{eq:two-phonon_ME_2}
    \frac{d\hat{\rho}(t)}{dt} = -i[g_2^*(\aop^2 - \alpha^2)\bopd + \hc, \hat{\rho}(t)] \\ + \kappa_b \mathcal{D}[\bop]\hat{\rho}(t) + \kappa_1\mathcal{D}[\aop]\hat{\rho}(t),
\end{multline}
where $\alpha^2 := -\epd/g_2^*$. After adiabatically eliminating the $\bop$ mode, this master equation becomes \cref{eq:two-phonon_ME_1}, with $\kappa_2 = 4|g_2|^2/\kappa_b$.

\subsection{Physical implementation of buffer and storage resonators} \label{subsec:buffer_and_storage_resonators}

To realize the dynamics described by \cref{eq:two-phonon_ME_2} in practice, previous demonstrations of two-phonon dissipation have relied on Josephson junctions~\cite{Leghtas2015, Touzard2018} or an ``asymmetrically-threaded SQUID'' (ATS)~\cite{Lescanne2020} as the source of nonlinearity. Other variations of the nonlinear elements exist, for instance the ``SNAIL''~\cite{Frattini2017, Grimm2020}, but in this proposal we adopt the ATS due to the advantages it has over other nonlinear elements. These advantages are outlined in Ref.~\cite{Lescanne2020}. 

The potential energy of an ATS has the form $\sin(\hat{\phi})$, where $\hat{\phi} = \varphi_a \aop + \varphi_b \bop + \hc$ is the superconducting phase difference across the ATS and $\varphi_a, \, \varphi_b$ are vacuum fluctuation amplitudes that quantify the contribution of the $\aop$ and $\bop$ modes to the phase $\hat{\phi}$. It is important to emphasize that here $\aop, \, \bop$ are the \emph{normal modes} of the combined storage and buffer resonators. Because these resonators are far-detuned, there is little mixing between them, so $\aop$ is ``storage-like'' and $\bop$ is ``buffer-like''. 

\etc{Terms of cubic and higher orders in the power-series expansion of $\sin(\hat{\phi})$ generate nonlinear couplings between the modes, provided that the required energy is injected with pumps tuned to the appropriate frequencies. The desired interaction $g_2^*\aop^2 \bopd + \hc$ can be resonantly activated by modulating the magnetic flux that threads the ATS at frequency $\omega_p = 2\omega_a - \omega_b$. This modulation, which from now on we refer to as the ``pump'', provides the missing energy in the conversion process --- two storage phonons get converted to a buffer photon and pump photon. To stimulate the reverse process --- the conversion of a buffer and a pump photon to two storage phonons --- a linear drive $\epd^* \bop e^{-i\omegad t} + \hc$ at frequency $\omega_d$ is applied to the buffer. From now on we refer to this simply as the ``drive''. For further details on the implementation and the calculation of $g_2$, see \cref{sec:single_mode_stabilization} and Ref.~\cite{Lescanne2020}.    }

For the storage oscillator, the three cited experiments have used either superconducting 3D microwave cavities~\cite{Leghtas2015, Touzard2018} or on-chip coplanar-waveguide (CPW) resonators~\cite{Lescanne2020}, and recent theoretical proposals have focused on similar implementations~\cite{Guillaud2019, GM2020}. Here we study the possibility of using nanomechanical resonators instead, and tailor our calculations specifically to the case of one-dimensional phononic-crystal-defect resonators (PCDRs) made of lithium niobate, a crystalline piezoelectric material. These devices support resonances at gigahertz frequencies, with modes that are localized inside a volume $<1 \, \mu\text{m}^3$ of a suspended nanostructure. They have been coupled to transmon qubits in recent experiments~\cite{Arrangoiz-Arriola2019, Mirhosseini2020} and may offer a number of advantages over electromagnetic resonators. 

First, a PCDR is a micron-scale nanostructured device, with an on-chip footprint (area) that is at least three orders of magnitude smaller than that of planar superconducting resonators, including lumped-element structures. This is not a significant advantage today, with the largest quantum computers only having a few dozen physical qubits, but it may become important in the future. 

A second consideration is that, unlike electromagnetic resonators, appropriately designed acoustic devices do not experience direct crosstalk (unwanted couplings) because acoustic waves do not propagate through vacuum. They can still couple through the circuitry that mediates interactions between them, but this can be mitigated with approaches such as filtering and a carefully chosen connectivity, both of which are important features of our proposal. 

The third and most important consideration is that there is recent experimental evidence that phononic-crystal-based devices can have very long coherence times as a result of the high degree of confinement of their modes and the quality of their materials. For example, devices fabricated from silicon and operating at a frequency of $5 \, \text{GHz}$ have been shown to have energy relaxation and pure dephasing times of $T_1 \approx 1.5 \, \text{s}$ and $T_\phi \approx 130 \, \mu\text{s}$, respectively~\cite{MacCabe2020}. These silicon devices cannot be easily coupled to superconducting circuits, but they offer insight into the decoherence mechanisms affecting nanomechanical resonators and suggest a roadmap for achieving similar levels of coherence with piezoelectric devices. For example, similar studies with lithium niobate PCDRs are already under way~\cite{Wollack2020}, and although their coherence times are currently limited to $\sim 1 \, \mu\text{s}$, it is possible their performance could approach that of the silicon devices after sufficient advances in materials and surface science. 

We remark that although we have tailored our calculations to the case of PCDRs, the results of this proposal are still applicable to a setting where the storage modes are electromagnetic. 

\subsection{Wiring and layout} \label{subsec:wiring_and_layout}

\begin{figure*}[th]
    \centering
    \includegraphics[width=\textwidth]{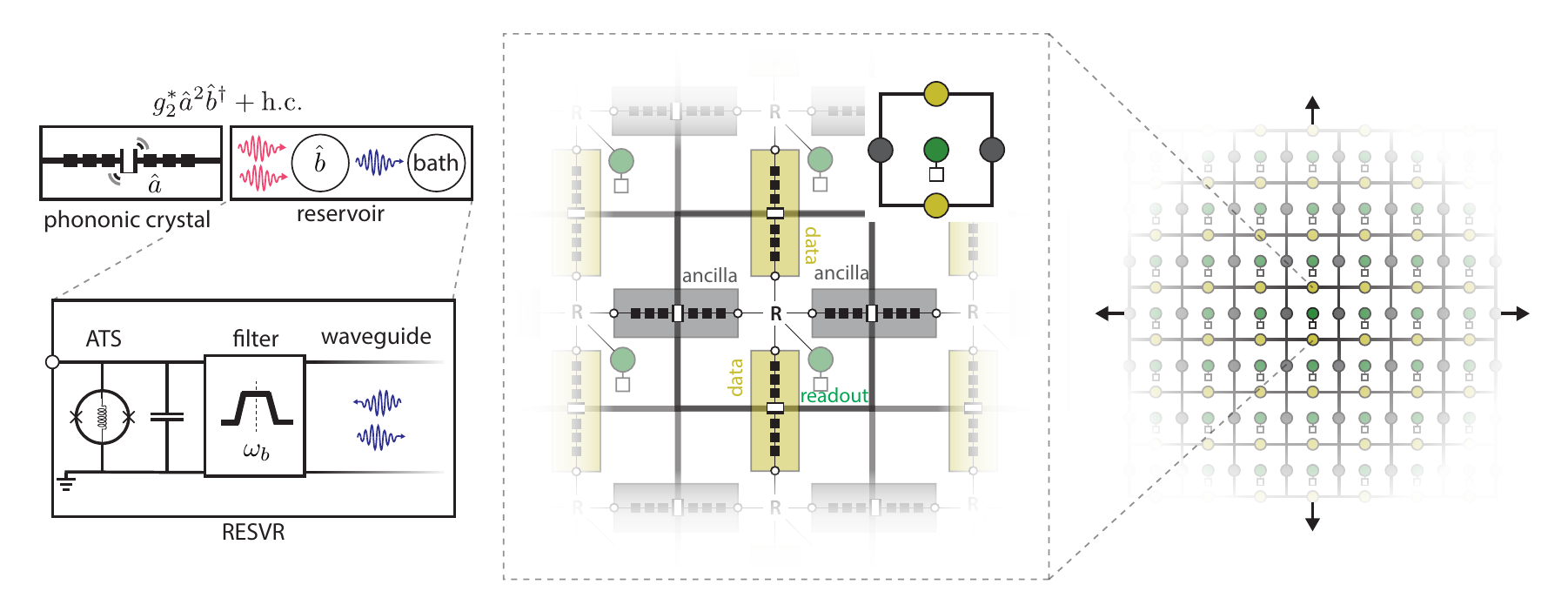}
    \caption{ Hardware implementation of the repetition- and surface-cat codes. In the 2D grid on the right, yellow circles represent data qubits where the logical information is encoded, and gray circles represent ancilla qubits which are used to measure the stabilizers and extract error syndromes. Both data and ancilla qubits are encoded as Schr{\" o}dinger cat states of localized acoustic modes of phononic-crystal-defect resonators (PCDRs), and are stabilized through a driven-dissipative two-phonon interaction with an engineered reservoir. This stabilization strongly biases the noise, suppressing $X$ errors and increasing $Z$ errors. The letter R in each plaquette represents the reservoir, which is implemented with a capacitively-shunted ATS (the ``buffer'' resonator), a bandpass filter, and an open waveguide. This circuit is shown inside the ``RESVR'' box in the left panel and has a single non-grounded terminal, marked with a white circle on the edge of the box. All resonators surrounding each reservoir in the layout diagram in the center panel connect to this one physical terminal. The green circles represent an additional acoustic mode used to measure the cat qubits in the $X$ basis with the aid of a transmon, which is represented by a white square. Altogether, five PCDRs are connected to each reservoir: four as active qubits, and one for readout.}
    \label{fig:hardware_cartoon}
\end{figure*}

We now describe a way to combine all of these building blocks to build a two-dimensional grid of cat qubits that form the basis for an outer code, such as the repetition code or the surface code. First, following Ref.~\cite{Lescanne2020} we form a buffer resonator with frequency $\omega_b$ by shunting an ATS with a capacitor.
\etc{This buffer mode is then coupled to the input of a bandpass filter that passes frequencies within a bandwidth $4J$ centered at $\omega_b$ and attenuates frequencies outside this range. The output of the filter is connected to an open waveguide (which can be accurately modeled as a resistive termination).} 
This filter configuration stands in contrast to the implementation in Ref.~\cite{Lescanne2020}, where a bandstop filter \etc{(that instead attenuates frequencies within some band and passes all others)} was used to protect the storage mode from radiatively decaying into the waveguide. In our proposal, the bandpass filter also serves this role, but it also plays a more fundamental role as a means of suppressing crosstalk mechanisms that arise as a result of our frequency-multiplexed scheme to stabilize (and perform gates between) multiple modes with a single ATS. From this point on, we refer to the combination of the buffer, filter, and waveguide as the ``reservoir''. 

We arrange reservoirs in a two-dimensional grid, as shown in \cref{fig:hardware_cartoon}, and connect neighboring reservoirs with a PCDR using each of the two terminals of the resonator. The reservoirs provide the connectivity between resonators and are located above, below, to the left, and to the right of each resonator. These four resonators serve as data and ancilla qubits in either the repetition or the surface code.  In addition, one more resonator coupled to each reservoir serves the purpose of an ancillary readout mode which is used to measure the cat qubits in the $X$ basis with the aid of a ordinary transmon. Alternatively, it is possible to omit this resonator altogether and perform the $X$ readout directly via the buffer --- see \cref{App:Alternative} for further details.  

There are two important considerations that motivate this architecture. The first is that present PCDR designs only have two available terminals, so each of them can be connected to at most two different reservoir circuits. This is simply a design choice --- it may be possible to add more terminals without a significant degradation of performance, and this would enable other variations of the 2D layout. The second consideration comes from our analysis of correlated errors in the frequency-multiplexed stabilization scheme, which we overview below and provide details of in \cref{sec:multimode_stabilization}. Our results show that the correlated error rates
increase rapidly with the number of modes connected to an ATS, and the error rates that come with choosing five modes per ATS are the largest that can be tolerated by the outer error-correcting codes. 

\subsection{Estimation of dissipation rates \texorpdfstring{$\kappa_1$}{kappa1} and \texorpdfstring{$\kappa_2$}{kappa2}} \label{subsec:calculation_of_loss}

The dissipation rates $\kappa_1$ and $\kappa_2$ are crucial parameters: they set the error rates of the gates, as well as the error rates during idling, state preparation, and measurement. Two-phonon loss is an engineered process, so the two-phonon loss rate $\kappa_2$ is a parameter that we can calculate. On the other hand, the single-phonon loss rate $\kappa_1$ is largely determined by intrinsic properties of the hardware. To construct the error model we use in this proposal, then, the starting points are to 1) calculate a prediction for the maximum achievable value of $\kappa_2$, and 2) infer the values of $\kappa_1$ that are required to reach various regimes of interest. We will consider three distinct regimes in this proposal, characterized by the magnitude of the ``dimensionless loss'' parameter $\kappa_1/\kappa_2$: $10^{-3}$ (\REGone), $10^{-4}$ (\REGtwo), and $10^{-5}$ (\REGthree). This parameter is particularly important because the $Z$-type error rates scale as $\sim \sqrt{\kappa_1/\kappa_2}$ for both the CNOT and Toffoli gates  (see \cref{sec:GatesMeas} 
and \cref{tab:Gateerrorrates} for further details). These regimes are summarized in \cref{tab:regimes}.

A summary of our $\kappa_2$ calculation is presented next, with further details contained in \cref{sec:single_mode_stabilization}. As described previously, the way the two-phonon loss is engineered is by inducing a nonlinear coupling $g_2^* \aop^2 \bopd + \text{h.c.}$ between the storage mode $\aop$ and a buffer mode $\bop$, which decays into the environment at rate $\kappa_b$. A key requirement is that $\kappa_b \gg 2 |\alpha| g_2$, so that the excited states of the buffer can be adiabatically eliminated to yield an effective description where the storage directly experiences two-phonon loss. In this adiabatic regime, $\kappa_2 \approx 4|g_2|^2 / \kappa_b$. We may write the adiabaticity constraint as $g_2 = \eta \kappa_b / 2 |\alpha|$ for some $\eta \ll 1$. 
We have observed numerically that $\eta = 1/5$ is sufficient to stabilize high-fidelity cat states. 
Putting this together, we find that the maximum achievable two-phonon dissipation rate scales linearly with the buffer decay rate $\kappa_b$ and inversely with the mean phonon number $|\alpha|^2$ (the ``distance'' of the cat code):
\begin{equation}
\label{eq:kappa_2_max}
\kappa_2 \approx \kappa_b \eta^2/|\alpha|^2.
\end{equation}
We note that the maximum achievable $\kappa_b$ is upper bounded by the filter bandwidth, $4J$ (see \cref{sec:single_mode_stabilization} for details). In this work, we fix $4J/2\pi = 100$MHz, which is sufficient to satisfy $\kappa_b \leq 4J$ for the values of $\kappa_b$ we consider.

We now move on to estimating the required values of the single-phonon loss rate $\kappa_1$. Before doing so, we first recall that $\aop$ and $\bop$ are the normal --- or hybridized --- modes of the system, and therefore $\kappa_1$ is given by
\begin{equation}
\label{eq:kappa_1_budget}
\kappa_1 \approx \kappa_{1, \text{i}} + \kappa_{1, \text{rad}} + (g/\delta)^2  \kappa_{b,\text{i}}.
\end{equation}
Here $g$ is the \emph{linear} coupling rate between the bare storage and buffer resonators, $\delta = \omega_b - \omega_a$ is their detuning, and $\kappa_{b,\text{i}}$ is the intrinsic decay rate of the bare buffer resonator. The first contribution $\kappa_{1, \text{i}}$ is the intrinsic loss rate of the \emph{bare} storage mode, an empirical quantity that depends, for example, on the quality of the resonator materials.  The second contribution $\kappa_{1, \text{rad}}$ is due to direct radiative decay into the buffer bath, which we make negligibly small by ensuring the storage frequency $\omega_a$ lies far outside of the filter passband, or in other words by ensuring the bath has a vanishing density of states at $\omega_a$. The third contribution $(g/\delta)^2 \kappa_{b,\text{i}}$ is due to the intrinsic loss of the bare buffer resonator, which the storage inherits due to their hybridization and which the filter cannot protect against. Usually $|g/\delta| \sim 10^{-2}$, so this last contribution is important when $\kappa_{1, \text{i}} \ll \kappa_{b,\text{i}}$.

\begin{figure}[ht]
\includegraphics[width=\linewidth]{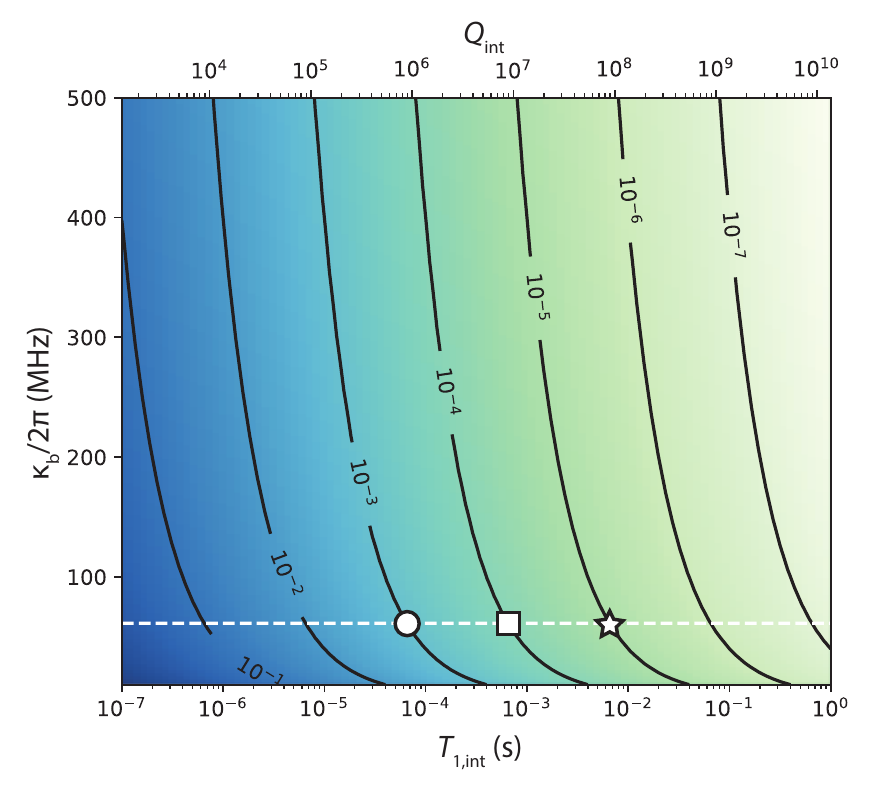}
\caption{\label{fig:loss_results_main_text} 
Dimensionless loss $\kappa_1/\kappa_2$, as given by \cref{eq:loss_main_text}, as a function of buffer decay rate $\kappa_b$ and storage intrinsic energy relaxation time $T_{1,\text{i}} = 1/\kappa_{1,\text{i}}$, assuming fixed $\omega_a/2\pi = 2.16 \, \text{GHz}$ and $|\alpha|^2 = 8$. We label the corresponding quality factor $Q_\text{i} = \omega_a/\kappa_{1,\text{i}}$ on the upper horizontal axis, and mark the $\kappa_b$ value used in our proposal with the white dashed line. The points corresponding to \REGone, \REGtwo, and \REGthree are indicated with a circle, a square, and a star, respectively. Future innovations in the multiplexed stabilization scheme may allow for larger bandwidths, which would relax these requirements proportionally.}
\end{figure}

Summing up, $\kappa_1 \approx \kappa_{1, \text{i}} + (g/\delta)^2 \kappa_{b, \text{i}}$ when the buffer bath has a vanishing density of states at $\omega_a$. A key result of our analysis is that $(g/\delta)^2 \kappa_{b, \text{i}}$ can be strongly suppressed by using a buffer resonator with a large characteristic impedance $Z_b$. This can be accomplished by increasing $|\delta|$ until $(g/\delta)^2 \kappa_{b,\text{i}}$ is suppressed to a value comparable to or smaller than $\kappa_{1,\text{i}}$. This comes at the cost of reducing the nonlinear interaction rate $g_2$, which also scales with the detuning as $g_2 \sim 1/\delta^2$. But one can offset this penalty by increasing $Z_b$, because $g_2 \sim Z_b^{5/2}$ as we show in \cref{sec:single_mode_stabilization}. We show that under certain assumptions of $\kappa_{b,\text{i}}$ and $\kappa_{1,\text{i}}$, once $Z_b \sim 1 \, \text{k}\Omega$ we can access a regime where $\kappa_1 \approx \kappa_{1, \text{i}}$ and therefore
\begin{equation}
\label{eq:loss_main_text}
    \kappa_1/\kappa_2 \approx \kappa_{1,\text{i}}|\alpha|^2/\eta^2 \kappa_b.
\end{equation}
This is a useful result, as it addresses the problems that arise when coupling a highly coherent, linear storage element to a much lossier superconducting circuit. In \cref{fig:loss_results_main_text}, we plot this simple expression for $\kappa_1/\kappa_2$ as a function of $\kappa_{1,\text{i}}$ and $\kappa_b$.  We assume $|\alpha|^2 = 8$, which is large enough to result in good performance of the outer codes.

In later sections, we analyze the performance of our architecture and find that there is an upper limit on $\kappa_b$, beyond which crosstalk begins to inhibit the performance of the architecture. Specifically, in \Cref{subsection:Surface code logical failure rates in the presence of crosstalk errors}, we show that the nonlinear coupling strength must satisfy $g_2/2\pi \leq 2$MHz, lest crosstalk degrade the logical lifetimes. As a result, we have the restriction that $\kappa_b = 2|\alpha| g_2 /\eta \leq 2\pi * 57$MHz. At this maximal value, the $\kappa_{1, \text{i}}$ values needed to reach the three different regimes we study in this paper are indicated in \cref{fig:loss_results_main_text} (see also \cref{tab:regimes}). 
Because crosstalk is thus a limiting factor for our architecture, we remark that there are several ways crosstalk could be mitigated in future designs. 
For example, in \Cref{App:Alternative}, we describe an alternative version of our architecture with 4 modes per unit cell as opposed to 5; this modification reduces crosstalk, thereby enabling larger $\kappa_2$ and easing the $\kappa_{1, \text{i}}$ requirements. 
Future approaches could reduce the number of modes per unit cell even further by increasing the number of terminals of each PCDR.

It is important to note that the value $\kappa_2 \approx 4|g_2|^2/\kappa_b\sim 2\pi * 280 \, \text{kHz}$ that we derive in this analysis, while theoretically possible, would require a larger values of $g_2$ (about 5 times larger) than those previously reported~\cite{Lescanne2020}. Because $\alpha^2 = -\epsilon_d/g_2^*$ (see \cref{subsec:cat_codes_overview}), this would require a larger drive amplitude on the buffer mode in order to maintain a fixed $\alpha$, which may cause unforeseen problems such as instabilities~\cite{Lescanne2019} or the excitation of spurious transitions~\cite{Sank2016, Zhang2019}. Furthermore, the large buffer impedance $Z_b$ required increases the size of the vacuum fluctuations of the superconducting phase $\phiop$, making the system more prone to instabilities. A detailed analysis of the power-handling capacity of our system is beyond the scope of this work. This is an area of active research, with promising advances such as the use of inductive shunts to suppress instabilities~\cite{Verney2019}.

\subsection{Multiplexed stabilization} \label{subsec:multiplexed_stabilization}

In our architecture, each reservoir is responsible for stabilizing multiple storage modes simultaneously, \ch{in contrast to prior proposals~\cite{Mirrahimi2014,Guillaud2019}.
This \emph{multiplexed stabilization} is both beneficial and necessary in the context of our architecture. 
Stabilizing multiple storage modes with a single reservoir is clearly beneficial from the perspective of hardware efficiency, as the required number of ATSs and control lines is reduced.  
Moreover, the use of PCDRs (as opposed to, e.g., electromagnetic resonators) actually necessitates multiplexed stabilization.
Current PCDR designs have only two terminals, meaning that a PCDR can couple to at most two different reservoir circuits, yet each reservoir must couple to at least four storage modes in order to achieve the required 2D-grid connectivity. 
Each reservoir must necessarily stabilize multiple storage modes as a result.}

Conveniently, we find that multiplexed stabilization can be implemented via a simple extension of the single-mode stabilization scheme demonstrated in Ref.~\cite{Lescanne2020}. The main idea  is to use \emph{frequency-division multiplexing} to stabilize different modes independently. Here, multiplexing refers to the fact that different regions of the filter passband are allocated to the stabilization of different modes. When the bandwidth allocated to each stabilization process is sufficiently large, multiple modes can be stabilized simultaneously and independently, as we now show. 

To stabilize the $n$-th mode coupled to a given reservoir, we apply a pump frequency $\omega_p^{(n)}=2\omega_a - \omega_b +\Delta_n$, and drive the buffer mode at frequency $\omega_d^{(n)}=\omega_b - \Delta_n$, where $\Delta_n$ denotes a detuning. To stabilize multiple modes simultaneously, we apply multiple such pumps and drives. Analogously to the single-mode stabilization case, the nonlinear mixing of the ATS  then gives rise to an interaction Hamiltonian of the form
\begin{equation}
    \label{eq:multimode_interaction}
    \hat H/\hbar = \sum_n g_2\left( \hat a_n^2 -\alpha^2 \right)\hat b^\dagger e^{i\Delta_n t} + \mathrm{H.c.},
\end{equation}
see \Cref{sec:multimode_stabilization} for derivation. 
Note that the sum does not run over all modes coupled to the ATS, but rather only over the modes stabilized by that ATS. 
In our architecture, though five modes couple to each ATS, only two must be stabilized simultaneously, so the sum contains only two terms. 
By adiabatically eliminating the lossy buffer mode, and assuming the detunings are chosen such that $|\Delta_n - \Delta_m|\gg 4|\alpha|^2\kappa_2$ for all $m\neq n$, one obtains an effective master equation describing the evolution of the storage modes,
\begin{equation}
\label{eq:multimode_incoherent}
\frac{d \hat{\rho}}{dt}\approx \sum_n \kappa_{2,n}\mathcal{D}\left[ \hat a_n^2 -\alpha^2 \right]\hat{\rho}(t),
\end{equation}
see \Cref{sec:multimode_stabilization} for derivation.
Here, $\kappa_{2,n}\approx 4 |g_2|^2/\kappa_b$ if the corresponding detuning falls inside the filter passband ($|\Delta_n| < 2J$), and $\kappa_{2,n}\approx 0$ otherwise, see~\Cref{sec:single_mode_stabilization}. The dynamics~(\ref{eq:multimode_incoherent}) stabilize cat states in different modes independently and simultaneously. 
Thus, by simply applying additional pumps and drives \emph{with appropriately chosen detunings}, multiple modes can be simultaneously stabilized by a single ATS.

\begin{figure}[t]
    \centering
    \includegraphics[width=\columnwidth]{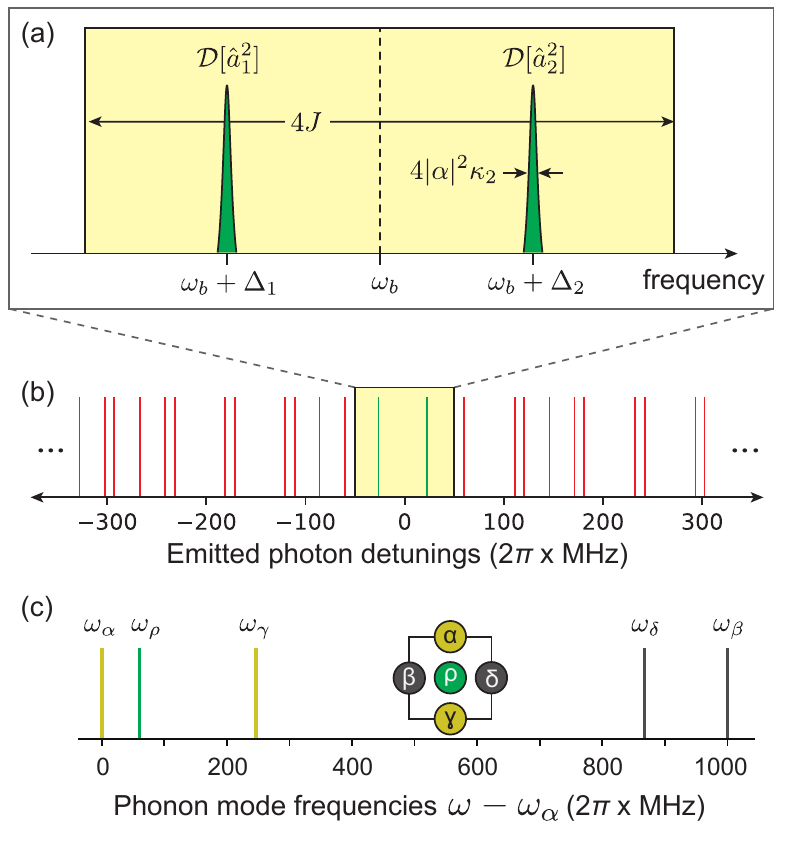}
    \caption{Multiplexed stabilization and crosstalk mitigation. (a) Frequency multiplexing. 
    Because the desired couplings $(g_2 \hat a_n^2 \hat b^\dagger e^{i\Delta_i t} +\text{H.c.})$ are detuned by different amounts, photons lost to the environment via the buffer have different frequencies. 
    When the corresponding emitted buffer mode photons (green lines) are spectrally well resolved, $|\Delta_n - \Delta_m| \gg 4|\alpha|^2\kappa_2$, the modes are stabilized independently. 
    Dissipation associated with photon emissions at frequencies inside the filter passband (the yellow box, with bandwidth $4J/2\pi = 100$MHz) is strong, while dissipation associated with emission at frequencies outside the passband is suppressed.
    (b),(c) Crosstalk suppression. 
    Red lines in (b) denote photon emission frequencies associated with various correlated errors, calculated for the specific storage mode frequencies plotted in (c). 
    The mode frequencies are deliberately chosen so that \emph{all} emissions associated with correlated errors occur at frequencies outside the filter passband (no red lines fall in the yellow box). In other words, \Cref{eq:crosstalk_constraint1,eq:crosstalk_constraint2} are simultaneously satisfied for any choices of the indices that lead to nontrivial errors in the cat qubits.
    See \Cref{sec:multimode_stabilization} for further details. 
    }
    \label{fig:multiplexing_and_crosstalk}
\end{figure}

The efficacy of this multiplexed stabilization scheme can be understood intuitively by considering the frequencies of photons that leak from the buffer mode to the filtered bath. 
In the case of $\Delta_n = 0$, a pump applied at frequency $2\omega_a-\omega_b$ facilitates the conversion of two phonons of frequency $\omega_a$ to a single photon of frequency $\omega_b$ (note that acoustic phonons are converted into buffer photons via piezoelectricity in our proposal). 
As a result, photons that leak from the buffer to the bath have frequency $\omega_b$.  If instead the pump is detuned by an amount $\Delta_n \neq 0$, it follows from energy conservation that the corresponding emitted buffer mode photons have frequency $\omega_b + \Delta_n$.  
When the differences in these emitted photon frequencies, $\Delta_n-\Delta_m$, are chosen to be much larger than the emitted photon linewidths, $4|\alpha|^2\kappa_2$ (see \Cref{appendix:Shifted Fock Basis}), emitted photons associated with different storage modes are spectrally resolvable by the environment. 
Therefore, when the stabilization of mode $n$ causes a buffer mode photon to leak to the environment, there is no back-action on modes $m\neq n$.
These ideas are illustrated pictorially in \Cref{fig:multiplexing_and_crosstalk}(a). 

\subsection{Crosstalk}

\ch{Our multiplexed stabilization scheme can induce undesired crosstalk among the cat qubits, and this crosstalk must be quantified in order to provide realistic performance estimates for our architecture.  
We now enumerate the different sources of crosstalk and show that the dominant sources can be largely suppressed through a combination of filtering and phonon-mode frequency optimization. 
Later on, in \Cref{sec:LogicalMemory}, we incorporate the residual crosstalk errors into calculations of the logical error rates for our architecture, finding that these small correlated errors can nevertheless be a limiting factor for overall performance. 
}

In acting as a nonlinear mixing element, the ATS not only mediates the desired $(g_2 \hat a_n^2 \hat b^\dagger +\mathrm{H.c.})$ interactions, but it also mediates spurious interactions between different storage modes. 
While most spurious interactions are far detuned and can be safely neglected in the rotating-wave approximation, there are others which cannot be neglected. Most concerning among these are interactions of the form
\begin{equation}
\label{eq:crosstalk_interactions}
g_2 \hat a_j \hat a_k \hat b^\dagger e^{i \delta_{ijk} t} +\mathrm{H.c.},
\end{equation} 
for $j\neq k$, where $\delta_{ijk} = \omega_p^{(i)} - \omega_j-\omega_k +\omega_b$. This interaction converts two phonons from different modes, $j$ and $k$, into a single buffer mode photon, facilitated by the pump that stabilizes mode $i$. These interactions cannot be neglected in general because they have the same coupling strength as the desired interactions~(\ref{eq:multimode_interaction}), and they can potentially be resonant or near-resonant, depending on the frequencies of the storage modes involved.

There are three different mechanisms through which the interactions~(\ref{eq:crosstalk_interactions}) can induce crosstalk among the cat qubits. These mechanisms are described in detail in~\Cref{sec:multimode_stabilization}, and we summarize them here. First, analogously to how the desired interactions ~(\ref{eq:multimode_interaction}) lead to two-phonon losses, the undesired interactions~(\ref{eq:crosstalk_interactions}) lead to correlated, single-phonon losses
\begin{equation}
\kappa_\mathrm{eff} \mathcal D[\hat a_j \hat a_k] \rightarrow \kappa_\mathrm{eff} |\alpha|^4 D[\hat Z_j \hat Z_k]
\label{eq:stochastic_crosstalk_1}
\end{equation}
where the rate $\kappa_\mathrm{eff}$ will be discussed shortly, and $\hat Z_i$ is the logical Pauli-$\hat Z$ operator for the cat qubit in mode $i$. The arrow denotes projection onto the code space, illustrating that these correlated losses manifest as \emph{stochastic}, correlated phase errors in the cat qubits.  

Second, the interplay between different interactions of the form~(\ref{eq:crosstalk_interactions}) gives rise to new effective dynamics~\cite{james2007,Gamel2010timeaveraged,Reiter2012} generated by Hamiltonians of the form
\begin{align}
\hat H_\mathrm{eff} = &\chi \hat a_i^\dagger \hat a_j^\dagger \hat a_m \hat a_n e^{i (\delta_{\ell m n} - \delta_{ijk})t} +\mathrm{H.c.}, \\
\label{eq:crosstalk_coherent}
\rightarrow &\chi |\alpha|^4 \hat Z_i \hat Z_j \hat Z_k \hat Z_l e^{i (\delta_{\ell m n} - \delta_{ijk})t} +\mathrm{H.c.},
\end{align}
where the coupling rate $\chi$ is defined in \Cref{sec:multimode_stabilization}. The projection onto the code space in the second line
reveals that $\hat H_\mathrm{eff}$ can induce undesired, \emph{coherent} evolution within the code space.

Third, $\hat H_\mathrm{eff}$ can also evolve the system out of the code space, changing the phonon-number parity of one or more modes in the process. 
Though the engineered dissipation subsequently returns the system to the code space, it does not correct changes to the phonon-number parity. The net result is that $\hat H_\mathrm{eff}$ also induces \emph{stochastic}, correlated phase errors in the cat qubits,
\begin{equation}
\gamma_\mathrm{eff} \mathcal D [ \hat Z_i \hat Z_j \hat Z_k \hat Z_\ell ],
\label{eq:stochastic_crosstalk_2}
\end{equation}
where the rate $\gamma_\mathrm{eff}$ will be discussed shortly.

Remarkably, all of the stochastic crosstalk errors, (\ref{eq:stochastic_crosstalk_1}) and (\ref{eq:stochastic_crosstalk_2}), can be suppressed to negligible levels through a combination of filtering and phonon-mode frequency optimization. In \Cref{sec:multimode_stabilization}, we show that both $\kappa_\mathrm{eff}\approx 0$ and $\gamma_\mathrm{eff}\approx 0$, provided
\begin{align}
\label{eq:crosstalk_constraint1}
|\delta_{ijk}|&>2J, \\
\label{eq:crosstalk_constraint2}
|\delta_{ijk}-\delta_{\ell mn}|&>2J,
\end{align}
respectively. This suppression can be understood as follows. The decoherence associated with $\kappa_\mathrm{eff} $ and $\gamma_\mathrm{eff}$ results from the emission of buffer mode photons at frequencies $\omega_b +\delta_{ijk}$ and $\omega_b\pm(\delta_{ijk}-\delta_{\ell mn})$, respectively. When the frequencies of these emitted photons lie outside the filter passband, their emission (and the associated decoherence) is suppressed. Crucially, we can arrange for all such errors to be suppressed \emph{simultaneously} by carefully choosing the frequencies of the storage modes, as shown in \Cref{fig:multiplexing_and_crosstalk}(b,c).
We note that the configuration of mode frequencies in \Cref{fig:multiplexing_and_crosstalk}(c) was found via a numerical optimization procedure described in \Cref{sec:multimode_stabilization} and is robust to realistic frequency fluctuations. 

\ch{The coherent crosstalk errors (\ref{eq:crosstalk_coherent}) can also be suppressed through phonon-mode frequency optimization, though the suppression is not sufficient to render them negligible. To suppress these errors, the phonon-mode frequencies have been chosen to maximize the detunings $\delta_{ijk}-\delta_{\ell m n}$, such that $\hat H_\mathrm{eff}$ is rapidly rotating and its damaging effects are mitigated to a large extent (see \Cref{sec:multimode_stabilization} for details). Even so, the residual crosstalk errors are not negligible and they must be accounted for when estimating the overall performance of the architecture. To this end, in \Cref{sec:multimode_stabilization} we precisely quantify the magnitude of these residual crosstalk errors,  and the impact of these errors on logical failure rates is calculated in \Cref{sec:LogicalMemory}. As described in that section, 
we must have $g_2 \lesssim 2\pi* 2$ MHz, lest these coherent errors degrade logical lifetimes. 
At the hardware level, this restriction limits the achievable $\kappa_2$ (see \cref{subsec:calculation_of_loss}), meaning that longer storage mode coherence times are required to reach a given $\kappa_1/\kappa_2$ because of these coherent crosstalk errors. 
}

\ch{Crosstalk also imposes another limitation on our architecture: though increasing the number of modes per unit cell would improve hardware efficiency and connectivity, crosstalk forces us to minimize the number of modes per unit cell. 
Indeed, as more modes are added to a unit cell, frequencies become increasingly crowded~\cite{naik2017,pechal2018,hann2019}, and magnitude of crosstalk errors increases. 
Accordingly, we have chosen four modes (plus one additional mode for readout) per unit cell because this is the minimum number consistent with our 2D square grid layout.} In \cref{App:Alternative}, we describe an alternative architecture that only uses four modes per unit cell, but requires a different approach to $X$ measurements that may be more challenging to implement.

\ch{Broadly speaking, these limitations illustrate the importance of accounting for crosstalk when designing and analyzing fault-tolerant quantum computing architectures. 
More specifically, these limitations reveal that finding further ways to mitigate crosstalk is an important direction for future research on dissipative cat qubits. 
In future designs, resonators with additional terminals, or tunable couplers~\cite{mundada2019,chen2014}, could be employed to further mitigate the effects of crosstalk, for example. Additionally, in \Cref{App:Alternative}, we describe an alternate version of our architecture that employs a different $X$-basis readout scheme in order to reduce the number of modes per unit cell and hence reduce crosstalk.  }

\section{Gates and Measurements}
\label{sec:GatesMeas}

In this section, we discuss the gates and measurements of the cat qubits. We first discuss the implementation of the $X$ gate via a rotating two-phonon dissipation; this will be helpful for understanding the $\cnot$ and Toffoli gates. We then review the fundamentals of the bias-preserving $\cnot$ and Toffoli gates acting on cat qubits \cite{Guillaud2019} and present several new analytical and numerical results. In particular, we explicitly characterize the extra geometric phase ($Z$ or CZ rotations) which must be taken into account in the implementation of the $\cnot$ and Toffoli gates if the average excitation number $|\alpha|^{2}$ is not an even integer. Moreover, we introduce the shifted Fock basis method and demonstrate that it is useful for the perturbative analysis of the $Z$ error rates of various cat-qubit gates. We then illustrate that the shifted Fock basis method also allows more efficient numerical simulation of large cat qubits (up to $|\alpha|^{2}=10$) than the usual Fock basis method. The numerical results on gate error rates are summarized in \cref{tab:Gateerrorrates} and detailed descriptions of the methods are given in \cref{appendix:Shifted Fock Basis,appendix:Perturbative analysis of cat qubit gates,app:Gate Error Simulations}. \kn{These results are fed into the simulations of the concatenated cat codes in \cref{sec:LogicalMemory,sec:BottomUp}.  }

\ch{We also describe schemes for $X$- and $Z$-basis readout. Our scheme for $X$-basis readout has only a small impact on the length of an error correction cycle, thanks to the use of an additional readout mode that is interrogated by a transmon in parallel with the next error correction cycle. We also present a fast $Z$-basis readout scheme which uses a coupling between the storage mode and buffer mediated by the ATS.} \jp{  This  achieves measurement error rates which improve exponentially as $|\alpha|^2$ increases.} \hp{Having hardware native $X$- and $Z$-basis readout schemes allows for higher fidelity surface code stabilizer measurements as explained in \cref{subsec:SurfaceCodeMem}.  A more detailed analysis of the readout schemes can be found in \cref{appendix:Measurement}.} \ch{ Additionally, in \cref{App:Alternative} we present an alternative $X$-basis readout scheme where the readout is performed directly using the ATS, obviating the need for the extra readout mode and transmon.   }

\subsection{X Gate}
The $X$ gate interchanges the cat-code computational basis states $|0\rangle$ and $|1\rangle$. For large values of $\alpha$ these cat-code states are approximately equal to the coherent states $| \alpha \rangle$ and $|-\alpha \rangle$, so the $X$ gate acts by rotating the coherent states by $\pi$ in \etc{the phase space representation}. \ji{The value of $\alpha$ for the stabilized cat state is given by $\alpha^2=-\epsilon_d/g_2^*$ (c.f. \cref{subsec:cat_codes_overview}), so that the phase of $\alpha$ is determined by the phase of the drive. Therefore, modulating the phase of the drive on the storage cavity such that the stabilized value of $\alpha$ rotates by $\pi$ over a time $T$ realizes an $X$ gate.} The code state evolves according to
\begin{align}
    \frac{d\hat{\rho}(t)}{dt} &= \kappa_2 \mathcal{D}[ \hat{a}^2 - \alpha^{2} e^{2i\frac{\pi}{T}t}]\hat{\rho}(t) .
\end{align}
This gives an adiabatic implementation of the $X$ gate. Furthermore, we can apply a compensating Hamiltonian given by 
\begin{align}
    \hat{H}_{X} &= -\frac{\pi}{T} \hat{a}^{\dagger}\hat{a} , \label{eq:X gate compensating Hamiltonian main text}
\end{align}
so that the code state rotates along with the fixed point of the dissipator. With this compensating Hamiltonian, the gate need not be adiabatic and will succeed for any $T$. When the $X$ gate is corrupted by phonon loss, gain, or by dephasing, the logical error rates during the $X$ gate are identical to the noise during idle. This is because in the rotating frame of the compensating Hamiltonian $\hat{H}_{X}$, the noise and the dissipator are identical to the case of idle. The error rates for idle are summarized in \cref{tab:Gateerrorrates}. 

\subsection{CNOT}
We can realize the bias-preserving CNOT gate from \cite{Guillaud2019} using an ATS coupled to a pair of acoustic modes. The CNOT gate rotates the cat-code states of the target mode just as for the $X$ gate, except that now the rotation is conditioned on the state of the control mode. Cavity mode 1 will be the control and cavity mode 2 the target. A time dependent dissipator that realizes this rotation is given by the Lindblad jump operator
\begin{equation}
    \hat{L}_2(t) = \hat{a}_2^2 - \alpha^2 + \frac{\alpha}{2}( e^{2 i \frac{\pi}{T}t} - 1) (\hat{a}_1 - \alpha) .
\end{equation}
When cavity mode 1 is in the $|1\rangle$ cat-code state, which is approximately equal to the $|-\alpha \rangle$ coherent state, the corresponding dissipator reduces approximately to the rotating dissipator for the $X$ gate on the second cavity mode. On the other hand when cavity mode 1 is in the $|0\rangle$ cat state, the operator $L_2$ reduces to the usual time-independent Lindblad operator. The control cavity mode is always  stabilized by the usual time-independent Lindblad operator:
\begin{equation}
    \hat{L}_{1} = \hat{a}_{1}^{2} - \alpha^{2} .
\end{equation}
When a cat-code state $\hat{\rho}(t)$ evolves according to 
\begin{equation}
    \frac{d \hat{\rho}(t)}{dt} = \kappa_2 \mathcal{D}[\hat{L}_1](\hat{\rho}) + \kappa_2 \mathcal{D}[\hat{L}_2(t)]\hat{\rho}(t) , 
\end{equation}
the encoded state undergoes a CNOT gate (up to an extra $Z$ rotation on the control qubit; see below), assuming the gate time $T$ is long compared to the stabilization rate $\kappa_2 |\alpha|^2$. This gate preserves the bias in the noise because the two cat-code states remain distantly separated during the conditional rotation. 
 
Just as for the $X$ gate the CNOT gate can be performed much faster with the help of a compensating Hamiltonian. \kn{In this case, an ideal compensating Hamiltonian would be $-(\pi/T)|-\alpha\rangle\langle-\alpha|_{1} \hat{a}_{2}^{\dagger}\hat{a}_{2}$. This Hamiltonian rotates the state of mode 2 conditioned on the state of mode 1, so that the two-mode system remains in the subspace stabilized by the static dissipator $\mathcal{D}[\hat{L}_1]$ and the rotating dissipator $\mathcal{D}[\hat{L}_2(t)]$. However, such a compensating Hamiltonian is highly nonlinear and would be hard to implement in practice. Hence, as in Ref. \cite{Guillaud2019}, we consider an approximate version of the above Hamiltonian which only requires at most third-order nonlinearities: In this case the compensating Hamiltonian has the form:}
\begin{equation}
\label{eq:CNOTContolH}
    \hat{H}_{\textrm{CNOT}} = \frac{\pi}{4\alpha T}(\hat{a}_{1}+\hat{a}_{1}^{\dagger}-2\alpha) ( \hat{a}_{2}^{\dagger}\hat{a}_{2}-\alpha^{2}) . 
\end{equation}
\kn{This Hamiltonian rotates the state of mode 2 conditioned on the state of mode 1, so that the two-mode system remains in the subspace stabilized by the dissipator $\mathcal{D}[\hat{L}_1]$ and the rotating dissipator $\mathcal{D}[\hat{L}_2(t)]$.}

The dissipators $\mathcal{D}[\hat{L}_1]$ and $\mathcal{D}[\hat{L}_2(t)]$ combined with the \kn{compromised version of the} compensating Hamiltonian $\hat{H}_{\textrm{CNOT}}$ in \cref{eq:CNOTContolH} implement a gate
\begin{align}
    CX' &\equiv \hat{Z}_{1}(-\pi\alpha^{2}) \cdot \textrm{CNOT}_{1\rightarrow 2}, 
\end{align}
in the $T \gg 1/(\kappa_{2}\alpha^{2})$ limit, which differs from the desired CNOT gate $\textrm{CNOT}_{1\rightarrow 2}$ by an extra $Z$ rotation on the control qubit $\hat{Z}_{1}(-\pi\alpha^{2})$ (see \cref{appendix:Perturbative analysis of cat qubit gates} for more details). Here, $\hat{Z}(\theta)$ is defined as $\hat{Z}(\theta)\equiv \exp[ i\theta |1\rangle\langle 1| ] $ and $|1\rangle$ is a computational basis state, the $-1$ eigenstate of the Pauli $Z$ operator. The extra $Z$ rotation is trivial if the average excitation number $|\alpha|^{2}$ is an even integer. We also remark that \kn{the extra $Z$ rotation is not present if an ideal compensating Hamiltonian $-(\pi/T)|-\alpha\rangle\langle-\alpha|_{1} \hat{a}_{2}^{\dagger}\hat{a}_{2}$ is used.}  

Since the compensating Hamiltonian in \cref{eq:CNOTContolH} is only an approximation of an ideal compensating Hamiltonian\kn{, e.g.,} (i.e., $-(\pi/T)|-\alpha\rangle\langle-\alpha|_{1} \hat{a}_{2}^{\dagger}\hat{a}_{2}$), it introduces a residual non-adiabatic error that scales like $1/T$, where $T$ is the gate time. Phonon loss, gain, and dephasing noise during the CNOT gate give rise to a $Z$ error rate on both cavities that is proportional to $T$. The balance between the non-adiabatic errors and the noise gives rise to an optimal gate time that maximizes the fidelity.  

In Ref.~\cite{Guillaud2019}, it was noticed that the residual non-adiabatic error scales as $c/(\kappa_{2}\alpha^{2}T)$ and found that the constant coefficient is given by $c \simeq 1/(2\pi)$ via a numerical fit. In \cref{appendix:Perturbative analysis of cat qubit gates}, we provide a first-principle perturbative analysis of the $Z$ error rates of the CNOT gate by using the shifted Fock basis as a main tool. The key idea of the shifted Fock basis is to use the displaced Fock states $\hat{D}(\pm\alpha)|\hat{n}=n\rangle$ as the (unorthonormalized) basis states, where $\hat n = \hat a^\dagger \hat a$ is the mode occupation number. In particular, for the perturbative analysis of the $Z$ error rates, it suffices to consider only the ground state manifold consisting of the coherent states $\hat{D}(\pm\alpha)|\hat{n}=0\rangle = |\pm\alpha\rangle$ and the first excited state manifold consisting of the displaced single-phonon Fock states $\hat{D}(\pm\alpha)|\hat{n}=1\rangle$. See \cref{appendix:Shifted Fock Basis} for a detailed description of the shifted Fock basis, including orthonormalization and matrix elements of the annihilation operator $\hat{a}$ in the shifted Fock basis. By taking the ground and the first excited state manifolds in the shifted Fock basis and using perturbation theory, we find that the $Z$ error rates (per gate) of the implemented $CX'$ gate are given by 
\begin{align}
    &\bar{p}_{Z_{1}} =   \kappa_{1}\alpha^{2}T + \frac{\pi^{2}}{64\kappa_{2}\alpha^{2}T}, 
    \nonumber\\
    &\bar{p}_{Z_{2}} = \bar{p}_{Z_{1}Z_{2}} = \frac{1}{2}\kappa_{1}\alpha^{2}T .   
\end{align}
Here, $\kappa_{1}$ is the single-phonon loss rate (per time) and we assumed no dephasing and gain for the moment. We use $\bar{p}$ for error rates predicted by the perturbation theory and $p$ for numerical results. Note that the coefficient $\pi^{2}/64 = 0.154$ in the non-adiabatic error term is close to the coefficient $1/(2\pi) = 0.159$ which was found earlier via a numerical fit \cite{Guillaud2019}. Hence, the optimal gate time that minimizes the total gate infidelity is given by
\begin{align}
    \bar{T}_{CX'}^{\star} &= \frac{ \pi }{8\alpha^{2}\sqrt{2\kappa_{1}\kappa_{2}}} , 
\end{align}
and at the optimal gate time, the $Z$ error rates are given by 
\begin{align}
    \bar{p}_{Z_{1}}^{\star} = 6\bar{p}_{Z_{2}}^{\star} = 6\bar{p}_{Z_{1}Z_{2}}^{\star} = \frac{3\pi}{8}\sqrt{\frac{\kappa_{1}}{2\kappa_{2}}} = 0.833\sqrt{\frac{\kappa_{1}}{\kappa_{2}}} .  \label{eq:Z error rates CX' main text}
\end{align}
These agree well with the numerical results (see \cref{tab:Gateerrorrates})
\begin{align}
    p_{Z_{1}}^{\star} = 6.067p_{Z_{2}}^{\star} = 6.067p_{Z_{1}Z_{2}}^{\star} = 0.91\sqrt{\frac{\kappa_{1}}{\kappa_{2}}} , 
\end{align}
within a relative error of $10\%$ (see \cref{appendix:Perturbative analysis of cat qubit gates} for the reasons for the discrepancy). Note that the perturbation theory predicts that the optimal $Z$ error rates of the $CX'$ gate (or the CNOT gate for even $|\alpha|^{2}$) are independent of the size of the cat code $|\alpha|^{2}$.  

We simulated the CNOT gate using the effective dissipators and Hamiltonian acting on two cavities. Our method was to use the shifted Fock basis as described in \cref{appendix:Shifted Fock Basis} to find the optimal gate time and perform tomography at the optimal gate. This allowed us to compute all of the two-qubit Pauli error rates. The shifted Fock basis approach allowed us to compute the $Z$ error rates with a small Hilbert space dimension that does not depend on $\alpha$. In the standard Fock basis the required Hilbert space dimension increases rapidly with $\alpha$. In contrast to the $Z$ error rates, to accurately resolve the full set of Pauli error rates a large dimension that increases with $\alpha$ is required even for the shifted Fock basis. However, even \jp{for the full set of Pauli error rates, our simulations are several times faster when we use the shifted Fock basis rather than the standard Fock space, because good accuracy can be attained using a smaller Hilbert space dimension.} 
 
Our code was written in Python using the QuTIP package to solve the master equation including the disspators and Hamiltonian terms. We ran the simulations using AWS EC2 C5.18xlarge instances with 72 virtual CPUs, and the total time required for the CNOT simulations was about 150 hours.

\subsection{Toffoli} \label{ToffoliGate}
The bias-preserving Toffoli or CCX gate is directly analogous to the CNOT gate. The two control \kn{modes} are stabilized by the usual jump operator $\hat{L}_{1} = \hat{a}_{1}^{2}-\alpha^{2}$ and $\hat{L}_{2} = \hat{a}_{2}^{2}-\alpha^{2}$, while the third \kn{mode} is stabilized by a jump operator that couples the three \kn{modes} and rotates the third conditioned on the state of the two controls,
\begin{align}
    \hat{L}_{3}(t) &= \hat{a}_{3}^{2}-\alpha^{2} - \frac{1}{4}(e^{2i\frac{\pi}{T}t} -1 )(\hat{a}_{1}-\alpha)(\hat{a}_{2}-\alpha) . 
\end{align}
When both \kn{modes} 1 and 2 are in the $|1\rangle \simeq |-\alpha\rangle$ cat-code state, this jump operator reduces to approximately $\hat{a}_{3}^2 - \alpha^2 e^{2i\frac{\pi}{T}t}$, which is the rotating jump operator that realizes the $X$ gate on the third \kn{mode}. When one of the control \kn{modes} is in the $|0\rangle \simeq |\alpha\rangle$ cat-code state, the jump operator is approximately equal to the usual $\hat{a}_{3}^2 - \alpha^2$ jump operator that stabilizes the cat-code states. In this way the jump operators $\hat{L}_{1}$, $\hat{L}_{2}$, and $\hat{L}_{3}(t)$ implement the Toffoli gate (up to a controlled-$Z$ rotation on the two control qubits). Also like the CNOT gate we can apply a Hamiltonian to drive the desired evolution and perform the gate much faster while canceling part of the non-adiabatic errors. For the Toffoli gate this Hamiltonian is given by 
\begin{align}
    \hat{H}_{\textrm{TOF}} &= - \frac{\pi}{8\alpha^{2} T}( (\hat{a}_{1}-\alpha)(\hat{a}_{2}^{\dagger}-\alpha) + \textrm{h.c.} ) (\hat{a}_{3}^{\dagger}\hat{a}_{3}-\alpha^{2}) . \label{eq:Toffoli compensating Hamiltonian main text}
\end{align}
This Hamiltonian is the natural extension of \cref{eq:CNOTContolH}. It does not cancel all non-adiabatic noise, and like the CNOT in the presence of noise, the trade-off between non-adiabatic errors and noise from loss or dephasing gives rise to an optimal gate time for each value of $\alpha$ and the noise parameters.

Similarly as in the case of the CNOT gate, we emphasize that the dissipators $\mathcal{D}[\hat{L}_{1}]$, $\mathcal{D}[\hat{L}_{2}]$, and $\mathcal{D}[\hat{L}_{3}(t)]$ combined with the compensating Hamiltonian $\hat{H}_{\textrm{TOF}}$ in \cref{eq:Toffoli compensating Hamiltonian main text} realize a gate 
\begin{align}
    CCX' \equiv CZ_{1,2}(-\pi\alpha^{2}) \cdot \textrm{TOF}_{1,2\rightarrow 3},  
\end{align}
which differs from the desired Toffoli gate $\textrm{TOF}_{1,2\rightarrow 3}$ by a CZ rotation on the two control qubits (see \cref{appendix:Perturbative analysis of cat qubit gates} for more details). Here, $CZ(\theta)$ is defined as $CZ(\theta) \equiv \exp[ i\theta |11\rangle\langle 11| ]$ and $|11\rangle$ is the \CC{simultaneous} $-1$ eigenstate of the Pauli $Z$ operators $\hat{Z}_{1}$ and $\hat{Z}_{2}$. \kn{The extra CZ rotation is not present if an ideal compensating Hamiltonian $-(\pi/T)|-\alpha,-\alpha\rangle\langle-\alpha,-\alpha|_{1,2} \hat{a}_{3}^{\dagger}\hat{a}_{3}$ is used}. Note that the extra CZ rotation $CZ_{1,2}(-\pi\alpha^{2})$ is trivial if $|\alpha|^{2}$ is an even integer.

We simulated the Toffoli gate subject to phonon loss, gain, and dephasing at different rates by solving the master equation given by the Hamiltonian $\hat{H}_{\textrm{TOF}}$, the dissipator on each \etc{mode}, and the Lindblad operators for the noise. These simulations were carried out using AWS EC2 c5.18xlarge instances and took about 170 hours running on instances with 72 virtual CPUs. Because we simulated three \etc{modes} for the Toffoli gate, we were able to resolve only the dominant $Z$-type error rates and not the other Pauli error rates that are exponentially small in $\alpha^2$. These simulations used the shifted Fock basis approach. With this method we are able to use a Hilbert space dimension of $8$ for each of the three \etc{modes} and simulate all of the $Z$ Pauli error rates with high precision. The numerical results for the optimal gate time and the 7 $Z$-type Pauli error rates  \ji{under a pure loss noise model are summarized in \cref{tab:Gateerrorrates}. The results including gain and dephasing can be found in \cref{Tab:ToffGateTimes}.} Our simulations match our perturbation theory calculations for the $Z$ error rates. 

\kn{Similarly as in the case of the CNOT gate, we can use the ground and the first excited state manifolds in the shifted Fock basis and perform a perturbative analysis. O}ur perturbation theory yields the following $Z$ error rates of the $CCX'$ gate, or the Toffoli gate \jp{when $|\alpha|^{2}$ is an even integer} (see \cref{appendix:Perturbative analysis of cat qubit gates}): 

\begin{align}
    &\bar{p}_{Z_{1}} = \bar{p}_{Z_{2}}  = \kappa_{1}\alpha^{2}T  + \frac{\pi^{2}}{128\kappa_{2}\alpha^{2}T}, 
    \nonumber\\
    &\bar{p}_{Z_{3}} = \frac{5}{8}\kappa_{1}\alpha^{2}T,  
    \nonumber\\
    &\bar{p}_{Z_{1}Z_{2}} =   \frac{\pi^{2}}{128\kappa_{2}\alpha^{2}T}, 
    \nonumber\\
    &\bar{p}_{Z_{1}Z_{3}} = \bar{p}_{Z_{2}Z_{3}} = \bar{p}_{Z_{1}Z_{2}Z_{3}} = \frac{1}{8}\kappa_{1}\alpha^{2}T. \label{eq:Z error rates TOF pert main text} 
\end{align} 
Note that, the optimal gate time that minimizes the total gate infidelity is given by $
    \bar{T}_{CCX'}^{\star} = ( \pi  / ( 8\alpha^{2} \sqrt{2\kappa_{1}\kappa_{2}} ) )$ which is identical to the optimal gate time of the $CX'$ gate (or the CNOT gate for even $|\alpha|^{2}$) predicted by the perturbation theory. At the optimal gate time, the $Z$ error rates (per gate) are given by 
\begin{align}
    &\bar{p}_{Z_{1}}^{\star} = \bar{p}_{Z_{2}}^{\star} = 3.2\bar{p}_{Z_{3}}^{\star} = 2\bar{p}_{Z_{1}Z_{2}}^{\star} 
    \nonumber\\
    &= 16\bar{p}_{Z_{1}Z_{3}}^{\star} =  16\bar{p}_{Z_{2}Z_{3}}^{\star} = 16\bar{p}_{Z_{1}Z_{2}Z_{3}}^{\star}  
    \nonumber\\
    &= \frac{\pi}{4} \sqrt{\frac{\kappa_{1}}{2\kappa_{2}}} = 0.555\sqrt{\frac{\kappa_{1}}{\kappa_{2}}}  .  \label{eq:Z error rates CCX' main text}
\end{align}
which agree well with the numerical results (see \cref{tab:Gateerrorrates}) 
\begin{align}
    &p_{Z_{1}}^{\star} = p_{Z_{2}}^{\star} = 3.05p_{Z_{3}}^{\star} = 1.81p_{Z_{1}Z_{2}}^{\star} 
    \nonumber\\
    &= 14.9p_{Z_{1}Z_{3}}^{\star} =  14.9p_{Z_{2}Z_{3}}^{\star} = 14.9p_{Z_{1}Z_{2}Z_{3}}^{\star}  = 0.58\sqrt{\frac{\kappa_{1}}{\kappa_{2}}} ,
\end{align} 
up to a relative error of $5\%$. Thus, 
as in the case of the CNOT gate, the perturbation theory predicts that the optimal $Z$ error rates of the $CCX'$ gate (or the Toffoli gate for even $|\alpha|^{2}$) are independent of the size $|\alpha|^{2}$ of the cat code. 

\begin{table*}
    \begin{center}
        \begin{tabular}{c c c c c}
            \toprule
            \multirow{2}{*}{} & \REGone & \REGtwo & \REGthree & \multirow{2}{*}{Formula} \\
            & ($\kappa_1/\kappa_2 = 10^{-3}$) & ($\kappa_1/\kappa_2 = 10^{-4}$)& ($\kappa_1/\kappa_2 = 10^{-5}$) & \\
            \midrule
            CNOT\\
            \midrule
            Optimal Gate Time & $3.9* 10^{-7} \, s$& $1.2 * 10^{-6} \, s$&$3.9 * 10^{-6} \, s$ &$0.31  |\alpha|^{-2} (\kappa_{1}\kappa_{2})^{-\frac{1}{2}}$ \\
            $Z_1$ & $2.9 * 10^{-2} $ & $9.1 * 10^{-3}$ & $2.9 * 10^{-3}$ & $0.91 \sqrt{\kappa_1/\kappa_2}$ \\
            $Z_2 \approx Z_1 Z_2$ & $4.7 * 10^{-3}$ & $1.5 * 10^{-3}$ & $4.7 * 10^{-4}$ &$0.15 \sqrt{\kappa_1/\kappa_2}$\\
            $X_1 \approx  X_2 \approx X_1 X_2$ & \multirow{2}{*}{$3.3 * 10^{-9}$} & \multirow{2}{*}{$1.0 * 10^{-9}$} & \multirow{2}{*}{ $3.3 * 10^{-10}$} & \multirow{2}{*}{$0.93 \exp(-2 |\alpha|^2) \sqrt{\kappa_1/\kappa_2}$}\\
            $\approx Y_1 \approx Y_1 X_2 \approx Z_1 X_2$ \\[3pt]
            $Y_2 \approx Y_1 Y_2 \approx X_1 Y_2$ & \multirow{2}{*}{$3.2 * 10^{-11}$} &\multirow{2}{*}{$3.2 * 10^{-12} $}& \multirow{2}{*}{$3.2* 10^{-13}$}& \multirow{2}{*}{$0.28 \exp(-2|\alpha|^2) \left(\kappa_1/\kappa_2 \right)$}\\
            $\approx X_1 Z_2 \approx Y_1 Z_2 \approx Z_1 Y_2$ \\
            \midrule
            Toffoli at CNOT optimal Time &\\
            \midrule
            $Z_1 \approx Z_2$ & $1.8 * 10^{-2}$ & $5.8 * 10^{-3}$ & $1.8 * 10^{-3}$ & $0.58 \sqrt{\kappa_1/\kappa_2}$\\
            $Z_3$ & $6.0 * 10^{-3}$ & $1.9 * 10^{-3}$ & $6.0 * 10^{-4}$ & $0.19 \sqrt{\kappa_1/\kappa_2}$\\
            $Z_1 Z_2$ & $1.0 * 10^{-2}$ & $3.2 * 10^{-3}$ & $1.0 * 10^{-3}$ &$0.32 \sqrt{\kappa_1/\kappa_2}$\\
            $Z_1 Z_3 \approx Z_2 Z_3$ & $1.2 * 10^{-3}$ & $3.9 * 10^{-4}$ & $1.2 * 10^{-4}$ & $0.039 \sqrt{\kappa_1/\kappa_2}$ \\
            $Z_1 Z_2 Z_3$ & $1.2 * 10^{-3}$ & $3.9 * 10^{-4}$ & $1.2 * 10^{-4}$ & $0.039 \sqrt{\kappa_1/\kappa_2}$\\
            \midrule
            $|0\rangle$ Prep &\\ 
            \midrule
            Time & $4.0 * 10^{-9}\, s$& $4.0 * 10^{-9}\, s$ & $4.0 * 10^{-9}\, s$  & $0.1*(\kappa_2\alpha^2)^{-1}$ \\
            $X$ Error Probability & $4.9 * 10^{-15}$& $4.9 * 10^{-15}$ & $4.9 * 10^{-15}$ &$0.39 \exp(-4 \alpha^2)$\\
            \midrule
            $|+\rangle$ Prep &&&\\
            \midrule
            Time & $4.0 * 10^{-7}\, s$ & $4.0 * 10^{-7}\, s$ & $4.0 * 10^{-9}\, s$ & $10*(\kappa_2\alpha^2)^{-1}$  \\
            $Z$ Error Probability &$7.5 * 10^{-3}$ & $7.5 * 10^{-4}$ & $7.5 * 10^{-5}$ & $7.5 \kappa_1/\kappa_2$\\
            \midrule
            Idle \\
            \midrule
            $X$ Error Probability & $6.2 * 10^{-17}$ & $2.0 * 10^{-17}$ & $6.2 * 10^{-18}$ & $0.5 \kappa_1 \alpha^2 T \exp(-4 \alpha^2)$ \\
            $Y$ Error Probability &$6.2 * 10^{-17}$ & $2.0 * 10^{-17}$& $6.2 * 10^{-18}$& $0.5 \kappa_1 \alpha^2 T \exp(-4 \alpha^2)$ \\
            $Z$ Error Probability &$1.0 * 10^{-2}$& $3.1 * 10^{-3}$ & $1.0 * 10^{-3}$ & $\kappa_1 \alpha^2 T$\\
            \bottomrule
        \end{tabular}
    \end{center}
    \caption{\ji{Table of gate error rates for CNOT and Toffoli from simulation for the three regimes defined in \cref{tab:regimes}. The noise model is pure loss with rate $\kappa_1$ and no gain or dephasing noise. $\alpha$ is the cat code parameter and $\kappa_2$ is the rate of two-phonon dissipation. The rightmost column gives the formula in terms of $\alpha$, $\kappa_1$ and $\kappa_2$. The Toffoli optimal gate time is similar to the CNOT optimal gate time (see \cref{app:Gate Error Simulations}). For simplicity and because the CNOT gate time sets the timescale of error correction, the Toffoli gate error rates are shown for a gate time equal to the CNOT optimal gate time. In the error probability formulas for Idling, $T$ is the length of time that the system idles. The error probabilities for idle for the three regimes are calculated using the CNOT optimal gate time.}}    
    \label{tab:Gateerrorrates}
\end{table*}

\begin{table*}[]
    \centering
    \begin{tabular}{cccc}
        \toprule
       & \REGone & \REGtwo & \REGthree\\
        \midrule
        $X$  Measurement &&& \\
        \cmidrule{1-1}
         Idling Time &$\longleftarrow$ & $3.1*10^{-6} s$& $\longrightarrow$ \\
         Infidelity Repetition Code &$7.2*10^{-3}$&$9.7*10^{-4}$&$3.6*10^{-4}$ \\
         Infidelity Surface Code &$7.2*10^{-3}$&$9.7*10^{-4}$&$1.0*10^{-4}$ \\
        \midrule
        $Z$  Measurement &&& \\
        \cmidrule{1-1}
        Infidelity &$\longleftarrow$ & $1.7*10^{-4}$& $\longrightarrow$ \\
        \bottomrule
    \end{tabular}
    \caption{
    Table of $X$-basis and $Z$-basis measurement error rates and measurement idling times used in the error correction simulations.   The error rates correspond to $|\alpha|^2=8$.  The number of measurements (up to 3(5) for repetition(surface) code) is chosen to maximize fidelity.   Plots with more datapoints and details on the simulations and assumed parameters can be found in \cref{appendix:Measurement}.}
    \label{tab:XMeasurementTimeAndErrorRate}
\end{table*}

\subsection{\texorpdfstring{$X$}{X} measurement}
\label{subsec:x_measurement}
$X$-basis readout entails determining the parity of a bosonic mode. Specifically this is readout in the basis of even and odd cat states, i.e., $|\pm\rangle \propto |\alpha\rangle \pm |-\alpha\rangle$. \etc{Here we describe our approach to $X$-basis readout.  In our scheme, we use an additional phononic readout mode in every unit cell that we do not stabilize with two phonon dissipation.  This readout mode is interrogated by a transmon qubit in parallel with the gates of the next error correction cycle.  This allows us to achieve high measurement fidelity and minimal idling time for the data qubits.  The additional readout mode (coloured green) and transmon are pictured in \cref{fig:hardware_cartoon}.  }

Here we outline the steps for the readout of an ancilla qubit in the $X$ basis.  First we ``deflate" the ancilla mode ($\hat{a}_1$) which maps the even parity cat state to the Fock state $|\hat{n} = 0\rangle$ and the odd cat state to the Fock state $|\hat{n}=1\rangle$ \cite{Grimm2020}.  This can be achieved by abruptly changing the engineered dissipation from $D[\hat{a}_1^2-\alpha^2]$ to $D[\hat{a}_1^2]$.  Pairs of phonons will be dissipated mapping the system to the $|\hat{n} = 0\rangle$ and $|\hat{n} = 1\rangle$ manifold while preserving parity.  The purpose of the deflation is to reduce susceptibility to single-phonon loss events which change the parity of the cat qubit.  After this deflation we turn off the two-phonon dissipation. Subsequent to the deflation the ancilla mode and readout mode ($\hat{a}_2$) evolve under the Hamiltonian $
\hat{H} = g (\hat{a}_1 ^\dagger \hat{a}_2 + \hat{a}_2 ^\dagger \hat{a}_1)$ which transfers the excitation \cite{Gao2018,Zhang2019,Gao2019} between the ancilla mode and readout mode in a time ${\pi}/{2 g}$. 

With the excitation in the readout mode, we perform repeated QND (quantum non-demolition) measurements of the readout mode \cite{sun2014tracking, HannRobustReadout, ElderHighFidelityMeasurement} and take a majority vote to get our final measurement outcome.  The individual measurements are standard QND bosonic parity measurements which are performed using a dispersive coupling between the readout mode and a transmon qubit ($\hat{q}$) described by $\hat{H}_{\textrm{dispersive}} = \chi \hat{a_2}^\dagger\hat{a_2}\hat{q}^\dagger\hat{q}$.  Evolution under the Hamiltonian for a time ${\pi}/{\chi}$ yields the controlled parity gate
$ U= I\otimes|g\rangle\langle g| + e^{i \hat{a}_2^\dagger \hat{a}_2 \pi} |e\rangle\langle e|$.
Combined with transmon state preparation and measurement this interaction can be used to realize parity measurements of the readout mode \cite{sun2014tracking}.

While this repeated parity measurement is taking place, the CNOT gates of the next error correction cycle can occur in parallel.  This enables us to reach high readout fidelity without affecting the length of an error correction cycle.  For our repetition code and surface code simulations we use up to 3 or 5 parity measurements respectively during the error correction gates of the next cycle. 

We simulated this measurement scheme to get a rough sense of the expected measurement fidelities.  The misassignment probabilities and measurement idling times can be found in \cref{tab:XMeasurementTimeAndErrorRate} for the three regimes considered in the paper.

\subsection{\texorpdfstring{$Z$}{Z} measurement}
In $Z$-basis measurement the goal is to distinguish $|0\rangle$ and $|1\rangle$, which are approximately the coherent states $|\alpha\rangle$ and $|-\alpha\rangle$.  We achieve \jp{this readout by engineering a coupling between the storage model $\hat{a}$ and the buffer mode $\hat{b}$ described by the ``beamsplitter'' Hamiltonian $\hat{H} = g (\hat{a}^\dagger \hat{b} + \hat{b}^\dagger \hat{a})$. The physical realization of this Hamiltonian is explained in \cref{appendix:Measurement}.}

\jp{If the state of the storage mode is $|\pm \alpha\rangle$, this coupling drives the buffer mode to a coherent state $|\pm \gamma\rangle$ whose phase is aligned with the initial phase of the storage mode. Hence a homodyne measurement of the buffer mode distinguishes the states $|\pm\alpha\rangle$ of the storage mode, as desired.}

In \cref{appendix:Measurement} we find that the \etc{signal to noise ratio (SNR)} for this readout scheme \hp{at time $\tau$ }is
\begin{align}
    \text{SNR}(\tau) &= \nonumber \\
    &\alpha \sqrt{8\kappab} \frac{\left[ 1-e^{-\kappa_b \tau/4} \left[\cosh{\frac{\beta \tau}{4}}+\frac{\kappa_b}{\beta}\sinh{\frac{\beta \tau}{4}} \right] \right]}{g\sqrt{\tau}}
\end{align}
in good agreement with numerics.  \jp{Here $\kappa_b$ is the single-phonon loss rate of the buffer mode,} and $\beta = \sqrt{\kappa_b^2-(4g)^2}$.  The measurement SNR scales as $\alpha$ which means there is an exponential improvement of the measurement fidelity with $|\alpha|^2$.  This measurement process is not quantum non-demolition (QND) and at long times the measurement SNR goes as ${1}/{\sqrt{\tau}}$ because the storage mode is emptied.  From the measurement SNR, the measurement separation error\hp{, which is the dominant contribution to the total measurement error $\epsilon(|\alpha|^2)$, } can be computed  using $\epsilon_\text{sep}(\tau) = \frac{1}{2}\text{Erfc}(\frac{\text{SNR}(\tau)}{2})$
\cite{krantz2019quantum}. 

\jp{For the values of $|\alpha|^2$ considered in our $Z$-basis measurement simulations, we found that the optimal measurement time is $\text{T}_{\text{measure}} \approx 850 \text{ ns}$, and that a conservative relation for the numerical probability of an incorrect readout as a function of $|\alpha|^2$ is 
\begin{align}
    \epsilon(|\alpha|^2) &= e^{-1.5-0.9|\alpha|^2}.
    \label{eq:ZMeasurementError}
\end{align}
For $\alpha^2 = 8$, used in much of our analysis of error correction, we have  $\epsilon \approx 2 * 10^{-4}$.  In contrast to gates, whose optimal fidelities depend only on the dimensionless ratio $\kappa_1/\kappa_2$,  readout fidelities and durations depend on additional parameter assumptions discussed in \cref{appendix:Measurement}.}

\subsection{Gain and Dephasing errors}

\etc{Here we summarize the impact of additional noise sources investigated in \cref{appendix:Perturbative analysis of cat qubit gates,app:Gate Error Simulations}.} \ji{Adding phonon gain to the storage mode in addition to loss only slightly enhances the error rates of all operations. If the thermal population is given by $n_{th} = 0.01$, then the error rates of the CZ gate are increased by about $1 \%$, for example. On the other hand, we expect that pure dephasing noise in the form of a Lindblad jump operator $a^\dagger a$ on the storage mode will substantially increase the $X$ error rate but only slightly increase the $Z$ rate. We numerically find that although the $Z$ error rates of the $Z$ and CZ gates are not measurably affected by pure dephasing noise, those of the \cnot and \tof gates are adversely affected. This is surprising given that pure dephasing consists of random rotations on the storage mode state and does not change the parity of the cat qubit. We provide a perturbative analysis to explain this behavior and attribute the enhanced $Z$ error rates of the \cnot and \tof gates to the fact that the stabilizing jump operators for the target cat qubits are not static and instead rotate conditioned on the state of the control qubits. Our perturbative analysis agrees well with our numerical results, and they predict that the optimal $Z$ error rates of the \cnot and \tof gates scale as $\sqrt{\kappa_{\phi}/\kappa_{2}}$, where $\kappa_{\phi}$ is the dephasing rate. These calculations can be found in \cref{appendix:Perturbative analysis of cat qubit gates}. Our simulations of gate error rates in the presence of gain and dephasing noise can be found in \cref{app:Gate Error Simulations}.}

\section{Logical failure rates for quantum memory}
\label{sec:LogicalMemory}

\begin{figure}[th]
	\centering
	\subfloat[\label{fig:RepetitionCodeStripExample}]{%
		\includegraphics[width=0.48\textwidth]{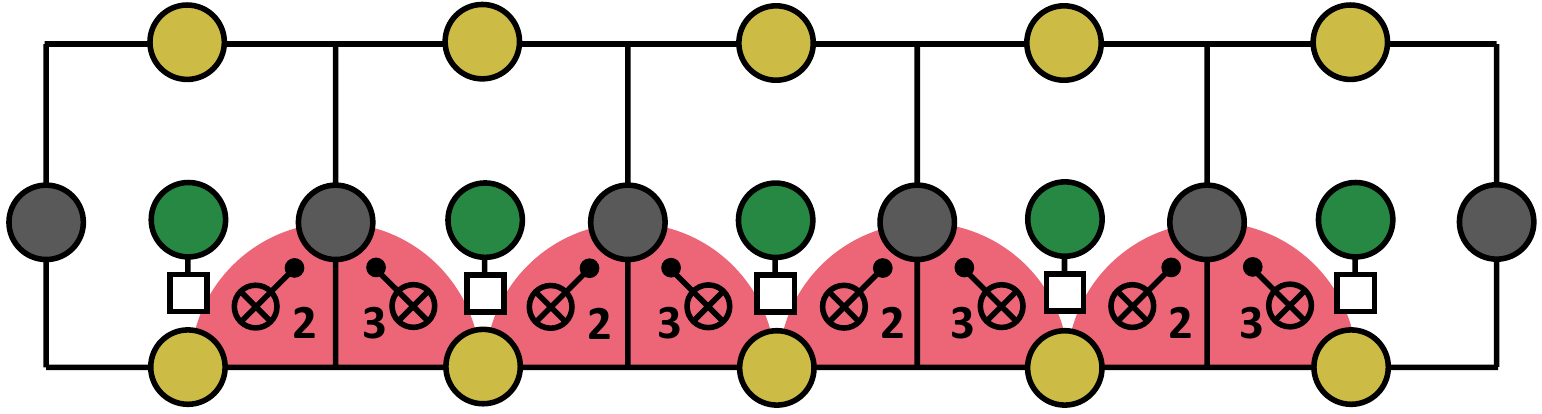}
	}
	\vfill
	\subfloat[\label{fig:SurfaceCode3by5StripPINK}]{%
		\includegraphics[width=0.48\textwidth]{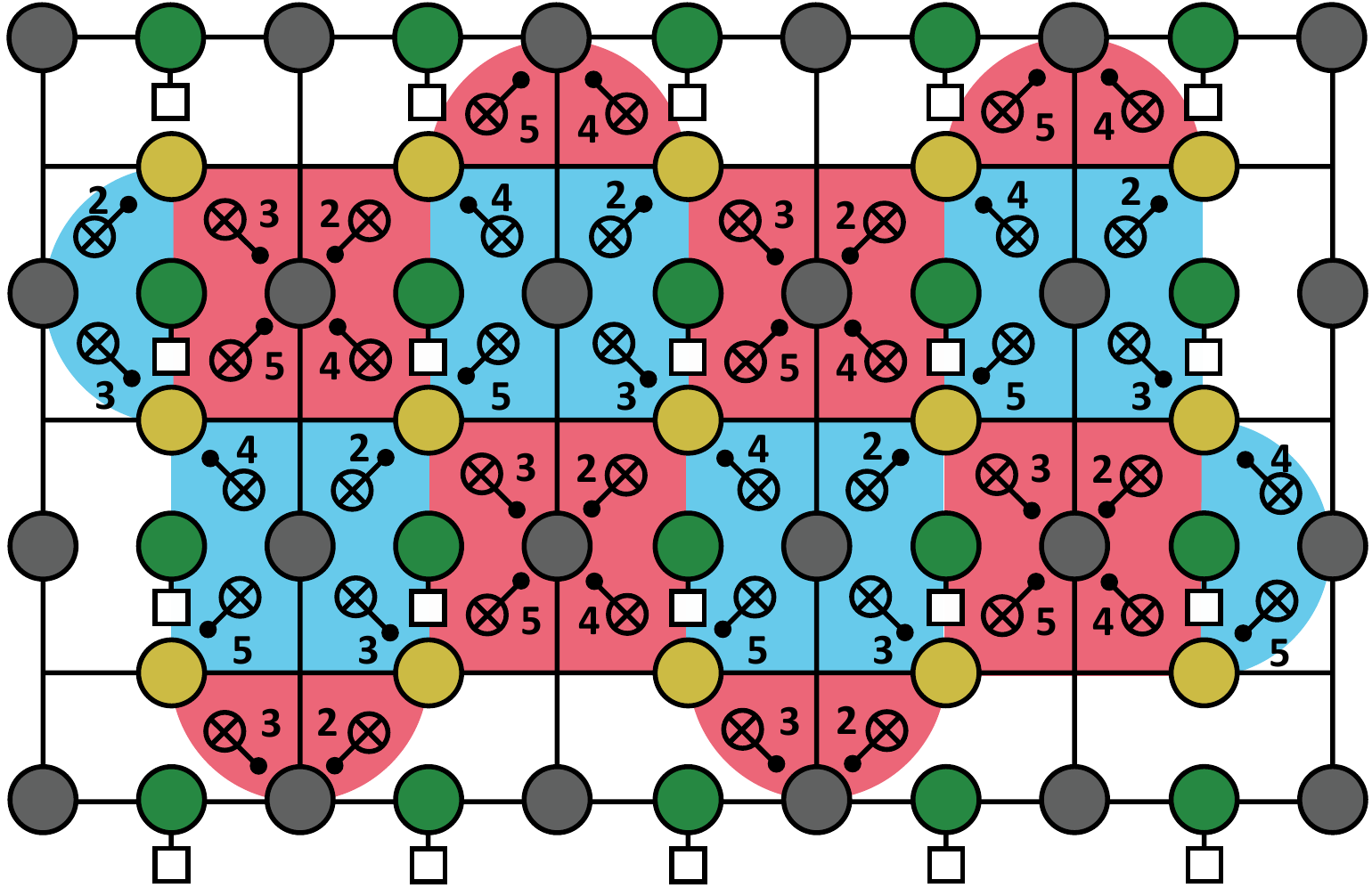}
	}
	\vfill
	\subfloat[\label{fig:KEYsurfaceRepLattice}]{%
		\includegraphics[width=0.48\textwidth]{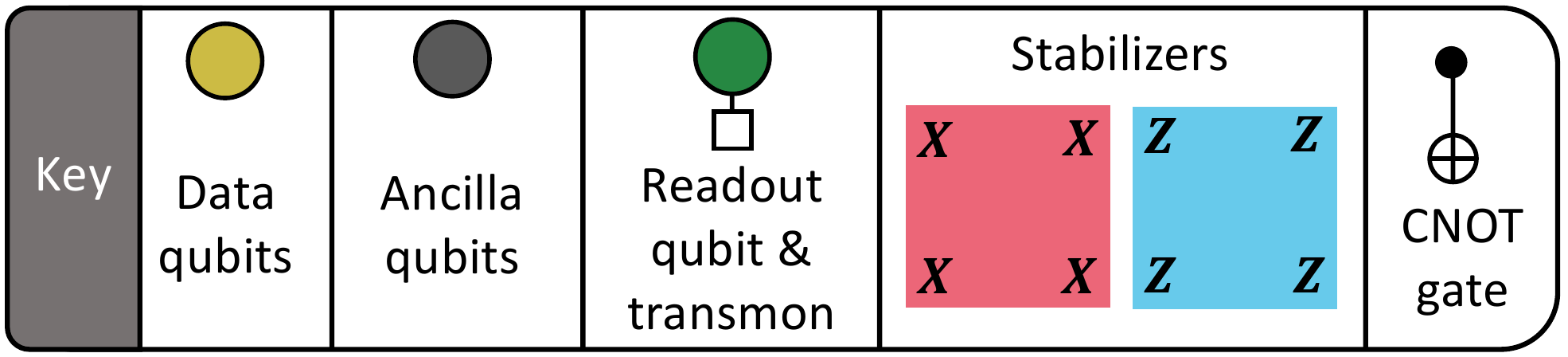}
	}

	\caption{(a) Circuit illustration of a $d=5$ repetition code embedded in our ATS layout. As explained in \cref{fig:hardware_cartoon}, the yellow vertices correspond to the data qubits and the gray vertices to the ancillas, whereas green circles are readout modes and the white squares are transmon qubits, which we use for $X$-basis measurements (see \cref{appendix:Measurement,fig:xMeasurementInfidelity} for more details). The pink semi-circles are used to illustrate the $X_iX_{i+1}$ stabilizer of the repetition code. \CC{We also label each CNOT gate by the corresponding time step in which it is applied.} (b) Circuit illustration of a $d_x = 3$ by $d_z = 5$ thin rotated surface code. The pink and blue plaquettes correspond to the $X$ and $Z$-type stabilizers respectively, with the numbers indicating the time steps in which the CNOT gates are applied. \CC{Measurements of $X$-type stabilizers detect $Z$ errors whereas measurements $Z$-type stabilizers detect $X$ errors.} (c) Key illustrating the different components of the repetition and surface code lattices. CNOT gates are used to couple qubits connected to the same ATS.}
	\label{fig:CircuitsRepSurf}
\end{figure}

\begin{figure}[th]
	\centering
	\includegraphics[width=0.48\textwidth]{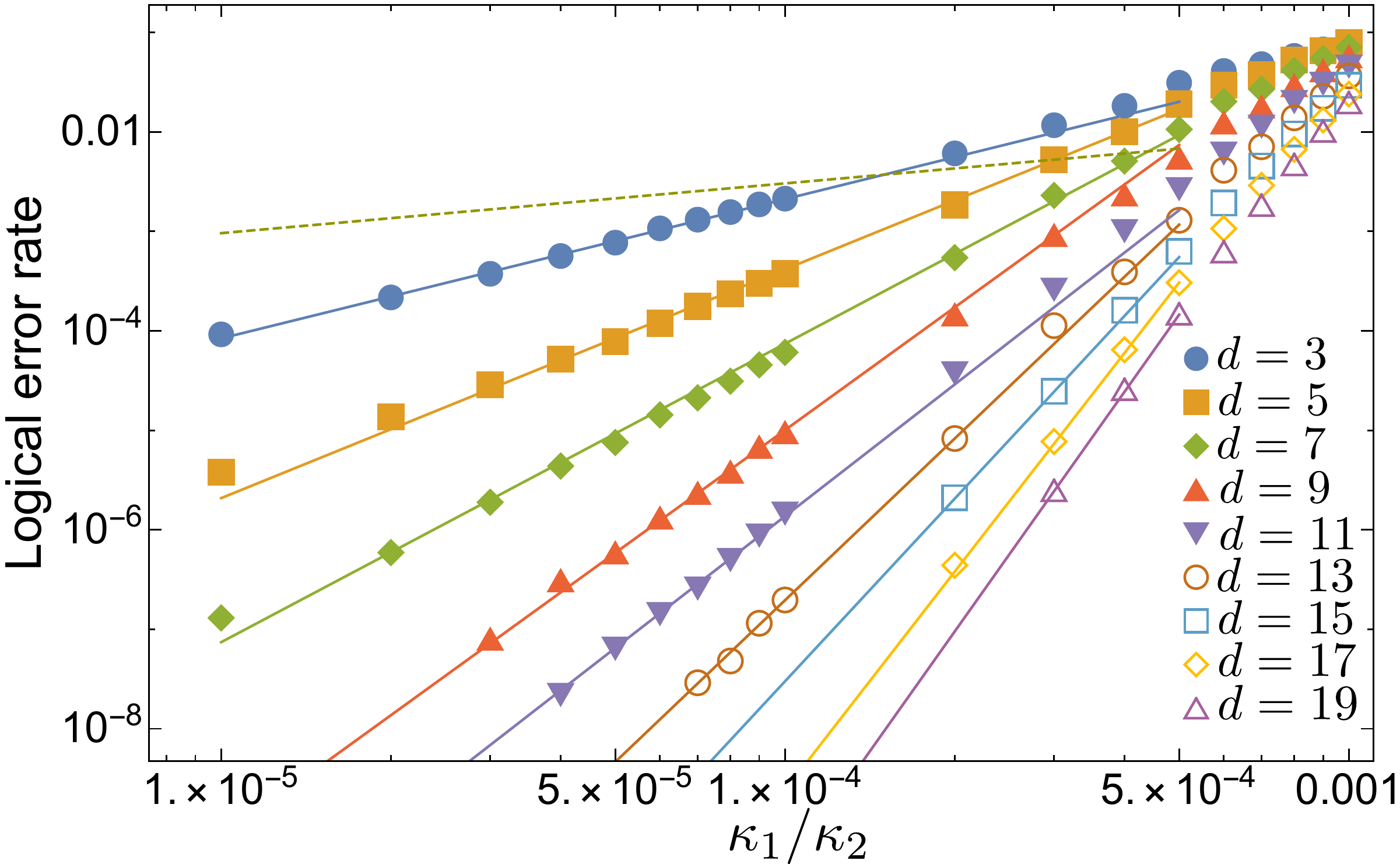}
	\caption{Logical $Z$ failure rates for the repetition code, for a variety of values of the code distance $d$. We use the circuit-level noise model described in \cref{sec:GatesMeas} with $\kappa_{\phi} = 0$ and $n_{th} = 0$. The $X$-basis measurement error rates are obtained from \cref{tab:XMeasurementTimeAndErrorRate} with three parity measurements. The number of syndrome measurement rounds $r$ for each distance is obtained using the \texttt{STOP} algorithm described in \cref{appendix:STOPdec}. The dashed green line is used as a stand-in for comparison with the logical memory error rates and corresponds to the function $ 0.3025 \sqrt{\kappa_1 / \kappa_2 } $ which is a quarter of the total $Z$ failure rate of a physical CNOT gate (see \cref{tab:Gateerrorrates}).}
	\label{fig:MemRepCode}
\end{figure}

\CC{Equipped with the noise model in \cref{sec:GatesMeas} for performing gates and measurements on stabilized cat qubits, we now transition to describing the outer level error correcting codes used to protect encoded logical qubits against phase-flip and bit-flip errors. Errors can be identified by measuring a codes stabilizer generators $g_i \in \mathcal{S}$ where $\mathcal{S}$ forms an abelian group, and the stabilizers act trivially on the encoded state. Detectable errors anticommute with a subset of the stabilizers in $\mathcal{S}$ and can be identified by their error syndrome. More details on the stabilizer formalism can be found in Ref.~\cite{Gottesman2010}.} 

\CC{The two codes} that we use in our architecture for implementing quantum algorithms are the repetition code and the rotated surface code \cite{TS14}. \CC{Illustrations for such codes, along with their corresponding syndrome extraction circuits (which measure the codes stabilizer generators), are given in \cref{fig:CircuitsRepSurf}}. As described in \cref{sec:BottomUp,Sec:TopDown}, the repetition code is used for preparing $\ket{\text{TOF}}$ magic states which will allow us to implement logical Toffoli gates. However, the repetition code alone is insufficient for universal quantum computation since, without the ability to correct at least one bit-flip error, the logical $X$-failure rates would be too high during the implementation of most quantum algorithms of interest for reasonable values of $\alpha^2$ (see \cref{fig:TotalRepetitionCodeFail}). As such, apart from the preparation of $\ket{\text{TOF}}$ states (which will be converted to $\ket{\text{TOF}}$ states encoded in the surface code using lattice surgery), all logical gates of quantum algorithms are performed in a $d_x = 3$ by $d_z$ rotated surface code lattice. Here $d_x$ and $d_z$ denote the minimum weight of the $X$- and $Z$-type logical operators of the rotated surface code. We fix $d_x = 3$ since as will be seen, we will only need to correct one bit-flip error at the surface code level to get the desired logical $X$ failure rates for the implementation of quantum algorithms of practical interest such as those considered in \cref{sec:Overhead}. 

In this section, we provide logical $Z$ failure rates for the repetition code and rotated surface code in the context of quantum memories using a minimum-weight perfect matching (MWPM) decoding algorithm with weighted edges described in \cref{sec:EdgeWeights} and the noise model described in \cref{sec:GatesMeas}. We also provide general logical $X$ failure rate polynomials of the rotated surface code as a function of the $d_z$ distance. 

\subsection{Repetition code logical failure rates}
\label{subsec:RepCodeMem}

The logical $Z$ failure rates of the repetition code for distance $3 \le d \le 19$ are provided in \cref{fig:MemRepCode}. All results were obtained from a Monte-Carlo simulation based on the circuit level noise model where each gate, state-preparation, idling qubits and measurements fail with probabilities given in \cref{tab:Gateerrorrates,tab:XMeasurementTimeAndErrorRate}. 

In error correction there are two settings of interest: where the logical information needs to be stored for some fixed period of time; and where there is flexibility to adapt the number of rounds before proceeding to the next stage of the computation.  Here we introduce the \texttt{STOP} algorithm, which is an adaptive policy for \CC{deciding} how many rounds to repeat the syndrome measurements.  In the limit of large code distances, \texttt{STOP} terminates (with high probability) in the same number of rounds as an algorithm using fixed $d$ rounds. For smaller code distances and low noise regimes, \texttt{STOP} provides an advantage over a fixed round decoder as  it requires $(d+1)/2$ rounds. Full details for the implementation of the \texttt{STOP} algorithm are provided in \cref{appendix:STOPdec}. We now give two important remarks.

\textit{Remark one:} Consider first the setting where the logical information is stored for a fixed period of time. The standard approach that is followed in the literature when obtaining numerical results for decoding such codes is to perform $d$ rounds of noisy syndrome measurements followed by one round of perfect syndrome measurement (where no additional errors are introduced). Errors are then corrected using the full syndrome history. The round of perfect syndrome measurement is added to ensure that the final error after correction is either in the stabilizer group or corresponds to a logical operator (i.e. we must ensure that we project to the code-space to declare success or failure). Furthermore, if the error syndrome was decoded based \textit{only} on $d$ noisy syndrome measurement rounds (i.e. without the round of perfect error correction), a single measurement error occurring in the $d$th round could result in a logical failure (a fact that is often not fully appreciated). However for many models of universal quantum computation, the data qubits are measured directly as part of the quantum algorithm or during the implementation of state injection for performing non-Clifford gates (see Refs.~\cite{DA07,CIP17,KBF+15} and \cref{fig:MeasDataRepEquiv}). As illustrated in \cref{fig:MeasDataRepEquiv} of \cref{appendix:STOPdec}, the direct measurement of the data qubits can be viewed as a round of perfect error correction since measurement errors in such a process are equivalent to data qubit errors occurring immediately prior to the measurement of the data. However for our purposes, the repetition code will be used during the preparation of $\ket{\text{TOF}}$ magic states where the circuits used in the preparation protocol contain non-Clifford gate locations (see \cref{fig:BottomUpTOFATS}). Prior to the application of these non-Clifford gates, errors on the encoded code-blocks need to be corrected without having access to a round of perfect syndrome measurements (since the data qubits cannot be measured directly prior to applying the non-Clifford gates). Hence, it is important to have a decoder which is robust to measurement errors occurring in the last round when rounds of perfect syndrome measurements cannot effectively be applied in the hardware. A solution is that instead of repeating the syndrome measurement $d$ times, one can repeat the syndrome $r$ times where $r$ is computed using the \texttt{STOP} algorithm mentioned above. Note that in this case, $r$ is not fixed but instead is a function of the observed syndrome history. For all logical $Z$ failure rates plotted in \cref{fig:MemRepCode}, the simulations were performed using the \texttt{STOP} algorithm for determining when to stop measuring the error syndrome. To ensure projection onto the codespace, we add 1 round of ideal syndrome measurements after the last round given by the \texttt{STOP} algorithm and implement MWPM over the full syndrome history.

\textit{Remark two:} The $x$-axis in \cref{fig:MemRepCode} is plotted as a function of $\kappa_1/\kappa_2$. It is important to note that some components of the hardware fail with probabilities proportional to $\kappa_1/\kappa_2$ whereas other components (such as the CNOT gates) \CC{fail} with probabilities proportional to $\sqrt{\kappa_1/\kappa_2}$ (see \cref{tab:Gateerrorrates}). \CC{In particular, in \REGthree, the noise is dominated by CNOT gates, whereas in \REGone}, some idling qubits during CNOT gate times are afflicted by errors with probabilities comparable with the CNOT failure rates, hence changing the slope of the logical failure rate curves. To be clear, in our simulations we took into account all different types of idling locations; for this reason, and also because we use the \texttt{STOP} algorithm for determining the number of syndrome measurement rounds instead of repeating a fixed $d$ times, our numerics should not be directly compared with previous works such as in \cite{GM2020}. Note further that for comparisons with other works (such as in \cite{GM2020}), the $x$-axis of our plots would need to be re-scaled as a function of $\sqrt{\kappa_1 /\kappa_2}$. 

\begin{figure}[th]
	\centering
	\includegraphics[width=0.48\textwidth]{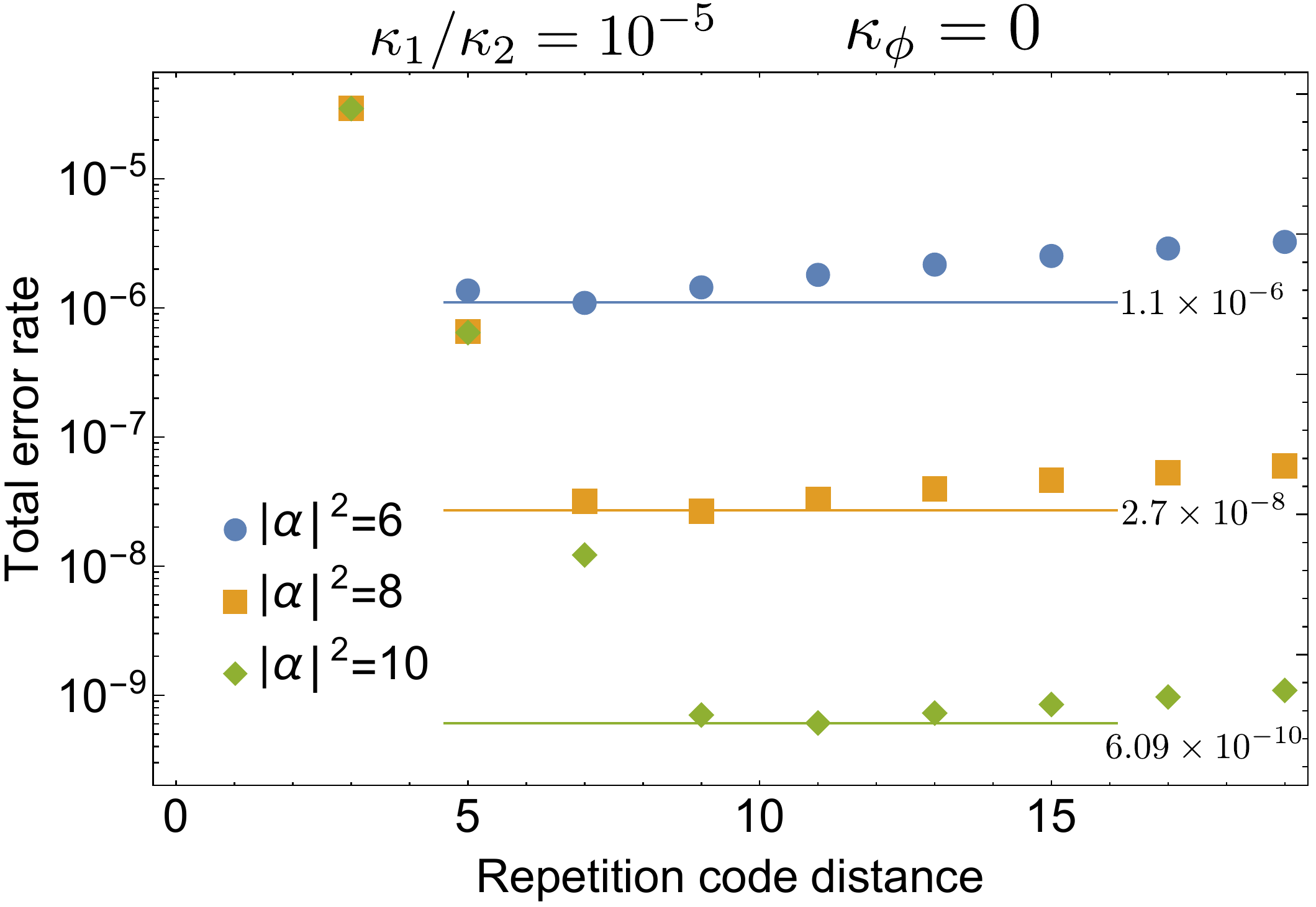}
	\caption{Total logical failure rate per code cycle for various repetition code distances and values of $|\alpha|^2$ with fixed $\kappa_1 / \kappa_2 = 10^{-5}$, $\kappa_{\phi} = 0$ and $n_{th} = 0$.  \etc{Therefore, the $|\alpha|^2=8$ data points correspond to \REGthree.}
	The logical $X$ and $Y$ failure rates were computed analytically (to leading order) using the noise model presented in \cref{sec:GatesMeas} while taking into account all malignant and benign fault locations. For $|\alpha|^2=8$, the lowest achievable total logical error rate is $2.7 * 10^{-8}$ per code cycle using $d=9$. }
	\label{fig:TotalRepetitionCodeFail}
\end{figure}

Given two strips of neighboring repetition codes, a logical CNOT gate can be implemented transversally between the two strips, and the failure probability of such a gate is approximately four times the values showed in \cref{fig:MemRepCode}. One possible interesting quantum error-correction experiment would be to demonstrate a logical CNOT gate with lower failure probability compared to a physical CNOT gate. As such, in \cref{fig:MemRepCode}, we also plotted a dashed green curve which corresponds to the function $ 0.3025 \sqrt{\kappa_1/ \kappa_2} $ which is a quarter of the total $Z$ failure rate of a CNOT gate (see \cref{tab:Gateerrorrates}). As can be seen, for $\kappa_1 / \kappa_2 < 4.5 * 10^{-4}$, the probability of failure of a CNOT gate encoded in a $d=7$ repetition code is lower than that of a physical CNOT gate. \CC{As such, experiments demonstrating a logical CNOT gate with a failure probability smaller than a physical CNOT could be achieved in \REGtwo.}
From the hardware analysis, we find that $\kappa_{2}/(2\pi) = 500\textrm{kHz}$ (or $\kappa_{2} = 3.14* 10^{6} s^{-1}$) is achievable for $|\alpha|^{2} = 8$ (see \cref{sec:HardwareImplementation,subsec:loss_results} for more details). In this case, $\kappa_{1}/\kappa_{2} =10^{-4}$ 
\CC{(\REGtwo)} corresponds to a lifetime of $3\textrm{ms}$. From \cref{fig:MemRepCode}, a logical CNOT gate implemented transversally with two $d=9$ repetition code strips \CC{in \REGtwo} fails with probability $3.7 * 10^{-5}$ which would correspond to the highest CNOT fidelities achieved to date \CC{(see for instance Refs.~\cite{HighCNOT1,HighCNOT2} where a two-qubit gate fidelity as high as $99.9(1) \%$ is achieved).} Furthermore, we find numerically that the general polynomial describing the logical $Z$ failure rate of a distance $d$ repetition code for $d$ rounds of syndrome measurements is given by
\begin{align}
    p^{(Z)}_{L}(d) = 0.014d \left(770 \frac{\kappa_1}{\kappa_2}\right)^{0.41d}.
\end{align}
The justifications for the chosen scaling of $p^{(Z)}_{L}(d)$ and the scaling of the logical failure rates for the rotated surface code in \cref{subsec:SurfaceCodeMem} are given in \cref{subsec:TimeLikeErrors}.

Lastly, in \cref{fig:TotalRepetitionCodeFail} we compute the total logical failure rate per code cycle (which includes contributions from logical $X$ and $Y$ failures) of the repetition code for distances in the range $3 \le d \le 19$ \CC{in \REGthree with} $\kappa_{\phi} = 0$ and $n_{th} = 0$. For $|\alpha|^2 = 8$, it can be seen that above $d=9$, contributions from bit-flip errors are the dominant factor in the total logical failure rate. As such, going to larger repetition code distances results in higher logical failure rates. Such features demonstrate the importance of taking into account contributions from bit-flip errors, even though they are exponentially suppressed. Further, such results demonstrate that the logical $X$ error rate when implementing a logical Toffoli gate using the piece-wise fault-tolerant construction of \cite{Guillaud2019,GM2020} would be too high for the algorithms considered in \cref{sec:Overhead} \CC{unless $|\alpha^2| \gg 8$}.

\subsection{Rotated surface code logical failure rates}
\label{subsec:SurfaceCodeMem}

\begin{figure}[th]
	\centering
	\includegraphics[width=0.48\textwidth]{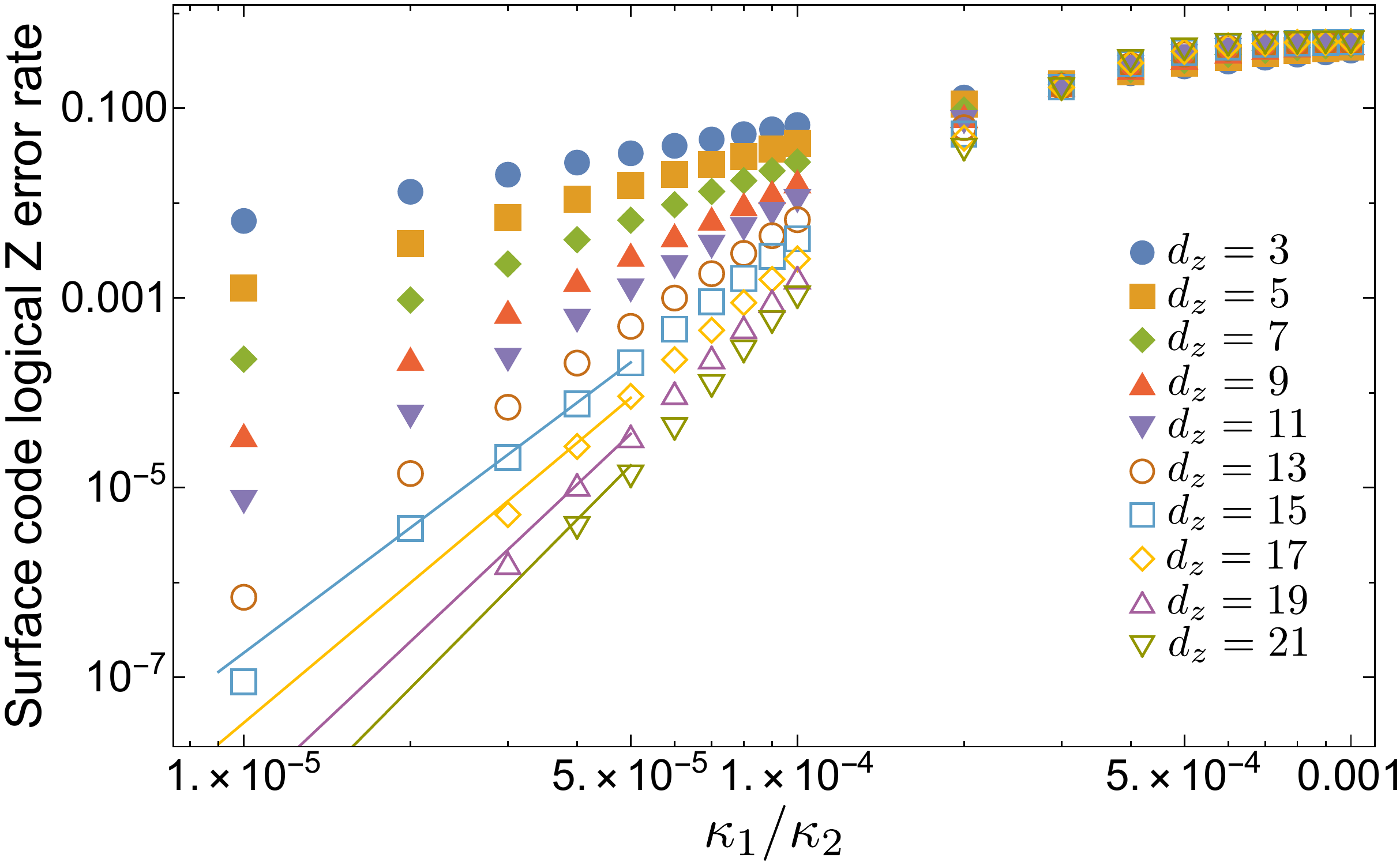}
	\caption{Logical $Z$ failure rates for the rotated surface code with  $d_x = 3$ and varying $d_z$. We use the circuit-level noise model described in \cref{sec:GatesMeas} with $\kappa_{\phi} = 0$ and $n_{th} = 0$. The $X$-basis measurement error rates are obtained from \cref{tab:XMeasurementTimeAndErrorRate} with five parity measurements. We point out that $\kappa_1 / \kappa_2 = 10^{-5}$, $\kappa_1 / \kappa_2 = 10^{-4}$ and $\kappa_1 / \kappa_2 = 10^{-3}$ correspond to CNOT failure rates of $3.8 * 10^{-3}$, $1.2 * 10^{-2}$ and $3.8 * 10^{-2}$. The simulations were done by performing $d_z$ rounds of noisy syndrome measurements followed by one round of perfect syndrome measurement. }
	\label{fig:SurfaceCodeMem}
\end{figure}

Using the circuit-level noise model described in \cref{tab:Gateerrorrates,tab:XMeasurementTimeAndErrorRate,eq:ZMeasurementError},  the logical $Z$ failure rates for the rotated surface code with $d_x = 3$ and varying  $d_z$ are given in \cref{fig:SurfaceCodeMem}. Note that the logical $X$ operator has minimum support on $d_x$ qubits along each column
of the lattice. The logical $Z$ operator has minimum support on $d_z$ qubits along each row of the lattice. Contrary to our repetition code simulation methodology, the simulation results were obtained by performing $d_z$ rounds of noisy syndrome measurements followed by one round of perfect syndrome measurement in order to guarantee projection onto the code-space. Throughout this paper, we use the surface code with only a fixed number of error correction rounds. Furthermore, in our proposal we never perform \CC{physical} non-Clifford gates directly on surface-code patches, rather non-Clifford gates are always achieved by \etc{gate teleportation using} a magic state. As such, all simulations are performed for \jp{$d_z$} rounds followed by one ideal round to project onto the codespace. 

\CC{
We also point out that all $Z$ stabilizers are measured in the $Z$-basis by using CNOT gates (as shown in \cref{fig:CircuitsRepSurf}) rather than measured in the $X$-basis using CZ gates. The reason is that, in addition to $Z$-basis measurements being more reliable, only $X$ or $Y$ errors (which are exponentially suppressed) on the target qubits of the CNOT's can result in measurement errors. If CZ gates were used, a CZ failure resulting in a $Z$ error on an ancilla qubit would flip the measurement outcomes. Hence \jp{if CZ gates paired with $X$-basis measurements were used instead of CNOT gates paired with $Z$-basis measurements, syndrome measurement errors in $Z$-type stabilizer measurements would be much more likely.}}  

\etc{Using the stabilizer measurement schedule of \cref{fig:CircuitsRepSurf} and detailed further in \cref{fig:xMeasurement}, the duration of each round of stabilizer measurements has three contributions: the optimal time for 4 CNOT gates; the time to deflate the cat qubit and the time to swap the ancilla qubit into the readout mode.  For example, for \REGthree this takes $31 \mu s$.}

As can be seen from \cref{fig:SurfaceCodeMem}, in order to obtain low logical $Z$ failure rates without requiring a very large $d_z$ distance (say $d_z > 40$), it is required that $\kappa_1 / \kappa_2 \le 5 * 10^{-5}$. \CC{Hence the hardware parameters must be in \REGthree.} Put another way, the total CNOT gate $Z$ failure rate should be less than $7.6 * 10^{-3}$ to achieve very low logical failure rates with reasonably small surface code distances.

\jp{Comparing the logical $Z$ error rates in} \cref{fig:MemRepCode,fig:SurfaceCodeMem}, one sees that the surface code significantly under-performs the repetition code. This is mainly 
\jp{because} a distance-$d$ repetition code requires a total of $2d-1$ data and ancilla qubits compared to the rotated surface code which requires $2d_xd_z - 1$ data and ancilla qubits. Further, the surface code requires weight-four stabilizer measurements compared to weight-two stabilizers for the repetition code and thus the syndrome measurement circuit has larger depth.

The logical $Z$ and $X$ failure rate polynomials for fixed $d_x = 3$ and arbitrary $d_z$ distances (with $d_z$ rounds of stabilizer measurements) were found numerically to be given by
\begin{align}
   & p^{(Z)}_{L}(d_z) = 0.028d_z\left(3559 \frac{\kappa_1}{\kappa_2}\right)^{0.292d_z}, \label{eq:SurfPolysNoDephZ}\\
   & p^{(X)}_{L}(d_z) = 3449d_z^2 e^{-4 |\alpha|^2}\left(\frac{\kappa_1}{\kappa_2}\right).
    \label{eq:SurfPolysNoDephX}
\end{align}
See \cref{subsec:TimeLikeErrors} for further details on the fitting procedure and additional results on errors during lattice surgery.

For the algorithms considered in \cref{sec:Overhead}, \CC{a detailed resource cost analysis shows that} we require $p^{(X)}_{L}(d_z) \le 10^{-10}$. As can be seen from \cref{eq:SurfPolysNoDephX}, if $|\alpha|^2 = 6$, $\kappa_1 / \kappa_2 = 10^{-5}$ (which requires $d_z = 31$), the logical $X$ error rate is approximately $1.3 * 10^{-9}$ which is an order of magnitude worse than the minimum requirements. However, setting $|\alpha|^2 = 8$, we obtain $p^{(X)}_{L}(d_z) = 4.2 * 10^{-13}$. \CC{Given the above,} the algorithms considered in \cref{sec:Overhead} require $|\alpha|^2 \ge 8$ \etc{or $d_x > 3.$}.

\CC{From the results obtained in this section and \cref{subsec:RepCodeMem}, we can make the following observations. In \REGone, the total logical failure rate of an idling qubit using the repetition code can be smaller than that of a physical qubit using a large enough repetition code since the parameters are below threshold. In order to obtained a logical CNOT gate with failure probability less than a physical CNOT, the hardware parameters must be in \REGtwo. Lastly, to implement the algorithms considered in \cref{sec:Overhead} using thin stripped surface codes, the hardware parameters must be in \REGthree. Note that our scheme for performing universal fault-tolerant quantum computing is described in \cref{Sec:LatticeSurgery,sec:BottomUp,Sec:TopDown}.}

\etc{We conclude this section by comparing the performance of our architecture (in \REGthree) with square surface codes subject to an unbiased, depolarizing noise model with CNOT gate infidelity of $p=10^{-3}$. In an algorithm example discussed further in \cref{sec:Overhead} (the $L=8$ and $u=4$ example), we require $d_x=3$ and $d_z=25$ to suppress $p^{(Z)}_{L} + p^{(X)}_{L}$ to $1.9 *10^{-11}$. This requires $d_x d_z = 75$ ATS and $3 d_x d_z =225 $ qubits (PCDRs) per surface code patch.  For depolarizing noise, the square surface code has been simulated~\cite{fowler2013surface} and fitted to a logical error rate  $p_{L} \approx d^2 (100 p )^{(d+1)/2}$ and so $d=26$ suffices to achieve a similar logical error rate ($p_L=2.1*10^{-11}$).  Most conventional architectures use $2d^2$ qubits per surface code patch (because they do not use readout qubits).  Therefore, our architecture require $\sim 6\times$ fewer qubits per surface code patch.}

\subsection{Surface code logical failure rates in the presence of crosstalk errors}\label{subsection:Surface code logical failure rates in the presence of crosstalk errors}

\begin{figure}[th]
    \centering
    \includegraphics[width=0.45\textwidth]{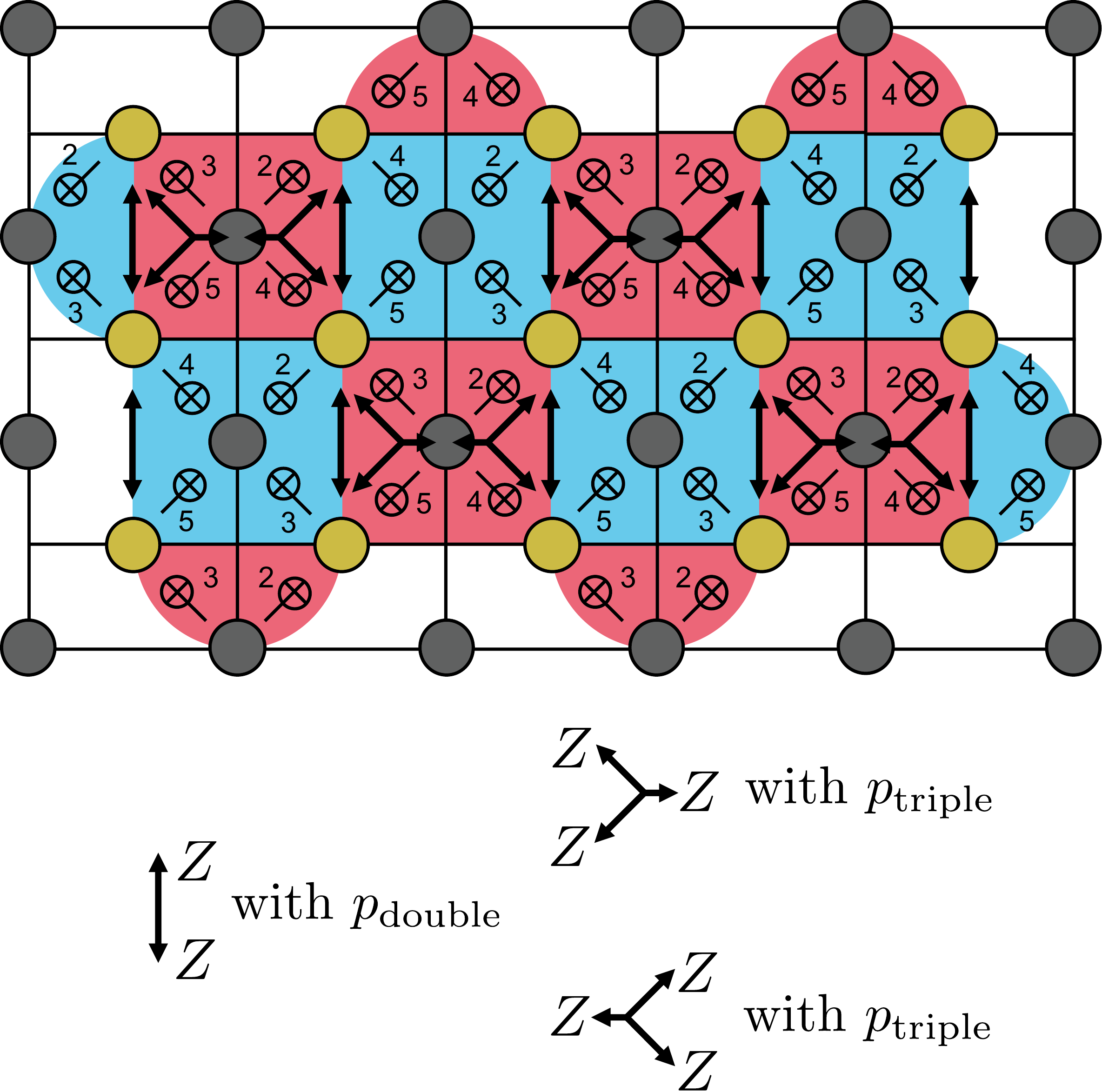}
    \caption{Crosstalk errors due to multiplexed stabilization. Phononic modes that are connected via a shared ATS experience correlated Pauli $Z$ errors due to micro-oscillation (see \cref{subsec:mitigation_optimization}). Every pair of two data qubits that are shared by the same ATS (hence aligned vertically) undergoes a correlated $Z$ error with a probability $p_{\textrm{double}}$. Also, every triple of two data qubits and an ancilla qubit that measures an $X$-type surface-code stabilizer undergoes a correlated $Z$ error with a probability $p_{\textrm{triple}}$, where the $Z$ error on the ancilla qubit manifests as a flipped outcome of the corresponding $X$-type stabilizer.   }
    \label{fig:Fig_surface_code_micro_oscillation_crosstalk_main_text}
\end{figure}

\begin{figure}
    \centering
    \includegraphics[width=\columnwidth]{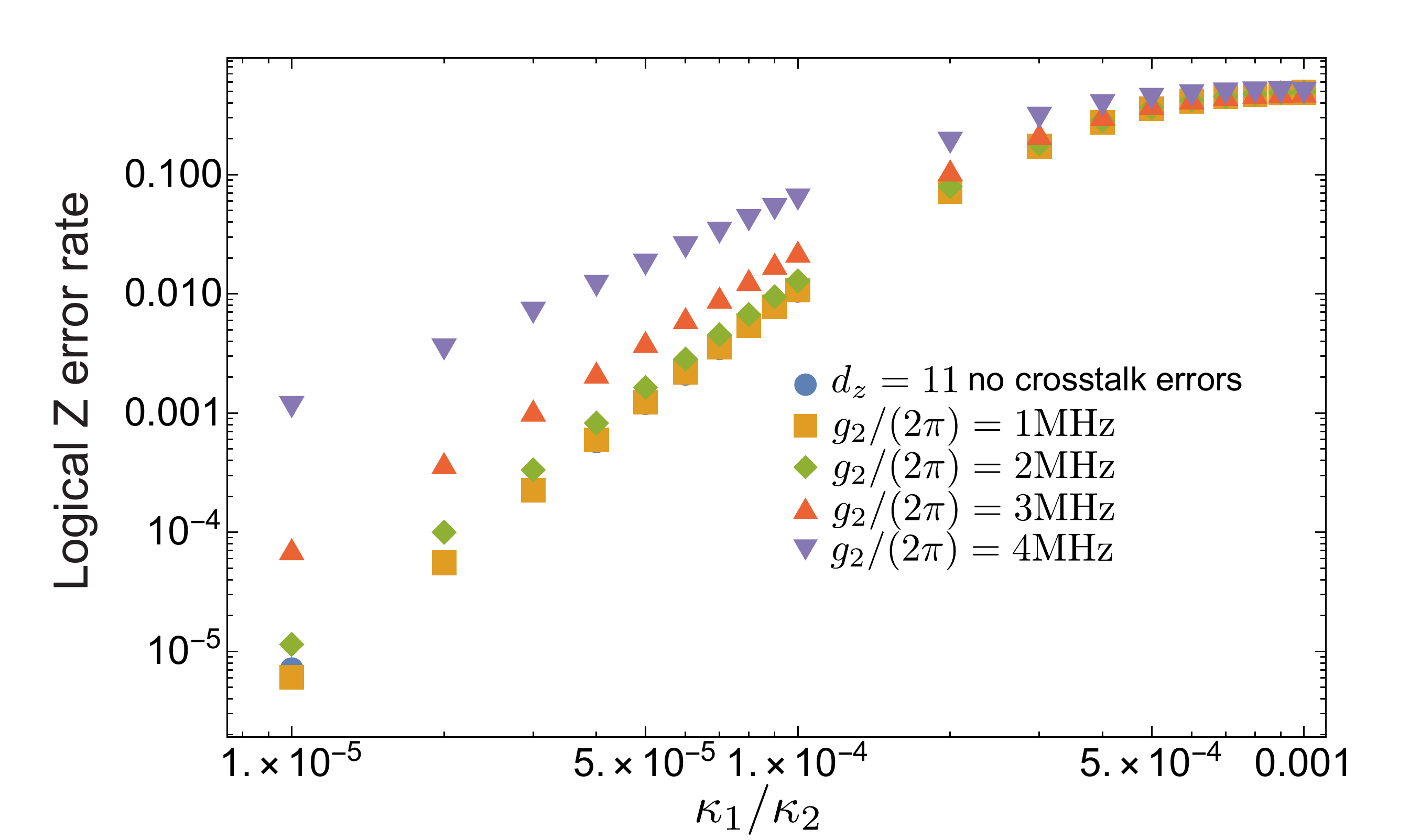}
    \caption{ Logical $Z$ failure rates for a $d_x=3$ and $d_z=11$ thin surface code in the presence of the residual crosstalk errors (given in \cref{eq:CrossPdoubs5Modes}) arising from the coherent micro-oscillations and the  circuit-level noise model of \cref{sec:GatesMeas} 
    The $X$-basis measurement error rates are obtained from \cref{tab:XMeasurementTimeAndErrorRate} with five parity measurements. We compute the logical $Z$ error rates for different values of $g_2$ shown in the legend, and compare such results to the case where the crosstalk errors are not present. 
    }
    \label{fig:CorrelatedNoiseLogZd11}
\end{figure}

Recall that in our architecture proposal, each ATS stabilizes multiple phononic modes. Since the ATS mediates various spurious interactions as well as desired interactions, phononic modes that are connected by the same ATS undergo crosstalk errors. While stochastic crosstalk errors can be strongly suppressed by filtering and careful choice of the frequencies of the phononic modes (see \cref{subsec:mitigation_filtering}), coherent micro-oscillation errors cannot be eliminated by the filters (see \cref{subsec:mitigation_optimization}). In particular, such residual crosstalk errors result in two non-trivial noise processes: every pair of data qubits that are connected by a shared ATS (hence aligned vertically) experiences a correlated $Z$ error with probability $p_{\textrm{double}}$ and every triple of two data qubits and an ancilla qubit that measures an $X$ stabilizer of the surface code experiences a correlated $Z$ error with probability \jp{$p_{\textrm{triple}}$}. In particular, the $Z$ error on the ancilla qubit is realized in the form of a flipped measurement outcome of the corresponding $X$-type stabilizer (see \cref{fig:Fig_surface_code_micro_oscillation_crosstalk_main_text}).

In \cref{subsec:mitigation_optimization}, we optimize the frequencies of the five phononic modes coupled to a shared ATS to minimize $p_{\textrm{double}}$ and \jp{$p_{\textrm{triple}}$}, assuming that the maximum frequency difference between different phononic modes is $2\pi * 1\textrm{GHz}$. With the optimal choice of phononic mode frequencies, we find that the correlated error rates $p_{\textrm{double}}$ and \CC{$p_{\textrm{triple}}$} are given by 
\begin{align}
    p_{\textrm{double}} &= 1.829* 10^{-8}|\alpha|^{8}\Big{(} \frac{g_{2}/(2\pi)}{1\textrm{MHz}} \Big{)}^{4}, 
    \nonumber\\
    p_{\textrm{triple}} &= 5.205* 10^{-10}|\alpha|^{8}\Big{(} \frac{g_{2}/(2\pi)}{1\textrm{MHz}} \Big{)}^{4} . 
    \label{eq:CrossPdoubs5Modes}
\end{align}
Here, $g_{2}$ is the strength of the desired interaction $\hat{a}^{2}\hat{b}^{\dagger}$ needed for the engineered two-phonon dissipation. See \cref{subsec:mitigation_optimization} for more details on why $p_{\textrm{double}}$ and \CC{$p_{\textrm{triple}}$} scale as $g_{2}^{4}$. Note that $p_{\textrm{triple}}$ is $35$ times smaller than $p_{\textrm{double}}$. For $g_{2}/(2\pi) = 1\textrm{MHz}$ and $|\alpha|^{2}=8$, $p_{\textrm{double}}$ is given by $p_{\textrm{double}} = 7.5* 10^{-5}$, which is negligible compared to the total error rate of the physical CNOT gate between two cat qubits. However, since $p_{\textrm{double}}$ scales as $p_{\textrm{double}}\propto g_{2}^{4}$, it increases rapidly as we use larger coupling \CC{strengths}. For instance, $p_{\textrm{double}}$ is given by $p_{\textrm{double}} = 1.2* 10^{-3}$ at $g_{2}/(2\pi) = 2\textrm{MHz}$ and $p_{\textrm{double}} = 1.9* 10^{-2}$ at $g_{2}/(2\pi) = 4\textrm{MHz}$.

In \cref{fig:CorrelatedNoiseLogZd11} we provide logical $Z$ failure rates of the thin surface code under the presence of the crosstalk errors described above for various values of $g_2$.
We note that in the presence of crosstalk errors with probabilities $p_{\textrm{double}}$ and $p_{\textrm{triple}}$ (which are given in \cref{eq:CrossPdoubs5Modes}), extra edges need to be added to the matching graphs of the surface code. Details of the modified graphs in addition to the edge weight calculations are provided in \cref{subsec:CorrErrEdges}. 

As can be seen from \cref{fig:CorrelatedNoiseLogZd11}, when $g_{2}/(2\pi) = 1\text{MHz}$, the effects from crosstalk errors are negligible (the logical error rate curves with and without crosstalk almost perfectly overlap). When $g_{2}/(2\pi) = 2\text{MHz}$, the effects are very small. However, if $g_{2}/(2\pi) \ge 3\text{MHz}$, the difference between logical $Z$ error rates of the surface code with and without crosstalk errors is large enough such that one would need to use larger code distances to achieve the target logical failure rates for the algorithms considered in \cref{sec:Overhead}. Hence, to maintain the overhead results obtained in \cref{sec:Overhead}, it would be preferable to use values of $g_{2}/2\pi \le 2\text{MHz}$, since in that case effects from crosstalk errors are very small. \ch{This bound on $g_2$ imposes a corresponding bound on $\kappa_2 = 4|g_2|^2/\kappa_b$ and hence dictates the maximum tolerable $\kappa_1$ to achieve a given ratio $\kappa_1/\kappa_2$ (see \Cref{tab:regimes}, as well as further discussion in \Cref{subsec:calculation_of_loss}). 
Thus, crosstalk errors are currently a limiting factor for our architecture because suppressing their impacts at the logical level  has required us to demand higher coherence at the physical level.} 

\ch{Ways to further mitigate crosstalk errors therefore represent an important direction for future research on dissipative cat qubits. 
One potential solution to this problem is described in \Cref{App:Alternative}. There, 
we consider an alternative version of our architecture where the limitations imposed by crosstalk errors are less severe. The alternative architecture employs a modified scheme for $X$-basis readout that allows us to reduce the number of modes per unit cell from 5 to 4, which, in turn, reduces crosstalk. With these modifications, we find the effects of crosstalk are negligible for $g_2/2\pi\leq 4$MHz, as opposed to $g_2/2\pi\leq 2$MHz for the five-mode case. This increase in the allowable range in $g_2$ would enable stronger engineered dissipation and hence ease requirements on the storage mode coherence times. See \Cref{App:Alternative} for details.  }

\begin{figure*}[t]
    \centering
    \includegraphics{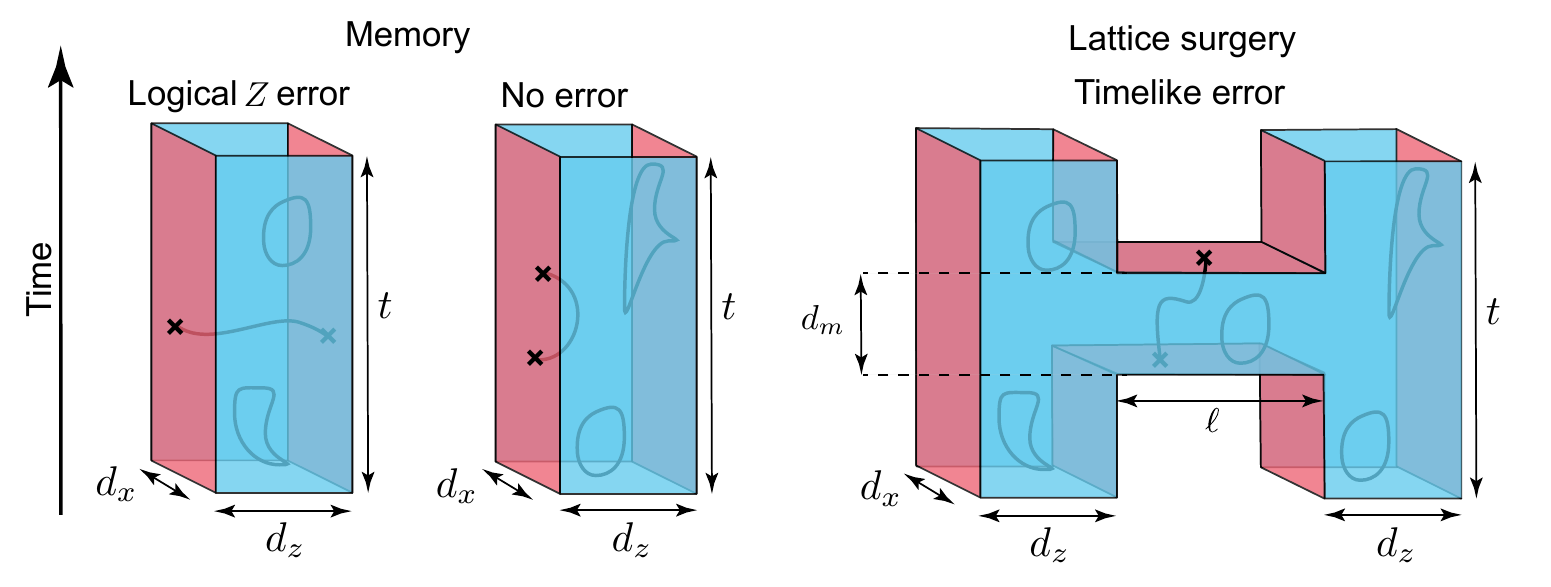}
    \caption{A spacetime diagram of memory and lattice surgery processes using a thin, rotated surface code with boundaries.  Pink (the left and right sides) represents boundaries where $Z$ strings can terminate. Blue (the fore and rear sides) represents boundaries where $X$ strings can terminate.  We show examples of $Z$ strings: when traveling in a spatial direction they represent physical $Z$ errors and when traveling in the vertical time direction, they represent errors on $X$ stabilizer measurements. We only show $Z$ strings that are closed loops or terminate on suitable boundaries. These can be regarded as the final $Z$ strings (after matching) including physical/measurement errors combined with recovery operations.  In the case of memory, a logical $Z$ error occurs whenever a $Z$ string propagates between two topologically disconnected red boundaries.  When performing lattice surgery to measure the $X_{L1} \otimes X_{L2}$ logical operator between two patches, an additional failure mechanism is possible.  If a $Z$ string propagates between two red boundaries disconnected in the time direction then we have a timelike $Z$-error. Computationally, this flips the outcome of the $X_{L1} \otimes X_{L2}$ measurement. Such processes are exponentially suppressed by increasing the measurement distance $d_m$.}
    \label{fig:TimeLikeCartoon}
\end{figure*}

\section{Computation by lattice surgery and timelike errors} \label{Sec:LatticeSurgery}

In both repetition and surface codes, the logical CNOT gate is transversal, which means $\cnot_L = \cnot^{\otimes n}$.  Therefore, a logical CNOT can be fault-tolerantly implemented whenever the hardware supports physical CNOTs between corresponding qubits in the code blocks.  For the repetition code, we can realize a transversal CNOT gate in a 2D layout between two repetition codes. However, for the surface code, a logical CNOT cannot be realized in a way that is both transversal and uses physical CNOT gates in a 2D hardware geometry. A well known solution is to use lattice-surgery between blocks of surface codes~\cite{HFDvM12,landahl14,LVO18,litinski2019game}. The simplest example of lattice surgery realizes a logical $X_{L}\otimes X_{L}$ or $Z_{L}\otimes Z_{L}$ measurement between two surface code patches separated by a distance $\ell$.  The two code blocks are merged into a single code block for $d_m$ rounds of surface code stabilizer measurement and then split apart. We illustrate this in ~\cref{fig:TimeLikeCartoon} with more fine-grained details in \cref{fig:LatticeSurgerySimple} of \cref{subsec:TimeLikeErrors}.  

During lattice surgery, certain types of logical errors can occur resulting in the wrong measurement outcome of multi-qubit logical Pauli operators. We call these timelike errors since in the spacetime picture they correspond to strings of errors in the time direction (see ~\cref{fig:TimeLikeCartoon}). As shown in \cref{subsec:TimeLikeErrors}, such logical failure modes are exponentially suppressed by increasing $d_m$, which comes at the price of increasing the execution time for this logical operation. A seemingly natural choice is to set $d_m=d_x=d_z$, but since our noise model is highly biased this leads to an asymmetry in the optimal choices. We discuss timelike errors in more detail in \cref{subsec:TimeLikeErrors} and present simulation results showing that for our noise model, the rate of timelike errors is comparable (even slightly lower) than logical $Z$ error rates. A detailed decoding scheme used for such simulations is described in \cref{subsec:DecodeTimeLikeErrors}.

Lattice surgery measurements combined with logical $\ket{0}$ and $\ket{+}$ preparations, and logical single-qubit $X$ and $Z$ measurements, can be used to perform logical $\cnot$, Hadamard and CZ gates~\cite{HFDvM12}.  Furthermore, the two codeblock lattice surgery sketched in \cref{fig:TimeLikeCartoon} can be generalized to act on multiple codeblocks to enable measurements of any tensor product of \jp{logical} $Z$ and $X$ operators. By making use of lattice twists, \etc{domain walls and lattice deformations}, any logical multi-qubit Pauli operator can be measured by lattice surgery~\cite{Litinski18}. 

\begin{figure}[th]
	\centering
	\includegraphics{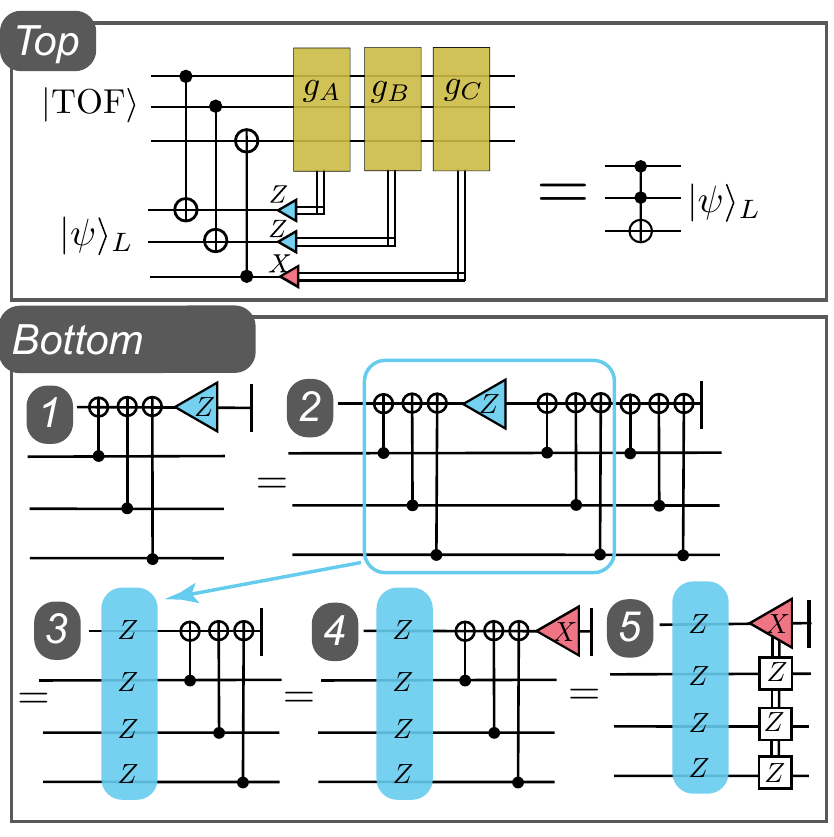}
	\caption{(\textit{Top}): \etc{Toffoli gate injection that implements a Toffoli gate by consuming a $\ket{\text{TOF}}$ resource state}. The Clifford corrections depend on the three measurement outcomes and are given in \cref{eq:g1,eq:g2,eq:g3}. All qubits and gates are implemented at the logical level. (\textit{Bottom}):
	\etc{These circuit identities illustrate how to convert a teleportation circuit into a Pauli-based computation. We present 
	five} equivalent circuits showing how to convert from a $\cnot$ followed by measurements with $m$ CNOT gates into a Pauli-based computation that can be realized by lattice surgery. Circuit 1 to 2: we insert the identity.  Circuit 2 to 3: we have replaced the highlighted box with a multi-qubit $Z^{\otimes m+1}$ measurement. Circuit 3 to 4: we add a single qubit $X$ measurement before we discard the qubits.  Circuit 4 to 5: we use the $X$ measurement to replace the $\cnot$ gates with classically controlled $Z$ gates.  A similar identity holds with the CNOT direction reversed and the roles of $X$ and $Z$ interchanged. Applying the identities of the bottom figure in the $m=1$ case to the top figure yields a Pauli-based Toffoli gate injection procedure. We make use of the the $m>1$ case in \cref{Sec:TopDown,app:GenCCinject}.}
	\label{fig:TOFCircuitInject}
\end{figure}

However, all these operations are either Clifford group gates or Pauli measurements, so some non-Clifford operation is required to complete a universal gate set. The model of Pauli-based computation~\cite{bravyi2016trading} shows that it is possible to perform universal quantum computation using just \jp{multi-qubit} \etc{Pauli measurements and access to suitable magic states and performing \textit{gate teleportation}. We denote the magic state for a Toffoli gate injection as}
\begin{align}
\ket{\text{TOF}} = \frac{1}{2} \sum_{a,b \in \mathbb{F}_2} \ket{a}\ket{b}\ket{a \wedge b},
\label{eq:TOFstate}
\end{align}
where $a \wedge b$ is the AND of bits $a$ and $b$. The $\ket{\text{TOF}}$ state is stabilized by the Abelian group $\mathcal{S}_{\text{TOF}} = \langle g_A, g_B, g_C  \rangle$ 
where 
\begin{align} \label{eq:g1}
g_A & = X_A \text{CNOT}_{B,C}, \\ \label{eq:g2}
g_B & = X_B \text{CNOT}_{A,C}, \\ \label{eq:g3}
g_C & = Z_C CZ_{A,B}.  
\end{align}
To simplify the notation used in \cref{Sec:TopDown}, we label the three qubits involved in a Toffoli gate by $A$, $B$ and $C$ instead of 1, 2 and 3. Given one copy of a $\ket{\text{TOF}}$ state, \etc{Toffoli gate teleportation} is performed using the circuit in \cref{fig:TOFCircuitInject} to realize a logical Toffoli gate. Notice that the circuit requires a Clifford correction $g_A^a g_B^b g_C^b$  for the binary measurement outcome $(a,b,c)$ of the single qubit Pauli measurements.

In a purely Pauli-based computation, rather than using lattice surgery to simulate the \cnot circuit for magic state injection, the CNOTs can be completely eliminated using the circuit identities shown in \cref{fig:TOFCircuitInject}. Furthermore, the Clifford corrections and Clifford gates in an algorithm do not necessarily need to be performed.  Rather we can keep a record of the accumulated Clifford gates so far into a Clifford \textit{frame} (see for instance Ref.~\cite{CIP17}). When we need to measure a Pauli $P$, we instead measure the Pauli $CPC^\dagger$ whenever the Clifford frame records $C$.  In such a Pauli-based computational model, Clifford gates do not contribute to an algorithms runtime.  Rather the runtime is determined by two factors: how fast we can prepare high fidelity $\tof$ states; and how fast they can be \etc{teleported} into the algorithm. The rate of \etc{gate teleportation} depends on how much routing space between qubits is budgeted for in the device. Using a fast data access structure~\cite{litinski2019game}, it is known that lattice surgery can perform a single arbitrary multi-qubit Pauli operator with approximately $\sim 2\times$ overhead in routing costs. Such a space overhead cost is pessimistic since not all qubits need to be involved in every lattice surgery operation, so considerable compression is possible. Ref.~\cite{paler2019opensurgery} assumed a $\sim 1.5\times$ overhead suffices and Refs.~\cite{FMMC12,Ogorman17,kivlichan2020improved} assumed this cost could be made negligible, so $\sim 1 \times$.  \etc{Furthermore, it has been shown that for biased noise architecture the routing overhead is lower than for depolarizing noise~\cite{chamberland2021universal}. In our later analysis of overheads, we assume a $\sim 1.3\times$ routing overhead cost (roughly midway between the commonly used $\sim 1.5\times$ and $\sim 1\times$ overheads) suffices to maintain this pace of \etc{teleportation}. Work in preparation will justify this routing overhead more rigorously.}  

One can also inject at a considerably faster pace than 
sequentially injecting magic states, up-to the limit of time-optimal quantum computation~\cite{fowler2012time}, though this approach incurs significantly higher routing overhead costs and is not practical for modest size quantum computers.  In the next two sections, we consider the pace and fidelity with which we can prepare \tof magic states. In what follows, we use $\ket{\text{TOF}}$ and \tof interchangeably when refering to the state in \cref{eq:TOFstate}.

\section{Toffoli distillation: Bottom-up scheme}
\label{sec:BottomUp}

\begin{figure*}[thp]
	\centering
	\includegraphics[width=0.8\textwidth]{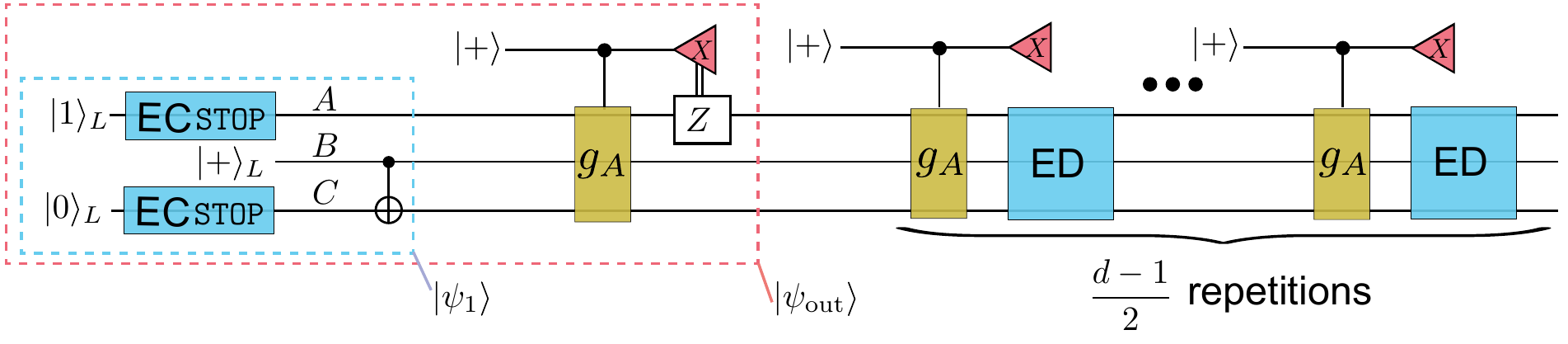}
	\caption{Circuit for our entire \BUTOFF protocol. The first step (shown in the dashed blue box) consists of preparing the state $\ket{\psi_1} = \frac{1}{\sqrt{2}}((\ket{100} + \ket{111})$. The preparation of the states $\ket{0}_L$ and $\ket{1}_L$ are described in \cref{subsec:compbasis}. The next step consists of measuring $g_A = X_A \text{CNOT}_{B,C}$. If the measurement outcome on the ancilla is $-1$, a $Z_A$ correction is applied to the output state. Note that at this stage, error correction is not applied to the data block. The first two steps are enclosed within the dashed red box. We label the output state of the first two steps as $\ket{\psi_{\text{out}}}$. Lastly, the measurement of $g_A$ is repeated $(d-1)/2$ times for a distance $d$ repetition code. The ED blocks correspond to one round of stabilizer measurements of the repetition code. If any of the measurement outcomes of ED or ancillas are non-trivial, the protocol is aborted and begins anew. }
	\label{fig:RepG2Meas}
\end{figure*}

In magic state distillation schemes, the goal is to distill magic states with circuits that require only stabilizer operations \cite{BraKit05,Bravyi12,Meier13}. The circuits used to distill such magic states are typically not fault-tolerant to all Clifford gate errors and thus must be implemented using a sufficiently large error-correcting code. Recently, with the advent of flag qubits and redundant ancilla encoding, scalable approaches to fault-tolerantly preparing magic states have been devised such that all stabilizer operations can be implemented directly at the physical level \cite{ChamberlandMagic,CNmagic20}. We refer to such methods as a bottom-up approach to preparing magic states. 

In this section, we provide a protocol to fault-tolerantly prepare \tof magic states encoded in the repetition code using a bottom-up approach (herein \BUTOFF). In \cref{Sec:TopDown}, we show how the scheme presented in this section can be supplemented by using a top-down approach to prepare \tof states with the very high fidelities required to implement the algorithms considered in \cref{sec:Overhead}. 

\begin{figure*}[th]
	\centering
	\subfloat[\label{fig:BottomUpTOF4ancillaATS}]{%
		\includegraphics[width=0.55\textwidth]{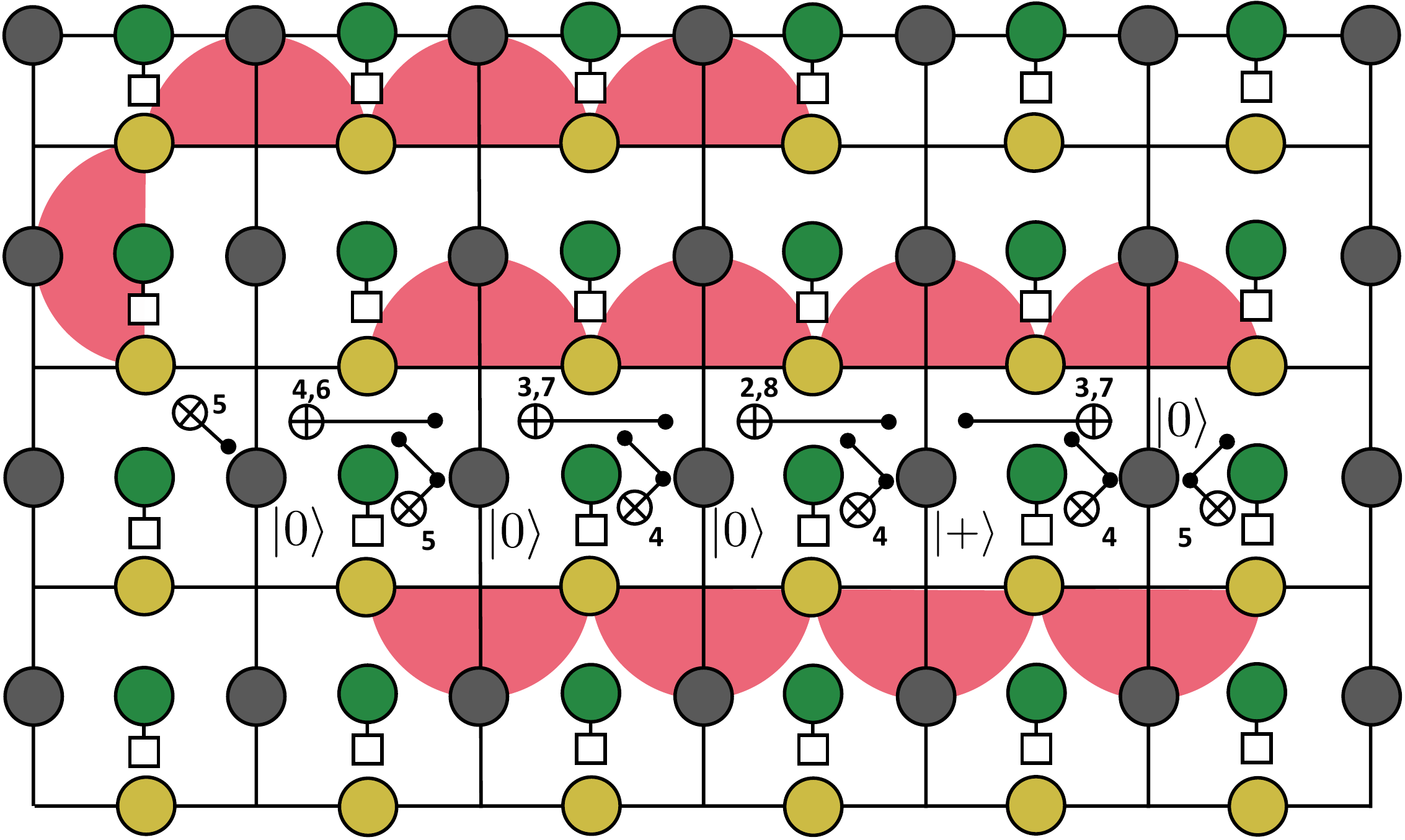}
	}
	\subfloat[\label{fig:TOFprepCircuitBottomUp}]{%
		\includegraphics[width=0.25\textwidth]{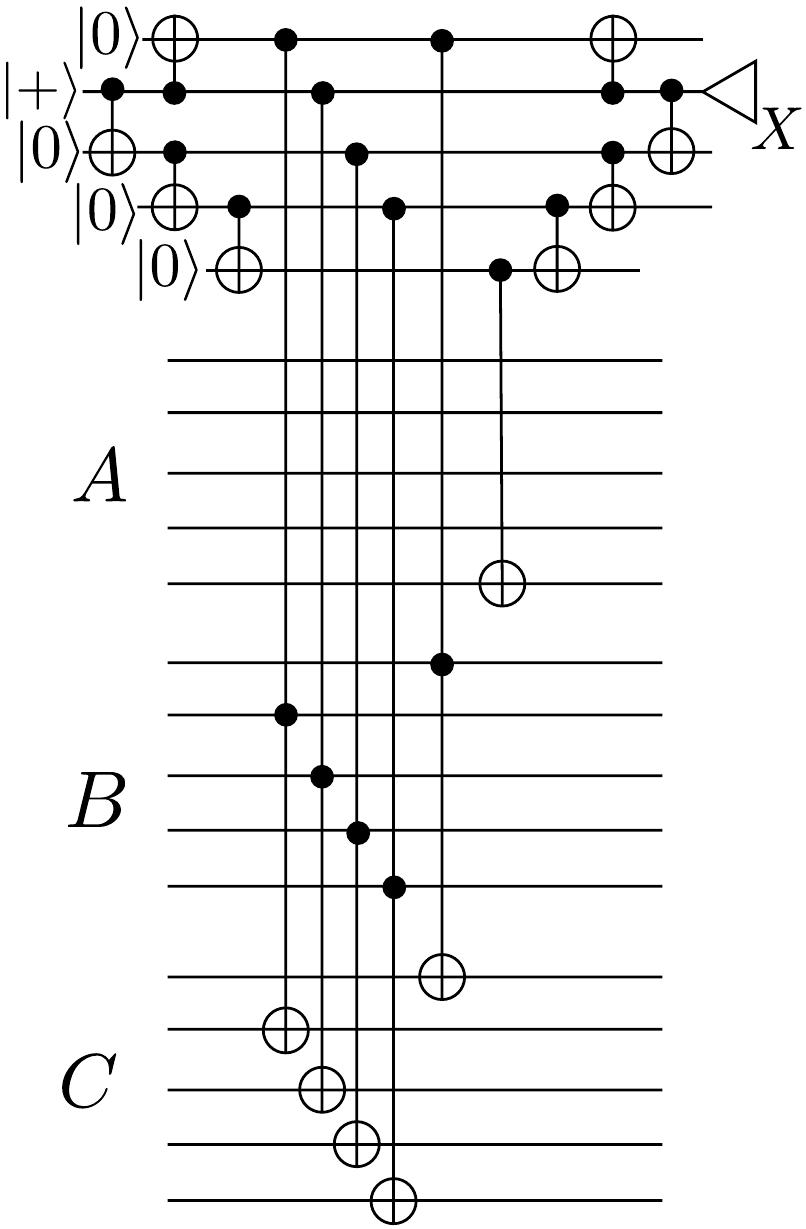}
	}

	\caption{\label{fig:BottomUpTOFATS}(a) Implementation of the $g_A$ measurement (for a distance $d=5$ repetition code) compatible with our ATS layout and lattice surgery implementation for universal quantum computation described in \cref{Sec:TopDown}. All operations are performed respecting the connectivity constraints of the ATS's and use the fewest possible ancilla qubits for preparing the GHZ state necessary for the fault-tolerant measurement of $g_A$. (b) Equivalent circuit for the implementation of (a). }
\end{figure*}

We now describe how to fault-tolerantly prepare the $\ket{\text{TOF}}$ state. First, note that the state $\ket{\psi_1} = \frac{1}{\sqrt{2}} (\ket{100} + \ket{111})$ is stabilized by $g_B$ and $g_C$. Such a state can straightforwardly be prepared using the circuit in the dashed blue box of \cref{fig:RepG2Meas}. In what follows, physical Toffoli gates will need to be applied between ancilla qubits and $\ket{\psi_1}$ prior to measuring the data. As such it is very important that the states $\ket{0}_L$ and $\ket{1}_L$ in the circuit of \cref{fig:RepG2Meas} (which are encoded in the repetition code) be prepared using the \texttt{STOP} algorithm since otherwise measurement errors in the last ancilla measurement round could lead to logical failures \footnote{Recently, it was also shown in Ref.~\cite{GM2020} that when implementing a logical Toffoli gate using a piece-wise fault-tolerant approach, one can track the $CZ$ errors that arise when $Z$ errors propagate through the target qubits of the physical Toffoli gates (see \cref{sec:BottomUpSimMethod}) and correct all errors at the final output of piece-wise circuit (instead of in between each blocks of physical Toffoli gates). Such an approach could potentially be used in our bottom-up $\ket{\text{TOF}}$ state preparation scheme, allowing us to avoid using the \texttt{STOP} algorithm to prepare $\ket{0}_L$ and $\ket{1}_L$. However we leave such an analysis to future work.}. An alternative to avoid using the \texttt{STOP} algorithm would be to prepare $\ket{0}_L$ and $\ket{1}_L$ using post selection. However, such an approach would reduce the acceptance probability of our scheme (see below) thus increasing its space-time overhead cost. Once $\ket{+}_L = \ket{+}^{\otimes n}$, $\ket{1}_L$ and $\ket{0}_L$ have been prepared, the CNOT gate in the dashed blue box of \cref{fig:RepG2Meas} is applied transversally. 

\begin{figure}
	\centering
	\subfloat[\label{fig:LogicalsZBUTOF}]{%
		\includegraphics[width=0.48\textwidth]{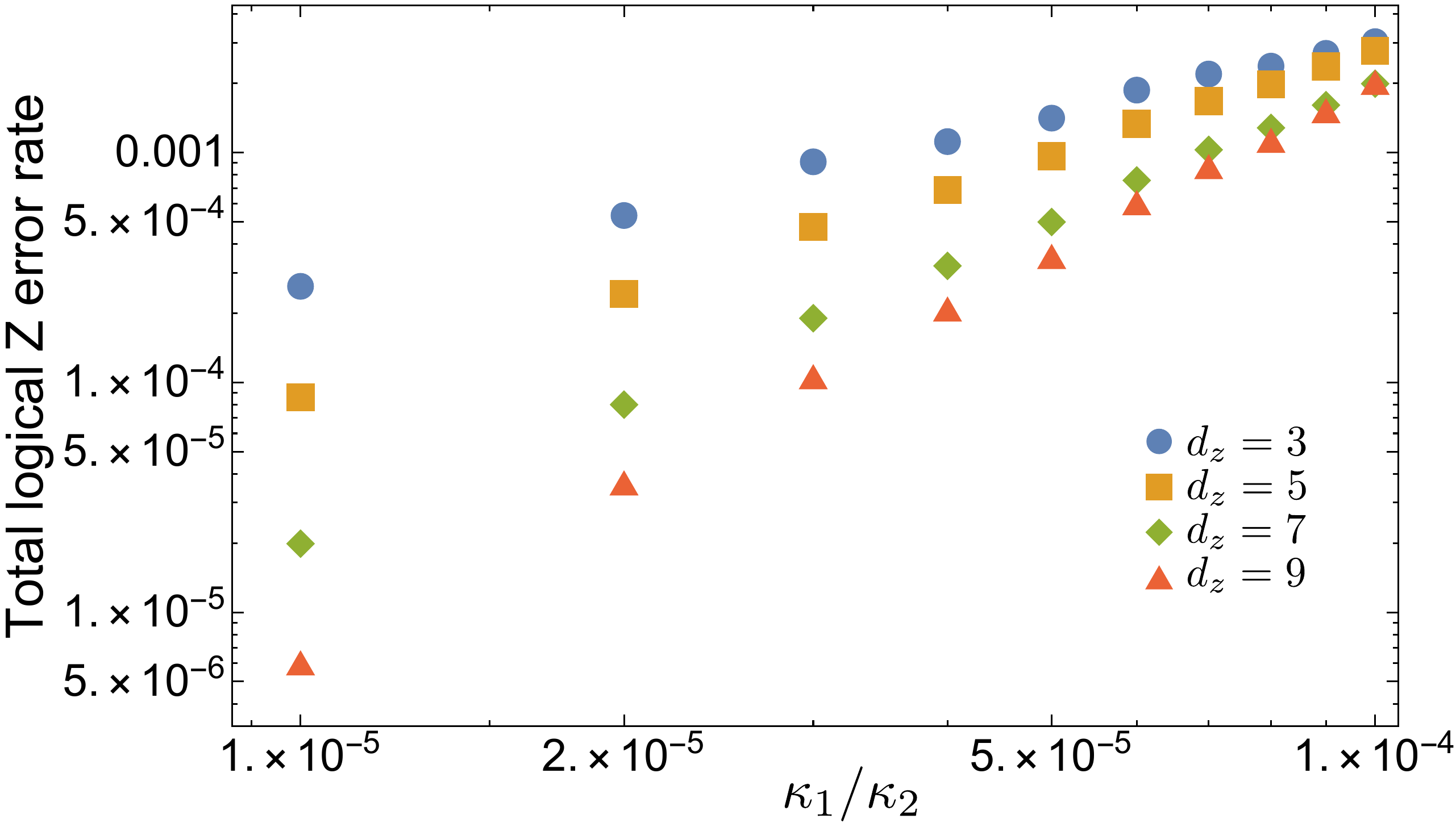}
	}
	\vfill
	\subfloat[\label{fig:AcceptanceProbsBUTOF}]{%
		\includegraphics[width=0.48\textwidth]{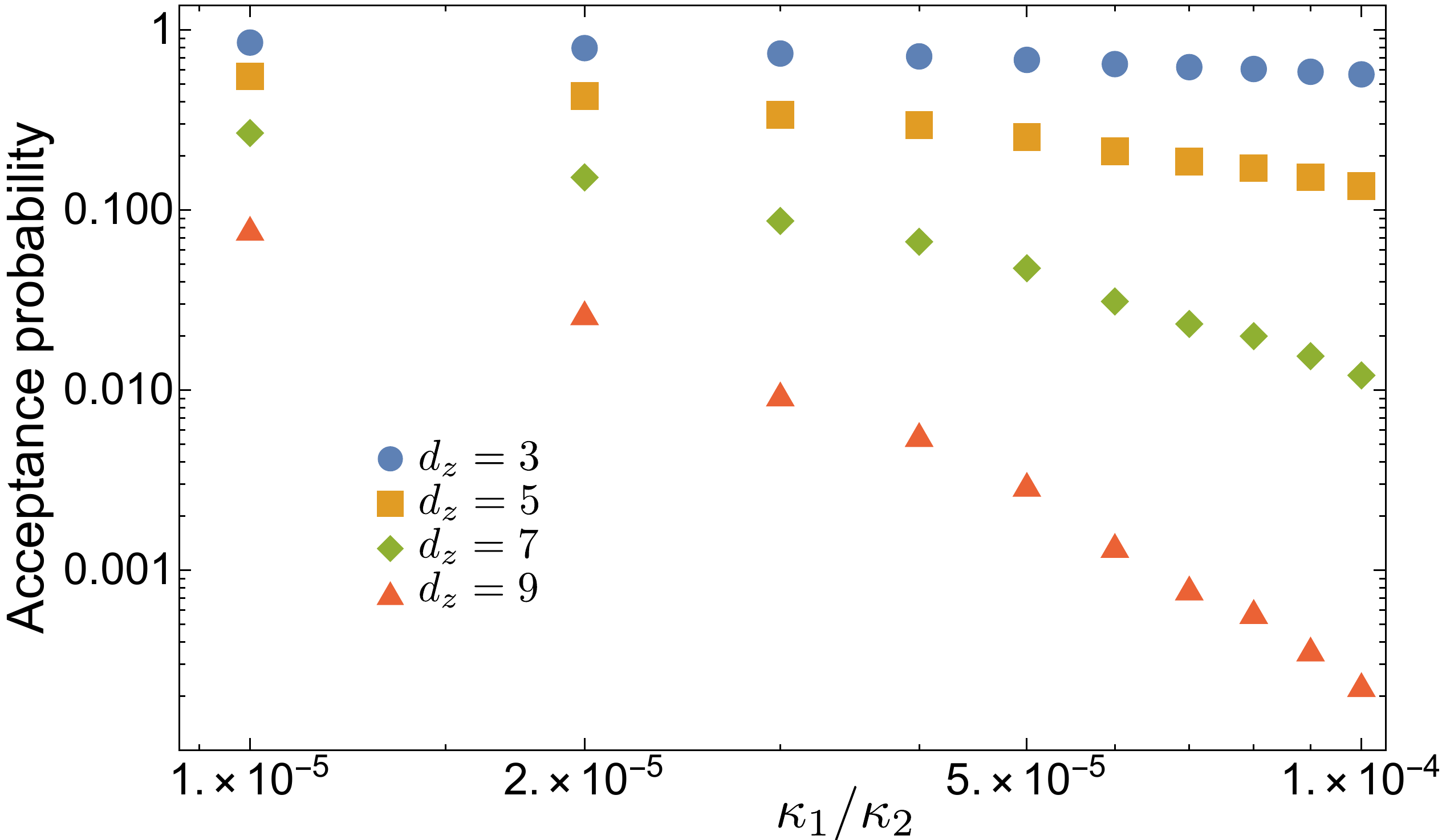}
	}
	\vfill
		\subfloat[\label{fig:DifferentZsBUTOF}]{%
		\includegraphics[width=0.48\textwidth]{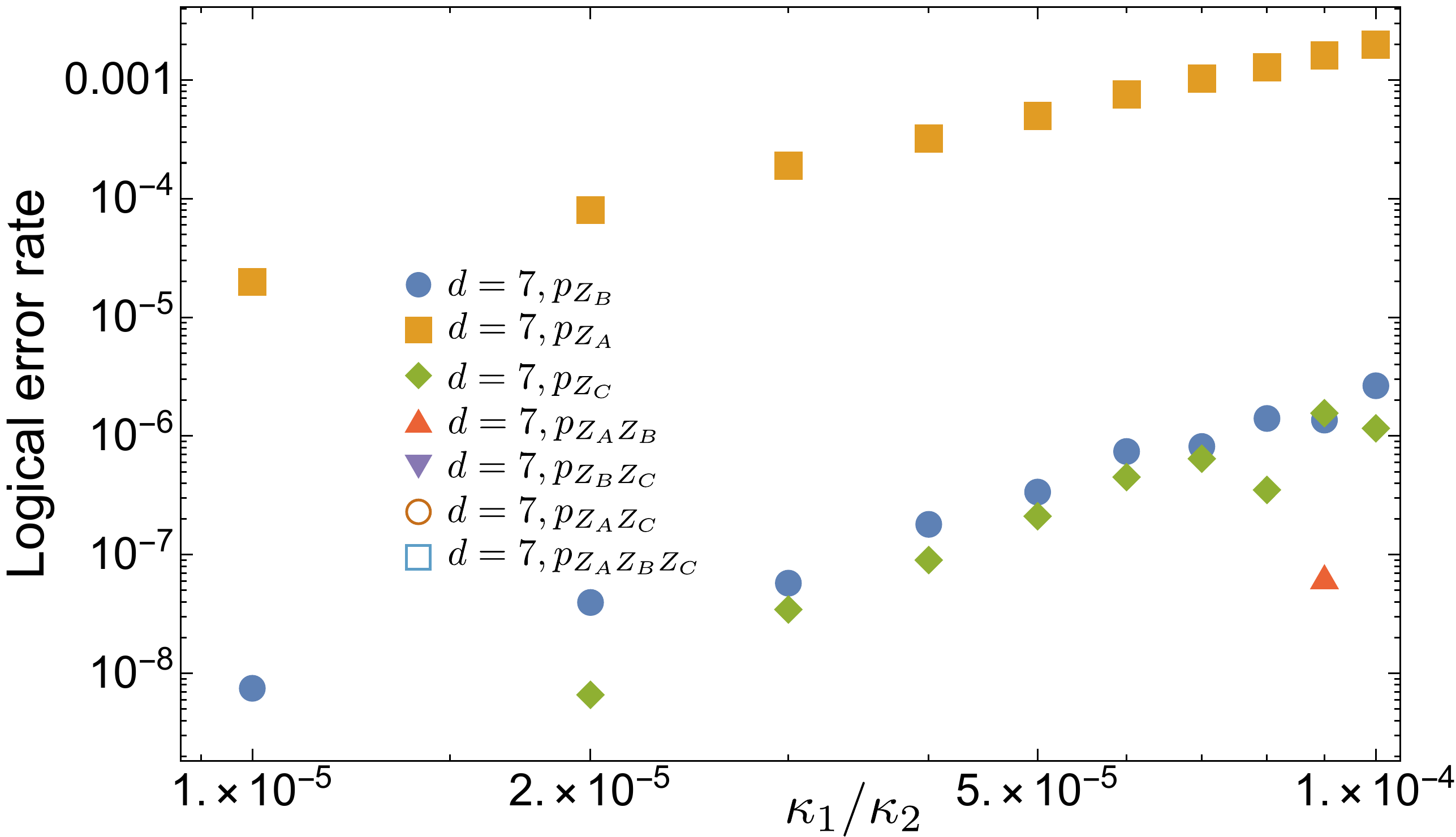}
	}

	\caption{\label{fig:PlotsBUTOF}(a) Total logical $Z$ failure rate for preparing a $\ket{\text{TOF}}$ state using the fault-tolerant \BUTOFF protocol described in this section. (b) Acceptance probabilities for preparing $\ket{\text{TOF}}$ states using the fault-tolerant protocols described in this section. (c) Decomposition of the logical $Z$ errors for a $d=7$ $\ket{\text{TOF}}$ state prepared using the fault-tolerant protocol described in this section. As can be seen, from all seven possible combinations of logical $Z$ errors, a logical $Z$ error on block $A$ is more likely by several orders of magnitude. All numerical simulations were performed by setting $n_{th} = 0$, $\kappa_{\phi} = 0$ and using the circuit level noise model described in \cref{sec:GatesMeas} \CC{with $|\alpha^2| = 8$}. }
\end{figure}

Now, given a copy of $\ket{\psi_1}$, we can prepare $\ket{\text{TOF}}$  by measuring $g_A$ using the circuit in the dashed red box of \cref{fig:RepG2Meas} resulting in the state $\ket{\psi}_{\text{out}}$. If the measurement outcome is $+1$, then $\ket{\psi}_{\text{out}} = \ket{\text{TOF}}$, and if it is $-1$, then $\ket{\psi}_{\text{out}} = Z_A\ket{\text{TOF}}$. Hence we apply a $Z_A$ correction given a $-1$ measurement outcome. Note that neither error detection nor error correction is applied to any of the data blocks at this stage. The reason is that it is not necessary for ensuring the fault-tolerance of our scheme. Further, we found numerically that adding error correction at this stage results in higher logical failure rates when preparing $\ket{\text{TOF}}$. Furthermore, adding unnecessary error detection units would lower the acceptance probability of our scheme. We provide a more detailed implementation of the controlled-$g_A$ gate in \cref{fig:BottomUpTOFATS} below.

A measurement error on the ancilla results in a logical $Z_A$ failure and so the measurement of $g_A$ needs to be repeated (similar repetitions are needed for the preparation of logical computational basis states, see \cref{appendix:StabOps}). This can be done using the \texttt{STOP} algorithm. However, due to the increasing circuit depth with increasing repetition code distance in addition to the high cost of the controlled-$g_A$ gate, such a scheme does not have a threshold and results in relatively high logical failure rates. As in Refs.~\cite{ChamberlandMagic,CNmagic20}, an alternative approach is to use an error detection scheme by repeating the measurement of $g_A$ exactly $(d-1)/2$ times for a distance $d$ repetition code. In between each measurement of $g_A$, one round of error detection is applied to the data qubits by measuring the stabilizers of the repetition code (see \cref{fig:RepG2Meas}). If any of the measurement outcomes are non-trivial, the \BUTOFF protocol is aborted and reinitialized.  In \cref{fig:BottomUpTOF4ancillaATS}, we provide an example of the two-dimensional layout and sequence of operations for measuring $g_A$ which is compatible with our ATS architecture for a distance-5 repetition code. To realize the protocol with local operations, we replace the $\ket{+}$ ancilla in \cref{fig:RepG2Meas} with 5 qubits that we prepare in a GHZ state. Subsequently, the required Toffoli and CNOT gates are applied, followed by a disentangling of the GHZ states and measurement of the $\ket{+}$ state ancilla. The equivalent circuit implementing the $g_A$ measurement for a $d=5$ repetition code is shown in \cref{fig:TOFprepCircuitBottomUp}.

As a remark, we point out that in general, it is possible to use one fewer ancilla in the circuit of \cref{fig:BottomUpTOF4ancillaATS} with a lattice that is no longer translationally invariant with respect to yellow and gray vertices. However, such a layout could not straightforwardly be used with our lattice surgery implementation of \cref{app:TopDown}.

In \cref{fig:LogicalsZBUTOF}, we provide the total $Z$ failure probability of our \BUTOFF protocol for various repetition code distances ranging from $d = 3$ to $d=9$. We note that given the increasing circuit depth of \BUTOFF with the repetition code distance $d$, our scheme does not have a threshold even though it is fault-tolerant. Further, as can be seen from \cref{fig:AcceptanceProbsBUTOF}, the acceptance probability for preparing such states (i.e. the probability that all measurement outcomes in \cref{fig:RepG2Meas} are trivial) decreases exponentially with increasing code distances. Hence, large repetition code distances should be avoided. However in \CC{\REGthree where $\kappa_1 / \kappa_2 \approx 10^{-5}$}, we can still obtain $\ket{\text{TOF}}$ states with total failure probabilities on the order of $6 * 10^{-6}$, which is orders of magnitude better than the failure probabilities that would be obtained by preparing $\ket{\text{TOF}}$ states using non-fault-tolerant methods. This drastically reduces the overhead requirements of the top-down approach of \cref{Sec:TopDown}. Also, as can be seen from \cref{fig:DifferentZsBUTOF}, logical $Z$ errors are highly concentrated on block $A$. The reason is that while the error detection units on each block can detect up to $d-1$ physical $Z$ errors, $(d-1)/2$ measurement errors on the GHZ ancilla will lead to a logical $Z$ error on block $A$. 

We note that the GHZ circuit in \cref{fig:TOFprepCircuitBottomUp}, which is used to measure $g_A$, is not fault-tolerant to $X$ or $Y$ errors \footnote{This should be compared to the circuits used in \cite{CNmagic20} for preparing $\ket{H}$ type magic states which are fault-tolerant to all types of Pauli noise given that a depolarizing circuit level noise model was assumed.}. However, since we are assuming that $X$ and $Y$ errors are exponentially suppressed, flag qubits for detecting $X$-type error propagation are unnecessary as long as $X$ or $Y$ error rates multiplied by the total number of fault locations are below the target levels for algorithms of interest. Indeed as is shown in \cref{Sec:TopDown,sec:Overhead} and for the parameters chosen in this work, $X$ error rates are low enough such that the desired failure rates can be achieved for implementing the quantum algorithms with over a million Toffoli gates (see \cref{tab:HubbardModelCossts}).

\CC{We also remark that all simulations for our \BUTOFF protocol were performed by setting $|\alpha^2|=8$. As can be seen in \cref{fig:TotalRepetitionCodeFail}, when setting $|\alpha^2|=6$, the total logical failure probability of a repetition code strip is roughly two orders of magnitude higher than the $|\alpha^2|=8$ results. Given such features and the fact that the GHZ circuit used to measured $g_A$ is not fault-tolerant to bit-flip errors, we thus require that $|\alpha^2| \ge 8$. Smaller values of $|\alpha^2|$ would not be suitable for implementing our \BUTOFF protocol with only one round of magic state distillation (described in \cref{Sec:TopDown}) since bit-flip failure probabilities would be too high. Therefore in settings where $|\alpha^2| \le 6$, a different approach would be needed to prepare high fidelity $\ket{\text{TOF}}$ magic states.}

Lastly, we note that simulating the circuit in \cref{fig:BottomUpTOFATS} can be challenging given the presence of physical Toffoli gates. In \cref{sec:BottomUpSimMethod}, we provide a method for performing a near exact simulation of such circuits (the simulation is exact if there are less than $d$ $Z$-type errors on block $C$ prior to applying the physical Toffoli gates). Also, when using the \texttt{STOP} algorithm to simulate the preparation of $\ket{0}_L$ and $\ket{1}_L$ prior to applying the physical Toffoli gates, we do not add one round of perfect error correction (since projecting to the codespace is not necessary at this stage). Residual errors at the output of the preparation of $\ket{0}_L$ and $\ket{1}_L$ using the \texttt{STOP} algorithm are propagated to the next stage of the protocol.

\section{Top-Down Scheme for higher fidelity Toffoli gates }
\label{Sec:TopDown}

\etc{
Previous sections established that surface codes and \REGthree are necessary to reach logical memory error rates needed for large scale quantum computation.  \cref{Sec:LatticeSurgery} reviewed lattice surgery and how it provided reliable Clifford operations through the Pauli-based model of computation. For large-scale, universal quantum computation we also need a very high fidelity non-Clifford gate, such as the Toffoli gate.  At the cat qubit level,  we proposed using an adiabatic bias-preserving Toffoli gate (see \cref{ToffoliGate}).  We then proposed using the adiabatic Toffoli in the \BUTOFF protocol to prepare $\tof$  states (see \cref{sec:BottomUp}). The lowest infidelity we reported for the \BUTOFF protocol was $6 * 10^{-6}$, which is insufficient for quantum  algorithms using over a million Toffoli gates.
}

\etc{This section completes our proposal for performing a very high fidelity Toffoli gate via the preparation of $\tof$ states. We propose a magic-state distillation protocol that utilizes the output of \BUTOFF.  We use thin surface code qubits wherever a potential bit-flip would lead to an error on the output $\tof$ state. We call this the top-down Toffoli protocol (herein \TDTOFF) because it assumes access to high-fidelity \jp{logical} Clifford gates, so we are attacking the problem with a view from the top of the stack.  Later, in \cref{sec:Overhead}, we show that using 1 round of \BUTOFF  concatenated with 1 round of \TDTOFF, achieves high enough fidelities to implement some quantum algorithms of interest.}

\etc{This section makes three significant contributions to the theory of magic state distillation.  Our first main contribution is a new set of quantum error correction codes that can be used for $\tof$-to-$\tof$ magic state distillation more efficiently than all previous proposals.  In the main text, we describe only the specific codes used in \TDTOFF, but \cref{sec:TD_Transversality} presents a general framework for code design and $\tof$ state distillation, of which \TDTOFF is just one example.  Our second main contribution is the idea of using Clifford symmetries of magic states to perform noise tailoring that enables us to exploit noise bias. Again, \cref{App:Noisetailoring} explores Clifford symmetries at a general level, and here we discuss the  consequences to \TDTOFF.  In summary, at high noise bias, exploiting Clifford symmetries enables cubic error reduction instead of quadratic error suppression.  Our third main contribution is numerical, we calculate the performance of \TDTOFF in our proposed hardware using \REGthree noise parameters.}

\begin{figure}
    \centering
    \includegraphics{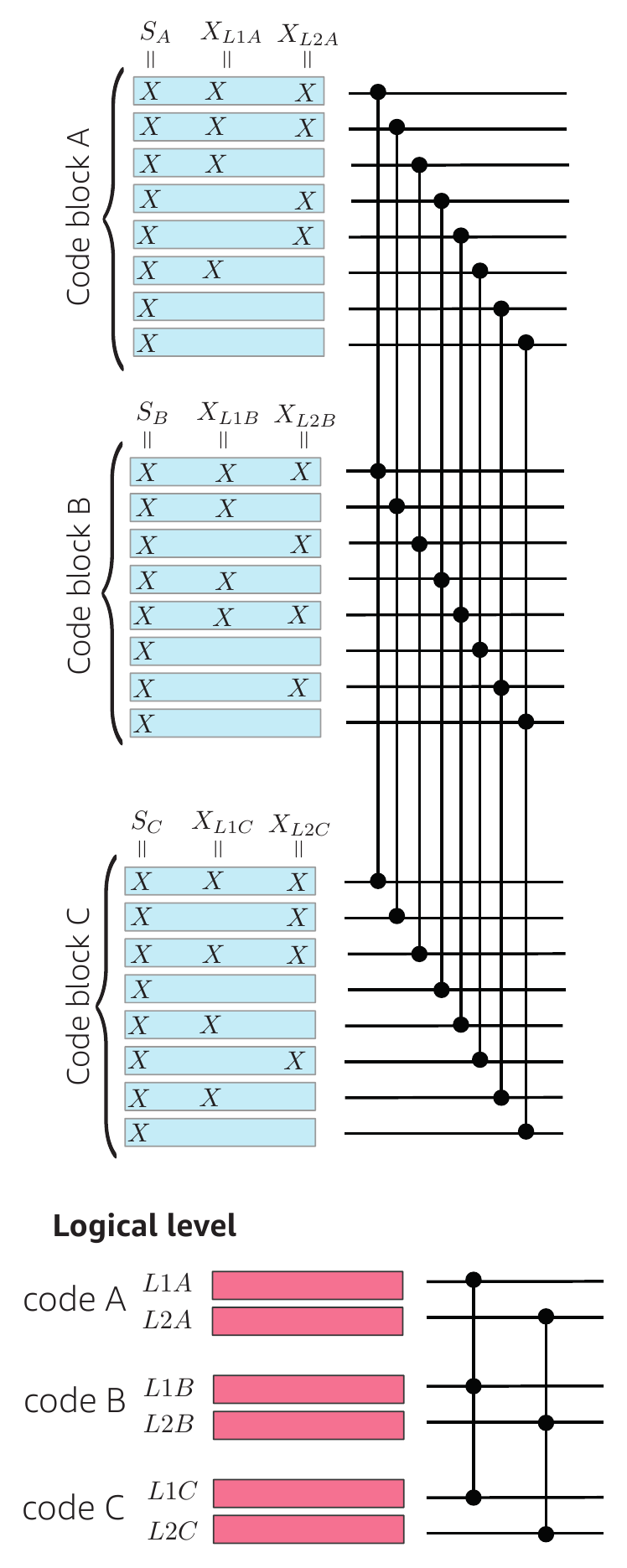}
    \caption{Our \TDTOFF protocol uses a trio of quantum error correction codes each using $n=8$ qubits to encode $k=2$ logical qubits.  The top circuit shows the $3n=16$ qubits and labels the $X$ stabilisers and logical operators for these codes.  Each of the $3k=6$ qubits is actually encoded again into a thin surface code. The bottom circuit shows the $3k$ logical qubits encoded by this code. The crucial and exotic property of these codes is that they have a transversal a CCZ gate: applying the 8 CCZ gates as shown in the top circuit, results in the 2 logical CCZ gates shown in the bottom circuit. See ~\cref{sec:TD_Transversality} for proof details.}
    \label{fig:Transversal}
\end{figure}

The Toffoli gate is equivalent, \jp{up to conjugation of the target qubit by a Hadamard transformation}, to the \jp{controlled-controlled-$Z$} ($\ccz$) gate, and for technical reasons \jp{we prefer to work with} $\ccz$ gates.  Throughout this section we work at an encoded level, so whenever we say qubit, we mean surface code qubit. Our starting point for design of \TDTOFF is identifying a trio of $n$-qubit codes each encoding $k$ logical qubits, which have a transversal, logical $\ccz$ gate.  By transversal, we mean that if $\ccz_j$ denotes a  $\ccz$ gate 
\jp{acting on the $j$th qubit in each of three code blocks},
then $\ccz^{\otimes n}=\prod_{j=1}^n \ccz_j$ realizes the logical $\ccz^{\otimes k}$ \jp{acting on} $3k$ logical qubits. The error correction codes we use for \TDTOFF are shown in ~\cref{fig:Transversal} and each block uses $n=8$ to encode $k=2$ logical qubits.  \cref{fig:Transversal} also illustrates what we mean by $\ccz$ transversality, with the transversality proof postponed until \cref{sec:TD_Transversality}.

Given such a code, a standard recipe for magic state distillation protocols goes as follows~\cite{Bravyi05}: prepare $\ket{+}_L^{\otimes 3k}$ encoded in the relevant codes; perform imperfect $\ccz^{\otimes n}$ by gate teleportation using noisy $\tof$ states; measure the $X$-stabilizers and post-select on ``+1" outcomes; and decode.  This would require $3n=24$ qubits plus workspace for Cliffords and routing.  However, one can make a space-time tradeoff~\cite{haah2018codes,litinski2019game,litinski2019magic} so that the full $24$-qubit code is never prepared; rather we work with $9$ qubits that we herein call the \textit{factory qubits}.  We label these 9 factory qubits with $(j,D)$ where  $D \in \{ A, B, C \}$ denotes the codeblock and $j \in \{ 1,2,3\}$ specifies the qubit within the codeblock. To achieve the space-time tradeoff, we can define an encoding Clifford $V$ such that for $D \in \{ A, B, C \}$ we have
\begin{align}
    V X_{1,D} V^\dagger & = X_{L1 D} ,\\ \nonumber
    V X_{2,D} V^\dagger & = X_{L2 D} ,\\ \nonumber
    V X_{3,D} V^\dagger & = (X^{\otimes 8})_D .
\end{align}
\jp{The logical operators  $X_{L1D}$ and $X_{L2D}$} are shown in \cref{fig:Transversal}.  Instead of encoding $\ket{+}_L^{\otimes 3k}$ and performing $\ccz^{\otimes n}$, we prepare  $\ket{+}^{\otimes 9}$ and perform $ V( \ccz^{\otimes n}) V^\dagger$.  At the end of the protocol, instead of measuring the $X$-stabilizers we need only measure the 3 check qubits labelled $X_{3,D}$.  

It is important that $ V( \ccz^{\otimes n}) V^\dagger $ acts non-trivially on only the 9 qubits identified and error correction properties of the protocol are unaffected (see App.~\ref{App:SpaceTimeTrade} or Refs.~\cite{haah2018codes,litinski2019game,litinski2019magic} for details). In a Pauli-based computation, each noisy gate $ V \ccz_j V^\dagger $ can be realized using a single noisy $\tof$ state (produced by \BUTOFF) followed by a sequence of multi-qubit Pauli measurements implemented through lattice surgery (recall \cref{Sec:LatticeSurgery} and see also \cref{App:PerformV}).  An explicit factory layout is given in \cref{fig:BigLayoutLabelled} of \cref{App:FactoryLayout} that provides ample routing space for lattice surgery to be executed rapidly, with \etc{4 Toffoli gate teleportations} happening in parallel.

To describe the fault-tolerance properties of \TDTOFF, let us first assume the underlying memory and lattice surgerxy operations are implemented perfectly. Since the protocol is based on a trio of codes, \jp{each of which} can detect a single error, we can detect any fault affecting a single noisy $\tof$ state.  Even if an error affects multiple qubits within a single $\tof$ state (e.g. a $Z \otimes Z \otimes Z$ error) we still call it a single fault-location error because it leads to no more than 1 error in each codeblock, and so is detectable.   Therefore, if the \jp{noisy input} $\tof$ states have infidelity $\epsilon$, then after postselection the \jp{output $\tof$ states} will have infidelity $O(\epsilon^2)$. In \cref{App:ErrorProp}, we show exactly how the output fidelity depends on the noise model of the input $\tof$ states. As a toy example,  in \cref{App:ErrorProp} we show that for depolarizing noise the output infidelity is $1.878\epsilon^2 + O(\epsilon^3)$ per $\tof$ state output.  

However, we saw in \cref{fig:PlotsBUTOF} that \BUTOFF outputs states with errors heavily dominated by $Z \otimes \id \otimes \id$.  Let us consider the case when the \tof states are generated by \BUTOFF with $d_{BU}=7$  and assume \REGthree parameters; we refer to this throughout as our benchmark example.  Assuming an ideal \etc{implementation} of \TDTOFF (without any further optimisation to the noise profile) gives an output error of $8*10^{-10} \sim 2 \epsilon^2$, so the noise correlations slightly degrade performance relative to a depolarizing noise model with the same total error.

\jp{If $C$ is a Clifford transformation such that $C\ket{\tof}=\ket{\tof}$, we say that $C$ is a \emph{Clifford symmetry} of the (error-free) TOF state. For example, the group of Clifford symmetries includes $\{ g_A, g_B, g_C \}$ of \cref{eq:g1,eq:g2,eq:g3}.
Using these symmetries, we can improve the fidelity of the output TOF states by tailoring the distillation protocol, exploiting the property that the noise on input TOF states is dominated by $Z$ errors on the first qubit. In the tailored protocol we apply a different Clifford symmetry to each of the 8 input TOF states. The Clifford transformation $C$ modifies the noise model, mapping an error $E$ to $CEC^\dagger$.}

We prove in \cref{App:Noisetailoring} the existence of a set of Clifford symmetries with the following property:  given an initial noise model dominated by a Pauli $Z \otimes \id \otimes \id$ error occurring with probability $\epsilon_1=p_{Z_{A}}$ and rarer $Z$ errors occurring with total probability $\epsilon_2=\sum_{E \neq Z_{A}} p_{E}$, the tailored protocol outputs $\tof$ states with infidelity $O(\epsilon_1^3)+ O(\epsilon_1 \epsilon_2)+O(\epsilon_2^2)$. Furthermore, performing the Clifford symmetries adds a mere 2 $\cnot$ gates to the protocol's gate complexity because most of the Clifford symmetries can be chosen as permutations of qubit labels. Having accounted for both space-time tradeoffs and noise tailoring, the full final protocol is described in \cref{tab:FactoryClock} of \cref{App:FactoryLayout}.

\begin{table*}[th]
\begin{tabular}{cccccc}  \toprule
infidelity & & & $P_{\mathrm{ACC}}$ & $T_{TD}$ Time ($\mu $s) & \\
 $\epsilon _{\mathrm{TD}}$ & $\#$ ATS & $\#$ PCDRs & ($\%$) & per $\tof$ & $d_{\text{BU}}$ \\ \midrule
 $2.4 * 10^{-9}$ & 2814 & 8442 & 97 & 2691 & 7 \\
 $2.6 * 10^{-9}$ & 2394 & 7182 & 97 & 2434 & 7 \\
 $9.0 * 10^{-9}$ & 2016 & 6048 & 99 & 2388 & 5 \\
 $2.8 * 10^{-8}$ & 1680 & 5040 & 99 & 2262 & 5 \\
 $5.6 * 10^{-8}$ & 1596 & 4788 & 99 & 1886 & 5 \\
 $2.6 * 10^{-7}$ & 1470 & 4410 & 98 & 1762 & 5 \\
 $9.9 * 10^{-7}$ & 1386 & 4158 & 98 & 1766 & 5 \\
 $1.5 * 10^{-6}$ & 1302 & 3906 & 94 & 1724 & 5 \\
 $7.6 * 10^{-6}$ & 1176 & 3528 & 93 & 1602 & 5 \\
  \bottomrule
\end{tabular}
    \caption{Resource costs for \TDTOFF generation of $\tof$ states using as input 8 noisy state produced by \BUTOFF using distance $d_{BU}$.  The protocol outputs 2 $\tof$ states with infidelity $2\epsilon _{\mathrm{TD}}$ and success probability $P_{\mathrm{ACC}}$ rounded to nearest integer.  We give the expected runtime per Toffoli as $T_{TD}$.  The whole factory (including \BUTOFF modules) has a footprint given in terms of the number of ATS components, or equivalently in terms of PCDRs (qubits) which is approximately three times the ATS count. Further details provided in \cref{tab:FactoryCost1}. We assume \REGthree hardware parameters.}
    \label{tab:FactorySummary}
\end{table*}

Returning to the previously discussed benchmark example, then $\epsilon_1=2*10^{-5}$ and $\epsilon_2=7.5*10^{-9}$ so $\epsilon_2 \ll \epsilon_1$ and we expect an improvement from noise tailoring.  Assuming an ideal \CC{implementation} of the noise-tailored \TDTOFF, then we have an output error of $1.2*10^{-12}$ that is dominated by a contribution $\sim 8 \epsilon_1 \epsilon_2$.  However, the protocol will not be implemented ideally. The protocol is realized with each qubit encoded into a memory: either a repetition code or a thin surface code.  We can exponentially suppress memory and lattice surgery errors by increasing the code distance, though this comes at increased resource cost.  The tuning of these code distances is one of the most important aspects of optimal factory design. Following an approach similar to prior work on code distance tuning~\cite{litinski2019magic}, we present our analysis of Clifford noise in \cref{App:CliffordNoise}.  We present a sample of our numerical results from \cref{App:CliffordNoise} in \cref{tab:FactorySummary}.

\jp{As we discuss in \cref{sec:Overhead}, the error rates in \cref{tab:FactorySummary} are sufficiently low for reliable implementation of quantum algorithms with up to $10^{8}$ Toffoli gates, at a quite low overhead cost. Note that the lowest error rate reported in the table is $2.4*10^{-9}$; this error rate is dominated by bit-flip errors in repetition code blocks, and could be surpassed by more extensive use of surface codes (see \cref{App:LimitingFactors}).} 
 
 Let us compare to the factory of Gidney and Fowler~\cite{gidney2019efficient} that concatenates $T$ state distillation with a protocol that distills a single $\tof$ state from a supply of $T$ states. Using a square surface code distance $d$, the factory requires $12 d \times 6 d$ qubits and takes $5.5 d$ surface code cycles.  They assume a superconducting transmon architecture with $p_{SC}=10^{-3}$ CNOT gate infidelity that can execute one cycle of surface code error correction in 1$\mu s$. For sample algorithms with $\sim 1-100$ million Toffoli gates, they considered a $d=31$ surface code which gives a $6.9 *10^{4}$ qubit footprint generating 1 $\tof$ state every $170 \mu s$.  This is a considerably larger size than our factory, mainly because we exploit \BUTOFF, thin surface codes, and where possible we use repetition codes. Note that \cref{tab:FactorySummary} assumed hardware parameters leading to surface code cycles of \etc{$31 \mu s$} rather than $1.1 \mu s$, so while our factory typically needs far fewer surface code cycles per $\tof$ state, \etc{our slower physical gate times mean that the overall factory runtime (per \tof state) are an order of magnitude slower.}

\begin{table*}[ht!]
\begin{tabular}{ccccccc}  \toprule
size & $\tof$ gates & T-gates &  &  & $RT$ & fac \\
 $L$ & $N_{\tof}$ & $N_{T}$ & $\#$ ATS & $\#$ PCDRs &
   \text{mins} & $\%$  \\ \midrule
   \multicolumn{6}{c}{$u/\tHubbard=4$} \\ \midrule
8  & $1.8 * 10^5$ & $1.7 * 10^6$ & $1.8 * 10^4$ & $5.4 * 10^4$ & 32 & 8.8 \\
16 & $1.9 * 10^5$ & $9.5 * 10^5$ & $6.5 * 10^4$ & $1.95 * 10^5$ & 23 & 2.5 \\
24 & $1.9 * 10^5$ & $8.5 * 10^5$ & $1.5 * 10^5$ & $4.5 * 10^5$ & 23 & 1.0 \\
32 & $2.0 * 10^5$ & $8.7 * 10^5$ & $2.7 * 10^5$ & $8.1 * 10^5$ & 24 & 0.6 \\
 \midrule
   \multicolumn{6}{c}{$u/\tHubbard=8$} \\ \midrule
8  & $4.3 * 10^5$ & $4.2 * 10^6$ & $1.8 * 10^4$ & $5.4 * 10^4$ & 89 & 9.5 \\
16 & $4.6 * 10^5$ & $2.3 * 10^6$ & $7.0 * 10^4$ & $2.1 * 10^5$ & 60 & 2.4 \\
24 & $4.7 * 10^5$ & $2.1 * 10^6$ & $1.5 * 10^5$ & $4.5 * 10^5$ & 57 & 1.0 \\
32 & $4.7 * 10^5$ & $2.1 * 10^6$ & $2.7 * 10^5$ & $8.1 * 10^5$ & 62 & 0.6 \\
 \bottomrule
\end{tabular}
    \caption{Column \text{$\#$ATS} refers to the total number of ATS components used.  The total \text{$\#$ATS} count includes: $2L^2$ logical qubits to represent the Hubbard model fermions; ancilla qubits for phase estimation, ancilla-assisted circuit synthesis~\cite{RUS}, Hamming weight phasing and catalysis~\cite{Gidney2018halvingcostof};  and the ATS space for 1 \TDTOFF factory ($\%$fac counts the percentage of this contribution rounded up to nearest integer); and we also include a generous $+30\%$ space overhead for routing and lattice surgery costs. } 
    \label{tab:HubbardModelCossts}
\end{table*}

\section{Overhead Estimates}
\label{sec:Overhead}

Here we consider how our architecture could be used to fault-tolerantly implement a quantum algorithm beyond the reach of classical computers. Throughout this section, \etc{we assume \REGthree hardware parameters and find competitive performance compared to other architectures. Since \REGone and \REGtwo incur much higher resource overhead costs and/or can not reach the required fidelities, a key conclusion is that \REGthree or better should be the long-term goal for the proposed architecture.}  Using a Pauli-based computation (recall \cref{Sec:LatticeSurgery}), the complexity is mainly determined by the number of qubits and Toffoli gates required for the algorithm.  Simulations of 100-qubit circuits are substantially beyond the reach of current classical methods unless they have low depth or are near-Clifford circuits.  \etc{Currently, the best} known classical simulation algorithm of near-Clifford circuits~\cite{bravyi2019simulation} for an $n$-qubit circuit with \jp{a total of $N_{\tof}$ Toffoli gates} has a runtime $O(\mathrm{poly}(n,N_{\tof}) 2^{0.83 N_{\tof}})$. \etc{Without substantial improvement of existing classical simulation algorithms,} for $N_{\tof}=1000$ the exponential component of the runtime is comfortably in the classically intractable regime.  

Let us consider a computation with $n=100$ and $N_{\tof}=1000$. A computation of this size could be executed reliably using only the repetition code; the protection against bit flips provided by thin-stripped surface codes is not needed. Using $d_{\mathrm{rep}}=9$ repetition codes for \BUTOFF, error rates of $6*10^{-6}$ per $\tof$ gate can be achieved and therefore an error probability of $0.6\%$ for the full algorithm arising from errors in  $\tof$ gates alone. However, logical failure rates for data qubits stored in memory must also be considered.  For data qubits encoded in the repetition code, the lowest achievable logical error rate is $2.7*10^{-8}$ using a $d_{\mathrm{rep}}=9$ repetition code (\cref{fig:TotalRepetitionCodeFail}).  With $1000* d_{\mathrm{rep}} = 9000$ repetition code cycles and $n=100$ logical qubits, the total probability of a memory error is $\sim 2.4 \%$. Hence the total failure probability of the algorithm due to the combination of memory TOF errors is $\sim 3 \%$. \etc{Since \BUTOFF is probabilistic, we can boost the success probability to near unity and produce Toffoli states effectively on demand by simply making many parallel attempts at \BUTOFF. The whole computation is therefore} achievable with 900 ATS components for memory and several hundred ATS components to parallelize \BUTOFF.

Additional resources are needed for routing and performing Clifford operations, so the entire device would require between 1 and 2 thousand ATS components, depending on routing and Clifford requirements. 

\etc{While algorithms using a thousand Toffoli gates are classically intractable with known methods, there are no known algorithms of this scale that offer a quantum advantage for useful problems.}  \jp{As a representative example of a problem where quantum advantage is reachable with a relatively modest quantum circuit, we consider the task of estimating the ground state energy density of the Hubbard model with Hamiltonian}~\cite{kivlichan2020improved,HubbardInPrep}

\begin{equation}
   H = u \sum_{i} a^{\dagger}_{i,\uparrow}a_{i,\uparrow}a^{\dagger}_{i,\downarrow}a_{i,\downarrow}  + \etc{ \tHubbard }  \sum_{i, j\in N(i)}  (a^{\dagger}_{i,\uparrow} a_{j,\uparrow}+ a^{\dagger}_{i,\downarrow} a_{j,\downarrow} ) ,
\end{equation}
\jp{which describes fermions hopping on an $L\times L$ square lattice with periodic boundary conditions; \etc{$\tHubbard$ is the coefficient} of the hopping term in $H$ ($N(i)$ denotes the set of lattice sites which are neighbors of site $i$), and $u$ is the coefficient of an on-site repulsive term. The fermion creation and annihilation operators $a^\dagger$, $a$ can be encoded using qubits by various methods.
The ratio $u/\tHubbard$ quantifies the interaction strength.}
We consider $u/ t= 4$ to enable an easier comparison with Ref.~\cite{kivlichan2020improved}.  However, a classical simulation of such a model is most difficult in the regime near \etc{$u/ \tHubbard = 8$}  \cite{zheng2017stripe} and so we also consider this choice. 

\etc{We use the plaquette Trotterization scheme and analysis of Ref.~\cite{HubbardInPrep} to count the non-Clifford gates for estimating the ground state energy density.  Overall, the gate complexity scales as $O(L^{3}/\epsilon^{3/2})$ where $\epsilon$ is the allowed error in the total energy.  Since we are interested in the energy density, we can consider a multiplicative (extensive) error of $5\%$ of the ground state energy.   Since the allowed energy error $\epsilon$ grows with the system size $L^2$, the overall runtime complexity is upper-bounded by a constant~\footnote{In contract, for an additive (intensive) error energy estimation, $\epsilon$ is constant in $L$, and so the gate complexity will grow with system size~\cite{kivlichan2020improved,HubbardInPrep}}.  We also require the algorithm to succeed with probability at least $90\%$. From \cref{tab:HubbardModelCossts}, we see this algorithm requires over 1 million Toffoli gates and over 100 logical qubits~\cite{kivlichan2020improved}.}  While simulating the Hubbard model is not feasible using just the repetition code to protect the logical data qubits, only very little bit-flip protection suffices, and so we can use a $d_x=3$ thin surface code as our primary storage and the \TDTOFF protocol for Toffoli states.   \cref{tab:HubbardModelCossts} separately presents the number of logical $\tof$ gates ($N_\tof$) and logical T gates ($N_{T}$) required by the algorithm.  We can catalyze 1 $\tof $ state into 2 T states~\cite{Gidney2018halvingcostof}, so that the algorithm consumes a total of
\begin{equation} \label{eqn:catalysis}
 \tau =N_\tof+(N_{T}/2)
\end{equation}
$\tof $ states. 

\textit{Caveats in architectural comparisons.-} Our results for $u/\tHubbard=4$ in \cref{tab:HubbardModelCossts} can be compared with the transmon architecture resource estimates of Table I of \cite{kivlichan2020improved}, though subject to several caveats that we list first.  Direct comparisons are difficult because the noise models are very different. Transmon architectures are typically considered with \cnot error probabilities of $p_{SC}=10^{-3}$ or $p_{SC}=10^{-4}$ \etc{and a depolarizing noise model.} \jp{In contrast,} for our \REGthree hardware parameters the \cnot infidelity is \etc{$3.6 * 10^{-3}$}, \jp{but with highly biased noise.} To perform a \cnot  with infidelity of $10^{-4}$ we would need $\kappa_1/\kappa_2 \sim 10^{-8}$ (see caption of \cref{fig:SurfaceCodeMem} for further discussions). So although we benefit greatly from bit-flip suppression due to cat-codes, our current projections for $Z$ error rates are far less optimistic than typically assumed for transmon qubits. Furthermore, transmon-architecture resource estimates are based on a toy depolarizing noise model, whereas our noise model has been derived from \jp{detailed} 
modeling of the hardware.  An additional important caveat is that we exploit the Hubbard model simulation analysis of \cite{HubbardInPrep}, which provides a $5.5 \times$ reduction in gate count for $L=8$ and a larger improvement for larger $L$ (compared to Ref.~\cite{kivlichan2020improved}).  \etc{These gate count reductions lead to a comparable $5.5 \times$ reduction in  runtime, but below we factor out these algorithmic improvements when making architectural comparisons with Ref.~\cite{kivlichan2020improved}.}

\textit{Qubit cost discussion:} \etc{\cref{tab:HubbardModelCossts} includes a column reporting the number of ATSs required. For an $L=8$ and $u=4$ Hubbard model, we estimate the cost at 18,000 ATS's, which corresponds to the headline figure in our abstract.  We need to triple this number to obtain the number of PCDRs (qubits)}. \jp{Comparing to a superconducting transmon-qubit architecture~\cite{kivlichan2020improved} with a \cnot infidelity $p$, we find that for} $p=10^{-3}$ we need $\sim 6 \times$ fewer qubits; and for $p=10^{-4}$ we use a comparable number of qubits.  \etc{At the end of \cref{subsec:SurfaceCodeMem}, we made a comparison of surface code overheads with a $p=10^{-3}$ depolarizing noise model and found we needed $\sim 6 \times$ fewer qubits, similar to the $\sim 6 \times$ improvement found here.} \etc{Given better  $\kappa_1/\kappa_2$ than assumed by \REGthree, there would be additional resource savings.} One loose assumption in our qubit counting is that we multiply our resource overhead costs by a factor of $1.3 \times$ to account for routing and lattice surgery costs (see discussion of \cref{Sec:LatticeSurgery} and Ref.~\cite{chamberland2021universal}) whereas we do not know what routing overhead was assumed in Ref.~\cite{kivlichan2020improved} but believe this was neglected.

\textit{Runtime discussion:} \etc{The total runtime of our architecture is in the practically reasonable range of \etc{23-89 minutes} for a classically challenging task}.  There are two important factors in the runtime analysis: (1)  the time it takes to prepare $\tau$ $\tof$ states, which is $T_{a}=\tau T_{TD}$ (see  \cref{tab:FactorySummary} for examples of values of $T_{TD}$); (2)  the time required to inject magic states and perform Toffoli uncomputations via lattice surgery \etc{$T_{b}=( 4 N_\tof+ N_{T} ) (d_m+1) T_{\mathrm{surf}}$ (see Ref.~\cite{chamberland2021universal} for further discussion)}, where $T_{\mathrm{surf}}$ is the time per surface code cycle and $d_m$ is the number of surface code cycles per lattice surgery operation (recall \cref{fig:TimeLikeCartoon}).  Note that we use $d_m + 1$ instead of $d_m$ to allow for ancilla qubits to be reinitialized between consecutive lattice surgery protocols. We take the runtime to be $RT=\mathrm{max}[T_{a},T_{b}]$.  We say the architecture is Clifford bottlenecked if $RT=T_{b}$ and magic-state bottlenecked if $RT=T_{a}$.  

Note that our estimate of $T_a$ assumes that we can only teleport 1 magic state qubit at a time, since faster injection rates could incur higher routing or Clifford gate costs.  For our hardware and factory design, \etc{we are Clifford bottlenecked} as the \TDTOFF factory is producing Toffoli states at about the same pace as they can be transported into the main algorithm. In contrast, estimates for superconducting transmon architectures~\cite{kivlichan2020improved} have assumed a single factory leading to them being significantly magic-state bottlenecked (with the algorithm often idle and waiting for the factory).  Let us consider the instance with $u/\tHubbard=4$ and $L=8$, for which we estimate a runtime of \etc{32 minutes}.  For a transmon architecture with $p_{SC}=10^{-3}$, one obtains a runtime estimate of 3 minutes, by reducing the results of Ref.~\cite{kivlichan2020improved} by a factor $5.5$ to account for recent algorithmic improvements \cite{HubbardInPrep}.  A similar runtime estimate (2.6 minutes) is obtained for the transmon architecture by assuming it generates 1 $\tof$ state per $170\mu s$ using the factory of Ref.~\cite{gidney2019efficient}. Overall, the transmon architecture runs about \etc{$11\times$} faster \jp{than our architecture, primarily due} to \etc{$28\times$} faster execution of each surface code cycle.

\section{Conclusion}

In this paper, we presented a comprehensive analysis of an architecture for a fault-tolerant quantum computer. At the lowest level, it is based on hybrid electro-acoustic devices to implement a stabilized cat code with highly biased noise, dominated by \ji{phase flips}. This cat code is then concatenated with an outer code that focuses mostly on correcting the \ji{phase-flip} errors. Our estimated overheads for performing fault-tolerant quantum algorithms showcase the promise of this approach \etc{if the appropriate parameter regime can be reached}. There are several interesting directions for future work to improve on our current proposal.  

On the hardware side, we would like to explore ways to increase the value of $\kappa_2$, which would allow us to achieve the desired ratio of $\kappa_1/\kappa_2$ with a less stringent constraint on $T_1 = 1/\kappa_1$ of the acoustic oscillators. Currently the value of $\kappa_2$ is upper bounded by the cross-talk error and the bandwidth of the filter. We believe similar set ups with tunable couplers, multiport resonators, and multiple buffer modes are promising for increasing substantially the attainable value of $\kappa_2$. Higher $\kappa_2$ would also give faster gates, allowing for a larger quantum advantage over classical computing. 

As was shown in this work, the magic state factory only accounts for \etc{at most $9.5 \%$ of the total resource overhead requirements. The other $90.5 \%$} of the overhead 
is largely dominated by the performance of the thin rotated surface code. Recently, an $XZZX$-type surface code which takes advantage of the noise bias for phase-flip errors was introduced and shown to have better thresholds compared to the rotated surface code~\cite{BonillaAtaides20}. An interesting avenue for future work would be to consider the implementation of the $XZZX$ surface code (or other topological codes which take advantage of the noise bias) in our architecture to determine if further reductions in overhead costs can be achieved. Further, one could use compass codes~\cite{LMNWB19,ChamberlandPRX,DLHB20,HB20compass2} which potentially require fewer resources compared to surface codes given the low-weight gauge operator measurements. However, details for implementing such codes in a lattice surgery scheme such as the one presented in this work remain to be addressed. We also note that better thresholds and lower logical failure rates for a given code distance doesn't necessarily correspond to lower resource costs for running algorithms. For instance, a more careful analysis of the $XZZX$ surface code shows that although it achieves lower logical failure rates than a thin-stripped surface code for the same $d_x$ and $d_z$ distances, the $XZZX$ code requires roughly double the amount of data and ancilla qubits compared to thin-stripped surface codes. As such, for desired target logical error rates required to implement certain algorithms, there are noise parameter regimes where the overall resource costs for running the algorithm using a thin-stripped surface code is less than using the $XZZX$ code. 

In this work we considered a standard model of Pauli-based computation with \jp{multi-qubit} Pauli operators measured via lattice surgery in order to inject magic states.  This approach comes with an additional qubit cost for data access and routing, and the choice of routing solution 
\jp{yields} a lower bound on runtime.
In previous resource analyses, these considerations were not especially important because algorithms were bottlenecked by the pace at which they could produce magic states.  In contrast, \jp{data routing} emerged as a bottleneck in our architecture, and so a more careful optimization of routing costs and speed of \etc{gate teleportation} 
\jp{might improve the runtime substantially.} Indeed, a rapid runtime is especially important in an architecture where bit flips are rare because it is desirable to execute the algorithm fast enough such that we can avoid needing a \jp{code with a higher $d_x$ distance}.

\begin{acknowledgments}

We thank Qian Xu for helping with the displaced Fock basis calculation and Alex Retzker for discussions. C.C. thanks Yunong Shi and Pierre-Yves Aquilanti for their help in setting up the AWS clusters where most of the error correction simulations were performed. We thank all the members of the AWS Center of Quantum Computing for our collaboration on building more powerful quantum technologies. We thank Richard Moulds, Nadia Carlsten, Eric Kessler, and all the members of the Amazon Braket and Quantum Solutions Lab teams. We thank Simone Severini for creating an environment where this research was possible in the first place. We thank Bill Vass, James Hamilton and Charlie Bell for their support and guidance throughout this project.

\end{acknowledgments}

\appendix

\section{Engineering two-phonon dissipation with piezoelectric nanostructures}
\label{sec:single_mode_stabilization}

In this Appendix we calculate the dimensionless loss parameter $\kappa_1/\kappa_2$ --- the ratio of the single-phonon and two-phonon dissipation rates --- and show how to minimize it to the lowest level allowable by the intrinsic loss of the hardware and the crosstalk constraints derived in \cref{sec:multimode_stabilization}. This Appendix is divided into four parts. First, in \cref{subsec:ats_implementation} we revisit an existing method to engineer two-photon (or in this case two-phonon) dissipation using an asymmetrically-threaded SQUID (ATS) device.~\cite{Lescanne2020}. Next we show in \cref{subsec:calculation_of_g2} how to calculate the interaction rate $g_2$ when the storage resonator is an arbitrary piezoelectric nanostructure, and explicitly calculate $g_2$ for the specific case of a one-dimensional phononic-crystal-defect resonator (PCDR)~\cite{Arrangoiz-Arriola2019}. Then in \cref{subsec:classical_filter_theory} we derive, using a classical description of the underlying superconducting circuits, a general expression for $\kappa_2$ when a bandpass filter is placed in between the output port of the buffer resonator and the external $50 \, \Omega$ environment and show how to design the filter to optimize $\kappa_2$. We include a filter in our analysis because filtering the output --- or engineering the density of states of the system's reservoir --- is crucial to the multiplexed stabilization protocol described in \cref{sec:multimode_stabilization}. Finally, in \cref{subsec:loss_results} we show that the loss $\kappa_1/\kappa_2$ can be minimized by utilizing a high-impedance buffer resonator and calculate a lower bound for this loss.

\subsection{\label{subsec:ats_implementation} Implementation of the required Josephson nonlinearity}

\begin{figure}[ht]
\includegraphics{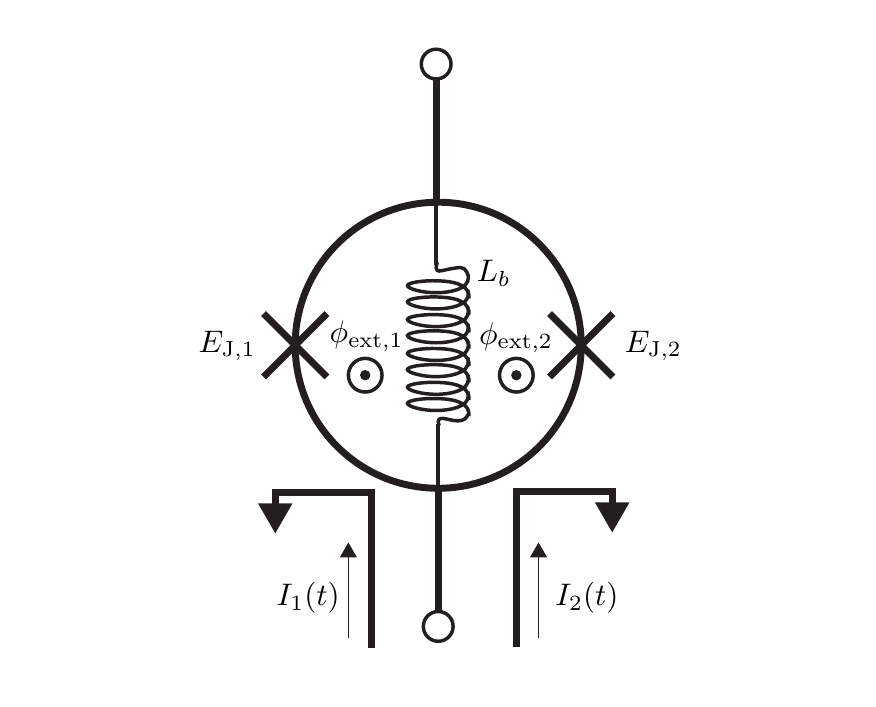}
\caption{\label{fig:ats_diagram} Schematic diagram of an ATS. Two junctions with Josephson energies $E_{J,1}$, $E_{J,2}$ are connected in parallel, forming a SQUID. The SQUID loop in turn is `split' in the middle by a linear inductor with inductance $L_b$, effectively forming two loops on either side of the inductor. The magnetic fluxes $\phi_{\text{ext}, 1}$ and $\phi_{\text{ext}, 2}$ threading the left and right loops, respectively, are controlled via externally applied, time-dependent currents $I_1(t)$, $I_2(t)$ that are buffered to ground in the vicinity of the loops using on-chip fluxlines.}
\end{figure}

In \cref{sec:HardwareImplementation} in the main text, we described at a high level how the two-phonon dissipation can be generated by engineering a nonlinear interaction $g_2^* \aop^2\bopd + \text{h.c.}$ between the storage mode $\aop$ and a very lossy ``buffer'' mode $\bop$. Here we describe in detail how this interaction can be engineered and calculate estimates of $g_2$ specifically for the hardware in this proposal. Following the method introduced in Ref.~\cite{Lescanne2020}, we propose implementing the required nonlinearity using an asymmetrically-threaded SQUID (``ATS'') device, which consists of an ordinary superconducting quantum interference device (SQUID) that is split in the middle by a linear inductor --- see \cref{fig:ats_diagram}. We reproduce some of the results of Ref.~\cite{Lescanne2020} here for convenience. 

In its most general form, the ATS potential is given by
\begin{multline}
\label{eq:ATS_potential_1}
U(\phiop) = \frac{1}{2}E_{L, b} \phiop^2 - 2\EJ \cos(\phi_\Sigma) \cos(\phiop + \phi_\Delta) \\ + 2\Delta \EJ \sin(\phi_\Sigma) \sin(\phiop + \phi_\Delta),
\end{multline}
where $\phiop$ is the phase difference across the ATS, $\phi_\Sigma := (\phi_{\text{ext},1} + \phi_{\text{ext},2})/2$, $\phi_\Delta := (\phi_{\text{ext},1} - \phi_{\text{ext},2})/2$, and $\phi_{\text{ext}, 1}$ ($\phi_{\text{ext}, 2}$) is the magnetic flux threading the left (right) loop, in units of the reduced flux quantum $\Phi_0 = \hbar/2e$. Here $E_{L,b} = \Phi_0^2/L_b$, $\EJ = (E_{J,1} + E_{J,2})/2$, and $\Delta \EJ = (E_{J,1} - E_{J,2})/2$ is the junction asymmetry. This ATS potential can be further simplified by tuning $\phi_\Sigma$ and $\phi_\Delta$ with two separate fluxlines, setting them to
\begin{align}
\phi_\Sigma &= \pi/2 + \epsilon_p(t), \\
\phi_\Delta &= \pi/2
\end{align}
where $\epsilon_p(t) = \epsilon_{p,0} \cos(\omegap t)$ is a small ac component added on top of the dc bias. At this bias point, and assuming that $|\epsilon_p(t)| \ll 1$, Eq.~(\ref{eq:ATS_potential_1}) reduces to
\begin{equation}
\label{eq:ATS_potential_2}
U(\phiop) = \frac{1}{2}E_{L,b}\phiop^2 - 2E_J \epsilon_p(t) \sin(\phiop) + 2\Delta \EJ \cos(\phiop). 
\end{equation}

\subsection{\label{subsec:calculation_of_g2} Calculation of nonlinear interaction rate \texorpdfstring{$g_2$}{g2}}
To make further progress, it is necessary to represent the nanomechanical element as an equivalent circuit that accurately captures its linear response. This can be done straightforwardly using the method of Foster synthesis, provided we know the admittance $Y_m(\omega)$ seen from the terminals of the mechanical resonator. This admittance can be accurately computed using modern FEM solvers. For further details on the piezoelectrics simulations, see Ref.~\cite{Arrangoiz-Arriola2016}.
\begin{figure}[ht]
\includegraphics[width=\linewidth]{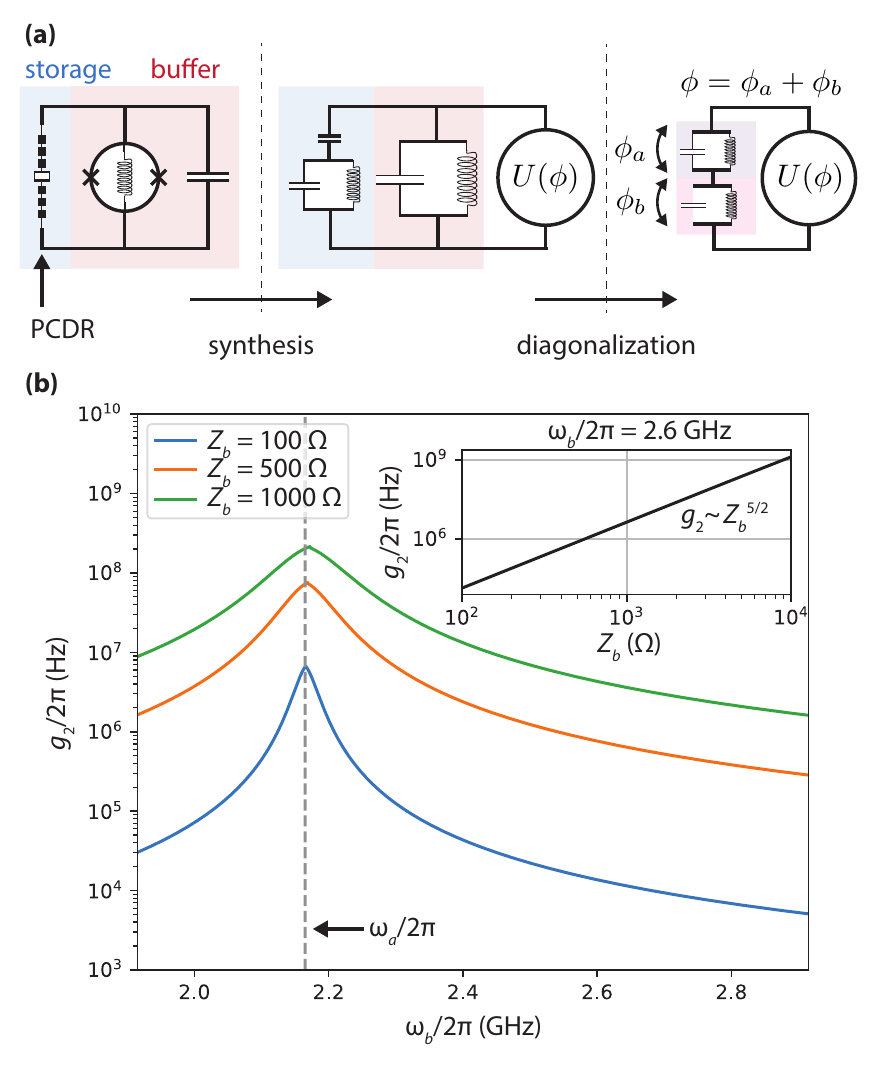}
\caption{\label{fig:g2_plots} Calculation of $g_2$. (a) Schematic summary of our method for calculating $g_2$. A PCDR, connected in parallel to a buffer resonator that is formed by shunting an ATS with a capacitance $C_b$, is synthesized as a simple Foster network with the same admittance function $Y_m(\omega)$ as the original piezoelectric structure. The Foster network consists of a parallel combination of an inductance $L_a$ and a capacitance $C_a$, in series with a `coupling capacitance' $C_g$. In turn, the linear components of the buffer resonator $L_b$ and $C_b$ are lumped together with the mechanical Foster circuit, leaving the nonlinear part of the ATS potential as an additional circuit element that we label by $U(\phi)$ in the diagram. The linear network is then diagonalized and the vacuum fluctuation amplitudes $\varphi_a$ and $\varphi_b$ of the storage-like and buffer-like eigenmodes are used to calculate $g_2$. (b) Dependence of $g_2$ on the buffer resonator frequency $\omega_b$ and impedance $Z_b$. The $g_2$ curves peak at the storage mode frequency $\omega_a$ where the modes are maximally hybridized. Inset: $g_2$ plotted as a function of $Z_b$ for a fixed $\omega_b$, showing the $5/2$ power law dependence.} 
\end{figure}

The equivalent circuit (or ``Foster network'') is shown in \cref{fig:g2_plots}(a) and in its simplest form consists of a `dc capacitance' in series with an LC block, with an additional resistor (not shown) inserted to include the effects of loss in the resonator. We note that this ``lossy Foster'' method is not exact but is accurate enough for our purposes provided that losses are sufficiently small~\cite{Nigg2012}. The linear part of the buffer resonator (including the inductor that splits the ATS) can also be represented as an LC block. In this representation the buffer and storage resonators are two linear circuits with a linear coupling and can therefore be diagonalized by a simple transformation of coordinates. The resulting ``storage-like'' ($\aop$) and ``buffer-like'' ($\bop$) eigenmodes both contribute to the total phase difference across the ATS, $\phiop = \phizpa (\aop + \aopd) + \phizpb (\bop + \bopd)$. These modes therefore mix via the ATS potential, which we redefine as $U(\phiop) \mapsto U(\phiop) - E_{L,b}\phiop^2/2$ because we already absorbed the inductor into the linear network. The vacuum fluctuation amplitudes of each mode mode are given by
\begin{equation}
\varphi_{k,j} = \sqrt{\frac{\hbar}{2\omega_k}}(C^{-1/2}U)_{jk},
\end{equation}
where $C$ is the Maxwell capacitance matrix of the circuit, $U$ is the orthogonal matrix that diagonalizes $C^{-1/2}L^{-1}C^{-1/2}$, and $L^{-1}$ is the inverse inductance matrix~\cite{Pechal2017}. The index $k \in \{a, b\}$ labels the mode and $j$ labels the node in question. Note that generally we omit the $j$ index in our notation because the node of interest is clear from context (it is the one where the ATS is located).

The way in which the ATS mixes the modes is now explicitly clear: the third-order term in the Taylor series expansion of the $\sin(\phiop)$ function in \cref{eq:ATS_potential_2} contains terms of the form $\aop^2 \bopd + \text{h.c.}$, which is precisely the required coupling. This is the key reason for using an ATS as opposed to an ordinary junction, which has a potential $\sim \cos(\phiop)$. Note also that a finite junction asymmetry $|\Delta \EJ| > 1$ partially eliminates the benefit of using an ATS, as this introduces additional self- and cross-Kerr terms. For the remainder of this analysis we assume we are operating in the ideal case $\Delta \EJ = 0$, noting that with state-of-the-art fabrication one can reliably achieve $\Delta E_J/E_J \sim 10^{-2}$~\cite{Kreikebaum2020}.

In order to select the desired terms one must set the pump frequency to $\omegap = 2\omegaa - \omegab$~\cite{Lescanne2020}. This brings the term $g_2^*\aop^2\bopd + \text{h.c.}$ into resonance and allows us to drop the other terms using a rotating-wave approximation (RWA). The coupling rate is given by $g_2 = (\EJ/\hbar)\epsilon_{p,0}\phizpa^2 \phizpb/2$. Additionally, a linear drive $\epsilon_d^* \bop + \hc$ at frequency $\omegad = \omegab$ is added to supply the required energy for the two-phonon drive.

We now explicitly calculate $g_2$ \textit{assuming that the storage resonator is a one-dimensional lithium niobate phononic-crystal-defect resonator (PCDR)} as reported in Ref.~\cite{Arrangoiz-Arriola2019}. We use for its Foster network parameters the values $C_g = 0.385 \, \text{fF}$, $C_a = 1.682 \, \text{fF}$, and $L_a = 2.614 \, \mu\text{H}$, which in previous work have produced accurate estimates of the linear coupling rate between the phononic mode and other electrical circuits~\cite{Arrangoiz-Arriola2019, Wollack2020}. These parameters set $\omega_a/2\pi \approx 2.17 \, \text{GHz}$ as the storage mode frequency, which will remain fixed for the reaminder of this Appendix. We further take $E_J/h = 90 \, \text{GHz}$ and $\epsilon_{p,0} = \pi/80$ as representative values that are experimentally realistic.~\cite{Lescanne2020}. We note that Ref.~\cite{Lescanne2020} did not explicitly report a value for $\epsilon_{p,0}$, but we inferred it by reproducing their reported value of $g_2$. In some instances we will set $\epsilon_{p,0}$ to an even smaller value, which we will indicate accordingly. In \cref{fig:g2_plots}(b) we show $g_2$ plotted as a function of the buffer mode's frequency $\omega_b \approx 1/\sqrt{L_b C_b}$ for three different values of the impedance $Z_b = \sqrt{L_b/C_b}$. The two parameters $\omega_b$ and $Z_b$ completely specify the properties of the buffer resonator for the purposes of this work. One salient feature is that $g_2$ scales as
\begin{equation}
\label{eq:g2_scaling}
    g_2 \sim Z_b^{5/2},
\end{equation}
which is due to the fact that $\varphi_b \sim \sqrt{Z_b}$ and $\varphi_a \sim Z_b$. This rapid scaling will prove useful later on, when we explore how to configure the system to minimize the dimensionless loss $\kappa_1/\kappa_2$.

\subsection{\label{subsec:classical_filter_theory} Classical filter theory and derivation of dissipation rates}

\begin{figure*}
\centering
\includegraphics[width=\linewidth]{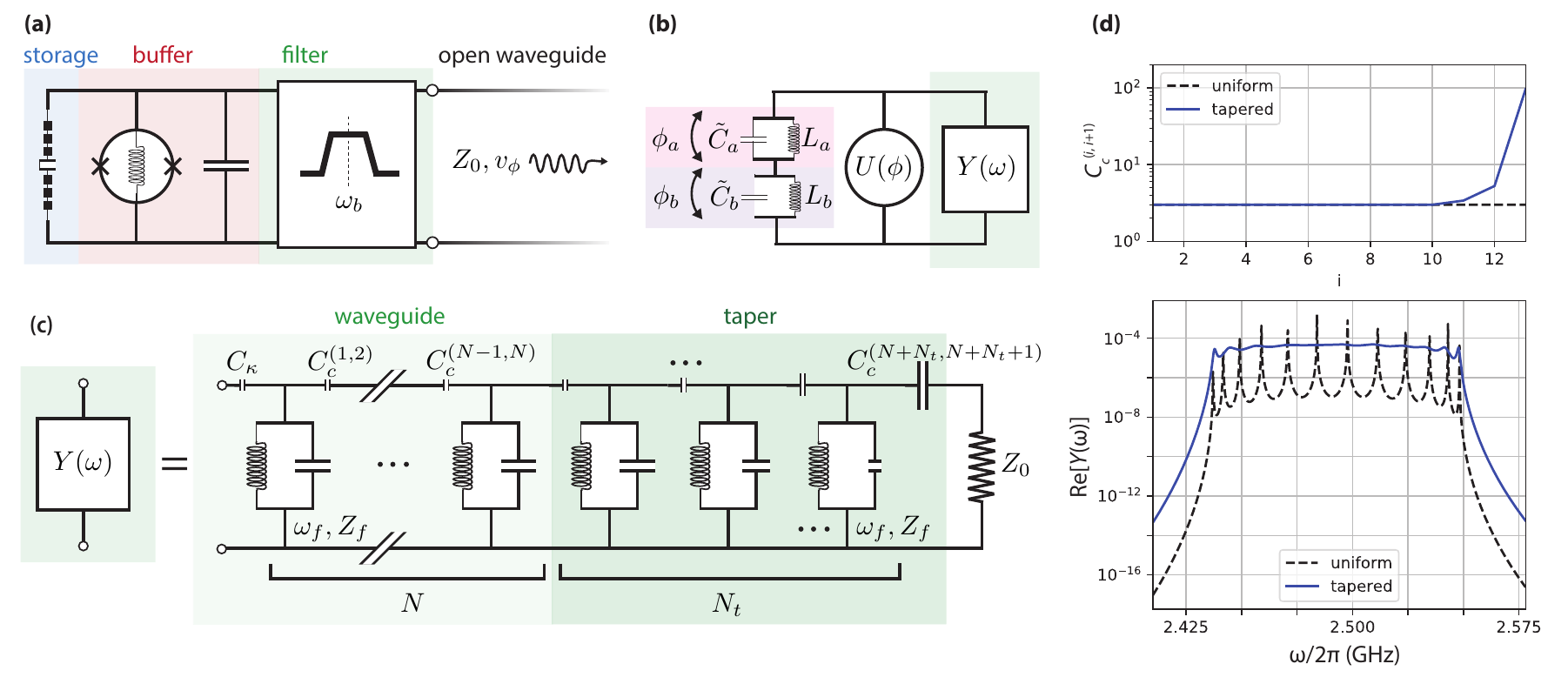}
\caption{\label{fig:filter_design} Filter design. (a) Schematic of the filtering setup. A bandpass filter centered at frequency $\omega_b$ is placed in between the output of the buffer resonator and an open waveguide with characteristic impedance $Z_0$ and phase velocity $v_\phi$. Photons that are transmitted through the filter enter the open waveguide and are irreversibly lost. (b) Circuit diagram showing the normal modes $a$ and $b$ and their connection to the filter described by an admittance function $Y(\omega)$. (c) Detailed circuit diagram for the filter structure, which consists of a main filter chain with $N$ ``unit cells'' followed by a taper section with $N_t$ cells, terminated at the end with a load resistance $Z_0$ that accurately models the open waveguide at the output port. Every cell of the filter has frequency $\omega_f$ and impedance $Z_f$, and neighboring cells are coupled capacitively with capacitances $C_c^{(i, i+1)}$. The coupling capacitance $C_\kappa$ between the buffer resonator and the first filter cell is defined separately for generality. (d) Top: coupling capacitances plotted as a function of cell index $i$ for tapered ($(N,N_t)=(10,3)$) and uniform ($(N,N_t)=(13,0)$) filters. The tapered structure, found automatically by a Nelder-Mead optimizer, is characterized by a rapid increase in $C_c$ near the end of the structure. Bottom: typical filter response, here shown as the real part of $Y(\omega)$ for tapered and uniform filters. The response of the uniform structure shows multiple sharp peaks, each corresponding to a standing-wave resonance of the structure, whereas the tapered response is relatively flat throughout the filter passband. In effect, the taper allows propagating waves to be transmitted to the open waveguide over a broad bandwidth.}
\end{figure*}

The above calculation of $g_2$ is only half the story, since we are ultimately interested in making accurate predictions of $\kappa_1/\kappa_2$. Indeed $\kappa_2 = 4g_2^2/\kappa_b$ in the simple two-mode model with the pump tuned perfectly on resonance $\omega_p = 2\omega_a - \omega_b$. However, as we show in \cref{sec:multimode_stabilization}, in order to stabilize multiple modes with a single ATS (which is necessary to achieve the required connectivity for the surface code), it is a critical requirement to utilize a bandpass filter between the buffer resonator and the open $50 \, \Omega$ port in order to protect the storage modes from radiative decay and to suppress unwanted correlated decay processes --- see \cref{fig:filter_design}(a) for a sketch of the device. We therefore need a more general expression for the two-phonon dissipation rate $\kappa_2$ in the case where the bath that the $b$ mode couples to is described by a general admittance function $Y(\omega)$. 
We begin with the Hamiltonian of the closed system comprising the storage mode $a$ and the buffer mode $b$, neglecting dissipation:
\begin{equation}
    H = \frac{1}{2}q^T C^{-1} q + \frac{1}{2}\Phi^T L^{-1} \Phi - 2E_J \epsilon_p(t)\sin(\phi_2),
\end{equation}
$q = (q_1, q_2)^T$, $\Phi = (\Phi_1, \Phi_2)^T$, $\Phi_j = \int dt V_j (t)$ is the node flux at node $j$ (with the voltage $V_j$ defined with respect to the ground node), and 
\begin{equation}
    C = \begin{pmatrix}
            C_a + C_g & -C_g \\
            -C_g & C_b + C_g
        \end{pmatrix}, \, \, \,
    L^{-1} = \begin{pmatrix}
            L_a^{-1} & 0 \\
            0 & L_b^{-1}
            \end{pmatrix}.
\end{equation}
We are also using the notation $\phi_j := \Phi_j / \Phi_0$ for the dimensionless flux, where $\Phi_0 = \hbar/2e$ is the reduced flux quantum. The equations of motion (EOMs) are
\begin{align}
    \label{eq:classical_eoms}
    \begin{split}
    \dot{\Phi} &= \partial_q H = C^{-1} q, \\
    \dot{q} &= -\partial_\Phi H = -L^{-1}\Phi + 2I_J \epsilon_p (t) \cos\phi_2 \begin{pmatrix} 0 \\ 1 \end{pmatrix},
    \end{split}
\end{align}
where we defined $I_J := E_J/\Phi_0$. Note that the charge EOM in \cref{eq:classical_eoms} is simply Kirchhoff's current law (KCL). To include the effect of the external admittance $Y(\omega)$, which describes both the filter and the $50 \, \Omega$ output line, we add an additional source of current $I_s(t)$ flowing into node 2:
\begin{align}
    I_s(t) &= \int_{-\infty}^{\infty} d\omega Y(\omega)\dot{\Phi}_{F,2}(\omega)e^{i\omega t} \\
    &= \int_{-\infty}^{\infty} d\omega Y(\omega) \left[ \frac{1}{2\pi}\int_{-\infty}^{\infty} dt' \dot{\Phi}_2(t')e^{-i\omega t'}\right] e^{i\omega t} \\
    &= \frac{1}{2\pi} \int_{-\infty}^{\infty} dt' \int_{-\infty}^{\infty} d\omega Y(\omega) \dot{\Phi}_2(t')e^{i\omega(t-t')},
\end{align}
where $\dot{\Phi}_{F,2}(\omega)$ is the Fourier transform of the voltage $\dot{\Phi}_2(t)$. Combining the EOMs \cref{eq:classical_eoms} and adding the source term, we find
\begin{equation}
    \label{eq:flux_eom}
    C\ddot{\Phi}(t) + L^{-1}\Phi(t) = F(t)\begin{pmatrix} 0 \\ 1 \end{pmatrix},
\end{equation}
where $F(t)$ is defined as
\begin{multline}
    F(t) \equiv 2I_J \epsilon_p(t)\cos\phi_2(t) - \\
    \frac{1}{2\pi}\int dt' \int d\omega Y(\omega)\dot{\Phi}_2(t') e^{i\omega(t - t')}. 
\end{multline}
Here both integrals run from $-\infty$ to $+\infty$. We will use this convention for the remainder of this section for notational simplicity, unless otherwise stated. Let $\Phi' = C^{1/2}\Phi$. Then \cref{eq:flux_eom} becomes
\begin{equation}
    \label{eq:flux_eom_2}
    \ddot{\Phi}'(t) + C^{-1/2}L^{-1}C^{-1/2}\Phi'(t) = F(t)C^{-1/2} \begin{pmatrix} 0 \\ 1 \end{pmatrix}.
\end{equation}
We now diagonalize the matrix $C^{-1/2}L^{-1}C^{-1/2}$ as
\begin{equation}
    C^{-1/2}L^{-1}C^{-1/2} = U \Omega^2 U^T,
\end{equation}
where $\Omega = \text{diag}(\omega_a, \omega_b)$ is a diagonal matrix containing the normal mode frequencies and $U$ is an orthogonal matrix. The normal modes are  
\begin{equation}
    \Phi'' = U^T \Phi' = U^T C^{1/2} \Phi = (\Phi''_1, \Phi''_2)^T.
\end{equation}
In terms of $\Phi''$, the flux EOM \cref{eq:flux_eom_2} is given by
\begin{align}
    \label{eq:normal_mode_eoms}
    \ddot{\Phi}''(t) + \Omega^2 \Phi''(t) &= F(t)U^T C^{-1/2}\begin{pmatrix} 0 \\ 1 \end{pmatrix} \\
    &= F(t) \begin{pmatrix} (U^T C^{-1/2})_{12} \\ (U^T C^{-1/2})_{22} \end{pmatrix} \\
    &= F(t) \begin{pmatrix} (C^{-1/2} U)_{21} \\ (C^{-1/2} U)_{22} \end{pmatrix},
\end{align}
where we have used the fact that $C$ (and therefore $C^{-1/2}$) is symmetric. If we define $\Phi_a := (C^{-1/2}U)_{21} \Phi''_1$ and $\Phi_b := (C^{-1/2}U)_{22} \Phi''_2$, \cref{eq:normal_mode_eoms} can be written more neatly as
\begin{equation}
    \tilde{C}_a\ddot{\Phi}_a + \tilde{C}_a\omega_a^2\Phi_a = \tilde{C}_b\ddot{\Phi}_b + \tilde{C}_b\omega_b^2\Phi_b = F(t),
\end{equation}
where 
\begin{equation}
    \label{eq:effective_capacitances}
    \tilde{C}_a := (C^{-1/2}U)^{-2}_{21}, \, \, \, \tilde{C}_b := (C^{-1/2}U)^{-2}_{22}
\end{equation}
are the effective capacitances of the $a$ and $b$ normal modes. \cref{eq:normal_mode_eoms} is KCL for a different network --- one where two $LC$ stages, one for each of the normal modes, are placed in series with each other. The series combination is in turn connected to the filtered environment $Y(\omega)$ and the ATS potential $U(\Phi)$ (see \cref{fig:filter_design}(b)). Note that this diagonalization procedure is completely equivalent to synthesizing a Foster network representing the coupled storage and buffer resonators, for example as done in black-box quantization~\cite{Nigg2012}.

Note that $\Phi = C^{-1/2} U \Phi''$, and in particular
\begin{align}
    \Phi_2 &= (C^{-1/2}U)_{21} \Phi''_1 + (C^{-1/2}U)_{22}\Phi''_2 \\
    &= \Phi_a + \Phi_b.
\end{align}
In terms of these normal mode amplitudes, $F(t)$ is given by
\begin{multline}
    F(t) = 2I_J \epsilon_p(t) \cos\left[\phi_a(t) + \phi_b(t)\right] \\ - \frac{1}{2\pi}\int dt' \int d\omega Y(\omega) \left[\dot{\Phi}_a(t') + \dot{\Phi}_b(t')\right]e^{i\omega(t - t')}.
\end{multline}
We now define the following dimensionless, time-varying mode amplitudes:
\begin{align}
    \begin{split}
    a(t) := \frac{1}{\sqrt{2\hbar}}\left[\sqrt{\tilde{C}_a \omega_a} \Phi_a(t) + i\frac{1}{\sqrt{\tilde{C}_a\omega_a}}\tilde{C}_a \dot{\Phi}_a(t)\right], \\
    b(t) := \frac{1}{\sqrt{2\hbar}}\left[\sqrt{\tilde{C}_b \omega_b} \Phi_b(t) + i\frac{1}{\sqrt{\tilde{C}_b\omega_b}}\tilde{C}_b \dot{\Phi}_b(t)\right].
    \end{split}
\end{align}
Defining $\varphi_j = \Phi_0^{-1}\sqrt{\hbar/2\omega_j\tilde{C}_j}$, where $j \in \{a, b\}$,  we have
\begin{equation}
    \phi_a = \varphi_a (a + a^\dagger), \, \, \,
    \phi_b = \varphi_b (b + b^\dagger).
\end{equation}
Here the $\dagger$ symbol indicates complex conjugation. \emph{We identify $\varphi_j$ as the amplitude of the vacuum fluctuations of the phase at node 2 due to mode $j$.} 

It is straightforward to show that the EOMs of these ``annihilation variables'' are
\begin{align}
    \label{eq:a_and_b_eoms}
    \begin{split}
    \dot{a}(t) &= -i\omega_a a(t) + i(\Phi_0/\hbar) \varphi_a F(t), \\
    \dot{b}(t) &= -i\omega_b b(t) + i(\Phi_0/\hbar) \varphi_b F(t).
    \end{split}
\end{align}
In terms of $a$ and $b$, the source term $F(t)$ is given by 
\begin{widetext}
\begin{multline}
    F(t) = 2I_J \epsilon_p (t) \cos\left[\varphi_a(a(t) + a^\dagger(t)) + \varphi_b(b(t) + b^\dagger(t))\right] \\
    -\frac{1}{2\pi}\int dt' \int d\omega Y(\omega)\left[\frac{i\hbar}{2\tilde{C}_a\Phi_0\varphi_a}(a^\dagger(t') - a(t')) + \frac{i\hbar}{2\tilde{C}_b\Phi_0\varphi_b}(b^\dagger(t') - b(t'))\right]e^{i\omega(t - t')}.
\end{multline}
\end{widetext}
We now invoke the rotating wave approximation (RWA) and neglect terms that are fast-rotating, namely $a^\dagger(t')$ and $b^\dagger(t')$ in both EOMs and $a(t')$ and $b(t')$ in the EOMs for $b$ and $a$, respectively. This is well-justified in the regime where $\omega_a$, $\omega_b$, and $|\omega_a - \omega_b|$ are all much larger than the dissipation rates $\text{Re}[Y]/2\tilde{C}_j, \, j \in \{a, b\}$. We will see shortly that indeed these quantities emerge as dissipation rates from our analysis, so this assumption is self-consistent. The EOMs \cref{eq:a_and_b_eoms} then become
\begin{widetext}
\begin{align}
    \begin{split}
    \dot{a}(t) &= -i\omega_a a(t) - \frac{1}{2\pi}\int dt' \int d\omega \frac{Y(\omega)}{2\tilde{C}_a} a(t') e^{i\omega(t-t')} + 2i (E_J/\hbar) \epsilon_p(t) \varphi_a \cos\left[\varphi_a(a(t) + a^\dagger(t)) + \varphi_b(b(t) + b^\dagger(t))\right] \\
    \dot{b}(t) &= -i\omega_b b(t) - \frac{1}{2\pi}\int dt' \int d\omega \frac{Y(\omega)}{2\tilde{C}_b} b(t') e^{i\omega(t-t')} + 2i (E_J/\hbar) \epsilon_p(t) \varphi_b \cos\left[\varphi_a(a(t) + a^\dagger(t)) + \varphi_b(b(t) + b^\dagger(t))\right].
    \end{split}
\end{align}
\end{widetext}
We now go to an ``interaction frame'' (or rotating frame) defined by the transformations
\begin{align}
    a(t) &\mapsto a(t)e^{i\omega_a t}, \\
    b(t) &\mapsto b(t)e^{i(\omega_b + \Delta)t},
\end{align}
and explicitly add the flux pump
\begin{equation}
    \epsilon_p(t) = \epsilon_{p,0} \cos\omega_pt, \, \, \, \omega_p = 2\omega_a - \omega_b - \Delta,
\end{equation}
which was introduced in \cref{subsec:ats_implementation}. We have also added a detuning $\Delta$ to keep the analysis general and also because finite $\Delta$ is a key requirement for multiplexed stabilization --- see \cref{sec:multimode_stabilization}.  Expanding the cosine term to second order and keeping only the resonant terms, we find:
\begin{widetext}
\begin{align}
\label{eq:time_nonlocal_eoms}
\begin{split}
    \dot{a}(t) &= -\frac{1}{2\pi}\int dt' \int d\omega \frac{Y(\omega)}{2\tilde{C}_a} a(t') e^{i(\omega + \omega_a)(t - t')} + 2ig_2 a^\dagger(t) b(t), \\
    \dot{b}(t) &= i\Delta b(t) - \frac{1}{2\pi}\int dt' \int d\omega \frac{Y(\omega)}{2\tilde{C}_b} b(t') e^{i(\omega + \omega_b + \Delta)(t - t')} + ig_2 a^2(t),
\end{split}
\end{align}
\end{widetext}
where $g_2 := (E_J/\hbar) \epsilon_{p,0} \varphi_a^2 \varphi_b/2$. 

The EOMs \cref{eq:time_nonlocal_eoms} do not have simple solutions in general because they are non-local in time. However, we can drastically simplify them --- and re-cast them into a form that is time-local --- under a specific regime of interest, which we describe next. First, note that
\begin{widetext}
\begin{equation}
    \int dt' \int d\omega Y(\omega) b(t')e^{i(\omega + \delta)(t - t')} = \int dt' Y_T(t - t') b(t')e^{i\delta (t - t')},
\end{equation}
\end{widetext}
where $\delta$ equals either $\omega_a$ or $\omega_b + \Delta$ depending on which EOM we are referring to, and $Y_T(t)$ is the Fourier transform of the admittance function $Y(\omega)$:
\begin{equation}
    Y_T(t) := \int d\omega Y(\omega) e^{i\omega t}.
\end{equation}
Now suppose for illustration that $Y(\omega)$ is a simple function
\begin{equation}
    Y(\omega) =
    \begin{cases}
    Y_0 & |\omega| \leq 2J \\
    0 & |\omega| > 2J,
    \end{cases}
\end{equation}
which describes an ``ideal'' filter with bandwidth $J$. We note this is not a physical admittance function and we are using this simply as an example --- in particular, it doesn't satisfy certain basic properties such as causality. Its Fourier transform is
\begin{equation}
    Y_T(t) = (2Y_0J) \frac{\sin(2Jt)}{Jt},
\end{equation}
so $|Y_T(t-t')e^{i\delta t}| = |Y_T(t - t')|$ is localized in the range defined by $J|t - t'| \sim 1$. Therefore, assuming $b(t')$ evolves much more slowly compared to the timescale $1/J$, the following approximation holds:
\begin{align}
    &\int dt' Y_T(t - t') e^{i\delta (t-t')}b(t') \\
    &\approx \int dt' Y_T(t - t') e^{i\delta (t-t')}b(t) \\
    &= \int dt' Y_T(t')e^{i\delta t'}b(t) \\
    &= 2\pi Y^*(\delta)b(t),
\end{align}
where in the last line we used $Y(-\delta) = Y^*(\delta)$. We shall verify shortly that this slowness assumption is self-consistent. For now, this approximation transforms the EOMs \cref{eq:time_nonlocal_eoms} to the following form:
\begin{align}
\label{eq:time_local_eoms}
\begin{split}
    \dot{a}(t) &= -\frac{\kappa_1}{2} a(t) + 2ig_2 a^\dagger(t) b(t), \\
    \dot{b}(t) &= \left[i\tilde{\Delta} - \frac{\kappa_{b, \text{eff}}(\Delta)}{2}\right] b(t) + ig_2 a^2(t) + \epsilon_d.
\end{split}
\end{align}
Here $\kappa_1 := \text{Re}\left[Y^*(\omega_a)\right]/\tilde{C}_a$ and $\kappa_{b, \text{eff}}(\Delta) := \text{Re}\left[Y^*(\omega_b + \Delta)\right]/\tilde{C}_b$ are the effective \emph{linear} dissipation rates of the $a$ and $b$ modes, respectively. We have also added an additional drive term $\epsilon_d$ (which rotates at frequency $\omega_b + \Delta$ in the lab frame and therefore here it is static), and defined $\tilde{\Delta} := \Delta - \text{Im}\left[Y^*(\omega_b + \Delta)\right]/2\tilde{C}_b$, which now includes the frequency shift of the $b$ mode due to its coupling to the filter. Note we have also neglected the corresponding shift $-\text{Im}\left[Y^*(\omega_a)\right]/2\tilde{C}_a$ of the $a$ mode, since this is negligibly small for the purposes of this analysis. 

Let us now find an effective description of the $a$ mode alone, valid in a regime where the linear dissipation rate $\kappa_{b,\text{eff}}$ is large (in a sense that will be made rigorous shortly). This procedure is the classical analogue of the formal adiabatic elimination procedure used in \cref{subsec:effective_operator}. Let us assume that $\dot{b}(t) = 0$, i.e. the $b$ mode is evolving sufficiently slowly that the time derivative can be neglected. Then \cref{eq:time_local_eoms} becomes
\begin{equation}
\label{eq:static_b_solution}
b(t) = \frac{ig_2a^2(t) + \epsilon_d}{-i\tilde{\Delta} + \kappa_{b,\text{eff}}(\Delta)/2},
\end{equation}
and
\begin{equation}
    \label{eq:final_eoms_2}
    \dot{a}(t) = -\frac{\kappa_1}{2}a(t) - \kappa_2 a^\dagger(t) a^2(t) +\alpha_d a^\dagger (t),
\end{equation}
where
\begin{align}
    \label{eq:kappa2_general_formula}
    \kappa_2(\Delta) &:= \text{Re}\left[\frac{4g_2^2}{-2i\tilde{\Delta} + \kappa_{b,\text{eff}}(\Delta)}\right] \\
    &= \frac{4g_2^2}{4\tilde{\Delta}^2 + \kappa_{b,\text{eff}}^2(\Delta)}\kappa_{b,\text{eff}}(\Delta),
\end{align}
and $\alpha_d := 2i g_2 \epsilon_d [-i\tilde{\Delta} + \kappa_b/2]^{-1}$. As a final step, let us linearize the EOMs around the static solutions $a=\pm \alpha$ given by setting $\dot{a}(t) = 0$. Assuming $4\kappa_2|\alpha|^2 \gg \kappa_1$, the solutions are $\alpha = \pm \sqrt{\epsilon_d/g_2}$. Defining $d := a - \alpha$ as the ``fluctuations'' around these fixed points, the linearized equation of motion for $d(t)$ becomes
\begin{equation}
\label{eq:linearized_d_eom}
    \dot{d}(t) = -\frac{\kappa_1}{2}d(t) - 2\kappa_2 |\alpha|^2 d(t) \approx -\frac{\kappa_\text{conf}}{2} d(t),
\end{equation}
where $\kappa_\text{conf} := 4|\alpha|^2 \kappa_2$. We call this rate the confinement rate in keeping with existing terminology~\cite{Lescanne2020}. Applying this linearization to \cref{eq:static_b_solution}, we find
\begin{equation}
\label{eq:linearized_b_eom}
    b(t) = \frac{2i g_2 \alpha}{-i\tilde{\Delta} + \kappa_{b, \text{eff}}(\Delta)/2}d(t) + const.
\end{equation}
The rate $\kappa_2$ we previously defined is now manifestly the two-phonon dissipation rate that we wanted to find, as it sets the rate $\kappa_\text{conf}$ at which fluctuations away from the fixed points $a = \pm \alpha$ decay back into the ``code space''. It reduces to the familiar form $\kappa_2 = 4g_2^2/\kappa_{b,\text{eff}}$ in the case of a perfectly resonant pump $\tilde{\Delta} = 0$, and to the form $\kappa_2 = (g_2/\tilde{\Delta})^2 \kappa_{b,\text{eff}}$ in the far off-resonant limit $|\tilde{\Delta}| \gg \kappa_{b,\text{eff}}$. This latter form is indeed equivalent to the expressions for $\kappa_2$ derived in \cref{sec:multimode_stabilization}, where the filter is modeled as a linear chain of oscillators with nearest-neighbor linear couplings. Here, the function $\kappa_{b,\text{eff}}(\Delta)$ contains all the information about the filtered environment, capturing effects such as the exponential suppression of $\kappa_2$ when $\Delta$ lies outside of the filter passband. Finally, we note that the straightforward linearization procedure above is the classical analogue of the shifted Fock basis technique described in \cref{appendix:Shifted Fock Basis}.

Let us go back and re-examine the two main assumptions that we have made so far: 1) that $b(t')$ evolves much more slowly compared to the filter response timescale $1/J$, and 2) the adiabatic assumption that $\dot{b}(t) = 0$ in \cref{eq:time_local_eoms}.

First, by inspecting equation \cref{eq:linearized_d_eom} we can extract the effective timescale of the dynamics of $d$ mode. We see that $d$ evolves on a timescale $1/|\alpha|^2\kappa_2$ (assuming $|\alpha|^2\kappa_2 \gg \kappa_1$, which is the regime we are interested in). Therefore, from the solution for $b(t)$ in \cref{eq:linearized_b_eom} we infer that the $b$ mode also evolves on this timescale. The slowness assumption that led to \cref{eq:time_local_eoms} is therefore self-consistent as long as $|\alpha|^2\kappa_2 \ll J$. Furthermore, even though we used a `toy model' for $Y(\omega)$ to illustrate the required hierarchy of timescales, we verified numerically using the simulations in \cref{subsec:filter_design} that this exact logic remains valid even when $Y(\omega)$ describes a real, appropriately designed filter.

Second, under which conditions is the adiabatic elimination $\dot{b}(t) = 0$ valid? The solution for $b(t)$ in \cref{eq:linearized_b_eom}, obtained by assuming $\dot{b}(t) = 0$, evolves on the same timescale $1/|\alpha|^2\kappa_2$ as $d(t)$. Therefore the adiabatic elimination step is self-consistent so long as $|\alpha|^2\kappa_2 \ll \kappa_b$, because $1/\kappa_b$ is the timescale in which $b(t)$, as described by the full EOM \cref{eq:time_local_eoms}, converges to its steady state. Since $\kappa_2(\Delta) \leq \kappa_2(0)$, this condition is equivalent to $2|\alpha| g_2 \ll \kappa_b$:
\begin{multline}
    |\alpha|^2\kappa_2 \ll \kappa_b \iff |\alpha|^2\kappa_2(0) = 4|\alpha|^2g_2^2/\kappa_b \ll \kappa_b \\ \iff 2|\alpha|g_2 \ll \kappa_b.
\end{multline}
For the purposes of this work we shall assume that $2|\alpha|g_2 = \eta \kappa_b$ is sufficient, for some small number $\eta < 1$. Using time-domain master equation simulations (not shown) we have verified that using $\eta = 1/5$ is sufficient to stabilize the storage mode.

\subsection{\label{subsec:filter_design} Filter design}

Here we turn to the problem of filter design. What should we use as the physical embodiment of the filtered environment described by $Y(\omega)$? We can start by outlining some general design principles based on the preceding analysis. First, recall that the effective dissipation rate of the $b$ mode is $\kappa_{b,\text{eff}}(\Delta) = \text{Re}\left[Y^*(\omega_b + \Delta)\right]/\tilde{C}_b$, and second, note that the two-phonon dissipation rate is given by \cref{eq:kappa2_general_formula}, which we repeat here for convenience: $\kappa_2(\Delta) = 4g_2^2 \kappa_{b,\text{eff}}(\Delta) \left[4\tilde{\Delta}^2 + \kappa_{b,\text{eff}}^2(\Delta)\right]^{-1}$. As discussed in \cref{sec:multimode_stabilization}, different values of $\Delta$ are required to stabilize multiple modes with a single ATS --- one value for each mode. Therefore, the function $\kappa_2(\Delta)$ should be constant --- and as large as possible --- over a certain band of frequencies $B = [\omega_b - \Delta_\text{max}, \omega_b + \Delta_\text{max}]$. In effect, there should be a finite density of states that the $b$ mode can radiate into within this band. Outside of this band, however, the density of states should vanish in order to suppress correlated phase-flip errors resulting from the multiplexed stabilization (see \cref{sec:multimode_stabilization}). These requirements translate to a simple design principle: the function $\text{Re}\left[Y(\omega)\right]$ should ideally be a constant in the band $\omega \in B$, and zero elsewhere, much like in the toy model discussed in \cref{subsec:classical_filter_theory} where we took $\Delta_\text{max} = 2J$. This is akin to a resistor that only absorbs radiation at certain frequencies.

\subsubsection{General properties}

\begin{figure*}
\centering
\includegraphics[width=\linewidth]{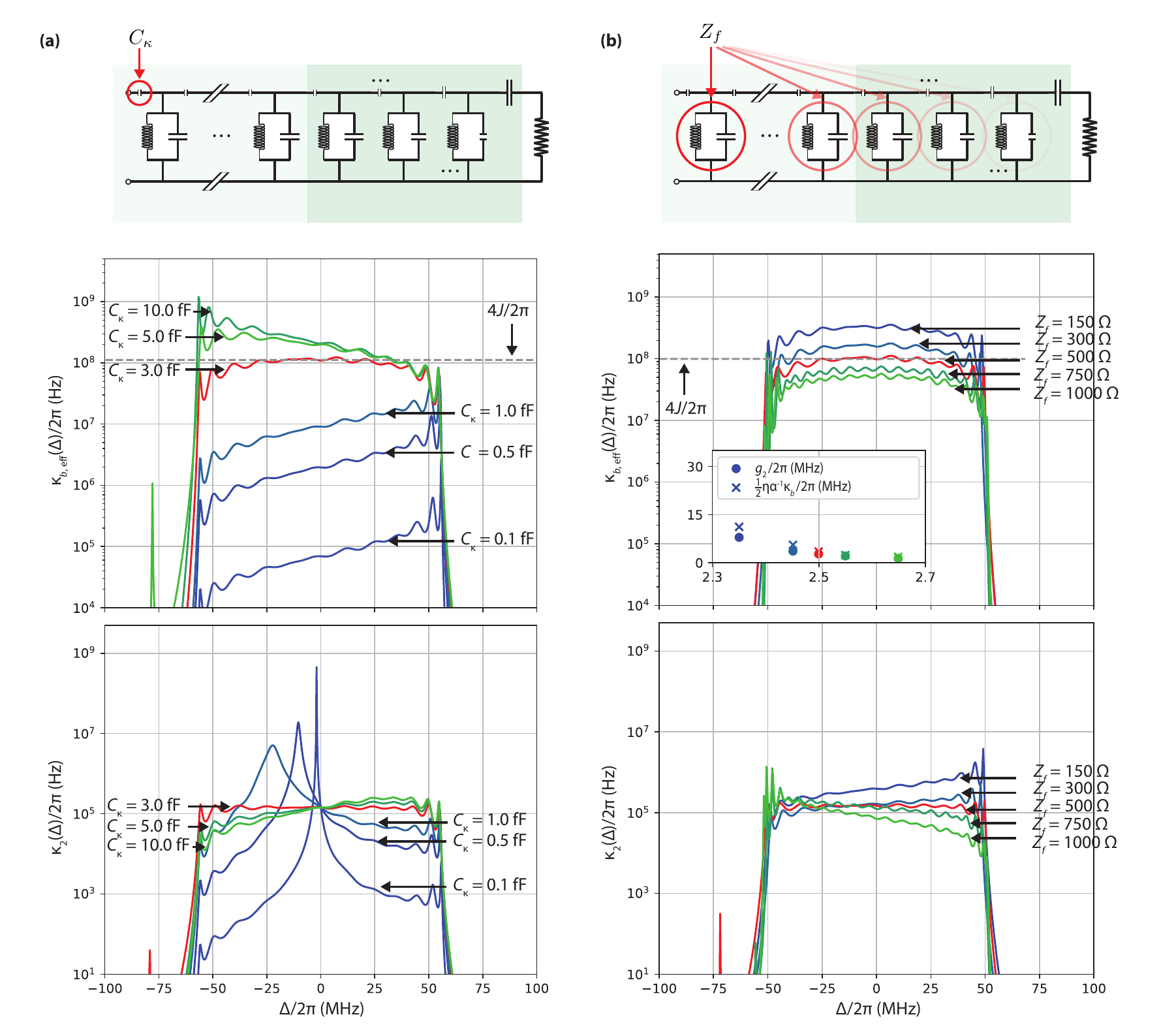}
\caption{\label{fig:filter_design_2} Exploring the filter design space. (a) Dependence of $\kappa_{b,\text{eff}}(\Delta)$ and $\kappa_2(\Delta)$ on the coupling capacitance $C_\kappa$ between the $b$ mode and the first filter resonator. Here we fix $\epsilon_{p,0} = 0.015$, $\omega_f/2\pi = 2.55 \, \text{GHz}$, $Z_f = 500 \, \Omega$, $\omega_b = \omega_f - 2J$, $Z_b = 1 \, \text{k}\Omega$, $C_c = 3.0 \, \text{fF}$, and $(N, N_t) = (10, 3)$. As $C_\kappa$ increases, we observe two regimes: an `undercoupled' regime $C_\kappa \ll C_c$ characterized by a sharply peaked $\kappa_2(\Delta)$, where the narrow $b$ mode filters the dissipation process, and an `overcoupled' regime $C_\kappa \gg C_c$ where $\kappa_2$ saturates and becomes asymmetric. In this latter regime the $b$ mode strongly hybridizes with the first filter cell. For large enough $C_\kappa$, their normal mode frequencies shift outside of the filter passband, forming bound resonances that are visible as sharp peaks to the left of the passband in some of the curves. The optimal value is $C_\kappa = C_c = 3.0 \, \text{fF}$, where $\kappa_2(\Delta)$ is maximized and flat, is shown in red. Note that at this optimal coupling, $\kappa_{b,\text{eff}} = 4J$ (gray dashed line). Note also that the adiabatic condition $g_2<\eta \kappa_b/2\alpha$ is not respected for several of the plots shown, as $g_2$ is fixed. (b) Dependence of $\kappa_{b,\text{eff}}(\Delta)$ and $\kappa_2(\Delta)$ on the characteristic filter impedance $Z_f$. In order to keep the filter bandwidth $4J$ constant, increasing $Z_f$ requires decreasing $C_c$, and to keep $g_2 < \eta \kappa_b/2\alpha$ (adiabatic threshold), increasing $Z_f$ requires increasing $\omega_f$ (which decreases $g_2$ due to the larger detuning between the $a$ and $b$ modes). The values used for the plotted curves are $\omega_f/2\pi = 2.4, \, 2.5, \, 2.55, \, 2.6, \, 2.7 \, \text{GHz}$, $C_c = 10, \, 4.5, \, 2.7, \, 1.7, \, 1.2 \, \text{fF}$, $\omega_b = \omega_f - 2J$, and $(N, N_t) = (10, 3), \, (10, 3), \, (10, 3), \, (14, 6), \, (14, 6)$. Larger values of $Z_f$ required larger $N_t$ to compensate for the larger impedance mismatch to the $50 \, \Omega$ line. We also fix $Z_b = 1 \, \text{k}\Omega$ here. The optimal value is $Z_f = Z_b/2 = 500 \, \Omega$, which produces a flat $\kappa_2(\Delta)$ curve (shown in red). Also note that at this optimal value, $\kappa_{b,\text{eff}} = 4J$ (gray dashed line). Inset: $g_2$ and $\eta \kappa_b/2\alpha$ corresponding to each of the simulations for the different values of $Z_f$, plotted as a function of $\omega_b$, showing the adiabatic constraint $g_2 < \eta \kappa_b/2\alpha$ is satisfied (here $\alpha = \sqrt{8}$ and $\eta = 1/5$).}
\end{figure*}

One of the simplest possible networks with these properties is a linear chain of $N$ LC resonators with capacitive couplings, as shown in \cref{fig:filter_design}(c). This resonator chain has a well-defined band with dispersion~\cite{Ferreira2020}
\begin{equation}
\label{eq:filter_dispersion}
\omega(k) = \omega_f + 2J[\cos(\pi k/N) - 1], \, \, k \in \{0, ..., N-1\}.
\end{equation}
Here $J$ is the coupling rate between neighboring resonators and is approximately given by
\begin{equation}
\label{eq:J_formula_1}
    J \approx \frac{\omega_f}{2}\frac{C_c}{C_f + 2C_c},
\end{equation}
where $\omega_f$ is the resonance frequency of each LC block, $C_c$ is the coupling capacitance, and $C_f$ is the shunt capacitance. This rate is directly tied to the filter bandwidth,
\begin{equation}
    (\text{bandwidth}) = 4J,
\end{equation}
and is controllable via $C_c$. Note also that we usually specify the frequency $\omega_f$ and impedance $Z_f$ of each LC block of the filter, which together with $C_c$ uniquely specify the shunt inductance $L_f = Z_f/\omega_f$ and shunt capacitance $C_f = 1/\omega_f Z_f$. Usually $C_f \gg C_c$, so
\begin{equation}
\label{eq:J_formula_2}
    J \sim \frac{1}{2}\omega_f^2 C_c Z_f.
\end{equation}
This means that for fixed values of $\omega_f$ and $C_c$, the filter bandwidth is directly proportional to $Z_f$. This formula will be useful shortly.

Normally the $N$ filter modes with dispersion relation \cref{eq:filter_dispersion} are standing waves that span the entire chain. These modes would therefore hybridize with the $b$ mode, effectively forming a ``multimode buffer'' with $N+1$ sharp resonances that the $a$ mode interacts with via the ATS. This is not the behavior we are interested in. Instead of a structure supporting standing resonances, we ideally seek a medium that is perfectly transparent to photons with frequency $\omega \in B$ and perfectly reflective otherwise. One way to achieve this is to add a small number of additional resonators at the end of the chain and rapidly ramp up the values of the coupling capacitances $C_c^{(i, i+1)}$ between neighboring cells (see \cref{fig:filter_design}(d)). We refer to this region as the `taper' in keeping with existing terminology~\cite{Ferreira2020}. The shunt capacitances are also adjusted in order to keep the frequency of each cell constant across the filter, including the taper cells. The effect of the taper is to significantly broaden the resonances of the structure so that the entire $B$ band is filled by their overlapping lineshapes, or alternatively, it allows the waves that propagate along the chain to be transmitted to the outside environment with negligible reflections.

We show in \cref{fig:filter_design}(d) the typical response of such a filter. The taper parameters (coupling capacitances and shunt capacitances) have been chosen to minimize the cost function $C = -\sum_{\omega \in B}\log \text{Re}[Y(\omega)]$, producing a relatively flat response over the band of interest $B$. We note that this choice of cost function is only a design heuristic that approximately produces the desired response.

\subsubsection{Optimizing the filter}

\begin{figure*}
\centering
\includegraphics[width=\linewidth]{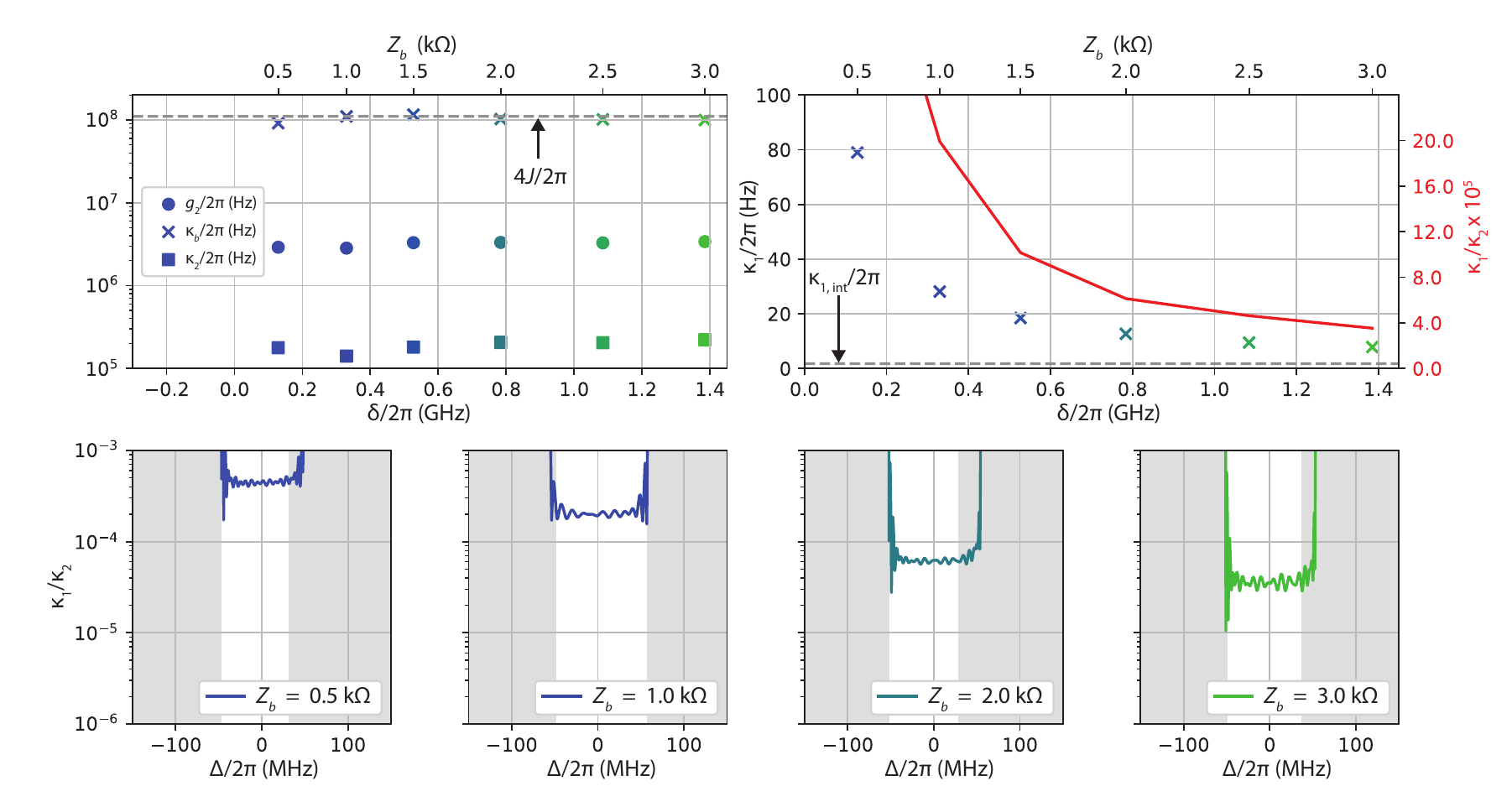}
\caption{\label{fig:loss_results} Behavior of $\kappa_1/\kappa_2$ in the large $Z_b$ limit. Top left panel: $g_2$, $\kappa_b$ and $\kappa_2$ plotted as a function of $Z_b$. The latter two rates are averages over the middle of filter passband, $\omega \in [\omega_b - J, \omega_b + J]$. The detunings $\delta = \omega_b - \omega_a$ corresponding to each similuation are also indicated on the horizontal axis. Each result is obtained by optimizing the filter for each value of $Z_b$, following the procedure outlined in \cref{subsec:filter_design}. We observe that all of these rates remain practically constant, an in particular $\kappa_{b} = 4J$. Top right panel: single-phonon relaxation rate $\kappa_1$ plotted as a function of $Z_b$ and $\delta$. This relaxation rate $\kappa_1 = \kappa_{1,\text{i}} + \kappa_{1,\text{p}}$ includes two contributions: the intrinsic loss $\kappa_{1,\text{i}}$ of the resonator, which here we assume has a fixed intrinsic quality factor $Q_{a,\text{i}} = \omega_a/\kappa_{1,\text{i}} = 10^9$, and the Purcell loss $\kappa_{1,\text{p}}$ due to its coupling to the buffer resonator. This latter rate has a contribution due to radiation into the waveguide (which is vanishingly small due to the strong filter suppression), and an important contribution $\sim (g/\delta)^2 \kappa_{b,\text{i}}$ due to the intrinsic decay of the buffer resonator itself, which we assume has $Q_{b,\text{i}} = 10^6$. This loss channel is \emph{not} suppressed by the filter. However, it can be mitigated by increasing the detuning $\delta$. Indeed, at large values of $\delta$, $\kappa_1$ asymptotes to $\kappa_{1,\text{i}}$ (gray dashed line). We also show the loss parameter $\kappa_1/\kappa_2$ plotted in red, where $\kappa_2$ is averaged over the filter band, which also asymptotes to a lower bound as $Z_b$ and $\delta$ become large. Bottom panels: loss spectra $\kappa_1/\kappa_2(\Delta)$ shown for a few selected values of $Z_b$. The gray shading indicates the regions where the adiabatic condition $g_2 < \eta \kappa_b/2\alpha$ is \emph{not} satisfied. These regions roughly correspond to the frequencies outside of the passband. Here $\alpha = \sqrt{8}$ and $\eta = 1/5$.}
\end{figure*}

Given fixed properties $\omega_a$, $\omega_b$, $Z_b$, etc. of the coupled storage-buffer system, what is the optimal choice of filter parameters? By now it should be self-evident what we mean by ``optimal'': those which maximize the two-phonon dissipation rate $\kappa_2(\Delta)$ across the filter band \{$\omega_b + \Delta \in B\}$ and make it as flat (constant) as possible within $B$. There are many parameters that describe the filter: $C_\kappa$, $C_c$, $\omega_f$, $Z_f$, $N$ (the number of ``unit cells''), $N_t$ (the number of ``taper cells''), and the set of coupling capacitances $\{C_c^{(i,i+1)}\}$ in the taper region. For fixed values of these first six parameters, the set $\{C_c^{(i,i+1)}\}$ is automatically optimized using the method described in the preceding paragraph, leaving six free parameters. What we show next is how to choose these parameters in order to optimize the function of interest $\kappa_2(\Delta)$ while simultaneously respecting the following constraints:
\begin{enumerate}
    \item $4J/2\pi = 100 \, \text{MHz}$ 
    \item $\omega_b = \omega_f - 2J$
    \item $g_2 < \eta \kappa_b/2\alpha$
\end{enumerate}
Constraint (1) is to ensure that photons created as a result of correlated decay of multiple storage modes during multiplexed stabilization have frequencies $\omega_{\text{corr. decay}} \not\in B$ outside of the passband. This prevents these photons from radiating into the environment and suppresses correlated phase-flip errors. The value $4J/2\pi = 100 \, \text{MHz}$ is approximately the largest possible bandwidth the filter can have while still satisfying this requirement --- for further detail see \cref{sec:multimode_stabilization}. Constraint (2) sets the $b$ mode frequency exactly in the middle of the passband, making the functions $\kappa_{b,\text{eff}}(\Delta)$ and $\kappa_2(\Delta)$ symmetric. This is not absolutely necessary but is rather a matter of convenience. Constraint (3) is to ensure that the system is in a regime where adiabatic elimination is valid, as found at the end of \cref{subsec:classical_filter_theory}. Here we fix $\alpha = \sqrt{8}$ and $\eta = 1/5$. Finally, we comment on what are reasonable values for $N$ and $N_t$. The number of taper cells $N_t$ depends on $Z_f$ and $Z_0 (= 50 \, \Omega)$, with $N_t$ needing to be larger the farther $Z_f$ deviates from $Z_0$. This agrees with the intuition that the taper is acting as an impedance-matching network. Once $Z_f$ and $N_t$ are chosen, we observe numerically that it is sufficient to choose a number of unit cells $N \gtrsim N_t/2$. Anything larger than this is unnecessary and does not change the results --- the chain being longer does not affect the dissipation rates we are interested in calculating. 

In \cref{fig:filter_design_2}(a) we show the effect of varying the capacitance $C_\kappa$, which sets the strength of coupling between the $b$ mode and the first resonator in the chain. Here $C_c = 3.0 \, \text{fF}$ is fixed, as well as $Z_f = 500 \, \Omega$. We observe two ``regimes'': a weak-coupling regime defined by $C_\kappa \ll C_c$, where $\kappa_b$ is small and $\kappa_2(\Delta)$ is sharply peaked near $\Delta=0$. This peak indicates that the $b$ mode is filtering the conversion process $g_2^* a^2 b^\dagger + \hc$, only allowing the emission of photons with frequencies inside its narrow bandwidth. Conversely, in the ``overcoupled'' regime $C_\kappa \gg C_c$ the $b$ mode decays rapidly, but interestingly $\kappa_2(\Delta)$ saturates and becomes asymmetric about $\Delta=0$. The optimal coupling $C_\kappa = C_c$, in between these two regimes, is where $\kappa_2(\Delta)$ is nearly perfectly symmetric and flat, and practically saturated. We remark that although $\kappa_b$ saturates to a value of around $\sim 4J$, it is possible to set $C_\kappa$ to a small enough value so that $\kappa_b$ is much smaller than this saturation value, assuming a flat $\kappa_2(\Delta)$ were not needed (which is not the case here).

Next, we show in \cref{fig:filter_design_2}(b) the effect of varying the characteristic filter impedance $Z_f$. Because of \cref{eq:J_formula_2}, in order to keep $J$ constant as $Z_f$ is varied we must adjust $C_c$ as well. Furthermore, we observe that $\kappa_b$ decreases with $Z_f$, and so in order to respect the adiabatic threshold $g_2 < \eta \kappa_b/2\alpha$ introduced earlier we reduce $g_2$ by increasing $\omega_b (=\omega_f - 2J)$ to further detune the $a$ and $b$ modes. The key observation is that the optimal value of $Z_f$, for which $\kappa_2(\Delta)$ is flat, is precisely $Z_f = Z_b/2$. This is true regardless of the chosen value of $Z_b$.

Together, these observations constrain $C_\kappa = C_c$ and $Z_f = Z_b/2$, and through \cref{eq:J_formula_2}, $C_c$ is constrained to the value $C_c \approx 4J/\omega_f^2 Z_b$. Once a value of $Z_b$ is chosen, the only remaining free parameter is $\omega_f$. But as we did in the preceding exercise, in what follows we will again use $\omega_f (= \omega_b + 2J)$ to fine-tune $g_2$ in order to satisfy the adiabaticity constraint. Therefore, with this design methodology, \textit{all of the (optimal) filter parameters are dictated by the properties of the storage and buffer resonators, with the exception of $N$ and $N_t \sim N/2$}.

\subsection{\label{subsec:loss_results} Optimization of the dimensionless loss \texorpdfstring{$\kappa_1/\kappa_2$}{k1/k2}}

We finally address the problem of optimizing the loss parameter $\kappa_1/\kappa_2$. For this we turn our attention back to \cref{eq:g2_scaling}, which we repeat here: $g_2 \sim Z_b^{5/2}$. Since $\kappa_2 \sim g_2^2 \sim Z_b^5$, the obvious question is, can we exploit this scaling to maximize $\kappa_2$? The answer is yes, but surprisingly this is not because of the obvious reason one would expect. In fact, as $Z_b$ increases, all of the filter parameters must be adjusted accordingly as described in \cref{subsec:filter_design}. We observe numerically that as this procedure is carefully repeated with different values of $Z_b$, \emph{the dissipation rate $\kappa_2$ remains practically constant and is independent of $Z_b$}. A semi-quantitative explanation is as follows: 1) because $C_\kappa = C_c$ and $Z_f = Z_b/2$ as found in the preceding section, the ``$b$'' resonator is hardly distinguishable from any other resonator in the main section of the filter chain. Its effective decay rate is therefore $\kappa_{b,\text{eff}} \sim J$, because the hopping rate $J$ is the rate that determines how quickly an excitation is transferred to the filter and out of the $b$ mode. Indeed, we observe numerically that this decay precisely matches the filter bandwidth, $\kappa_{b,\text{eff}} = 4J$, as we increase $Z_b$ while re-optimizing all of the filter parameters every time $Z_b$ changes. 2) Since $g_2 = \eta \kappa_b/2\alpha = 2\eta J/\alpha$ (to satisfy the adiabaticity constraint), $\kappa_{2, \text{max}} = 4 g_2^2/\kappa_b \approx 4 \eta^2 J/\alpha^2 \approx 4J/25\alpha^2$. \emph{Therefore, $\kappa_2$ only depends on the filter bandwidth, which is upper-bounded by the crosstalk analysis of \cref{sec:multimode_stabilization}, and the mean phonon number $|\alpha|^2$}. This result has important implications for our proposal and, as we will see shortly, imposes a lower bound on the phonon relaxation rate $\kappa_1$ required to reach the low values of $\kappa_1/\kappa_2$ that are necessary for our architecture.

Even though $\kappa_2$ depends solely on $J$ and $|\alpha|^2$, there is still something to be gained by increasing $Z_b$. In \cref{fig:loss_results} we show the ``loss spectrum'' $\kappa_1/\kappa_2(\Delta)$ for different values of $Z_b$. We observe that this loss does indeed decrease as $Z_b$ increases, but only relatively slowly and eventually asymptotes to a fixed value. This is because as $Z_b$ increases, $g_2$ increases as well, so the optimization procedure pushes $\omega_b$ further away from $\omega_a = \omega_b + \delta$ to compensate and keep $g_2$ below the adiabatic threshold $\eta \kappa_b/2\alpha$. In doing so, the loss $\sim (g/\delta)^2 \kappa_{b,\text{i}}$ that originates from the hybridization of the buffer and storage modes (here $g$ is the linear coupling between them) decreases as well. Note that only the intrinsic loss $\kappa_{b,\text{i}}$ of the buffer resonator enters this formula, because the radiative contribution is strongly suppressed since $\omega_a$ lies far outside the filter passband. Nevertheless this intrinsic contribution is still important, because in this proposal we operate under the assumption that the intrinsic decay rate $\kappa_{b,\text{i}}$ of the buffer mode (which is a superconducting circuit that suffers from several loss channels including two-level systems, quasiparticles, etc.) is at least two orders of magnitude larger than that of the storage mode, $\kappa_{1, \text{i}}$. In the limit $g/\delta \ll 1$, this contribution becomes negligibly small, and the phonon relaxation rate is purely intrinsic: $\kappa_1 \approx \kappa_{1, \text{i}}$. This causes the loss $\kappa_1/\kappa_2$ to asymptote to 
\begin{equation}
\label{eq:loss_terminal_value}
    \kappa_1/\kappa_2 \xrightarrow[Z_b \to \infty]{} \kappa_{1, \text{i}}|\alpha|^2/4 \eta^2 J. 
\end{equation}
This is of course only a theoretical exercise: one cannot build a device with arbitrarily large $Z_b$, and $\omega_b$ cannot be arbitrarily large. In particular, increasing the impedance will increase the size of the vacuum phase fluctuations, making the system more prone to instabilities, and a detailed analysis of this physics is left for future work. However, as we show in \cref{fig:loss_results}, there is a feasible range of values of $Z_b$ with which we could begin to approach the limiting value of loss in \cref{eq:loss_terminal_value}, depending on what assumptions we make for the intrinsic losses of the buffer and storage modes. These limiting values are plotted in \cref{fig:loss_results_main_text} in the main text as a function of $\kappa_{1, \text{i}}$ and for different filter bandwidths. It is important to emphasize that it maybe be possible to increase $J$ beyond its presently constrained value $4J/2\pi = 100 \, \text{MHz}$ through further innovations in the stabilization protocols, or by reducing the number of resonators coupled to each ATS. This is why we plot these curves for different bandwidths.

\section{\label{sec:multimode_stabilization} Multiplexed stabilization and crosstalk}

In this Appendix, we show how multiple storage resonators coupled to a common ATS can be stabilized simultaneously.  Coupling to a common ATS leads to crosstalk, and we discuss how this crosstalk can be quantified and mitigated. The main result of this Appendix is that the predominant sources of crosstalk can be effectively mitigated when up to five modes are coupled to a common ATS, so that the five-mode unit cells of our architecture are largely free of crosstalk. 

In~\Cref{subsec:effective_operator}, we begin by reviewing the effective operator formalism described in Ref.~\cite{Reiter2012}, which is the main tool we employ to analyze the dynamics of these multimode systems. In~\Cref{subsec:multiplex}, we describe our procedure for stabilizing multiple modes with a single ATS, and in~\Cref{subsec:crosstalk_sources} we discuss the resulting sources of crosstalk. Finally, in~\Cref{subsec:mitigation_filtering,subsec:mitigation_optimization} we show how this crosstalk can be effectively mitigated through a combination of filtering and storage mode frequency optimization. Throughout this appendix, we take $\hbar = 1$ to simplify notation.

\subsection{Effective operator formalism}
\label{subsec:effective_operator}

In this Appendix, we frequently employ adiabatic elimination as a tool to extract the effective dynamics of an open quantum system within some stable subspace. The purpose of this subsection is to describe the effective operator formalism that we employ in order to perform this adiabatic elimination. While adiabatic elimination has been described in a variety of prior works (see, e.g.,~\cite{Reiter2012,azouit2016,azouit2017}), we privilege the treatment in Ref.~\cite{Reiter2012} due to its simplicity and ease of application. We briefly review the relevant results. 
  
Consider an open quantum system evolving according to the master equation
\begin{equation}
\dot {\hat \rho}  = -i [\hat H,\hat \rho] + \sum_i \mathcal D[\hat L_i](\hat \rho),
\end{equation}
with Hamiltonian $\hat H$, jump operators $\hat L_i$, and where $\mathcal D[\hat L](\hat \rho) = \hat L \hat \rho \hat L^\dagger -\frac{1}{2}\left( \hat L^\dagger \hat L \hat \rho + \hat \rho \hat L^\dagger \hat L \right)$. We suppose that the system can be divided into two subspaces: a stable ground subspace, and a rapidly-decaying excited subspace, defined by the projectors $\hat P_g$ and $\hat P_e$, respectively. The Hamiltonian can be written in block form with respect to these subspaces as
\begin{equation}
\hat H =   \begin{pmatrix}
   \hat H_g & \hat V_-   \\
   \hat V_+ & \hat H_e  \\
  \end{pmatrix}
\end{equation}
where $\hat H_{g,e} =\hat  P_{g,e} \hat H \hat P_{g,e} $, and $\hat V_{+,-} = \hat P_{e,g} \hat H \hat P_{g,e}$. We also suppose that the jump operators take the system from the excited to the ground subspace, i.e., $ \hat L_i = \hat P_g \hat L_i \hat P_e $, and we define the non-Hermitian Hamiltonian
\begin{equation}
\hat H_{\mathrm{NH}} = \hat H_e -\frac{i}{2}\sum_i \hat L_i^\dagger \hat L_i.
\end{equation}
$\hat H_\text{NH}$ describes the evolution within the excited subspace; unitary evolution is generated by $\hat H_e$, while the remaining term describes the non-unitary, deterministic ``no jump'' evolution induced by the dissipators $\mathcal D [{\hat L_i}]$.  

The authors of Ref.~\cite{Reiter2012} consider the case where the evolution between the subspaces induced by $\hat V_{+,-}$ is perturbatively weak relative to the evolution induced by $\hat H_0 \equiv \hat H_g +\hat  H_{\mathrm{NH}}$. Because the excited subspace is barely populated due to the rapid decays, the dynamics of the system are well-approximated by those within the ground subspace, governed by the effective master equation
\begin{equation}
\dot {\hat \rho}  = -i [\hat H_\mathrm{eff},\hat \rho] + \sum_i \mathcal D[\hat L_{\mathrm{eff},i}](\hat \rho),
\end{equation} 
where
\begin{equation}
\hat H_\mathrm{eff} = -\frac{1}{2} \hat V_- \left[ \hat H_\mathrm{NH}^{-1} + \left(\hat H_\mathrm{NH}^{-1}\right)^\dagger \right] \hat V_+ + \hat H_g,
\end{equation}
and
\begin{equation}
\hat L_{\mathrm{eff},i} = \hat L_i \hat H_\mathrm{NH}^{-1} \hat V_+.
\end{equation}
These expressions apply for time-independent Hamiltonians. However, we will also be interested in situations where the perturbations $\hat V_{+,-}$ are time-dependent and take the form
\begin{align}
\hat V_+(t)  &= \sum_n \hat V_{+,n} e^{i \delta_n t}, \\
\hat V_-(t)  &= \sum_n \hat V_{-,n} e^{-i \delta_n t} .
\end{align}
In this case, the effective Hamiltonian and jump operators are given by
\begin{align}
\label{eq:Heff}
&\hat H_\mathrm{eff} = \hat H_g \nonumber \\
&-\frac{1}{2} \sum_{m,n}  \hat V_{-,n}  \left[ \hat H_{\mathrm{NH},m} ^{-1} + \left(\hat H_{\mathrm{NH},n}^{-1} \right)^\dagger \right]  
 \hat V_{+,m}  e^{i(\delta_m - \delta_n) t},
\end{align}
and
\begin{equation}
\hat L_{\mathrm{eff},i} = \hat  L_i \sum_n \hat H_{\mathrm{NH},n}^{-1} \hat V_{+,n} e^{i \delta_n t},
\end{equation}
where $\hat H_{\mathrm{NH},n} = \hat H_\mathrm{NH} +\delta_n$.

\subsection{Simultaneous stabilization of multiple cat qubits with a single ATS }
\label{subsec:multiplex}

We consider a collection of $N$ storage modes mutually coupled to a common reservoir. For the moment, we take reservoir to be a capacitively-shunted ATS (buffer resonator) with a large decay rate. The Hamiltonian of the system is
\begin{align}
    \hat H &= \hat H_d + \omega_b \hat b^\dagger \hat b + \sum_{n=1}^N \omega_n \hat a_n^\dagger \hat a_n \nonumber \\
    &-2\EJ \epsilon_p(t) \sin \left( \hat \phi_b + \sum_{n=1}^N \hat \phi_{n}  \right),
\end{align}
where $\hat H_d$ is a driving term (defined below), $\hat a_n$ ($\hat{b}$) is the annihilation operator for the $n$-th storage mode (buffer mode) with frequency $\omega_n$ ($\omega_b$), and $\hat \phi_n = \varphi_n (\hat a_n + \hat a_n^\dagger)$ is the phase across the ATS due to mode $n$, with vacuum fluctuation amplitudes $\varphi_n$. To stabilize multiple storage modes simultaneously, we apply separate pump and drive tones for each mode. Explicitly,
\begin{equation}
    \epsilon_p(t) = \sum_n \epsilon_p^{(n)} \cos\left(\omega_p^{(n)} t\right),
\end{equation}
and
\begin{equation}
    \hat H_d = \sum_n \left( \epsilon_d^{(n)} \hat b\, e^{i \omega_d^{(n)} t } + \mathrm{H.c.}\right).
\end{equation}
We choose the frequencies of the $n$-th pump and drive tones, respectively, as
\begin{align}
\label{eq:nth_pump}\omega_p^{(n)} &= 2\omega_n - \omega_b + \Delta_n,  \\
\label{eq:nth_drive}\omega_d^{(n)} & = \omega_b - \Delta_n,
\end{align}
where $\Delta_n$ denote detunings whose importance will be made clear shortly. Note that, in the architecture proposed in the main text, only a subset of the modes coupled to a given reservoir are stabilized by that reservoir. Accordingly, only the corresponding subset of the drives and pumps above need actually be applied.

\begin{figure*}
\begin{center}
\includegraphics[width=0.9\linewidth]{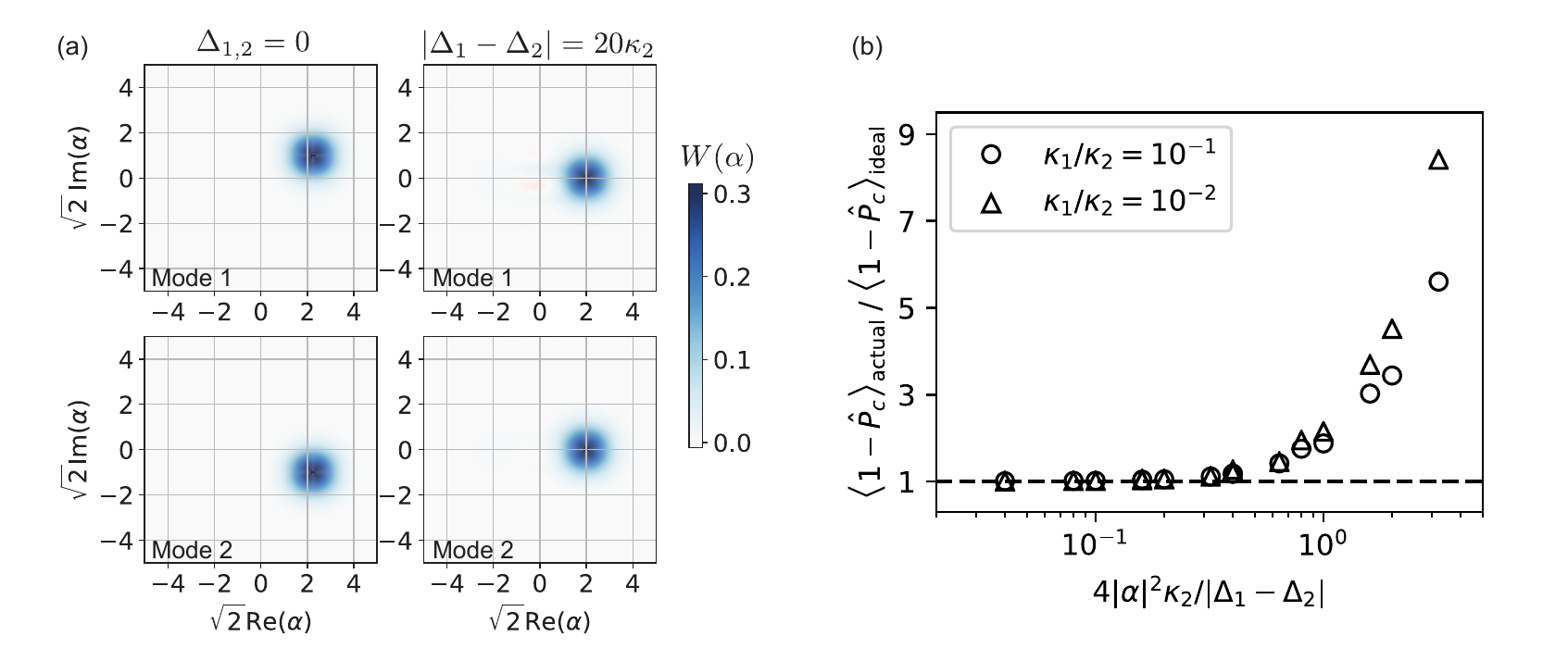}
\caption{Multiplexed stabilization. (a) Comparison of stabilization for $\Delta_n = 0$ and $|\Delta_n-\Delta_m|\gg 4|\alpha|^2\kappa_2$. Wigner plots are shown of two storage modes after evolution under the master equation $\dot{\hat \rho } = -i[\hat H,\hat \rho] +\kappa_b \mathcal{D}[\hat b]$, with $\hat H$ given by~(\ref{eq:desired_terms}). The storage modes are initialized in a product state $\ket{\beta_1}\ket{\beta_2}$ that does not lie in the code space but which is a steady state of (\ref{eq:coherent_dissipator}). Thus, when $\Delta_n = 0$ (left plots), the evolution is (approximately) trivial. The left two plots thus also serve as Wigner plots of the initial state $\ket{\beta_1}\ket{\beta_2}$. However, when $|\Delta_1-\Delta_2|\gg 4|\alpha|^2\kappa_2$ (right plots), the system evolves to the code space, defined here by $\alpha = \sqrt{2}$. 
(b) Validity of approximating \Cref{eq:coherent_dissipator} by \Cref{eq:incoherent_dissipator}. Master equations (\ref{eq:coherent_dissipator},\ref{eq:incoherent_dissipator}) are simulated (with decoherence added to each mode via the dissipators $\kappa_1\mathcal{D}[\hat a]$ and $\kappa_1\mathcal{D}[\hat a^\dagger \hat a]$), and the expectation value of $1-\hat P_c$ is computed once the system reaches its steady state. Here $\hat P_c$ denotes the projector onto the cat code space, and the subscripts ``actual'' and ``ideal'' denote expectation with respect to the steady states of (\ref{eq:coherent_dissipator}) and (\ref{eq:incoherent_dissipator}), respectively. The ratio of expectations, plotted on the vertical axis, quantifies the relative increase in population outside the code space. A ratio $\sim 1$ indicates the approximation works well. Parameters are chosen from the ranges $|\alpha|^2\in [1,4]$ and $|\Delta_1-\Delta_2|/\kappa_2\in [5,100]$.
}
\label{fig:multiplexed_stabilization}
\end{center}
\end{figure*}

To proceed, we expand the sine to third order and move to the frame where each mode rotates at its respective frequency. The resultant Hamiltonian is
\begin{align}
\label{eq:multimode_ham}
\hat H &\approx \sum_n \left( \epsilon_d^{(n)} \hat b\,  e^{-i \Delta_n t } + \mathrm{H.c.} \right) \nonumber \\
&-2 E_J \epsilon_p(t)\Bigg [ \varphi_b  \hat b\, e^{-i\omega_b t}  + \sum_n \varphi_n  \hat a_n\, e^{-i\omega_n t}  + \mathrm{H.c.}   \Bigg ] \nonumber \\
&+\frac{E_J}{3} \epsilon_p(t) \Bigg [ \varphi_b  \hat b\, e^{-i\omega_b t}  + \sum_n \varphi_n  \hat a_n\, e^{-i\omega_n t}  + \mathrm{H.c.}   \Bigg ]^3
\end{align}
This Hamiltonian contains terms that lead to the required two-phonon dissipators for each storage mode, 
\begin{equation}
\label{eq:desired_terms}
	\sum_n \left[ g_{2,n} \left( \hat a_n^2 -\alpha_n^2 \right)\hat b^\dagger e^{i\Delta_n t}  + \mathrm{H.c.} \right],
\end{equation}
with 
\begin{align}
g_{2,n} &= E_J \epsilon_p^{(n)}\varphi_n^2 \varphi_b/2, \\
\alpha_n^2 &=  - \left(\epsilon_d^{(n)}\right)^*/g_{2,n}.
\end{align}
However, the Hamiltonian~(\ref{eq:multimode_ham}) contains numerous other terms. While many of these other terms are fast-rotating and can be neglected in the rotating wave approximation (RWA), others can have non-trivial effects. For example, the interplay between the terms in the second and third lines of~(\ref{eq:multimode_ham}) gives rise to effective frequency shifts (a.c.~Stark shifts) of the buffer and storage modes, which modify the resonance conditions~(\ref{eq:nth_pump}) and (\ref{eq:nth_drive}).
One can calculate the magnitudes of these shifts (and hence compensate for them) by applying the effective operator approach of Refs.~\cite{james2007,Gamel2010timeaveraged}, in which case the Stark shifts are given by the coefficients of the $\hat b^\dagger \hat b$ and $\hat a^\dagger \hat a$ terms that arise in the effective Hamiltonian. Alternatively, the shifts can be calculated by moving to a displaced frame with respect to the linear terms on the second line of~(\ref{eq:multimode_ham}), as is done in Ref.~\cite{Lescanne2020}.  The Hamiltonian~(\ref{eq:multimode_ham}) also contains terms which lead to crosstalk, but we defer the discussion of these terms to the next section. For now, we keep only the desired terms~(\ref{eq:desired_terms}).

We proceed by adiabatically eliminating the lossy buffer mode $\hat b$, following the approach described in \cref{subsec:effective_operator}. Specifically, we designate the the ground subspace as the subspace where the buffer mode is in the vacuum state, and the excited subspace as the subspace where the buffer mode contains at least one excitation.      
We find that the effective dynamics of the storage modes within the ground subspace are described by the master equation
\begin{equation}
\label{eq:ME1}
 \dot {\hat \rho}  = -i[\hat H_\mathrm{eff},\hat \rho] + \mathcal D\left [ \sum_n \frac{\sqrt{\kappa_{b}}g_{2,n }}{\Delta_n - i \kappa_b/2}\left(  \hat a_n^2 -\alpha^2_n \right) e^{i\Delta_n t}  \right](\hat \rho),
\end{equation}
where
\begin{align}
\hat H_\mathrm{eff} &= -\frac{1}{2} \sum_{m,n} \Bigg\{  g_{2,n}^* g_{2,m}  (\hat a_n^2 - \alpha_n^2 )^\dagger  (\hat a_m^2 - \alpha_m^2 )  \nonumber \\ 
&\times \left[ \frac{1}{\Delta_m-i\kappa_b/2} + \frac{1}{\Delta_n + i\kappa_b/2} \right] e^{i(\Delta_m-\Delta_n) t}  \Bigg \}.
\end{align}
To understand these dynamics, let us first consider the simple case where $\Delta_n = 0$. The above master equation reduces to
\begin{equation}
\label{eq:coherent_dissipator}
 \dot {\hat \rho}  = \kappa_2 D\left [ \sum_n \left(  \hat a_n^2 - \alpha_n^2 \right)   \right](\hat \rho),
\end{equation}
where $\kappa_2 = 4 |g_2|^2/\kappa_b$. Any product of coherent states
\begin{equation}
\ket{\beta_1}\otimes \ket{\beta_2} \otimes \ldots \otimes \ket{\beta_N}
\end{equation}
that satisfies $\sum_n \beta_n^2 = \sum_n \alpha_n^2$ is a steady state of~(\ref{eq:coherent_dissipator}). The subspace of steady states includes states in the code space, for which $\beta_n^2 = \alpha_n^2$, but it also includes states outside of the code space. Because a strictly larger space is stabilized, when noise pushes the system outside of the code space, the stabilization is not guaranteed to return the system to the code space. The coherent dissipation in Eq.~(\ref{eq:coherent_dissipator}) is thus not sufficient for our purposes. 

Consider instead the case where the detunings are chosen to be distinct, satisfying $|\Delta_n - \Delta_m| \gg 4|\alpha|^2\kappa_2$. In this limit, we can drop the now fast-rotating cross terms in the dissipator in Eq.~(\ref{eq:ME1}), and the effective master equation becomes
\begin{equation}
\label{eq:incoherent_dissipator}
 \dot {\hat \rho}  = \sum_n \mathcal \kappa_{2,n}D\left [ \hat a_n^2 -\alpha^2_n  \right](\hat \rho),
\end{equation}
where
\begin{equation}
\kappa_{2,n} = \frac{\kappa_b|g_{2,n }|^2}{\Delta_n^2 + \kappa_b^2/4}.
\end{equation}
The incoherent dissipator~\cref{eq:incoherent_dissipator} stabilizes cat states in each mode, as desired. Thus, by simply detuning the pumps and drives used to stabilize each mode, multiple modes can be stabilized simultaneously and independently by a single ATS.

Two remarks about the approximation of \Cref{eq:coherent_dissipator} by \Cref{eq:incoherent_dissipator} are necessary. 
First, the condition $|\Delta_n - \Delta_m| \gg 4|\alpha|^2\kappa_2$ can be derived by expressing the operators in \Cref{eq:coherent_dissipator} in the displaced Fock basis (\Cref{appendix:Shifted Fock Basis}). Roughly speaking, the condition dictates that $|\Delta_n - \Delta_m|$ be much larger than the rate at which photons are lost from the stabilized modes. 
Second, we have neglected $\hat H_\mathrm{eff}$; the rotating terms in $\hat H_\mathrm{eff}$ can be dropped in the RWA in the considered limit, and the non-rotating terms provide an additional source of stabilization~\cite{puri2017} that we neglect for simplicity. 
It is also worth noting that the two-phonon dissipation rate, $\kappa_{2,n}$, decreases monotonically with $\Delta_n$. To avoid significant suppression of this engineered dissipation, one can choose $\Delta_n \lesssim  \kappa_b$ so that $\kappa_{2,n}$ remains comparable to $\kappa_2$, or alternatively one can exploit the filtering procedure described in \Cref{sec:single_mode_stabilization} and further analyzed in \Cref{subsec:mitigation_filtering} which enables strong effective dissipation even for $\Delta_n>\kappa_b$.

We demonstrate our scheme for multiplexed stabilization numerically in Fig.~\ref{fig:multiplexed_stabilization}. Through master equation simulations, we observe good stabilization for $|\Delta_1 -\Delta_2| \gg 4|\alpha|^2 \kappa_2$, but not $\Delta_{1,2} =0$, as expected. Moreover, we also quantify the validity of approximating \Cref{eq:coherent_dissipator} by \Cref{eq:incoherent_dissipator}. Strictly speaking, the approximation is valid only in the regime $ |\Delta_n - \Delta_m| \gg 4|\alpha|^2\kappa_2$, but we find that even for $ |\Delta_n - \Delta_m| \sim 4|\alpha|^2\kappa_2$ the stabilization works reasonably well, by which we mean that the population that leaks out of the code space is comparable for the two dissipators~(\ref{eq:coherent_dissipator}) and~(\ref{eq:incoherent_dissipator}), see \Cref{fig:multiplexed_stabilization}(b). The approximation breaks down beyond this point, and accounting for the additional terms in \Cref{eq:coherent_dissipator} becomes increasingly important.

We conclude this subsection by providing some physical intuition as to why detuning the pumps and drives allows one to stabilize multiple cat qubits simultaneously. When $\Delta_n = 0$, excitations lost from different storage modes via the buffer cannot be distinguished by the environment. As a result, we obtain a single coherent dissipator $\hat L \propto \sum_n (\hat a_n^2 -\alpha_n^2)$.  When distinct detunings are chosen for each mode, however, excitations lost from different modes via the buffer are emitted at different frequencies. When these buffer mode photons are spectrally resolvable, the environment can distinguish them, resulting in a collection of independent, incoherent dissipators $\hat L_n \propto  (\hat a_n^2 -\alpha_n^2)$ instead. The emitted photon linewidth is $4|\alpha|^2\kappa_2$, which can be seen by expressing $\kappa_2\mathcal{D}[\hat a^2 - \alpha^2]$ in the displaced Fock basis (\Cref{appendix:Shifted Fock Basis}).
Thus, the emitted photons are well-resolved when $|\Delta_n - \Delta_m| \gg 4|\alpha|^2\kappa_2$, which is the same condition assumed in the derivation of (\ref{eq:incoherent_dissipator}). We illustrate this idea pictorially in \Cref{fig:multiplexing_and_crosstalk}(a) of the main text.

\subsection{Sources of crosstalk}
\label{subsec:crosstalk_sources}

In this subsection we describe how undesired terms in the Hamiltonian~(\ref{eq:multimode_ham}) lead to crosstalk among modes coupled to the same ATS. In particular, we show that these undesired terms lead to effective dissipators and effective Hamiltonians that can cause correlated phase errors in the cat qubits. 

The predominant sources of crosstalk are undesired terms in the Hamiltonian~(\ref{eq:multimode_ham}) of the form
\begin{equation}
\label{eq:unwanted}
g_2\, \hat a_i \hat a_j \hat b^\dagger e^{i\delta_{ijk}  t} +\mathrm{H.c.},
\end{equation}
where
\begin{equation}
\delta_{ijk} = \omega_k^{(p)} - \omega_i - \omega_j + \omega_b,
\end{equation}
and we have neglected the dependence of $g_2$ on the indices $i,j$ for simplicity. 
In contrast to the other undesired terms in~(\ref{eq:multimode_ham}), these terms have the potential to induce large crosstalk errors because they both (i) have coupling strengths comparable to the desired terms (\ref{eq:desired_terms}), and (ii) can be resonant or near-resonant.   
In particular, the undesired term is resonant $(\delta_{ijk} = 0)$ for $2\omega_k +\Delta_k = \omega_i + \omega_j$. This resonance condition can be satisfied, for example, when the storage modes have near uniformly-spaced frequencies.

These unwanted terms may not be exactly resonant in practice, but we cannot generally guarantee that they will be rotating fast enough to be neglected in the RWA either. In contrast, all other undesired terms in~(\ref{eq:multimode_ham}) are detuned by at least $\min_n |\omega_n - \omega_b|$, which is on the order of $\sim 2\pi * 1 \, \text{GHz}$ for the parameters considered in this work. We therefore focus on crosstalk errors induced by the terms~(\ref{eq:unwanted}).

The terms~(\ref{eq:unwanted}) can lead to three different types of correlated errors:
\begin{itemize}
\item{Type I: Stochastic errors induced by effective dissipators}
\item{Type II: Stochastic errors induced by effective Hamiltonians}
\item{Type III: Coherent errors induced by effective Hamiltonians}
\end{itemize}
We describe each type of error in turn. Without mitigation (see \Cref{subsec:mitigation_filtering,subsec:mitigation_optimization}), these correlated phase errors could be a significant impediment to performing high-fidelity operations.

\subsubsection*{Type I: stochastic errors induced by effective dissipators}
The terms~(\ref{eq:unwanted}) can lead to correlated phonon losses at rates comparable to $\kappa_2$, resulting in  significant correlated phase errors in the cat qubits. 
These deleterious effects manifest when one adiabatically eliminates the buffer mode. Explicitly, we apply the effective operator formalism described in Subsecton~\ref{subsec:effective_operator} to the operators
\begin{align}
\hat H^{(1)} &= g_2\, \hat a_i \hat a_j \hat b^\dagger e^{i\delta_{ijk}  t} +\mathrm{H.c.}, \\
\hat L^{(1)} &= \sqrt{\kappa_b}\, \hat b
\end{align}
and obtain the effective operators
\begin{align}
\hat H_\mathrm{eff}^{(1)} &= -\frac{|g_2|^2 \delta_{ijk}}{\delta_{ijk}^2+\kappa_b^2/4} (\hat a_i \hat a_j)^\dagger (\hat a_i \hat a_j) +\mathrm{H.c.}, \\
\label{eq:Leff1}
\hat L_\mathrm{eff}^{(1)} &= \frac{g_2 \sqrt{\kappa_b}}{\delta_{ijk} - i \kappa_b/2} \hat a_i \hat a_j  e^{i\delta_{ijk} t}.
\end{align}
The effective Hamiltonian preserves phonon-number parity and thus does not induce phase flips. The effective jump operator $\hat L_\mathrm{eff}$ describes correlated single-phonon losses in modes $i$ and $j$ at a rate
\begin{equation}
\kappa_\mathrm{eff} = \frac{\kappa_b |g_2|^2 }{\delta_{ijk}^2 + \kappa_b^2/4}  
\end{equation}
which is comparable to $\kappa_2$ for $\delta_{ijk} \lesssim \kappa_b$. These correlated single-phonon losses induce correlated phase flips in the cat qubits, which can be seen by projecting $\hat L_\mathrm{eff}$ into the code space, 
\begin{equation}
\hat L_\mathrm{eff}^{(1)} \rightarrow \sqrt{\kappa_\mathrm{eff}}\, \alpha^2  \hat Z_i \hat Z_j e^{i\delta_{ijk} t}.
\end{equation}

\subsubsection*{Type II: stochastic errors induced by effective Hamiltonians}
The interplay between different terms of the form~(\ref{eq:unwanted}) can lead to further correlated errors. As an example, consider the operators
\begin{align}
\hat H^{(2)} &= g_2\, \hat a_i \hat a_j \hat b^\dagger e^{i\delta_{ijk}  t} + g_2\, \hat a_\ell \hat a_m \hat b^\dagger e^{i\delta_{\ell m n}  t} +\mathrm{H.c.}, \\
\hat L^{(2)} &= \sqrt{\kappa_b}\, \hat b.
\end{align}
Adiabatically eliminating the buffer mode yields,
\begin{align}
\label{eq:Heff2}
\hat H^{(2)}_\mathrm{eff} &=  
\left[ \chi (\hat a_i \hat a_j )^\dagger  (\hat a_\ell \hat a_m )   e^{i(\delta_{\ell m n}-\delta_{i j k}) t}  +\mathrm{H.c.} \right] + \ldots,  \\
\hat L^{(2)}_\mathrm{eff} &= \frac{g_2 \sqrt{\kappa_b}}{\delta_{ijk} - i \kappa_b/2} \hat a_i \hat a_j  e^{i\delta_{ijk} t} \nonumber \\ 
&+ \frac{g_2 \sqrt{\kappa_b}}{\delta_{\ell m n} - i \kappa_b/2} \hat a_\ell \hat a_m  e^{i\delta_{\ell m n} t}.
\end{align}
where
\begin{equation}
\chi = -\frac{|g_2|^2}{2} \left[ \frac{1}{\delta_{ijk}-i\kappa_b/2} + \frac{1}{\delta_{\ell m n} + i\kappa_b/2} \right]   \nonumber \\ 
\end{equation}
and ``$\ldots$'' denotes additional terms in the effective Hamiltonian that preserve phonon-number parity.  Note that the effective dissipator $\hat L^{(2)}_\mathrm{eff}$ leads to Type I correlated phase errors.  Indeed, for sufficiently large $|\delta_{ijk}-\delta_{\ell m n}|$, the action of $\hat L^{(2)}_\mathrm{eff}$ can be approximated by replacing it with two independent dissipators of the form~(\ref{eq:Leff1}). 

What is different about this example is that the effective Hamiltonian $\hat H_\mathrm{eff}^{(2)}$ contains terms $\propto (\hat a_i \hat a_j )^\dagger (\hat a_\ell \hat a_m )$ that generally do not preserve phonon-number parity. Such terms can unitarily evolve the system out of the code space, changing the parity in the process. In turn, the engineered dissipation returns the system to the code space, but it does so without changing the parity. Therefore, the net effect of such excursions out of the code space and back is to induce \emph{stochastic} parity-flips in the storage modes, which manifest as correlated phase errors on the cat qubits. The errors are stochastic even though the evolution generated by $\hat H^{(2)}_\mathrm{eff}$ is unitary because the stabilization itself is stochastic. Specifically, the errors are of the form $\mathcal{D}[\hat Z_i \hat Z_j \hat Z_\ell \hat Z_m]$, which one can show by adiabatically eliminating the excited states of the storage modes (see \Cref{appendix:Shifted Fock Basis}). 

\subsubsection*{Type III: coherent errors induced by effective Hamiltonians}

The parity-non-preserving effective Hamiltonian $\hat H^{(2)}_\mathrm{eff}$ also induces non-trivial coherent evolution within the code space. This can be seen by projecting $\hat H_\mathrm{eff}^{(2)}$ into the code space
\begin{equation}
\label{eq:Heff_2}
\hat H^{(2)}_\mathrm{eff} \rightarrow  ( |\alpha|^4 \chi \hat Z_i \hat Z_j \hat Z_\ell \hat Z_m e^{i(\delta_{\ell m n}-\delta_{i j k}) t}  + \mathrm{H.c.}).
\end{equation}
This undesired evolution does not decohere the system but can nevertheless degrade the fidelity of operations. See further discussion in \Cref{subsec:mitigation_optimization}.

\subsection{Crosstalk mitigation: filtering}
\label{subsec:mitigation_filtering}

In this subsection, we show how Type I and Type II crosstalk errors can be suppressed by placing a bandpass filter at the output port of the buffer mode (see \Cref{subsec:classical_filter_theory} for additional discussion of filtering). The purpose of the filter is to allow photons of only certain frequencies to leak out of the buffer, such that the desired engineered dissipation remains strong but spurious dissipative processes are suppressed. A crucial requirement of this approach is that the desired dissipative processes be spectrally resolvable from the undesired ones, and we show that adequate spectral resolution is achievable in the next section (\Cref{subsec:mitigation_optimization}).

\begin{figure*}
\begin{center}
\includegraphics[width=\linewidth]{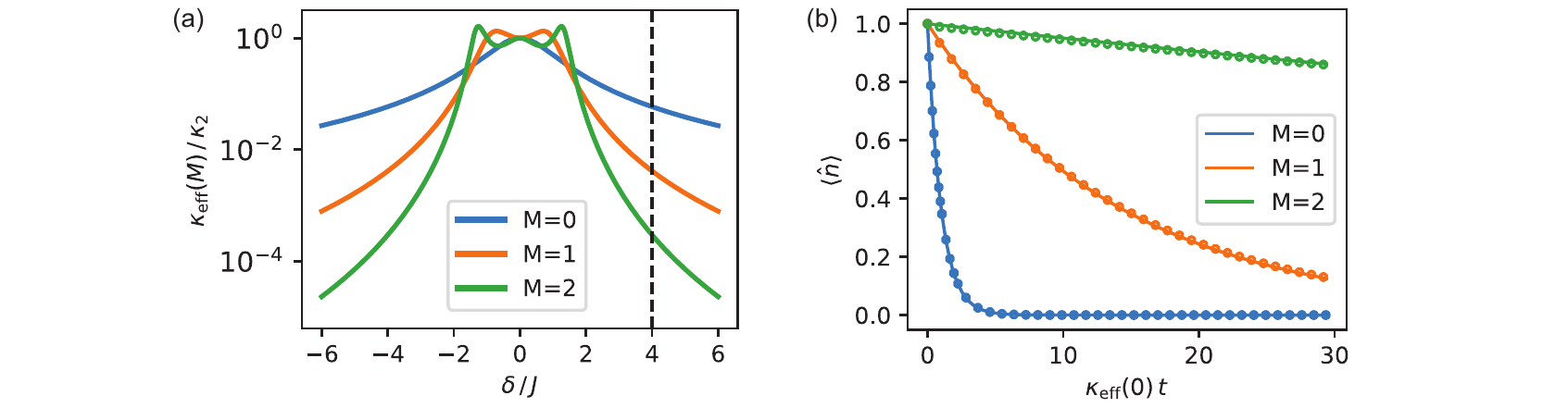}
\caption{Suppression of Type I errors. (a) Plots of $\kappa_{\mathrm{eff}}(M)$ as a function of the detuning, $\delta$, of the unwanted term. (b) Master equation simulations.  The system is initialized with a single excitation in the storage mode and evolved according to the dynamics $\dot{\hat \rho } = -i[(g_2 \hat a \hat b^\dagger e^{i\delta t}+\mathrm{H.c.})+\hat H_{\mathrm{buffer+filter}},\hat \rho ]+\mathcal{D}[\hat L^{(3)}](\hat\rho)$. These dynamics are analogous to those generated by $\hat H^{(3)}$ and $\hat L^{(3)}$; in both cases the unwanted term induces losses at rates $\kappa_\mathrm{eff}(M)$. Simulation results are indicated by open circles, and the analytical expressions for $\kappa_{\mathrm{eff}}(M)$ are plotted as solid lines. Parameters: $\alpha = \sqrt{2}, \kappa_c/g_2 = 10, J/g_2 = 5.$ For (b), $\delta = 4J$, as indicated by the dashed line in (a).}
\label{fig:decay_rates}
\end{center}
\end{figure*}

We begin by providing a quantum mechanical model of a bandpass filter~\cite{sete2015,Ferreira2020}. While a detailed classical analysis of the filter is given in \Cref{subsec:classical_filter_theory}, here we employ a complementary quantum model. The quantum model not only allows us to study the filter's effects numerically via master equation simulations, but it is also sufficiently simple so as to enable a straightforward analytical treatment via the effective operator formalism described in \Cref{subsec:effective_operator}. 

Motivated by the filter designs described in \cref{subsec:filter_design}, we employ a tight-binding model where the filter consists of a linear chain of $M$ bosonic modes with annihilation operators $\hat c_i$, and each with the same frequency $\omega_b$. Modes in the chain are resonantly coupled to their nearest neighbors with strength $J$. The first mode in the chain couples to the buffer mode $\hat b$, which is no longer coupled directly to the open waveguide. Instead, the $M$-th mode is now the one which couples strongly to the waveguide, such that its single-photon loss rate is given by $\kappa_c$. The buffer-filter system is described by the Hamiltonian (in the rotating frame)
\begin{equation}
\hat H_{\mathrm{buffer + filter}} = J(\hat c_1^\dagger \hat b + \hat c_1 \hat b^\dagger) + \sum_{i = 1}^{M-1} J(\hat c_{i+1}^\dagger  \hat c_i   +\hat c_{i+1}  \hat c_i^\dagger),
\end{equation}
together with the dissipator $\kappa_c\mathcal{D}[\hat c_M]$. We show below that these additional modes act as a bandpass filter, with center frequency $\omega_b$ and bandwidth $4J$, and they suppresses the emission of photons with frequencies outside of this passband. 

\subsubsection*{Suppression of Type I errors}
 To illustrate the suppression of Type I errors, we consider the operators
\begin{align}
\hat H^{(3)} &= \left(g_2\, \hat a_i \hat a_j \hat b^\dagger e^{i\delta_{ijk}  t} +\mathrm{H.c.} \right) + \hat H_{\mathrm{buffer + filter}}, \\
\hat L^{(3)} &= \sqrt{\kappa_c}\, \hat c_M
\end{align}
where the first term in $\hat H^{(3)}$ is the same as the unwanted term $\hat H^{(1)}$ from \Cref{subsec:crosstalk_sources}. We adiabatically eliminate \emph{both the buffer and filter modes} in order to obtain an effective dynamics for only the storage modes. We note that adiabatically eliminating the buffer and filter modes together is not fundamentally different from adiabatically eliminating the buffer; both calculations are straightforward applications of the methods in Subsection~\ref{subsec:effective_operator}. 
We obtain the effective dissipator
\begin{equation}
\hat L^{(3)}_\mathrm{eff} = \sqrt{\kappa_\mathrm{eff} (M)}\, \hat a_i \hat a_j  e^{i\delta_{ijk} t}
\end{equation}
where the rates for the first few values of $M$ are
\begin{align}
\label{eq:eff_loss_rate_M0}
\kappa_\mathrm{eff} (0) &=  \frac{\kappa_c |g_2|^2 }{\delta_{ijk}^2 + \kappa_c^2/4} \approx \kappa_c \frac{|g_2|^2}{\delta_{ijk}^2}\\
\label{eq:eff_loss_rate_M1}
\kappa_\mathrm{eff} (1) &=  \frac{\kappa_c |g_2|^2 J^2 }{(J^2-\delta_{ijk}^2)^2 + \delta_{ijk}^2\kappa_c^2/4}  \nonumber \\ 
&\approx \kappa_\mathrm{eff} (0) \left(\frac{J}{\delta_{ijk}}\right)^2 \\ \label{eq:eff_loss_rate_M2}
\kappa_\mathrm{eff} (2)  &=  \frac{\kappa_c |g_2|^2 J^4 }{(2J^2\delta_{ijk}-\delta_{ijk}^3)^2 + (J^2-\delta_{ijk}^2)^2\kappa_c^2/4} \nonumber \\
&\approx \kappa_\mathrm{eff} (0) \left(\frac{J}{\delta_{ijk}}\right)^4,
\end{align} 
where the approximations assume that $\delta_{ijk}\gg J,\kappa_c$.  In this regime, $\kappa_\mathrm{eff}(M)$ is exponentially suppressed with increasing $M$ via the factor $(J/\delta_{ijk})^{2M}$. 

We plot these rates as a function of $\delta_{ijk}$ in \Cref{fig:decay_rates}(a), where the exponential suppression of the decoherence rates outside the filter band is evident. \Cref{fig:decay_rates}(a) should be understood as analogous to \cref{fig:filter_design_2} in \cref{sec:single_mode_stabilization}, though we emphasize that here the rates are derived from a fully quantum model of the filter. We also remark that unlike in \cref{sec:single_mode_stabilization}, where the emphasis was on detailed classical filter design, here we do not taper the filter. This explains the ``ripples'' in $\kappa_\text{eff}$ within the filter passband. \Cref{fig:decay_rates}(b) shows the results of analogous master equation simulations; good quantitative agreement with the analytical expressions is observed. 
Thus we conclude that Type I errors are indeed suppressed by the filter, provided $|\delta_{ijk}|> 2J $.

\begin{figure*}
\begin{center}
\includegraphics[width=\linewidth]{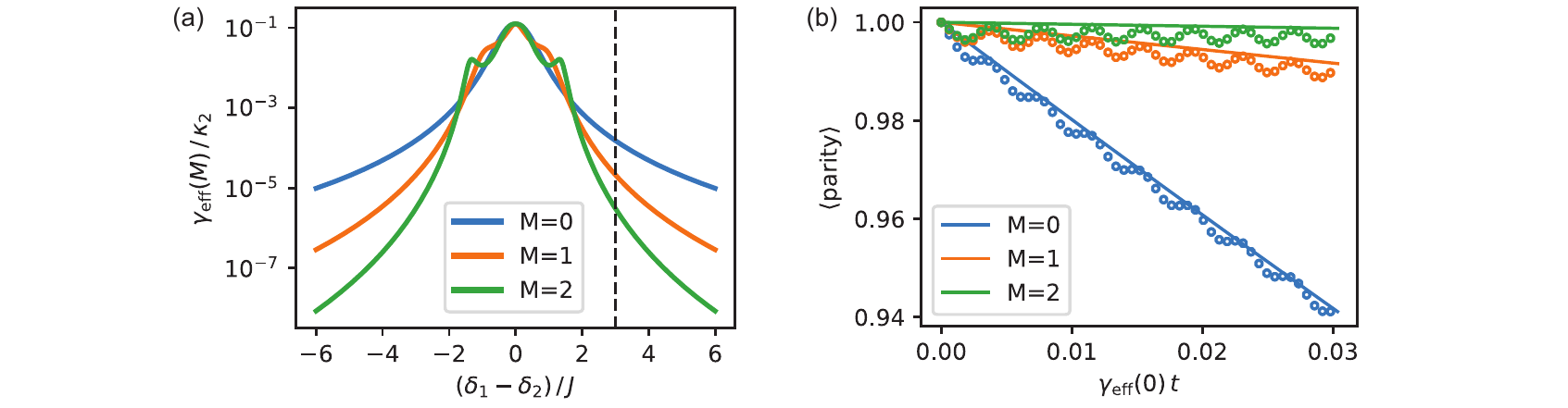}
\caption{Suppression of Type II errors. (a) Plots of $\gamma_{\mathrm{eff}}(M)$ as a function of the detuning, $\delta_1-\delta_2$, of the effective Hamiltonian. (b) Master equation simulations.  The storage mode is initialized in the even parity cat state and evolved according to the dynamics $\dot{\hat \rho } = -i[\hat{\bar H}^{(4)},\hat \rho ]+\mathcal{D}[\hat L^{(4)}](\hat\rho)$. Simulation results are indicated by open circles, and the analytical expressions for $\gamma_{\mathrm{eff}}(M)$ are plotted as solid lines.  Parameters: $\alpha = \sqrt{2}, \kappa_c/g_2 = 10, J/g_2 = 5$. Rather than specify values for $g$ and $\delta_{1,2}$, we simply fix $\chi_\mathrm{eff}(M)/g_2 = 0.2$. For (b), $\delta = 3J$, as indicated by the dashed line in (a).
}
\label{fig:typeii_suppression}
\end{center}
\end{figure*}

\subsubsection*{Suppression of Type II errors}
To illustrate the suppression of Type II errors, we construct a simple toy model that both captures the relevant physics and is easy to study numerically. Consider the operators
\begin{align}
\label{eq:H4}
\hat H^{(4)} &=
\left( g\, \hat a \hat b^\dagger e^{i\delta_{1}  t} + g \, \hat b^\dagger e^{i\delta_{2}  t} +\mathrm{H.c.}\right)\nonumber\\
&+\left[g_2 (\hat a^2 - \alpha^2) \hat b^\dagger +\mathrm{H.c.}\right]  + \hat H_\mathrm{buffer + filter}  \\
 \hat L^{(4)} &= \sqrt{\kappa_c}\, \hat c_M.
\end{align}
where $\hat a$ is the annihilation operator for the single storage mode that we consider in this model.  
In this toy model, the first line of $\hat H^{(4)}$ should be understood as analogous to $\hat H^{(2)}$. Indeed we obtain the former from the latter by replacing $\hat a_i \hat a_j \rightarrow \hat a$ and $\hat a_\ell \hat a_m \rightarrow 1$.

Adiabatically eliminating the buffer and filter modes yields the effective operators
\begin{align}
\hat H^{(4)}_\mathrm{eff} &= \left[ \chi_\mathrm{eff}(M)\, \hat a\, e^{i (\delta_1 - \delta_2) t} +\mathrm{H.c.} \right] +\ldots\\
\hat L^{(4)}_\mathrm{eff} &= \sqrt{\kappa_\mathrm{eff}^{( \delta_1)}(M)} \, \hat a\, e^{i \delta_1 t} + \sqrt{\kappa_{\mathrm{eff}}^{(0)}(M)}(\hat a^2-\alpha^2).
\end{align}
Here, ``$\ldots$'' denotes a parity-preserving term $(\propto \hat a^\dagger \hat a)$ that we neglect, $\kappa_\mathrm{eff}^{(\delta)}(M)$ denotes the effective loss rate [\cref{eq:eff_loss_rate_M0,eq:eff_loss_rate_M1,eq:eff_loss_rate_M2}] with the replacement $\delta_{ijk}\rightarrow\delta$, and 
\begin{equation}
\chi_\mathrm{eff}(M) \approx -\frac{|g|^2}{2}\left(\frac{1}{\delta_1} + \frac{1}{\delta_2} \right)
\end{equation}
is independent of $M$ in the limit $\delta_{1,2} \gg J,\kappa_b$. The first term in $\hat L_\mathrm{eff}^{(4)}$ gives rise to the Type I errors that are suppressed by the filter, as already discussed. Our present interest is the Type II errors induced by the interplay of $\hat H_\mathrm{eff}^{(4)}$, the stabilization, and the filter. 

Unfortunately, the effective operators $\hat H_\mathrm{eff}^{(4)}$ and $\hat L_\mathrm{eff}^{(4)}$ do not properly capture this interplay. In particular, it follows from energy conservation that Type II errors induced by $\hat H^{(4)}_\mathrm{eff}$ result in photon emissions at frequency $\omega_b + \delta_2-\delta_1$. Intuitively, such emissions should be exponentially suppressed when this frequency lies outside the filter band. However, this suppression is not apparent in the operators $\hat H^{(4)}_\mathrm{eff},\hat L_\mathrm{eff}^{(4)}$ because, in the course of deriving $\hat H^{(4)}_\mathrm{eff}$, we already eliminated the filter. After adiabatic elimination the only vestige of the filter is the term $\sqrt{\kappa_{\mathrm{eff}}^{(0)}(M)}(\hat a^2-\alpha^2)$, which embodies the behavior of the filter at frequency $\omega_b$, \emph{but not at frequency} $\omega_b + \delta_2-\delta_1$. As such, proceeding to calculate the Type II error rate from these operators is not valid, and an alternate approach is required. 

In order to properly capture the subtle interplay between the effective Hamiltonian, the stabilization, and filter, we defer adiabatic elimination and instead begin by calculating an effective Hamiltonian that describes the time-averaged dynamics generated by $\hat H^{(4)}$. We restrict our attention to a regime where the terms on the first line of \Cref{eq:H4} are rapidly rotating, so that evolution generated by $\hat H^{(4)}$ is well approximated by its time average. We calculate the time-averaged effective Hamiltonian $\hat{\bar H}^{(4)}$ following the approach described in Refs.~\cite{Gamel2010timeaveraged,james2007},
\begin{align}
&\hat{\bar H}^{(4)} = \left[g_2 (\hat a^2 - \alpha^2) \hat b^\dagger +\mathrm{H.c.}\right]  + \hat H_\mathrm{buffer + filter} \nonumber \\ 
    &-\frac{|g|^2}{2}\left(\frac{1}{\delta_1}+\frac{1}{\delta_2}\right)\left(2\hat b^\dagger \hat b + 1\right)
 \left(\hat ae^{i(\delta_1-\delta_2)t} +\mathrm{H.c.}\right) 
\end{align}
where we have neglected a parity-preserving term $(\propto \hat a^\dagger \hat a)$, and terms rotating at the fast frequencies $\delta_{1,2}$. 
Notice that 
\begin{equation}
    \hat{\bar H}^{(4)} \approx  \left[g_2 (\hat a^2 - \alpha^2) \hat b^\dagger +\mathrm{H.c.}\right]  + \hat H_\mathrm{buffer + filter} + \hat{H}_\mathrm{eff}^{(4)},
\end{equation}
where the approximation is obtained by preemptively replacing $\hat b^\dagger \hat b$ with its expected value of $0$. Doing so reveals that $\hat H_\mathrm{eff}^{(4)}$ can be understood as arising from the time-averaged dynamics of the the unwanted terms in $\hat{H}^{(4)}$ in the limit of large $\delta_{1,2}$. 
In effect, time averaging provides a way of introducing $\hat H_\mathrm{eff}^{(4)}$ into the dynamics without having to eliminate the filter, thereby allowing us to study the interplay of the filter and effective Hamiltoninan. 

We proceed by taking the operators $\hat{\bar H}^{(4)}$ and $\hat L ^{(4)}$ and adiabatically eliminating the buffer, the filter, and all excited states of the storage mode, i.e.~all states that do not lie in the code space. Adiabatically eliminating the storage mode excited states is valid in the regime where the engineered dissipation is strong relative to couplings that excite the storage mode ($\hat H_\mathrm{eff}^{(4)}$ in this case), such that these excited states are barely populated. See \Cref{appendix:Shifted Fock Basis} for further details. We obtain 
\begin{align}
\hat {\bar H}^{(4)}_\mathrm{eff} &=   \chi_\mathrm{eff}(M)\, \alpha \hat Z\, e^{i(\delta_1 - \delta_2) t} + \mathrm{H.c.}, \\
\hat {\bar L}_\mathrm{eff}^{(4)} &=\sqrt{ \gamma_\mathrm{eff}(M) } \hat Z.
\end{align}
\begin{widetext}
The rates for the first few values of $M$ are
\begin{align}
\gamma_\mathrm{eff}(0) &= \frac{
4\kappa_c |2 g_2 \alpha\, \chi_\mathrm{eff}(0)|^2
}{
4\left(|2g_2\alpha|^2-\delta_{12}^2  \right)^2+ \delta_{12}^2 \kappa_c^2}, \\
\gamma_\mathrm{eff}(1) &=
\frac{
4 J^2 \kappa_c |2 g_2 \alpha\, \chi_\mathrm{eff}(1)|^2
}{
4\delta_{12}^2\left(J^2 + |2g_2\alpha|^2 -\delta_{12}^2 \right)^2 + \left(|2g_2\alpha|^2-\delta_{12}^2 \right)^2 \kappa_c^2} \approx \gamma_{\mathrm{eff}}(0)\left(\frac{J}{\delta_{12}}\right)^2,
\\
\gamma_\mathrm{eff}(2) &=
\frac{
4 J^4 \kappa_c |2 g_2 \alpha\, \chi_\mathrm{eff}(2)|^2
}{
4\left( |2g_2\alpha|^2(J-\delta_{12})(J+\delta_{12}) + \delta_{12}^4 - 2J^2 \delta_{12}^2   \right)^2 + \delta_{12}^2\left(|2g_2\alpha|^2+J^2-\delta_{12}^2 \right)^2 \kappa_c^2} \approx \gamma_{\mathrm{eff}}(0)\left(\frac{J}{\delta_{12}}\right)^4,
\end{align}
\end{widetext}
where we have used the shorthand $\delta_{12}\equiv \delta_1-\delta_2$ to simplify the expressions, and the approximations are obtained in the in the limit of large $|\delta_{1}-\delta_{2}|$.
In this limit, we find that the phase flip rate is exponentially suppressed by the filter,
\begin{equation}
\gamma_\mathrm{eff}(M)\approx  \gamma_\mathrm{eff}(0) \left(\frac{J}{\delta_1 - \delta_2}\right)^{2 M},
\end{equation}
as expected.

We plot the rates $\gamma_\mathrm{eff}(M)$ as a function of $\delta_{1}-\delta_{2}$ in \Cref{fig:typeii_suppression}(a), where the exponential suppression of the decoherence rates outside the filter band is again evident. \Cref{fig:typeii_suppression}(b) shows the results of corresponding master equation simulations. Good quantitative agreement with the analytical expressions is observed. (Note that the small parity oscillations in the simulation results are Type III errors---coherent micro-oscillations due to evolution generated by the effective Hamiltonian within the code space. These errors are not suppressed by the filter.)
Thus we find that Type II errors are also suppressed by the filter, provided the effective Hamiltonian detuning lies outside the filter passband.

\subsection{Crosstalk mitigation: mode frequency optimization}
\label{subsec:mitigation_optimization}

We have shown that stochastic correlated phase errors (Types I and II) can be suppressed by a filter if the corresponding emitted buffer mode photons have frequencies outside the filter passband.
We now show that it is possible to suppress \emph{all} such errors simultaneously
by carefully choosing the frequencies of the storage modes. In doing so, the effects of Type III errors can also be simultaneously minimized. Importantly, the storage mode frequencies are chosen to be compatible with error correction in the surface code, and we begin this section by describing how the surface code architecture constrains the choice of storage mode frequencies. 

\begin{figure*}
    \centering
    \includegraphics[width=0.75\textwidth]{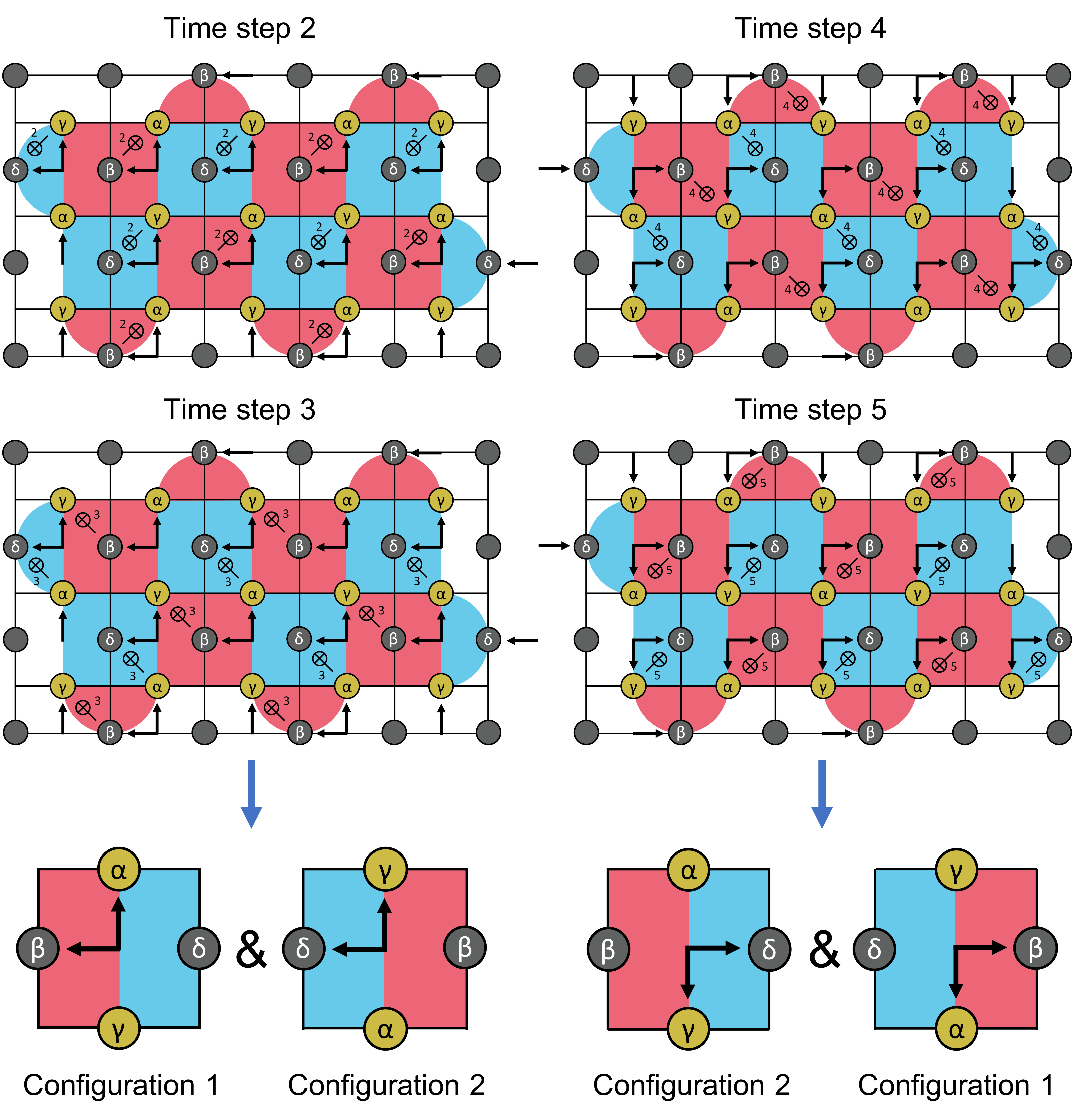}
    \caption{Cat-qubit stabilization in the surface-code architecture. Each ATS is coupled to two data modes $\alpha,\gamma$ and two ancilla modes $\beta,\delta$. In practice, ATSs are also coupled to a fifth readout mode (not shown here because it is not stabilized by any ATS). Each ATS is responsible for performing four CNOT gates (at different time steps) and stabilizing two phononic modes in the cat-code manifold during each time step.  In the top panel, we show configurations of the cat-qubit stabilization which respect the constraint discussed in the main text: at each time step, a CNOT's target mode must be stabilized by an ATS that also couples to its control mode. Each phononic mode, pointed by a black arrow, is stabilized by an ATS where the black arrow originates from. In the bottom panel, we show two stabilization configurations in the perspective of each host ATS. In configuration $1$ ($2$), modes $\alpha,\beta$ ($\gamma,\delta$) are stabilized by the host ATS and the remaining modes $\gamma,\delta$ ($\alpha,\beta$) are stabilized by some other neighboring ATSs.  }
    \label{fig:surface_code_stabilization_configuration}
\end{figure*}

\begin{figure}
    \centering
    \includegraphics[width=0.45\textwidth]{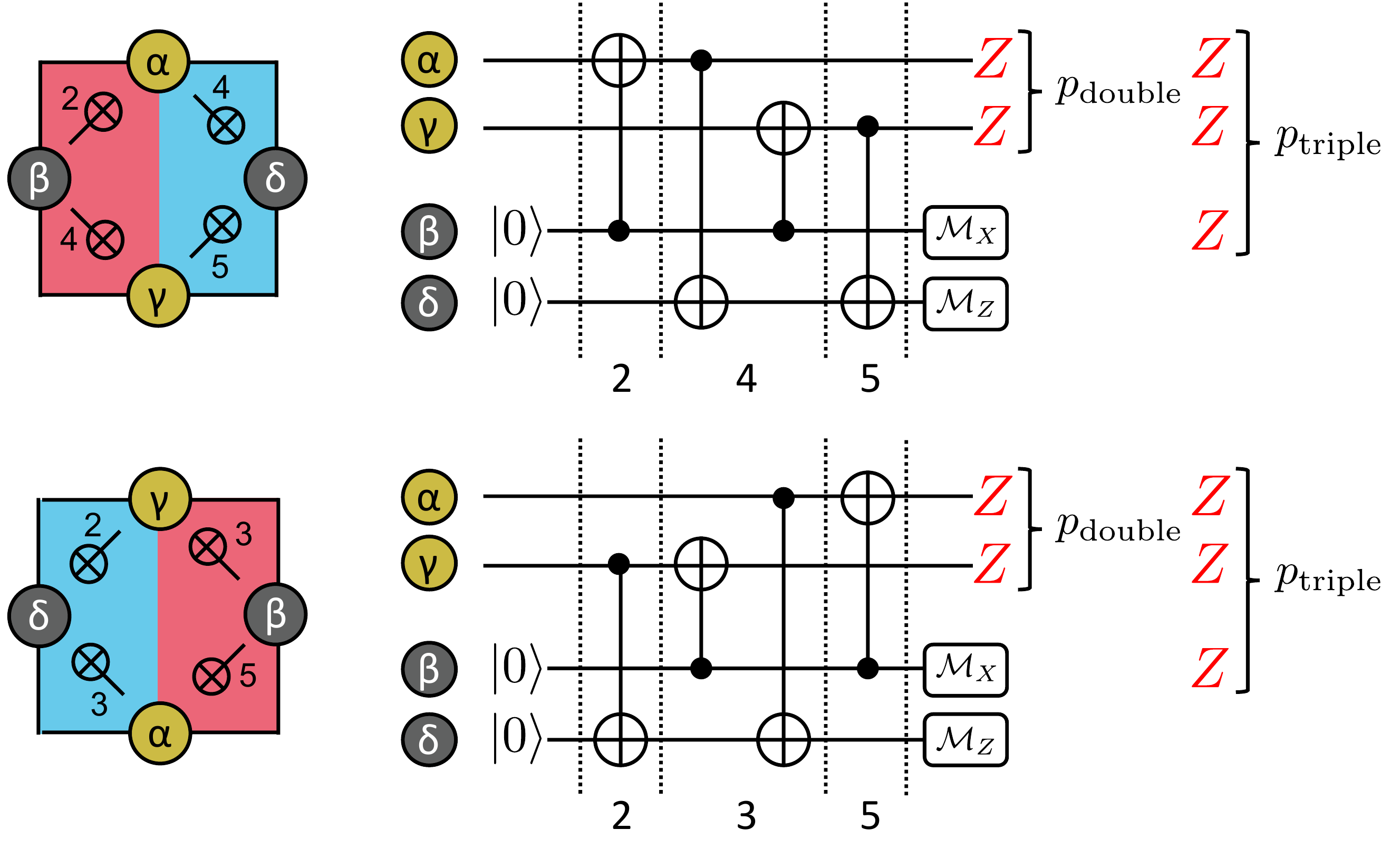}
    \caption{Type III crosstalk errors in the surface-code architecture. We define $p_{\textrm{double}}$ as the probability of getting a Type III error $\propto \hat Z_\alpha \hat Z_\gamma \hat I_\beta$, and $p_{\textrm{triple}}$ as the probability of getting a Type III error $\propto \hat Z_\alpha \hat Z_\gamma \hat Z_\beta$.}
    \label{fig:correlated_error_due_to_micro_oscillation_surface_code}
\end{figure}

We consider the surface-code architecture and optimize the storage mode frequencies such that they are compatible with the surface-code stabilizer measurement. To understand the constraints imposed by the implementation of the surface code, recall that each ATS is coupled to five phononic modes in our proposal (see \cref{fig:hardware_cartoon}). Among the five modes, four modes (two data and two ancilla modes for the surface code) are stabilized in the cat-code manifold by an ATS. Another mode (readout mode) is dedicated to measuring cat qubits in the $X$ basis and is not stabilized by any ATS. Since every data or ancilla mode couples to two ATSs, each ATS is only responsible for stabilizing two of the five phononic modes to which it couples. Thus, for each given ATS, we must determine which two phononic modes should be stabilized. 

An important consideration in deciding which phononic modes should be stabilized by a given ATS is that each ATS is used to realize four CNOT gates (performed in four different time steps) to measure the stabilizers of the surface code. While a CNOT gate is being performed, the target mode of the CNOT gate is stabilized by a rotating jump operator $\hat{L}_{2}(t) = \hat{a}_{2}^{2}-\alpha^{2} + (\alpha/2)(\exp[2i\pi t/T] - 1)(\hat{a}_{1}-\alpha) $ that acts non-trivially both on the target mode ($\hat{a}_{2}$) and the control mode ($\hat{a}_{1}$). Thus, while a CNOT gate is being performed, the target mode must be stabilized by the ATS that also couples to the control mode.

In Fig.\ \ref{fig:surface_code_stabilization_configuration} we show how these stabilization constraints can be satisfied. 
In the top panel of the figure, we show four (out of six, state preparation and measurement not show) time steps of the surface-code stabilizer measurement. During each time step, different CNOT gates between data and ancilla cat qubits are applied. We label data modes as $\alpha$ and $\gamma$ and ancilla modes as $\beta$ and $\delta$. Ancilla modes labelled as $\beta$ ($\delta$) are used to measure the $X$-type ($Z$-type) stabilizers of the surface code. 
We use black arrows to indicate which phononic modes are stabilized by each ATS at each time step; each phononic mode at the tip of a black arrow is stabilized by the ATS at the arrow's tail. Importantly, every target mode of a CNOT gate is stabilized by an ATS that also couples to the corresponding control mode at all time steps.
Note, however, that a given ATS stabilizes different modes at different time steps, as summarized in the bottom panel of \cref{fig:surface_code_stabilization_configuration}. In particular, there are two stabilization configurations: in configuration $1$ ($2$) modes $\alpha,\beta$ ($\gamma,\delta$) are stabilized by the given ATS, and the remaining modes $\gamma,\delta$ ($\alpha,\beta$) are stabilized by some other neighboring ATSs.           

Now, our goal is to choose the frequencies of the storage modes and detunings of the pumps in order to minimize crosstalk. In order to ensure that the choice of mode frequencies is compatible with the surface-code stabilizer measurement, we assign modes with the same label in \cref{fig:surface_code_stabilization_configuration} to have the same frequency. Thus, there are only five mode frequencies that must be chosen: the frequencies $\omega_\alpha, \omega_\beta, \omega_{\gamma}, \omega_{\delta}$ corresponding to the four labels in \cref{fig:surface_code_stabilization_configuration}, plus the frequency of the readout mode (not shown in \cref{fig:surface_code_stabilization_configuration}), which we take to be the same in each unit cell and denote by $\omega_\rho$. 
Similarly, there are four pump detunings, $\Delta_\alpha, \Delta_\beta, \Delta_{\gamma}, \Delta_{\delta}$, that must be chosen. Here, as above, $\Delta_i$ denotes the detuning of the pump (and buffer drive) used to stabilize mode $i$. 
In the following, we construct a cost function $C$ that quantifies crosstalk as a function of these nine parameters (five mode frequencies and four pump detunings). Numerically minimizing $C$ allows us to find the choices of the frequencies and detunings that minimize crosstalk. 

First, $C$ should be large if any emitted buffer mode photons associated with Type I and II errors lie inside the filter's bandwidth $4J$. We thus take $C=1$ if any of the following conditions are met for either of the two stabilization configurations shown in \cref{fig:surface_code_stabilization_configuration}:
\begin{itemize}
    \item $|\delta_{ijk}| < 2J$ (Type I errors not suppressed)
    \item $|\delta_{ijk}-\delta_{\ell m n}|< 2J$ (Type II errors not suppressed)
    \item $|\delta_{iii}| > 2J$ (desired dissipation suppressed)
\end{itemize}
In other words, we set $C=1$ if any Type I or II errors are not suppressed by the filter, or if any of the desired engineered dissipation is suppressed by the filter. We emphasize that these conditions must be checked for both stabilization configurations in \cref{fig:surface_code_stabilization_configuration}; checking both configurations is necessary in order to ensure that Type I and II crosstalk is suppressed by the filter at \emph{all} time steps.

Second, $C$ should be large if the coherent Type III errors have significant damaging effects, and we now quantify these effects in the context of the surface code.
Recall that these errors are generated by effective Hamiltonian terms of the form (\ref{eq:Heff_2}), which we repeat for convenience,
\begin{equation}
|\alpha|^4 \chi \hat Z_i \hat Z_j \hat Z_\ell \hat Z_m e^{i(\delta_{\ell m n}-\delta_{i j k}) t}  + \mathrm{H.c.}.
\end{equation}
When these terms are rapidly rotating, i.e., when $|\alpha^4 \chi| \ll |\delta_{ijk}-\delta_{\ell m n}|$, their effects are suppressed. Indeed, these terms effectively induce detuned Rabi oscillations between states of different parity, and the magnitude of these oscillations is small in the far-detuned limit. 
To quantify this suppression, note that these micro-oscillation errors remain coherent during gates but can be converted to incoherent, correlated $\hat Z$ errors when the $X$-type stabilizers are measured.
The probability $p_{ijk\ell mn}$ of inducing a correlated phase error upon a such a measurement scales quadratically in the ratio of the coupling strength and detuning,
\begin{equation}
p_{ijk\ell mn} = \left(\frac{|\alpha^4 \chi|}{\delta_{ijk}-\delta_{\ell m n}}\right)^2.
\end{equation} 
Among the various Type III errors, we focus on those that induce phase errors in both of the data modes $\alpha$ and $\gamma$ since such errors are specific to our architecture and not taken into account in the standard surface-code analysis. In particular, we define $p_{\textrm{double}}$ as the total probability at least one Type III error $\propto \hat Z_\alpha\hat Z_\gamma \hat I_\beta$, and $p_{\textrm{triple}}$ as the total probability of at least one Type III error $\propto \hat Z_\alpha \hat Z_\gamma \hat Z_\beta$. Explicitly,
\begin{align}
    p_{\textrm{double}} &= \sum_{\{ijk\ell m n\}\in \mathcal{D}} p_{ijk\ell m n} ,\\ 
    p_{\textrm{triple}} &= \sum_{\{ijk\ell m n\}\in \mathcal{T}} p_{ijk\ell m n},
\end{align}
where $\mathcal D$ and $\mathcal T$ denote sets of indices that give rise to errors $\propto \hat Z_\alpha\hat Z_\gamma \hat I_\beta$ and $\propto \hat Z_\alpha \hat Z_\gamma \hat Z_\beta$, respectively, see \cref{fig:correlated_error_due_to_micro_oscillation_surface_code}. 
Note that the $\hat Z$ error on the ancilla mode $\beta$ manifests as a flipped $ X$-basis measurement outcome. On the other hand, $\hat Z$ errors on the other ancilla mode $\delta$ do not flip the measurement outcomes. This is because the mode $\delta$ is measured in the $ Z$ basis, and $Z$-basis measurements commute with $\hat Z$ errors. 

\begin{table*}[t]
  \centering
  \begin{tabular}{ |c|c|c|c|c| } 
    \hline
    $4J$  &
    $\omega_\alpha, \omega_\beta,\omega_\gamma,\omega_\delta, \omega_\rho$  &
    $\frac{1}{2}(p_\text{double}^{(1)}+p_\text{double}^{(2)})$  & $\frac{1}{2}(p_\text{triple}^{(1)}+p_\text{triple}^{(2)})$ &
    $C$
    \\\hline
    100 & 0, 1000, 242, 879, 61 & 
    $1.83*10^{-8}\left[\frac{|\alpha|^2 g_2}{2\pi\text{MHz}}\right]^4$
    & 
    $5.20*10^{-10}\left[\frac{|\alpha|^2 g_2}{2\pi\text{MHz}}\right]^4$
    & $ 1.88*10^{-8} \left[\frac{|\alpha|^2 g_2}{2\pi\text{MHz}}\right]^4$ \\\hline
  \end{tabular}
  \caption{Frequency optimization results. The parameters $4J$ and $\omega$ are given in units of $2\pi \times $ MHz.  The Type III error probabilities and the cost $C$ are expressed in terms of $\alpha$ and $g_2$. For realistic choices of $|\alpha| = \sqrt{8}$ and $g_2/2\pi = 2$ MHz, the cost function evaluates to $C = 1.23* 10^{-3}$. We fix $-\Delta_{\alpha} = \Delta_{\beta} = -\Delta_{\gamma} = \Delta_{\delta} = J$.}
  \label{tab:frequency_opt}
\end{table*}

We incorporate these Type III errors into the cost function as follows.
We take $C=1$ if Type I or II errors are not suppressed by the filter (see aforementioned conditions on the $\delta_{ijk}$), and otherwise we take
\begin{equation}
\label{eq:typeIII_cost}
C = \frac{1}{2}\left(p_{\mathrm{double}}^{(1)} + p_{\mathrm{triple}}^{(1)}+p_{\mathrm{double}}^{(2)} + p_{\mathrm{triple}}^{(2)}\right),
\end{equation}
where $p_{\mathrm{double}}^{(i)}$ and $p_{\mathrm{triple}}^{(i)}$ denote the values of $p_{\mathrm{double}}$ and $p_{\mathrm{triple}}$ for the $i$-th stabilization configuration. \Cref{eq:typeIII_cost} thus represents the average probability of a Type III error occurring during one time step. 
Costs $C\ll1$ are thus only achieved when both the probability of Type III errors is small, and all Type I and II errors are suppressed by the filter.

Having defined the cost function $C$, we perform a numerical search for the values of the mode frequencies and pump detunings which minimize the cost. In performing this optimization, we place two additional restrictions on allowed frequencies and detunings. 
First, we restrict the mode frequencies to lie within a $1$ GHz bandwidth. This is done because the modes are supported by phononic-crystal-defect resonators (PCDRs), and as such all mode frequencies must lie within the phononic bandgap, or at least within the union of two separate bandgaps each associated with different PCDRs. These bandgaps are typically not more than $500$ MHz wide for the devices we consider~\cite{Arrangoiz-Arriola2019}. 
Second, we restrict the values of the detunings to $\Delta = \pm J$. This is done to maximize use of the filter bandwidth; emitted buffer mode photons are detuned from one another by $2J$ and from the nearest band edge by $J$, see \cref{fig:multiplexing_and_crosstalk}(a). Additionally, we choose $4J/2\pi = 100$MHz because this value is both small enough to ensure that all stochastic crosstalk errors are suppressed by the filter, and large enough so that the buffer decay rate $\kappa_b/2\pi = 57$MHz is not limited by the filter bandwidth.  

\begin{figure}
\begin{center}
\includegraphics[width=\linewidth]{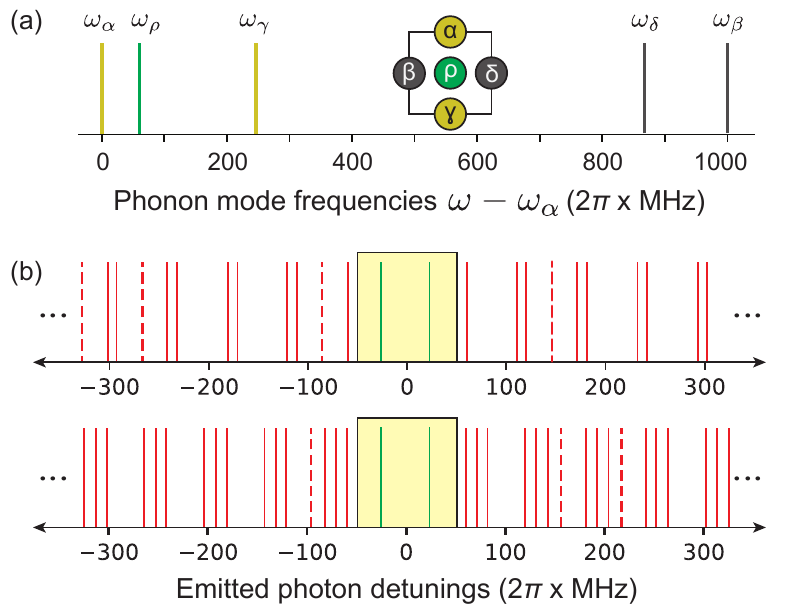}
\caption{Optimized mode frequencies. (a) Plot of the optimized frequencies of the five storage modes. (b) Emitted buffer mode photon detunings. Red dashed (solid) lines indicate photons emitted via parity-non-preserving Type~I (Type II) processes. The yellow box covers the region $[-50,50]\, (2\pi \times \text{MHz})$, representing a bandpass filter with center frequency $\omega_b$ and a $4J = 2\pi * 100$ MHz passband. The fact that no lines lie inside the yellow box indicates that all Type I and II processes are sufficiently far detuned so as to be suppressed by the filter. The top (bottom) plot in (b) is for the case where modes $\alpha$ and $\beta$ ($\gamma$ and $\delta$) are stabilized simultaneously. }
\label{fig:CTH_freq_opt}
\end{center}
\end{figure}

The optimization results are listed in \Cref{tab:frequency_opt} and illustrated in \Cref{fig:CTH_freq_opt}. For the optimal configurations, \emph{all} Type I and Type II errors are simultaneously suppressed by the filter. Note also that all emitted photon frequencies associated with Type I or II errors lie at least $10$ MHz outside the filter passband. As a result, the optimized configuration is robust to deviations in the mode frequencies of the same order, and larger deviations can be tolerated by decreasing the filter bandwidth. Moreover, for realistic values of $|\alpha|$ and $g_2$, we have $C\ll 1$, indicating that Type III errors are strongly suppressed. Therefore, all dominant sources of crosstalk are strongly suppressed.

\section{Shifted Fock basis}
\label{appendix:Shifted Fock Basis}

Simulating a large cat qubit (with large $|\alpha|^{2} \gg 1$) by using the usual Fock basis becomes quickly inefficient. Here, we introduce a shifted Fock basis method which can describe large cat states in a more efficient way (i.e., using a smaller Hilbert space dimension) than the usual Fock basis. Specifically, we will explain how to construct the annihilation operator $\hat{a}$ in the shifted Fock basis.  

Recall that a cat state is composed of two coherent state components $| \pm \alpha\rangle$ which can be understood as displaced vacuum states $\hat{D}(\pm\alpha)|\hat{n}=0\rangle$, where $\hat{D}(\alpha)\equiv \exp[\alpha\hat{a}^{\dagger}-\alpha^{*}\hat{a}]$ is the displacement operator. In the shifted Fock basis, we use $2d$ displaced Fock states $\hat{D}(\pm \alpha)|\hat{n}=n\rangle$ as basis states, where $n\in \lbrace 0,\cdots, d-1 \rbrace$. Note that while displaced Fock states in each $\pm \alpha$ branch are orthonormalized, displaced Fock states in different branches are not necessarily orthogonal to each other. We thus need to orthonormalize the displaced Fock states. 

We first define the non-orthonormalized basis states as follows: 
\begin{align}
|\phi_{n,\pm}\rangle \equiv \frac{1}{\sqrt{2}} \big{[} \hat{D}(\alpha) \pm (-1)^{n} \hat{D}(-\alpha) \big{]} |\hat{n}=n\rangle,  \label{eq:displaced Fock cat basis states}
\end{align}
where $|\phi_{n,+}\rangle$ and $|\phi_{n,-}\rangle$ have even and odd excitation number parity, respectively. Note that we grouped the non-orthonormalized states into the even and odd branches instead of the $\pm\alpha$ branches. As a result, in the ground state manifold ($n=0$), the normalized basis states $|\phi_{0,\pm}\rangle$ are equivalent to the complementary basis states of the cat qubit $|\pm\rangle$, not the computational basis states $|0/1\rangle$, i.e., 
\begin{align}
    |\pm\rangle \propto |\phi_{0,\pm}\rangle = \frac{1}{\sqrt{2}}( | \alpha\rangle \pm | -\alpha\rangle ) . 
\end{align} 
We use the even/odd branching convention so that any two basis states in different branches are orthogonal to each other and hence the orthonormalization can be done separately in each parity sector. Note that 
\begin{align}
\Phi^{\pm}_{m,n}\equiv \langle \phi_{m,\pm}|\phi_{n,\pm}\rangle &= \delta_{m,n} \pm (-1)^{m}D_{m,n}(2\alpha),  \label{eq:Phi matrix}
\end{align} 
where $D_{m,n}(\alpha) \equiv \langle \hat{n}=m| \hat{D}(\alpha) |\hat{n}=n\rangle$ are the matrix elements of the displacement operator $\hat{D}(\alpha)$ in the usual Fock basis: 
\begin{align}
D_{m,n}(\alpha)  &= e^{-\frac{|\alpha|^{2}}{2} } \sqrt{ \frac{ \min(m,n)! }{ \max(m,n)! } } L_{\min(m,n)}^{( |m-n| )}( |\alpha|^{2} )
\nonumber\\
&\qquad\times   \begin{cases}
\alpha^{m-n}  & m\ge n \\
(-\alpha^{*})^{n-m} & m< n
\end{cases}  . 
\end{align}
 Here, $L_{n}^{(\alpha)}(x)$ is the generalized Laguerre polynomial. Since $ |D_{m,n}(2\alpha)|  = \mathcal{O}(  |\alpha|^{m+n} e^{-2|\alpha|^{2}} )$, $D_{m,n}(2\alpha)$ is negligible if $m+n \ll |\alpha|^{2}$. In this regime, the basis states $|\phi_{n,\pm}\rangle$ are almost orthonormal. For the purpose of estimating the phase-flip (or $Z$) error rates within a small multiplicative error, it is often permissible to neglect the non-orthogonality of the states $|\phi_{n,\pm}\rangle$. However, this is generally not the case if we want to evaluate the $Z$ error rates with a very high precision or if we want to estimate the bit-flip (or $X$) error rates because the bit flip error rates decrease exponentially in $|\alpha|^{2}$. In these cases, taking into account the non-orthogonality of the states $|\phi_{n,\pm}\rangle$ is essential. 

We orthonormalize the basis states $|\phi_{n,\pm}\rangle$ by applying the Gram-Schmidt orthonormalization procedure. Specifically, given the non-orthonormalized basis states $|\phi_{n,\pm}\rangle$, we construct $d$ orthonormalized basis states in each parity sector starting from the ground state $|\phi_{0,\pm}\rangle$:
\begin{align}
|\psi_{n,\pm} \rangle &= \sum_{m=0}^{d-1} c^{\pm}_{m,n} |\phi_{m,\pm}\rangle . 
\end{align}
The coefficients $c^{\pm}_{m,n}$ ($0 \le m,n \le d-1$) are determined inductively. In the base case ($k=0$), 
\begin{align}
c^{\pm}_{0,0} &= \frac{1}{\sqrt{ \Phi^{\pm}_{0,0} }}, \quad c^{\pm}_{m,0} =0 \,\,\, \textrm{for all}\,\,\, 1\le m \le d-1,  \label{eq:cmatrix 1} 
\end{align}
and thus the logical $|\pm\rangle$ states of the cat qubit are given by 
\begin{align}
    |\pm\rangle \equiv |\psi_{0,\pm} \rangle = \frac{1}{\sqrt{ \Phi^{\pm}_{0,0} }}|\phi_{0,\pm}\rangle = \frac{ |\alpha\rangle \pm |-\alpha\rangle }{\sqrt{ 2(1\pm e^{-2|\alpha|^{2}}) }}. 
\end{align}
In the general case with $1 \le k \le d-1$, suppose we are given with $c^{\pm}_{mn}$ for all $0\le m \le d-1$ and $0\le n \le k-1$. Thus, at this point, the first $k$ columns of $c^{\pm}$ are known. Let $c^{\pm}_{:,0:k-1}$ be the $d\times k$ matrix which is obtained by taking the first $k$ columns of the matrix $c^{\pm}$. Given $c^{\pm}_{:,0:k-1}$, we assign the $k+1^{\textrm{th}}$ column of $c^{\pm}$ as follows.  
\begin{align}
c^{\pm}_{m,k} &= -\frac{ ( c^{\pm}_{:,0:k-1} (c^{\pm}_{:,0:k-1})^{\dagger} \Phi^{\pm} )_{ m , k } }{ \sqrt{ \Phi^{\pm}_{k,k} -  (  (\Phi^{\pm})^{\dagger}  c^{\pm}_{:,0:k-1} (c^{\pm}_{:,0:k-1})^{\dagger} \Phi^{\pm} )_{k,k}   }  }   ,\label{eq:cmatrix 2}
\end{align}
for $0\le m \le k-1$,  
\begin{align} 
c^{\pm}_{k,k} &=  \frac{1 }{ \sqrt{ \Phi^{\pm}_{k,k} -  (  (\Phi^{\pm})^{\dagger}  c^{\pm}_{:,0:k-1} (c^{\pm}_{:,0:k-1})^{\dagger} \Phi^{\pm} )_{k,k}   }  }, \label{eq:cmatrix 3} 
\end{align}
and $c^{\pm}_{m,k} =0$ for all $m > k$. 

Having constructed the $2d$ orthonormalized shifted Fock basis states $|\psi_{n,\pm}\rangle$, we now need to find the matrix elements of an operator $\hat{O}$ (e.g., $\hat{O} = \hat{a}$) in the orthonormalized basis. Let $|\phi_{n}\rangle = |\phi_{n,+}\rangle$ and $|\phi_{n+d}\rangle = |\phi_{n,-}\rangle$ for $n \in \lbrace 0,\cdots, d-1 \rbrace$ and also define $|\psi_{n}\rangle$ and $|\psi_{n+d}\rangle$ similarly. Suppose that the operator $\hat{O}$ transforms the non-orthonormalized basis states $|\phi_{n}\rangle$ as follows
\begin{align}
    \hat{O}|\phi_{n}\rangle = \sum_{m=0}^{2d-1}O_{m,n}|\phi_{m}\rangle. 
\end{align} 
We call $O_{m,n}$ the matrix elements of the operator $\hat{O}$ in the non-orthonormalized basis $|\phi_{n}\rangle$. Then, in the orthonormalized basis, the matrix elements of the operator $\hat{O}$ are given by 
\begin{align}
    O'_{m,n} &\equiv \langle \psi_{m} |\hat{O} |\psi_{n}\rangle = (c^{\dagger}\Phi O c)_{m,n}, \label{eq:matrix element transformation from nonortho to ortho}
\end{align}
where $\Phi$ and $c$ are $2d\times 2d$ matrices which are defined as 
\begin{align}
    \Phi = \begin{bmatrix}
    \Phi^{+} & 0\\
    0 & \Phi^{-}
    \end{bmatrix}, \quad c = \begin{bmatrix}
    c^{+} & 0\\
    0 & c^{-}
    \end{bmatrix} .   
\end{align}
The matrix elements of the $d\times d$ matrices $\Phi^{\pm}$ and $c^{\pm}$ are given in \cref{eq:Phi matrix,eq:cmatrix 1,eq:cmatrix 2,eq:cmatrix 3}. 

Consider the annihilation operator $\hat{O}=\hat{a}$ and note that it transforms the non-orthonormalized basis states $|\phi_{n,\pm}\rangle$ as follows: 
\begin{align}
    \hat{a}|\phi_{n,\pm}\rangle &= \sqrt{n}|\phi_{n-1,\mp}\rangle + \alpha|\phi_{n,\mp}\rangle. 
\end{align}
Note that the annihilation operator $\hat{a}$ flips the $\pm$ parity to the $\mp$ parity. 
Thus, in the non-orthonormalized basis, the matrix elements of the annihilation operator are given by 
\begin{align}
    \begin{bmatrix}
    0 & \hat{b}+\alpha\\
    \hat{b}+\alpha & 0
    \end{bmatrix} = \hat{X}\otimes ( \hat{b} + \alpha ),  
\end{align}
where $\hat{X}$ is the Pauli $X$ operator and $\hat{b}$ is the truncated annhilation operator of size $d\times d$. Then, the matrix elements of the annihilaton operator in the orthonormalized basis $|\psi_{n,\pm}\rangle$ can be obtained via the transformation given in \cref{eq:matrix element transformation from nonortho to ortho}. 

Recall that $|\psi_{n,\pm}\rangle$ are complementary basis states. To find the matrix elements of an operator in the computational basis states, we should conjugate the matrix by the Hadamard operator $\hat{H}$. Thus, in the orthonormalized computational basis, the annihilation operator is given by
\begin{align}
    \hat{a}&\equiv (\hat{H}\otimes \hat{I}) \cdot c^{\dagger}\Phi ( \hat{X}\otimes ( \hat{b} + \alpha ) ) c \cdot  (\hat{H}\otimes \hat{I}) 
    \nonumber\\
    &\xrightarrow{|\alpha|^{2} \gg d } \hat{Z}\otimes (\hat{b}+\alpha) . 
\end{align}
The approximate expression $\hat{a} \simeq \hat{Z}\otimes (\hat{b}+\alpha)$ is useful for analyzing the $Z$ error rates of large cat qubits (with $|\alpha|\gg 1$) in the perturbative regime where the cat qubit states may sometimes be excited to the first excited state manifold ($n=1$) but quickly decay back to the ground state manifold ($n=0$). In particular, the engineered two-phonon dissipator $\kappa_{2}\mathcal{D}[\hat{a}^{2}-\alpha^{2}]$ is given by 
\begin{align}
    \kappa_{2}\mathcal{D}[ \hat{I}\otimes (\hat{b}^{2} + 2\alpha\hat{b}) ] &\simeq 4\kappa_{2}\alpha^{2}\mathcal{D}[\hat{I}\otimes \hat{b}]
\end{align}
by using the approximate expression $\hat{a} \simeq \hat{Z}\otimes (\hat{b}+\alpha)$ and disregarding higher than second excited states (i.e., $\hat{b}^{2}=0$). Hence, the linewidth of the engineered two-phonon dissipation is approximately given by $4\kappa_{2}\alpha^{2}$, which from now on we refer to as the confinement rate $\kappa_\text{conf}$. For numerical simulations (\cref{app:Gate Error Simulations}), we thoroughly take into account the orthonormalization and use the orthonormalized shifted Fock basis obtained by the Gram-Schmidt process. We lastly remark that the parity operator $e^{i\hat{\pi}\hat{a}^{\dagger}\hat{a}}$ is exactly given by $\hat{X}\otimes \hat{I}$ in the shifted Fock basis (with the orthonormalization accounted for) because of the way we define the basis states, i.e., $|\psi_{n,+}\rangle$ ($|\psi_{n,-}\rangle$) has an even (odd) excitation number parity.  

\section{Perturbative analysis of the Z error rates of the cat qubit gates }
\label{appendix:Perturbative analysis of cat qubit gates}

Here, we analyze the $Z$ error rates of the cat qubit gates (idling, Z rotations, CZ rotations, CNOT, and Toffoli) by using the shifted Fock basis (\cref{appendix:Shifted Fock Basis}) and adiabatic elimination or effective operator formalism (\cref{subsec:effective_operator}). 

\subsection{Idling}
\label{subsection:Idling appendix}

Consider an idling single cat qubit which is stabilized by the two-phonon dissipation $\kappa_{2}\mathcal{D}[\hat{a}^{2}-\alpha^{2}]$ and is subject to single-phonon loss $\kappa_{1}\mathcal{D}[\hat{a}]$: 
\begin{align}
    \frac{d\hat{\rho}(t)}{dt} &= \kappa_{2}\mathcal{D}[\hat{a}^{2}-\alpha^{2}]\hat{\rho}(t) + \kappa_{1}\mathcal{D}[\hat{a}]\hat{\rho}(t). 
\end{align}
Assuming $|\alpha|\gg 1$, the above master equation is given by 
\begin{align}
    \frac{d\hat{\rho}(t)}{dt} &= \kappa_{2}\mathcal{D}[  \hat{I} \otimes ( \hat{b}^{2} + 2\alpha\hat{b} ) ]\hat{\rho}(t) 
    \nonumber\\
    &\quad + \kappa_{1}\mathcal{D}[ \hat{Z} \otimes (\hat{b} + \alpha ) ]\hat{\rho}(t) \label{eq:master eq idling shifted Fock basis}
\end{align}
in the shifted Fock basis, where we used the mapping $\hat{a}\rightarrow \hat{Z}\otimes (\hat{b} + \alpha)$. Suppose that the system is initially in the cat qubit manifold, i.e., $\hat{\rho}(0) = \hat{\rho}_{g}(0)\otimes |0\rangle'\langle 0|'$, where $\hat{\rho}_{g}(0)$ is a density operator of size $2\times 2$ and $|0\rangle' \equiv |\hat{b}^{\dagger}\hat{b}=0\rangle$ (not to be confused with the computational basis state $|0\rangle$). When the system is idling, the states are never excited to the excited state manifold and thus $\hat{\rho}(t) = \hat{\rho}_{g}(t)\otimes |0\rangle'\langle 0|'$. Projecting the master equation in \cref{eq:master eq idling shifted Fock basis} to the ground state manifold, we find    
\begin{align}
    \frac{d\hat{\rho}_{g}(t)}{dt} &= \kappa_{1}\alpha^{2}\mathcal{D}[\hat{Z}]\hat{\rho}_{g}(t) , 
\end{align}
and hence 
\begin{align}
    \hat{\rho}_{g}(T) \simeq (1-\bar{p}_{Z})\hat{\rho}_{g}(t) + \bar{p}_{Z}\hat{Z}\hat{\rho}_{g}(t)\hat{Z}, 
\end{align}
provided that the idling $Z$ error rate (per gate) $\bar{p}_{Z} \equiv \kappa_{1}\alpha^{2}T$ is small (i.e., $\bar{p}_{Z} \ll 1$) where $T$ is the idling time. Note that we used the notation $\bar{p}$ with a bar to indicate that the presented expression is obtained via a perturbative analysis. We use $p$ without bar to refer to numerical results.

\subsection{Z rotations}
\label{subsection:Z rotations appendix}

Assume that $\alpha$ is real and positive. To implement a Z rotation 
$Z(\theta) \equiv \exp[i\theta |1\rangle\langle 1| ]$ on a cat qubit (where $|1\rangle\simeq |-\alpha\rangle$ is a cat-code computational basis state), we need to apply a linear drive $\epsilon_{Z}(\hat{a}+\hat{a}^{\dagger})$:
\begin{align}
    \frac{d\hat{\rho}(t)}{dt} &= \kappa_{2}\mathcal{D}[\hat{a}^{2}-\alpha^{2}]\hat{\rho}(t) + \kappa_{1}\mathcal{D}[\hat{a}]\hat{\rho}(t) 
    \nonumber\\
    &\quad -i[ \epsilon_{Z}(\hat{a}+\hat{a}^{\dagger}) , \hat{\rho}(t) ] . 
\end{align}
In the shifted Fock basis, the linear drive $\epsilon_{Z}(\hat{a}+\hat{a}^{\dagger})$ is given by $\epsilon_{Z}\hat{Z}\otimes (\hat{b}+\hat{b}^{\dagger}+2\alpha)$. Thus, in the ground state manifold, it induces a Z rotation via the term $2\epsilon_{Z}\alpha\hat{Z}$. At the same time, the term $\epsilon_{Z}\hat{Z}\otimes \hat{b}^{\dagger}$ excites the cat qubit to its first excited state, which then quickly decays back to the ground state manifold due to the engineered dissipaton $\kappa_{2}\mathcal{D}[\hat{a}^{2}-\alpha^{2}] \leftrightarrow \kappa_{2}\mathcal{D}[ \hat{I} \otimes(  \hat{b}^{2} + 2\alpha \hat{b}) ]$. Thus, to capture the first order effects, we only consider the ground state manifold and the first excited state manifold ($n=0,1$), hence ignoring $\hat{b}^{2}$ in $\kappa_{2}\mathcal{D}[ \hat{I} \otimes( \hat{b}^{2}+2\alpha\hat{b} )]$. Also, assuming $\kappa_{2} \gg \kappa_{1}$, we ignore the intrinsic decay due to the single phonon loss in the excited state manifold, i.e., $\kappa_{1}\mathcal{D}[\hat{a}] \leftrightarrow \kappa_{1}\mathcal{D}[ \hat{Z} \otimes (\hat{b} + \alpha) ] \simeq \kappa_{1}\alpha^{2} \mathcal{D}[\hat{Z} \otimes \hat{I} ]$, where we used $\mathcal{D}[c\hat{A}] = |c|^{2}\mathcal{D}[\hat{A}]$. 
Then, the master equation is given by
\begin{align}
    \frac{d\hat{\rho}(t)}{dt} &= 4\kappa_{2}\alpha^{2}\mathcal{D}[\hat{I}\otimes \hat{b}]\hat{\rho}(t) + \kappa_{1}\alpha^{2}\mathcal{D}[ \hat{Z} \otimes \hat{I}  ]\hat{\rho}(t) 
    \nonumber\\
    &
    -2i\alpha\epsilon_{Z} [\hat{Z} \otimes \hat{I} , \hat{\rho}(t) ] 
    -i[ \epsilon_{Z} \hat{Z} \otimes (\hat{b}+\hat{b}^{\dagger}) , \hat{\rho}(t) ]
\end{align}
The second term on the right-hand side of this master equation describes a $Z$ error acting on the encoded cat qubit due to single-phonon loss, occurring at the rate (per time) $\kappa_1\alpha^2$. The third term rotates the cat qubit about the $Z$ axis. The fourth term excites the cat qubit from its ground-state manifold to its first-excited-state manifold, with a coupling strength  $g = \epsilon_{Z}$, and at the same time inflicts a $Z$ error on the cat qubit. This excitation decays back to the cat code ground-state manifold with a decay rate $\kappa = 4\kappa_{2}\alpha^{2}$ due to the engineered dissipation described by the first term.
Assuming $\kappa \gg g$ the creation and decay of this excitation results in an additional $Z$ error in the ground state manifold with an effective error rate (per time) $4g^{2} / \kappa  =  \epsilon_{Z}^{2} / (\kappa_{2}\alpha^{2})$, augmenting the $Z$ error rate due to single-phonon loss. 
The effective master equation therefore becomes
\begin{align}
    \frac{d\hat{\rho}_{g}(t)}{dt} = \Big{(} \kappa_{1}\alpha^{2} + \frac{\epsilon_{Z}^{2}}{\kappa_{2}\alpha^{2}} \Big{)} \mathcal{D}[\hat{Z}]\hat{\rho}_{g}(t) -i[ 2\epsilon_{Z}\alpha \hat{Z} , \hat{\rho}_{g}(t) ] , 
\end{align}
where we have used the subscript $g$ to indicate that $\hat{\rho}_{g}(t)$ is the density operator in the ground-state manifold of the cat state.

Given this effective master equation, we can analyze the effective Hamiltonian and the effective phase-flip error separately because they commute with each other. The effective Hamiltonian $\hat{H}_{\textrm{eff}} = 2\epsilon_{Z}\alpha\hat{Z}$ induces a $Z$ rotation $\hat{Z}(\theta)$ with $\theta = 4\epsilon_{Z}\alpha T $ after the gate time $T$, i.e., $\epsilon_{Z} = \theta / (4\alpha T)$. Then, the $Z$ error rate (per gate) due to the effective phase-flip is given by
\begin{align}
    \bar{p}_{Z} = \kappa_{1}\alpha^{2}T + \frac{\epsilon_{Z}^{2}}{\kappa_{2}\alpha^{2}}T = \kappa_{1}\alpha^{2}T + \frac{\theta^{2}}{16\kappa_{2}\alpha^{4}T}, 
\end{align}
provided that $\bar{p}_{Z}\ll 1$. This $Z$ error rate is minimized at the optimal gate time 
\begin{align}
    \bar{T}^{\star}_{Z(\theta)} = \frac{|\theta|}{4\alpha^{3} \sqrt{\kappa_{1}\kappa_{2}} } , 
\end{align}
and the corresponding optimal $Z$ error rate is given by
\begin{align}
    \bar{p}_{Z}^{\star} = \frac{|\theta|}{2\alpha} \sqrt{\frac{\kappa_{1}}{\kappa_{2}}} .
\end{align}

\subsection{CZ rotations}
\label{subsection:CZ rotations appendix}

A $ZZ$ interaction between two cat qubits can be implemented by using a beam-splitter coupling $\epsilon_{ZZ}(\hat{a}_{1}\hat{a}_{2}^{\dagger}+\hat{a}_{1}^{\dagger}\hat{a}_{2})$, which is given by $2\epsilon_{ZZ}\alpha^{2}\hat{Z}_{1}\hat{Z}_{2}$ in the ground state manifold of the cat qubits. To implement a controlled Z rotation $CZ(\theta) \equiv \exp[ i\theta |11\rangle\langle 11| ]$, we should add single-qubit Z rotations so that only the state $|11\rangle$ accumulates a non-trivial phase. More specifically, we need 
\begin{align}
    \hat{H} &= \epsilon_{ZZ}( \hat{a}_{1}\hat{a}_{2}^{\dagger} + \hat{a}_{1}^{\dagger}\hat{a}_{2} ) -\epsilon_{ZZ}\alpha(\hat{a}_{1}+\hat{a}_{1}^{\dagger}) 
    \nonumber\\
    &\qquad\qquad\qquad\qquad\quad - \epsilon_{ZZ}\alpha(\hat{a}_{2}+\hat{a}_{2}^{\dagger}), 
\end{align}
and the master equation is given by
\begin{align}
    \frac{d\hat{\rho}(t)}{dt} &= \kappa_{2} \Big{[} \mathcal{D}[\hat{a}_{1}^{2}-\alpha^{2}] + \mathcal{D}[\hat{a}_{2}^{2}-\alpha^{2}] \Big{]} \hat{\rho}(t) 
    \nonumber\\
    &\quad + \kappa_{1} \Big{[} \mathcal{D}[\hat{a}_{1}] + \mathcal{D}[\hat{a}_{2}] \Big{]} \hat{\rho}(t)   -i[ \hat{H}  , \hat{\rho}(t)  ]. 
\end{align}
Similarly as in the case of single-qubit Z rotations, the engineered dissipation induces a strong decay from the first excited state manifold to the cat qubit manifold with a decay rate (per time) $\kappa = 4\kappa_{2}\alpha^{2}$. Also, the single-phonon loss causes local phase-flip errors in each cat qubit manifold with an error rate $\kappa_{1}\alpha^{2}$. In the shifted Fock basis, the Hamiltonian $\hat{H}$ is given by 
\begin{align}
     \hat{H} &= 2\epsilon_{ZZ}\alpha^{2} (\hat{Z}_{1}\hat{Z}_{2} - \hat{Z}_{1}-\hat{Z}_{2}) \otimes \hat{I}
     \nonumber\\
     &\quad +\epsilon_{ZZ}\alpha (\hat{Z}_{1}\hat{Z}_{2}-\hat{Z}_{1}) \otimes (\hat{b}_{1}+\hat{b}_{1}^{\dagger})
     \nonumber\\
     &\quad +\epsilon_{ZZ}\alpha (\hat{Z}_{1}\hat{Z}_{2}-\hat{Z}_{2}) \otimes (\hat{b}_{2}+\hat{b}_{2}^{\dagger})
     \nonumber\\
     &\quad +\epsilon_{ZZ}\hat{Z}_{1}\hat{Z}_{2}\otimes (\hat{b}_{1}\hat{b}_{2}^{\dagger} + \hat{b}_{1}^{\dagger}\hat{b}_{2}). 
\end{align}
The first term generates an effective Hamiltonian $\hat{H}_{\textrm{eff}} = 8\epsilon_{ZZ}\alpha^{2}|11\rangle\langle 11|$ in the ground state manifold. Due to the second (third) term, the first (second) cat qubit is excited to its first excited state manifold with a coupling strength $g = \epsilon_{ZZ}\alpha$ while the encoded cat qubits are subjected to 
a $\hat{Z}_{1}\hat{Z}_{2}-\hat{Z}_{1}$ ($\hat{Z}_{1}\hat{Z}_{2}-\hat{Z}_{2}$) error.
The excited state decays back to the ground-state manifold at the rate $\kappa= 4 \kappa_2\alpha^2$ due to the engineered dissipation; as a result the cat qubits experience effective $\hat{Z}_{1}\hat{Z}_{2}-\hat{Z}_{1}$ and $\hat{Z}_{1}\hat{Z}_{2}-\hat{Z}_{2}$ errors, each with rate (per time) $4g^{2}/\kappa = \epsilon_{ZZ}^{2}/\kappa_{2}$. Note that the last term in the effective Hamiltonian can in principle induce excitation exchange between the two modes but we may neglect this effect because the excited states decay very quickly back to the ground state manifold (i.e., $\epsilon_{ZZ}\ll 4\kappa_{2}\alpha^{2}$ which is indeed the case in the parameter regime we focus on). Putting all this together, we find the following effective master equation in the ground-state manifold of two cat qubits: 
\begin{align}
    \frac{d\hat{\rho}_{g}(t)}{dt} &=  \kappa_{1}\alpha^{2}\Big{[} \mathcal{D}[\hat{Z}_{1}] + \mathcal{D}[\hat{Z}_{2}]  \Big{]} \hat{\rho}_{g}(t)
    \nonumber\\
    &\quad + \frac{\epsilon_{ZZ}^{2}}{\kappa_{2}} \Big{[} \mathcal{D}[ \hat{Z}_{1}\hat{Z}_{2} - \hat{Z}_{1}] + \mathcal{D}[\hat{Z}_{1}\hat{Z}_{2} - \hat{Z}_{2}]  \Big{]} \hat{\rho}_{g}(t)
    \nonumber\\
    &\quad-i [8\epsilon_{ZZ}\alpha^{2} |11\rangle\langle 11| , \hat{\rho}_{g}(t) ] . 
\end{align}
The effective Hamiltonian (which commutes with the $Z$-type effective jump operators) generates a CZ rotation $CZ(\theta)$ with $\theta = -8\epsilon_{ZZ}\alpha^{2}T$ where $T$ is the gate time. Hence, $\epsilon_{ZZ} = -\theta / (8\alpha^{2}T)$. The remaining effective jump operators induce an error channel 
\begin{align}
    \mathcal{N}_{CZ(\theta)}(\hat{\rho}) &\simeq \hat{\rho} + \kappa_{1}\alpha^{2}T\Big{[} \mathcal{D}[\hat{Z}_{1}] + \mathcal{D}[\hat{Z}_{2}]  \Big{]} \hat{\rho}
    \nonumber\\
    &\quad + \frac{\theta^{2}}{64\kappa_{2}\alpha^{4}T} \Big{[} \mathcal{D}[ \hat{Z}_{1}\hat{Z}_{2} - \hat{Z}_{1}] 
    \nonumber\\
    &\qquad\qquad\qquad + \mathcal{D}[\hat{Z}_{1}\hat{Z}_{2} - \hat{Z}_{2}]  \Big{]} \hat{\rho}, 
\end{align}
provided that the error rates (per gate) $\kappa_{1}\alpha^{2}T$ and $\theta^{2}/(64\kappa_{2}\alpha^{4}T)$ are much smaller than unity. Ignoring the off-diagonal terms like $\hat{Z}_{1}\hat{Z}_{2}\hat{\rho}\hat{Z}_{1}$, we get Pauli $Z$ error rates 
\begin{align}
    &\bar{p}_{Z_{1}} = \bar{p}_{Z_{2}} = \kappa_{1}\alpha^{2}T  + \frac{\theta^{2}}{64\kappa_{2}\alpha^{4}T},  
    \nonumber\\
    &\bar{p}_{Z_{1}Z_{2}} = \frac{\theta^{2}}{32\kappa_{2}\alpha^{4}T} . 
\end{align}
The total gate infidelity $1-\bar{p}_{Z_{1}}-\bar{p}_{Z_{2}}-\bar{p}_{Z_{1}Z_{2}}$ is minimized at the optimal gate time 
\begin{align}
    \bar{T}^{\star}_{CZ(\theta)} &= \frac{|\theta|}{4\alpha^{3} \sqrt{2\kappa_{1}\kappa_{2}} } , 
\end{align}
and the $Z$ error rates (per gate) at this optimal gate time are given by
\begin{align}
    \bar{p}_{Z_{1}}^{\star} = \bar{p}_{Z_{2}}^{\star} = \frac{3}{2}p_{Z_{1}Z_{2}}^{\star} = \frac{3|\theta|}{8\alpha}\sqrt{\frac{\kappa_{1}}{2\kappa_{2}}} . 
\end{align}
Note that the optimal $Z$ error rates for $Z$ and CZ rotations decrease as $\alpha$ increases. Below, we show that this is not the case for the CNOT and Toffoli gates.      

\subsection{CNOT}
\label{subsection:CNOT appendix}

The CNOT gate between two cat qubits can be realized by 
\begin{align}
    \frac{d\hat{\rho}(t)}{dt} &= \kappa_{2}\Big{[} \mathcal{D}[\hat{a}_{1}^{2}-\alpha^{2}]  + \mathcal{D}[\hat{L}_{2}(t)]  \Big{]} \hat{\rho}(t)  
    \nonumber\\
    &\quad + \kappa_{1} \Big{[} \mathcal{D}[\hat{a}_{1}] + \mathcal{D}[\hat{a}_{2}] \Big{]} \hat{\rho}(t)  -i [ \hat{H}  , \hat{\rho}(t) ], 
\end{align}
where $\hat{a}_{1}$ and $\hat{a}_{2}$ are the annihilation operators of the control and the target modes, respectively, and $\hat{L}_{2}(t)$ and $\hat{H}$ are given by  
\begin{align}
    \hat{L}_{2}(t) &= \hat{a}_{2}^{2}-\alpha^{2} + \frac{\alpha}{2}(e^{2i\frac{\pi}{T}t} -1 )(\hat{a}_{1}-\alpha) , 
    \nonumber\\
    \hat{H} &= \frac{\pi}{4\alpha T}(\hat{a}_{1}+\hat{a}_{1}^{\dagger}-2\alpha) ( \hat{a}_{2}^{\dagger}\hat{a}_{2}-\alpha^{2}) . 
\end{align}
How and whether this master equation can be physically implemented is discussed in \cref{appendix:physical implementation of cat qubit gates}. Here, we focus on analyzing the effective $Z$ error rates on the cat qubits under this master equation.

Note that the time-dependent engineered jump operator $\hat{L}_{2}(t)$ stabilizes the target mode in the $|\pm\alpha\rangle$ (or $|\pm \alpha e^{i\frac{\pi}{T}t}\rangle$) manifold if the control cat qubit is in the $|0\rangle \simeq |\alpha\rangle$ (or $|1\rangle \simeq |-\alpha\rangle$) state. As a result, the target cat qubit is rotated by $180\degree$ at time $t=T$ only if the control qubit is in the $|1\rangle$ state. That is, an $\hat{X}$ gate is applied to the target cat qubit (i.e., $|\pm\alpha\rangle \rightarrow |\mp \alpha\rangle$) conditioned on the control cat qubit being in the $|1\rangle$ state, hence the desired CNOT gate. Note that for this conditional stabilization to work, the engineered jump operator $\hat{L}_{2}$ should be modulated adiabatically (i.e., $T\gg 1/(\kappa_{2}\alpha^{2})$) such that the target mode does not leak out of the $|\pm \alpha e^{i\frac{\pi}{T}t}\rangle$ manifold if the control qubit is in the $|1\rangle$ state. Adverse effects due to the non-adiabaticity can be partially (but not fully) compensated for by the compensating Hamiltonian $\hat{H}$. See more on this below.  

To analyze this master equation, we first use a hybrid basis where the control and the target modes are described by the shifted and usual Fock basis, respectively. In the hybrid basis, assuming $|\alpha|\gg 1$ and using an approximate expression $\hat{a} \simeq \hat{Z}\otimes (\hat{b} + \alpha)$, the compensating Hamiltonian is given by
\begin{align}
    \hat{H} &= - \frac{\pi}{T}|1\rangle\langle 1|_{1}\otimes (\hat{a}_{2}^{\dagger}\hat{a}_{2}-\alpha^{2}) 
    \nonumber\\
    &\quad +\frac{\pi}{4\alpha T} \hat{Z}_{1}\otimes( \hat{b}_{1}+\hat{b}_{1}^{\dagger})( \hat{a}_{2}^{\dagger}\hat{a}_{2}-\alpha^{2}) . \label{eq:CNOT compensating Hamiltonian good and bad}
\end{align}
Since we are using the shifted Fock basis for the control mode and the usual Fock basis for the target mode at this point, $\hat{b}_{1}$ is a $d\times d$ matrix whereas $\hat{a}_{2}$ is a $2d\times 2d$ matrix, where $d$ is defined in \cref{appendix:Shifted Fock Basis}. 

Note that the first term in \cref{eq:CNOT compensating Hamiltonian good and bad}, which is a desired term, rotates the target mode conditioned on the control mode being in the $|1\rangle$ state branch. Hence, this term actively brings the target mode to the $|\pm\alpha e^{i\frac{\pi}{T}t}\rangle$ manifold (if the control qubit is in the $|1\rangle$ state) and thus makes it unnecessary for the system to adiabatically relax under the engineered jump operator $\hat{L}_{2}$. In particular, conditioned on the control qubit being in the $|1\rangle$ state, this term makes the target mode rotate by $180\degree$ at $t=T$, implementing an X gate (i.e., $|\pm\alpha\rangle \rightarrow |\mp\alpha\rangle$) to the target cat qubit. 

While the first term compensates for the adverse effects of the non-adiabaticity, the second term induces an undesirable back-action to the control mode which, as we show below, turns out to be a significant error source for the CNOT gate. Intuitively, the reason why the second term is detrimental is because the cat states in the target mode are not eigenstates of the excitation number operator $\hat{a}_{2}^{\dagger}\hat{a}_{2}$ and rather follow a Poissonian-like distribution with mean excitation number $\alpha^{2}$. Due to such fluctuations in the excitation number of the target mode, the undesired second term makes the control mode leak out of its ground state manifold and at the same time causes a $Z$ error on the control qubit space. How this undesired term degrades the CNOT gate fidelity can be best described in a rotating frame and in the full shifted Fock basis which we describe below.  

Now, we go to a rotating frame with respect to the desired compensating Hamiltonian 
\begin{align}
    \hat{H}'\equiv - \frac{\pi}{T}|1\rangle\langle 1|_{1}\otimes (\hat{a}_{2}^{\dagger}\hat{a}_{2}-\alpha^{2}), 
\end{align}
that is, we consider the time evolution of $\hat{\rho}_{I}(t) \equiv e^{i\hat{H}'t} \hat{\rho}(t)e^{-i\hat{H}'t}$ which should ideally be idling. In the rotating frame (assuming $|\alpha|\gg 1$), the annihilation operator of the control mode $\hat{Z}_{1}\otimes (\hat{b}_{1}+\alpha)$ is unchanged since $\hat{Z}_{1}$ commutes with $|1\rangle\langle 1|_{1}$ in $\hat{H}'$ (this is not the case when the orthonormalization is taken into account as there are exponentially small time-dependent corrections to $\hat{a}_{1}$ in the rotating frame). On the other hand, $\hat{a}_{2}$ is transformed as
\begin{align}
    \hat{a}_{2} &\rightarrow e^{i\hat{H}'t} \hat{a}_{2} e^{-i\hat{H}'t} 
    \nonumber\\
    &= |0\rangle\langle 0|_{1}\otimes \hat{a}_{2} + |1\rangle\langle 1|_{1}\otimes \hat{a}_{2}e^{i\frac{\pi}{T}t} = \hat{Z}_{1}\Big{(} \frac{\pi}{T}t \Big{)}\otimes \hat{a}_{2},  
\end{align}
where we define $\hat{Z}_{k}(\theta)$ as $\hat{Z}_{k}(\theta)\equiv \exp[i\theta |1\rangle\langle 1|_{k}]$. Having moved to the rotating frame, we finally use the shifted Fock basis for the target mode and replace $\hat{a}_{2}$ by $\hat{Z}_{2}\otimes (\hat{b}_{2}+\alpha)$. 

In the rotating frame (and in the full shifted Fock basis), the master equation is given by
\begin{align}
    \frac{d\hat{\rho}_{I}(t)}{dt} &= \kappa_{2}\Big{[} \mathcal{D}[ \hat{I}_{1,2}\otimes (\hat{b}_{1}^{2}+2\alpha\hat{b}_{1}) ]   + \mathcal{D}[ \hat{L}'_{2}(t)  ]  \Big{]} \hat{\rho}_{I}(t)  
    \nonumber\\
    &\!\!\!\!\!\! + \kappa_{1} \Big{[} \mathcal{D}[ \hat{Z}_{1} \otimes (\hat{b}_{1} + \alpha) ] 
    \nonumber\\
    &\qquad + \mathcal{D}[ \hat{Z}_{1}\Big{(} \frac{\pi}{T}t \Big{)}\hat{Z}_{2} \otimes (\hat{b}_{2}+\alpha) ] \Big{]} \hat{\rho}_{I}(t) 
    \nonumber\\
    &\!\!\!\!\!\! -i\Big{[}\frac{\pi}{4\alpha T} \hat{Z}_{1}\otimes( \hat{b}_{1}+\hat{b}_{1}^{\dagger})( \hat{b}_{2}^{\dagger}\hat{b}_{2} +\alpha(\hat{b}_{2}+\hat{b}_{2}^{\dagger}) ) ,\hat{\rho}_{I}(t) \Big{]}, 
\end{align}
where the jump operator $\hat{L}'_{2}(t)\equiv e^{i\hat{H}'t} \hat{L}_{2}(t) e^{-i\hat{H}'t} $ in the rotating frame is given by
\begin{align}
    \hat{L}'_{2}(t) =  \hat{Z}_{1}\Big{(}  \frac{2\pi}{T}t \Big{)} \otimes (\hat{b}_{2}^{2}+2\alpha\hat{b}_{2}) + \frac{\alpha}{2}(e^{2i\frac{\pi}{T}t} -1 ) \hat{Z}_{1}\otimes \hat{b}_{1}  . \label{eq:CNOT target stabilization before approximation}
\end{align}
Similarly as in the case of Z and CZ rotations, we only consider the first excited state in each mode ($\hat{b}_{1}^{2}=\hat{b}_{2}^{2}=0$) and ignore weak internal couplings and dissipations within the excited state manifold assuming that the engineered dissipation rate $\kappa_{2}$ dominates. Lastly, we ignore the second term in the jump operator $\hat{L}_{2}(t)$ to not complicate the analysis and convey the main idea more easily. This approximation can have a minor quantitative impact as the second term in $\hat{L}_{2}(t)$ is only four times weaker than the first term in the worst case ($t = T/2$). However, the key qualitative features (e.g., scaling) are not affected by this simplification. 

With the above simplifications, the master equation is given by 
\begin{align}
    \frac{d\hat{\rho}_{I}(t)}{dt} &= 4\kappa_{2}\alpha^{2}\Big{[} \mathcal{D}[ \hat{I}_{1,2}\otimes \hat{b}_{1}  ]   + \mathcal{D}[ \hat{Z}_{1}\Big{(}  \frac{2\pi}{T}t \Big{)}\otimes \hat{b}_{2}  ]  \Big{]} \hat{\rho}_{I}(t) 
    \nonumber\\
    &\quad +  \kappa_{1}\alpha^{2} \Big{[} \mathcal{D}[ \hat{Z}_{1} \otimes \hat{I}] + \mathcal{D}[ \hat{Z}_{1}\Big{(} \frac{\pi}{T}t \Big{)}\hat{Z}_{2} \otimes \hat{I} ] \Big{]} \hat{\rho}_{I}(t) 
    \nonumber\\
    &\quad -i\Big{[}\frac{\pi}{4 T} \hat{Z}_{1}\otimes( \hat{b}_{1}\hat{b}_{2} + \hat{b}_{1}^{\dagger}\hat{b}_{2}^{\dagger}) ,\hat{\rho}_{I}(t) \Big{]}. \label{eq:CNOT equation shifted Fock basis before AE}
\end{align}
Note that the undesired term in the compensating Hamiltonian $\hat{H}-\hat{H}' = \frac{\pi}{4 T} \hat{Z}_{1}\otimes( \hat{b}_{1}\hat{b}_{2} + \hat{b}_{1}^{\dagger}\hat{b}_{2}^{\dagger})$ jointly excites both the control and the target modes with a coupling strength $g= \pi/(4T)$ and at the same time causes a $\hat{Z}_{1}$ error on the control qubit. The excited state $|11\rangle'$ (defined as $|\hat{b}_{1}^{\dagger}\hat{b}_{1} = 1\rangle \otimes |\hat{b}_{2}^{\dagger}\hat{b}_{2} = 1\rangle$, not to  be confused with the computational basis state $|11\rangle$) eventually decays back to the code space through either $|11\rangle' \rightarrow |01\rangle' \rightarrow |00\rangle'$ or $|11\rangle' \rightarrow |10\rangle' \rightarrow |00\rangle'$ with a total decay rate (per time) $\kappa = 8\kappa_{2}\alpha^{2}$. Note that whichever way the excited state decays, the decay is accompanied by a Z rotation on the control mode, i.e., $\hat{Z}_{1}( \frac{2\pi}{T}t )$. Thus, after adiabatically eliminating the excited states, we get an effective jump operator $\hat{Z}_{1}\hat{Z}_{1}( \frac{2\pi}{T}t )$ with a decay rate (per time) $4g^{2}/\kappa = \pi^{2}/(32\kappa_{2}\alpha^{2}T^{2})$ in the ground state manifold. Thus, we have the following master equation. 
\begin{align}
    \frac{d\hat{\rho}_{I,g}(t)}{dt} &= \kappa_{1}\alpha^{2} \Big{[} \mathcal{D}[ \hat{Z}_{1}] + \mathcal{D}[ \hat{Z}_{1}\Big{(} \frac{\pi}{T}t \Big{)}\hat{Z}_{2} ] \Big{]} \hat{\rho}_{I,g}(t) \nonumber\\
    &\quad + \frac{\pi^{2}}{32\kappa_{2}\alpha^{2}T^{2}}\mathcal{D}\Big{[} \hat{Z}_{1}\hat{Z}_{1}\Big{(} \frac{2\pi}{T}t \Big{)} \Big{]} \hat{\rho}_{I,g}(t), \label{eq:CNOT equation shifted Fock basis after AE in the interaction picture}
\end{align}
where the dissipators in the first line are due to the single phonon loss projected to the ground state manifold. By integrating and ignoring higher order terms, we find 
\begin{align}
    \hat{\rho}_{I,g}(T) &\simeq \hat{\rho}_{g}(0) + \int_{0}^{T}dt \Big{(} \kappa_{1}\alpha^{2} \Big{[} \mathcal{D}[ \hat{Z}_{1}] + \mathcal{D}[ \hat{Z}_{1}\Big{(} \frac{\pi}{T}t \Big{)}\hat{Z}_{2} ] \Big{]}
    \nonumber\\
    &\qquad\qquad + \frac{\pi^{2}}{32\kappa_{2}\alpha^{2}T^{2}}\mathcal{D}\Big{[} \hat{Z}_{1}\hat{Z}_{1}\Big{(} \frac{2\pi}{T}t \Big{)} \Big{]} \Big{)} \hat{\rho}_{g}(0) 
\end{align}
at the gate time $T$.

To go back to the original frame (i.e., $\hat{\rho}(T) = e^{-i\hat{H}' T} \hat{\rho}_{I}(T)e^{i\hat{H}' T}$), note that $e^{-i\hat{H}' T}$ is given by
\begin{align}
    e^{-i\hat{H}' T} &= |0\rangle\langle 0|_{1} \otimes \hat{I} + |1\rangle\langle 1|_{1}\otimes e^{i\pi \hat{a}_{2}^{\dagger}\hat{a}_{2} } e^{-i \pi \alpha^{2} }
\end{align}
in the hybrid basis. In the shifted Fock basis, $e^{i\pi\hat{a}^{\dagger}\hat{a}}$ is exactly given by $\hat{X}\otimes \hat{I}$ and thus we have 
\begin{align}
    e^{-i\hat{H}' T} &= ( \hat{Z}_{1}(-\pi\alpha^{2}) \cdot \textrm{CNOT}_{1\rightarrow 2} ) \otimes \hat{I} 
\end{align}
in the full shifted Fock basis. Thus, projecting $e^{-i\hat{H}' T}$ to the ground state manifold of the cat qubits, we find $\hat{\rho}_{g}(T) = CX' \hat{\rho}_{I,g}(T) CX'^{\dagger}$ where $CX' \equiv \hat{Z}_{1}(-\pi\alpha^{2}) \cdot \textrm{CNOT}_{1\rightarrow 2}$. Therefore, we can understand $\hat{\rho}_{g}(T)$ as a state that results from applying a unitary operation $CX'$ to the input state $\hat{\rho}_{g}(0)$ which is then corrupted by an error channel
\begin{align}
    \mathcal{N}_{CX'}(\hat{\rho}) &\simeq \hat{\rho} + \int_{0}^{T}dt \Big{(} \kappa_{1}\alpha^{2} \Big{[} \mathcal{D}[ \hat{Z}_{1}] + \mathcal{D}[ \hat{Z}_{1}\hat{Z}_{1}\Big{(} \frac{\pi}{T}t \Big{)}\hat{Z}_{2} ] \Big{]}
    \nonumber\\
    &\qquad\qquad + \frac{\pi^{2}}{32\kappa_{2}\alpha^{2}T^{2}}\mathcal{D}\Big{[} \hat{Z}_{1}\hat{Z}_{1}\Big{(} \frac{2\pi}{T}t \Big{)} \Big{]} \Big{)} \hat{\rho} ,
\end{align}
where we used the fact that $\hat{Z}_{2}$ is transformed via $\textrm{CNOT}_{1\rightarrow 2}$ into $\hat{Z}_{1}\hat{Z}_{2}$. Performing the integration explicitly and ignoring off-diagonal terms
similarly as in the analysis of the controlled Z rotations, we find that the $Z$ error rates (per gate) of the $CX'$ gate are given by
\begin{align}
    &\bar{p}_{Z_{1}} =   \kappa_{1}\alpha^{2}T + \frac{\pi^{2}}{64\kappa_{2}\alpha^{2}T}, 
    \nonumber\\
    &\bar{p}_{Z_{2}} = \bar{p}_{Z_{1}Z_{2}} = \frac{1}{2}\kappa_{1}\alpha^{2}T . 
\end{align}
Hence, the optimal gate time that minimizes the total gate infidelity is given by 
\begin{align}
    \bar{T}_{CX'}^{\star} &= \frac{ \pi }{8\alpha^{2}\sqrt{2\kappa_{1}\kappa_{2}}} , 
\end{align}
and at the optimal gate time, the $Z$ error rates (per gate) of the $CX'$ gate are given by
\begin{align}
    \bar{p}_{Z_{1}}^{\star} = 6\bar{p}_{Z_{2}}^{\star} = 6\bar{p}_{Z_{1}Z_{2}}^{\star} = \frac{3\pi}{8}\sqrt{\frac{\kappa_{1}}{2\kappa_{2}}} = 0.833\sqrt{\frac{\kappa_{1}}{\kappa_{2}}} . \label{eq:Z error rates CX'}
\end{align}
Note that the $Z$ errors (per gate) due to the single phonon loss only account for half the total $CX'$ gate error rate at the optimal gate time. The remaining half comes from the $Z$ error due to the undesired term in the compensating Hamiltonian (see the discussion below \cref{eq:CNOT compensating Hamiltonian good and bad}). Numerically, we find that the optimal $Z$ error rates (per gate) of the CNOT gate are given by (see \cref{tab:Gateerrorrates})
\begin{align}
    p_{Z_{1}}^{\star} = 6.067p_{Z_{2}}^{\star} = 6.067p_{Z_{1}Z_{2}}^{\star} = 0.91\sqrt{\frac{\kappa_{1}}{\kappa_{2}}} , 
\end{align}
which agree well with the perturbative prediction in \cref{eq:Z error rates CX'} within a relative error of $10\%$. Note that the quantitative differences are mostly due to the fact that we neglected the second term in \cref{eq:CNOT target stabilization before approximation} to make the analysis simpler and also that we only consider the first excited state manifold in each mode.     

We emphasize that to really implement the desired $\textrm{CNOT}_{1\rightarrow 2}$ gate, one should apply a Z rotation $\hat{Z}_{1}(\pi\alpha^{2})$ to the control cat qubit to compensate for the extra Z rotation in the $CX'$ gate and such an extra operation will result in additional $Z$ errors (see \cref{subsection:Z rotations appendix}). However, if the average excitation number $\alpha^{2}$ is an even integer, the extra Z rotation is not needed and thus the $Z$ error rates of the CNOT gate are simply given by the ones in \cref{eq:Z error rates CX'}.  

It is often said that bosonic dephasing $\kappa_{\phi}\mathcal{D}[\hat{a}^{\dagger}\hat{a}]$ does not cause any $Z$ errors on cat qubits because it preserves the parity. While this is true for idling, Z and CZ rotations, this is not the case for the CNOT and Toffoli gates. To see why this is the case, note that $\kappa_{\phi}\mathcal{D}[\hat{a}^{\dagger}\hat{a}]$ is given by 
\begin{align}
    \kappa_{\phi}\mathcal{D}[\hat{a}^{\dagger}\hat{a}] &= \kappa_{\phi}\mathcal{D}[ \hat{I} \otimes (\hat{b}^{\dagger}+\alpha)(\hat{b}+\alpha)]
    \nonumber\\
    &= \kappa_{\phi}\mathcal{D}[ \hat{I} \otimes (\hat{b}^{\dagger}\hat{b}+\alpha(\hat{b}+\hat{b}^{\dagger}))]
\end{align}
in the shifted Fock basis, where we assumed $|\alpha|\gg 1$ and used the fact that $\mathcal{D}[\hat{O}+c\hat{I}] = \mathcal{D}[\hat{O}]$ for all hermitian operators $\hat{O}^{\dagger} = \hat{O}$ and a scalar $c$. If the cat qubit is in its ground state manifold, $\hat{b}^{\dagger}\hat{b}+\alpha\hat{b}$ acts trivially and thus the dominant effect due to the dephasing is the heating caused by the term $\alpha\hat{b}^{\dagger}$, i.e., 
\begin{align}
     \kappa_{\phi}\mathcal{D}[\hat{a}^{\dagger}\hat{a}] &\simeq \kappa_{\phi}\alpha^{2}[\hat{I}\otimes \hat{b}^{\dagger}].  \label{eq:dephasing induces heating}
\end{align}
Such heating, however, does not induce any $Z$ errors on the qubit space, as indicated by the identity operator in the first slot of the tensor product; this is consistent with the fact that the bosonic dephasing alone cannot change the excitation number parity.  

In the case of the CNOT gate, dephasing in each mode independently causes heating, resulting in direct population transfer from the ground state manifold associated with $|00\rangle'$ to the excited states manifolds with $|10\rangle'$ and $|01\rangle'$. As shown in the first line of \cref{eq:CNOT equation shifted Fock basis before AE}, the excited states $|10\rangle'$ and $|01\rangle'$ decay back to the code space via the engineered dissipation $4\kappa_{2}\alpha^{2} \mathcal{D}[ \hat{I}_{1,2}\otimes \hat{b}_{1}  ]  $ and $  4\kappa_{2}\alpha^{2}\mathcal{D}[ \hat{Z}_{1} (  \frac{2\pi}{T}t )\otimes \hat{b}_{2}  ] $, respectively. While the former engineered dissipation (corresponding to the control mode) is parity preserving, the latter (corresponding to the target mode) induces a Z rotation of the control mode, i.e., $\hat{Z}_{1} (  \frac{2\pi}{T}t )$. This is because the engineered jump operator on the target mode $\hat{L}_{2}(t)$ rotates conditioned on the state of the control mode. Consequently, while the process $ |10\rangle' \rightarrow |00\rangle'$ is parity preserving in overall, the other process $|01\rangle'\rightarrow |00\rangle'$ induces $Z$ errors on the qubit degree of freedom. More explicitly, the heating followed by the fast relaxation in the target mode induces a new noise process 
\begin{align}
     \kappa_{\phi}\alpha^{2}\mathcal{D}\Big{[} \hat{Z}_{1} \Big{(}  \frac{2\pi}{T}t \Big{)} \Big{]}\hat{\rho}_{I,g}(t)
\end{align}
in addition to the noise processes described in the right hand side of \cref{eq:CNOT equation shifted Fock basis after AE in the interaction picture}. Integrating over the time window $t \in [0,T]$ and ignoring off-diagonal terms, such a noise process adds an error rate (per gate) $\kappa_{\phi}\alpha^{2}T/2$ to $p_{Z_{1}}$, i.e., 
\begin{align}
    &\bar{p}_{Z_{1}} =   \kappa_{1}\alpha^{2}T + \frac{1}{2}\kappa_{\phi}\alpha^{2}T  + \frac{\pi^{2}}{64\kappa_{2}\alpha^{2}T}, 
    \nonumber\\
    &\bar{p}_{Z_{2}} = \bar{p}_{Z_{1}Z_{2}} = \frac{1}{2}\kappa_{1}\alpha^{2}T .
\end{align}
That is, even in the lossless case (i.e., $\kappa_{1}=0$), the CNOT gate is not free from $Z$ errors and is instead limited by $\bar{p}_{Z_{1}}^{\star} \propto \sqrt{ \kappa_{\phi}/\kappa_{2} }$ at the optimal gate time. In contrast, dephasing does not induce any additional $Z$ errors in the case of idling, Z rotations, and CZ rotations because in these cases the engineered dissipation is always static (i.e., $\kappa_{2}\mathcal{D}[\hat{a}^{2}-\alpha^{2}]$ in the usual Fock basis or approximately $4\kappa_{2}\alpha^{2}\mathcal{D}[\hat{I}\otimes \hat{b}]$ in the shifted Fock basis) and thus preserves the parity when it brings the excited states back to the cat code manifold. We also remark that in the presence of non-zero thermal population $n_{\textrm{th}}$, we simply need to replace $\kappa_{1}$ by $\kappa_{1}(1+2n_{\textrm{th}})$.  

We reinforce that the above perturbative approach based on an approximate expression $\hat{a} \simeq \hat{Z}\otimes (\hat{b}+\alpha)$ is not capable of capturing non-$Z$-type errors which decrease exponentially in $|\alpha|^{2}$. Numerically, however, we simulate the master equation in the shifted Fock basis without making any approximations to capture the exponentially small error rates and get accurate $Z$ error rates. In particular, we use an exact expression of the annihilation operator in the shifted Fock basis (obtained via the procedure described in \cref{appendix:Shifted Fock Basis}) and perform the frame transformations similarly as in this section (i.e., hybrid basis, rotating frame, and then full shifted Fock basis) but in a way that takes into account exponentially small corrections in $|\alpha|^{2}$. See \cref{app:Gate Error Simulations} for numerical results.

\subsection{Toffoli}
\label{subsection:Toffoli appendix}

A Toffoli gate among three cat qubits can be implemented by
\begin{align}
    \frac{d\hat{\rho}(t)}{dt} &= \kappa_{2}\Big{[} \mathcal{D}[\hat{a}_{1}^{2}-\alpha^{2}] + \mathcal{D}[\hat{a}_{2}^{2}-\alpha^{2}]  +  \mathcal{D}[ \hat{L}_{3}(t) ]  \Big{]} \hat{\rho}(t)  
    \nonumber\\
    &\quad+ \kappa_{1} \Big{[} \mathcal{D}[\hat{a}_{1}]  + \mathcal{D}[\hat{a}_{2}] + \mathcal{D}[\hat{a}_{3}] \Big{]} \hat{\rho}(t)  -i [ \hat{H} , \hat{\rho}(t) ]
\end{align}
where the engineered dissipation $\hat{L}_{3}(t)$ and the compensating Hamiltonian $\hat{H}$ are given by
\begin{align}
    \hat{L}_{3}(t) &= \hat{a}_{3}^{2}-\alpha^{2} - \frac{1}{4}(e^{2i\frac{\pi}{T}t} -1 )(\hat{a}_{1}-\alpha)(\hat{a}_{2}-\alpha), 
    \nonumber\\
    \hat{H} &= - \frac{\pi}{8\alpha^{2} T}( (\hat{a}_{1}-\alpha)(\hat{a}_{2}^{\dagger}-\alpha) + \textrm{h.c.} ) (\hat{a}_{3}^{\dagger}\hat{a}_{3}-\alpha^{2}) .
\end{align}
Similarly as in the case of the CNOT gate, the time-dependent engineered jump operator $\hat{L}_{3}(t)$ stabilizes the target mode $\hat{a}_{3}$ in the $|\pm\alpha e^{i\frac{\pi}{T}t} \rangle$ manifold if the control modes $\hat{a}_{1}$ and $\hat{a}_{2}$ are in the ``trigger'' state $|11\rangle \simeq |-\alpha,-\alpha\rangle$ or in the usual cat code manifold $|\pm\alpha\rangle$ otherwise. Hence, the target mode is rotated by $180\degree$ (i.e., X gate on the cat qubit) at the gate time $t=T$ only if the control qubits are in the trigger state $|11\rangle$, realizing the controlled-controlled-X gate, or the Toffoli gate on the three cat qubits. Moreover, the compensating Hamiltonian $\hat{H}$ mitigates the adverse effects due to the non-adiabaticity by actively bringing the target mode in the desired manifold $|\pm\alpha e^{i\frac{\pi}{T}t} \rangle$ when the control qubits are in the trigger state $|11\rangle \simeq |-\alpha,-\alpha\rangle$. 

To analyze the $Z$ error rates of the Toffoli gate perturbatively, we first use to the hybrid basis system where the control modes are described by the shifted Fock basis and the target mode is described by the usual Fock basis. In the hybrid basis, the compensating Hamiltonian is given by 
\begin{align}
    \hat{H} &= -\frac{\pi}{T}|11\rangle\langle 11|_{1,2}\otimes (\hat{a}_{3}^{\dagger}\hat{a}_{3}-\alpha^{2})
    \nonumber\\
    &\quad -\frac{\pi}{8\alpha T}(\hat{Z}_{1}-\hat{Z}_{1}\hat{Z}_{2})\otimes (\hat{b}_{1}+\hat{b}_{1}^{\dagger})(\hat{a}_{3}^{\dagger}\hat{a}_{3}-\alpha^{2}) 
    \nonumber\\
    &\quad -\frac{\pi}{8\alpha T}(\hat{Z}_{2}-\hat{Z}_{1}\hat{Z}_{2})\otimes (\hat{b}_{2}+\hat{b}_{2}^{\dagger})(\hat{a}_{3}^{\dagger}\hat{a}_{3}-\alpha^{2}) 
    \nonumber\\
    &\quad +\frac{\pi}{8\alpha^{2} T}\hat{Z}_{1}\hat{Z}_{2}\otimes (\hat{b}_{1}\hat{b}_{2}^{\dagger}+\hat{b}_{1}^{\dagger}\hat{b}_{2})(\hat{a}_{3}^{\dagger}\hat{a}_{3}-\alpha^{2}), 
\end{align}
where we used $\hat{a}_{k} \simeq \hat{Z}_{k}\otimes (\hat{b}_{k}+\alpha)$ for $k\in \lbrace 1,2\rbrace$. Note that the first term is a desired term that rotates the target mode by $180\degree$ over the gate time $T$ only if the two control qubits are in the trigger state. The fourth term acts trivially if the system is in the ground state manifold. The second and the third terms, on the other hand, make the system excited and leak out of the ground state manifold. 

Similarly as in the case of the CNOT gate, we go to a rotating frame with respect to the desired compensating Hamiltonian
\begin{align}
    \hat{H}' \equiv -\frac{\pi}{T}|11\rangle\langle 11|_{1,2}\otimes (\hat{a}_{3}^{\dagger}\hat{a}_{3}-\alpha^{2}), 
\end{align}
i.e., $\hat{\rho}_{I}(t)\equiv e^{i\hat{H}'t}\hat{\rho}(t)e^{-i\hat{H}'t}$. In this frame, the annihilation operators of the control modes $\hat{a}_{1}$ and $\hat{a}_{2}$ are unchanged but the annihilation operator of the target mode $\hat{a}_{3}$ is transformed as 
\begin{align}
    \hat{a}_{3} \rightarrow e^{i\hat{H}'t}\hat{a}_{3}e^{-i\hat{H}'t} = CZ_{1,2}\Big{(} \frac{\pi}{T}t \Big{)} \otimes \hat{a}_{3}, 
\end{align}
where $CZ_{1,2}(\theta) \equiv \exp[ i\theta |11\rangle\langle 11|_{1,2} ]$. Lastly, by using the shifted Fock basis for the target mode as well (i.e., $\hat{a}_{3}\simeq \hat{Z}_{3}\otimes (\hat{b}_{3} + \alpha)$), we find the following equation of motion for $\hat{\rho}_{I}(t)$: 
\begin{align}
    \frac{d\hat{\rho}_{I}(t)}{dt} &= \kappa_{2}\Big{[} \mathcal{D}[\hat{I}_{1,2,3} \otimes (\hat{b}_{1}^{2} + 2\alpha \hat{b}_{1})] 
    \nonumber\\
    &\qquad + \mathcal{D}[\hat{I}_{1,2,3} \otimes (\hat{b}_{2}^{2} + 2\alpha \hat{b}_{2})]  +  \mathcal{D}[ \hat{L}'_{3}(t) ]  \Big{]} \hat{\rho}_{I}(t)  
    \nonumber\\
    &\quad+ \kappa_{1} \Big{[} \mathcal{D}[ \hat{Z}_{1} \otimes (\hat{b}_{1}+\alpha) ]  + \mathcal{D}[ \hat{Z}_{2} \otimes (\hat{b}_{2}+\alpha) ] 
    \nonumber\\
    &\qquad\quad + \mathcal{D}[ CZ_{1,2}\Big{(} \frac{\pi}{T}t \Big{)}\hat{Z}_{3} \otimes (\hat{b}_{3}+\alpha)] \Big{]} \hat{\rho}_{I}(t)  
    \nonumber\\
    &\quad -i [ \hat{H}-\hat{H}' , \hat{\rho}_{I}(t) ]. 
\end{align}
Here, $\hat{L}'_{3}(t) \equiv e^{i\hat{H}'t}\hat{L}_{3}(t) e^{-i\hat{H}'t}$ is given by 
\begin{align}
    \hat{L}'_{3}(t) &= CZ_{1,2}\Big{(} \frac{2\pi}{T}t \Big{)} \otimes (\hat{b}_{3}^{2}+ 2\alpha\hat{b}_{3}) 
    \nonumber\\
    &\,\,\, -\frac{1}{4}(e^{2i\frac{\pi}{T}t}-1)\Big{[} \hat{Z}_{1}\hat{Z}_{2}\otimes \hat{b}_{1}\hat{b}_{2}  - \alpha(\hat{Z}_{1}-\hat{Z}_{1}\hat{Z}_{2})\otimes \hat{b}_{1}
    \nonumber\\
    &\qquad\qquad\qquad\quad  - \alpha(\hat{Z}_{2}-\hat{Z}_{1}\hat{Z}_{2})\otimes \hat{b}_{2}  \Big{]} . 
\end{align}
We neglect all the other terms than the first term in the right hand side because they are much smaller than the first term. Also, we only consider the first excited states and set $\hat{b}_{1}^{2} = \hat{b}_{2}^{2} = \hat{b}_{3}^{2}=0$. 

In the full shifted Fock basis, $\hat{H}-\hat{H}'$ is given by
\begin{align}
    &\hat{H}-\hat{H}' 
    \nonumber\\
    &= -\frac{\pi}{8\alpha T}(\hat{Z}_{1}-\hat{Z}_{1}\hat{Z}_{2})\otimes (\hat{b}_{1}+\hat{b}_{1}^{\dagger})(\hat{b}_{3}^{\dagger}\hat{b}_{3} +\alpha(\hat{b}_{3}+\hat{b}_{3}^{\dagger}) ) 
    \nonumber\\
    &\quad -\frac{\pi}{8\alpha T}(\hat{Z}_{2}-\hat{Z}_{1}\hat{Z}_{2})\otimes (\hat{b}_{2}+\hat{b}_{2}^{\dagger})(\hat{b}_{3}^{\dagger}\hat{b}_{3} +\alpha(\hat{b}_{3}+\hat{b}_{3}^{\dagger}) ) 
    \nonumber\\
    &\quad +\frac{\pi}{8\alpha^{2} T}\hat{Z}_{1}\hat{Z}_{2}\otimes (\hat{b}_{1}\hat{b}_{2}^{\dagger}+\hat{b}_{1}^{\dagger}\hat{b}_{2})(\hat{b}_{3}^{\dagger}\hat{b}_{3} +\alpha(\hat{b}_{3}+\hat{b}_{3}^{\dagger}) ).
\end{align}
As explained above, the third term acts trivially on the code space and thus we focus on the first two terms. In particular, we only consider the dominant driving effects due to the first two terms and approximate $\hat{H}-\hat{H}'$ as 
\begin{align}
    \hat{H}-\hat{H}' &\simeq -\frac{\pi}{8 T}(\hat{Z}_{1}-\hat{Z}_{1}\hat{Z}_{2})\otimes (\hat{b}_{1}\hat{b}_{3} +  \hat{b}_{1}^{\dagger}\hat{b}_{3}^{\dagger}) 
    \nonumber\\
    &\quad -\frac{\pi}{8 T}(\hat{Z}_{2}-\hat{Z}_{1}\hat{Z}_{2})\otimes (\hat{b}_{2}\hat{b}_{3} +  \hat{b}_{2}^{\dagger}\hat{b}_{3}^{\dagger}) . \label{eq:TOF compensating Hamiltonian approximate}
\end{align}
Putting everything together, we find the following master equation: 
\begin{align}
    \frac{d\hat{\rho}_{I}(t)}{dt} &= 4\kappa_{2}\alpha^{2}\Big{[} \mathcal{D}[\hat{I}_{1,2,3} \otimes  \hat{b}_{1}] + \mathcal{D}[\hat{I}_{1,2,3} \otimes  \hat{b}_{2}] 
    \nonumber\\
    &\qquad\qquad +  \mathcal{D}[ CZ_{1,2}\Big{(} \frac{2\pi}{T}t \Big{)} \otimes  \hat{b}_{3} ]  \Big{]} \hat{\rho}_{I}(t)  
    \nonumber\\
    &\!\!\!\! + \kappa_{1}\alpha^{2} \Big{[} \mathcal{D}[ \hat{Z}_{1} \otimes \hat{I} ]  + \mathcal{D}[ \hat{Z}_{2} \otimes \hat{I} ] 
    \nonumber\\
    &\!\!\!\!\qquad\qquad + \mathcal{D}[ CZ_{1,2}\Big{(} \frac{\pi}{T}t \Big{)}\hat{Z}_{3} \otimes \hat{I}] \Big{]} \hat{\rho}_{I}(t)  
    \nonumber\\
    &\!\!\!\! +i \Big{[} \frac{\pi}{8 T}(\hat{Z}_{1}-\hat{Z}_{1}\hat{Z}_{2})\otimes (\hat{b}_{1}\hat{b}_{3} +  \hat{b}_{1}^{\dagger}\hat{b}_{3}^{\dagger}) 
    \nonumber\\
    &  +  \frac{\pi}{8 T}(\hat{Z}_{2}-\hat{Z}_{1}\hat{Z}_{2})\otimes (\hat{b}_{2}\hat{b}_{3} +  \hat{b}_{2}^{\dagger}\hat{b}_{3}^{\dagger})    , \hat{\rho}_{I}(t) \Big{]}. 
\end{align}
The undesired terms of the compensating Hamiltonian in \cref{eq:TOF compensating Hamiltonian approximate} make the system excited to the manifold associated with $|101\rangle'$ ($|011\rangle'$) via $\hat{b}_{1}^{\dagger}\hat{b}_{3}^{\dagger}$ ($\hat{b}_{2}^{\dagger}\hat{b}_{3}^{\dagger}$) and at the same time cause an error $\hat{Z}_{1}-\hat{Z}_{1}\hat{Z}_{2}$ ($\hat{Z}_{2}-\hat{Z}_{1}\hat{Z}_{2}$) on the qubit space at a rate (per time) $g=\pi/(8T)$. These excited states decay back to the code space via the engineered dissipation. For instance, $|101\rangle'$ decays back to the code space through either $|101\rangle' \rightarrow |001\rangle' \rightarrow |000\rangle'$ or $|101\rangle' \rightarrow |100\rangle' \rightarrow |000\rangle'$ with a total decay rate (per time) $\kappa = 8\kappa_{2}\alpha^{2}$. In both decay routes, the annihilation of the excitation in the target mode (i.e., $\hat{b}_{3}$) is accompanied by an additional error $CZ_{1,2}( \frac{2\pi}{T}t )$ on the control qubits. The same is true for the other excited state $|011\rangle'$ which decays back to the code space either via $|011\rangle' \rightarrow |001\rangle' \rightarrow |000\rangle'$ or $|011\rangle' \rightarrow |010\rangle' \rightarrow |000\rangle'$. Consequently, by using adiabatic elimination, we find that these driven-dissipative processes induce two independent decay processes with jump operators $(\hat{Z}_{1}-\hat{Z}_{1}\hat{Z}_{2})CZ_{1,2}( \frac{2\pi}{T}t )$ and $(\hat{Z}_{2}-\hat{Z}_{1}\hat{Z}_{2})CZ_{1,2}( \frac{2\pi}{T}t )$ with an effective decay rate (per time) $4g^{2}/\kappa = \pi^{2}/(128\kappa_{2}\alpha^{2}T^{2})$. Hence, the effective master equation in the ground state manifold is given by
\begin{align}
    &\frac{d\hat{\rho}_{I,g}(t)}{dt} 
    \nonumber\\
    &= \kappa_{1}\alpha^{2} \Big{[} \mathcal{D}[ \hat{Z}_{1} ]  + \mathcal{D}[ \hat{Z}_{2} ] + \mathcal{D}\Big{[} CZ_{1,2}\Big{(} \frac{\pi}{T}t \Big{)}\hat{Z}_{3} \Big{]} \Big{]} \hat{\rho}_{I,g}(t)
    \nonumber\\
    &\quad+ \frac{\pi^{2}}{128\kappa_{2}\alpha^{2}T^{2}}\Big{[} \mathcal{D}\Big{[}(\hat{Z}_{1}-\hat{Z}_{1}\hat{Z}_{2})CZ_{1,2}\Big{(} \frac{2\pi}{T}t \Big{)}\Big{]}
    \nonumber\\
    &\qquad\quad + \mathcal{D}\Big{[}(\hat{Z}_{2}-\hat{Z}_{1}\hat{Z}_{2})CZ_{1,2}\Big{(} \frac{2\pi}{T}t \Big{)}\Big{]} \Big{]}\hat{\rho}_{I,g}(t), \label{eq:TOF equation shifted Fock basis after AE in the interaction picture}
\end{align}
where $\hat{\rho}_{I,g}(t) \equiv \langle 000|' \hat{\rho}_{I} |000\rangle'$ is the projected density matrix (of size $2^{3}\times 2^{3}$) to the ground state manifold of the three cat qubits.  

To go back to the original frame (i.e., $\hat{\rho}(T) = e^{-i\hat{H}'T}\hat{\rho}_{I}(T)e^{i\hat{H}'T}$), note that $e^{-i\hat{H}'T}$ is given by 
\begin{align}
    e^{-i\hat{H}'T} &= (\hat{I}_{1,2,3}-|11\rangle\langle 11|_{1,2})\otimes \hat{I} 
    \nonumber\\
    &\quad + |11\rangle\langle 11|_{1,2} \otimes e^{i\pi (\hat{a}_{3}^{\dagger}\hat{a}_{3}-\alpha^{2})}  
\end{align}
in the hybrid basis, and since $e^{i\pi\hat{a}^{\dagger}\hat{a}}$ is given by $\hat{X}\otimes \hat{I}$ in the shifted Fock basis, we have
\begin{align}
    e^{-i\hat{H}'T} &= (CZ_{1,2}(-\pi\alpha^{2}) \cdot \textrm{TOF}_{1,2\rightarrow 3}) \otimes \hat{I}
\end{align}
in the full shifted Fock basis, where $\textrm{TOF}_{1,2\rightarrow 3}$ is the desired Toffoli gate. Thus, projecting $e^{-i\hat{H}' T}$ to the ground state manifold, we find $\hat{\rho}_{g}(T) = CCX' \hat{\rho}_{I,g}(T) CCX'^{\dagger}$ where $CCX' \equiv CZ_{1,2}(-\pi\alpha^{2}) \cdot \textrm{TOF}_{1,2\rightarrow 3}$. Therefore, we can understand $\hat{\rho}_{g}(T)$ as a state that results from applying a unitary operation $CCX'$ to the input state $\hat{\rho}_{g}(0)$ which is then corrupted by an error channel
\begin{align}
    \mathcal{N}_{CCX'}(\hat{\rho}) &\simeq \hat{\rho} + \int_{0}^{T}dt \Big{(} \kappa_{1}\alpha^{2}\Big{[}\mathcal{D}[\hat{Z}_{1}]
    \nonumber\\
    &\qquad\quad + \mathcal{D}[\hat{Z}_{2}] + \mathcal{D}\Big{[} CZ_{1,2}\Big{(} \frac{\pi}{T}(t+T) \Big{)}\hat{Z}_{3} \Big{]}
    \nonumber\\
    & + \frac{\pi^{2}}{128\kappa_{2}\alpha^{2}T^{2}}\Big{[} \mathcal{D}\Big{[}(\hat{Z}_{1}-\hat{Z}_{1}\hat{Z}_{2})CZ_{1,2}\Big{(} \frac{2\pi}{T}t \Big{)}\Big{]} 
    \nonumber\\
    &\,\,\,\, + \mathcal{D}\Big{[}(\hat{Z}_{2}-\hat{Z}_{1}\hat{Z}_{2})CZ_{1,2}\Big{(} \frac{2\pi}{T}t \Big{)}\Big{]}  \Big{]}\Big{)}\hat{\rho}. 
\end{align}
Here, we used the fact that $\hat{Z}_{3}$ is transformed via $\textrm{TOF}_{1,2\rightarrow 3}$ into $CZ_{1,2}\hat{Z}_{3}$. Evaluating the integral explicitly and ignoring off-diagonal Pauli errors, we find the following $Z$ error rates (per gate) of the $CCX'$ gate: 
\begin{align}
    &\bar{p}_{Z_{1}} = \bar{p}_{Z_{2}}  = \kappa_{1}\alpha^{2}T + \frac{\pi^{2}}{128\kappa_{2}\alpha^{2}T}, 
    \nonumber\\
    &\bar{p}_{Z_{3}} = \frac{5}{8}\kappa_{1}\alpha^{2}T, \quad \bar{p}_{Z_{1}Z_{2}} = \frac{\pi^{2}}{128\kappa_{2}\alpha^{2}T}, 
    \nonumber\\
    &\bar{p}_{Z_{1}Z_{3}} = \bar{p}_{Z_{2}Z_{3}} = \bar{p}_{Z_{1}Z_{2}Z_{3}} = \frac{1}{8}\kappa_{1}\alpha^{2}T. 
\end{align}
Hence, the optimal gate time that minimizes the total gate infidelity is given by 
\begin{align}
    \bar{T}_{CCX'}^{\star} &= \frac{ \pi }{8\alpha^{2}\sqrt{2\kappa_{1}\kappa_{2}}} , 
\end{align}
and at the optimal gate time, the $Z$ error rates (per gate) of the $CCX'$ gate are given by
\begin{align}
    &\bar{p}_{Z_{1}}^{\star} = \bar{p}_{Z_{2}}^{\star} = 3.2\bar{p}_{Z_{3}}^{\star} = 2\bar{p}_{Z_{1}Z_{2}}^{\star} 
    \nonumber\\
    &= 16\bar{p}_{Z_{1}Z_{3}} =  16\bar{p}_{Z_{2}Z_{3}} = 16\bar{p}_{Z_{1}Z_{2}Z_{3}}  
    \nonumber\\
    &= \frac{\pi}{4} \sqrt{\frac{\kappa_{1}}{2\kappa_{2}}} = 0.555\sqrt{\frac{\kappa_{1}}{\kappa_{2}}}  .  \label{eq:Z error rates CCX'}
\end{align}
Numerically, we find that the optimal $Z$ error rates (per gate) are given by (see \cref{tab:Gateerrorrates}) 
\begin{align}
    &p_{Z_{1}}^{\star} = p_{Z_{2}}^{\star} = 3.05p_{Z_{3}}^{\star} = 1.81p_{Z_{1}Z_{2}}^{\star}
    \nonumber\\
    &= 14.9p_{Z_{1}Z_{3}} =  14.9p_{Z_{2}Z_{3}} = 14.9p_{Z_{1}Z_{2}Z_{3}}  = 0.58\sqrt{\frac{\kappa_{1}}{\kappa_{2}}} , 
\end{align}
which agree well with the perturbative prediction in \cref{eq:Z error rates CCX'}. 

Similarly as in the case of the CNOT gate, we remark that the implemented gate $CCX'$ differs from the desired Toffoli gate $\textrm{TOF}_{1,2\rightarrow 3}$ by a CZ rotation $CZ_{1,2}(-\pi\alpha^{2})$. Thus, unless the average excitation number $\alpha^{2}$ is given by an even integer, one should apply $CZ_{1,2}(\pi\alpha^{2})$ to compensate for the extra phase shift. Lastly, note that dephasing can induce direct heating in each mode with a heating rate (per time) $\kappa_{\phi}\alpha^{2}$ (see \cref{eq:dephasing induces heating}). The excited states due to the heating decay back to the code space via the engineered dissipation. The engineered jump operators in the control modes are static and thus the excitations in the control modes decay back to the code space in a parity-preserving way, i.e., $4\kappa_{2}\alpha^{2}\mathcal{D}[\hat{I}_{1,2,3}\otimes \hat{b}_{1}]$ and $4\kappa_{2}\alpha^{2}\mathcal{D}[\hat{I}_{1,2,3}\otimes \hat{b}_{2}]$. On the other hand, the engineered jump operator on the target mode is time-dependent and thus the the relaxation of the excitation in the target mode is accompanied by a CZ rotation in the control qubits, i.e., $4\kappa_{2}\alpha^{2}\mathcal{D}[CZ_{1,2}(\frac{2\pi}{T}t)\otimes \hat{b}_{1}]$. Consequently, such a heating-relaxation process in the target mode generates a new noise process 
\begin{align}
    \kappa_{\phi}\alpha^{2}\mathcal{D}\Big{[} CZ_{1,2}\Big{(} \frac{2\pi}{T}t \Big{)} \Big{]} \hat{\rho}_{I,g}(t)
\end{align}
in addition to the noise processes described in the right hand side of \cref{eq:TOF equation shifted Fock basis after AE in the interaction picture} and adds $\kappa_{\phi}\alpha^{2}T/8$ to $p_{Z_{1}}$, $p_{Z_{2}}$, and $p_{Z_{1}Z_{2}}$, i.e., 
\begin{align}
    &\bar{p}_{Z_{1}} = \bar{p}_{Z_{2}}  = \kappa_{1}\alpha^{2}T + \frac{1}{8}\kappa_{\phi}\alpha^{2}T + \frac{\pi^{2}}{128\kappa_{2}\alpha^{2}T}, 
    \nonumber\\
    &\bar{p}_{Z_{3}} = \frac{5}{8}\kappa_{1}\alpha^{2}T, \quad \bar{p}_{Z_{1}Z_{2}} =  \frac{1}{8}\kappa_{\phi}\alpha^{2}T + \frac{\pi^{2}}{128\kappa_{2}\alpha^{2}T}, 
    \nonumber\\
    &\bar{p}_{Z_{1}Z_{3}} = \bar{p}_{Z_{2}Z_{3}} = \bar{p}_{Z_{1}Z_{2}Z_{3}} = \frac{1}{8}\kappa_{1}\alpha^{2}T. 
\end{align}
Hence, even in the lossless case (i.e., $\kappa_{1}=0$), the Toffoli gate has non-zero $Z$ error rates which scale as $\bar{p}^{*}_{Z_{1}} = \bar{p}^{*}_{Z_{2}} = \bar{p}^{*}_{Z_{1}Z_{2}} \propto \sqrt{ \kappa_{\phi}/\kappa_{2} }$ at the optimal gate time. Lastly, in the presence of non-zero thermal population $n_{\textrm{th}}$, we simply need to replace $\kappa_{1}$ by $\kappa_{1}(1+2n_{\textrm{th}})$.

\section{Simulations of gate error rates}
\label{app:Gate Error Simulations}

\subsection{CNOT}

\begin{figure}
    \centering
    \includegraphics[width=0.48\textwidth]{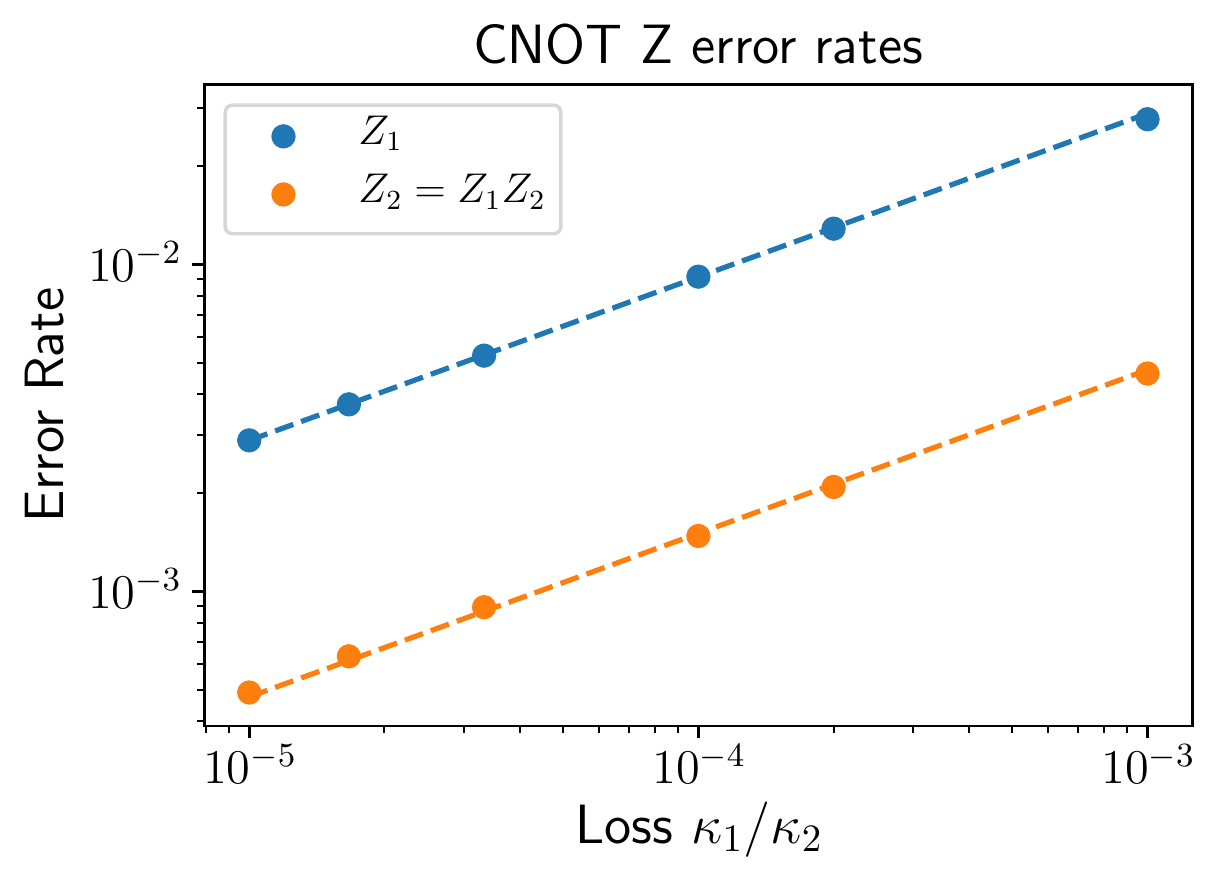}
	\caption{Log-log plot of Pauli $Z$-type error rates for the CNOT gate at optimal gate time with mean phonon number $n=8$ in the presence of pure loss at rate $\kappa_1$. The fits are performed over the range $\kappa_1/\kappa_2$ from $10^{-4}$ to $10^{-5}$. The error rates $Z_2$ and $Z_1 Z_2$ differ by no more than $10^{-5}$.}
    \label{fig:CNOTZVsLoss}
\end{figure}

\begin{figure}
    \centering
    \includegraphics[width=0.48\textwidth]{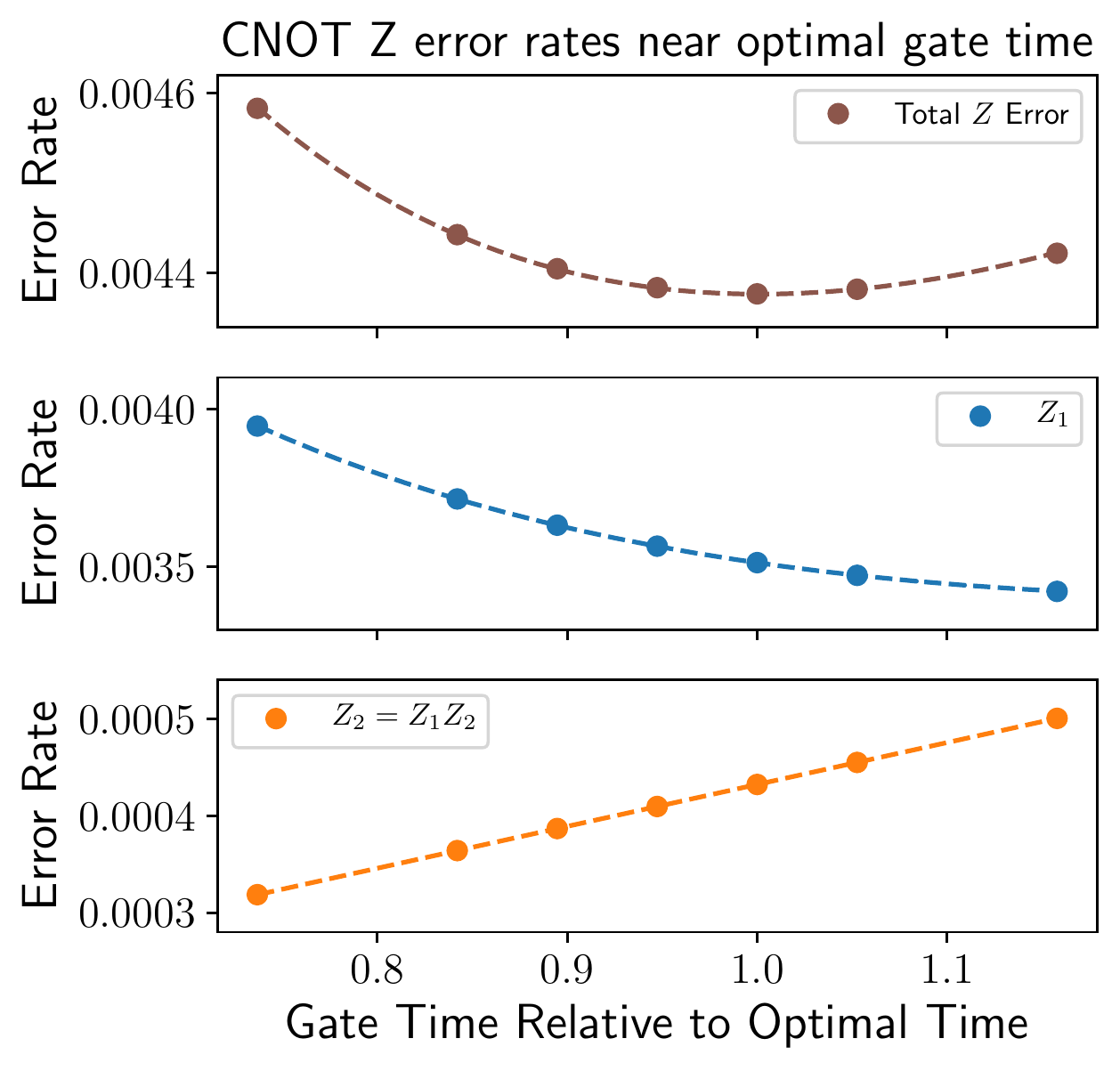}
	\caption{Plot of Pauli $Z$-type error rates for the CNOT gate with mean phonon number $n=10$ at various values of the gate time. The noise model is at rate $\kappa_1 = 10^{-5} \kappa_2$, dephasing at a rate $\kappa_\phi = \kappa_1$, and gain with $n_{th} = 1/100$. The gate time is plotted relative to the optimal gate time for these parameters. The optimal gate time minimizes the total error. The dotted curves are a linear fit for the $Z_2$ error rate and a sum of a linear term and a $1/T$ term for the $Z_1$ and total error rates, representing the contributions of loss and non-adiabatic errors.}
    \label{fig:CNOTZVsT}
\end{figure}

\begin{figure}
    \centering
    \includegraphics[width=0.48\textwidth]{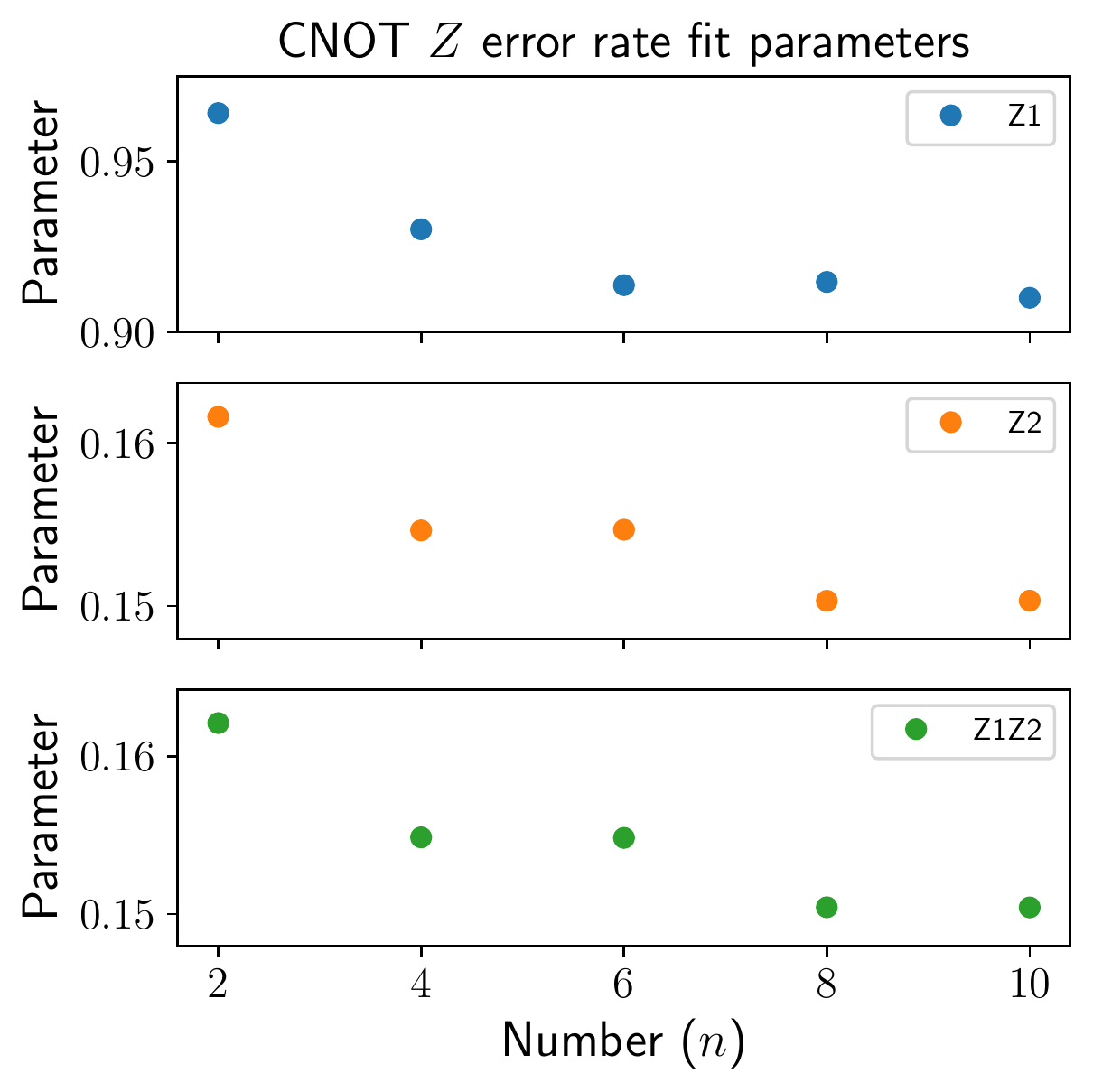}
	\caption{Plot of the fit parameters of the square root fit as shown in figure \ref{fig:CNOTZVsLoss} for the Pauli $Z$-type error rates of the CNOT gate for different values of mean phonon number $n$. The noise model in this figure is pure phonon loss.}
    \label{fig:CNOTZVsn}
\end{figure}

We simulated the CNOT gate as described in \cref{subsection:CNOT appendix} using the shifted Fock basis approach on AWS EC2 instances. Our code is written in Python using the QuTiP package. The results presented here took approximately 150 hours to run on an AWS EC2 C5.18xlarge instance with 72 virtual CPUs. To compute the Pauli error rates for the CNOT gate, we use two types of simulation. One set of simulations is aimed at the Z-type Pauli error rates and also determined the optimal gate time. These simulations require only a small dimension in the shifted Fock basis. The second type of simulation uses a much larger Hilbert space dimension to perform full tomography of the CNOT gate at the optimal gate time for relatively small values of the cat-code size $\alpha$.

We consider four noise models: first pure phonon loss at a number of different rates. We are most interested in the range of loss ($\kappa_1/\kappa_2$) from $10^{-4}$ to $10^{-5}$. Next, we consider phonon loss at rate $\kappa_1$, phonon gain at a rate such that the thermal occupation is given by $n_{th} = 1/100$, and dephasing noise at three different rates $\kappa_\phi = 1$, $2.5$, and $10$ times $\kappa_1$. This value of the thermal occupation number is larger than what we expect in acoustic cavities. We chose $n_{th} = 1/100$  so that we could resolve the contribution of phonon gain on the gate error rates. With $n_{th} = 1/100$ the gate error rates are enhanced by a factor of about $1.01$ relative to the error rates with no phonon gain. Dephasing noise is more significant; it increases the dominant error rate, $Z$ error on the control qubit and decreases the optimal gate time. \ji{The CNOT error rates including different values of dephasing are shown in \cref{tab:FullCNOTerrorrates}.} \ji{While idling, \jp{the bosonic dephasing term $\kappa_{\phi}\mathcal{D}[\hat{a}^{\dagger}\hat{a}]$ in the master equation} does not induce any additional $Z$ errors since dephasing preserves the excitation number parity. Thus, one might be tempted to conclude that dephasing only affects non-$Z$-type error rates of the CNOT gate and leaves the $Z$ error rates unchanged. However, surprisingly, we numerically find that this is not the case. In particular, as shown in \cref{tab:FullCNOTerrorrates}, we observe that the optimal gate time decreases noticeably and the total optimal $Z$ error rate (per gate) of the CNOT gate increases as the dephasing rate (per time) $\kappa_{\phi}$ increases. In \cref{appendix:Perturbative analysis of cat qubit gates}, we show that the enhanced $Z$ error rates of the CNOT gate due to dephasing are attributed to the fact that the target stabilization operator $\hat{L}_{2}(t)$ is not static and instead rotates conditioned on the state of the control mode. More specifically, dephasing in each mode causes direct population transfer from the ground state manifold of a cat qubit to its first excited state manifold. While such a heating itself does not cause a phase-flip error since dephasing preserves the excitation number parity, the rotating target stabilization operator $\hat{L}_{2}$ does cause a $Z$ error on the control qubit \jp{when} it brings the excited states of the target mode back to the ground state manifold.}

\begin{table*}
    \begin{center}
        \begin{tabular}{c c c c}
            \toprule
            \multirow{2}{*}{CNOT} & $\kappa_\phi=\kappa_1$, & $\kappa_\phi=2.5 \kappa_1$, & $\kappa_\phi=10 \kappa_1$,\\
            & $n_{th} = 1/100$ & $n_{th} = 1/100$ & $n_{th} = 1/100$\\
            \midrule
            Optimal Gate Time & $0.27|\alpha|^{-2} (\kappa_1 \kappa_2)^{-\frac{1}{2}}$ & $0.24|\alpha|^{-2} (\kappa_1 \kappa_2)^{-\frac{1}{2}}$ & $0.16 |\alpha|^{-2} (\kappa_1 \kappa_2)^{-\frac{1}{2}}$\\
            $Z_1$ & $1.10 \sqrt{\kappa_1/\kappa_2}$ & $1.33 \sqrt{\kappa_1/\kappa_2}$ & $2.14 \sqrt{\kappa_1/\kappa_2}$ \\
            $Z_2 \approx Z_1 Z_2$ & $0.14 \sqrt{\kappa_1/\kappa_2}$ & $0.12 \sqrt{\kappa_1/\kappa_2}$ & $0.079 \sqrt{\kappa_1/\kappa_2} $\\
            $X_1 \approx X_2 \approx X_1 X_2$ & \multirow{2}{*}{$1.07 \exp(-2 |\alpha|^2) \sqrt{\kappa_1/\kappa_2}$} & \multirow{2}{*}{$1.28 \exp(-2 |\alpha|^2) \sqrt{\kappa_1/\kappa_2}$} & \multirow{2}{*}{$2.01 \exp(-2|\alpha|^2) \sqrt{\kappa_1/\kappa_2}$} \\
            $\approx Y_1 \approx Y_1 X_2 \approx Z_1 X_2$ \\[3pt]
            $Y_2 \approx Y_1 Y_2 \approx X_1 Y_2$ & \multirow{2}{*}{$0.29 \exp(-2|\alpha|^2) \left(\kappa_1/\kappa_2 \right)$} & \multirow{2}{*}{$0.30 \exp(-2|\alpha|^2) \left(\kappa_1/\kappa_2 \right)$} & \multirow{2}{*}{$0.28 \exp(-2|\alpha|^2) \left(\kappa_1/\kappa_2 \right)$}\\
            $\approx X_1 Z_2 \approx Y_1 Z_2 \approx Z_1 Y_2$ \\
            \bottomrule
        \end{tabular}
    \end{center}
    \caption{\ji{Table showing the CNOT optimal gate time and error rates for nonzero thermal gain ($n_{\mathrm{th}}$) and dephasing ($\kappa_\phi$) along with loss ($\kappa_1$). In each case phonon gain is with $n_{th} = 1/100$, while the rate of dephasing noise relative to loss varies across the three columns.}}
    \label{tab:FullCNOTerrorrates}
\end{table*}

\begin{figure}
    \centering
    \includegraphics[width=0.48\textwidth]{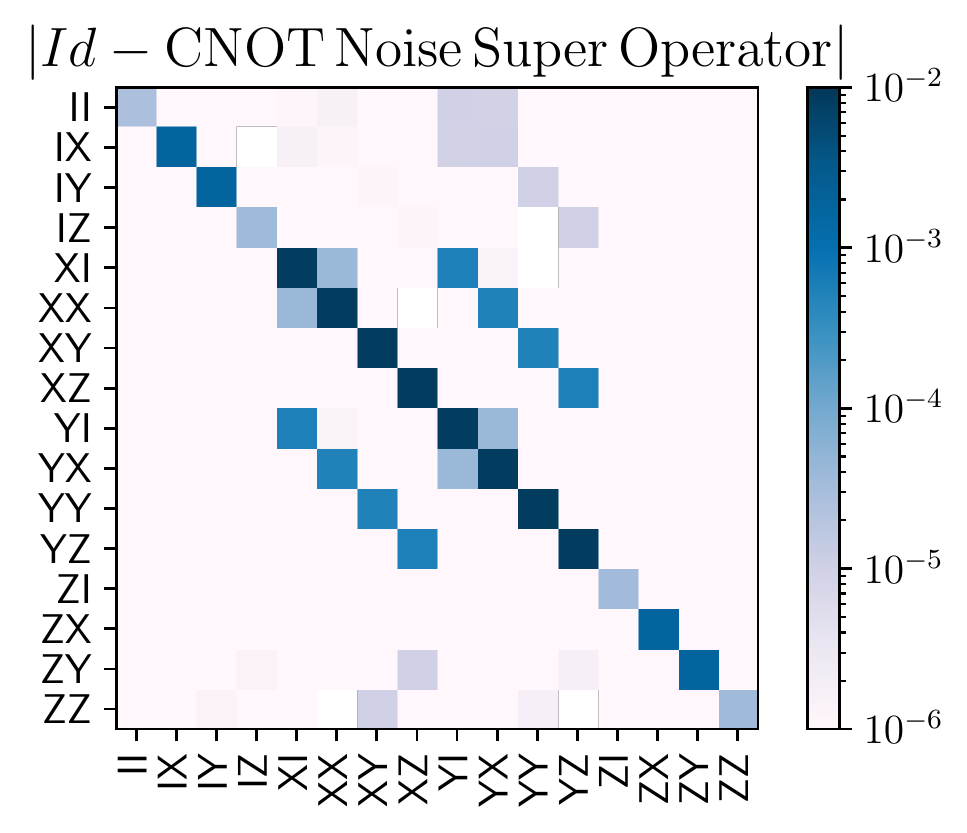}
	\caption{Plot of identity minus the super operator for the CNOT noise channel in the Pauli basis. The CNOT parameters are $n=4$, $\kappa_1/\kappa_2 = 10^{-5}$, $n_{th} = 1/100$, and $\kappa_\phi = \kappa_1$. The diagonal components of the matrix are the Pauli infidelities, in other words, one minus the probabilities that the CNOT noise channel maps a given Pauli operator back to itself. The off-diagonal components represent the coherent part of the noise channel. These terms are orders of magnitude smaller than the dominant noise terms. The dominant $Z_1$ error rate manifests itself as the relatively larger diagonal terms that are sensitive to a $Z_1$ error, i.e. Pauli operators with $X$ or $Y$ on the first qubit.}
    \label{fig:CNOTTom}
\end{figure}

\begin{figure}
    \centering
    \includegraphics[width=0.48\textwidth]{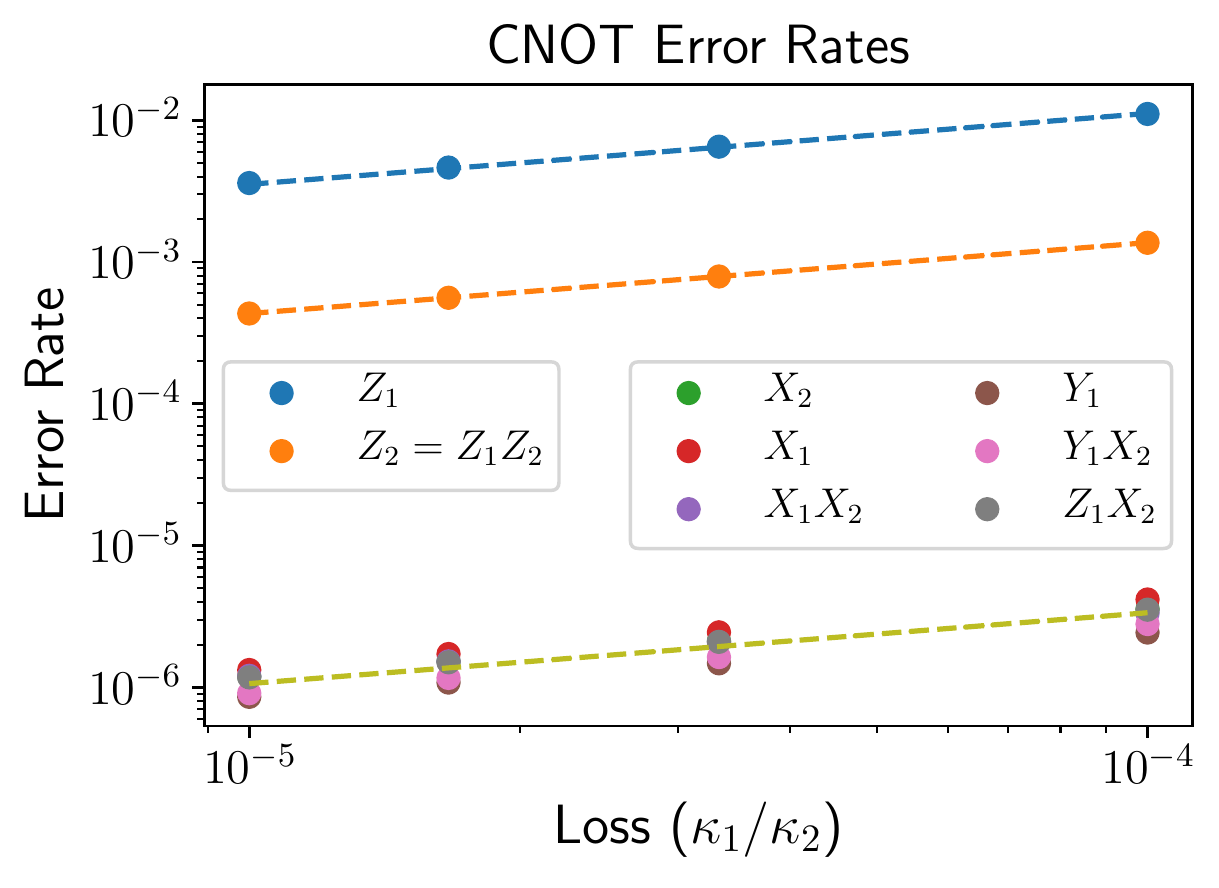}
	\caption{Log-log plot of the Pauli error rates for the CNOT gate with parameters, $n=4$, $\kappa_1 = \kappa_\phi$ and $n_{th} = 1/100$. Each of these error rates scale like $\sqrt{\kappa_1/\kappa_2}$.}
    \label{fig:CNOTTomError}
\end{figure}

\begin{figure}
    \centering
    \includegraphics[width=0.48\textwidth]{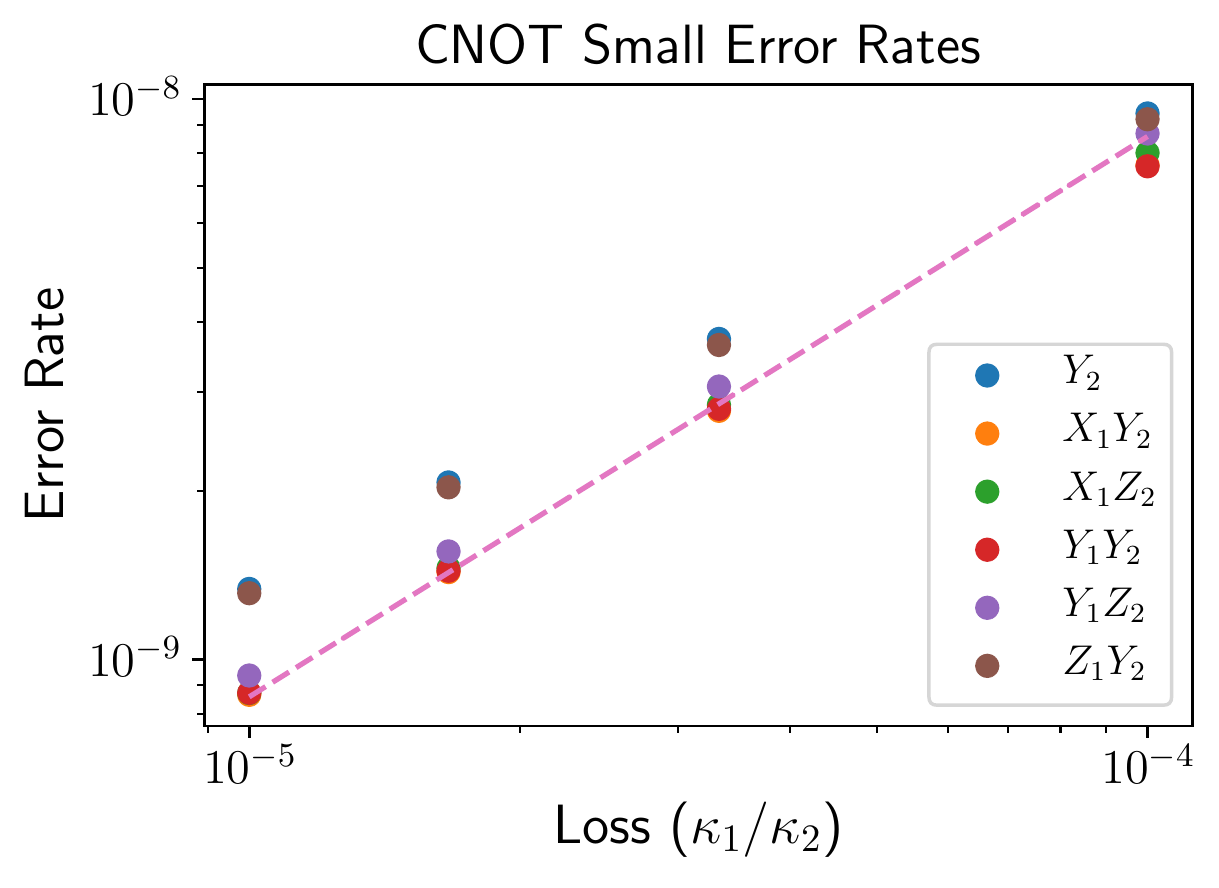}
	\caption{Log-log plot of the smallest error rates for the CNOT gate with parameters, $n=4$, $\kappa_1 = \kappa_\phi$ and $n_{th} = 1/100$. These error rates are proportional to $\kappa_1/\kappa_2$ rather than the square root scaling of the other error rates in \cref{fig:CNOTTomError}.}
    \label{fig:CNOTTomSmall}
\end{figure}

\begin{figure}
    \centering
    \includegraphics[width=0.48\textwidth]{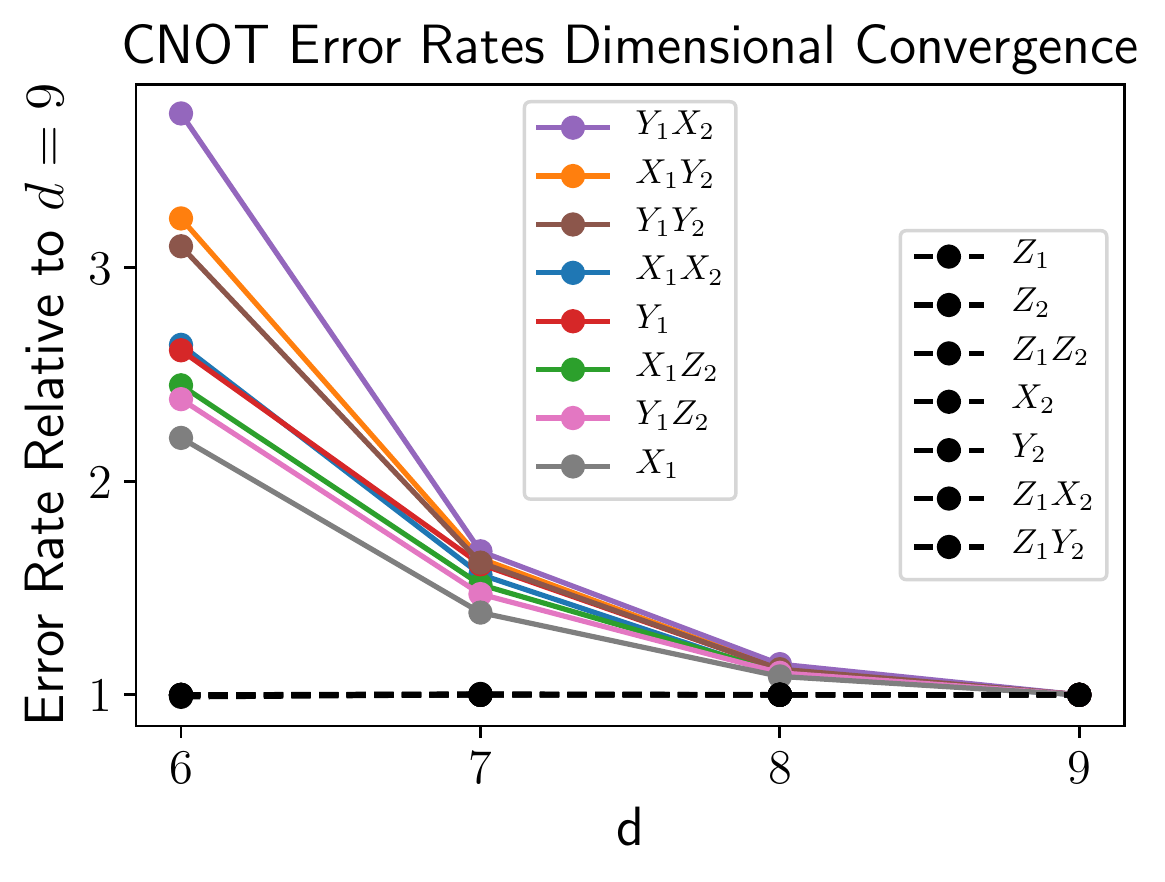}
	\caption{Plot of the Pauli error rates at a fixed value of the noise parameters and different values of the shifted Fock basis dimension $d$. Each error rate is scaled by its value at largest value of dimension $d=9$ to show the convergence as $d$ increases. The parameters are set to $n=3$, $\kappa_1 = 10^{-5}$, $\kappa_\phi = 0$, and $n_{th} = 0$. One set of Pauli error rates converges rapidly as $d$ increases. This includes all Pauli errors where $Z$ or $Id$ act on the control qubit. Another set of Pauli errors with $X$ or $Y$ acting on the control qubit require much higher Hilbert space dimension to capture accurately. This implies that these error rates include significant contributions from highly excited states.}
    \label{fig:CNOTTomDimConv}
\end{figure}

The $Z$ error rates for the CNOT gate are well-captured by the shifted Fock basis with small dimension, indicating that the $Z$ error rates are dominated by dephasing resulting from the excitation of the cat qubit to the lowest energy excited states. The results plotted in \cref{fig:CNOTZVsLoss,fig:CNOTZVsT,fig:CNOTZVsn} were obtained with $d=7$, or a total Hilbert space dimension of 14. The simulations converge rapidly as the dimension increases. The relative difference in the error rates shown in \cref{fig:CNOTZVsLoss} between the simulations with $d=6$ and $d=7$ is about $10^{-6}$, and this gives a bound on how closely these simulations reflect the true error rates in an infinite-dimensional cavity. We call the control cavity 1 and the target 2. As described in \cref{subsection:CNOT appendix}, the non-adiabatic error contribution to the $Z_1$ error rate of the CNOT gate scales with $1/T$, where $T$ is the gate time, while the error due to single-phonon loss scales with $T$. As a result of the tradeoff between non-adiabatic error, the optimal gate time scales like $1/\sqrt{T}$. As shown in \cref{fig:CNOTZVsT}, around the optimal gate time the $Z_1$ error rate is decreasing with $T$, whereas the $Z_2$ and $Z_1 Z_2$ error rates are increasing. This is because the non-adiabatic errors affect only the control cavity, i.e. $Z_1$. We find an optimal gate time that differs only slightly from the prediction in \cref{subsection:CNOT appendix}. In \cref{fig:CNOTZVsLoss} we find the expected square root scaling of the $Z$ error rates with loss rate over a wide range of loss rates. We do observe 
that the points corresponding to larger values of loss near $\frac{\kappa_1}{\kappa_2} = 10^{-3}$ tend to lie below the square root best-fit curve. For this reason, we perform our fits over the range of loss from $10^{-4}$ to $10^{-5}$, which is our range of interest for our error correction simulations. This leads to slightly larger error rate fit parameters than if we fit over the full range of loss. \cref{fig:CNOTZVsn} shows the dependence of the $Z$ error rate coefficients on the mean phonon number of the cat $n = \alpha^2$. These coefficients come from fits of each error rate to $c \sqrt{\kappa_1/\kappa_2}$ for each value of $n$. There is variation over the range $n=2, \dots 10$, but for $n=8$ and $n=10$ the variation is quite small. The values quoted in \cref{tab:Gateerrorrates} represent this large-$n$ value.

\begin{figure}
    \centering
    \includegraphics[width=0.48\textwidth]{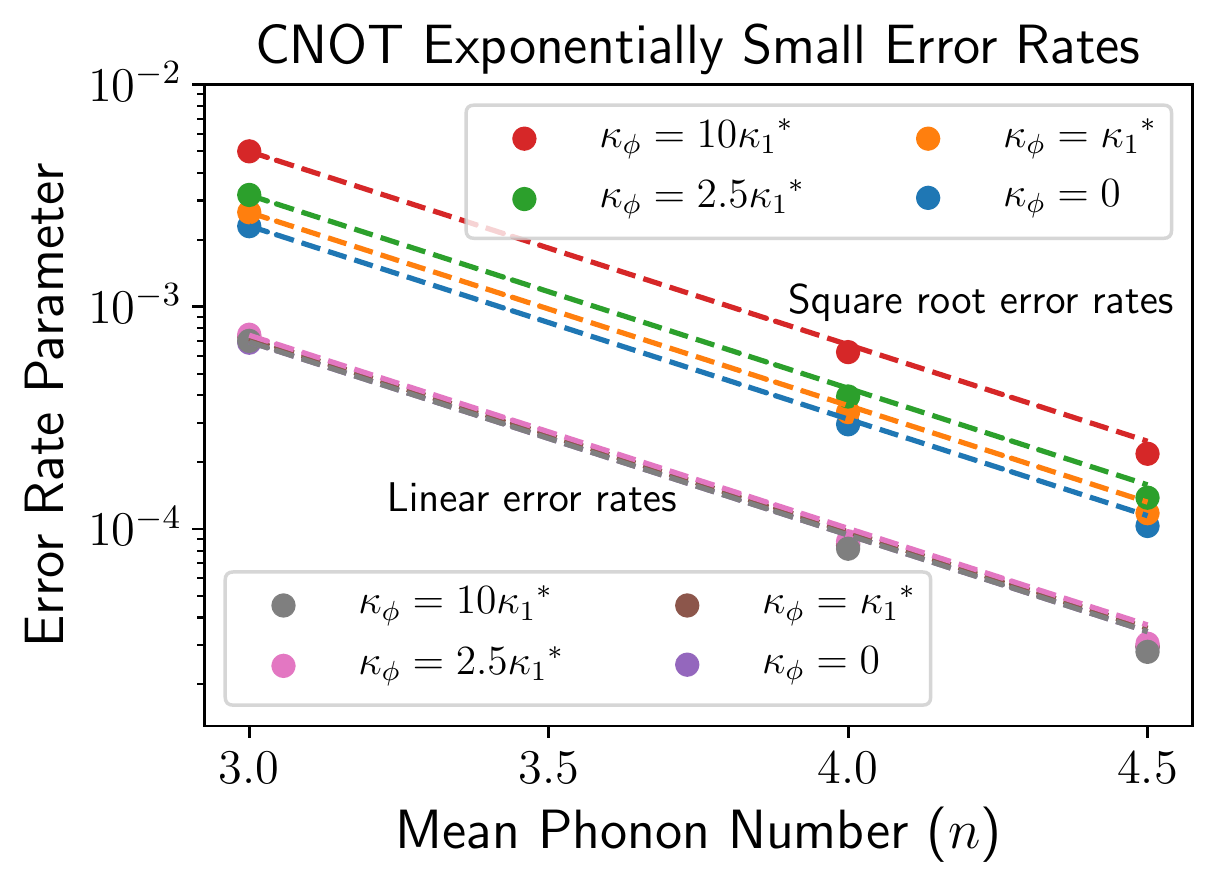}
	\caption{Log-linear plot showing exponential decay of non-$Z$ error rates as mean phonon number $n=|\alpha|^2$ increases. Four values of dephasing are plotted, $\kappa_\phi = 0$, $1$, $2.5$, and $10$ times $\kappa_1$. The asterisk that appears for the non-zero values of dephasing represents that these points include phonon gain at a rate $n_{th} = 1/100$. No gain is present in the $\kappa_\phi = 0$ points. For each set of noise parameters, the upper set of points represents the Pauli error rates from  \cref{fig:CNOTTomError} that scale with $\sqrt{\kappa_1/\kappa_2}$ and exponentially with $n$. The lower set of points are the Pauli error rates from  \cref{fig:CNOTTomError} that scale linearly with $\kappa_1/\kappa_2$ and exponentially with $n$. The parameters of the exponential fits can be found in \cref{tab:Gateerrorrates}.}
    \label{fig:CNOTTomExp}
\end{figure}

Once the optimal gate time is found using the $Z$ error rate simulation, we performed tomography for the CNOT gate at several values of loss, dephasing, and $n$ to compute the full noise channel. The noise channel for $n=4$, $n_{th} = 1/100$, and $\kappa_\phi = \kappa_1 = 10^{-5}$ is illustrated in \cref{fig:CNOTTom}. The noise channel is largely incoherent with small off-diagonal elements. The diamond distance from identity is equal to about $2.5$ times the average infidelity of the channel across all values of $\alpha$, loss, and dephasing that we simulated. The Hilbert-Schmidt norm of the off-diagonal elements of the super operator in the Pauli basis is $10^{-2}$--$10^{-3}$ times the norm of the diagonal elements. Neglecting the off-diagonal components, we are able to read off the full set of 15 two-qubit Pauli error rates. For the values of $n=\alpha^2$ that are not even integers, we must cancel the extra $Z_1$ rotation by angle $\pi \alpha^2$ that comes with our implementation of the CNOT gate. In practice this would entail additional error, but we do not include the effect of the noisy $Z$ rotation because we are interested in the error intrinsic to the CNOT gate and we expect to operate with even $n$ as much as possible. Besides the dominant $Z$ error rates, each of the other Pauli error rates is exponentially small in $\alpha$. However, we observe that these exponentially small error rates are divided into two classes---six of them scale like the square root of $\kappa_1/\kappa_2$ just like the $Z$ error rates and the remaining six error rates scale linearly with $\kappa_1/\kappa_2$. The error rates with square root scaling are plotted for one choice of parameters in \cref{fig:CNOTTomError}. The error rates scaling linearly are much smaller and are shown in \cref{fig:CNOTTomSmall}. A large dimension is required to accurately recover some of the Pauli error rates. As shown in \cref{fig:CNOTTomDimConv} when $n=3$ the Pauli errors that involve $X$ or $Y$ acting on the control qubit require a much larger value of $d$ than the other error rates. The \ji{relative} difference between the error rates with $d=8$ and $d=9$ in this case was as much as $15\%$. We used a dimension of $d=9$, $10$, and $11$ for $n=3$, $4$, and $4.5$, respectively, and the total Hilbert space dimension is $2d$ in the shifted Fock basis. We did not go to larger values of $n$ because the required Hilbert space dimension required an unreasonably long time to simulate. Across all values of loss, dephasing, and mean phonon number, the error rates for the largest dimension $d$ that we used and the error rates at $d-1$ differed by several percent. This provides a sense of the difference we expect between the largest dimension we used and the $d\rightarrow \infty$ limit. Because of this uncertainty of perhaps several percent in certain of the Pauli error rates and for simplicity, we have chosen to report a single fit for each of the two groups of exponentially small error rates. These include both the small error rates that scale with the square root of loss in \cref{fig:CNOTTomError} and those that scale linearly in \cref{fig:CNOTTomSmall}. This is why only a single best fit curve appears over the clusters of small error rates in those plots. We have taken the average within each of the two classes of exponentially small error rates and fit the square root or linear curve to those averages. Correspondingly, in our simulations of error correction we assume that $p_{X_1} = p_{X_2} = p_{X_1 X_2} = p_{Y_1} = p_{Y_1 X_2} = p_{Z_1 X_2}$ and $p_{Y_1} = p_{X_1 Y_2} = p_{X_1 Z_2} = p_{Y_1 Y_2} = p_{Y_1 Z_2} = p_{Z_1 Y_2}$, and the error probabilities are given by the average fits. Both classes of small error rates exhibit the expected exponential scaling with the mean phonon number $n$ of the cat code as shown in \cref{fig:CNOTTomExp}.

\subsection{Toffoli}

\begin{figure}
    \centering
    \includegraphics[width=0.48\textwidth]{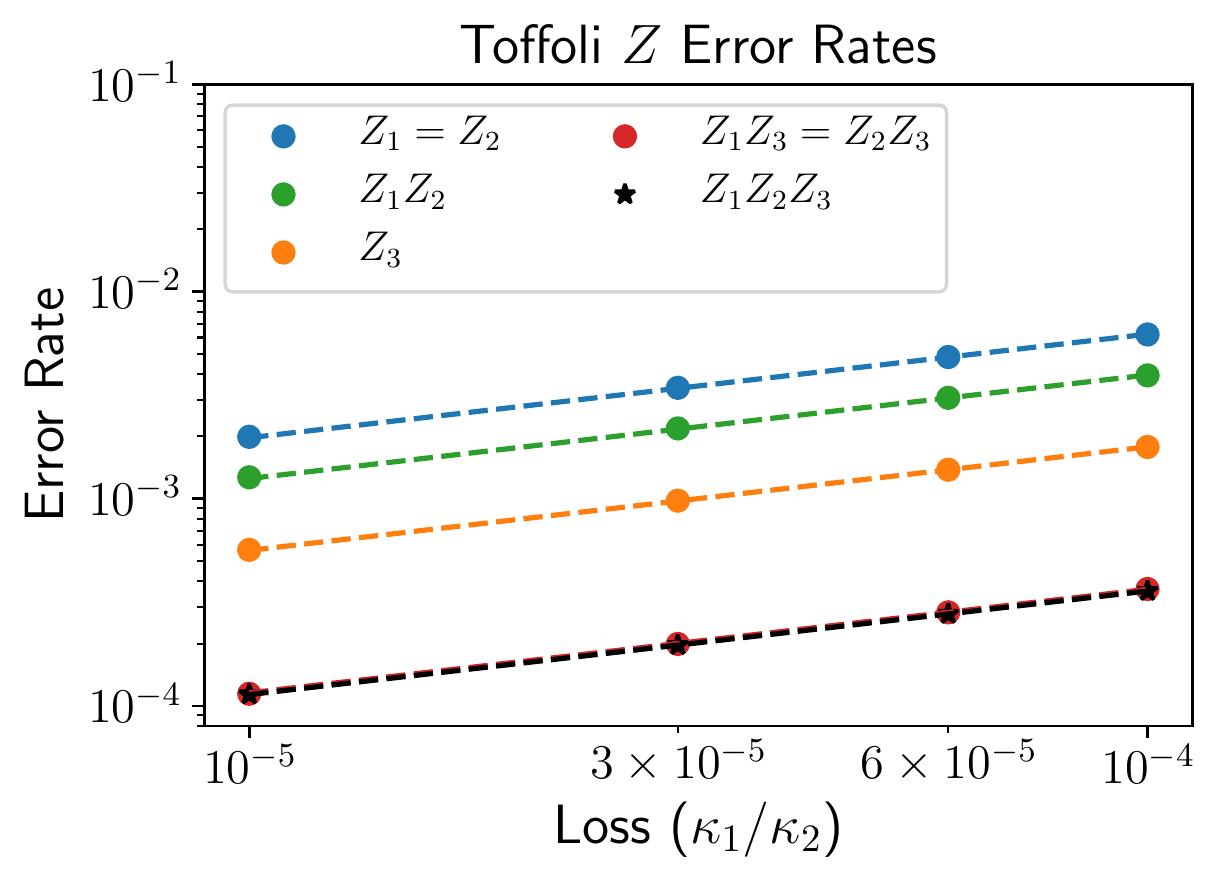}
	\caption{Log-log plot of the various $Z$-type error rates for the Toffoli gate at optimal gate time with parameters $n=8$, $\kappa_\phi = \kappa_1$, and $n_{th} = 1/100$. These error rates were obtained in a shifted Fock basis simulation using $d=4$ for a total Hilbert space dimension of 8 for each of the three cavities involved in the Toffoli gate. Qubits 1 and 2 are the controls and 3 is the target.}
    \label{fig:ToffVsLoss}
\end{figure}

\begin{figure}
    \centering
    \includegraphics[width=0.48\textwidth]{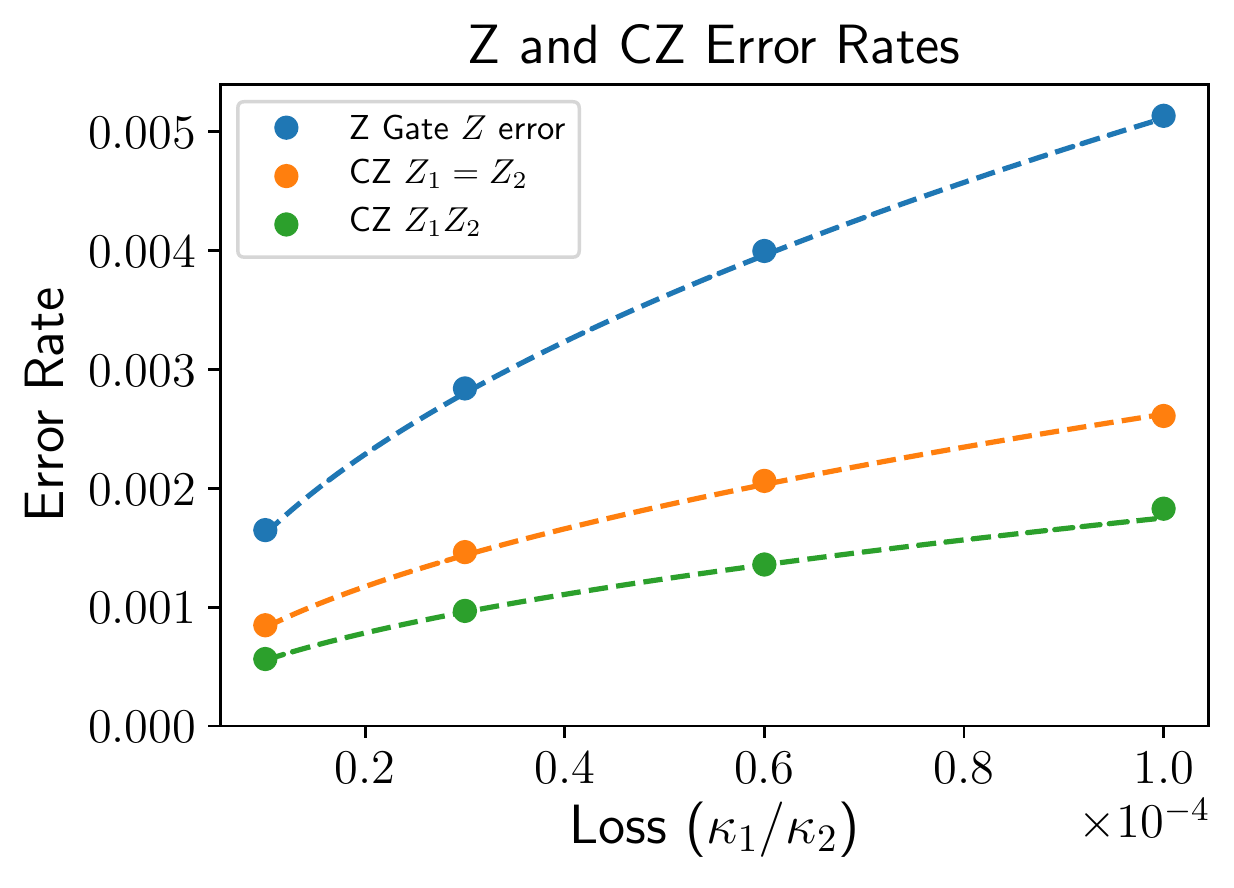}
	\caption{Plot of the $Z$ error rates for the $Z$ and CZ gates as a function of loss for $n=10$. The noise model for this plot is pure phonon loss with no gain. Gain will have a small effect of these error rates, while dephasing noise will have only a negligible effect. The dotted curves are best fits in the form $c*\sqrt{\kappa_1/\kappa_2}$.}
    \label{fig:ZandCZVsLoss}
\end{figure}

    We simulate the Toffoli gate using the shifted Fock basis as we did for the CNOT gate. In this case we solve the master equation for three cavities. This leads to a much larger total Hilbert space dimension, and for this reason we are unable to use the large values of $d$ necessary to resolve all $63$ Pauli error probabilities. Instead we focus on the dominant errors, which are the $Z$-type Pauli errors. These errors do not require a large value of $d$ to calculate with good precision. We used $d=4$ for each of the three cavities in these Toffoli simulations, which required a total of about $170$ hours running on an AWS EC2 c5.18xlarge instance with 72 virtual CPUs. We simulated the noise channel on a complete set of $X$ eigenstates and averaged over the initial states. We simulated a range of gate times\ji{, and the optimal gate time is the one} that minimizes the total error rate. For loss without gain or dephasing, we found that this gate time matched the optimal gate time for the CNOT gate. \ji{The $Z$ error rates in this case are summarized in \cref{tab:Gateerrorrates}.} 
    
    With dephasing noise added we found a small difference in optimal gate time. \ji{These error rates are found in \cref{Tab:ToffGateTimes}.} \ji{As in the case of the CNOT gate, dephasing noise increases the $Z$ error rates and shortens the optimal gate time. The dominant $Z$ error on the control qubits 1 and 2 increases from $0.58 \sqrt{\kappa_1/\kappa_2}$ to $0.91 \sqrt{\kappa_1/\kappa_2}$ as the dephasing rate increases from $0$ to $10 \kappa_1$. Dephasing noise primarily affects the $Z_1$, $Z_2$ and $Z_1 Z_2$ error rates. Many of the other Pauli $Z$ error rates decrease because of the reduction in optimal gate time. Also, with dephasing noise in addition to loss, the optimal gate time for the Toffoli gate differs from the optimal gate time for the CNOT gate. With large dephasing $\kappa_\phi = 10 \kappa_1$ the optimal gate time for the Toffoli gate is about $1.18$ times the optimal gate time for CNOT. For simplicity, we have chosen to always operate the Toffoli gate using a gate time equal to the CNOT optimal gate time. This has a small effect on the total fidelity of the Toffoli gate and on the relative size of the different Pauli $Z$ error probabilities.} We chose to use the optimal gate times for the CNOT gate throughout the paper. The Toffoli error rates at the true optimal gate time and at the CNOT optimal gate time are shown in \cref{Tab:ToffGateTimes}. The difference in the total fidelity of the Toffoli gate is small, however the relative size of individual Pauli $Z$ error rates does differ by several percent when $\kappa_\phi = 10 \kappa_1$.
    
    \cref{fig:ToffVsLoss} shows the seven $Z$-type error probabilities for the Toffoli gate at optimal gate time as a function of the loss rate with $n=8$, $\kappa_\phi = \kappa_1$ and $n_{th} = 1/100$. We see the expected square root scaling with $\kappa_1/\kappa_2$ for each of the error rates and perform best fits. We simulate Toffoli with $n=4$, $6$, $8$, and $10$ and for four sets of noise parameters: only phonon loss and then phonon loss, gain and dephasing at three different rates, $\kappa_\phi = 1$, $2.5$, and $10$ times $\kappa_1$. Similar to the CNOT example in \cref{fig:CNOTZVsn}, the parameters of the square root fits depend on $n$ but reach a plateau around $n=8$ or $10$. For our error correction simulations we are most interested in values of $n$ in this regime. To produce the numbers in \cref{Tab:ToffGateTimes} we have average the values for $n=8$ and $n=10$. The relative difference between these two is only order $10^{-2}$ or less.

\begin{table*}
    \centering
    \begin{tabular}{c c c c}
        \toprule
        \multirow{2}{*}{Toffoli at optimal gate times} & $\kappa_\phi = \kappa_1$ & $\kappa_\phi = 2.5 \kappa_1$ & $\kappa_\phi = 10 \kappa_1$ \\ & $n_{th} = 1/100$ & $n_{th} = 1/100$ & $n_{th} = 1/100$\\
        \midrule
        Gate Time & $0.28 |\alpha|^{-2} (\kappa_1 \kappa_2)^{-\frac{1}{2}}$ & $0.25 |\alpha|^{-2} (\kappa_1 \kappa_2)^{-\frac{1}{2}}$ & $0.18 |\alpha|^{-2} (\kappa_1 \kappa_2)^{-\frac{1}{2}}$\\
        $Z_1 = Z_2$& $0.62 \sqrt{\kappa_1/\kappa_2}$ & $0.68 \sqrt{\kappa_1/\kappa_2}$ & $0.90 \sqrt{\kappa_1/\kappa_2}$ \\
        $Z_3$& $0.18 \sqrt{\kappa_1/\kappa_2}$ & $0.16 \sqrt{\kappa_1/\kappa_2}$ & $0.12 \sqrt{\kappa_1/\kappa_2}$ \\
        $Z_1 Z_2$& $0.40 \sqrt{\kappa_1/\kappa_2}$ & $0.48 \sqrt{\kappa_1/\kappa_2}$ & $0.79 \sqrt{\kappa_1/\kappa_2}$ \\
        $Z_1 Z_3 = Z_2 Z_3$ & $0.036 \sqrt{\kappa_1/\kappa_2}$ & $0.033 \sqrt{\kappa_1/\kappa_2}$ & $0.024 \sqrt{\kappa_1/\kappa_2}$ \\
        $Z_1 Z_2 Z_3$& $0.035 \sqrt{\kappa_1/\kappa_2}$ & $0.032 \sqrt{\kappa_1/\kappa_2}$ & $0.024 \sqrt{\kappa_1/\kappa_2}$ \\
        \midrule
        Toffoli at CNOT optimal times &&\\
        \midrule
        Gate Time & $0.27 |\alpha|^{-2} (\kappa_1 \kappa_2)^{-\frac{1}{2}} $ & $0.24 |\alpha|^{-2} (\kappa_1 \kappa_2)^{-\frac{1}{2}}$ & $0.16 |\alpha|^{-2} (\kappa_1 \kappa_2)^{-\frac{1}{2}}$ \\
        $Z_1 = Z_2$& $0.62 \sqrt{\kappa_1/\kappa_2}$ & $0.68 \sqrt{\kappa_1/\kappa_2}$ & $0.91 \sqrt{\kappa_1/\kappa_2}$ \\
        $Z_3$& $0.17 \sqrt{\kappa_1/\kappa_2}$ & $0.15 \sqrt{\kappa_1/\kappa_2}$ & $0.098 \sqrt{\kappa_1/\kappa_2}$ \\
        $Z_1 Z_2$& $0.41 \sqrt{\kappa_1/\kappa_2}$ & $ 0.50 \sqrt{\kappa_1/\kappa_2}$ & $0.84 \sqrt{\kappa_1/\kappa_2}$ \\
        $Z_1 Z_3 = Z_2 Z_3$& $0.035 \sqrt{\kappa_1/\kappa_2}$ & $0.031 \sqrt{\kappa_1/\kappa_2}$ & $0.020 \sqrt{\kappa_1/\kappa_2}$ \\
        $Z_1 Z_2 Z_3$& $0.034 \sqrt{\kappa_1/\kappa_2}$ & $0.030 \sqrt{\kappa_1/\kappa_2}$ & $0.020 \sqrt{\kappa_1/\kappa_2}$\\
        \bottomrule
    \end{tabular}
    \caption{\ji{Table comparing Toffoli $Z$ Pauli error rates, on the one hand, at the optimal gate time and, on the other hand, using a gate time equal to the optimal gate time for the CNOT gate. The error rates from our numerical simulations were fit to $\sqrt{\kappa_1/\kappa_2}$ to produce the coefficients that appear in the table. Three different values of dephasing are included. The gate times for CNOT and Toffoli match in the case of no dephasing, and the difference between the two increases as the dephasing rate $\kappa_\phi$ increases. Qubits 1 and 2 are the controls, and qubit 3 is the target.}}
    \label{Tab:ToffGateTimes}
\end{table*}

\begin{table*}
    \centering
    \begin{tabular}{c c c}
    \toprule
         $Z$ Gate & Loss, no gain & Loss and gain $n_{th} = 1/100$\\
         \midrule
         Opt. Time & $0.61 (\alpha^3 \sqrt{ \kappa_1 \kappa_2})^{-1}$ & $0.61 (\alpha^3 \sqrt{ \kappa_1 \kappa_2})^{-1}$\\
         $Z$ & $1.63 \sqrt{\kappa_1/\kappa_2}/\alpha$ & $1.64 \sqrt{\kappa_1/\kappa_2}/\alpha$ \\
         \midrule
         CZ Gate & & \\
         \midrule
         Opt. Time & $0.56 (\alpha^3 \sqrt{ \kappa_1 \kappa_2})^{-1}$ & $0.56 (\alpha^3 \sqrt{ \kappa_1 \kappa_2})^{-1}$ \\
         $Z_1 = Z_2$ & $0.83 \sqrt{\kappa_1/\kappa_2}/\alpha$ & $0.84 \sqrt{\kappa_1/\kappa_2}/\alpha$\\
         $Z_1 Z_2$ & $0.56 \sqrt{\kappa_1/\kappa_2}/\alpha$ & $0.56 \sqrt{\kappa_1/\kappa_2}/\alpha$ \\
         \bottomrule
    \end{tabular}
    \caption{\ji{Table of $Z$ gate and CZ gate optimal times and $Z$ error rates from numerical simulations. Dephasing noise has a negligible effect on the $Z$ error rates.}}
    \label{tab:ZandCZRates}
\end{table*}

\begin{figure}
    \centering
    \includegraphics[width=0.48\textwidth]{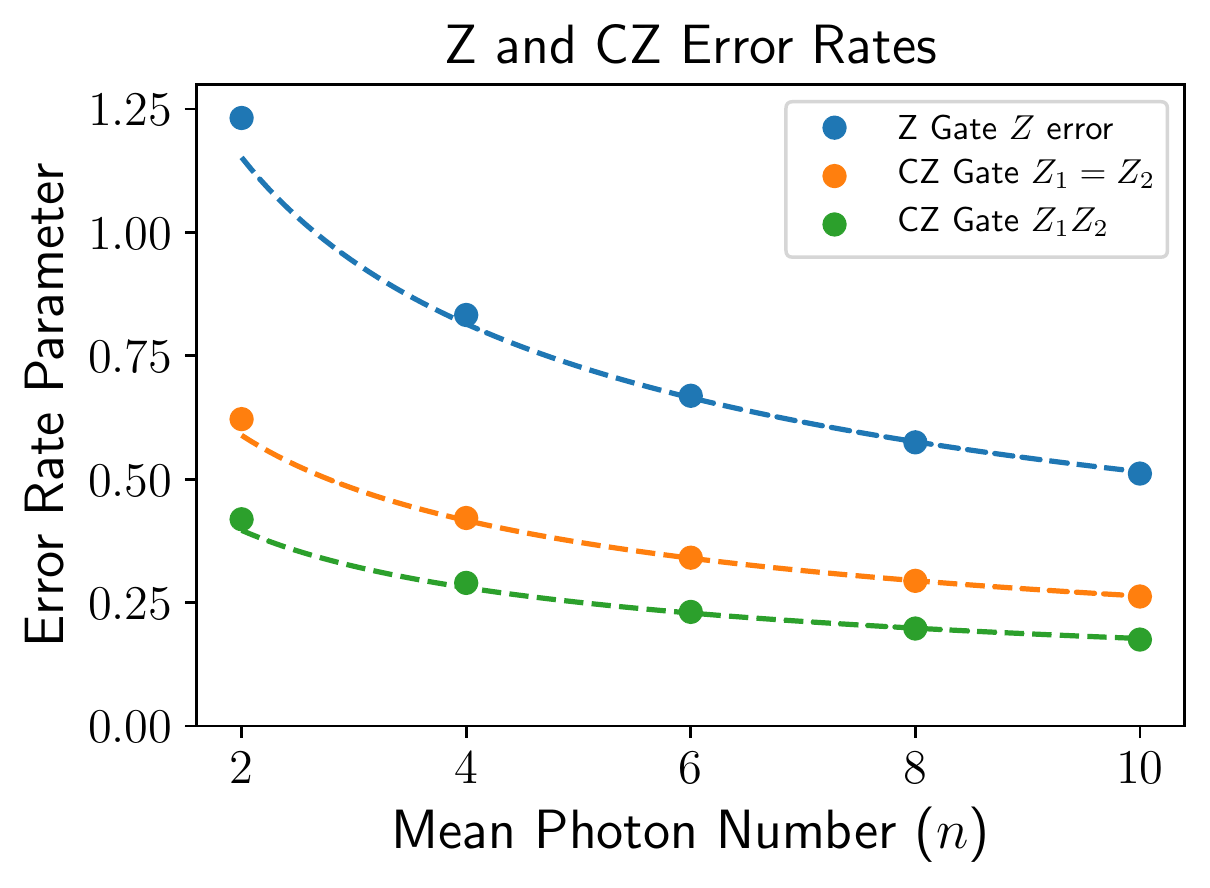}
	\caption{Plot of the $Z$ error rate parameters from best fits like the ones in \cref{fig:ZandCZVsLoss} as a function of mean phonon number $n$. The error rates are the product of these fit parameters and the loss rate $\sqrt{\kappa_1/\kappa_2}$. The dotted best fit curves in this plot are fits to $c/\alpha$ where $\alpha = \sqrt{n}$. For small values of $\alpha$ the scaling differs somewhat from the $1/\alpha$ scaling in the large $\alpha$ limit. For this reason the fits were performed over the range $n=6$ to $10$.}
    \label{fig:ZandCZVsn}
\end{figure}

\subsection{Z and CZ}
To implement our CNOT and Toffoli gates with values of $\alpha$ such that $n=\alpha^2$ is not an even integer, we need to apply an additional $Z$ or CZ rotations on the control cavity or cavities. These rotations can be implemented as described in \cref{sec:GatesMeas}. The dominant error rates are the $Z$ error rates, and at the optimal gate time the $Z$ error rates scale with $\sqrt{\kappa_1/\kappa_2}$. Unlike the case of CNOT or Toffoli, the error rates decrease with $\alpha$ for the $Z$ and CZ rotations. We simulate the $Z$ error rates for the $Z$ and CZ gates, in other words $Z$ and CZ rotations by angle $\pi$. Once again we use the shifted Fock basis as described in \cref{appendix:Shifted Fock Basis} to simulate the $Z$ error rates using a small Hilbert space dimension. \cref{fig:ZandCZVsLoss} shows the $Z$ error rates for both the $Z$ and CZ gates when $n=10$ and the noise model is phonon loss. We fit the error rates to $\sqrt{\kappa_1/\kappa_2}$ for each value of $n$ and for each noise model. Then \cref{fig:ZandCZVsn} shows the scaling of the coefficients of the $\sqrt{\kappa_1/\kappa_2}$ fits as a function of $n$ when the noise model is phonon loss. We fit these curves to $1/\alpha$. The results of the fits that give the $Z$ error rates as functions of $\alpha$ and $\kappa_1/\kappa_2$ are summarized in \cref{tab:ZandCZRates}. We also simulated the $Z$ and CZ gates subject to dephasing noise and confirmed that dephasing noise does not contribute significantly to the $Z$ error rates. Including phonon gain with $n_{th} = 1/100$ has a small effect as shown in \cref{tab:ZandCZRates}.

\section{Physical implementation of cat qubit gates}
\label{appendix:physical implementation of cat qubit gates}

Here, we discuss physical realization of the cat qubit gates. Note that engineering static two-phonon dissipations in a multiplexed setting has been extensively discussed in the previous section. Also implementation of the rotating dissipators for the CNOT and Toffoli gates are discussed in detail in Ref.~\cite{Guillaud2019}. We thus focus on engineering Hamiltonian interactions needed to implement the cat-qubit gates. In particular, we discuss realization of the linear drive in $\hat{H}_{Z}$, beam-splitter coupling in $\hat{H}_{CZ}$, selective frequency shift in $\hat{H}_{X}$, cubic optomechanical coupling in $\hat{H}_{\textrm{CNOT}}$, and the quartic interaction in $\hat{H}_{\textrm{TOF}}$ in the stated order.  

Recall the Hamiltonian of the system consisting of multiple phononic modes $\hat{a}_{k}$ coupled to a shared ATS mode $\hat{b}$:       
\begin{align}
    \hat H &= \sum_{k=1}^N \omega_k \hat{a}_k^\dagger \hat{a}_k +  \omega_b \hat{b}^\dagger \hat{b} -2E_{J} \epsilon_{p}(t) \sin \Big{(} \sum_{k=1}^N  \hat{\phi}_{k} + \hat{\phi}_{b} \Big{)}. 
\end{align}
Here, $\hat{\phi}_{k} \equiv \varphi_{k}(\hat{a}_{k} + \hat{a}_{k}^{\dagger})$ and $\hat{\phi}_{b} \equiv \varphi_{b}(\hat{b}+\hat{b}^{\dagger})$. Also, $\varphi_{k}$ and $\varphi_{b}$ quantify zero-point fluctuations of the modes $\hat{a}_{k}$ and $\hat{b}$. To simplify the discussion, we neglect small frequency shifts due to the \ji{flux} pump $\epsilon_{p}(t)$ for the moment and assume that the frequency of a mode is given by its bare frequency (in practice, however, the frequency shifts need to be taken into account; see below for the frequency shift due to pump). Then, in the rotating frame where every mode rotates with its own frequency, we have 
\begin{align}
    \hat{H}_{\textrm{rot}} &= -2E_{J} \epsilon_{p}(t) \sin \Big{(} \sum_{k=1}^N  \varphi_{k}\hat{a}_{k}e^{-i\omega_{k}t}  + \textrm{h.c.}
    \nonumber\\
    &\qquad\qquad\qquad\qquad\quad +  \varphi_{b}\hat{b}e^{-i\omega_{b}t} + \textrm{h.c.} \Big{)} . \label{eq:Sine potential rotating frame gate implementation}
\end{align}

Linear drive on a phononic mode, say $\hat{a}_{k}$, can be readily realized by using a \ji{flux} pump $\epsilon_{p}(t)=\epsilon_{p}\cos(\omega_{p}t)$ and choosing the pump frequency $\omega_{p}$ to be the frequency of the mode we want to drive, that is, $\omega_{p} = \omega_{k}$. Then, by taking only the leading order linear term in the sine potential (i.e., $\sin(\hat{x}) \simeq \hat{x}$), we get the desired linear drive
\begin{align}
    \hat{H}_{\textrm{rot}} &= -E_{J}\epsilon_{p}\varphi_{k}(\hat{a}_{k}+\hat{a}_{k}^{\dagger}) + \hat{H}', 
\end{align}
i.e., $\epsilon_{Z} = -E_{J}\epsilon_{p}\varphi_{k}$, where $\hat{H}'$ contains fast-oscillating terms such as $-E_{J}\epsilon_{p}(\varphi_{l}\hat{a}_{l}e^{-i(\omega_{l}-\omega_{k})t}+\textrm{h.c.})$ with $l\neq k$ and $-E_{J}\epsilon_{p}(\varphi_{b}\hat{b}e^{-i(\omega_{b}-\omega_{k})t}+\textrm{h.c.})$ as well as other terms that rotate even faster, e.g., $-E_{J}\epsilon_{p}\varphi_{k}(\hat{a}_{k}e^{-2i\omega_{k}t}+\textrm{h.c.})$. Since the frequency differences between different modes are on the order of $100$MHz but $|\epsilon_{Z}|/(2\pi)$ is typically not required to be larger than $1$MHz, the fast-oscillating terms can be ignored by using a rotating wave approximation (RWA). For instance, the strength of the linear drive needed for the compensating Hamiltonian for the CNOT gate $\hat{H}_{\textrm{CNOT}}$ is given by  
\begin{align}
    \frac{\pi\alpha}{4T^{\star}_{\textrm{CNOT}}} &= \frac{\pi\alpha^{3}}{1.24}\sqrt{\kappa_{1}\kappa_{2}} 
    \nonumber\\
    &= \begin{cases}
    2\pi * 2.89\textrm{MHz}  & \kappa_{1}/\kappa_{2} = 10^{-3} \\
    2\pi * 912\textrm{kHz}  & \kappa_{1}/\kappa_{2} = 10^{-4} \\
    2\pi * 289\textrm{kHz} & \kappa_{1}/\kappa_{2} = 10^{-5}
    \end{cases} 
\end{align}
at the optimal CNOT gate time $T^{\star}_{\textrm{CNOT}} = 0.31/(\sqrt{\kappa_{1}\kappa_{2}}\alpha^{2})$ assuming $\alpha^{2} =8$ and $\kappa_{2}=10^{7}s^{-1}$. Note that the subleading cubic term in the sine potential is also neglected here. These unwanted cubic terms are smaller than the desired linear term by a factor of $\varphi_{k}^{2}$. We remark that to avoid driving unwanted higher order terms, one may alternatively drive the phononic mode directly, at the expense of increased hardware \etc{complexity}, instead of using the pump $\epsilon_{p}(t)$ at the ATS node.    

Let us now consider a beam-splitter interaction between two phononic modes, e.g., $\epsilon_{ZZ}(\hat{a}_{1}^{\dagger}\hat{a}_{2} + \hat{a}_{1}\hat{a}_{2}^{\dagger})$, which is needed for implementing a CZ rotation between two cat qubits. It is also used to realize the compensating Hamiltonian for the Toffoli gate $\hat{H}_{\textrm{TOF}}$ and to realize the SWAP operation for the $X$ readout of a cat qubit. Note that the beam-splitter interaction is quadratic and even. Hence, it cannot be directly driven with a single pump tone since the sine potential has an odd parity. We thus jointly apply one pump tone and another drive tone to off-resonantly drive two odd terms and choose the detunings such that these two odd terms realize a resonant beam-splitter interaction when they are combined together. Since average Hamiltonian theory is useful for the analysis of the above scheme as well as many other schemes we propose below, we briefly state a key result of average Hamiltonian theory \cite{James2007effective,Gamel2010timeaveraged}: given a time-dependent Hamiltonian
\begin{align}
    \hat{H} = \hat{H}_{0} + \sum_{n}\Big{[}\hat{V}_{n}e^{-i\Delta_{n}t} + \textrm{h.c.} \Big{]}
\end{align}
with fast-oscillating time-dependent terms, one gets the following effective Hamiltonian by averaging out fast-oscillating terms
\begin{align}
    \hat{H}_{\textrm{eff}} &= \hat{H}_{0} + \frac{1}{2}\sum_{m,n}\Big{(} \frac{1}{\Delta_{m}} + \frac{1}{\Delta_{n}} \Big{)}[\hat{V}_{m}^{\dagger},\hat{V}_{n}]e^{i(\Delta_{m}-\Delta_{n})t}. 
\end{align}

To realize the beam-splitter interaction $\hat{a}_{1}^{\dagger}\hat{a}_{2} +\textrm{h.c.}$, we drive the two terms $\hat{a}_{1}^{\dagger}\hat{a}_{2}\hat{b}^{\dagger}$ and $\hat{b}$ off-resonantly. In particular, we use a pump $\epsilon_{p}(t) = \epsilon_{p}\cos(\omega_{p}t)$ with a pump frequency $\omega_{p} = \omega_{2}-\omega_{1}-\omega_{b} -\Delta$ to off-resonantly drive the term $\hat{a}_{1}^{\dagger}\hat{a}_{2}\hat{b}^{\dagger}$ and directly drive the $\hat{b}$ mode via 
\begin{align}
    \hat{H}_{d} &= \epsilon_{d}(\hat{b}^{\dagger}e^{-i\omega_{d}t} + \textrm{h.c.})
    \label{eq:BeamSplitterDerivationStart}
\end{align}
with a drive frequency $\omega_{d} = \omega_{b} + \Delta$ to off-resonantly drive the linear term $\hat{b}^{\dagger}$. Note that the size of the detuning $|\Delta|$ must not be larger than half the filter bandwidth $2J$ so that the drive is not filtered out. Then, by taking up to the third order terms in the sine potential (i.e., $\sin(\hat{x})\simeq \hat{x} - \hat{x}^{3}/6$) in \cref{eq:Sine potential rotating frame gate implementation}, we find  
\begin{align}
    \hat{H}_{\textrm{rot}} &= E_{J}\epsilon_{p}\varphi_{1}\varphi_{2}\varphi_{b}\hat{a}_{1}^{\dagger}\hat{a}_{2}\hat{b}^{\dagger}e^{-i\Delta t}  + \textrm{h.c.}
    \nonumber\\
    &\quad +\epsilon_{d}\hat{b}^{\dagger}e^{-i\Delta t} + \textrm{h.c.} + \hat{H}', \label{eq:sine potential explicit beam splitter main text}
\end{align}
where $\hat{H}'$ contains fast-oscillating terms, which we ignore for the moment. Let $\chi_{1} \equiv E_{J}\epsilon_{p,1}\varphi_{1}\varphi_{2}\varphi_{b}$ and $\chi_{2} \equiv \epsilon_{d}$. Then, neglecting $\hat{H}'$, the average Hamiltonian theory yields 
\begin{align}
    \hat{H}_{\textrm{eff}} &= \frac{1}{\Delta}[ (\chi_{1}\hat{a}_{1}\hat{a}_{2}^{\dagger} + \chi_{2})\hat{b} , (\chi_{1}\hat{a}_{1}^{\dagger}\hat{a}_{2} + \chi_{2})\hat{b}^{\dagger} ]
    \nonumber\\
    &= \frac{1}{\Delta}\Big{[} (\chi_{1}\hat{a}_{1}\hat{a}_{2}^{\dagger} + \chi_{2})(\chi_{1}\hat{a}_{1}^{\dagger}\hat{a}_{2} + \chi_{2})
    \nonumber\\
    &\qquad + [ (\chi_{1}\hat{a}_{1}\hat{a}_{2}^{\dagger} + \chi_{2}) , (\chi_{1}\hat{a}_{1}^{\dagger}\hat{a}_{2} + \chi_{2}) ]\hat{b}^{\dagger}\hat{b} \Big{]}
    \nonumber\\
    &\xrightarrow{ \hat{b}^{\dagger}\hat{b} \ll 1 } \frac{1}{\Delta} (\chi_{1}\hat{a}_{1}\hat{a}_{2}^{\dagger} + \chi_{2})(\chi_{1}\hat{a}_{1}^{\dagger}\hat{a}_{2} + \chi_{2})
    \nonumber\\
    &= \frac{\chi_{1}\chi_{2}}{\Delta}( \hat{a}_{1}^{\dagger}\hat{a}_{2} + \hat{a}_{1}\hat{a}_{2}^{\dagger} ) + \frac{\chi_{1}^{2}}{\Delta}(\hat{a}_{1}^{\dagger}\hat{a}_{1}+1)\hat{a}_{2}^{\dagger}\hat{a}_{2}  . \label{eq:average Hamiltonian theory beam splitter main text}
\end{align}
Note that we assumed that the population in the $\hat{b}$ mode is negligible (i.e., $\hat{b}^{\dagger}\hat{b} \ll 1$) and dropped the constant energy shift $\chi_{2}^{2}/ \Delta$ in the last line. The first term in the last line is the desired beam-splitter interaction $\epsilon_{ZZ}( \hat{a}_{1}^{\dagger}\hat{a}_{2} + \hat{a}_{1}\hat{a}_{2}^{\dagger} )$ with a coupling strength 
\begin{align}
    \epsilon_{ZZ} = \frac{\chi_{1}\chi_{2}}{\Delta} = E_{J}\epsilon_{p,1}\varphi_{1}\varphi_{2}\varphi_{b}  \beta, 
\end{align}
where $\beta \equiv \chi_{2}/\Delta = \epsilon_{d}/\Delta$ can be understood as an effective displacement in the $\hat{b}$ mode. For the population of the $\hat{b}$ mode to be negligible, we need $|\beta|\ll 1$. Assuming $\beta=0.1$ and noting that $E_{J}\epsilon_{p,1}\varphi_{1}\varphi_{2}\varphi_{b}\sim g_{2} \lesssim 2\pi \times  5\textrm{MHz}$, we find that $\epsilon_{ZZ} \sim 2\pi \times 500\textrm{kHz}$ is achievable. The strength of the beam-splitter interaction in the compensating Hamiltonian for the Toffoli gate $\hat{H}_{\textrm{TOF}}$ is given by (see \cref{eq:compensating Hamiltonian TOF expanded appendix})
\begin{align}
    \frac{\pi}{8T^{\star}_{\textrm{TOF}}} &= \frac{\pi \alpha^{2}}{2.48}\sqrt{\kappa_{1}\kappa_{2}} 
    \nonumber\\
    &= \begin{cases}
    2\pi * 1.02\textrm{MHz}  & \kappa_{1}/\kappa_{2} = 10^{-3} \\
    2\pi * 323\textrm{kHz}  & \kappa_{1}/\kappa_{2} = 10^{-4} \\
    2\pi * 102\textrm{kHz} & \kappa_{1}/\kappa_{2} = 10^{-5}
    \end{cases} 
\end{align}
at the optimal Toffoli gate time $T^{\star}_{\textrm{TOF}} = 0.31/(\sqrt{\kappa_{1}\kappa_{2}}\alpha^{2})$ assuming $\alpha^{2}=8$ and $\kappa_{2}=10^{7}s^{-1}$. We also remark that the second term in the last line of \cref{eq:average Hamiltonian theory beam splitter main text} gives rise to undesired cross-Kerr interaction and energy shift of the $\hat{a}_{2}$ mode. The unwanted cross-Kerr interaction $\hat{a}_{1}^{\dagger}\hat{a}_{1}\hat{a}_{2}^{\dagger}\hat{a}_{2}$ can in principle be cancelled by off-resonantly driving the term $\hat{a}_{1}\hat{a}_{2}\hat{b}^{\dagger}$ with a detuning $\Delta'$ different from $\Delta$. The frequency shift of the mode $\hat{a}_{2}$ (i.e., $(\chi_{1}^{2}/\Delta)\hat{a}_{2}^{\dagger}\hat{a}_{2}$) can either be incorporated into the frequency matching condition or physically cancelled by off-resonantly driving the term $\hat{a}_{2}\hat{b}^{\dagger}$ (see below for more details). 

Note that we have so far ignored fast-oscillating terms (i.e., $\hat{H}'$ in \cref{eq:sine potential explicit beam splitter main text}). These fast-oscillating terms include unwanted cubic terms, e.g., $E_{J}\epsilon_{p,1}\varphi_{2}\varphi_{3}\varphi_{b}\hat{a}_{2}^{\dagger}\hat{a}_{3}\hat{b}^{\dagger}e^{i(2\omega_{2}-\omega_{1}-\omega_{3} - \Delta) t} +\textrm{h.c.}$ which would give rise to an unwanted beam-splitter interaction $\hat{a}_{2}^{\dagger}\hat{a}_{3} + \textrm{h.c.}$. If the frequencies of the modes $\hat{a}_{1}$, $\hat{a}_{2}$, and $\hat{a}_{3}$ are equally spaced, $2\omega_{2}-\omega_{1}-\omega_{3}$ vanishes and the unwanted term $\hat{a}_{2}^{\dagger}\hat{a}_{3}\hat{b}^{\dagger}$ interferes with the desired term $\hat{a}_{1}^{\dagger}\hat{a}_{2}\hat{b}^{\dagger}$ as they rotate with the same frequency. However, in practice, equal frequency spacing is avoided in the optimization of the frequencies of the phononic modes. Hence, unwanted beam-splitter interactions are far detuned from the desired beam-splitter interaction. We remark that remaining fast-rotating terms in $\hat{H}'$ (different from the above beam-splitter type) are of less concern as their rotating frequencies are farther away from the frequencies of the desired terms. 

Let us now move on to the selective frequency shift which is needed, e.g., for removing non-adiabatic errors of the $X$ gate if we were to implement the $X$ gate physically (see \cref{eq:X gate compensating Hamiltonian main text}). In practice, the $180\degree$ rotation $e^{i\pi\hat{a}^{\dagger}\hat{a}}$ (or $\hat{a}\rightarrow -\hat{a}$) for the $X$ gate can be performed via software by adapting the phases of subsequent drives. However, we still discuss the selective frequency shift because it is conceptually useful for understanding our proposal for implementing the compensating Hamiltonians for the CNOT and Toffoli gates. 

We first consider frequency shifts due to a pump $\epsilon_{p}(t) = \epsilon_{p}\cos(\omega_{p}t)$. Note that the terms $\hat{a}_{k}^{\dagger}\hat{a}_{k}\hat{b}^{\dagger}$ and $\hat{b}^{\dagger}$ in the sine potential are off-resonantly driven by the pump with the same detuning $\Delta = \omega_{p}-\omega_{b}$ and with coupling strengths $E_{J}\epsilon_{p}\varphi_{k}^{2}\varphi_{b}$ and $-E_{J}\epsilon_{p}\varphi_{b}$, respectively. Hence, through the average Hamiltonian theory, we find that the frequency of the $\hat{a}_{k}$ mode is shifted by
\begin{align}
    \delta\omega_{k} &= -\frac{E_{J}^{2}\epsilon_{p}^{2}\varphi_{k}^{2}\varphi_{b}^{2}}{\omega_{p}-\omega_{b}} . 
\end{align}
Similarly as in the case of beam-splitter interaction, the frequency shift is accompanied by undesirable quartic terms such as self-Kerr $(\hat{a}_{k}^{\dagger})^{2}\hat{a}_{k}^{2}$ and cross-Kerr $\hat{a}_{k}^{\dagger}\hat{a}_{k}\hat{a}_{l}^{\dagger}\hat{a}_{l}$ nonlinearities. While we have ignored the frequency shifts due to pump in the discussions so far, they need to be carefully taken into account in practice. 

Note that the size of frequency shift can be modulated by changing the pump amplitude $\epsilon_{p}$ (i.e., $|\delta\omega_{k}| \propto \epsilon_{p}^{2}$). However, we cannot engineer the frequency shifts due to $\hat{a}_{k}^{\dagger}\hat{a}_{k}\hat{b}^{\dagger}$ and $\hat{b}^{\dagger}$ in a mode-selective manner since $\hat{a}_{l}^{\dagger}\hat{a}_{l}\hat{b}^{\dagger}$ with $l\neq k$ rotates with the same frequency as those of $\hat{a}_{k}^{\dagger}\hat{a}_{k}\hat{b}^{\dagger}$ and $\hat{b}^{\dagger}$. In particular, since $\delta\omega_{k} / \delta\omega_{l} = \varphi_{k}^{2}/\varphi_{l}^{2}$ and the zero-point fluctuations of phononic modes are almost identical, the frequency shifts of the phononic modes $\delta\omega_{k}$ are approximately independent of the mode index $k$. Thus, we cannot rely on frequency shifts due to $\hat{a}_{k}^{\dagger}\hat{a}_{k}\hat{b}^{\dagger}$ and $\hat{b}^{\dagger}$ to exclusively shift the frequency of a specific mode $\hat{a}_{k}$. 

Selective frequency shift the mode $\hat{a}_{k}$ can nevertheless be realized by off-resonantly driving the term $\hat{a}_{k}\hat{b}^{\dagger}$: if we are given with a Hamiltonian $\hat{H} = \chi \hat{a}_{k}\hat{b}^{\dagger}e^{-i\Delta t} + \textrm{h.c.}$, the average Hamiltonian theory yields (assuming $\hat{b}^{\dagger}\hat{b}\ll 1$ similarly as in \cref{eq:average Hamiltonian theory beam splitter main text})  \begin{align}
    \hat{H}_{\textrm{eff}} &= \frac{\chi^{2}}{\Delta}\hat{a}_{k}^{\dagger}\hat{a}_{k}, 
\end{align} 
i.e., frequency shift of the mode $\hat{a}_{k}$. In practice, the pumps used to off-resonantly drive the term $\hat{a}_{k}\hat{b}^{\dagger}$ may also drive $\hat{a}_{l}\hat{b}^{\dagger}$ with $l\neq k$ which will lead to the frequency shift of another mode $\hat{a}_{l}$. However, $\hat{a}_{l}\hat{b}^{\dagger}$ is detuned from $\hat{a}_{k}\hat{b}^{\dagger}$ by $\omega_{l}-\omega_{k}$ so the relevant detuning $\Delta'$ of the unwanted term $\hat{a}_{l}\hat{b}^{\dagger}$ is given by $\Delta' = \Delta + \omega_{l}-\omega_{k}$. Hence, the unwanted frequency shift in another mode $\hat{a}_{l}$ can in principle be suppressed by ensuring $|\Delta'|\gg |\Delta|$.  

Building up on the intuitions gained from the discussion of selective frequency shift, we now discuss implementation of the compensating Hamiltonian for the CNOT gate in \cref{eq:CNOTContolH}. Without loss of generality, we focus on the CNOT gate between the modes $\hat{a}_{1}$ (control) and $\hat{a}_{2}$ (target). Note that $\hat{H}_{\textrm{CNOT}}$ consists of an optomechanical coupling $(\pi/(4\alpha T))(\hat{a}_{1}+\hat{a}_{1}^{\dagger})\hat{a}_{2}^{\dagger}\hat{a}_{2}$ between two phononic modes, a linear drive on the control mode $-(\pi\alpha/(4T))(\hat{a}_{1}+\hat{a}_{1}^{\dagger})$, and a selective frequency shift of the target mode $-(\pi/(2T))\hat{a}_{2}^{\dagger}\hat{a}_{2}$. Similarly as the $180\degree$ rotation for the $X$ gate needs not be implemented physically, the selective frequency shift of the target mode can be taken care of via software. That is, instead of using $\hat{H}_{\textrm{CNOT}}$ in \cref{eq:CNOTContolH}, one may use a different compensating Hamiltonian 
\begin{align}
    \hat{H}'_{\textrm{CNOT}} &= \frac{\pi}{4\alpha T} (\hat{a}_{1}+\hat{a}_{1}^{\dagger})(\hat{a}_{2}^{\dagger}\hat{a}_{2}-\alpha^{2}) 
\end{align}
as well as an appropriately modified rotating jump operator $\hat{L}'_{2}(t)$ such that the cat states $|0\rangle \simeq |\alpha\rangle$ and $|1\rangle \simeq |-\alpha\rangle$ in the target mode are mapped to $|-i\alpha\rangle$ and $|i\alpha\rangle$ if the control mode is in the state $|0\rangle \simeq |\alpha\rangle$, and to $|i\alpha\rangle$ and $|-i\alpha\rangle$ if the control mode is in the trigger state $|1\rangle \simeq |-\alpha\rangle$. Hence, one may simply redefine the cat-code computational basis states of the target mode as $|0\rangle \leftarrow |-i\alpha\rangle$ and $|1\rangle \leftarrow |i\alpha\rangle$ and adjust the phases of subsequent drives accordingly. 

Note that the optomechanical coupling and the linear drive on the control mode still need to be implemented physically. Implementation of the linear drive is already discussed above. To realize the optomechanical coupling, one might be tempted to directly drive the cubic term $\hat{a}_{1}\hat{a}_{2}^{\dagger}\hat{a}_{2} + \textrm{h.c.}$ in the sine potential via a pump $\epsilon_{p}(t) = \epsilon_{p}\cos(\omega_{p}t)$. However, the direct driving scheme is not suitable for a couple of reasons: since the term $\hat{a}_{1}\hat{a}_{2}^{\dagger}\hat{a}_{2}$ rotates with frequency $\omega_{1}$, the required pump frequency is given by $\omega_{p}=\omega_{1}$ which is the same pump frequency reserved to engineer a linear drive on the $\hat{a}_{1}$ mode. Moreover, the term $\hat{a}_{1}\hat{a}_{2}^{\dagger}\hat{a}_{2}$ rotates at the same frequency as those of undesired cubic terms such as $\hat{a}_{1}\hat{a}_{3}^{\dagger}\hat{a}_{3}$, $\hat{a}_{1}\hat{a}_{4}^{\dagger}\hat{a}_{4}$, and also $\hat{a}_{1}^{\dagger}\hat{a}_{1}^{2}$. Hence, even if the linear drive is realized via a direct driving of the phononic mode, one still cannot selectively drive the desired optomechanical coupling by using the pump frequency $\omega_{p}=\omega_{1}$ due to the frequency collision with other unwanted cubic terms. This issue is analogous to the one we had earlier that the selective frequency shift of the $\hat{a}_{1}$ mode is not possible via the synthesis of two terms $\hat{a}_{1}^{\dagger}\hat{a}_{1}\hat{b}^{\dagger}$ and $\hat{b}^{\dagger}$. 

To circumvent the above frequency-collision issue, we propose to realize the optomechanical coupling $(\hat{a}_{1}+\hat{a}_{1}^{\dagger})\hat{a}_{2}^{\dagger}\hat{a}_{2}$ by off-resonantly driving the term $(\hat{a}_{1} + \lambda)\hat{a}_{2}\hat{b}^{\dagger}$. That is, given a Hamiltonian $\hat{H} = \chi(\hat{a}_{1} + \lambda)\hat{a}_{2}\hat{b}^{\dagger}e^{-i\Delta t} + \textrm{h.c.}$, we get the following effective Hamiltonian through the time averaging
\begin{align}
    \hat{H}_{\textrm{eff}} &= \frac{\chi^{2}\lambda}{\Delta}\Big{(}  \hat{a}_{1}+\hat{a}_{1}^{\dagger} + \lambda +  \frac{1}{\lambda}\hat{a}_{1}^{\dagger}\hat{a}_{1} \Big{)}\hat{a}_{2}^{\dagger}\hat{a}_{2} , 
\end{align}
where we again assumed that the population of the $\hat{b}$ mode is negligible (i.e., $\hat{b}^{\dagger}\hat{b}\ll 1$). In particular, by choosing $\lambda = -2\alpha$, we can realize the optomechanical coupling as well as the selective frequency shift of the $\hat{a}_{2}$ mode, i.e., $\hat{H}_{\textrm{eff}} \propto (\hat{a}_{1}+\hat{a}_{1}^{\dagger}-2\alpha)\hat{a}_{2}^{\dagger}\hat{a}_{2}$ up to an undesired cross-Kerr term $- \hat{a}_{1}^{\dagger}\hat{a}_{1}\hat{a}_{2}^{\dagger}\hat{a}_{2}/(2\alpha)$ (which can in principle be cancelled by off-resonantly driving the term $\hat{a}_{1}\hat{a}_{2}\hat{b}^{\dagger}$). Hence, if we realize $\hat{H}_{\textrm{CNOT}}$ this way, we need not rely on software to keep track of the phase of the target mode as the phase shift is physically realized. We also remark that the term $(\hat{a}_{1} + \lambda)\hat{a}_{2}\hat{b}^{\dagger}$ is detuned from other undesired terms such as $(\hat{a}_{1} + \lambda)\hat{a}_{k}\hat{b}^{\dagger}$ with $k\ge 3$ by a frequency difference $\omega_{2}-\omega_{k}$. Thus, the unwanted optomechanical coupling $(\hat{a}_{1}+\hat{a}_{1}^{\dagger})\hat{a}_{k}^{\dagger}\hat{a}_{k}$ can be suppressed by a suitable choice of the detuning $\Delta$ similarly as in the case of selective frequency shift. 

Note that while the cubic term $\hat{a}_{1}\hat{a}_{2}\hat{b}^{\dagger}$ in $(\hat{a}_{1} + \lambda)\hat{a}_{2}\hat{b}^{\dagger}$ can be realized by using the sine potential, the other quadratic term $\hat{a}_{2}\hat{b}^{\dagger}$ cannot be directly realized from the sine potential which has an odd parity. The quadratic interaction $\hat{a}_{2}\hat{b}^{\dagger}$ can in principle be realized by synthesizing (using the average Hamiltonian theory) two odd terms $\hat{a}_{2}(\hat{b}^{\dagger})^{2}$ and $\hat{b}^{\dagger}$. 
To put everything together and get the desired optomechanical coupling, however, the results of average Hamiltonian theory need to be concatenated. In other words, to analyze the full scheme for the desired optomechanical coupling, a higher-order average Hamiltonian theory is needed. We leave it as a future work to thoroughly analyze such a scheme.

Lastly, let us consider the compensating Hamiltonian $\hat{H}_{\textrm{TOF}}$ for the Toffoli gate in \cref{eq:Toffoli compensating Hamiltonian main text}. $\hat{H}_{\textrm{TOF}}$ is explicitly given by  
\begin{align}
    \hat{H}_{\textrm{TOF}} &= -\frac{\pi}{8\alpha^{2}T}(\hat{a}_{1}^{\dagger}\hat{a}_{2} + \hat{a}_{1}\hat{a}_{2}^{\dagger})(\hat{a}_{3}^{\dagger}\hat{a}_{3} - \alpha^{2}) 
    \nonumber\\
    &\qquad + \frac{\pi}{8\alpha T}(\hat{a}_{1}+\hat{a}_{1}^{\dagger}-\alpha)(\hat{a}_{3}^{\dagger}\hat{a}_{3} - \alpha^{2}) 
    \nonumber\\
    &\qquad + \frac{\pi}{8\alpha T}(\hat{a}_{2}+\hat{a}_{2}^{\dagger}-\alpha)(\hat{a}_{3}^{\dagger}\hat{a}_{3} - \alpha^{2}) . \label{eq:compensating Hamiltonian TOF expanded appendix}
\end{align}
Note that the terms in the second and the third lines are in the same form as the compensating Hamiltonian for the CNOT gate. Thus, they can be realized in a similar way as described above. The terms in the first line contain a beam-splitter interaction $(\hat{a}_{1}^{\dagger}\hat{a}_{2} + \hat{a}_{1}\hat{a}_{2}^{\dagger})$, which we have already discussed above, as well as a quartic term $(\hat{a}_{1}^{\dagger}\hat{a}_{2} + \hat{a}_{1}\hat{a}_{2}^{\dagger})\hat{a}_{3}^{\dagger}\hat{a}_{3}$. Since the sine potential has an odd parity, it is not possible to drive the quartic term directly. The quartic term can nevertheless be realized by off-resonantly driving the term $(\hat{a}_{1}+\hat{a}_{2})\hat{a}_{3}\hat{b}^{\dagger}$: given $\hat{H} = \chi(\hat{a}_{1}+\hat{a}_{2})\hat{a}_{3}\hat{b}^{\dagger}e^{-i\delta t} + \textrm{h.c.}$, we get 
\begin{align}
    \hat{H}_{\textrm{eff}} &= \frac{\chi^{2}}{\Delta}( \hat{a}_{1}^{\dagger}\hat{a}_{2} + \hat{a}_{1}\hat{a}_{2}^{\dagger}  )\hat{a}_{3}^{\dagger}\hat{a}_{3}  + \frac{\chi^{2}}{\Delta}( \hat{a}_{1}^{\dagger}\hat{a}_{1} + \hat{a}_{2}\hat{a}_{2}^{\dagger}  )\hat{a}_{3}^{\dagger}\hat{a}_{3} , 
\end{align}
i.e., the desired quartic interaction and unwanted cross-Kerr interactions between a control and the target modes. The undesired cross-Kerr terms, which are as strong as the desired quartic term, can in principle be cancelled by off-resonantly driving the terms $\hat{a}_{1}\hat{a}_{3}\hat{b}^{\dagger}$ and $\hat{a}_{2}\hat{a}_{3}\hat{b}^{\dagger}$ with detunings $\Delta_{1}$ and $\Delta_{2}$ which are different from each other and also from $\Delta$. 

The required coupling strength of the quartic interaction $(\hat{a}_{1}^{\dagger}\hat{a}_{2} + \hat{a}_{1}\hat{a}_{2}^{\dagger})\hat{a}_{3}^{\dagger}\hat{a}_{3}$ is given by 
\begin{align}
    \frac{\pi}{8\alpha^{2}T_{\textrm{TOF}}} &= \frac{\pi}{2.48}\sqrt{\kappa_{1}\kappa_{2}} 
    \nonumber\\
    &= \begin{cases}
    2\pi * 128\textrm{kHz}  & \kappa_{1}/\kappa_{2} = 10^{-3} \\
    2\pi * 40.3\textrm{kHz}  & \kappa_{1}/\kappa_{2} = 10^{-4} \\
    2\pi * 12.8\textrm{kHz} & \kappa_{1}/\kappa_{2} = 10^{-5}
    \end{cases} 
\end{align}
at the optimal Toffoli gate time $T^{\star}_{\textrm{TOF}} = 0.31/(\sqrt{\kappa_{1}\kappa_{2}}\alpha^{2})$ assuming $\kappa_{2} = 10^{7}s^{-1}$. Note that the coupling strength of the term $\hat{a}_{1}\hat{2}\hat{b}^{\dagger}$ and $\hat{a}_{1}\hat{3}\hat{b}^{\dagger}$ are comparable to $g_{2} \lesssim 2\pi \times 5\textrm{MHz}$. Incorporating the bosonic enhancement factor due to the average excitation number $\alpha^{2}$, we require the detuning $\Delta$ to be much larger than $g_{2}\alpha^{2}$, e.g., $\Delta = 10g_{2}\alpha^{2}$. Then the achievable coupling strength of the quartic interaction is given by $g_{2}^{2}/(10g_{2}\alpha^{2}) = g_{2}/(10\alpha^{2}) \lesssim 2\pi \times 60\textrm{kHz}$ assuming $\alpha^{2}=8$.

\section{Measurement}
\label{appendix:Measurement}
In this appendix we discuss measurement schemes for high fidelity readout in both the $X$ and $Z$ basis.  \hp{Compared to gates where the optimal gate errors can be cleanly described in terms of dimensionless constants like $\kappa_1/\kappa_2$, for readout absolute timescales matter more.  We enumerate the parameter choices for the different schemes at the end of the corresponding sections.}
\subsection{\texorpdfstring{$X$}{X}-basis measurement}
\label{subsection: X measurement}
Here we discuss in more detail the $X$-basis readout scheme used to generate the infidelities used in most of the error correction simulations.  Note that throughout this appendix, when we refer to measurement infidelities or error probabilities, we are referring to misassignment probabilities 
\begin{align}
    \epsilon_s = 1-\text{P}(s|s)
\end{align} where $\text{P}(s|s)$ is the probability of reading out the state s given that the cavity was in state $s$.

\begin{figure*}
	\centering
	\includegraphics[width=\textwidth]{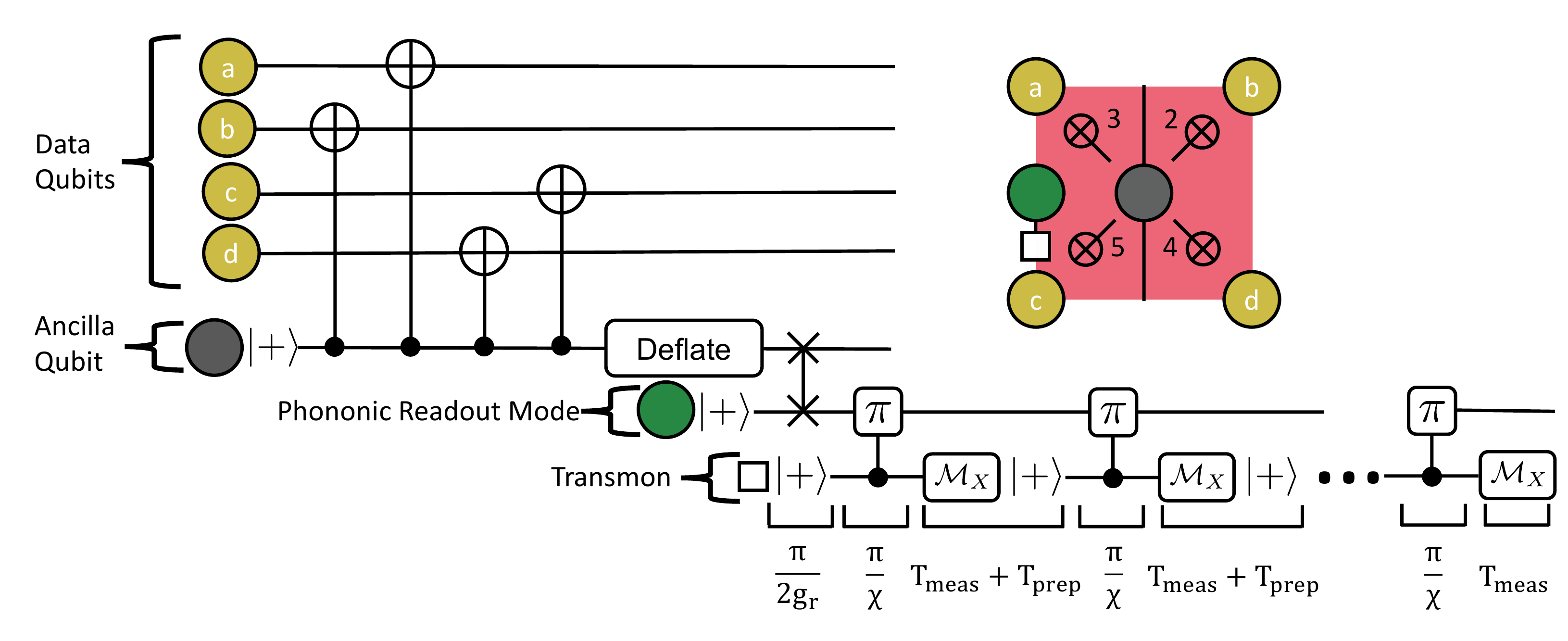}
	\caption{Circuit used for an $X$-basis measurement in the context of an $X$-type stabilizer measurement.  The first step consists of entangling the ancilla qubit with the data qubits. Afterwords, the ancilla qubit is deflated followed by a SWAP with a readout mode.   Lastly, the readout mode is repeatedly measured using a transmon qubit. The duration's for the parts of the measurement procedure are labeled at the bottom of the figure below each circuit element. While these repeated parity measurements are occurring, the CNOT gates of the next error correction cycle can begin.  Also included is a diagram of the physical layout of the stabilizer to give context to the measurement circuit.}
	\label{fig:xMeasurement}
\end{figure*}

$X$-basis measurement refers to determining the parity of a phononic mode or equivalently readout in the basis of even and odd cat states, i.e., $|\pm\rangle \propto |\alpha\rangle \pm |-\alpha\rangle$.  To realize such a measurement with minimal impact on the length of an error correction cycle we utilize an additional phononic mode which we refer to as the readout mode.  This mode is interrogated by a transmon in parallel with the next error correction cycle.  As is pictured in \cref{fig:hardware_cartoon}, every unit cell contains this additional readout mode connected to a transmon.  Pictured in \cref{fig:xMeasurement} is the circuit we use for measuring an $X$ stabilizer which we now walk through in more detail.  
To perform an $X$ stabilizer measurement first the ancilla qubit $\hat{a}_1$ is entangled with the data qubits.  Subsequently we ``deflate'' the ancilla qubit on a timescale comaparable $1/\kappa_2$, mapping the even parity state to $|\hat{n} = 0\rangle$ and the odd parity state to $|\hat{n} = 1\rangle$ \cite{Grimm2020}.  Deflation is achieved under evolution with the two-phonon dissipator, 
\begin{align}
   \frac{d\hat{\rho}(t)}{dt} &= \kappa_2 \mathcal{D}[ \hat{a}_1^2 - \alpha(t)^2]\hat{\rho}(t),
\end{align}
by taking $\alpha(t)$ from $\alpha_0$ to $\alpha_1 < \alpha_0$.  In our case we rapidly take $\alpha(t)$ from its initial value to $\alpha_1=0$ where we evolve for a time on the order of $1/\kappa_2$.  The deflation is not required to be adiabatic since we do not need to maintain phase coherence between the even and odd parity states.  The utility of the deflation is that it makes the state of the cavity less susceptible to single phonon loss events which change its parity.  

After the deflation we perform a SWAP between the ancilla phononic mode ($\hat{a}_1$) and the readout mode ($\hat{a}_2$).  The SWAP is performed using the Hamiltonian
\begin{align}
    \hat{H}_{rot} &= g_r (\hat{a}_1^\dagger \hat{a}_2 + \hat{a}_1^\dagger \hat{a}_2)
\end{align}
Evolution under this Hamiltonian for a duration ${\pi}/{2 g_r}$ realizes a SWAP gate (there is a rotation of the swapped state by 90 degrees).  Physical implementation of this Hamiltonian is discussed in \cref{eq:BeamSplitterDerivationStart}.

An advantage of this readout scheme is that after the exchange has occurred the next cycle of quantum error correction can continue in parallel with the measurement of the readout mode.  With the idling time now set only by the deflation + SWAP steps, we can spend more time measuring the readout mode without compromising on idling error.  This simple layout choice could be generally useful in other architectures.  In the specific case of this proposal, we perform repeated QND parity measurements which we majority vote to get our final measurement outcome \cite{HannRobustReadout, ElderHighFidelityMeasurement, sun2014tracking}.  In general more advanced methods than majority voting will give higher fidelity \cite{HannRobustReadout}.

Measurement of the readout mode parity is done using a dispersive coupling with a transmon qubit \cite{sun2014tracking} $H = -\chi \hat{\sigma}_z\hat{a}^\dagger \hat{a}$ where $\hat{a}$ corresponds to a bosonic mode and $\hat{\sigma}_z$ corresponds to a transmon qubit.  Evolution under this Hamiltonian for a time $t = {\pi}/{\chi}$ realizes the unitary
$\hat{U} = I\otimes|g\rangle\langle g| + e^{i \hat{a}_2^\dagger \hat{a}_2\pi} |e\rangle\langle e|$ in  which is a controlled parity gate.  As pictured in \cref{fig:xMeasurement}, when the controlled parity gate is placed between an initialization in $|+\rangle$ and measurement of the transmon in the $X$-basis this realizes a QND measurement of the readout mode parity.  \hp{Importantly our use of repeated measurement suppresses the effect of transmon error mechanisms (i.e. transmon readout error, transmon T1...) on the final readout fidelity.  The readout is still QND in the presence of such errors as they commute with the dispersive coupling}.  

\begin{figure}
	\centering
	\includegraphics[width=.48\textwidth]{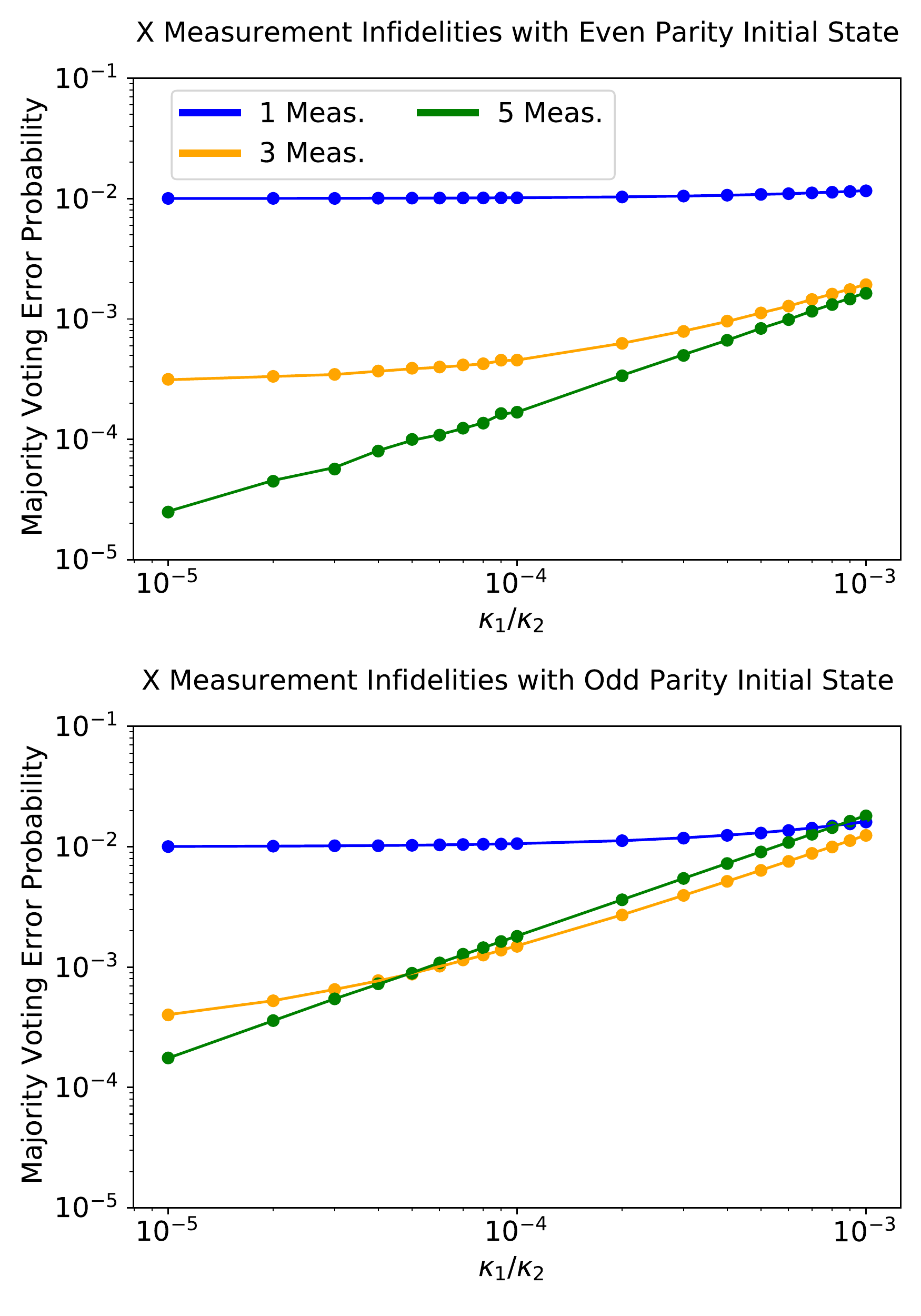}
	\caption{Simulated (markers) and analytical (curves) error probabilities for $X$ measurement for the cases of both even parity and odd parity initial states.  We have taken the parity measurement to be QND and projective.  The dependence on $\kappa_1/{\kappa_2}$ is stronger for the case of an odd initial state since the cavity is mapped to $|1\rangle$ after the deflation.  The plotted curve is the leading order analytic model for the case of $\kappa_\phi = 0$ \cref{eq:XMeasurementModelLeadingOrder}.  Numerical imprecision predominantly due to the deflation simulation can have about a 10 percent effect on the simulated infidelities for the smaller $\kappa_1/\kappa_2$.}
	\label{fig:xMeasurementInfidelity}
\end{figure}

We have performed simulations of this measurement scheme to determine the rough infidelities for different single phonon loss rates $\kappa_1 = \kappa_2 *(\kappa_1/\kappa_2)$.  We start by performing master equation evolution under the deflation and swap to determine $P(\text{even})$ ($P(\text{odd})$) which is the probability that the final state after the deflation and swap steps is even (odd).  We include the effects of single phonon loss during the deflation and swap.  Then we sample from these probabilities to determine the state of the readout mode after the first measurement.  After the measurement the readout mode is in the $|\hat{n} = 0\rangle$ and $|\hat{n} = 1\rangle$ manifold which is a good approximation since after the deflation step the population of the readout mode is very close to 0 or 1 depending on the initial state.  Starting from the state the readout mode is projected onto after the first measurement we perform master equation evolution to include the effects of the single phonon loss, gain, and dephasing on the readout mode during the inter measurement period and during measurement ($T_\mathrm{entangle}+T_\mathrm{meas} + T_\mathrm{reset}$).  We then repeat this projection and evolution for the remaining number of measurements that are used, giving us one sequence of projections.  To include the effect of transmon errors such as loss, dephasing, and incorrect measurement we add additional randomness associated with a fixed transmon error probability ($\epsilon_q$) giving us the final measurement sequence.  We have performed Monte Carlo sampling of these measurement sequences to determine the infidelities of this measurement process.  A plot of the infidelities are pictured in \cref{fig:xMeasurementInfidelity}.  The assumed numbers in the simulations are listed at the end of this section.

In other circumstances, such as stabilization of four component cat codes, decay during entanglement with the transmon is problematic because it induces dephasing of the cavity.  In our case since we are only concerned about measuring the parity this dephasing is not important.  As a result we are justified in lumping the effect of this transmon decay into our fixed parameter representing the transmon infidelity mechanisms.  We also note that recent advances in transmon measurement would allow more aggressive transmon measurement fidelities than what we assume \cite{ElderHighFidelityMeasurement}.  This would allow us to use fewer measurements to achieve the same fidelities we currently expect to achieve. 

We can also get an approximate form for the measurement infidelities to expect with this repeated measurement procedure.  Defining $N$ (odd) to be the total number of measurements and $k\equiv (N+1)/2$, to leading order the error probability for the majority voting of the repeated measurements for initial even and odd cat states in the case of no gain are
\begin{align}
    \epsilon_\text{even} &= \epsilon_\text{(deflate + SWAP)} + \binom{N}{k}\epsilon_q^k (1-\epsilon_q)^{N-k} \nonumber \\
    \epsilon_\text{odd} &= \epsilon_\text{(deflate + SWAP)} + \binom{N}{k}\epsilon_q^k (1-\epsilon_q)^{N-k}+ \kappa_1 \text{T}_p
    \label{eq:XMeasurementModelLeadingOrder}
\end{align}
where $\text{T}_p$ is the amount of time after the SWAP and before the kth measurement and $\epsilon_\text{(deflate + SWAP)}$ is the error from the deflation and SWAP for the given initial state.  In the above expressions the first term is the contribution to the error from the deflation and SWAP steps.  The second term is due to transmon error where k measurements are incorrect.  The last term in the case of an odd initial state is the probability of a $T_1$ event before the kth measurement which will lead with high probability to all the remaining measurements giving 0. Note that this is the reason that for an odd initial state and larger $\kappa_1/\kappa_2$ values majority voting 5 measurements underperforms majority voting 3 measurements.  Using a more advanced procedure than majority voting would mitigate this problem.

\hp{\textit{Assumptions--}The properties of the measurement were chosen to make the error probabilities dependent only on $\kappa_1/\kappa_2$ and the measurement times scale as $1/\kappa_2$ to follow the gates which have the same dependence.  This is convenient for the error correction simulations because it means that the logical failure rates are independent of the absolute scale of $\kappa_2$. }

\hp{The properties that feed into our measurement error probabilities and times are a transmon related error probability ($\epsilon_q$) of $1\%$, a deflation time of $3/{\kappa_2}$,  transmon related times of $\text{T}_\text{entangle}=\text{T}_\text{readout}=\text{T}_\text{reset} = 200 \mathrm{ ns}*(1/((100\ \mathrm{ ns})*\kappa_2) $, $\alpha^2 = 8$, and $g/{2\pi} = 1\mathrm{ MHz}*((100\ \mathrm{  ns})*\kappa_2)$\cite{HeinsooHighFidelity2018, JeffreyFastAccurate2014}.  For simplicity, in the error correction simulations we have used up to 3 measurements in the repetition code and up to 5 measurements in the surface code.  This choice is well justified in the important regime for \REGthree.  For example, for the surface code the duration of 4 CNOT gates is $\sim 28 \mu s$, larger than the relevant measurement duration of $5*T_\mathrm{entangle}+5*T_\mathrm{measure}+5*T_\mathrm{reset}\sim 17.05 \mu s$.   In \REGone and \REGtwo this simplification breaks down and fewer measurements or more time for the measurements would be needed.  However in these regimes fault tolerant quantum computation is infeasible even with this optimistic measurement error model.  In the overhead calculations the idling duration corresponding to the $X$-basis measurement is $3/\kappa_2 + \pi/2g$.}

\subsection{\texorpdfstring{$Z$}{Z}-basis measurement}
\label{subsection: Z measurement}
For $Z$ measurement we use a beamsplitter interaction between the buffer mode and a phononic storage mode 
\begin{align}
    \hat{H}_{r} = g_{r} (\hat{a}^{\dagger} \hat{b} + \hat{b}^{\dagger} \hat{a})
\end{align}
where $\hat{a}$ is an annihilation operator on a storage mode and $\hat{b}$ is an annihilation operator on the buffer mode.  By homodyning the output of the buffer mode we determine the state of the storage mode.  We perform this readout scheme with the two-phonon dissipation off.  A similar scheme has been realized for Kerr-Cat qubits in \cite{Grimm2020}.

To realize this interaction we drive the two terms $\hat{a}^{\dagger}\hat{b}^{2}$ and $\hat{b}$ off-resonantly. In particular, we use a pump $\epsilon_{p}(t) = \epsilon_{p}\cos(\omega_{p}t)$ with frequency $\omega_{p} = 2\omega_b -\omega_a+\Delta$ to off-resonantly drive the term $\hat{a}\hat{b}^{\dagger 2}$ and directly drive the $\hat{b}$ mode at frequency $\omega_{d} = \omega_{b} + \Delta$ to produce the term
\begin{align}
    \hat{H}_{d} &= \epsilon_{d}(\hat{b}^{\dagger}e^{-i\omega_{d}t} + \textrm{h.c.}).
\end{align}
Then the complete Hamiltonian in the rotating frame of all of the modes is 
\begin{align}
    \hat{H}_{\textrm{rot}} &=\frac{1}{2} E_{J}\epsilon_{p}\varphi_{a}\varphi_{b}^2\hat{a}^{\dagger}\hat{b}^{2}e^{i\Delta t}  + \textrm{h.c.}
    \nonumber\\
    &\quad +\epsilon_{d}\hat{b}^{\dagger}e^{-i\Delta t} + \textrm{h.c.} + \hat{H}'
\end{align}
where $\hat{H}'$ contains rapidly rotating terms. Now let $\chi_{1} \equiv E_{J}\epsilon_{p}\varphi_{a}\varphi_{b}^2/2$. Then, neglecting $\hat{H}'$ and constants, average Hamiltonian theory yields \cite{Gamel2010timeaveraged, james2007}
\begin{align}
    \hat{H}_{\textrm{eff}} &= \frac{1}{\Delta}[\chi_1 \hat{a}^\dagger \hat{b}^2 + \epsilon_d \hat{b}, \chi_1 \hat{a}\hat{b}^{\dagger 2} + \epsilon_d \hat{b}^\dagger]
    \nonumber\\
    &= \frac{1}{\Delta}\Big{[} 
    \chi_1^2 [2(1+2\hat{b}^\dagger \hat{b})\hat{a}^\dagger\hat{a} -\hat{b}^{\dagger 2}\hat{b}^2] + 2\chi_1\epsilon_d (\hat{a}^\dagger \hat{b} + \hat{a}\hat{b}^\dagger) \Big{]}
    \nonumber\\
    &\xrightarrow{ \hat{b}^{\dagger}\hat{b} \ll 1 } g (\hat{a}^\dagger \hat{b} + \hat{a}\hat{b}^\dagger) + g_b \hat{a}^\dagger\hat{a}.
\end{align}
The coupling constant is given by $g = \EJ\epsilon_p\beta\varphi_a\varphi_b^2$ where $\beta = \epsilon_d/\Delta$ and there is an energy shift.  The strength of the coupling is on the order of $g_2$ since it depends twice on $\varphi_b > \varphi_a$ and $ \beta<1$ to ensure $\hat{b}^\dagger\hat{b}<1$.  
\begin{figure*}
    \centering
    \subfloat[Error vs. Time]{\includegraphics[width=0.48\textwidth]{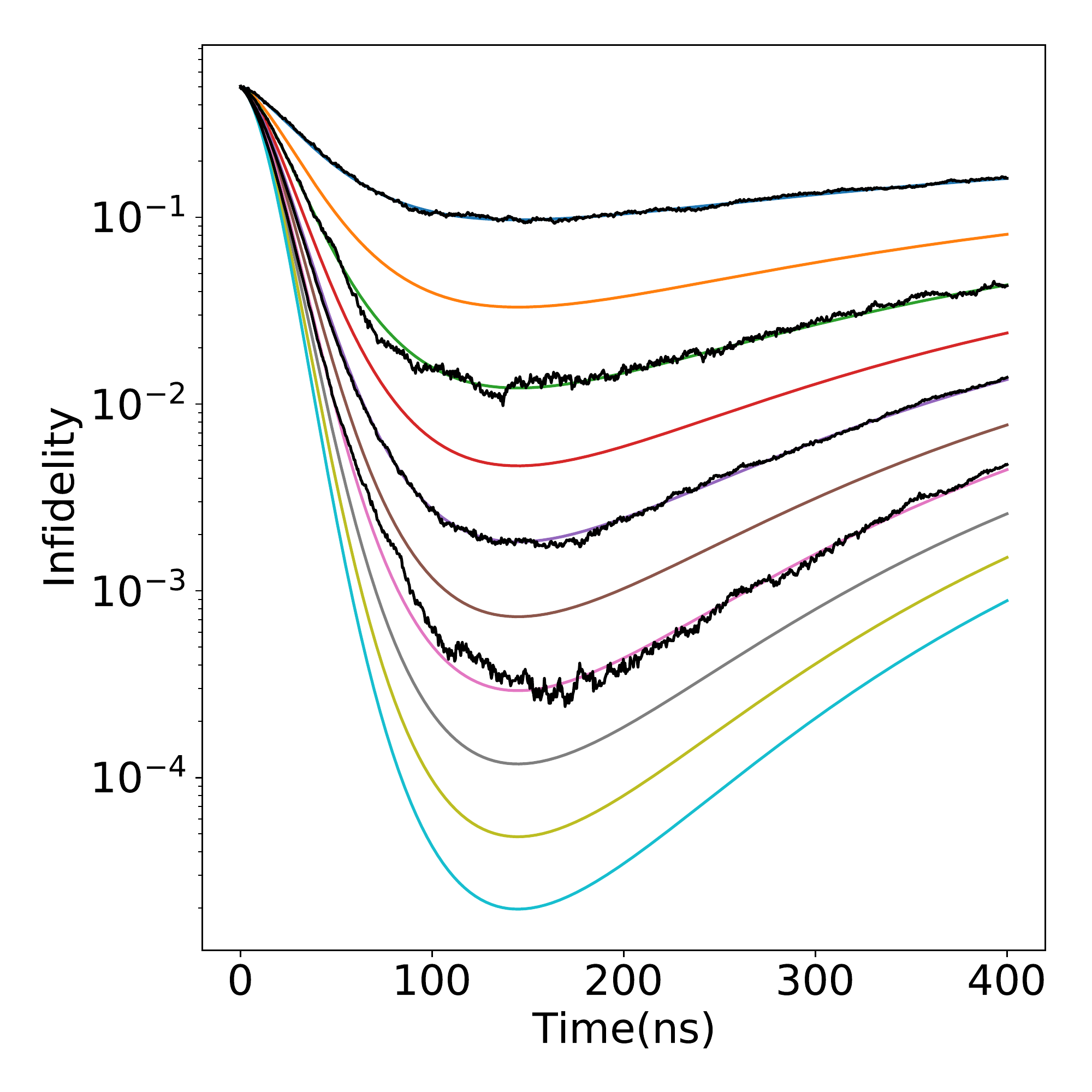}}
    \subfloat[Error Fit]{\includegraphics[width=0.48\textwidth]{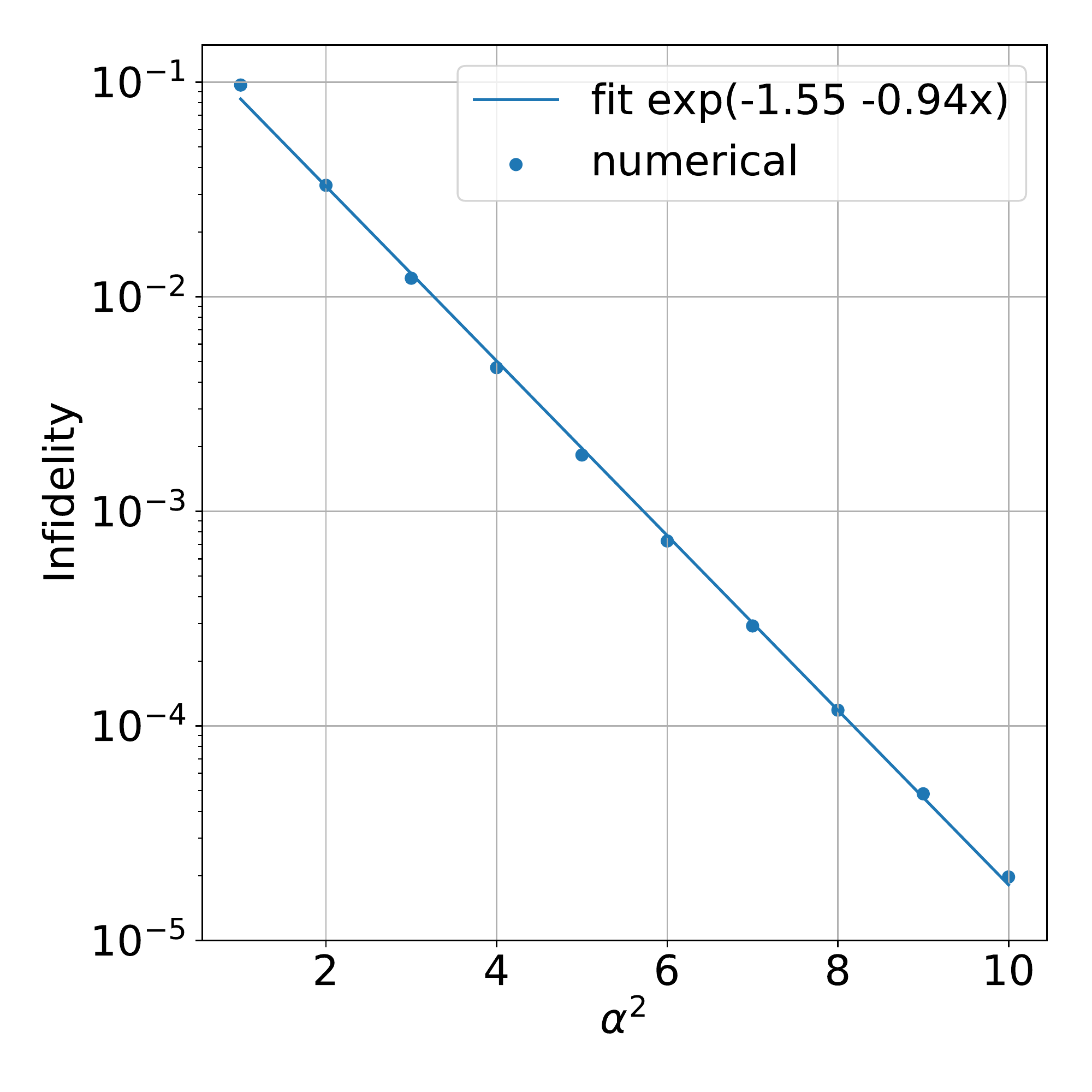}}
    \caption{a.) Measurement error probability for $Z$-basis readout as a function of time.  The colored lines correspond to the analytic formula for the separation error \cref{eq:snrtofidelityZ} for $\alpha^2 = 1,2,...,10$.  The black lines correspond to infidelities from simulations of the corresponding stochastic master equation \cref{eq:stochastic_master_equation_evolution} (QuTiP) in the interaction picture for a few thousand trajectories with the initial condition $|\alpha\rangle$ for the cases $\alpha^2 = 1, \alpha^2 = 3, \alpha^2 = 5$, and $\alpha^2 = 7$.  The simulated curved which include $\kappa_1/2\pi=1\ \mathrm{KHz}$ and analytic curves agree well indicating the small effect of the additional single phonon loss.  In order to get concrete time numbers the simulations use $1/\kappa_2 = 100$ ns but as discussed this can be scaled to any $\kappa_2$.  b.) Plot of the minimum infidelities vs. $\alpha^2$ and the fit line.}
	\label{fig:zMeasurementError}
\end{figure*}

Next we proceed by finding the measurement SNR for this readout scheme.  The coupled Langevin equations governing the evolution of the storage and buffer modes in the interaction picture are
\begin{align}
    \dot{\hat{a}} &= -i[\hat{a}, \hat{H}_r]=-i g_{r} \hat{b},
    \nonumber\\
    \dot{\hat{b}} &= -i[\hat{b}, \hat{H}_r] - \frac{\kappab}{2}\hat{b} - \sqrt{\kappa_{b}}\hat{b}_{in}=-i g_{r} \hat{a }- \frac{\kappab}{2}\hat{b} - \sqrt{\kappa_{b}}\hat{b}_{in}
\end{align}
Here $\kappab$ is the single photon loss rate of the buffer mode and we have neglected the single phonon loss rate of the storage mode under the assumption that it is far slower than the readout timescale.  These equations can be straightforwardly integrated \cite{Pfaff2017Catapult} to give
\begin{align}
    \hat{a}(t) &= \frac{\hat{a}(0)}{\beta} e^{-\frac{\kappa_b t}{4}} (\beta \cosh{\frac{\beta t}{4}}+\kappa_b \sinh{\frac{\beta t}{4}}),
    \nonumber\\
    \hat{b}(t) &= -i \frac{4 g \hat{a}(0)}{\beta}e^{-\kappa_b t/4} \sinh{\frac{\beta t}{4}}
\end{align}
where $\beta=\sqrt{\kappa_b^2 -(4 g_r)^2}$.  Here we have not included the mean zero terms with $\hat{b}_\text{in}$ since they are not relevant for computing the signal.  
The measurement operator with a uniform readout window is defined to be \cite{ClerkNoiseMeasurement, DidierLongitudinal}
\begin{align}
    \hat{M}(\tau) = \sqrt{\kappa_b} \int_{0}^{\tau} dt[\hat{b}_{out} ^ {\dagger} (t)e^{i \phi_h} + \hat{b}_{out} (t)e^{-i \phi_h}]
\end{align}
Using the input-output boundary condition that $\hat{b}_{out} = \hat{b}_{in} + \sqrt{\kappab}\hat{b}$. We can determine the average of the measurement operator to be
\begin{align}
    \langle\hat{M}(\tau)\rangle &=
    \nonumber\\ 
    &\frac{2\kappab\langle\hat{a}(0)\rangle\sin{\phi_h}}{g_r} * \nonumber \\
    &\left[1-e^{-\kappa_b \tau/4}\left[\cosh{\frac{\beta \tau}{4}}+\frac{\kappa_b}{\beta}\sinh{\frac{\beta \tau}{4}}\right]\right].
\end{align}
We have also taken the input to be the vacuum with the property $\langle \hat{b}_{in} (t') \hat{b}_{in}^\dagger (t)\rangle = \delta(t-t')$.  There is a weak drive on $\hat{b}$ to realize the Hamiltonian which we neglect and could be replaced with a flux pump.  From the average of the measurement signal $\langle\hat{M}(\tau)\rangle$ we can determine the measurement SNR using
\begin{align}
    \text{SNR}^2 = \frac{|\langle \hat{M} \rangle _\alpha - \langle \hat{M} \rangle _{-\alpha}|^2}{\langle \hat{M}_{N(\alpha)}^2 \rangle + \langle \hat{M}_{N(-\alpha)}^2 \rangle}.
\end{align}
We take the noise terms to be given by \begin{align}
    \langle \hat{M}_{N(\pm \alpha)}^2 \rangle=\langle (\hat{M}(\tau)-\langle\hat{M}(\tau) \rangle_{\pm \alpha}) \rangle = \kappab \tau
\end{align}  
which we have checked in numerics.

Solving for the SNR and optimizing the phase we get 
\begin{align}
    \text{SNR}_\alpha(\tau) &= \nonumber \\
    &\alpha \sqrt{8\kappab} \frac{\left[ 1-e^{-\kappa_b \tau/4} \left[\cosh{\frac{\beta \tau}{4}}+\frac{\kappa_b}{\beta}\sinh{\frac{\beta \tau}{4}} \right] \right]}{g\sqrt{\tau}}.
\end{align}

As is expected since this readout scheme is not QND and does not preserve the state of the cavity at long times the readout SNR goes as $1/\sqrt{\tau}$ as we are only integrating noise.  The readout separation error, which is the dominant source of error for this readout scheme, will be given by 
\begin{align}
    \epsilon_{\text{sep},\alpha}(\tau) = \frac{1}{2} \text{Erfc}(\frac{\text{SNR}_\alpha(\tau)}{2}).
    \label{eq:snrtofidelityZ}
\end{align}
From these equations we can then determine the fidelity as a function of time for different alpha for our measurement scheme as is shown in the colored line in \cref{fig:zMeasurementError} a.).  In \cref{fig:zMeasurementError} b.) we fit the optimal readout error to an exponential decay.  We used a more conservative relation $\epsilon = e^{-1.5-0.9|\alpha|^2}$ for the error correction simulations.

\begin{figure*}
	\centering
	\includegraphics[width=0.75\textwidth]{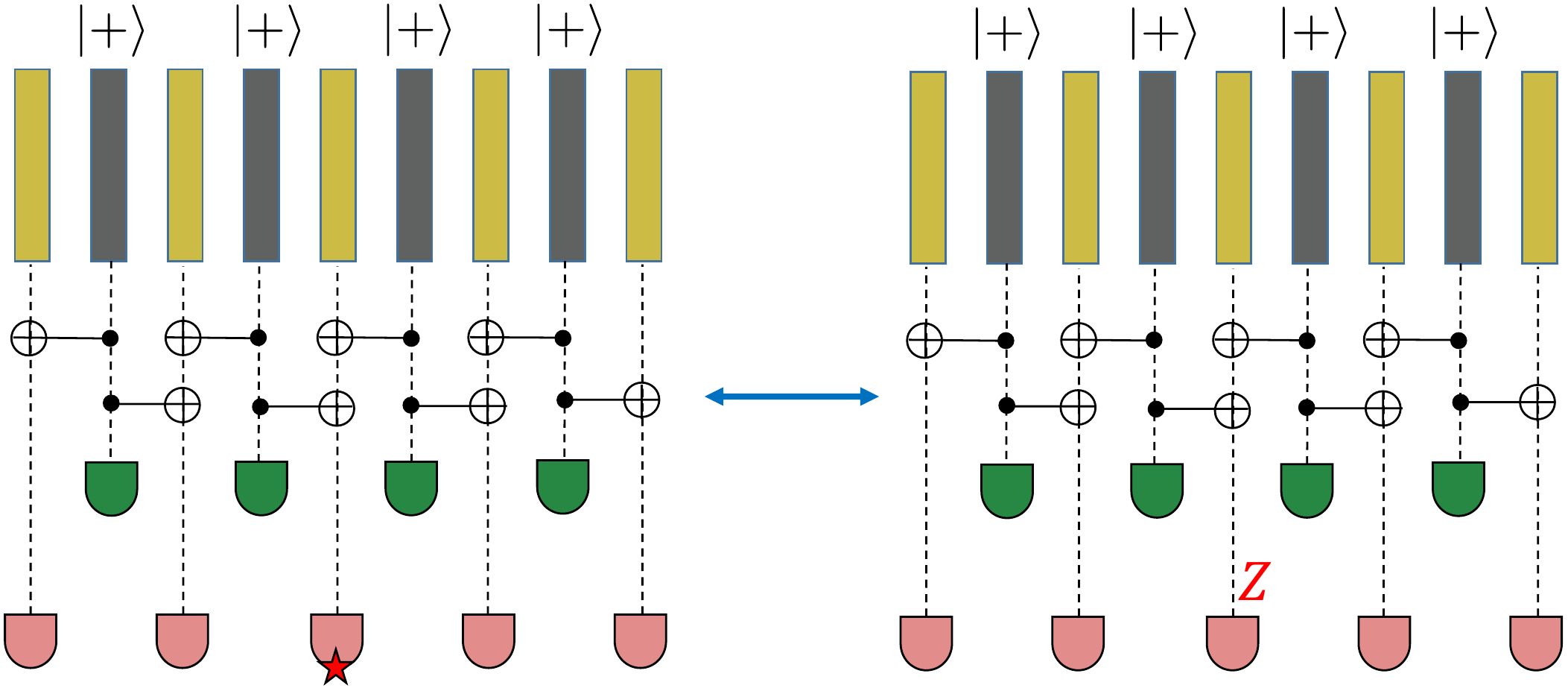}
	\caption{Distance-5 repetition code, where one round of stabilizer measurements is performed (green measurements) followed by a direct measurement of the data qubits (red measurements). The data qubits are the yellow rectangles, and the ancilla qubits (prepared in $\ket{+}$) are the grey rectangles. On the left, a measurement error on the third data qubit occurs during the direct measurement of the data, which is equivalent to having a $Z$ data qubit error immediately before the measurement (shown on the right). Such settings illustrate the importance of the \texttt{STOP} algorithm, where one might have to correct errors prior to applying a non-Clifford gate, and a round of perfect error correction (which in practice is achieved by directly measuring the data) cannot be performed. In such settings, a single measurement error during the last round of stabilizer measurements (green measurements in the figure) can lead to a logical failure if the syndrome measurement is repeated a fixed number of times (say $d$) rather than using the \texttt{STOP} algorithm.}
	\label{fig:MeasDataRepEquiv}
\end{figure*}

We have simulated trajectories for this measurement procedure using a stochastic master equation for confirmation.  The stochastic master equation simulations include $\kappa_1/2\pi = 1$ KHz.  The integrated and classified and measurement results from the stochastic master equation compared to the analytic expression are pictured in \cref{fig:zMeasurementError} a.).  The stochastic master equation simulations were performed with the evolution
\begin{align}
    \frac{d\hat{\rho}(t)}{dt} =& -i[g_r (\hat{a}^\dagger \hat{b} + \hat{b}^\dagger \hat{a}, \hat{\rho}(t)]  \nonumber \\ 
    &+\kappa_b (1-\eta) D[\hat{b}]\hat{\rho}(t) + \kappa_b \eta D[\hat{b}]\hat{\rho}(t) \nonumber \\
    &+\kappa_1 D[\hat{a}]\hat{\rho}(t) + n_{\mathrm{th}}\kappa_1 D[\hat{a}^\dagger]\hat{\rho}(t) + \kappa_\phi D[\hat{a}^\dagger \hat{a}]\hat{\rho}(t)
    \label{eq:stochastic_master_equation_evolution}
\end{align}
where the second term on the second line is the detection part of the master equation that is unraveled.  We see exponential suppression of the error probability with increasing $\alpha^2$.  In the future we expect to be able to improve the performance by optimizing the window function for the readout and using the confidence of the measurement result to feed back and improve the matching.  These advances in addition to the robustness of the EC to larger measurement errors than those currently assumed would allow us to make looser assumptions.

\hp{\textit{Assumptions--} The chosen parameters for this work are $\kappa_b/2\pi = 20 * ((100\ \mathrm{ ns})*\kappa_2)$ MHz and $g/2\pi = 4 * ((100\ \mathrm{ ns})*\kappa_2)$ MHz.  These rates depend on $\kappa_2$ so the readout time will scale as $1/\kappa_2$.  We also include the effect of a non-unity quantum efficiency $\eta = .5$ ($\eta$ is the proportion of the signal detected).  For $\kappa_2=2\pi*280e3$ used in the paper the optimal duration of the measurement is roughly $850\ ns$.  Note that $\kappa_b$ can be made larger with the main effect of lengthening the readout time. }

\section{Alternative architecture with ATS-based basis readout}
\label{App:Alternative}

\begin{figure}
    \centering
    \includegraphics[width=\columnwidth]{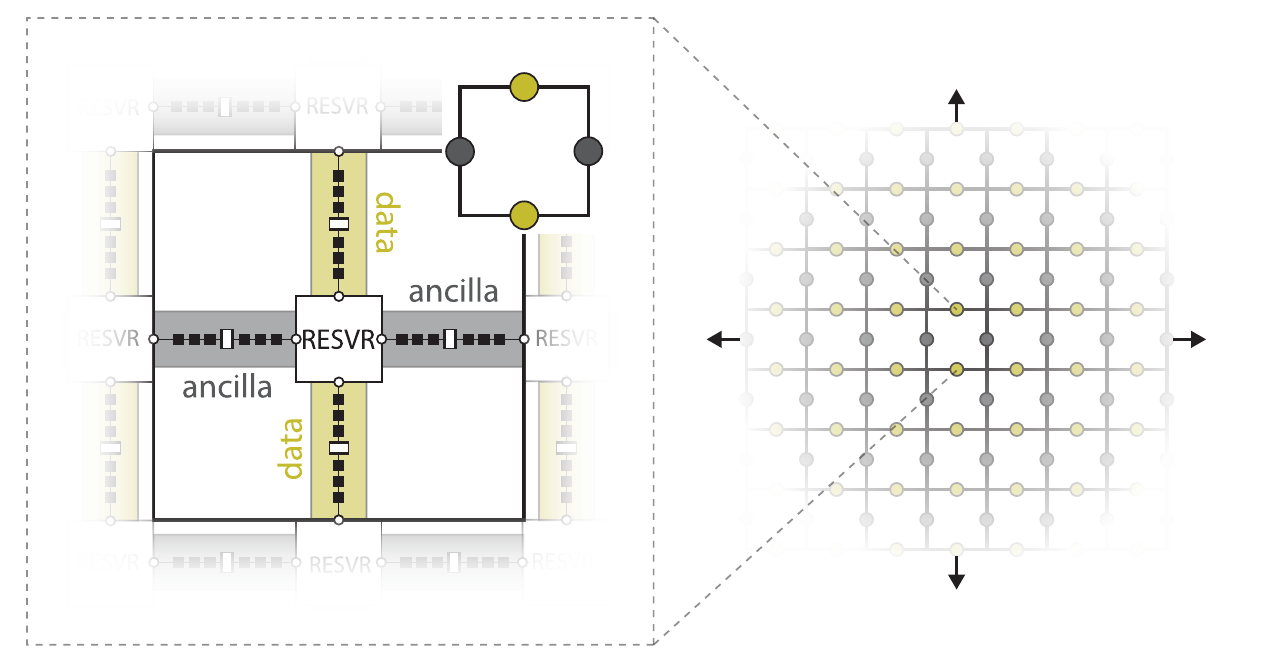}
    \caption{Alternative architecture. The $X$-basis readout is performed directly using the ATS [contained within the reservoir (RESVR)], as opposed to using an ancillary readout mode and transmon qubit, cf.~\Cref{fig:hardware_cartoon}. By eliminating the readout mode, the number of modes per unit cell is reduced from five to four. }
    \label{fig:alternative_architecture}
\end{figure}

\ch{In this appendix, we describe an alternative version of our architecture where the $X$-basis readout is performed directly using the ATS. 
As discussed below, this modification favorably impacts practicality and performance across several different levels of our architecture.  
For example, this modification obviates the need for a transmon in each unit cell, and allows us to reduce the number of modes coupled to each reservoir from five to four, see \Cref{fig:alternative_architecture}. 
As a result, crosstalk is reduced, easing limitations on logical lifetimes posed by correlated errors. 
Though such benefits are highly desirable, the ATS-based X-basis readout that underlies this alternative architecture may be difficult to perform in practice. 
Because the practical feasibility of the readout scheme is more speculative, we do not use the ATS-based readout scheme in any of the analyses in the main text. 
Instead, we describe the ATS-based readout scheme and enumerate its favorable impacts here in this appendix. The ATS-based readout scheme is presented and analyzed in \Cref{sec:modified_X_readout}, while \Cref{sec:modified_X_benefits} catalogs the beneficial impacts of this modification across different levels of the architecture. }

\subsection{ATS-based {$X$}-basis measurement scheme}
\label{sec:modified_X_readout}
\ch{
Here we present a modified $X$-basis readout scheme that can be implemented using only the ATS and the buffer mode. The readout scheme is implemented by engineering a coupling Hamiltonian of the form
\begin{align}
    \label{eq:modified_X_Ham}
    \hat{H}_{r} = i g_{r} \hat{a}^{\dagger} \hat{a} (\hat{b}^{\dagger}-\hat{b}).
\end{align}
Here $\hat{a}$ is the annihilation operator for a storage mode and $\hat{b}$ is the annihilation operator for the buffer mode. Note that this Hamiltonian is equivalent to the longitudinal readout discussed with transmons \cite{DidierLongitudinal}. 
Practically speaking, engineering this Hamiltonian may be challenging because it rotates at the frequency of the buffer mode.  That is, we cannot engineer this Hamiltonian simply by pumping the system at the resonance frequency of the buffer mode, because the same pump would also bring corresponding Hamiltonians for other storage modes on resonance simultaneously (e.g., $\hat{a}_2^\dagger \hat{a}_2 (\hat{b} + \hat{b}^\dagger)$). } 

\ch{To circumvent this frequency-selectivity problem, we sketch how a Hamiltonian of the form (\ref{eq:modified_X_Ham}) could be engineered via off-resonant pumping.  Consider the following Hamiltonian (in the rotating frame),
\begin{align}
    \label{eq:modified_X_interaction}
    \hat{H} &= (g_a \hat{a} \hat{b}^{\dagger 2} + i g_b \hat{a} \hat{b}^{\dagger}) e^{-i\Delta t} +h.c.
\end{align}
This Hamiltonian can be realized in a number of ways. For example, the $\hat{a} {{\hat{b}}^{\dagger 2}}$ term can be realized by pumping the ATS at frequency $2 \omega_b - \omega_a + \Delta$, while the $\hat{a} \hat{b}^{\dagger}$ term can be realized by synthesizing two odd terms (as described in the vicinity of \Cref{eq:BeamSplitterDerivationStart}). For simplicity, we assume access to Hamiltonians of the form (\ref{eq:modified_X_interaction}) and leave the precise details of how they are to be engineered to future work.}
\ch{For $g_{a,b}\ll \Delta$, evolution generated by the Hamiltonian (\ref{eq:modified_X_interaction}) is well described by the corresponding time-averaged effective Hamiltonian~\cite{Gamel2010timeaveraged},
\begin{align}
    \label{eq:modified_X_eff}
    \hat{H}_{\text{eff}} 
    &= \frac{g_a^2}{\Delta} [\hat{a}^\dagger\hat{b}^2,\hat{a} {{\hat{b}}^{\dagger 2}}]+ \frac{g_b^2}{\Delta}[\hat{a}^\dagger \hat{b},\hat{a}\hat{b}^\dagger]+\nonumber \\ 
    &\frac{i g_a g_b}{\Delta} ([\hat{a}^\dagger\hat{b}^2,\hat{a}\hat{b}^\dagger]-
    [\hat{a}^\dagger \hat{b}, \hat{a} {{\hat{b}}^{\dagger 2}}]) \nonumber \\
    &= \frac{g_a^2}{\Delta} (2\hat{a}^\dagger\hat{a}(1+2\hat{b}^\dagger\hat{b}) - {\hat{b}^{\dagger 2}}\hat{b}^2)+
    \frac{g_b^2}{\Delta} (\hat{a}^\dagger\hat{a} -\hat{b}^\dagger\hat{b})+\nonumber\\
    &\frac{i g_a g_b}{\Delta}(2 \hat{a}^\dagger \hat{a} - \hat{b}^\dagger \hat{b})(\hat{b}-\hat{b}^\dagger)
    \nonumber\\
    &\xrightarrow{ \left<\hat{b}^{\dagger}\hat{b}\right> \ll 1 } \frac{2 i g_a g_b}{\Delta} \hat{a}^\dagger \hat{a} (\hat{b}-\hat{b}^\dagger) + \frac{2 g_a^2 + g_b^2}{\Delta}\hat{a}^\dagger\hat{a}
\end{align}
where in the last line we have neglected many of the terms due to the small occupation of the buffer mode.  }

\ch{To implement an $X$-basis readout using this Hamiltonian, we first deflate the storage mode.  As described in \Cref{appendix:Measurement}, this deflation is achieved by abruptly setting the two-phonon dissipator for the storage mode equal to $\mathcal{D}[\hat{a}^2]$, as opposed to $\mathcal{D}[\hat{a}^2-\alpha^22]$, and waiting for a timescale comparable to $1/\kappa_2$. When the storage mode begins in an even (odd) parity state, this deflation procedure maps it to the $\ket{0}$ ($\ket{1}$) phonon Fock state. Subsequent to this deflation, we perform a homodyne measurement of the buffer mode while the system evolves under the  Hamiltonian (\ref{eq:modified_X_eff}).  In effect, this Hamiltonian drives the buffer mode conditioned on whether the parity of the storage mode was initially even or odd.} 

\ch{We compute the fidelity of this homodyne readout, where we aim to distinguish $|\hat{n}_a=0\rangle$ from $|\hat{n}_a=1\rangle$. We closely follow the derivation of SNR in \cite{DidierLongitudinal}.  The Langevin equation for the evolution of the buffer mode in the interaction picture is
\begin{align}
    \dot{\hat{b}} &= -i[\hat{b}, \hat{H}_{\text{eff}}] - \frac{\kappab}{2}\hat b - \sqrt{\kappa_{b}}\hat{b}_{in}\nonumber \\
    &= g_{r} \hat{a}^\dagger \hat{a} - \frac{\kappab}{2}\hat{b} - \sqrt{\kappa_{b}}\hat{b}_{in}
\end{align}
where $\hat{b}_{in}$ is the input field and $g_r \equiv 2 g_a g_b/\Delta$.  In the following calculations we will neglect the single phonon loss of the ancilla mode which in this simple case will add an average readout error probability of roughly $\kappa_a t/{4}$.  We integrate this equation to get the expected value of the buffer mode
\begin{align}
    \langle \hat{b}(t) \rangle = \frac{2 g_{r}}{\kappa_b} \langle\hat{a}^\dagger \hat{a}\rangle (1-e^{-\frac{\kappab t}{2}}).
\end{align}
The measurement operator for integration up to time $\tau$ and with homodyne angle $\phi_h$ is defined as \cite{ClerkNoiseMeasurement, DidierLongitudinal}
\begin{align}
    \hat{M}(\tau) = \sqrt{\kappa_b} \int_{0}^{\tau} dt[\hat{b}_{out} ^ {\dagger} (t)e^{i \phi_h} + \hat{b}_{out} (t)e^{-i \phi_h}].
\end{align}
Evaluating the average of this integral with the optimal phase gives
\begin{align}
    \langle\hat{M}(t)\rangle = \frac{4 g_r \langle\hat{a}^\dagger\hat{a}\rangle}{\kappab} (-2 + 2 e^{-\frac{\kappab t}{2}}+\kappab t).
\end{align}
Here we have used the standard input-output condition that $\hat{b}_{out} = \hat{b}_{in}+\sqrt{\kappab}\hat{b}$ and the conditions on $\hat{b}_{in}$ that $\langle\hat{b}_{in}\rangle=0$ and $\langle\hat{b}_{in}(t)\hat{b}_{in}^\dagger(t')\rangle = \delta(t-t')$.  There is a drive on $\hat{b}$ to realize the Hamiltonian which we neglect and could be replaced by an appropriate pump.  Next we compute the SNR which is defined as
\begin{align}
    \text{SNR}^2 = \frac{|\langle \hat{M} \rangle _1 - \langle \hat{M} \rangle _0|^2}{\langle \hat{M}_{N(1)}^2\rangle+\langle \hat{M}_{N(0)}^2\rangle}.
\end{align}
where $\hat{M}_{N(x)} = \hat{M} - \langle\hat{M}\rangle_{x}$ so  $\langle\hat{M}_{N(0)}^2\rangle=\langle\hat{M}_{N(1)}^2\rangle=\kappab t$. Thus the SNR is
\begin{align}
    \text{SNR}(\tau) = \frac{4 g_r}{\kappab \sqrt{2\kappab t}} (-2 + 2 e^{-\frac{\kappab t}{2}}+\kappab t).
\end{align}
The separation error for this readout, which will be in addition to the effect of the single phonon loss mentioned earlier, will be given by \cite{krantz2019quantum}
\begin{align}
    \epsilon_{\text{sep}}(\tau) = \frac{1}{2} \text{Erfc}(\frac{\text{SNR}(\tau)}{2}).
\end{align}
There will also be an additional contribution to the readout error associated with the deflation procedure (see \Cref{appendix:Measurement})}

\subsection{Benefits of ATS-based X-basis readout}
\label{sec:modified_X_benefits}
\ch{
The analysis of the previous section demonstrates that, in principle, high-fidelity $X$-basis readout can be performed directly using the ATS. We now show how this readout scheme can be exploited to improve the practicality and performance of our architecture. The following analysis demonstrates that improved $X$-basis readout represents one potential path to improved future designs. At the same time, it also illustrates a more general point: improvements at low levels of the architecture can propagate into significant savings at higher levels. These findings thus underscore the importance and utility of comprehensive architectural analyses. }

\ch{At the hardware level, the modified $X$-basis readout scheme enables us to simplify the unit cell of our architecture. 
Specifically, in comparison to the unit cell described in the main text (\Cref{fig:hardware_cartoon}), the unit cell of \Cref{fig:alternative_architecture} has no transmon and has four phononic modes instead of five. Removing the transmon reduces the number of control lines and is helpful for device layout because the transmon requires significant additional space in comparison to the phononic modes}

\ch{Reducing the number of modes per unit cell from five to four significantly reduces crosstalk within each unit cell. We can  quantify this reduction following the approach described in \Cref{sec:multimode_stabilization}. To briefly summarize: we optimize the frequencies of the storage modes within each unit cell in order to minimize the effects of crosstalk. More specifically, the frequencies are chosen so that the effects of coherent \emph{crosstalk} errors are minimized subject to the constraint that all \emph{incoherent} crosstalk errors must be exponentially suppressed by the filter. The impacts of the residual coherent errors are quantified via two parameters, $p_{\text{double}}$ and $p_{\text{triple}}$, that respectively describe the probabilities that either two data qubits, or two qubits and an ancilla qubit, suffer correlated $Z$ errors. }

\ch{In \Cref{tab:frequency_opt_four_mode}, we compare the results of this mode-frequency optimization for four- and five-mode unit cells. There are two important takeaways from these optimization results. The first takeaway is simply that the magnitude of the residual coherent errors is substantially reduced. Indeed, the total correlated error   probability, $p_{\text{double}} + p_{\text{triple}}$, is over an order of magnitude smaller for the four-mode case in than the five-mode case. As we discuss below, this reduction can significantly impact logical error rates. 
}

\begin{table*}[t]
  \centering
  \begin{tabular}{ |c|c|c|c|c|c| } 
    \hline
    \# modes & 
    $4J$  &
    $\omega_\alpha, \omega_\beta,\omega_\gamma,\omega_\delta, \omega_\rho$  &
    $p_\text{double}$  & $p_\text{triple}$ &
    $p_\text{double}+p_\text{triple}$
    \\\hline
    4 & 180 & 0, 1000, 798, 101, - &
    $1.22* 10^{-9}\left[\frac{|\alpha|^2 g_2}{2\pi\text{MHz}}\right]^4$
    & $3.87* 10^{-10}\left[\frac{|\alpha|^2 g_2}{2\pi\text{MHz}}\right]^4$
    & $1.60* 10^{-9} \left[\frac{|\alpha|^2 g_2}{2\pi\text{MHz}}\right]^4$ \\\hline
    5 & 100 & 0, 1000, 242, 879, 61 & 
    $1.83*10^{-8}\left[\frac{|\alpha|^2 g_2}{2\pi\text{MHz}}\right]^4$
    & 
    $5.20*10^{-10}\left[\frac{|\alpha|^2 g_2}{2\pi\text{MHz}}\right]^4$
    & $ 1.88*10^{-8} \left[\frac{|\alpha|^2 g_2}{2\pi\text{MHz}}\right]^4$ \\\hline
  \end{tabular}
  \caption{Frequency optimization results. The parameters $4J$ and $\omega$ are given in units of $2\pi \times $ MHz.  The correlated error probabilities in the last three columns are expressed in terms of $\alpha$ and $g_2$. For realistic choices of $|\alpha| = \sqrt{8}$ and $g_2/2\pi = 2$ MHz, the probabilities evaluate to $C = 1.05* 10^{-4}$ and $C = 1.23* 10^{-3}$ for the four- and five-mode configurations respectively. See \Cref{sec:multimode_stabilization} for further details.}
  \label{tab:frequency_opt_four_mode}
\end{table*}

 \ch{The second important takeaway from \Cref{tab:frequency_opt_four_mode} is that the four-mode unit cell can accommodate a larger filter bandwidth, which ultimately eases requirements on the lifetimes of the phononic resonators to achieve a desired value of $\kappa_1/\kappa_2$. As described in \Cref{sec:multimode_stabilization}, the constraint that all incoherent crosstalk errors be suppressed by the filter places an upper bound on the filter bandwidth, and this upper bound decreases as the number of modes in the unit cell increases. Reducing the number of modes from five to four thus enables us to increase the filter bandwidth. 
 In particular, we find that the bandwidth can be increased by nearly a factor of 2 (from $4J= 2\pi * 100$ MHz to $4J= 2\pi * 180$ MHz). As described in \Cref{sec:HardwareImplementation}, the filter bandwidth constrains the achievable value of $\kappa_2$, meaning that an increased bandwidth enables an increased $\kappa_2$. Equivalently, an increased filter bandwidth eases the coherence requirement on the phononic resonators to achieve a desired value of $\kappa_1/\kappa_2$. From \Cref{fig:loss_results_main_text}, we see that the factor of $\sim 2$ increase in bandwidth correspondingly lowers the $T_1$ requirement on the phononic resonators by roughly a factor of 2. }

\begin{figure}[th]
	\centering
	\includegraphics[width=0.48\textwidth]{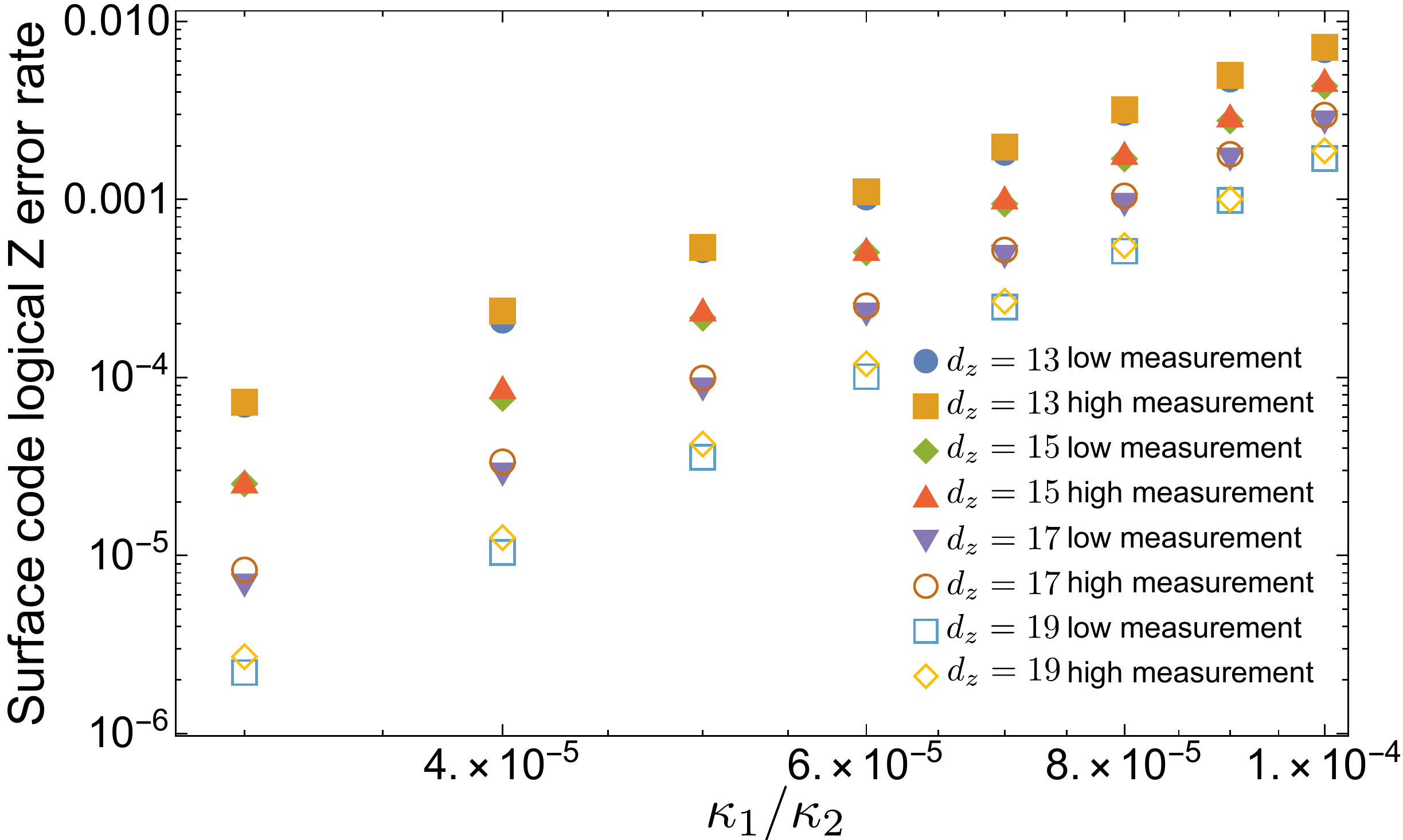}
	\caption{
	Logical $Z$ error rates of the thin rotated surface code with five modes coupled to an ATS (which includes a transmon qubit and an additional readout mode in each unit cell and corresponds to the data in \cref{fig:SurfaceCodeMem}) and four modes coupled to an ATS (which excludes the transmon qubit and performs direct $X$-basis measurements). We labeled measurements with the transmon qubit as \textit{low measurement} and the direct $X$-basis measurement as \textit{high measurement}. For the direct $X$-basis measurement, the measurement error rate is fixed at $2 * 10^{-3}$ for all values of $\kappa_1 / \kappa_2$. Measurement error rates with the transmon qubit were obtained from \cref{tab:XMeasurementTimeAndErrorRate} with five parity measurements.  }
	\label{fig:CompareHighLowMeasSurf_app}
\end{figure}

 \ch{Having described the effects of the ATS-based $X$-basis readout scheme at the hardware level, we now consider effects at the logical level. We begin by quantifying the impact of the increased measurement error rate of the ATS-based readout scheme (see \Cref{tab:XMeasurementTimeAndErrorRate}) on the logical error rates. 
In \cref{fig:CompareHighLowMeasSurf_app}, we plot the logical $Z$ error rate of the thin rotated surface code, both for the main-text architecture of (five-mode unit cell, transmon-based readout) and for the alternative architecture of this appendix (four-mode unit cell, ATS-based readout). For the former architecture, we used the same data as shown in \cref{fig:SurfaceCodeMem}, while for the latter architecture (labeled \textit{high measurement}), we fixed the measurement error rate to $2* 10^{-3}$. As can be seen, even though the measurement error rate can be more than an order of magnitude larger for the ATS-based readout, the logical $Z$ error rates increase by a small amount in the low $\kappa_1 / \kappa_2$ regime. The reason the logical failure rate is not greatly affected by the large increase in measurement failure rates is that CNOT failures are the dominant source of noise. As such, we do not expect the overhead results of \cref{sec:Overhead} to increase when using the alternative architecture of \Cref{fig:alternative_architecture} because the same code distances can be used for implementing the algorithms of interest.}

\begin{figure}[th]
	\centering
	\subfloat[\label{fig:CorrelatedNoiseLogZd11_app}]{%
		\includegraphics[width=0.48\textwidth]{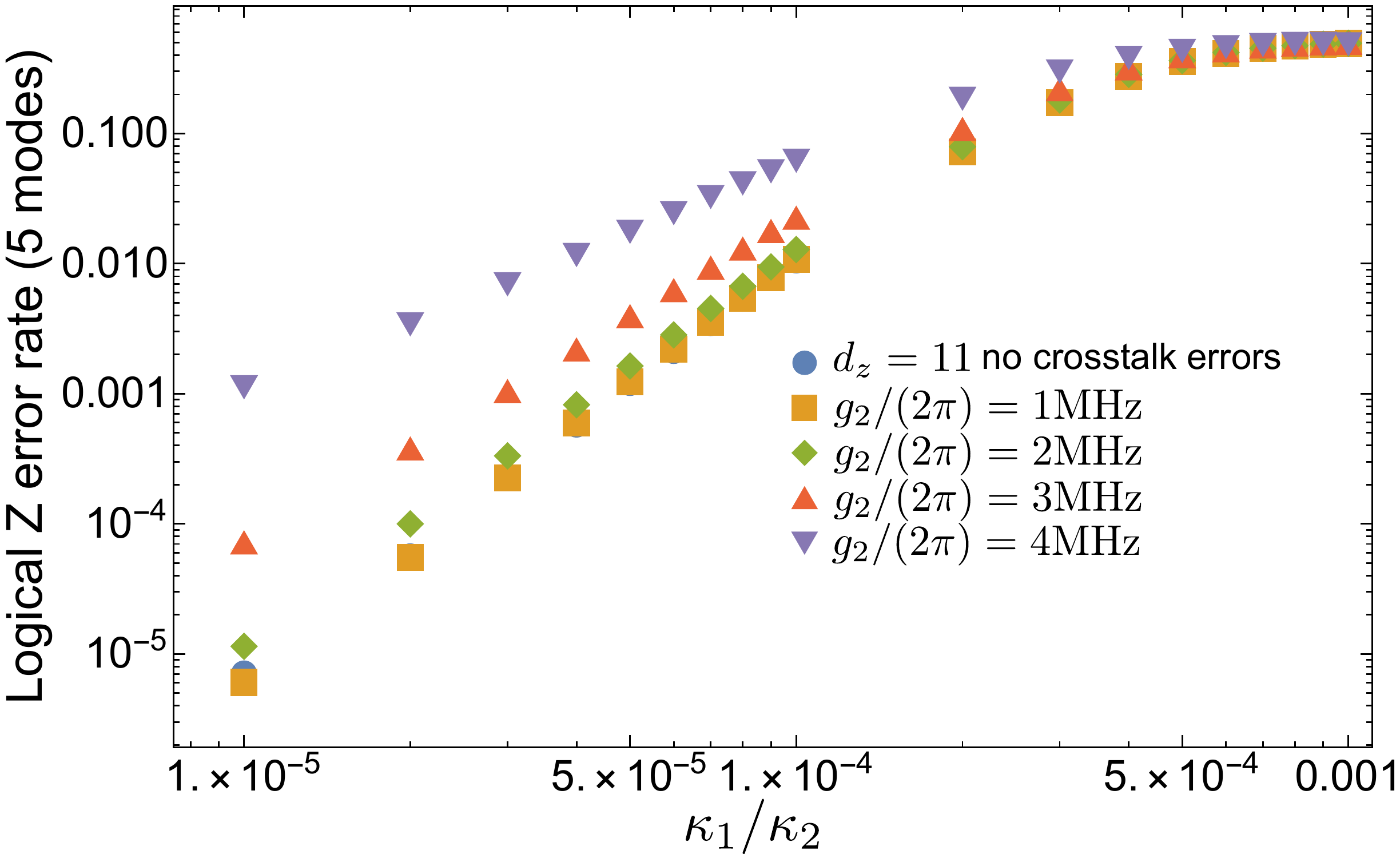}
	}
	\vfill
	\subfloat[\label{fig:CorrelatedNoiseLogZd114Modes_app}]{%
		\includegraphics[width=0.48\textwidth]{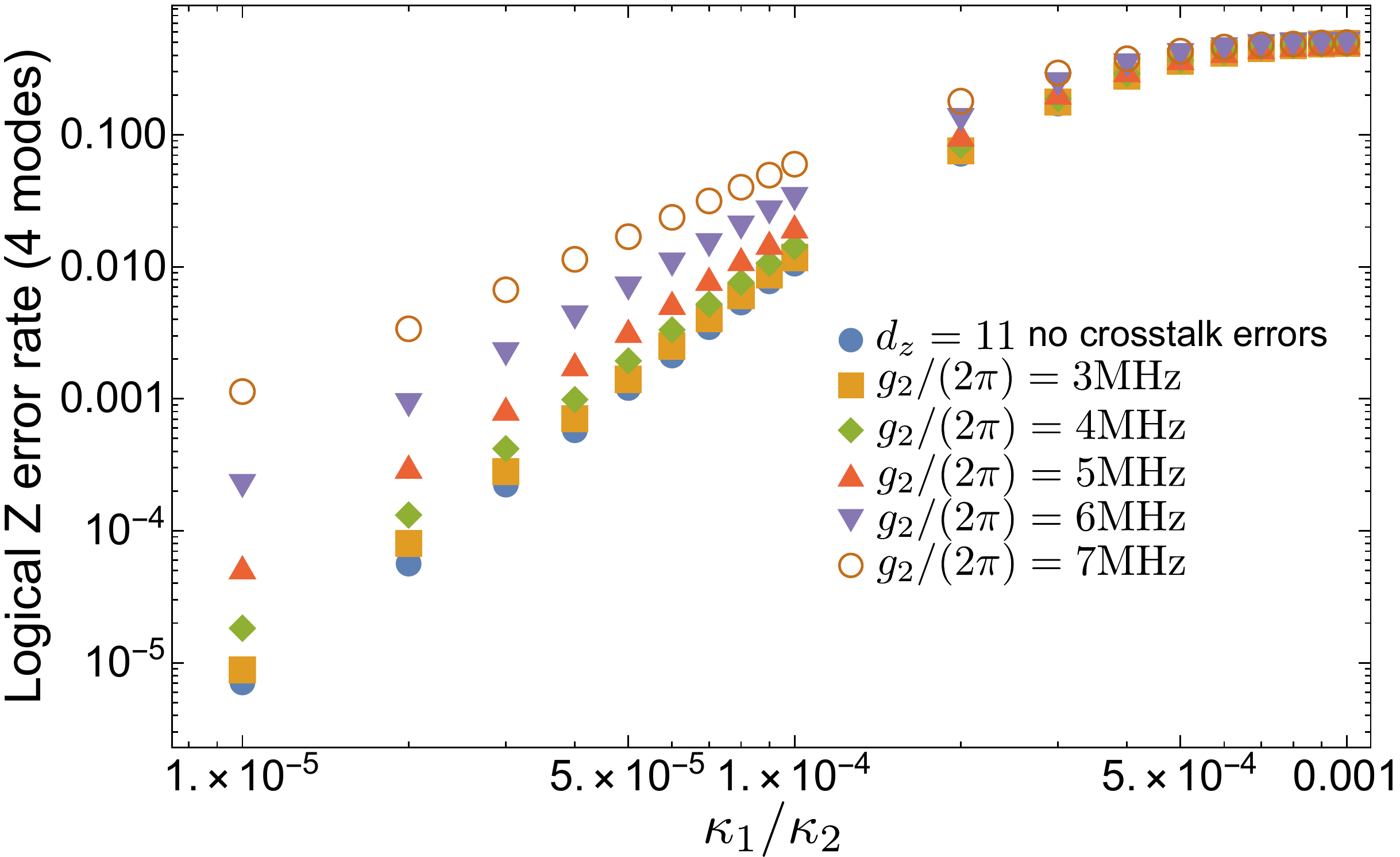}
	}

	\caption{Logical $Z$ failure rates for a $d_x=3$ and $d_z=11$ thin surface code in the presence of the residual crosstalk errors (given in \cref{tab:frequency_opt_four_mode}). The $X$-basis measurement error rates are obtained from \cref{tab:XMeasurementTimeAndErrorRate} with five parity measurements. We compute the logical $Z$ error rates for different values of $g_2$ shown in the legend, and compare such results to the case where the crosstalk errors are not present. The results in panel (a) [resp.~(b)] are obtained with crosstalk error rates taken from the five-mode (four-mode) row of \cref{tab:frequency_opt_four_mode} with $|\alpha|^2 = 8$. }
	\label{fig:d11CorrelatedSurfLogZ_app}
\end{figure}

\ch{ Finally, we consider how the reduction in crosstalk (\Cref{tab:frequency_opt_four_mode}) impacts logical error rates. Logical $Z$ error rates of the thin rotated surface code for the main-text architecture and the alternative architecture are plotted in \Cref{fig:d11CorrelatedSurfLogZ_app} (a) and (b), respectively, where correlated errors are simulated with probabilities given in \Cref{tab:frequency_opt_four_mode}. The results reveal that the effects of crosstalk on logical errors are less severe in the alternative architecture. In particular, the main-text architecture can accommodate values of $g_2/(2\pi)$ up to $\sim 2$ MHz before the impacts of correlated errors become significant, while the alternative architecture can accommodate values twice as large $g_2 \sim 4$ MHz before the impacts of correlated errors become significant. This increase in the maximum allowable $g_2$ can enable larger stabilization rates $\kappa_2$, or equivalently can reduce the requirement on the phonon-mode lifetime required to reach a given $\kappa_1/\kappa_2$.}
 
\section{\STOP ~algorithm}
\label{appendix:STOPdec}

When performing physical non-Clifford operations in between rounds of error correction (EC), in order to maintain the full effective code distance, it is crucial to use a fault-tolerant error correction protocol which satisfies the following definition (taken from \cite{Gottesman2010,CB17}):

\begin{definition}{\underline{Fault-tolerant error correction}}
	
	For $t = \lfloor (d-1)/2\rfloor$, an error correction protocol using a distance-$d$ stabilizer code $C$ is $t$-fault-tolerant if the following two conditions are satisfied:
	\begin{enumerate}
		\item For an input codeword with error of weight $s_{1}$, if $s_{2}$ faults occur during the protocol with $s_{1} + s_{2} \le t$, ideally decoding the output state gives the same codeword as ideally decoding the input state.
		\item For $s$ faults during the protocol with $s \le t$, no matter how many errors are present in the input state, the output state differs from a codeword by an error of at most weight $s$.
	\end{enumerate}
	\label{Def:FaultTolerantDef}
\end{definition}

Apart from being useful for proving thresholds of fault-tolerant error correction schemes based on code concatenation \cite{AGP06}, such a definition of fault-tolerant error correction is also relevant when performing physical non-Clifford operations on encoded qubits \textit{before} directly measuring the data qubits. 
In particular, if one implements a minimum weight perfect matching (MWPM) decoder (see Ref.~\cite{Edmonds65}) with $\mathcal{O}(d)$ rounds of stabilizer measurements (where $d$ is the code distance of the error correcting code protecting the data), a measurement error in the last round can lead to a logical failure and \cref{Def:FaultTolerantDef} would not be satisfied. 
In many fault-tolerant implementations, such a problem can be avoided by implementing non-Clifford operations via \etc{gate injection} and stabilizer operations, followed by direct measurement of the data qubits (hence physical non-Clifford gates are never directly applied to the data qubits). An example of direct measurements of the data qubits after performing one round of stabilizer measurements for a $d=5$ repetition code is given in \cref{fig:MeasDataRepEquiv}. 
By measuring the data qubits, measurement errors can be treated as data qubit errors arising prior to performing the measurement \cite{KBF+15}. 
As such, measuring the data directly acts as a round of perfect error correction. 

As was shown in \cref{sec:BottomUp}, post-selection can be avoided when preparing the logical $\ket{0}_L$ and $\ket{1}_L$ states (used to obtain the state $\ket{\psi}_1$) if we have a decoder that is robust to measurement errors in the last syndrome measurement round prior to applying the physical Toffoli gates. 
For the \BUTOFF protocol, we cannot directly measure the data prior to applying the physical Toffoli gates. 
Using ideas from Ref.~\cite{CB17}, in this section we propose an algorithm which tells us when to terminate the sequence of error syndrome measurements, which we call the \STOP ~algorithm, and which satisfies \cref{Def:FaultTolerantDef} when using the syndrome measurement from the last round to correct errors. 
Further, in \cref{appendix:StabOps}, we show how the \texttt{STOP} algorithm can be used with \etc{gate injection} to perform all stabilizer operations of the repetition code. 

\begin{figure}
	\centering
	\includegraphics[width=0.4\textwidth]{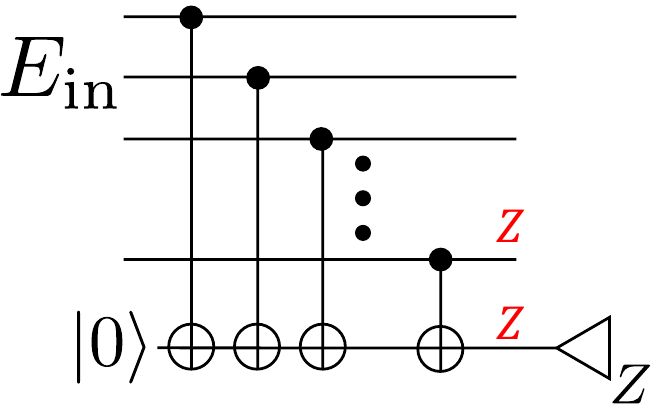}
	\caption{Example of a single controlled-$Z$ failure resulting in the error $Z_{n}\otimes Z_{n+1}$ (where $Z_{n+1}$ acts on the ancilla qubit) when measuring the operator $Z^{\otimes n}$. 
	Here $n$ is the number of data qubits. 
	This single fault can cause three consecutive syndrome measurements to yield three distinct outcomes. 
	Here $E_{\text{in}}$ is an input error with syndrome $s(E_{\text{in}}) = s_1$.}
	\label{fig:ExampleSTOP1}
\end{figure}

The goal of the \texttt{STOP} algorithm is to track consecutive syndrome outcomes $s_1, s_2, \cdots, s_r$ and to compute the \textit{minimum} number of faults which could have caused this sequence of syndromes. 
In particular, let $n_{\text{diff}}$ be a counter which tracks the minimum number of faults causing changes in syndrome outcomes, and consider the consecutive syndromes $s_{k-1}, s_k$ and $s_{k+1}$. 
Given that a single fault can lead to two syndrome changes as in the example below, suppose we obtain different syndromes in rounds $k$ and $k+1$ (so that $s_k \neq s_{k+1}$). 
In order to decide whether to increment $n_{\text{diff}}$ by one, we must first check whether $n_{\text{diff}}$ was incremented after measuring the $k$'th error syndrome. 
If $n_{\text{diff}}$ didn't increase after the $k$'th round, then we increment $n_{\text{diff}}$ by one. 
Otherwise, $n_{\text{diff}}$ remains unchanged.

As an example, suppose a single fault occurs during the second round of stabilizer measurements of an EC protocol adding a weight-one error to the data qubits while also flipping the measurement outcome of one of the stabilizers (in this case $Z^{\otimes n}$ as shown in \cref{fig:ExampleSTOP1}). 
Further, suppose the input error to the second round of the EC protocol $E_{\text{in}}$ has the error syndrome $s(E_{\text{in}}) = s_1$, and that the error $Z_n E_{\text{in}}$ has error syndrome $s_3 \neq s_1$ (here $Z_{n}$ is the $Z$ error added to the data qubit arising from the two-qubit gate failure). Since the $Z$ error flipped the measurement outcome of $Z^{\otimes n}$, the syndrome $s_2$ measured during the second round can differ from both $s_1$ and $s_3$.

\begin{algorithm}[t]
\SetAlgoLined
\KwResult{Final syndrome $s_r$ for $r$ repetitions of the syndrome measurement.}
 \textbf{initialize:} $t = (d-1)/2$; $n_{\text{diff}} = 0$; $\text{countSyn} = 1$; $\text{SynRep} = 1$; $n_{\text{diff}}\text{Increase} = 0$; $\text{test} = 0$\;
 \While{$\text{test} = 0$}{
  \If{$n_{\text{diff}} = t$}{
      $\text{test} = 1$
  }
  \textbf{Measure the error syndrome $s_j$.} Store the error syndrome $s_{j-1}$ from the previous round in $\textbf{synPreviousRound}$ and the current syndrome $s_j$ in $\textbf{synCurrentRound}$.\;
  \If{$\text{countSyn} > 1$}{
      \eIf{$\text{synPreviousRound} = \text{synCurrentRound}$}{
           $\text{SynRep} = \text{SynRep} + 1$\;
           $n_{\text{diff}}\text{Increase} = 0$\;
   }{
           $\text{SynRep} = 0$\;
           \eIf{$n_{\text{diff}}\text{Increase} = 0$}{
               $n_{\text{diff}} = n_{\text{diff}} + 1$\;
               $n_{\text{diff}}\text{Increase} = 1$\;
           }{
               $n_{\text{diff}}\text{Increase} = 0$\;
           }
  }
  }
  \If{$\text{SynRep} = t - n_{\text{diff}} + 1$}{
      $\text{test} = 1$\;
  }
  $\text{countSyn} = \text{countSyn} + 1$\;
 }
 \caption{\texttt{STOP} algorithm}
 \label{Alg:STOP}
\end{algorithm}

With the above example in mind, the \texttt{STOP} algorithm is given by \cref{Alg:STOP}. 
To see why a decoding algorithm based on \cref{Alg:STOP} satisfies \cref{Def:FaultTolerantDef}, consider the case where the total number of input errors and faults during the EC is $t=(d-1)/2$ for a distance $d$ error correcting code. 
If at any time during the EC the same syndrome $s_j$ is measured $t-n_{\text{diff}}+1$ times in a row, then it must have been the correct syndrome (with very high probability). 
The reason is that given the value of $n_{\text{diff}}$, which counts the minimum number of faults compatible with the syndrome history since the beginning of the current cycle of error correction, there would need to be more than $t$ faults to cause all $t-n_{\text{diff}}+1$ consecutive syndromes to be incorrect due to failures resulting in flipped measurement outcomes. 
As such one could use the syndrome $s_j$ to correct errors and terminate the protocol. 
Doing so, there could only be $\le t$ residual leftover errors that went undetected in the last measurement round. 

Similarly, if after measuring the $r-1$'th syndrome $n_{\text{diff}}$ gets incremented to $n_{\text{diff}}=t$, then we know that at least $t$ faults must have occurred during the EC. 
As such, by repeating the syndrome measurement one more time (resulting in the syndrome $s_r$) and using that syndrome to decode, there would need to have been more than $t$ faults for $s_r$ to be the wrong syndrome (due to faults flipping some of the stabilizer measurement outcomes in the last round). 
Hence using $s_r$ to decode would result in residual errors with weight $v \le t$ (where, as stated at the beginning of the previous paragraph, the total number of input errors and faults during the EC is $t$). 

Given the above, we conclude that when using \cref{Alg:STOP}, the sequence of syndrome measurements will terminate if one of the following conditions is satisfied:
\begin{enumerate}
	\item The syndrome $s_r$ is repeated $t-n_{\text{diff}}+1$ times in a row.
	\item The counter $n_{\text{diff}}$ gets incremented to $n_{\text{diff}}=t$. Measure the syndrome one more time resulting in the syndrome $s_r$. Use $s_r$ to decode. 
\end{enumerate}
Decoding will succeed if the total number of input errors and faults during the EC cycle is $\le t$.

We now provide a few remarks. 
Firstly, given a particular error correcting code and decoder along with the \texttt{STOP} algorithm for repeating the syndrome measurement, one can satisfy \cref{Def:FaultTolerantDef} by using the last measured syndrome $s_r$ to decode while ignoring the entire syndrome history. 
Hence in such settings, one can use a simple code-capacity-type decoder to decode with $s_r$ (i.e.\ a decoder which ignores measurement and space-time correlated errors). 
As an example, one can decode with the surface code using a MWPM or Union Find decoder (see Ref. \cite{LinTimeDec17}) on a two-dimensional graph instead of a three-dimensional graph tracking the entire syndrome history. 
Doing so could significantly reduce the overall decoding time. 
In general however, the approach where the \texttt{STOP} algorithm is used to ignore the entire syndrome history apart the last syndrome $s_r$ does not have a threshold~\footnote{There are cases where families of error correcting codes have thresholds when using the \texttt{STOP} algorithm to decode with the syndrome from the last round. One such example includes concatenated codes using the methods of \cite{AGP06}}. 
To see this, consider a distance $d$ repetition code and a stochastic noise model where fault locations fail with probability $p$. 
After $d$ rounds of repeating the syndrome measurement, there will be approximately $pd^2$ failures. 
For large distances $d$, with high probability the error syndrome will change in every round. 
Hence the probability of a measurement error in the final round will not depend on the past syndrome history and the decoder will fail to correct the errors with high probability. 

On the other hand, tracking the entire syndrome history and using \cref{Alg:STOP} to decide when to terminate the rounds of repeated syndrome measurement generally leads to lower failure rates and has a threshold. 
Indeed, when computing the memory failure rates of the repetition code using \cref{Alg:STOP} for deciding when to terminate the syndrome measurements), we found that performing MWPM on the entire syndrome history leads to lower logical failure rates compared to performing MWPM on a one-dimensional graph using only the final syndrome $s_r$ (note that the logical $Z$ failure rates for the repetition code in \cref{fig:MemRepCode} were computed by applying MWPM to the full syndrome history of the measured syndromes using the \texttt{STOP} algorithm). 
As such, the EC protocols used in this work when considering repetition codes implement MWPM on the entire syndrome history in conjunction with \cref{Alg:STOP} to decide when to stop measuring the error syndrome.

We conclude this section by providing a lower and upper bound on the maximum number of syndrome measurement repetitions that can be performed using the \texttt{STOP} algorithm. 
For the case where there are no failures, it is straightforward to see that the syndrome measurement will be repeated $t+1$ times. 
To find the upper bound, we consider the worst case scenario, where (starting with $n_{\text{diff}} = 0$) there are no failures in the first $t$ rounds of syndrome measurement, so that the same syndrome is repeated $t$ times. 
However, in round $t+1$, a measurement error occurs and  $n_{\text{diff}}$ gets incremented to $n_{\text{diff}} = 1$. 
Now again, suppose there are no failures in the next $t - 1$ rounds (so the same syndrome is repeated $t-1$ times) and a measurement error occurs in the $t$'th round, so that $n_{\text{diff}}$ is incremented to $n_{\text{diff}} = 2$. 
Suppose the same pattern repeats itself until all $t$ faults are exhausted resulting in $n_{\text{diff}} = t$. 
By the protocol of the \texttt{STOP} algorithm, we must repeat the syndrome measurement one more time. 
For such a fault pattern, the total number of syndrome measurements $s_{\text{tot}}$ is then given by
\begin{align}
    s_{\text{tot}} = \sum_{k = 0}^{t-1} (t-k) + t + 1 = \frac{1}{2}(t^2 + 3t + 2) = \binom{t+2}{2}.
\end{align}

For low code distances and low noise rate regimes, the average number of repetitions will approach $t+1$. 
However for large code distances, with high probability, the syndrome measurement outcome will change every round and thus $n_{\text{diff}}$ changes every other round. 
Thus after $2t$ rounds, $n_{\text{diff}} = t$ and the syndrome must be repeated one more time resulting in a total number of $2t + 1 = d$ rounds. 
It should then be expected that for large code distances, the performance of MWPM decoders based on a fixed $d$ rounds will perform similarly to a decoder which uses the \texttt{STOP} algorithm to terminate while implementing MWPM over the full syndrome history.

\section{Stabilizer operations with the repetition code }
\label{appendix:StabOps}

In this section we describe how to do all stabilizer operations with the repetition code. 
However, the methods we provide apply to any family of Calderbank-Shor-Steane (CSS) codes. 

\subsection{Computational basis states  }
\label{subsec:compbasis}

We begin by describing how to prepare the logical computational basis states of the repetition code. 
Doing so, we provide two schemes for preparing $\ket{0}_{L}$. 

\begin{figure}
	\centering
	\includegraphics[width=0.48\textwidth]{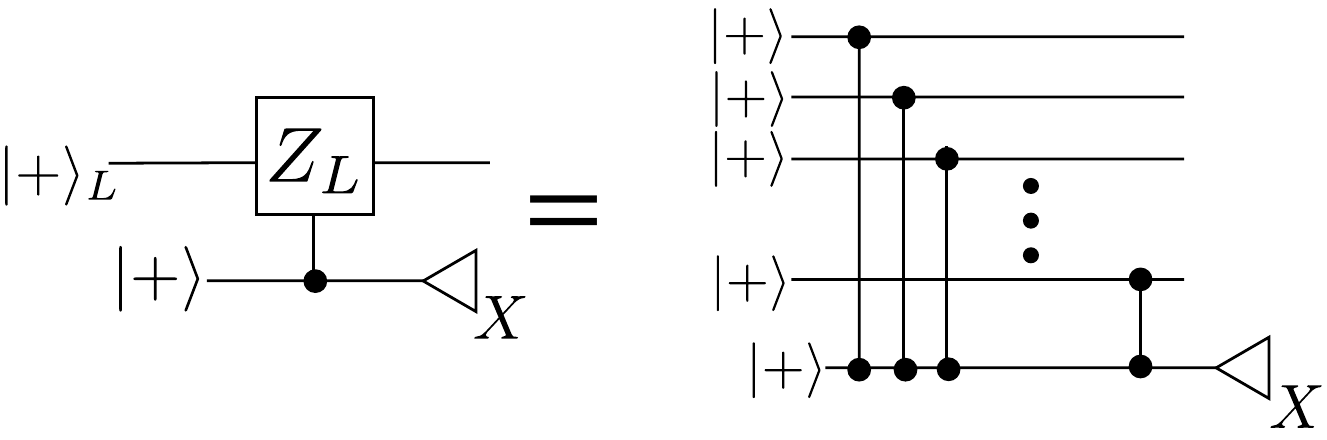}
	\caption{Circuit for preparing logical computational basis states of the repetition code. 
	A measurement error results in a logical $X$ error applied to the data. Fault-tolerance is achieved by repeating the measurement using the \texttt{STOP} algorithm.}
	\label{fig:PrepOone}
\end{figure}

\textbf{Scheme 1:} Using the fact that for an $n$-qubit repetition code $\ket{+}_{L} = \ket{+}^{\otimes n}$, preparing $ \ket{+}^{\otimes n}$ followed by a logical $Z_{L} = Z^{\otimes n}$ measurement (see \cref{fig:PrepOone}) projects the state to $\ket{0}_L$ given a $+1$ outcome and $\ket{1}_{L}$ given a $-1$ outcome. 
Since a measurement error on the ancilla results in a logical $X_L = X_1$ error applied to the data, fault-tolerance can be achieved by repeating the measurement of $Z_L$ using the \texttt{STOP} algorithm (where the syndrome corresponds to the ancilla measurement outcome) and applying the appropriate $X_L$ correction given the final measurement outcome. 
For instance, if $\ket{0}_L$ is the desired state and the final measurement outcome at the termination of the \texttt{STOP} algorithm is $-1$, $X_1$ would be applied to the data. 
Lastly, note that only $X$ errors can propagate from the ancilla to the data but these are exponentially suppressed by the cat-qubits. 

\textbf{Scheme 2:} Here we present a more conventional approach for preparing the computational basis states which only involves stabilizer measurements (see for instance Refs.~\cite{brooks2013fault,ADP14,VLCABT19}). 
Starting with the state $\ket{\psi_1} = \ket{0}^{\otimes n}$ which is a $+1$ eigenstate of $Z_L$, measure all stabilizers of the repetition code (each having a random $\pm 1$ outcome) resulting in the state
\begin{align}
\ket{\psi_2} = \prod_{i=1}^{n-1} \Big ( \frac{I \pm X_{i}X_{i+1}}{2} \Big )\ket{0}^{\otimes n}.
\end{align}
If the measurement outcome of $X_{k}X_{k+1}$ is $-1$, the correction $\prod_{j=1}^{k}Z_{j}$ can be applied to the data to flip the sign back to $+1$. 
However given the possibility of measurement errors, the measurement of all stabilizers $\langle X_{1}X_{2}, X_{2}X_{3}, \cdots , X_{n-1}X_{n} \rangle$ must be repeated. 
If physical non-Clifford gates are applied prior to measuring the data, then the \texttt{STOP} algorithm can be used to determine when to stop measuring the syndrome. 
Subsequently, MWPM is applied to the full syndrome history to correct errors and apply the appropriate $Z$ corrections to fix the code-space given the initial stabilizer measurements. 
After performing numerical simulations, we found that \textbf{scheme 2} achieves lower logical failure rates compared to \textbf{scheme 1}. Further, since physical Toffoli gates are applied to the data qubits in order to prepare a $\ket{\text{TOF}}$ magic state (see \cref{sec:BottomUp}) and given the constraints imposed by our ATS architecture (which make performing global $Z$ measurements very challenging using a single ancilla qubit), we always use \textbf{scheme 2} along with the \texttt{STOP} algorithm when preparing logical computational basis states. 

Lastly, we remark that although the logical component of an uncorrectable error $E^{(z)}Z_{L}$ (where $E^{(z)}$ is correctable) can always be absorbed by $\ket{0}_L$ resulting in an output state $\ket{\psi_{\text{out}}} = E^{(z)} \ket{0}_{L}$, it is still important to have a fault-tolerant preparation scheme for $\ket{0}_L$ (and thus to repeat the measurement of all stabilizers enough times). 
For instance, if a single fault results in a weight-two correctable $Z$ error (assuming $n\ge 5$), a second failure adding one or more data qubit errors during a subsequent part of the computation can combine with the weight-two error resulting in an uncorrectable data qubit error. 
Hence, such a preparation protocol would not be fault-tolerant up to the full code distance. 

\subsection{Implementation of logical Clifford gates}
\label{subsec:logCliffGateImp}

\begin{figure}
	\centering
	\includegraphics[width=0.45\textwidth]{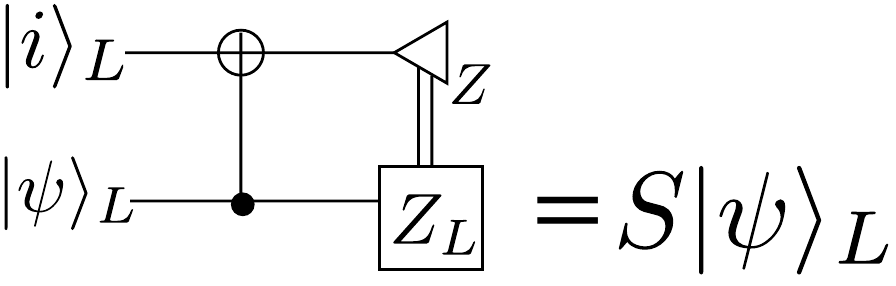}
	\caption{Circuit for implementing a logical $S$ gate. The circuit requires the preparation of $\ket{i}_{L}$, and the CNOT gate is transversal. A logical $\ket{Z}_L$ operator is applied when the measurement outcome of the ancilla is $-1$.}
	\label{fig:SgateCirc}
\end{figure}

Since the CNOT gate is transversal for the repetition code, we focus on implementing a generating set of single-qubit Clifford operations.
Recall that the Clifford group is generated by $\mathcal{P}_{n}^{(2)} = \langle H_i, S_i, \text{CNOT}_{ij} \rangle$, where
\begin{align}
H = \frac{1}{\sqrt{2}} 
\left(
\begin{array}{c}
\begin{array}{ccc}
 1 & 1 \\
 1 & -1\
\end{array}
 \\
\end{array}
\right),\ S=\left(
\begin{array}{c}
\begin{array}{ccc}
 1 & 0 \\
 0 & i  \\
\end{array}
 \\
\end{array}
\right),
\end{align}
are the Hadamard and phase gate operators. 
In what follows we show how to implement $S$ and $Q = SHS$ which also forms a generating set for single-qubit Clifford operations. 
A key to the implementation of such gates will be the injection of the state $\ket{i} = \frac{1}{\sqrt{2}} (\ket{0} + i\ket{1})$, which is a $+1$ eigenstate of the Pauli $Y$ operator. 
The logical state $\ket{i}_{L}$ can be prepared using \textbf{scheme 1} of \cref{subsec:compbasis} by replacing $Z_L$ with $Y_{L} = Y_1Z_2 \cdots Z_n$. 

\begin{figure}
	\centering
	\includegraphics[width=0.45\textwidth]{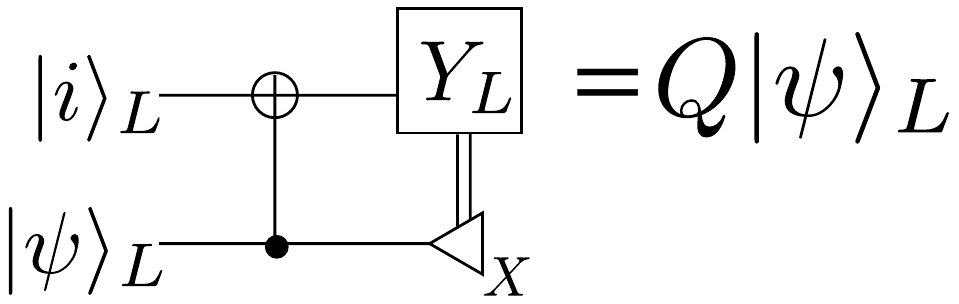}
	\caption{Circuit for implementing a logical $Q = SHS$ gate. The circuit requires the preparation of $\ket{i}_{L}$, and the CNOT gate is transversal. A logical $\ket{Y}_L$ operator is applied when the measurement outcome of the ancilla is $-1$.}
	\label{fig:Qcirc}
\end{figure}

In \cref{fig:SgateCirc} we provide a circuit for implementing $S_L$ which requires $\ket{i}_L$ as an input state, a transversal CNOT gate, and a logical $Z$-basis measurement. 
If a $-1$ measurement outcome is obtained, we apply a $Z_L$ correction to the data. 
Note however that a measurement error can result in a logical $Z_L$ being applied incorrectly to the data. 
As such, to guarantee fault-tolerance, one can repeat the circuit of \cref{fig:SgateCirc} and use the \texttt{STOP} algorithm to decide when to terminate.
The final measurement outcome is then used to determine if a $Z_L$ correction is necessary. 
The implementation of $S_L$ can thus be summarized as follows:

\begin{figure}
	\centering
	\includegraphics[width=0.45\textwidth]{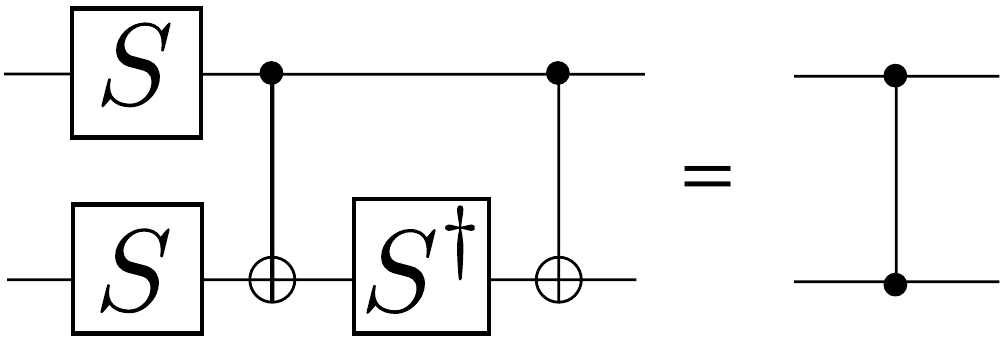}
	\caption{Efficient circuit for implementing a $CZ$ gate given the higher cost of logical $H$ gates compared to logical $S$ gates.}
	\label{fig:CZcirc}
\end{figure}

\textbf{$S_L$ gate implementation:}

\begin{enumerate}
   \item Implement the circuit in \cref{fig:SgateCirc} and let the measurement outcome be $s_1$.
   \item Repeat the circuit in \cref{fig:SgateCirc} and use the \texttt{STOP} algorithm to decide when to terminate.
   \item If the final measurement outcome $s_r = +1$, do nothing, otherwise apply $Z_L = Z_1Z_2 \cdots Z_n $ to the data.
\end{enumerate}

The circuit for implementing the logical $Q=SHS$ gate is given in \cref{fig:Qcirc}. 
The circuit consists of an injected $\ket{i}_L$ state, a transversal CNOT gate and a logical $X$-basis measurement is applied to the input data qubits. 
If the measurement outcome is $-1$, $Y_L$ is applied to the data. 
As with the $S_L$ gate, we repeat the application of the circuit in \cref{fig:Qcirc} to protect against measurement errors. 
The full implementation of $Q_L$ is given as follows:

\textbf{$Q_L$ gate implementation:}

\begin{enumerate}
   \item Implement the circuit in \cref{fig:Qcirc} and let the measurement outcome be $s_1$.
   \item Repeat the circuit in \cref{fig:Qcirc} and use the \texttt{STOP} algorithm to decide when to terminate.
   \item If the final measurement outcome $s_r = +1$, do nothing, otherwise apply $Y_L = Y_1Z_2 \cdots Z_n $ to the data.
\end{enumerate}

Note that the logical Hadamard gate can be obtained from the $S_L$ and $Q_L$ protocols using the identity $H = S^{\dagger} SHS S^{\dagger} = S^{\dagger} Q S^{\dagger}$. 
Hence ignoring repetitions of the circuits in \cref{fig:SgateCirc,fig:Qcirc}, the implementation of $H_L$ requires three logical CNOT gates, two $\ket{-i}_L$ and one $\ket{i}_L$ state, two logical $Z$ basis measurements, and one logical $X$ basis measurement. 
Instead of using two logical Hadamard gates and one CNOT gate to obtain a $CZ$ gate, we provide a more efficient circuit in \cref{fig:CZcirc}.

Lastly, we point out that since the circuits in \cref{fig:SgateCirc,fig:Qcirc} contain only stabilizer operations and injected $\ket{i}_L$ states, using the \texttt{STOP} algorithm to repeat the measurements is not strictly necessary. 
For instance, one could repeat the measurement a fixed number of times and majority vote instead of using the \texttt{STOP} algorithm. 
However in low noise rate regimes, the \texttt{STOP} algorithm can potentially be much more efficient since the average number of repetitions for the measurements can approach $t+1$ where $t = (d-1)/2$.

\section{Growing encoded data qubits to larger code distances with the repetition code}
\label{sec:Growing}

In this section, we provide a simple protocol for growing a state $\ket{\overline{\psi}}_{d_1} = \alpha \ket{0}_{d_1} + \beta \ket{1}_{d_1}$ encoded in a distance $d_1$ repetition code to a state $\ket{\overline{\psi}}_{d_2} = \alpha \ket{0}_{d_2} + \beta \ket{1}_{d_2}$ encoded in a distance $d_2 > d_1$ repetition code. We emphasize that the protocol presented in this section is applicable to arbitrary states and will be used for growing $\ket{\text{TOF}}$ states prepared using the fault-tolerant methods of \cref{sec:BottomUp} to larger code distances. 

Let $\mathcal{S}_{d_1} = \langle X_1X_2, X_2X_3, \cdots, X_{d_1-1}X_{d_1} \rangle$ be the stabilizer group for the distance $d_1$ repetition code with cardinality $|\mathcal{S}_{d_1}| = d_1 - 1$. Similarly, we define $\mathcal{S}_{d'_1} = \langle X_{d_1+1}X_{d_1 + 2}, \cdots, X_{d_2-1}X_{d_2} \rangle$ with $|\mathcal{S}_{d'_1}| = d_2 - d_1 - 1$. Furthermore, the stabilizer group for the distance $d_2$ repetition code is given by $\mathcal{S}_{d_2} = \langle X_1X_2, X_2X_3, \cdots, X_{d_2-1}X_{d_2} \rangle$.

In the remainder of this section, we define $g_i^{(d_1)}$ to be the $i$'th stabilizer in $\mathcal{S}_{d_1}$ and $g_i^{(d'_1)}$ to be the $i$'th stabilizer in $\mathcal{S}_{d'_1}$, so that $g_i^{(d_1)} = X_{i}X_{i + 1}$ and $g_i^{(d'_1)} = X_{d_1 + i}X_{d_1 + i + 1}$. 

\textbf{Protocol for growing $\ket{\overline{\psi}}_{d_1}$ to $\ket{\overline{\psi}}_{d_2}$:}

\begin{enumerate}
   \item Prepare the state $\ket{\psi_1} = \ket{0}^{\otimes (d_2 - d_1)}$.
   \item Measure all stabilizers in $\mathcal{S}_{d'_1}$ resulting in the state $\ket{\psi_2}_{d'_1} = \prod_{i=1}^{d_2-d_1-1} \Big ( \frac{I \pm g_i^{(d'_1)}}{2} \Big ) \ket{0}^{\otimes (d_2 - d_1)}$.
   \item Repeat the measurement of stabilizers in $\mathcal{S}_{d'_1}$ and apply MWPM to the syndrome history to correct errors and project to the code-space. If $g_i^{(d'_1)}$ is measured as $-1$ in the first round, apply the correction $\prod_{k= d_1 + 1}^{d_1 + i}Z_k$ to the data.
   \item Prepare the state $\ket{\psi_3} = \ket{\psi}_{d_1} \otimes \ket{\psi_2}_{d'_1}$ and measure $X_{d_1}X_{d_1+1}$.
   \item Repeat the measurement of all stabilizers if $\mathcal{S}_{d_2}$ and use MWPM over the syndrome history to correct errors. If in the first round the stabilizer $X_{d_1}X_{d_1+1}$ is measured as $-1$, apply the correction $\prod_{i = 1}^{d_1}Z_i$.
\end{enumerate}

As remark, the corrections stated in step 3 and 5 can be postponed to a later time after the growing protocol is completed. The reason is that one can use the entire syndrome history from each step, in addition to the syndromes measured after the states have merged to apply the appropriate corrections.

\begin{figure}
	\centering
	\includegraphics[width=0.48\textwidth]{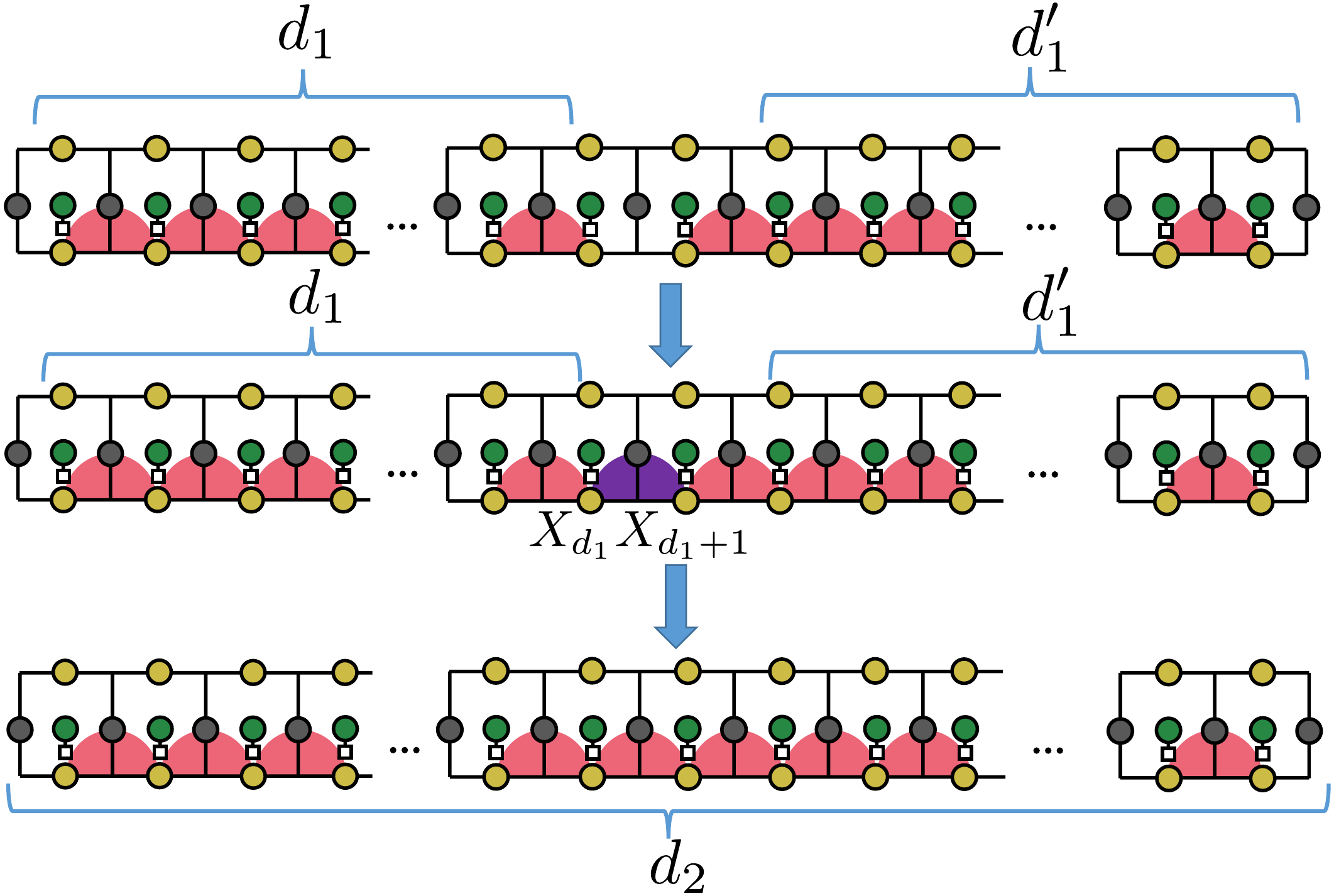}
	\caption{Diagram illustrating our protocol for growing the state $\ket{\psi}_{d_1}$ to  $\ket{\psi}_{d_2}$ with the ATS layout by starting with the two blocks stabilized by $\mathcal{S}_{d_1}$ and $\mathcal{S}_{d'_1}$ . The yellow vertices are the data qubits, and the gray vertices correspond to the ancilla qubits used to measure the stabilizers of the repetition code. The measurement of $X_{d_1}X_{d_1+1}$ (with random $\pm 1$ outcome) is highlighted by the purple semi-circle. After performing the appropriate corrections, the final block is stabilized by $\mathcal{S}_{d_2}$. }
	\label{fig:GrowingPic}
\end{figure}

The growing scheme involves two blocks, the first being the state  $\ket{\overline{\psi}}_{d_1}$ which we want to grow to  $\ket{\overline{\psi}}_{d_2}$. The second block involves the set of qubits which are prepared in the state $\ket{\psi_2}_{d'_1}$ and stabilized by $\mathcal{S}_{d'_1}$ (steps 1-3). The key is to measure the boundary operator $X_{d_1}X_{d_1 + 1}$ between the two blocks, which effectively merges both blocks into the encoded state $\ket{\overline{\psi}}_{d_2}$ and constitutes a simple implementation of lattice surgery \cite{HFDvM12,LR14,LVO18,FGLatticeSurg18}. To see this, consider the state prior to step 4:

\begin{align}
\ket{\psi_3} &= \ket{\overline{\psi}}_{d_1} \otimes \ket{\psi_2}_{d'_1} \nonumber\\
&= \alpha\ket{0}_{d_1} \otimes \ket{\psi_2}_{d'_1} + \beta \ket{1}_{d_1} \otimes \ket{\psi_2}_{d'_1} \nonumber\\
&= \alpha \prod_{i = d_1 + 1}^{d_2 - 1} \Big( \frac{I + g_{i}^{(d'_1)}}{2} \Big ) \ket{0}_{d_1} \otimes \ket{0}^{\otimes (d_2 - d_1)} \nonumber\\
&+ \beta X_1  \prod_{i = d_1 + 1}^{d_2 - 1} \Big( \frac{I + g_{i}^{(d'_1)}}{2} \Big ) \ket{0}_{d_1} \otimes \ket{0}^{\otimes (d_2 - d_1)},
\end{align}
where we used $\ket{1}_{d_1} = X_1 \ket{0}_{d_1}$. When measuring $X_{d_1}X_{d_1 + 1}$ and performing the correction $\prod_{i = 1}^{d_1}Z_i$ if the measurement outcome is $-1$, $\ket{\psi}_3$ is projected to
\begin{widetext}
\begin{align}
\ket{\psi}_f &= \alpha \prod_{i = d_1 + 1}^{d_2  - 1} \Big( \frac{I + g_{i}^{(d'_1)}}{2} \Big ) \Big ( \frac{I + X_{d_1}X_{d_1 + 1}}{2}   \Big ) \prod_{j = 1}^{d_1 - 1} \Big( \frac{I + g_{j}^{(d_1)} }{2} \Big ) \ket{0}^{\otimes d_2}  \nonumber\\ 
&+ \beta X_1 \prod_{i = d_1 + 1}^{d_2 - 1} \Big( \frac{I + g_{i}^{(d'_1)}}{2} \Big ) \Big ( \frac{I + X_{d_1}X_{d_1 + 1}}{2}   \Big ) \prod_{j = 1}^{d_1 - 1} \Big( \frac{I + g_{j}^{(d_1)} }{2} \Big ) \ket{0}^{\otimes d_2} \nonumber\\
&= \alpha \prod_{i = 1}^{d_2 - 1}\Big ( \frac{I + X_{i}X_{i + 1}}{2}   \Big )\ket{0}^{\otimes d_2} \nonumber\\
&+ \beta X_1 \prod_{i = 1}^{d_2 - 1}\Big ( \frac{I + X_{i}X_{i + 1}}{2}   \Big )\ket{0}^{\otimes d_2} \nonumber\\
&= \alpha \ket{0}_{d_2} + \beta X_1 \ket{0}_{d_2} \nonumber\\
&= \ket{\psi}_{d_2}.
\end{align}
\end{widetext}

The rounds of repeated stabilizer measurements in steps 3 and 5 are required due to the random outcomes and measurement errors which can occur when performing the appropriate projections. A pictorial representation for the growing scheme is shown in \cref{fig:GrowingPic}.

\section{Toffoli simulation twirling approximation}
\label{sec:BottomUpSimMethod}

To simulate the fault-tolerant preparation of the $\ket{\text{TOF}}$ state taking into account all fault-locations, we implement Monte-Carlo methods using a Gottesman-Knill type simulation \cite{GottesmanHeisenberg99} to avoid running into scalability issues. However, since the circuit in \cref{fig:TOFprepCircuitBottomUp} contains physical Toffoli gates, some type of approximation is necessary to perform a Gottesman-Knill type simulation with such circuits. 

In order to determine the most appropriate type of approximation, writing a Toffoli gate as \text{CCX}, we first observe that 
\begin{align}
(\text{CCX})(I \otimes I \otimes Z)\ket{\psi} =(CZ_{A,B} \otimes Z) (\text{CCX}) \ket{\psi},
\end{align}
for some arbitrary state $\ket{\psi}$. In other words, propagating a $Z$ error through the target qubit of the Toffoli gate results in a $CZ$ error on the two control qubits. Recall that we label the three logical qubits by $\{A,B,C\}$.

In what follows, we will consider the transformation of the $\ket{\text{TOF}}$ state with input data qubit $Z$ errors on the third block when using a single $\ket{+}$ ancilla to measure $g_A$. Note that all conclusions remain unchanged if instead we used the GHZ state of \cref{fig:TOFprepCircuitBottomUp}. 

Let $A_k$ be a subset of $k$ qubits and consider $k \ge 1$ data qubit errors on the third block expressed as $E^{(C)} = I \otimes I \otimes \prod_{j \in A_k} Z_{j} \equiv \prod_{j = 1}^{k} Z^{(C)}_{j}$. We have that
\begin{align}
\ket{\psi}_{\text{in}} = \ket{+} \prod_{j = 1}^{k} Z_j^{(C)} \ket{\text{TOF}}.
\label{eq:BeforeProp}
\end{align}
After applying \etc{$g_A$} and propagating the $Z$ errors through the Toffoli gates, $\ket{\psi}_{\text{in}}$ becomes 
\begin{align}
&\prod_{j = 1}^{k} Z_j^{(C)} \Big (\ket{0}\ket{\text{TOF}} + \ket{1} \prod_{j=1}^{k}Z_{j}^{(A)} \ket{\text{TOF}}     \Big) \nonumber \\
&= \prod_{j = 1}^{k} Z^{(C)}_{j} \Big [ \ket{+}\Big (\frac{I + \prod_{j = 1}^{k} Z^{(A)}_{j}}{\sqrt{2}} \Big )\ket{\text{TOF}} \nonumber \\
&+ \ket{-} \Big (\frac{I - \prod_{j = 1}^{k} Z^{(A)}_{j}}{\sqrt{2}} \Big )\Big ],
\end{align}
where $\prod_{j=1}^{k}Z_{j}^{(A)}$ are products of $Z$ errors on the first data block which have identical support with the $Z$ errors on the third block. After measuring the ancilla in the $X$ basis, a $\pm 1$ measurement outcome results in the state $\ket{\psi}_{f}$ given by
\begin{align}
\ket{\psi}_{f} = \prod_{j=1}^{k}Z_{j}^{(C)} \Big (\frac{I \pm \prod_{j = 1}^{k} Z^{(A)}_{j}}{\sqrt{2}} \Big )\ket{\text{TOF}} .
\label{eq:StateBeforeEC}
\end{align}
From \cref{eq:StateBeforeEC}, we see that when performing one round of error detection of the first block $A$, the error $\Big (\frac{I \pm \prod_{j = 1}^{k} Z^{(A)}_{j}}{\sqrt{2}} \Big )$ will project either to $I$ or $\prod_{j = 1}^{k} Z^{(A)}_{j}$ with $50 \%$ probability each unless $\prod_{j = 1}^{k} Z^{(A)}_{j} = Z^{(A)}_{L}$ in which case the state remains unchanged. 

Given the above, when performing our Gottesman-Knill type simulations when measuring $g_A$, if the input $Z$ errors to the third block are $\prod_{j = 1}^{k} Z^{(C)}_{j}$, we flip the GHZ ancilla measurement outcome with $50 \%$ probability and do the following: If $k < d$, we add the $Z$ errors  $\prod_{j = 1}^{k} Z^{(A)}_{j}$ to the first block with $50 \%$ probability. If $k = d$, we add $Z_{L}$ to the first block with $100 \%$ probability. 

Note that such a simulation method is exact when $k < d$ and only introduces a discrepancy when $k = d$. Since such events are rare, our approximation method differs from an exact simulation of the bottom-up $\ket{\text{TOF}}$ state preparation scheme only by a small amount.

\section{Fitting procedure for memory and lattice surgery}
\label{subsec:TimeLikeErrors}

Here we extend the discussion of lattice surgery presented in \cref{Sec:LatticeSurgery} as well as describe and justify the fitting procedures used in our error correction simulations. These fits enable us to reliably extrapolate to larger code sizes than simulated, which is required for our analysis of resource costs for large scale quantum computations (see ~\cref{sec:Overhead}).

\begin{figure*}[t]
    \centering
    \includegraphics{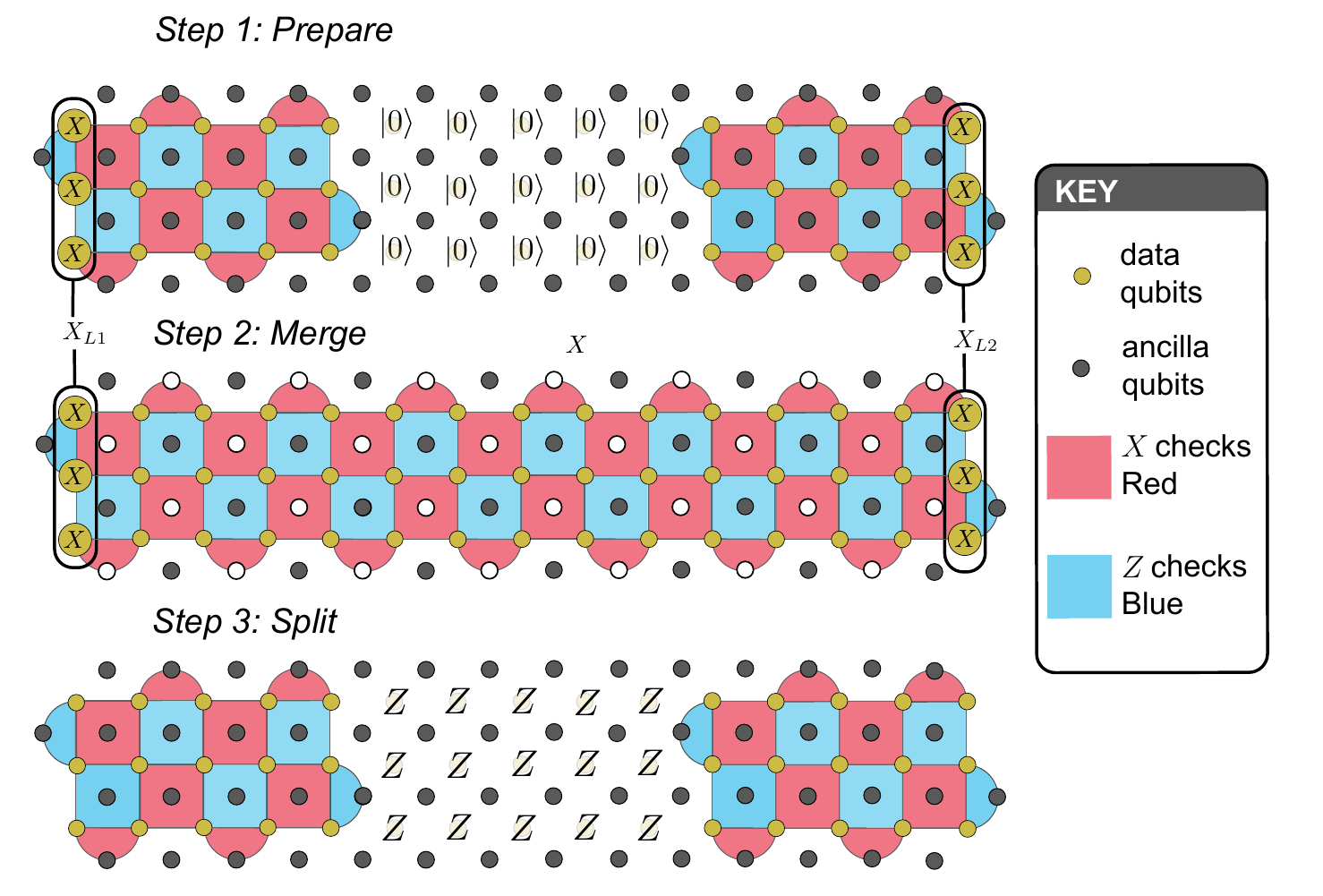}
    \caption{The three stages of lattice surgery corresponding to cross sections (time slices) of the lattice surgery spacetime diagram in \cref{fig:TimeLikeCartoon}. \textit{Step 1 Prepare:} data qubits between the surface code blocks are prepared in the $\ket{0}$ state.  \textit{Step 2 Merge:} start measuring the $Z$ and $X$ stabilizers indicated. The product of the $X$ stabilizers (highlighted with white vertices) yields the outcome $X_{L1} X_{L2}$. However, a measurement error on a white vertex will flip the outcome and so these stabilizer measurements must be repeated $d_m$ times, with $d_m$ chosen sufficiently large to suppress time-like errors to the desired probability. \textit{Step 3 Split:} The qubits between the initial surface code blocks are measured in the $Z$ basis. Note that it is not possible to use $X$ basis measurements to disentangle as this would measure $X_{L1}$ and $X_{L2}$. If the parity of the single-qubit $Z$ measurements is $``-1"$ then we must apply a Pauli correction $X_{L1}$ (or equivalently $X_{L2}$) as a correction. Both the measurement of $X_{L1} X_{L2}$ and the estimated Pauli correction must be done fault-tolerantly after having decoded the syndrome. In the case of $X_{L1} X_{L2}$, we choose 1 particular timeslice $t_p$ and make an initial guess by multiplying all the white vertices at time $t_p$. If the decoder assigns a measurement error to any white vertex at time $t_p$, then we must account by flipping the $X_{L1} X_{L2}$. If the accumulated physical $Z$ errors before time $t_p$ anticommute with $X_{L1} X_{L2}$ then we flip the outcome. For a similar discussion of lattice surgery see Ref.~\cite{FGLatticeSurg18}. Compared to \cref{fig:hardware_cartoon,fig:CircuitsRepSurf} we use a similar graphical representation but for simplicity omit the location of the transmon, readout qubit and ATS. 
    \label{fig:LatticeSurgerySimple}}
\end{figure*}

In addition, to presenting results for memory errors we also consider lattice surgery errors. Lattice surgery is the primary technique we consider for performing Clifford gates and magic state injection.  It is a procedure for measuring multi-qubit logical Pauli operators such as $X_L^{\otimes m}$ with $m \geq 2$. It can be regarded as a code deformation where the $m$ logical qubits are temporally merged into a code of $m-1$ logical qubits, and then split into their constituent $m$ logical qubits.  For the simple $m=2$ case, we illustrate the space-time diagram for this process in \cref{fig:TimeLikeCartoon} of the main text.  Here we present a more detailed in \cref{fig:LatticeSurgerySimple}.  

An incredibly powerful and beautiful feature of lattice surgery is that decoding via matching naturally extends over this 3D spacetime structure without being interrupted by lattice surgery.  However, some care is needed to correctly account for boundaries and assess different failure modes.  For a planar surface code, it is well known that one must allow defects to match with the appropriate boundaries in the space direction.  When performing lattice surgery, it is also important to match to appropriate boundaries in the time directions.  

To understand boundary effects, consider the more detailed explanation of lattice surgery in \cref{fig:LatticeSurgerySimple}. The procedure starts and ends with $Z$ basis state preparations and measurements.  A bit-flipped single qubit measurement or preparation will yield a pair of $Z$ syndrome defects. That is, the initial and final rounds of $Z$ stabilizer measurements are semi-ideal as they are reconstructed from single-qubit information so that any defects occur in pairs.  These short $X$ strings are then easily matched.  In contrast, an $X$ syndrome measurement error (at the start/end of lattice surgery) can lead to an isolated defect and is potentially harmful as it flips the outcome of the lattice surgery operation.  However, for such an isolated defect near a time boundary, the best explanation is clearly an isolated measurement error.  Therefore, we match these defects to red boundaries in the time direction.  

As a warm-up to discussing the probability of time-like errors, we first recap the error scaling properties of memory and logical $Z$-errors.  Consider a $d_x$ by $d_z$ surface code patch storing a logical qubit for $t$ surface code cycles.  We expect the total logical error probability to scale as $(1-\exp(-\lambda t))/2$ for some constant rate $\lambda$, which for small lambda is approximately $\sim \lambda t / 2$.  Furthermore, as $d_x$ increases the number of paths across the code increases linearly, so we expect that $\lambda \propto d_x$ and the total $Z$-logical probability to scale as 
\begin{equation}
    P_{Z} = d_x t F(d_z, p_1, \ldots , p_k ) ,
\end{equation}
for some function $F$ of $d_z$ and relevant hardware parameters $(p_1, \ldots , p_k)$. Note that $d_x t$ corresponds to the area of the vertical red boundaries in \cref{fig:TimeLikeCartoon}. For fixed parameters $(d_z, p_1, \ldots , p_m )$ the value of $F(d_z, p_1, \ldots , p_m )$ can be estimated by Monte Carlo simulation and evaluating $P_{Z}/(d_x t)$. For simulation purposes, standard practice is to assume: at time zero, the system is in a $``+1"$ eigenstates of all stabilizers; at time $t$, the round of stabilizer measurements is ideal.  This assumption introduces a finite size effect error into $P_{Z}/(d_x t)$.  This is suppressed by taking $t$ large, and community folklore suggests that $t=\mathrm{max}[d_z, d_x]$ will suffice though one could push higher.  The exact form of function $F$ can be quite involved, though we know it will be exponentially suppressed by the relevant distance $d_z$.  Taking our sole experimental parameter to be $\kappa_1 / \kappa_2$ we find good fits of the form
\begin{equation}
    P_{Z} = d_x t  a_z ( b_z  \kappa_1 / \kappa_2 )^{c_z d_z} ,
    \label{eq:LeadingOrderLogicalFail}
\end{equation}
where $a_z,b_z,c_z$ are fitted parameters. For small $d_x$, there will be a finite size effect so the scaling is not linear in $d_x$.  However, we can still use such a fit when $d_x$ is held constant provided we do not attempt to extrapolate to larger $d_x$.  Note that \cref{eq:LeadingOrderLogicalFail} is not necessarily a leading order fit of the classical form $O(p^{(d-1)/2})$. Since the probability of logical failures has a entropic/combinatorial component, it is dominated by errors with a weight much larger than $(d-1)/2$. As such, we do not attempt a leading order fit but rather it is appropriate to fit the scaling exponent $c_z$. 

\begin{figure}
    \centering
    \includegraphics[width=0.45 \textwidth]{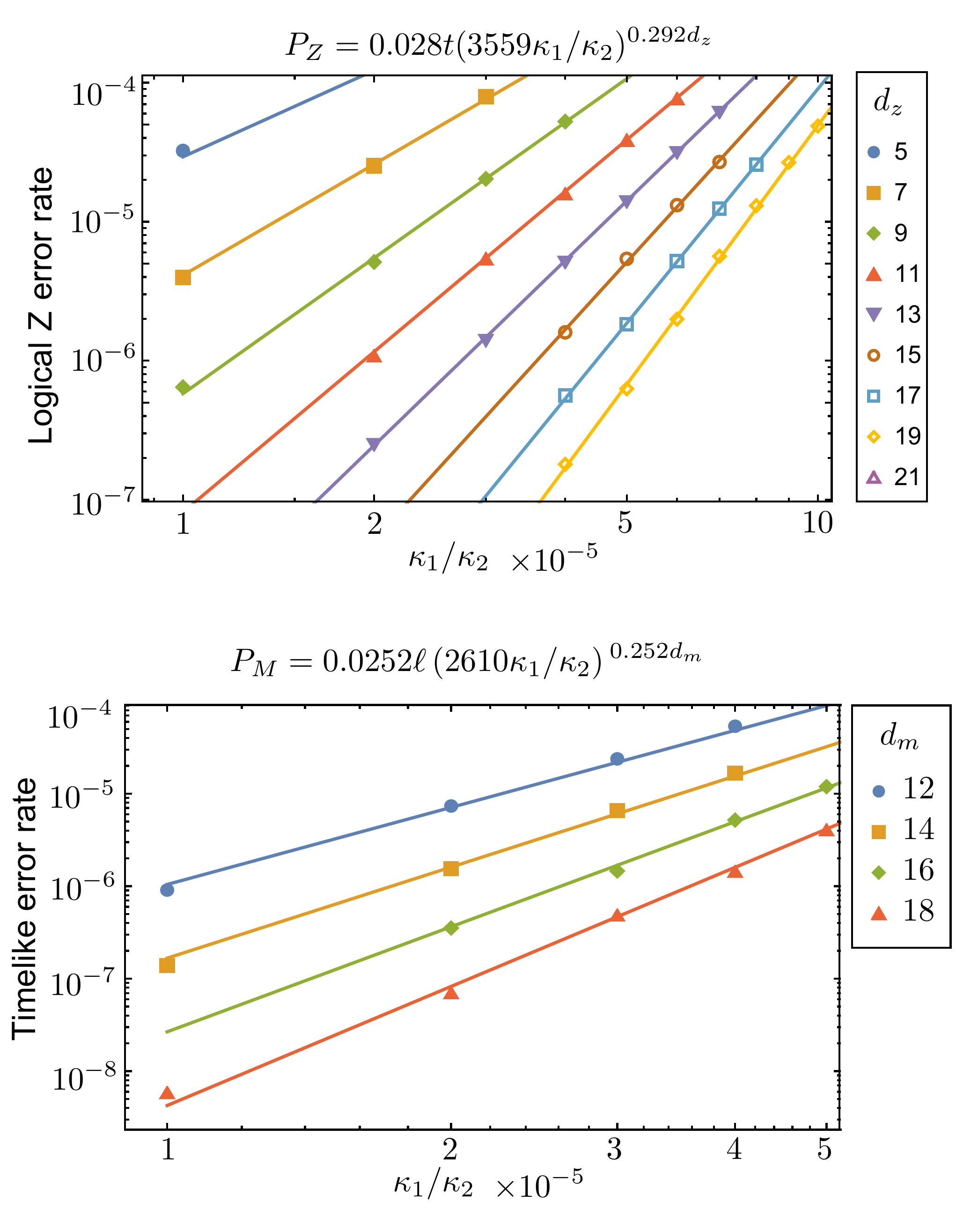}
    \caption{Fitted results for simulation of $d_x=3$ surface code for logical $Z$ and timelike errors. We fit according to the anstaz of \cref{eq:LeadingOrderLogicalFail,eq:PMscaling}. (Top) The logical $Z$ simulations for which we set $t=d_z$ and plot the error probability divided by $t$. (Bottom) the probability of a timelike error during lattice surgery for which we set $\ell=d_m-1$. All data points shown are used in fitting.  This is a truncated data set eliminating points above $10^{-4}$ on the error rate axis and eliminating points outside the relevant range of $\kappa_1/\kappa_2$.}
    \label{fig:SurfaceCodeErrorsGenFit}
\end{figure}

We present the result of this fitting procedure in Fig.~\ref{fig:SurfaceCodeErrorsGenFit} and observe that it works well over the interval $10^{-5} \leq \kappa_1 / \kappa_2 \leq 10^{-4}$.  At higher values of $\kappa_1 / \kappa_2$,   higher order contributions to a  logical $Z$ failure become important. Similarly, at lower values of $\kappa_1 / \kappa_2$, lower order  contributions become important. Even if a more sophisticated fitting function of $\kappa_1 / \kappa_2$ is assumed, we expect a finite range of applicability since there are other relevant experimental parameters in the noise model.

Similar reasoning can be applied to timelike errors.  The relevant boundary has an area $\ell d_x$ where $\ell$ is the distance between the codeblocks. As with $Z$-logical errors, the exponential decay of timelike errors follows from a percolation theory analysis~\cite{dennis02,kovalev2012fault} of a strings connecting the timelike boundaries. As always in percolation problems, the probability of a percolation event is exponentially suppressed in the distance between the boundaries (whenever below some threshold). The relevant boundaries are separated by a distance $d_m$, which we call the \textit{measurement distance}, and physically corresponds to the number of repeated rounds of syndrome measurements during the merge step. Therefore, we fit to the ansatz
\begin{equation}
    P_{M} = \ell d_x  a_z ( b_m  \kappa_1 / \kappa_2 )^{c_m d_m} ,
    \label{eq:PMscaling}
\end{equation}
where $a_m,b_m,c_m$ are fitted parameters. To obtain an estimate of $P_{M}$ we simulate the middle group of qubits in \cref{fig:LatticeSurgerySimple}. We wish to isolate the timelike errors and so freeze out $Z$-logical errors by assuming that the left-most and right-most qubits are ideal and error-free.  This is analogous to the assumption of ideal measurements in a memory simulation. Furthermore, since the $d_z$ distance is temporally extended during lattice surgery, such errors will be rare in comparison.  Again, this idealization introduces a finite size effect that vanishes as $\ell$ grows relative to $d_m$. 

We present the result of this fitting procedure for thin surface codes in Fig.~\ref{fig:SurfaceCodeErrorsGenFit} and observe that it works well over the interval $10^{-5} \leq \kappa_1 / \kappa_2 \leq 5 * 10^{-4}$.  We did not collect data for $\kappa_1 / \kappa_2 \geq 5* 10^{-4}$ as we had already identified that the surface code overhead will be prohibitively large in this regime.

To the best of our knowledge, there have not been previous simulations that investigate time-like errors in codes with boundaries and/or using circuit-level noise.  For instance, timelike errors were accounted for by Raussendorf and Harrington~\cite{raussendorf2007fault} but using a toy, phenomenological noise model and periodic boundary conditions in both space and time. 

Widespread practice is to set $d_m=d_x=d_z$ but there is no \textit{a prior} reason to believe this is optimal.  Indeed, just as physical bias in $X$ and $Z$ noise leads to an asymmetry in our choice of $d_x$ and $d_z$, a realistic noise model will influence the optimal choice of $d_m$. In later calculations we find that $d_m=d_z-2$ is the most common optimal choice for the main algorithm.  Furthermore, in the design of magic state distillation factories, the time-like errors are not critically important (see \cref{tab:CliffordFaults}) and so inside the factory $d_m$ can be set much smaller (by about a factor 1/2) than one would otherwise expect.

\section{Edge weights and decoding graphs for the repetition and surface codes }
\label{sec:EdgeWeights}

In this section we provide the decoding graphs used to implement MWPM with the repetition and surface codes considered in this paper. We also provide details for computing the edge weights of all edges in a given graph. 

\subsection{Repetition code decoding graphs}
\label{subsec:RepCodeGraphs}

\begin{figure}
	\centering
	\subfloat[\label{fig:RepCodeCircuit}]{%
		\includegraphics[width=0.17\textwidth]{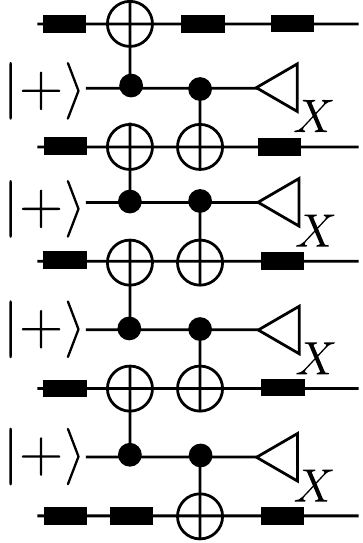}
	}
	\subfloat[\label{fig:RepCodeGraph}]{%
		\includegraphics[width=0.3\textwidth]{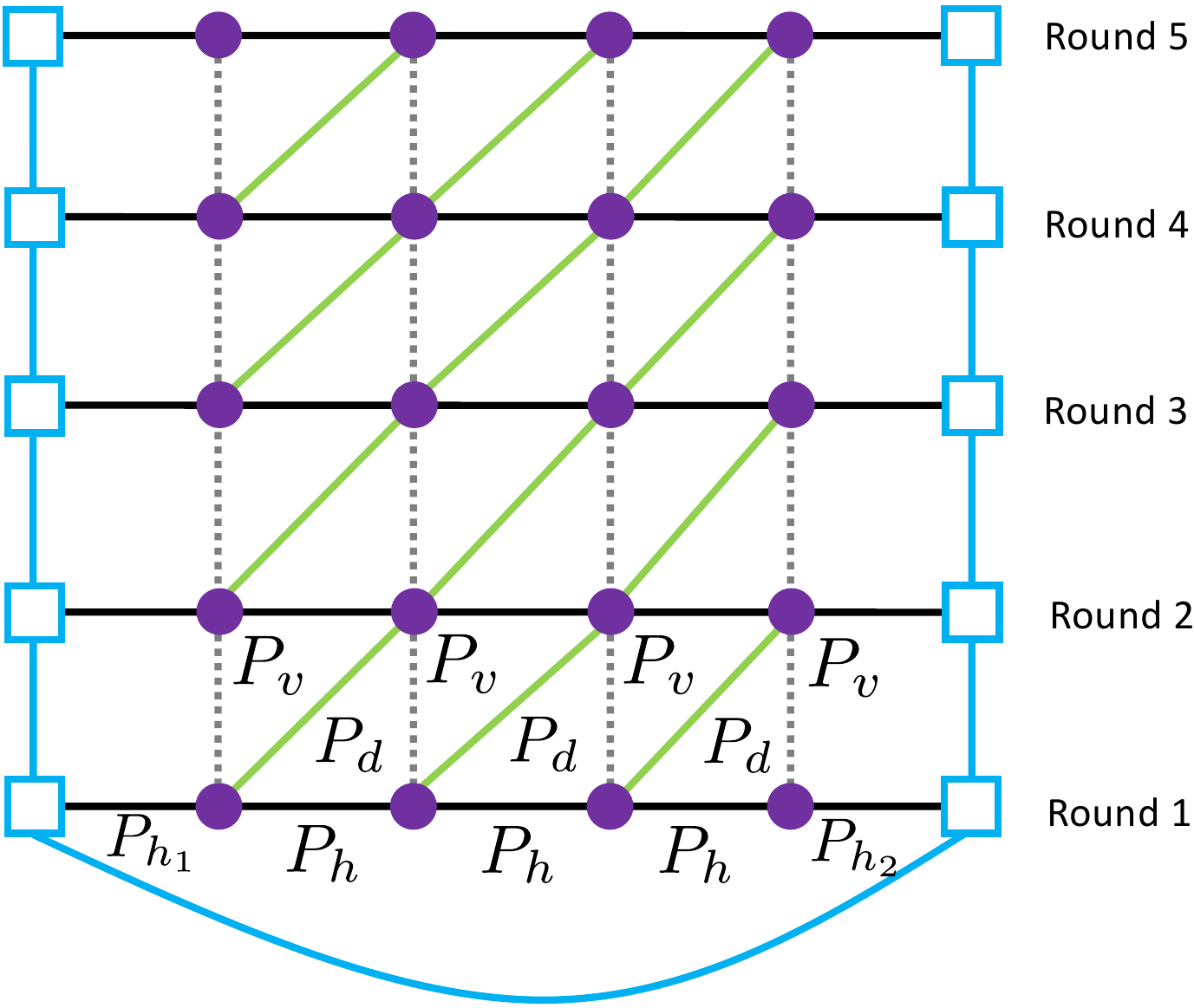}
	}

	\caption{\label{fig:RepCodeCircGraph}(a) Circuit for measuring the stabilizers of the $d=5$ repetition code. The dark rectangular boxes correspond to idling qubit locations. (b) MWPM decoding graph for the $d=5$ repetition code where the syndrome measurement is repeated five times. }
\end{figure}

The circuit for measuring the stabilizer of the $d=5$ repetition code is shown in \cref{fig:RepCodeCircuit} and can straightforwardly be generalized to arbitrary code distances. The corresponding graph for decoding the $d=5$ repetition code using MWPM with five rounds of syndrome measurements is shown in \cref{fig:RepCodeGraph}. The purple vertices correspond to the measurement outcome of each ancilla qubit (prepared in $\ket{+}$ and measured in the $X$-basis), and the horizontal edges correspond to the physical data qubits. A vertex is highlighted if the measurement outcome of the corresponding ancilla is non-trivial. We also add space-like boundary vertices and edges (shown in red). For a given syndrome measurement round (corresponding to a one-dimensional slice of the graph in \cref{fig:RepCodeGraph}), a boundary vertex is highlighted if on odd number of bulk vertices in the corresponding one-dimensional slice are highlighted. To deal with measurement errors, dashed grey vertical edges are added and connect vertices of two one-dimensional graphs. Lastly, cross-diagonal edges (shown in green) are added to deal with space-time correlated errors arising from failures at CNOT gate locations (see below for explicit examples). More details for implementing graph based decoding using MWPM can be found in Refs.~\cite{FowlerEdgeWeights,FowlerMatching,FMMC12}.

We now describe how to compute the edge weights for each edge type of the graph in \cref{fig:RepCodeGraph}. For a given edge $e$, we must first compute the probability $P_{e}$ of all failure events resulting in $e$ being highlighted. The weight $w_{e}$ for the edge $e$ is then given by $w_e = -\log (P_e)$ (see for instance Refs.~\cite{FowlerEdgeWeights,ChamberlandPRX,CKYZ20}). Such a prescription ensures that edges arising from more likely failure events are chosen with higher probability when finding the lowest weight path between two highlighted vertices. In what follows, we will refer to $P_e$ as the edge weight probability for the edge $e$.

The first and last data qubits in \cref{fig:RepCodeCircuit} have an extra idling location compared to all other data qubits, and their edge weight probabilities are labeled $P_{h_1}$ and $P_{h_2}$, whereas the other data qubits have edge weight probabilities labeled by $P_h$. The dashed grey vertical edges connecting are labeled $P_v$ and the green space-time correlated egdes are labeled $P_d$. Next we define $P_s$ to be the probability of a $\ket{+}$ state preparation error, $P_m$ the probability of a measurement error, $P_i$ the probability of an idle error and $P_{Z_1}$, $P_{Z_2}$ and $P_{Z_1Z_2}$ the probability of a $Z \otimes I$, $I \otimes Z$ and $Z \otimes Z$ CNOT failure (where the first qubit is the control qubit of the CNOT). 

We now show how to compute $P_h$ and $P_d$ to leading order (the other edge weight probabilities can be obtained using analogous methods). In the case of a single failure, a bulk horizontal edge (say corresponding to an error on the data qubit $q_j$) can be highlighted if either a $Z$ error occurs at the idle location during the preparation of $\ket{+}$, a $Z\otimes Z$ failure on the CNOT gate at the second time step with $q_j$ as a target qubit, or an $I \otimes Z$ failure on the CNOT gate on the third time step occuring in the previous syndrome measurement round. Hence we have
\begin{align}
P^{(t_1)}_{h} = P_i(1-P_{Z_1Z_2}) + P_{Z_1Z_2}(1-P_i),
\end{align}
and
\begin{align}
P^{(t>t_1)}_{h} &= 2P_i(1-P_{Z_1Z_2})(1-P_{Z_2})(1-P_i) \nonumber \\
&+  P_{Z_1Z_2}(1-P_i)^2(1-P_{Z_2}) \nonumber \\
&+ P_{Z_2}(1-P_i)^2(1-P_{Z_1Z_2}),
\end{align}
where $P^{(t_1)}_{h} = P_h$ in the first syndrome measurement round, and $P^{(t>t_1)}_{h} = P_h$ in all subsequent syndrome measurement rounds. 

Now suppose a $Z \otimes Z$ error occurs on a CNOT gate in the third time step of the syndrome measurement round $t$ resulting in a $Z$ data qubit error on qubit $q_j$ while also flipping the measurement outcome of the ancilla $a_k$. Note that if a $Z$ error had occurred on qubit $q_j$ prior to applying the two CNOT gates, both ancillas $a_k$ and $a_{k+1}$ would be measured non-trivially. Hence in round $t+1$ (and assuming no other failures), the measurement outcome of $a_k$ will not change whereas the measurement outcome of $a_{k+1}$ will change. To ensure that such an event is treated to leading order, we introduce a green cross-diagonal edge as seen in \cref{fig:RepCodeGraph}. Also note that a $I \otimes Z$ error on a CNOT in the second time step also results in such an edge. Hence we have that
\begin{align}
P_d = P_{Z_1Z_2}(1-P_{Z_2}) + P_{Z_2}(1-P_{Z_1Z_2}).
\end{align}

A similar analysis results in the following expressions for the remaining edge weight probabilities
\begin{align}
    P_v &= P_m(1-P_s)(1-P_{Z_1})^2 + P_s(1-P_m)(1-P_{Z_1})^2 \nonumber \\
    &+ 2P_{Z_1}(1-P_{Z_1})(1-P_s)(1-P_m), 
\end{align}
\begin{align}
    P^{(t_1)}_{h_1} = P^{(t_1)}_{h},
\end{align}
\begin{align}
    P^{(t > t_1)}_{h_1} &= 3P_i(1-P_i)^2(1-P_{Z_1Z_2})(1-P_{Z_2}) \nonumber \\
    &+ P_{Z_1Z_2}(1-P_i)^3(1-P_{Z_2}) \nonumber \\
    &+ P_{Z_2}(1-P_i)^3(1-P_{Z_1Z_2}),
\end{align}
\begin{align}
    P^{(t_1)}_{h_2} = 2P_i(1-P_i)(1-P_{Z_1Z_2}) + P_{Z_1Z_2}(1-P_i)^2,
\end{align}
and
\begin{align}
    P^{(t > t_1)}_{h_2} = P^{(t > t_1)}_{h_1}.
\end{align}

\subsection{Surface code decoding graphs}
\label{subsec:SurfaceCodeGraphs}

\begin{figure*}
	\centering
	\subfloat[\label{fig:Surface5by7}]{%
		\includegraphics[width=0.61\textwidth]{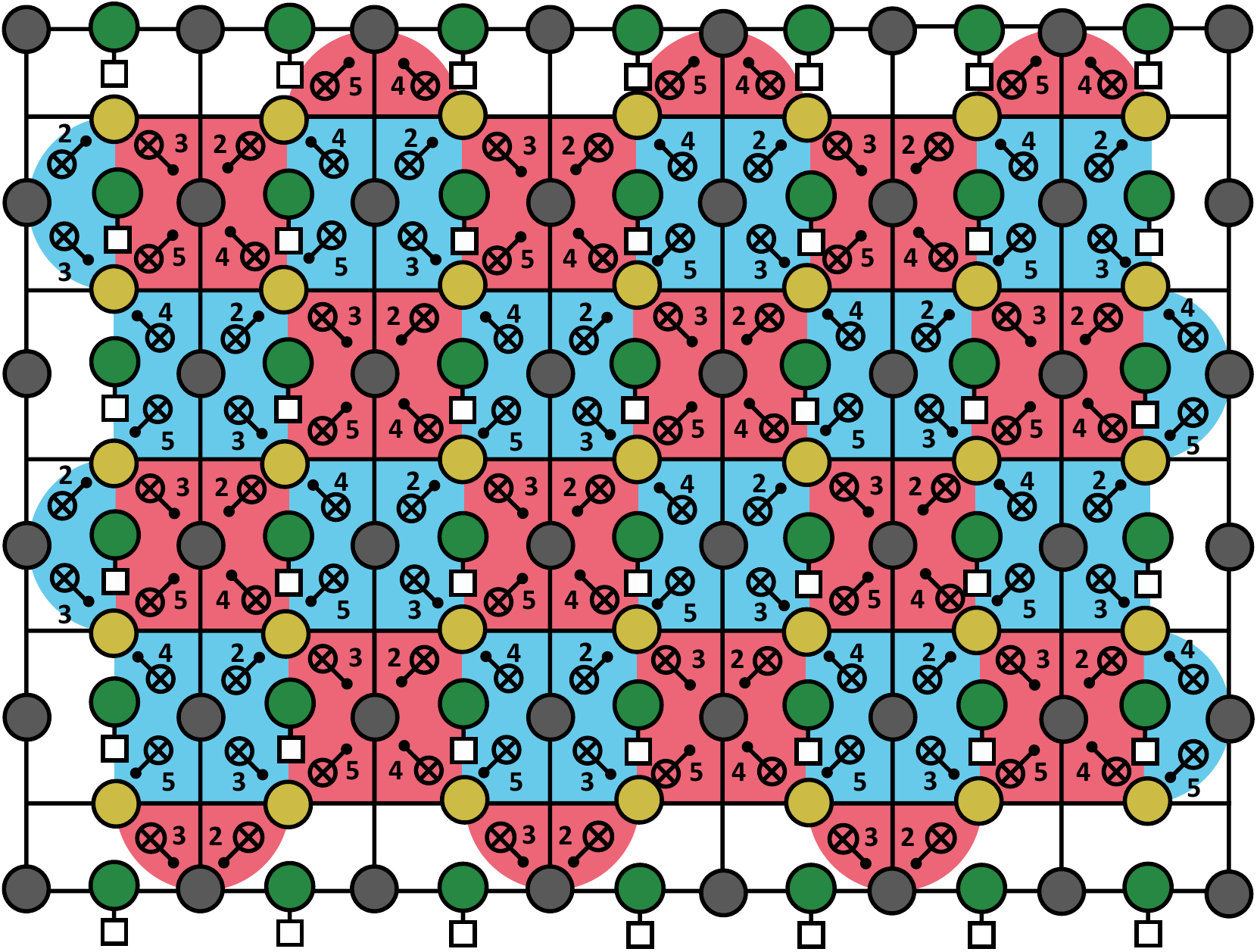}
	}
	\vfill
	\subfloat[\label{fig:Surface2DXgraph}]{%
		\includegraphics[width=0.61\textwidth]{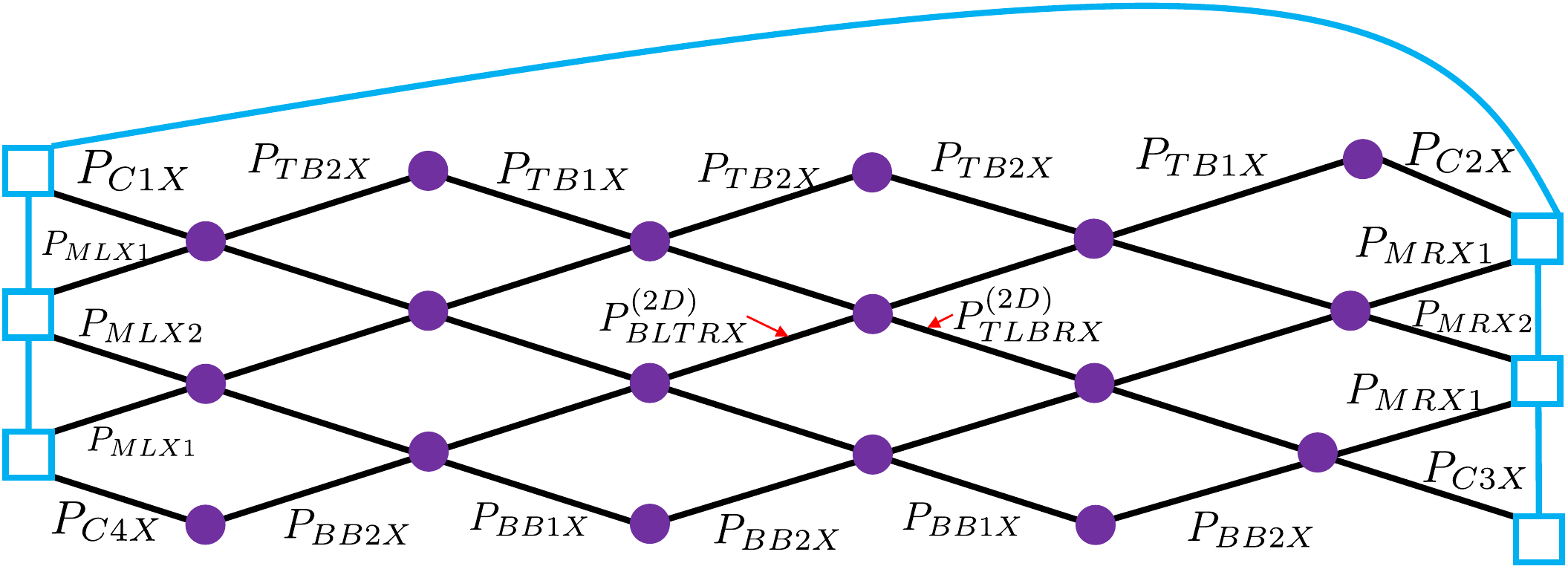}
	}
	\vfill
	\subfloat[\label{fig:Surface2DZgraph}]{%
		\includegraphics[width=0.61\textwidth]{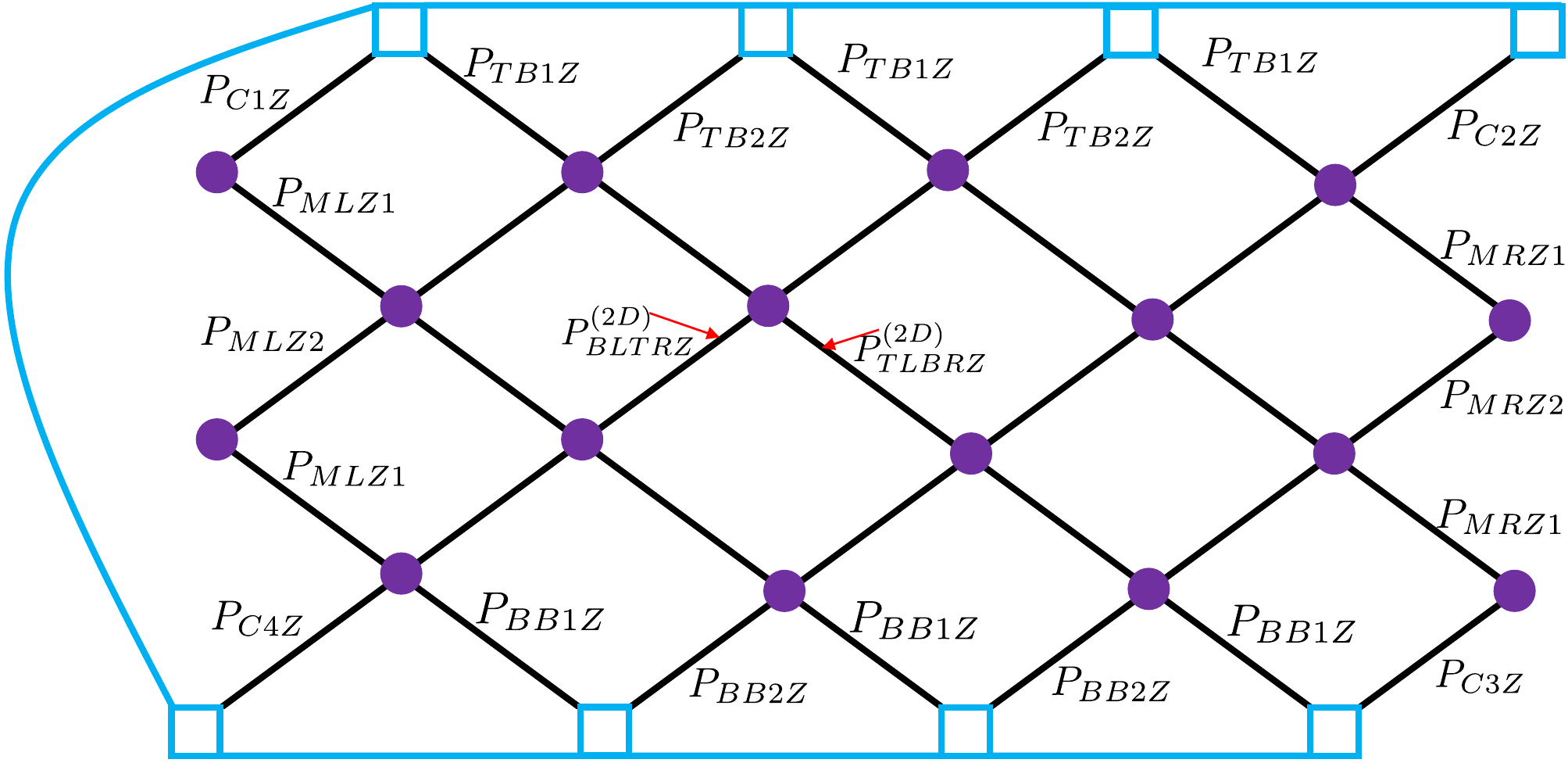}
	}

	\caption{\label{fig:SurfaceGraphs2D}(a) Surface code lattice with $d_x=5$ and $d_z=7$. (b) Graph used for decoding $X$ stabilizer measurement outcomes with both bulk and boundary edge weight probability labels. (c) Graph used for decoding $Z$ stabilizer measurement outcomes with both bulk and boundary edge weight probability labels. }
\end{figure*}

The two-dimensional graphs for decoding the $X$ and $Z$ stabilizer measurement outcomes of a $d_x=5$ and $d_z=7$ surface code, along with their corresponding edge weight probability labels, are shown in \cref{fig:SurfaceGraphs2D}. We will show below the edges that need to be added when considering measurement errors and space-time correlated errors arising from CNOT gate failures. However, we first provide edge weight probabilities for the edges of the two-dimensional graphs.

Let $G^{(2D)}_{(d_x)}$ and $G^{(2D)}_{(d_z)}$ be the two-dimensional graphs corresponding to the $X$ and $Z$ stabilizer measurement outcomes. For the graph $G^{(2D)}_{(d_x)}$, we label the bulk edge weight probabilities by $P^{(2D)}_{BLTRX}$ and $P^{(2D)}_{TLBRX}$. All other labels in \cref{fig:Surface2DXgraph} are used for boundary edges. Similarly, for the graph $G^{(2D)}_{(d_z)}$, we label the bulk edge weight probabilities by $P^{(2D)}_{BLTRZ}$ and $P^{(2D)}_{TLBRZ}$ with all other labels in \cref{fig:Surface2DZgraph} representing boundary edge weight probabilities. In order to simplify the expressions for the edge weight probabilities, we define the following function
\begin{align}
    &\Gamma(P_1,P_2, \cdots ,P_j; n_1,n_2, \cdots ,n_j) \equiv \nonumber \\
    &\sum_{k=1}^{j}n_kP_k(1-P_k)^{n_k - 1}\prod_{i=1,i \neq k}^{j}(1-P_i)^{n_i}.
    \label{eq:GammaEdgeProbs}
\end{align}

In what follows, we define $P^{(P_iP_j)}_{\text{CNOT}}$ to be the probability that a CNOT gate failure results in a two-qubit Pauli error of the form $P_i \otimes P_j$. We also define $P^{(P_i)}_{\text{Id}}$ to be the probability that a single-qubit idling location results in a $P_i$ Pauli error on that qubit. To further simplify the edge weight probability polynomials, we define the following probabilities:
\begin{align}
    P^{(1)}_{ZZCX} = P^{(ZZ)}_{\text{CNOT}} + P^{(ZY)}_{\text{CNOT}} + P^{(YZ)}_{\text{CNOT}} + P^{(YY)}_{\text{CNOT}},
    \label{eq:EdgeCX1}
\end{align}
\begin{align}
    P^{(1)}_{IZCX} = P^{(IZ)}_{\text{CNOT}} + P^{(XZ)}_{\text{CNOT}} + P^{(IY)}_{\text{CNOT}} + P^{(XY)}_{\text{CNOT}},
    \label{eq:EdgeCX2}
\end{align}
\begin{align}
    P^{(1)}_{ZICX} &= P^{(ZI)}_{\text{CNOT}} + P^{(ZX)}_{\text{CNOT}} + P^{(ZY)}_{\text{CNOT}} + P^{(ZZ)}_{\text{CNOT}} + P^{(YI)}_{\text{CNOT}} \nonumber \\
    &+ P^{(YX)}_{\text{CNOT}} + P^{(YZ)}_{\text{CNOT}} + P^{(YY)}_{\text{CNOT}},
    \label{eq:EdgeCX3}
\end{align}
\begin{align}
    P^{(2)}_{IZCX} &= P^{(IZ)}_{\text{CNOT}} + P^{(XZ)}_{\text{CNOT}} + P^{(IY)}_{\text{CNOT}} + P^{(XY)}_{\text{CNOT}} + P^{(ZI)}_{\text{CNOT}} \nonumber \\
    &+ P^{(ZX)}_{\text{CNOT}} + P^{(YI)}_{\text{CNOT}} + P^{(YX)}_{\text{CNOT}},
    \label{eq:EdgeCX4}
\end{align}
\begin{align}
    P^{(3)}_{IZCX} &= P^{(IZ)}_{\text{CNOT}} + P^{(IY)}_{\text{CNOT}} + P^{(ZZ)}_{\text{CNOT}} + P^{(ZY)}_{\text{CNOT}} + P^{(XZ)}_{\text{CNOT}} \nonumber \\
    &+ P^{(XY)}_{\text{CNOT}} + P^{(YZ)}_{\text{CNOT}} + P^{(YY)}_{\text{CNOT}},
    \label{eq:EdgeCX5}
\end{align}
and
\begin{align}
    P^{(2)}_{ZICX} = P^{(ZI)}_{\text{CNOT}} + P^{(YI)}_{\text{CNOT}} + P^{(ZX)}_{\text{CNOT}} + P^{(YX)}_{\text{CNOT}},
    \label{eq:EdgeCX6}
\end{align}

\begin{align}
    P^{(1)}_d = P^{(Z)}_{\text{Id}} + P^{(Y)}_{\text{Id}}.
    \label{eq:Edgedata1}
\end{align}

Using \cref{eq:GammaEdgeProbs,eq:EdgeCX1,eq:EdgeCX2,eq:EdgeCX3,eq:EdgeCX4,eq:EdgeCX5,eq:EdgeCX6,eq:Edgedata1} and the same methods as in \cref{subsec:RepCodeGraphs}, the leading order edge weight probabilities for the graph $G^{(2D)}_{(d_x)}$ are given by:
\begin{align}
    P^{(2D)}_{BLTRX} = \Gamma(P^{(1)}_{ZZCX},P^{(1)}_{IZCX}, P^{(1)}_d; 1,1,1),
\end{align}
\begin{align}
    &P^{(2D)}_{TLBRX} = \nonumber \\ &\Gamma(P^{(1)}_{ZZCX},P^{(1)}_{IZCX},P^{(1)}_{ZICX},P^{(2)}_{IZCX}, P^{(1)}_d; 2,2,1,1,1),
\end{align}
\begin{align}
        P_{C1X} = \Gamma(P^{(3)}_{IZCX},P^{(2)}_{ZICX}, P^{(1)}_d; 1,1,1),
\end{align}
\begin{align}
        P_{TB2X} = P^{(2D)}_{BLTRX},
\end{align}
\begin{align}
        P_{TB1X} = \Gamma(P^{(1)}_{ZZCX},P^{(1)}_{IZCX},P^{(2)}_{IZCX}, P^{(1)}_d; 2,1,1,1),
\end{align}
\begin{align}
        P_{C2X} = \Gamma(P^{(3)}_{IZCX},P^{(2)}_{IZCX},P^{(1)}_{ZZCX}, P^{(1)}_d; 1,1,1,1),
\end{align}
\begin{align}
        P_{MRX1} = \Gamma(P^{(3)}_{IZCX},P^{(2)}_{ZICX},P^{(1)}_d; 1,2,1),
\end{align}
\begin{align}
        &P_{MRX2} = \nonumber \\
        &\Gamma(P^{(3)}_{IZCX},P^{(1)}_{IZCX},P^{(1)}_{ZICX},P^{(2)}_{IZCX},P^{(1)}_d; 1,2,1,1,1),
\end{align}
\begin{align}
        P_{C3X} = \Gamma(P^{(3)}_{IZCX},P^{(1)}_{IZCX},P^{(1)}_{ZICX},P^{(1)}_d; 1,1,1,1),
\end{align}
\begin{align}
        P_{BB2X} = P^{(2D)}_{BLTRX},
\end{align}
\begin{align}
        P_{BB1X} = \Gamma(P^{(1)}_{IZCX},P^{(1)}_{ZZCX},P^{(1)}_{ZICX},P^{(1)}_d; 2,1,1,1),
\end{align}
\begin{align}
        P_{C4X} = P_{C3X},
\end{align}
\begin{align}
        &P_{MLX1} = \nonumber \\
        &\Gamma(P^{(3)}_{IZCX},P^{(1)}_{IZCX},P^{(2)}_{ZICX},P^{(1)}_{ZICX},P^{(1)}_d; 1,1,1,1,1),
\end{align}
and 
\begin{align}
        P_{MLX2} = P_{MLX1}.
\end{align}

For the graph $G^{(2D)}_{(d_z)}$, we first define the following probabilities:
\begin{align}
    P^{(1)}_{XXCX} = P^{(XX)}_{\text{CNOT}} + P^{(XY)}_{\text{CNOT}} + P^{(YX)}_{\text{CNOT}} + P^{(YY)}_{\text{CNOT}},
    \label{eq:EdgeCZ1}
\end{align}
\begin{align}
    P^{(1)}_{XICX} = P^{(XI)}_{\text{CNOT}} + P^{(YI)}_{\text{CNOT}} + P^{(XZ)}_{\text{CNOT}} + P^{(YZ)}_{\text{CNOT}},
    \label{eq:EdgeCZ2}
\end{align}
\begin{align}
    P^{(1)}_{IXCX} = P^{(IX)}_{\text{CNOT}} + P^{(ZX)}_{\text{CNOT}} + P^{(IY)}_{\text{CNOT}} + P^{(ZY)}_{\text{CNOT}},
    \label{eq:EdgeCZ3}
\end{align}
\begin{align}
    &P^{(2)}_{IXCX} = P^{(IX)}_{\text{CNOT}} + P^{(IY)}_{\text{CNOT}} + P^{(ZX)}_{\text{CNOT}} + P^{(ZY)}_{\text{CNOT}} + P^{(XX)}_{\text{CNOT}} \nonumber \\ &+ P^{(XY)}_{\text{CNOT}} + P^{(YX)}_{\text{CNOT}} + P^{(YY)}_{\text{CNOT}},
    \label{eq:EdgeCZ4}
\end{align}
\begin{align}
    &P^{(3)}_{IXCX} = P^{(IX)}_{\text{CNOT}} + P^{(IY)}_{\text{CNOT}} + P^{(ZX)}_{\text{CNOT}} + P^{(ZY)}_{\text{CNOT}} + P^{(XI)}_{\text{CNOT}} \nonumber \\ &+ P^{(XZ)}_{\text{CNOT}} + P^{(YI)}_{\text{CNOT}} + P^{(YZ)}_{\text{CNOT}},
    \label{eq:EdgeCZ5}
\end{align}
\begin{align}
    &P^{(2)}_{XICX} = P^{(XI)}_{\text{CNOT}} + P^{(YI)}_{\text{CNOT}} + P^{(XX)}_{\text{CNOT}} + P^{(YX)}_{\text{CNOT}} + P^{(XZ)}_{\text{CNOT}} \nonumber \\ &+ P^{(YZ)}_{\text{CNOT}} + P^{(XY)}_{\text{CNOT}} + P^{(YY)}_{\text{CNOT}},
    \label{eq:EdgeCZ6}
\end{align}
and

\begin{align}
    P^{(2)}_{d} = P^{(X)}_{\text{Id}} + P^{(Y)}_{\text{Id}}.
    \label{eq:EdgeCZ7}
\end{align}

Using \cref{eq:EdgeCZ1,eq:EdgeCZ2,eq:EdgeCZ3,eq:EdgeCZ4,eq:EdgeCZ5,eq:EdgeCZ6,eq:EdgeCZ7}, the leading order edge weight probabilities for the graph $G^{(2D)}_{(d_z)}$ are given by:
\begin{align}
    P^{(2D)}_{BLTRZ} = \Gamma(P^{(1)}_{XXCX},P^{(1)}_{XICX}, P^{(2)}_d; 1,1,1),
\end{align}
\begin{align}
    &P^{(2D)}_{TLBRZ} = \nonumber \\ &\Gamma(P^{(1)}_{XXCX},P^{(1)}_{XICX},P^{(2)}_{IXCX},P^{(3)}_{IXCX}, P^{(2)}_d; 2,2,1,1,1),
\end{align}
\begin{align}
    P_{C1Z} = \Gamma(P^{(2)}_{XICX},P^{(1)}_{IXCX}, P^{(2)}_d; 1,1,1),
\end{align}
\begin{align}
    &P_{TB1Z} = \nonumber \\ &\Gamma(P^{(2)}_{XICX},P^{(1)}_{XICX},P^{(2)}_{IXCX},P^{(3)}_{IXCX},P^{(1)}_{XXCX}, P^{(2)}_d \nonumber \\
    &;1,1,1,1,1,1),
\end{align}
\begin{align}
    &P_{TB2Z} = \Gamma(P^{(2)}_{XICX},P^{(1)}_{IXCX}, P^{(2)}_d; 1,2,1),
\end{align}
\begin{align}
    P_{C2Z} = P_{C1Z},
\end{align}
\begin{align}
    P_{MRZ1} = \Gamma(P^{(1)}_{XXCX},P^{(1)}_{XICX},P^{(3)}_{IXCX}, P^{(2)}_d; 1,2,1,1),
\end{align}
\begin{align}
    P_{MRZ2} = P^{(2D)}_{BLTRZ},
\end{align}
\begin{align}
    P_{C3Z} = P_{C1Z},
\end{align}
\begin{align}
    &P_{BB1Z} = \nonumber \\
    &\Gamma(P^{(2)}_{XICX},P^{(1)}_{IXCX},P^{(2)}_{IXCX},P^{(1)}_{XICX}, P^{(2)}_d; 1,1,1,1,1),
\end{align}
\begin{align}
    P_{BB2Z} = P_{BB1Z},
\end{align}
\begin{align}
    P_{C4Z} = \Gamma(P^{(2)}_{XICX},P^{(1)}_{XICX},P^{(2)}_{IXCX}, P^{(2)}_d; 1,1,1,1),
\end{align}
\begin{align}
    P_{MLZ1} = \Gamma(P^{(1)}_{XXCX},P^{(1)}_{XICX},P^{(2)}_{IXCX}, P^{(2)}_d; 1,2,1,1),
\end{align}
and
\begin{align}
    P_{MLZ2} = P^{(2D)}_{BLTRZ}.
\end{align}

\begin{figure*}
	\centering
	\subfloat[\label{fig:Xgraph3D}]{%
		\includegraphics[width=0.62\textwidth]{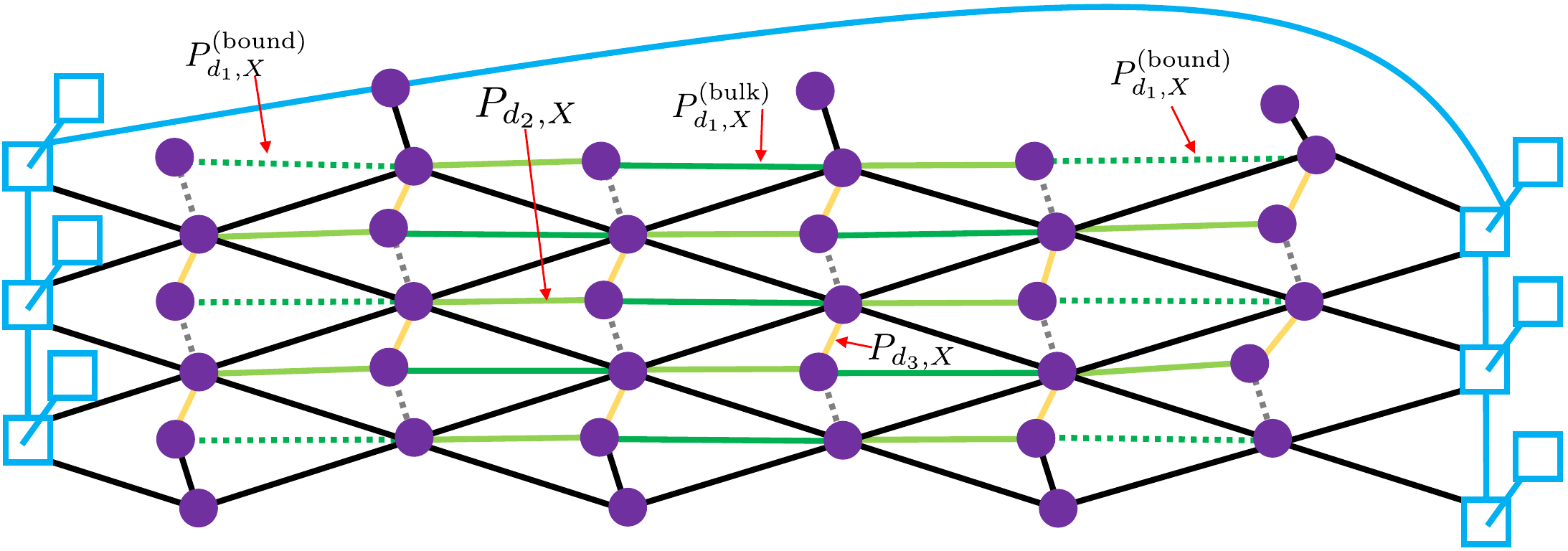}
	}
	\vfill
	\subfloat[\label{fig:Zgraph3D}]{%
		\includegraphics[width=0.55\textwidth]{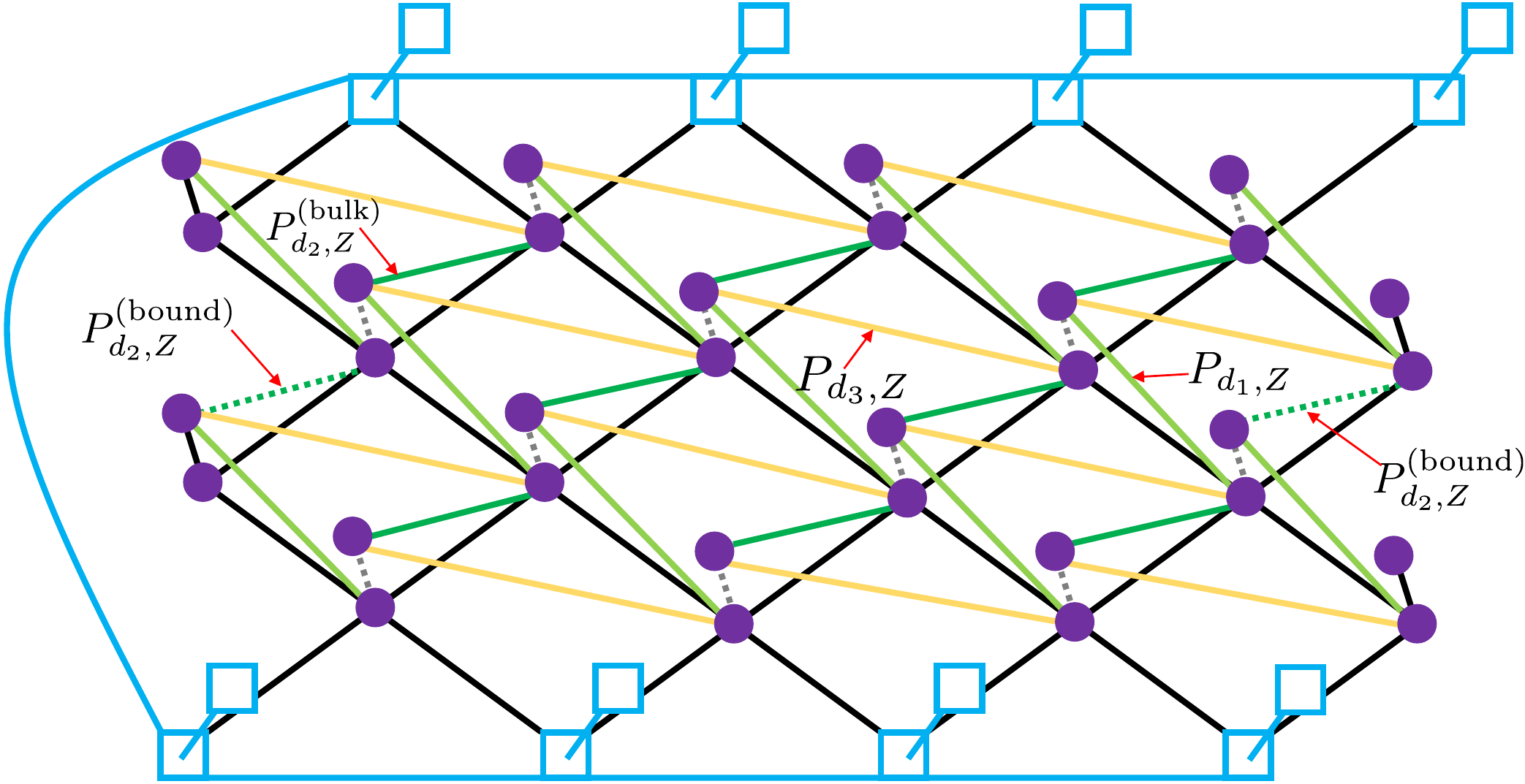}
	}
	\caption{\label{fig:3DgraphsSurfaceCode}(a) Graph used for decoding $X$-type stabilizer measurements which include vertical edges (dashed gray edges) for dealing with measurement errors and space-time correlated edges for correcting errors arising from CNOT gate failures causing two different syndrome measurement outcomes in consecutive rounds. (b) Same as in (a) but for $Z$-type stabilizer measurements.}
\end{figure*}

We now consider the three-dimensional version of the graphs in \cref{fig:Surface2DXgraph,fig:Surface2DZgraph} (which we label $G^{(3D)}_{(d_x)}$ and $G^{(3D)}_{(d_z)}$) to deal with measurement errors in addition to space-time correlated errors arising from CNOT gate failures. As an example, consider an $I \otimes Z$ error arising from a CNOT gate failure in the second time-step of an $X$-type (red) plaquette during the $k$'th syndrome measurement round. Such a failure adds a $Z$ data-qubit error which propagates through the CNOT in the fifth time-step of the top right red $X$-type plaquette. Let $v_j$ and $v_k$ be the vertices corresponding to the measurement outcomes of the two ancilla qubits which would detect the $Z$ error. Assuming there were no other failures, only one of the two vertices (say $v_j$) changes from rounds $k-1$ to round $k$. In the next syndrome measurement round, both $X$-type plaquettes will detect the $Z$ data qubit error the ancilla qubits in both $X$-type plaquettes will be highlighted. Hence only the vertex $v_k$ changes from round $k$ to $k+1$. In order to ensure that the highlighted ancillas arising from failures as in the example considered here can be reached by a single edge when implementing MWPM, the dark green edges in the graph of \cref{fig:Xgraph3D} (labeled $P^{(\text{bulk})}_{d_1,X}$) are added to the graph of \cref{fig:Surface2DXgraph}. The other types of space-time correlated edges are distinguished by their color and associated label (all edges of the same color have identical edge-weight probabilities). Similarly, we add the dashed grey vertical edges in \cref{fig:Xgraph3D,fig:Zgraph3D} connecting identical vertices from two consecutive syndrome measurement rounds to deal with measurement errors. The edge weight probabilities of such edges are labeled $P^{X}_{V}$ and $P^{Z}_{V}$. Note that there are also solid dark vertical edges at some of the boundaries of the graphs where weight-two $X$-type and $Z$-type stabilizers occur in \cref{fig:Surface5by7}. These vertical edges have different edges weight probabilities which are labeled $P^{X,\text{bound}}_{V}$ and $P^{Z,\text{bound}}_{V}$.

In order to avoid making the visualization of the three-dimensional graphs too cumbersome, in \cref{fig:Xgraph3D,fig:Zgraph3D} we only included vertices corresponding to the first two syndrome measurement rounds. Further, the two-dimensional edges from the second round were omitted in order to maintain focus on the vertical and space-time correlated edges connecting vertices from two consecutive syndrome measurement rounds. 

Let
\begin{align}
    P_{VCX} = P^{(ZI)}_{\text{CNOT}} + P^{(ZX)}_{\text{CNOT}} + P^{(YI)}_{\text{CNOT}} + P^{(YX)}_{\text{CNOT}},
    \label{eq:VerticalXgraph}
\end{align}
and
\begin{align}
    P_{VCZ} = P^{(IX)}_{\text{CNOT}} + P^{(IY)}_{\text{CNOT}} + P^{(ZX)}_{\text{CNOT}} + P^{(ZY)}_{\text{CNOT}}.
    \label{eq:VerticalZgraph}
\end{align}
Further, let $P_s$ be the probability of preparing $\ket{-}$ instead of $\ket{+}$ and $P_{m}$ be the probability that a $X$-basis measurement outcome is flipped. The edge weight probabilities corresponding to the dashed grey edges in \cref{fig:Xgraph3D,fig:Zgraph3D} (i.e. the vertical edges of $G^{(3D)}_{(d_x)}$ and $G^{(3D)}_{(d_z)}$) are given by
\begin{align}
     P^{X}_{V} = \Gamma(P_{VCX},P_s,P_m; 4,1,1),
\end{align}
\begin{align}
     P^{X,\text{bound}}_{V} = \Gamma(P_{VCX},P_s,P_m; 2,1,1),
\end{align}
\begin{align}
    P^{Z}_{V} = \Gamma(P_{VCZ},P_s,P_m; 4,1,1),
\end{align}
and
\begin{align}
    P^{Z,\text{bound}}_{V} = \Gamma(P_{VCZ},P_s,P_m; 2,1,1).
\end{align}

\begin{figure*}
	\centering
	\subfloat[\label{fig:CorrelatedGates}]{%
		\includegraphics[width=0.62\textwidth]{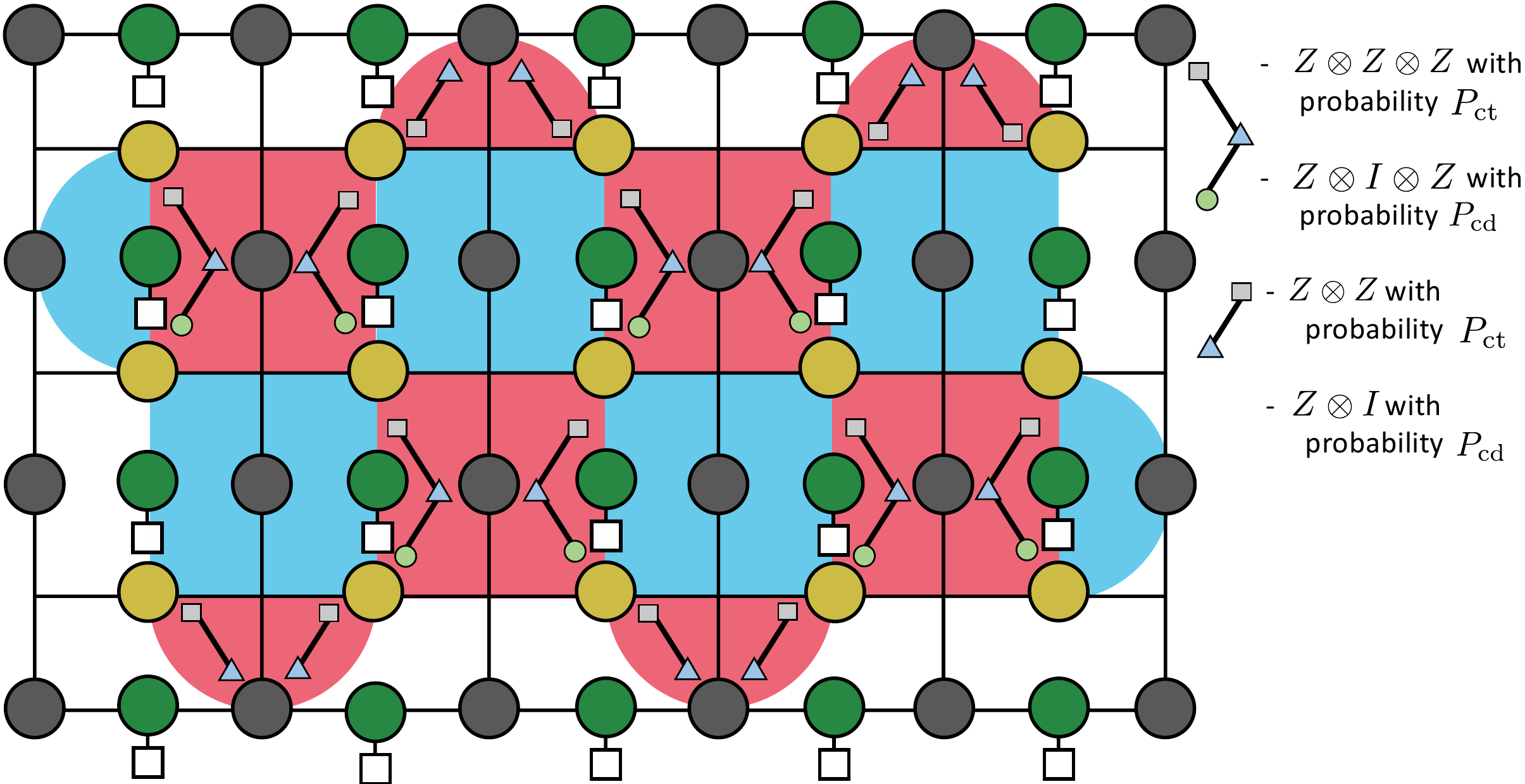}
	}
	\vfill
	\subfloat[\label{fig:XGraph3DCorr}]{%
		\includegraphics[width=0.65\textwidth]{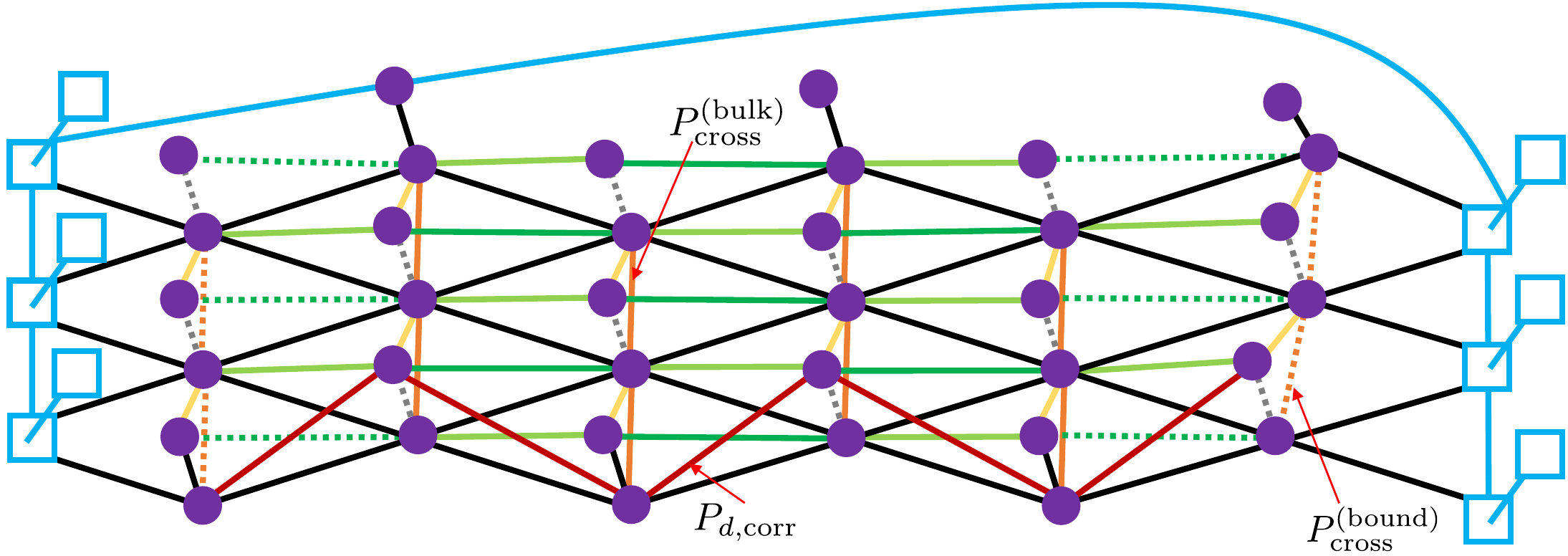}
	}
	\caption{\label{fig:CorrelatedErrGraph}(a) Fictitious identity gates illustrating the possible correlated errors arising before the $X$-basis measurement of the $X$-type ancilla qubits. Grey squares correspond to the first qubit, blue triangles to the second qubit and green circles to the third qubit. (b) $X$-type decoding graph with added edges to correct correlated errors. The edge-weight probabilities of the orange cross-edges are labeled $P_{\text{cross}}$. We also add red edges with edge-weight probabilities labelled $P_{d,\text{corr}}$ at the bottom row of the graph.}
\end{figure*}

Next we consider the edge-weight probabilities for the space-time correlated edges of $G^{(3D)}_{(d_x)}$. The dark green edges labeled by $P^{(\text{bulk})}_{d_1,X}$ have different values at the boundaries (dashed dark green edges in the first and last column of \cref{fig:Xgraph3D}) and are labeled by $P^{(\text{bound})}_{d_1,X}$. We have that
\begin{align}
    P^{(\text{bulk})}_{d_1,X} = \Gamma(P^{(1)}_{IZCX},P^{(1)}_{ZZCX},P^{(2)}_{ZICX}; 1,1,2),
\end{align}
and
\begin{align}
    P^{(\text{bound})}_{d_1,X} = \Gamma(P^{(1)}_{IZCX},P^{(1)}_{ZZCX},P^{(2)}_{ZICX}; 1,1,1).
    \label{eq:PboundD1}
\end{align}
The edge weight probability $P_{d_2,X}$ (represented by the light green edges in \cref{fig:Xgraph3D}) is given by
\begin{align}
    P_{d_2,X} = \Gamma(P^{(1)}_{IZCX},P^{(1)}_{ZZCX}; 1,1).
    \label{eq:PboundD2}
\end{align}
Lastly, the edge weight probability $P_{d_3,X}$ (represented by the yellow edges in \cref{fig:Xgraph3D}) is given by
\begin{align}
    P_{d_3,X} = P_{d_2,X}.
\end{align}

Similarly, for the graph $G^{(3D)}_{(d_z)}$, the edge weight probability $P_{d_1,Z}$ (represented by the light green edges) is given by
\begin{align}
    P_{d_1,Z} = \Gamma(P^{(1)}_{XICX},P^{(1)}_{XXCX}; 1,1).
\end{align}
The bulk and boundary edge weight probabilities $P^{(\text{bulk})}_{d_2,Z}$ (dark green edges) and $P^{(\text{bound})}_{d_2,Z}$ (dashed dark green edges) are given by
\begin{align}
    P^{(\text{bulk})}_{d_2,Z} = \Gamma(P^{(1)}_{XICX},P^{(1)}_{XXCX},P^{(1)}_{IXCX}; 1,1,2),
\end{align}
and
\begin{align}
    P^{(\text{bound})}_{d_2,Z} = \Gamma(P^{(1)}_{XICX},P^{(1)}_{XXCX},P^{(1)}_{IXCX}; 1,1,1).
\end{align}
Lastly, the edge weight probability $P_{d_3,Z}$ (represented by the yellow edges) is given by
\begin{align}
    P_{d_3,Z} = P_{d_1,Z}.
\end{align}

\subsection{Adding edges for dealing with correlated errors}
\label{subsec:CorrErrEdges}

In this section we provide a modified version of the graph $G^{(3D)}_{(d_x)}$ (described in \cref{subsec:SurfaceCodeGraphs}) which includes extra edges to deal with two-qubit and three-qubit correlated errors arising from the micro oscillations described in \cref{subsec:mitigation_optimization}. 

For the purposes of the edge weight analysis, in \cref{fig:CorrelatedGates}, we illustrate fictitious two-qubit and three-qubit gates which act as the identity and which are applied immediately prior to the $X$-basis measurements of the red plaquettes. The two-qubit correlated errors can be viewed as an $Z \otimes I \otimes Z$-type error at a three-qubit gate location, where the $Z$ errors act on the qubits adjacent to the grey squares and green circles of such gates. Such errors occur with probability $P_{\text{cd}}$. Similarly, the three-qubit correlated errors can be viewed as an $Z \otimes Z \otimes Z$-type error at a three-qubit gate location. Such errors occur with probability $P_{\text{ct}}$. Additionally, there can be correlated errors occurring between the ancilla and data qubits at the top and bottom boundaries of the lattice in \cref{fig:CorrelatedGates}. Hence, we add fictitious two-qubit gate locations at such boundaries as shown in the figure. 

In order to incorporate the different types of correlated errors mentioned above into our MWPM decoding protocol, extra edges are added to the graph $G^{(3D)}_{(d_x)}$ as shown in \cref{fig:XGraph3DCorr}. The first type of extra edges are two-dimensional cross edges shown in orange that deal with two and three-qubit correlated errors arising at the three-qubit fictitious gate locations of \cref{fig:CorrelatedGates}. The edge-weight probabilities of such edges are labeled $P^{(\text{bulk})}_{\text{cross}}$. Due to boundary effects, we also add dashed orange edges with edge-weight probabilities labeled $P^{(\text{bound})}_{\text{cross}}$. Additionally, extra space-time correlated edges (shown in red) are added at the bottom row of the graph in \cref{fig:XGraph3DCorr} with edge weight probabilities labeled by $P_{d,\text{corr}}$. Note that the two-qubit correlated errors arising at the top boundary of \cref{fig:CorrelatedGates} result in space-time correlated edges which are already included in $G^{(3D)}_{(d_x)}$.

 \begin{figure}
	\centering
	\includegraphics[width=0.35\textwidth]{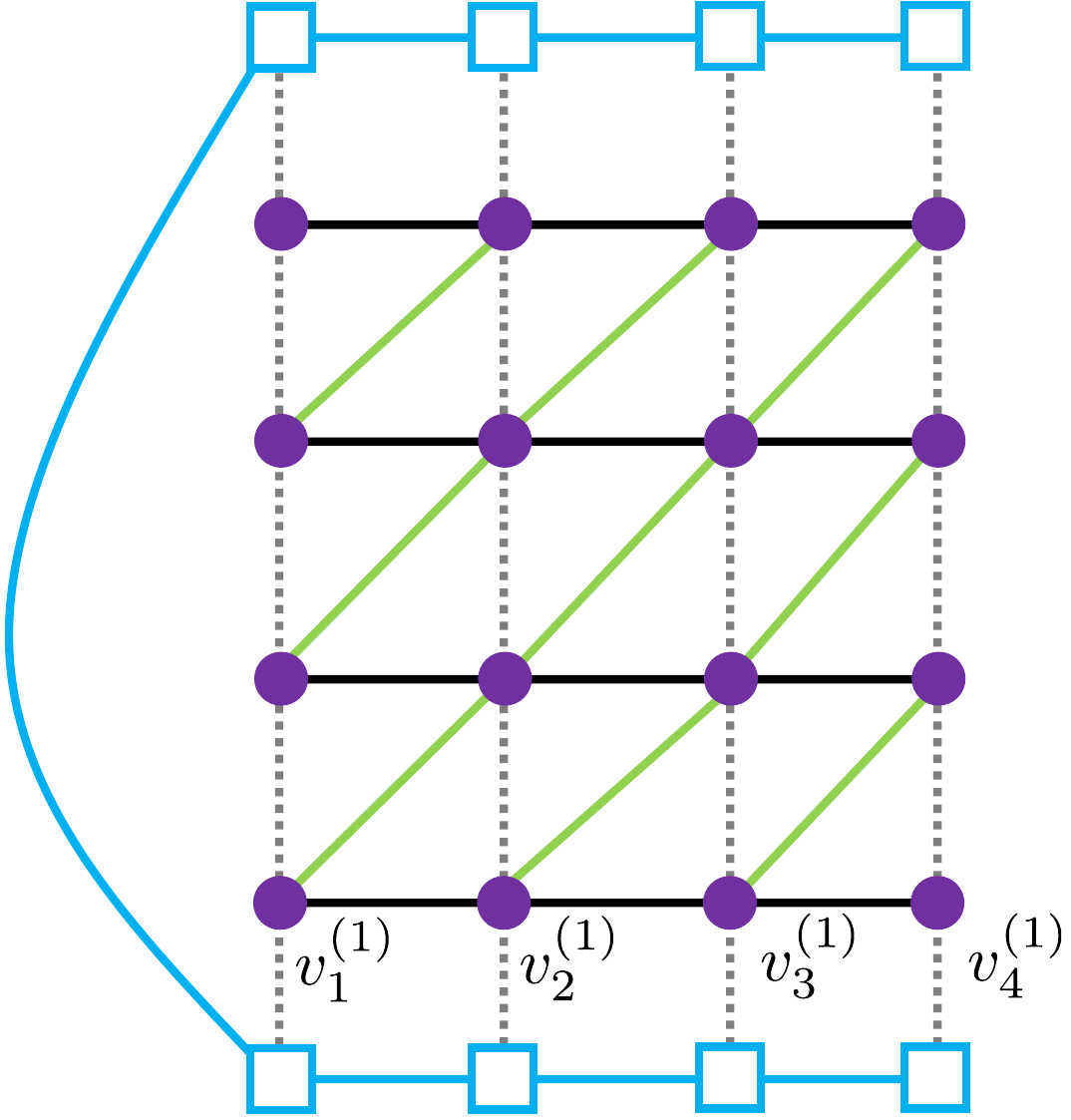}
	\caption{Example of a decoding graph for correcting timelike errors using a $d=5$ repetition code with $d_m = 4$. The top and bottom boundary edges (with zero weight) and vertices are shown in blue and are connected by a blue edge with zero weight. As explained in \cref{subsec:TimeLikeErrors}, we have removed the left and two-dimensional black edges (which correspond to the left and rightmost qubits) to isolate timelike errors.}
	\label{fig:TimeLikeGraph}
\end{figure}

\begin{figure}
	\centering
	\subfloat[\label{fig:ExampleTimeLikeCorr1}]{%
		\includegraphics[width=0.3\textwidth]{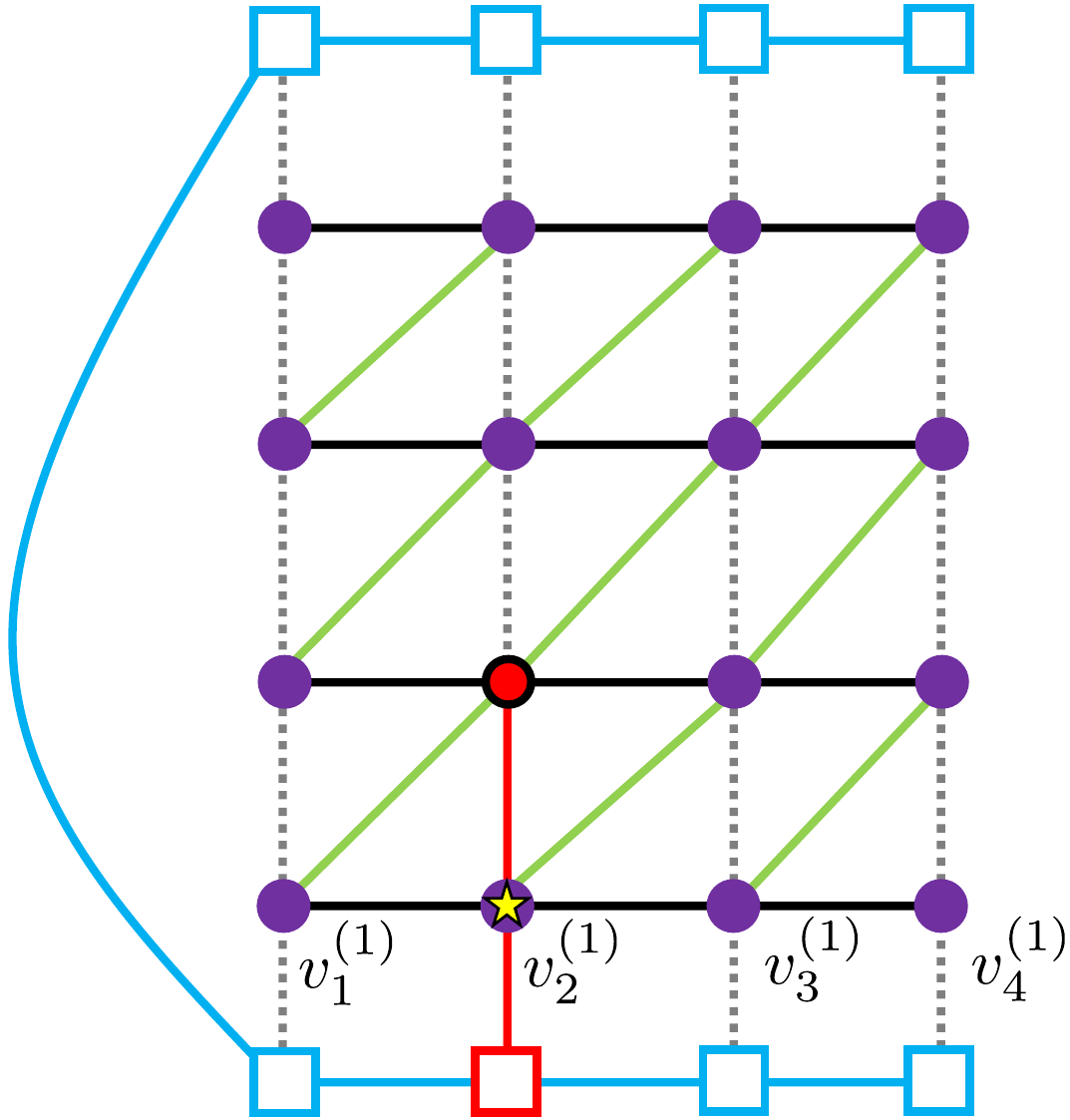}
	}
	\vfill
	\subfloat[\label{fig:ExampleTimeLikeCorr2}]{%
		\includegraphics[width=0.3\textwidth]{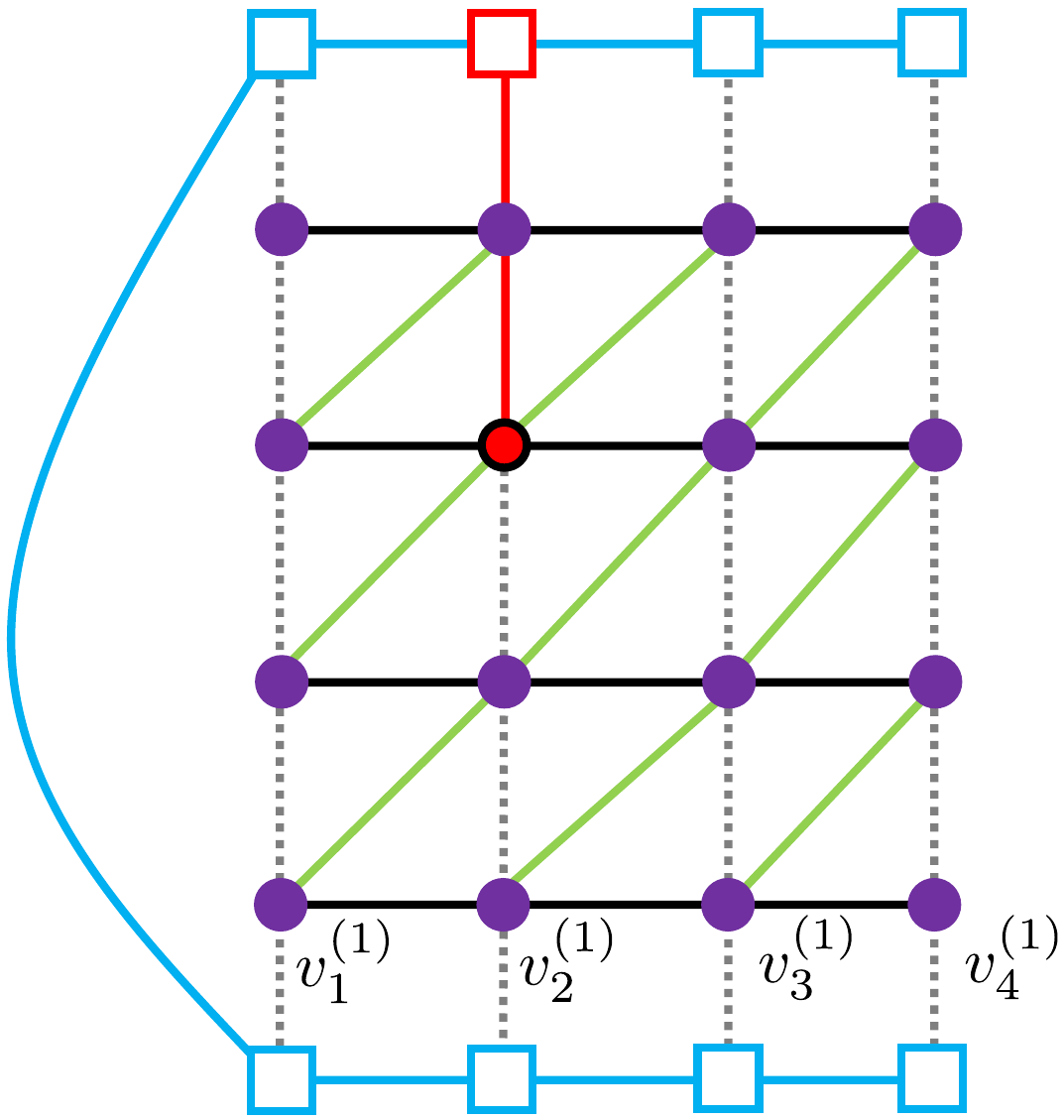}
	}
	\caption{(a) Implementation of the timelike decoding protocol in the presence of a single measurement error when measuring the stabilizer $X_2X_3$ during the first round. The minimum weight path matches to the bottom boundary going through the vertex $v^{(1)}_2$ whose outcome is correctly flipped (illustrated by the yellow star). (b) Same as in (a) but with an additional measurement error occurring in the second round when measuring $X_2X_3$. In this case, the minimum weight path matches to the top boundary and fails to flip the measurement outcome of $v^{(1)}_2$ (which is incorrect given the measurement error in the first round) resulting in a logical failure.}
	\label{fig:ExamplesTimeLikeErrs}
\end{figure}

In addition to the extra edges added to $G^{(3D)}_{(d_x)}$, the edge-weight probabilities of a subset of the edges already included in $G^{(3D)}_{(d_x)}$ need to be renormalized. The edge-weight probabilities of the added edges in addition to the renormalized edges are given by:
\begin{align}
    P^{(\text{bulk})}_{\text{cross}} = \Gamma(P_{\text{ct}}P_{\text{cd}}; 2,2),
\end{align}
\begin{align}
    P^{(\text{bound})}_{\text{cross}} = \Gamma(P_{\text{ct}}P_{\text{cd}}; 1,1),
\end{align}
\begin{align}
    P_{d,\text{corr}} = P_{\text{ct}},
\end{align}
\begin{align}
    P_{TB2X} = \Gamma(P^{(1)}_{ZZCX},P^{(1)}_{IZCX}, P^{(1)}_d,P_{\text{cd}}; 1,1,1,1),
\end{align}
\begin{align}
    P_{TB1X} = \Gamma(P^{(1)}_{ZZCX},P^{(1)}_{IZCX},P^{(2)}_{IZCX}, P^{(1)}_d,P_{\text{cd}}; 2,1,1,1,1),
\end{align}
\begin{align}
    &P_{C2X} = \nonumber \\
    &\Gamma(P^{(3)}_{IZCX},P^{(2)}_{IZCX},P^{(1)}_{ZZCX}, P^{(1)}_d,P_{\text{cd}},P_{\text{ct}}; 1,1,1,1,1,1),
\end{align}
\begin{align}
    P_{BB2X} = \Gamma(P^{(1)}_{ZZCX},P^{(1)}_{IZCX},P^{(1)}_d,P_{\text{cd}}; 1,1,1,1),
\end{align}
\begin{align}
    P_{BB1X} = \Gamma(P^{(1)}_{IZCX},P^{(1)}_{ZZCX},P^{(1)}_{ZICX},P^{(1)}_d,P_{\text{cd}}; 2,1,1,1,1),
\end{align}
\begin{align}
        &P_{C4X} = \nonumber \\
        &\Gamma(P^{(3)}_{IZCX},P^{(1)}_{IZCX},P^{(1)}_{ZICX},P^{(1)}_d,P_{\text{cd}},P_{\text{ct}}; 1,1,1,1,1,1),
\end{align}
\begin{align}
     P^{X}_{V} = \Gamma(P_{VCX},P_s,P_m,P_{\text{ct}}; 4,1,1,1),
\end{align}

For the space-time correlated edges, at the top row of the graph in \cref{fig:XGraph3DCorr}, we have
\begin{align}
    P^{(\text{bound,top})}_{d_1,X} = \Gamma(P^{(1)}_{IZCX},P^{(1)}_{ZZCX},P^{(2)}_{ZICX},P_{\text{ct}}; 1,1,1,1),
\end{align}
whereas at the bottom boundary $P^{(\text{bound,bottom})}_{d_1,X}$ is given by \cref{eq:PboundD1}. Similarly, at the top row of \cref{fig:XGraph3DCorr}, we have
\begin{align}
    P^{(\text{top})}_{d_2,X} = \Gamma(P^{(1)}_{IZCX},P^{(1)}_{ZZCX},P_{\text{ct}}; 1,1,1),
\end{align}
whereas anywhere else in the graph $P_{d_2,X}$ is given by \cref{eq:PboundD2}.

\subsection{Decoding time-like errors}
\label{subsec:DecodeTimeLikeErrors}

In this section, we show how the decoding graphs in addition to the MWPM decoding protocols need to be modified for correcting timelike errors discussed in \cref{subsec:TimeLikeErrors}. Since visualizing three-dimensional graphs can be challenging, we focus on correcting timelike errors in the context of the repetition code, even though timelike errors occur in surface code patches when implementing our lattice surgery schemes. However the main techniques discussed in the context of the repetition can straightforwardly be applied to the rotated surface code. 

An example of a decoding graph for timelike errors occurring in a $d=5$ repetition code with $d_m=4$ is given in \cref{fig:TimeLikeGraph}. Note that unlike \cref{fig:RepCodeGraph}, the boundary edges and vertices (shown in blue) are at the top and bottom of the graph since we follow the matching protocol explained in \cref{fig:TimeLikeCartoon}. In particular, we are considering a setting analogous to \cref{fig:LatticeSurgerySimple}, where data qubits between two repetition code patches are initially prepared in the $\ket{0}$ state, and the product of the $X$-stabilizers yields the outcome $X_{L1}X_{L2}$. Note that although the measurement of each $X$-stabilizer in the first round is random, the parity of the product of all measurement outcomes gives the outcome of $X_{L1}X_{L2}$. Due to possibility of measurement errors, measurements of the $X$-stabilizers are repeated $d_m$ times. MWPM is then performed over the entire syndrome history in order to determine if measurement errors occurred during the measurement of $X$-stabilizers in the first round. We thus summarize the decoding protocol as follows:
\begin{enumerate}
    \item Repeat the measurement of all $X$-stabilizers $d_m$ times.
    \item Implement MWPM using a timelike decoding graph (such as the one in \cref{fig:TimeLikeGraph}). If there is an odd number of highlighted vertices (purple vertices in \cref{fig:TimeLikeGraph}), highlight a boundary vertex (the particular choice is irrelevant). 
    \item Let $v^{(1)}_j$ correspond to the $j$'th $X$-stabilizer measurement outcome in the first round (represent by the $j$'th purple vertex, starting from the left, in the first layer of the graph in \cref{fig:TimeLikeGraph}). If there are highlighted timelike edges (i.e. vertical edges) incident to $v^{(1)}_j$, flip the measurement outcome of $v^{(1)}_j$.
    \item Let $\tilde{v}^{(1)}_j$ correspond to the values of $v^{(1)}_j$ after implementing MWPM and performing the appropriate measurement flips described in the previous step. The outcome $v_f$ of $X_{L1}X_{L2}$ is given by $v_f = \sum^{d-1}_{j = 1} \tilde{v}^{(1)}_j \text{mod}(2)$.
\end{enumerate}

In \cref{fig:ExamplesTimeLikeErrs}, we provide two examples for the implementation of the timelike decoder. In \cref{fig:ExampleTimeLikeCorr1}, we consider the case where a single measurement error occurs in the first round when measuring the stabilizer $X_2X_3$. Since the syndrome changes between the first and second round, the second vertex (starting from the left to right) of the second two-dimensional layer is highlighted. A boundary vertex is also highlighted to ensure the total number of highlighted vertices is even. After implementing MWPM, the minimum weight path connecting the two highlighted vertices correctly passes through $v^{(1)}_2$ in the timelike direction. The decoder then flips the measurement outcome of $X_2X_3$ in the first round resulting in the correct parity for the he outcome of $X_{L1}X_{L2}$. In \cref{fig:ExampleTimeLikeCorr2}, we consider a similar setting but with two consecutive measurement errors of the stabilizer $X_2X_3$ occurring in the first and second round. In this case, the syndrome only changes between the second and third round resulting in the red highlighted vertex shown in \cref{fig:ExampleTimeLikeCorr2}. After implementing MWPM, the minimum weight path connect to the top boundary and so the measurement outcome of $X_2X_3$ in the first round is incorrectly left unchanged resulting in a logical failure. 

We conclude this section with an important remark. Suppose a measurement error occurs in the first round when measuring the $X$-type stabilizer $S^{(x)}_j$ of a given code. In order to prevent highlighted timelike edges from being incident to the vertex $v^{(1)}_j$, one requires additional measurement errors such that minimum weight paths are matched in the top timelike portion of the decoding graph (as in \cref{fig:ExampleTimeLikeCorr2}). By increasing $d_m$ to $d_m + 2$, one requires an additional measurement error to guarantee that the minimum weight path is not incident to $v^{(1)}_j$, thus explaining the scaling in \cref{eq:PMscaling}.

\section{Toffoli state distillation (\TDTOFF)} 
\label{app:TopDown}
 
\subsection{Prior state of the art}
\label{sec:TD_PriorArt}

Here we give a high-level comparison of how our \TDTOFF protocol compares to the prior art in terms of magic state conversion rates.

Early protocols for fault-tolerant quantum computation focused on $\tof$ state preparation in concatenated codes~\cite{Shor96} or they protected against 1 type of error~\cite{Dennis01}.  However, none of these protocols are suitable for protecting against generic noise in topological (e.g. repetition or surface) codes.

A more modern approach to magic state distillation uses a supply of low fidelity $T$ magic states. 
There are many protocols for distillation of noisy $T$-states to purer $T$-states~\cite{BraKit05,Bravyi12,Jones13,Meier13,fowler13,hastings2018distillation}. 
One can also use $T$-states as input to protocols that output other types of magic states, including \tof states~\cite{jones13b,eastin13,Ogorman16,campbell2017unified,campbell2017unifying,haah2017magic}. 
For instance, there were parallel discoveries of protocols~\cite{jones13b,eastin13} that distill 1 \tof state from 8 noisy $T$ states, which we will write as $8 T \rightarrow 1 \tof$.  
This was later generalized using synthillation ~\cite{campbell2017unified,campbell2017unifying} to a family of protocols $(6k+2)T \rightarrow  k \tof $ for any integer $k$. 
However, in some settings, the supply of noisy \tof states can be prepared with better fidelities than the noisy $T$ states. For instance, in this paper we have shown that in system with highly-biased noise we can use a repetition encoding and the \BUTOFF protocol to realize \tof state at better fidelities than physical \tof gates, with only a mild additional resource cost.

It has been previously noted~\cite{paetznick2013universal} that triorthogonal codes enabling $(6k+8)T \rightarrow (2k) T $ state distillation can also be lifted to perform $(6k+8)\tof \rightarrow (2k) \tof$.  The conversion rate of these protocols is $2k/(6k+8)$, which is poor when $k$ is small (starting at $1/7$ for $k=1$) but improving when $k$ is larger (approaching $1/3$ for $k \rightarrow \infty$). However, the ratio of inputs to outputs is not the sole metric of importance; also crucial is the space-time complexity of the Clifford circuit implementing the distillation protocol.  Previous analysis has found that the space-time complexity of Clifford distillation circuits tends to be more favorable for simpler protocols using smaller block sizes~\cite{fowler13,litinski2019game,litinski2019magic} and that this effect can outweigh the improvement of conversion rate in the asymptotic regime. In other words, the desiderata for distillation protocols converting $n \rightarrow k$ magic states, are that: the protocol has a good rate, so $k/n$ is large; the protocol is compact so $n$ is as small as possible. These desiderata are in tension since rates tends to improve asymptotically as block size $n$ is increased. A protocol satisfying these desiderata, will likely have a small space-time footprint when compiled down to physical qubits and gates.   In this work, we present a $8 \tof \rightarrow 2 \tof$ protocol that protects against any single location fault (of $X$, $Y$ or $Z$ type), so it has a relatively high conversion rate of $1/4$ without needing to scale to large blocks.  In contrast, to achieve the same conversion rate using the ideas of Ref.~\cite{paetznick2013universal} would require a much larger $32 \tof \rightarrow 8 \tof$ protocol. 

\subsection{Transversality proofs}
\label{sec:TD_Transversality}

Here we prove that the trio of $[[8,2,2]]$ codes introduced in \cref{Sec:TopDown} have the required $\ccz $ tranversality properties. Recall that \ccz is a 3-qubit gate that adds a ``$-1$" phase to the state $\ket{111}$ and ``$+1$" to all other computational basis states.  The corresponding magic state $\ket{\ccz}$ differs from $\ket{\tof}$ by a single Hadamard gate.  For reasons of mathematical elegance, it is simpler to work mostly in terms of $\ket{\ccz}$ state distillation, but our final description of the distillation protocol will be presented in terms of $\ket{\tof}$ states.

For our trio of codes, each block encodes $k=2$ logical qubits into $n=8$ physical qubits and can detect an error on any single qubit. In the main text, we define the code using \cref{fig:Transversal} and for completeness we give an algebraic definition here.  All three code blocks one $X$-stabilizer $X^{\otimes 8}$ but different logical $X$ operators
\begin{align}  \label{eqs:LogicalOps}
    X_{L1A} & = 
    (X \otimes X \otimes X \otimes \id \otimes \id \otimes X \otimes \id \otimes \id)_A, \\ \nonumber
    X_{L2A} & = 
    (X \otimes X \otimes  \id \otimes \id \otimes X \otimes X \otimes \id \otimes \id)_A , \\ \nonumber
    X_{L1B} & = 
   ( X \otimes X \otimes  \id \otimes \id \otimes X \otimes X \otimes \id \otimes \id)_{B} , \\ \nonumber
    X_{L2B} & = 
   ( X \otimes \id \otimes X \otimes \id \otimes X \otimes \id \otimes X \otimes \id )_{B} , \\ \nonumber 
    X_{L1C} & =  ( X \otimes \id \otimes X \otimes \id \otimes X \otimes \id \otimes X \otimes \id )_{C}, \\ \nonumber 
    X_{L2C}  &= ( X \otimes X \otimes X \otimes \id \otimes \id \otimes X \otimes \id \otimes \id )_{C},
\end{align}
where the index $\{A, B, C \}$ labels the three different codeblocks and the numerical index labels the 2 logical qubits in this code block. We write $( \ldots )_{D=A,B,C}$ to emphasize that the operator acts non-trivially on codeblock $D$ and trivially on other codeblocks. While the code blocks share the same $X$-stabilizer, they will have different $Z$-stabilizers as a consequence of having different logical $X$ operators.

We say a set of  $[[n,k,d]]$ codes is $\ccz$ transversal whenever $\ccz^{\otimes n}$ performs a logical $\ccz^{\otimes k}$ gate.  Note that if we take three copies of a CSS code that has a transversal $T$ gate (so that $T^{\otimes n}=T_{L}$ or similar, then it must also be $\ccz$ transversal).  This is simply because CSS codes have transversal $\cnot$ gates and we can synthesize $\ccz$ gates from $\cnot$ and $T$ gates.  Essentially, this is the observation exploited to construct $(6k+8)\tof \rightarrow (2k) \tof$ protocols~\cite{paetznick2013universal}.  However, it is possible for a trio of codes to be $\ccz$ transversal, but not be $T$ transversal.  To the best of our knowledge this was first shown for the 3D surface codes by showing an equivalence (via unfolding) to 3D colour codes~\cite{kubica2015unfolding}.  Later, Vasmer and Brown gave a more direct proof that the 3D surface codes are $\ccz$ transversal~\cite{vasmer2018universal}.  

Here, we use similar proof techniques to Vasmer and Brown~\cite{vasmer2018universal}, though generalized (to $k>1$) and with a new code construction that code not appear to be a surface code.
From just the $X$-stabilizer and logical operator information, we will prove that our trio of codes are $\ccz$ transversal. The key element of the proof is a lemma relating tranversality to the support of logical $X$ operators and $X$ stabilizers.  The lemma requires that for $j=1,2$, the operators $X_{LjA}$, $X_{LjB}$ and $X_{LjC}$ share support on an odd number of qubit indices.  Furthermore, we need that for any other choice of three $X$ operators (either logical $X$ or $X$ stabilizer) with one selected from each code block, they must share support on an even number of qubit indices. It is easy to verify the operators provided above have this property.

We define codes here using slightly different notation from the main text.  Given an $n$-qubit bit string $\mathbf{s}=(s_1, s_2,  \ldots , s_n)$, we use $X[\mathbf{s}]:= \otimes_j X^{s_j}$.  For example, if
\begin{equation}
    \mathbf{u} = (1,1,1,0,0,1,0,0) ,
\end{equation}
then
\begin{equation}
    X[\mathbf{u}] = X \otimes X \otimes X \otimes \id \otimes \id \otimes X \otimes \id \otimes \id  .
\end{equation}
With this notation we can define an $[[n,k,d]]$ CSS code using a binary $G$-matrix representation as follows. 

Let $G$ be a binary matrix that is row-wise linearly independent and
partitioned as follows
\begin{equation}
    G = \left(  \begin{array}{c}
    G_1  \\ \hline
    G_0
    \end{array}
    \right) ,
\end{equation}
where $G$ has $n$ columns and $G_1$ has $k$ rows.  Letting $m$ denote the number of rows in $G_0$, then for a non-trivial ($ d\geq 2$) code we know $m \geq 1$. Here, we review the relevant facts for $G$-matrices, but for additional details and proofs refer the reader to Refs.~\cite{haah2017magic,campbell2017unifying,campbell2017unified,haah2018codes}.  This allows us to define a CSS code with all-zero logical state
\begin{equation}
    \ket{ (0,\ldots, 0) }_L = 2^{-m/2} \sum_{\mathbf{u} \in \mathbb{F}_2^m} \ket{ \mathbf{u} G_0} .
\end{equation}
Note we use bold-font for row vectors.  The notation $\mathbf{u} G_0$ represents left multiplication of matrix $G_0$ by the row vector $\mathbf{u}$, performed modulo 2, which will produce a length $n$ row-vector describing a physical, computational basis state. The set of all $\mathbf{u} G_0$ corresponds to the row-span of $G_0$ and form a group under addition modulo two.

Furthermore, logical computation basis states can be represented by a $k$-bit string $\mathbf{x}=(x_1, \ldots, x_k)$ as follows
\begin{equation} \label{eq:compdefined}
    \ket{ \mathbf{x} }_L = \frac{1}{\sqrt{|\mathcal{G}_0|}} \sum_{\mathbf{u} \in \mathbb{F}_2^m } \ket{\mathbf{u} G_0 + \mathbf{x} G_1} ,
\end{equation}
where $\mathbf{x} G_1$ is again obtained by matrix multiplication (modulo 2) and is a constant shift identifying a coset of the group generated by addition (modulo 2) of rows of $G_0$.  We can compress this notation slightly by noting
\begin{equation}
\label{eq:neaterGform}
    \mathbf{u} G_0 + \mathbf{x} G_1 = (\mathbf{x},\mathbf{u})G,
\end{equation}
where $(\mathbf{x},\mathbf{u})$ is the row-vector resulting from joining $\mathbf{u}$ and $\mathbf{x}$.  Again, note that \cref{eq:neaterGform} should be read as modulo two and this will be the convention for such expressions throughout the remainder of this appendix.

The $j^{\mathrm{th}}$ logical $X$ operator, denoted $X_{Lj}$, ought to flip the $\ket{\mathbf{0}}_L$ state to $\ket{(\hat{\mathbf{o}}_j)}_L$ state, where $\hat{\mathbf{o}}_j$ is a unit vector with a single ``1" entry at the $j^{\mathrm{th}}$ location.  It is straightforward to verify that $X_{Lj}=X[\hat{\mathbf{o}}_j G_1]$ performs the required flip and that $\hat{\mathbf{o}}_j G_1$ is equal to the  $j^{\mathrm{th}}$ row of $G_1$. Therefore, the logical operators of the code are given by the row vectors of $G_1$. Furthermore, for every $\mathbf{g}$ in the row-span of $G_0$, the operator $X[\mathbf{g}]$ is an $X$-stabilizer of the codespace, and this enumerates all the $X$-stabilizers. 

As a final notational preliminary, we will make use of a triple dot product between triples of vectors. If $\mathbf{a}$, $\mathbf{b}$ and $\mathbf{c}$ are binary vectors of equal length, we define
\begin{equation} \label{eq:wedge}
    | \mathbf{a} \wedge \mathbf{b} \wedge \mathbf{c} |= \sum_j a_j b_j c_j \pmod{2},
\end{equation}
which we again evaluate modulo 2.  It is useful to note that this counts the parity of the number of locations where operators $X[\mathbf{a}]$, $X[\mathbf{b}]$ and $X[\mathbf{c}]$ all act non-trivially.

This $G$-matrix representation was also used for triorthogonal codes~\cite{Bravyi12} and quasi-triorthogonal codes~\cite{campbell2017unifying,campbell2017unified} except we are interested in different transversality properties and so we will require different constraints on the weight of rows in $G_0$ and $G_1$. The additional constraints determine the transversality properties that we summarise with the following result, which is a slight generalization (beyond $k=1$) of the proof techniques used by Vasmer and Browne~\cite{vasmer2018universal}
\begin{lemma} \label{lem:TransversalityLemma}
Let $\{ G^{A} , G^{B}, G^{C} \}$ be a trio of $G$-matrices that represent a trio of $[[n,k,d]]$ codes. Additionally, assume the following triple dot product conditions (recall Eq.~\ref{eq:wedge}) 
\begin{equation} \label{eq:Condition}
      | \hat{\mathbf{o}}_p G^A \wedge  \hat{\mathbf{o}}_q G^B   \wedge  \hat{\mathbf{o}}_r G^C | = \begin{cases} 
      1 \mbox{ if } p=q=r \leq k \\
      0 \mbox{ otherwise } \\
      \end{cases}
\end{equation}
where $\hat{\mathbf{o}}_p$ is a binary unit vector with $1$ in location $p$ and 0 everywhere else.  Then it follows that a physical $\ccz^{\otimes n}$ realizes a transveral, logical $\ccz^{\otimes k}$.
\end{lemma}
Let us remark on what \cref{eq:Condition} means in terms of operators. Observe that when $p \leq k$, the operator $X[\hat{\mathbf{o}}_p G^D]$ is the $p^{\mathrm{th}}$ logical operator for codeblock $D \in \{A,B,C \}$.  Therefore, the condition of \cref{eq:Condition} tells us that the $X_{Lp}$ logical operators must share an odd number of qubit indices where they all act non-trivially.  All other combinations of logical operators and stabilizers have an even number of such locations.  

\begin{proof}  To determine the phase acquired from acting on the codespace with $\ccz^{\otimes n}$, we first ask how this operator acts on an arbitrary computational basis state.  Recall that $\ccz^{\otimes n} = \prod_{j=1}^n \ccz_j$ where $\ccz_j$ acts on qubit $j$ in each block.  Given a triple of $n$-qubit binary vectors $\mathbf{a}$, $\mathbf{b}$ and  $\mathbf{c}$ we have
\begin{equation}
    \ccz_j \ket{\mathbf{a}} \ket{\mathbf{b}} \ket{\mathbf{c}} = (-1)^{a_j b_j c_j } \ket{\mathbf{a}} \ket{\mathbf{b}} \ket{\mathbf{c}} ,
\end{equation}
and so
\begin{equation} \label{eq:EachTermTrans}
    \ccz^{\otimes n} \ket{\mathbf{a}} \ket{\mathbf{b}} \ket{\mathbf{c}} = (-1)^{| \mathbf{a} \wedge \mathbf{b} \wedge \mathbf{c} | } \ket{\mathbf{a}} \ket{\mathbf{b}} \ket{\mathbf{c}} ,
\end{equation}
where $ | \mathbf{a} \wedge \mathbf{b} \wedge \mathbf{c} |=\sum_j a_j b_j c_j$ as we introduced earlier.  Next, we ask how this acts on the codespace.

Consider a trio of computational basis states $\ket{\mathbf{x}}_L\ket{\mathbf{y}}_L\ket{\mathbf{z}}_L$ encoded in blocks $A$, $B$ and $C$ respectively.  Using \cref{eq:compdefined,eq:neaterGform}, we see that
\begin{align}
    & \ket{\mathbf{x}}_L\ket{\mathbf{y}}_L\ket{\mathbf{z}}_L  \\ \nonumber
    & = 2^{-3m/2} \sum_{\mathbf{u},\mathbf{v},\mathbf{w} \in \mathbb{F}_2^m} \ket{ (\mathbf{x},\mathbf{u})G^A} \ket{(\mathbf{y},\mathbf{v})G^B } \ket{(\mathbf{z},\mathbf{w}) G^C}.
\end{align}
 To determine the phase acquired from acting on $\ket{\mathbf{x}}_L\ket{\mathbf{y}}_L\ket{\mathbf{z}}_L$ with $\ccz^{\otimes n}$, we consider its action on each term in the superposition using \cref{eq:EachTermTrans}.  Each term acquires a phase
\begin{align} \label{eq:ActionEachTerm}
  & \ccz^{\otimes n} \ket{ (\mathbf{x},\mathbf{u})G^A} \ket{(\mathbf{y},\mathbf{v})G^B } \ket{(\mathbf{z},\mathbf{w}) G^C} \\ \nonumber
   & = (-1)^{\lambda} \ket{ (\mathbf{x},\mathbf{u})G^A} \ket{(\mathbf{y},\mathbf{v})G^B } \ket{(\mathbf{z},\mathbf{w}) G^C},
\end{align}
where the phase exponent is
\begin{align}
 \lambda & =  |( \mathbf{x},\mathbf{u} ) G^A \wedge (\mathbf{y},\mathbf{v})G^B \wedge (\mathbf{z},\mathbf{w}) G^C |.
\end{align}
Using linearity of the triple dot-product and expanding the vectors in terms of unit-vectors, e.g $(\mathbf{x}, \mathbf{u} )= \sum_p \hat{\mathbf{o}}_p (\mathbf{x}, \mathbf{u} )_p $, we have
\begin{align}
 \lambda & =  \sum_{p,q,r}  (\mathbf{x},\mathbf{u} )_p  (\mathbf{y},\mathbf{v})_q (\mathbf{z},\mathbf{w})_r  | \hat{\mathbf{o}}_p G^A \wedge  \hat{\mathbf{o}}_q G^B \wedge \hat{\mathbf{o}}_r G^C |.
\end{align}
Next, using the assumption of \cref{eq:Condition}, we see almost all these terms vanish except a few when $p=q=r\leq k$
\begin{align}
 \lambda & =  \sum_{p \leq k} (   \mathbf{u},\mathbf{x} )_p  (\mathbf{v},\mathbf{y})_p  (\mathbf{w},\mathbf{z})_p  .
\end{align}
Notice that if $p \leq k$, $( \mathbf{x}  , \mathbf{u})_p = ( \mathbf{x})_p$ since $\mathbf{x}$ is length $k$. Therefore,
\begin{align}
 \lambda & =  \sum_{p \leq k} ( \mathbf{x} )_p  (\mathbf{y})_p (\mathbf{z})_p  \\ \nonumber
 & = | \mathbf{x} \wedge \mathbf{y} \wedge \mathbf{z} | ,
\end{align}
where in the last line we have noted that the summation is exactly the triple dot-product between these vectors.  Substituting this back into \cref{eq:ActionEachTerm} we have
\begin{align}
 &  \ccz^{\otimes n} \ket{ (\mathbf{x},\mathbf{u})G^A} \ket{(\mathbf{y},\mathbf{v})G^B } \ket{(\mathbf{z},\mathbf{w}) G^C} \\ \nonumber
   & = (-1)^{| \mathbf{x} \wedge \mathbf{y} \wedge \mathbf{z} |}  \ket{ (\mathbf{x},\mathbf{u})G^A} \ket{(\mathbf{y},\mathbf{v})G^B } \ket{(\mathbf{z},\mathbf{w}) G^C}. 
\end{align}
Since the dependence on $\mathbf{u}$, $\mathbf{v}$ and $\mathbf{w}$ has vanished, $\ccz^{\otimes n}$ acts identically on every term in the superposition comprising the logical computation basis states so we have
\begin{align}
   \ccz^{\otimes n} \ket{\mathbf{x}}_L\ket{\mathbf{y}}_L\ket{\mathbf{z}}_L 
   & = (-1)^{| \mathbf{x} \wedge \mathbf{y} \wedge \mathbf{z} |}  \ket{\mathbf{x}}_L\ket{\mathbf{y}}_L\ket{\mathbf{z}}_L .
\end{align}
This is precisely the phase expected from $\ccz^{\otimes k}=\prod_j \ccz_{Lj}$ since each $\ccz_{Lj}$ contributes one term $x_j y_j z_j$ to the phase exponent. \end{proof}

We remark that the above proof closely follows previous work on 3D surface codes~\cite{vasmer2018universal} but generalised to arbitrary $k$.  This approach could be further extended using a proof technique similar to Refs.~\cite{campbell2017unifying,campbell2017unified} to cover cases where:  the logical unitary is not $\ccz^{\otimes k}$ but some other non-Clifford unitary; and/or the full codespace is not necessarily divisible into 3 equal sized blocks.  However, this more sophisticated approach is not required for our present purposes.

Rather, we are interested in the special case
\begin{lemma} \label{lem:examples}
Consider a trio of $G$-matrices as follows
  \begin{equation} \label{Trio}
      G^{A}=\left( \begin{array}{c}
         \mathbf{u}_1 \\  \mathbf{u}_2    \\ \hline
          \mathbf{1}  
      \end{array} \right) , G^{B}=\left( \begin{array}{c}
         \mathbf{u}_2  \\  \mathbf{u}_3    \\ \hline
          \mathbf{1}  
      \end{array} \right) , G^{C}=\left( \begin{array}{c}
         \mathbf{u}_3  \\ \mathbf{u}_1   \\ \hline
          \mathbf{1}  
      \end{array} \right),
  \end{equation}
where $\mathbf{1}=(1,1,\ldots,1)$.  Assume that
\begin{enumerate}
    \item $\forall t$:  $|\mathbf{u}_t|=\sum_{j=1}^n (\mathbf{u}_t)_j \pmod{2}=0$  ;
    \item $\forall t, t'$: $|\mathbf{u}_t \wedge \mathbf{u}_{t'} |=\sum_{j=1}^n (\mathbf{u}_t)_j (\mathbf{u}_{t'})_j \pmod{2}=0$  ;
    \item $|\mathbf{u}_1 \wedge \mathbf{u}_{2} \wedge \mathbf{u}_{3} |=1$.
\end{enumerate}
Then the corresponding codes are $[[n,2,2]]$ codes with a tranversal logical $\ccz^{\otimes n} = \ccz_L^{\otimes 2}$.   For instance, these conditions are met by setting
\begin{align} \label{eq:Uvector1}
    \mathbf{u}_1 &=  (1,1,1,0,0,1,0,0), \\  \label{eq:Uvector2}
    \mathbf{u}_2 &= (1, 1, 0, 1, 1, 0, 0, 0),  \\ \label{eq:Uvector3}
    \mathbf{u}_3 &= (1, 0, 1, 0, 1, 0, 1, 0) , 
\end{align}
to produce a trio of $[[8,2,2]]$ codes with $\ccz$ transversality as above.
\end{lemma}
 The above lemma provides an example trio of $[[8,2,2]]$ codes with the desired transversality property. To be more concrete, by combining \cref{Trio} and \cref{eq:Uvector1,eq:Uvector2,eq:Uvector3} the trio of codes have $G$-matrix representation
\begin{align} \label{eq:Galpha}
    G^A & = \left( \begin{array}{cccccccc}
        1&1&1&0&0&1&0&0   \\
        1&1&0&1&1&0&0&0 \\ \hline 
         1&1&1&1&1&1&1&1
    \end{array} \right) , \\  \label{eq:Gbeta}
     G^B & = \left( \begin{array}{cccccccc}
        1&1&0&1&1&0&0&0 \\ 
        1&0&1&0&1&0&1&0 \\ \hline 
         1&1&1&1&1&1&1&1
    \end{array} \right) , \\  \label{eq:Ggamma}
         G^C & = \left( \begin{array}{cccccccc}
         1&0&1&0&1&0&1&0 \\ 
         1&1&1&0&0&1&0&0 \\ \hline 
         1&1&1&1&1&1&1&1
    \end{array} \right)  ,
\end{align} 
By translating these matrices into $X$-stabilizers (all have the  $X[\mathbf{1}]=X^{\otimes 8}$ stabilizer) and logical $X$ operators (which differ), we verify that these are the same codes as specified by the operators given the main text (see e.g. \cref{eqs:LogicalOps}).  However, the lemma provides some general conditions under which transversality is satisfied to provide a better insight into the proof technique.

\begin{proof}The proof of \cref{lem:examples} follows quickly from \cref{lem:TransversalityLemma} by simply verifying all the cases. For instance, for $p=q=r$ we have
\begin{align}
| \hat{\mathbf{o}}_1 G^A \wedge  \hat{\mathbf{o}}_1 G^B   \wedge  \hat{\mathbf{o}}_1 G^C |& =| \mathbf{u}_1 \wedge \mathbf{u}_2 \wedge \mathbf{u}_3 | = 1 , \\ \nonumber
| \hat{\mathbf{o}}_2 G^A \wedge  \hat{\mathbf{o}}_2 G^B   \wedge  \hat{\mathbf{o}}_2 G^C |& =| \mathbf{u}_2 \wedge \mathbf{u}_3 \wedge \mathbf{u}_1 | = 1, \\ \nonumber| \hat{\mathbf{o}}_3 G^A \wedge  \hat{\mathbf{o}}_3 G^B   \wedge  \hat{\mathbf{o}}_3 G^C |& =| \mathbf{1} \wedge \mathbf{1} \wedge \mathbf{1} | = 0 .
\end{align}
In the second line, we have used that the triple dot product is invariant under permutation of vectors, for instance $| \mathbf{a} \wedge \mathbf{b} \wedge \mathbf{c} | =  | \mathbf{b} \wedge \mathbf{c} \wedge \mathbf{a} |$. The last equality in each line comes from the assumptions in \cref{lem:examples}.  Since $k=2$, we see that we indeed get unity when $p=q=r\leq k$ and zero for $p=q=r > 2$ (as required by \cref{eq:Condition}). Let us consider a case when $p,q,r \leq k$ but $p \neq q$, such as
\begin{align}
| \hat{\mathbf{o}}_2 G^A \wedge  \hat{\mathbf{o}}_1 G^B   \wedge  \hat{\mathbf{o}}_1 G^C |& =| \mathbf{u}_2 \wedge \mathbf{u}_2 \wedge \mathbf{u}_3 | \\ \nonumber
& = | \mathbf{u}_2  \wedge \mathbf{u}_3 | = 0 .
\end{align}
We have used the simple identity that in $\mathbb{F}_2$ we have $a^2b= ab$ and the natural extension to vectors that $| \mathbf{a} \wedge \mathbf{a} \wedge \mathbf{b} |= | \mathbf{a}  \wedge \mathbf{b} |$. The last equality comes from the assumptions in \cref{lem:examples} and gives the result required by \cref{eq:Condition}. By inspecting \cref{Trio}, we find that for any triple of rows (except for the special case when $p=q=r$) from the upper block $G_1$, two of the selected rows will be equal and so the triple dot product will again give zero, therefore satisfying \cref{eq:Condition}.   

Next, let us consider a case when one row comes from $G_0$, for instance $q=3$ and so
\begin{align}
| \hat{\mathbf{o}}_1 G^A \wedge  \hat{\mathbf{o}}_2 G^B   \wedge  \hat{\mathbf{o}}_3 G^C |& =| \mathbf{u}_1 \wedge \mathbf{u}_3 \wedge \mathbf{1} | \\ \nonumber
& =|  \mathbf{u}_1 \wedge \mathbf{u}_3 | \\ \nonumber
& =0 .
\end{align}
In the second line, we have use that $ a \cdot 1 =a$ extends to vectors so that in general $| \mathbf{a} \wedge \mathbf{b} \wedge \mathbf{1} | =| \mathbf{a} \wedge \mathbf{b} |$.  The last line uses assumption 2 of \cref{lem:examples}.  Indeed, whenever one (or more) of the rows is $\mathbf{1}$, we will be able to deploy assumption 1 (or 2) of \cref{lem:examples}.  This enumerates all possible cases and confirms that \cref{eq:Condition} always holds, therefore proving the main transversality statement of
 \cref{lem:examples}.  
 
 Lastly, that \cref{eq:Uvector1,eq:Uvector2,eq:Uvector3} satisfy assumptions 1-3 of \cref{lem:examples} is easily verified.  For example, 
\begin{align} \nonumber
    |\mathbf{u}_1 \wedge \mathbf{u}_{2} \wedge \mathbf{u}_{3} |& =  (\mathbf{u}_1 )_1 (\mathbf{u}_2 )_1 (\mathbf{u}_3 )_1 + \sum_{j=2}^8  (\mathbf{u}_1 )_j (\mathbf{u}_2 )_j (\mathbf{u}_j )_j \\ \nonumber
    & =  1 + \sum_{j=2}^8  0 = 1 ,
\end{align}
where in the first line we split off the $j=1$ case from the rest of the summation to highlight that this is the only non-zero term.   A deeper explanation is provided by noticing that the example vectors $\mathbf{u}_t$ correspond to generators of a  Reed-Muller code for which these properties are well-known~\cite{Sloane}.\end{proof}

\begin{figure*}
    \centering
    \includegraphics[width=2 \columnwidth]{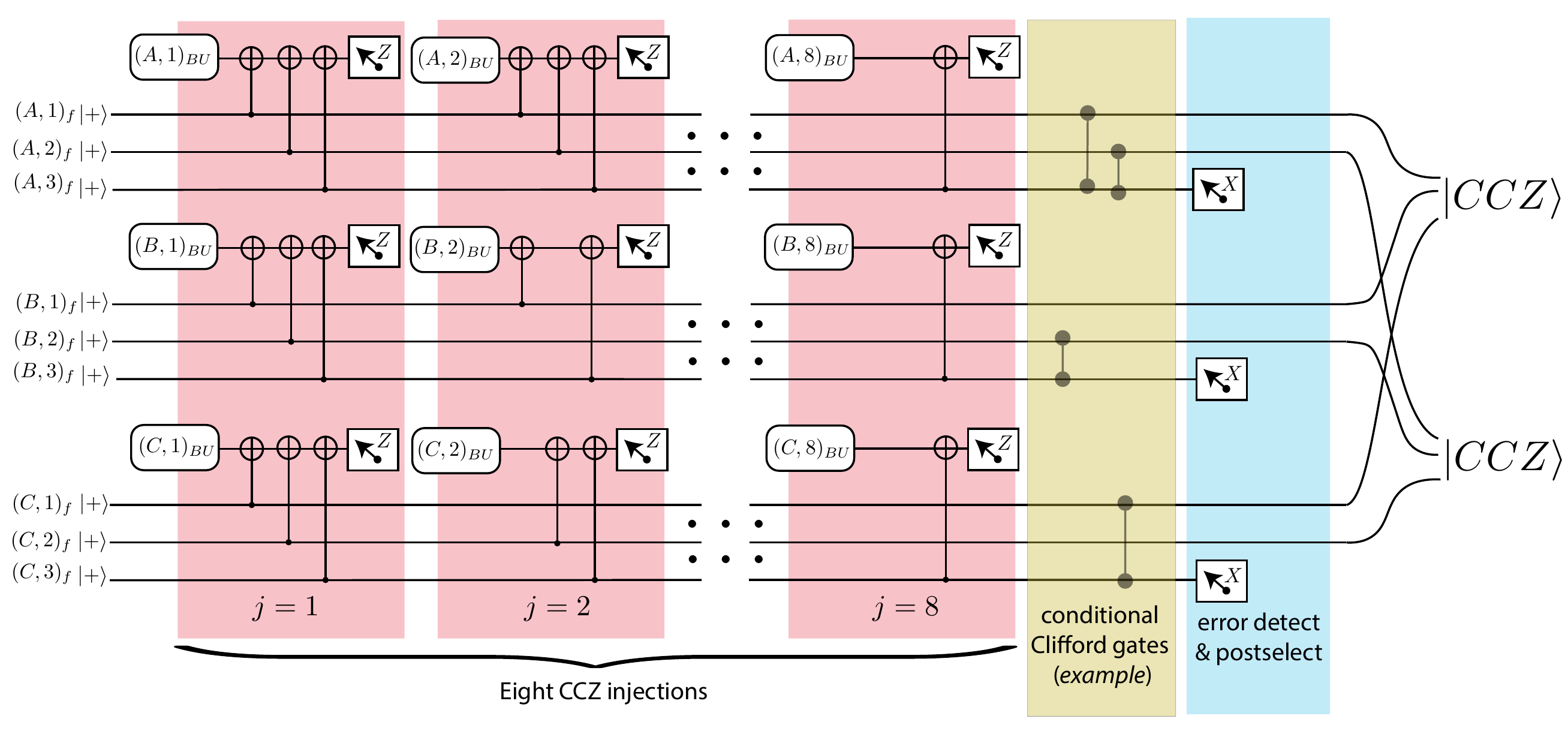}
    \caption{A magic state distillation protocol for $8 \ccz \rightarrow 2 \ccz$ with the eight $\ccz$ injections performed using \cref{algo:CNOTcircuit}. Qubit labels of form $(D,i)_f$ and $(D,j)_{BU}$ follow notation of \cref{def:QubitLabels}.  For each $j$, the triple of qubits $(A,1)_{BU}$, $(B,2)_{BU}$ and $(C,3)_{BU}$ are prepared in a noisy $\ccz$ states (e.g. using \BUTOFF) but this preparation is not shown. We show explicitly the $\cnot$ gates for the first 2 steps and the last step, but omit the middle steps for brevity.  The full circuit is reproducible using \cref{eq:Galpha,eq:Gbeta,eq:Ggamma} to specify the $\cnot$ pattern as outlined in \cref{algo:CNOTcircuit}.}
    \label{fig:CNOTcircuitforTDTOFF}
\end{figure*}

\subsection{Trading space and time}
\label{App:SpaceTimeTrade}

Here we construct a magic state distillation protocol from the $G$-matrix representation of \cref{sec:TD_Transversality} that minimizes space requirements.  The intuition is that one never encodes into the full codespace but rather converts the $\ccz^{\otimes n}$ gate into a product of $n$ conjugated $\ccz$ gates that we can think of as being conjugated by some partial encoding unitary. 

In particular, consider some $G^{D}$-matrix representing an $[[n,k,d]]$ code and a unitary $V^{D}$ such that
\begin{equation} \label{eq:encodingClifford}
    V^{D} \ket{\mathbf{x}} \ket{\mathbf{u}} \ket{\mathbf{0}} = \ket{(\mathbf{x},\mathbf{u} ) G^D} ,
\end{equation}
where $\mathbf{x}$ is a length $k$ bit-string and $\mathbf{u}$ is length $m$ (recall $m$ the number of rows in $G_0^D$). Furthermore, it is known that such a unitary $V^{D}$ can be found that is Clifford and composed solely of $\cnot$ gates~\cite{campbell2017unifying}. It follows that
\begin{align}
    V^{D} \ket{\mathbf{x}} \ket{+^{\otimes m}} \ket{\mathbf{0}} & = 2^{m/2} \sum_{\mathbf{u} \in \mathbb{F}_2^m} V^{D} \ket{\mathbf{x}} \ket{\mathbf{u}} \ket{\mathbf{0}}  \\ \nonumber
    & =  2^{m/2} \sum_{\mathbf{u} \in \mathbb{F}_2^m} \ket{(\mathbf{x},\mathbf{u} ) G^D}  \\ \nonumber
        & = \ket{\mathbf{x}}_L ,
\end{align}
where the second line uses \cref{eq:encodingClifford} and the last line uses \cref{eq:compdefined}.  This confirms that $V^{D}$ is an encoding unitary for the code associated with $G^D$.  To encode the logical state $\ket{+^{\otimes k}}_L$, we simply use linearity so that
\begin{align}
    V^{D} \ket{+^{\otimes k}} \ket{+^{\otimes m}} \ket{\mathbf{0}}   & = \ket{+^{\otimes k}}_L.
\end{align}
Given three codeblocks, we can encode simultaneously with $V=(V^A \otimes V^B \otimes V^C)$. The all $\ket{+}$ state encoded across three codeblocks is then
\begin{align}
    V (\ket{+^{\otimes k}} \ket{+^{\otimes m}} \ket{\mathbf{0}})^{\otimes 3}   & = \ket{+^{\otimes 3k}}_L .
\end{align}
A standard recipe for magic state distillation protocols~\cite{BraKit05,Bravyi12,campbell2017unified,campbell2017unified} is to encode into logical $\ket{+}$ states, perform tranversal non-Cliffords as follows
\begin{align}
  \ccz^{\otimes n} V  (\ket{+^{\otimes k}} \ket{+^{\otimes m}} \ket{\mathbf{0}})^{\otimes 3} & = \nonumber (\ccz\ket{+^{\otimes 3}}_L)^{\otimes k} \\
  & = \ket{\ccz}_L^{\otimes k} ,
\end{align}
which produces $k$ logical $\ccz$ states.  Decoding gives
\begin{align}
\label{eq:encodedCCZ}
 & (V^\dagger\ccz ^{\otimes n} V)  (\ket{+^{\otimes k}} \ket{+^{\otimes m}} \ket{\mathbf{0}})^{\otimes 3} \\
 & = V^\dagger  \ket{\ccz}_L^{\otimes k} \\ \nonumber
 & = \ket{\ccz ^{\otimes k}} \ket{+^{\otimes 3m}} \ket{\mathbf{0}^{\otimes 3}},
\end{align}
where in the last line we have slightly abused qubit ordering to collect together the physical $\ket{\ccz}$ state output. In the case of a detectable error, at least one of the $\ket{+^{\otimes 3m}}$ qubits will be phase flipped and detected by an $X$-measurement.

To reduce space overhead, we observe that the $\ket{\mathbf{0}^{\otimes 3}}$ qubits effectively play no role here. Furthermore, the unitary $V^\dagger\ccz ^{\otimes n} V$ acts non-trivially only on the first $3(k+m)$ qubits, so that the $\ket{\mathbf{0}^{\otimes 3}}$ qubits (a total of $3(n-m-k)$ qubits) are truly surplus to requirement.  Using our earlier notation $V^\dagger\ccz ^{\otimes n} V= \prod_j V^\dagger\ccz _j V$ where $\ccz_j$ acts on qubit $j$ of each block, one then has that
\begin{align} \label{eq:GenCCZ}
  & ( V^\dagger\ccz _j V )  \ket{(\textbf{x},\textbf{u})}\ket{(\textbf{y},\textbf{v})}\ket{(\textbf{z},\textbf{w})} \\ \nonumber
  & =(-1)^{[ (\textbf{x},\textbf{u}) G^A ]_j  [ (\textbf{y},\textbf{v}) G^B ]_j (\textbf{z},\textbf{w}) G^C ]_j } \ket{(\textbf{x},\textbf{u})}\ket{(\textbf{y},\textbf{v})}\ket{(\textbf{z},\textbf{w})} \\ \nonumber
   & =(-1)^{ (\textbf{x},\textbf{u}) [G^A ]_j  \cdot (\textbf{y},\textbf{v}) [G^B ]_j \cdot (\textbf{z},\textbf{w}) [G^C ]_j } \ket{(\textbf{x},\textbf{u})}\ket{(\textbf{y},\textbf{v})}\ket{(\textbf{z},\textbf{w})} ,
\end{align}
where $[\ldots]_j$ denotes the $j^{\mathrm{th}}$ element of the vector inside or the $j^{\mathrm{th}}$ column of a matrix. Notice,  we have suppressed the presence of the redundant $\ket{\mathbf{0}}$ qubits. 

In \cref{App:PerformV} we describe two concrete implementations of the $V \ccz V^\dagger$ gates.
Of course, it is crucial that the space reduction and $V \ccz V^\dagger$ implementation does not distort the way errors propagate and that error correction properties are retained, which is proven from first principles in \cref{App:ErrorProp}.

\subsection{Implementing Conjugated-CCZ gates}
\label{App:PerformV}

Here we give explicit implementations for the 8 conjugated-$\ccz $ gates described in \cref{eq:GenCCZ}. Any such gate can be realized using a single $\ccz $ magic state and we give further details for two different implementations: the first implementation uses $\cnot$ gates and single qubit measurements (\cref{App:PerformVwithCNOT}); the second implementation uses only multi-qubit Pauli measurements via lattice surgery (\cref{app:GenCCinject}).

Herein, we label qubits as follows. 
\begin{definition}[Qubit labels] \label{def:QubitLabels}
Consider a magic state distillation protocol using $n$ noisy $\ccz $ magic states and $G^D$ matrices with $k+m$ rows. We label each input magic states by $(D, j)_{BU}$ where $j \in [1, n]$ labels which $\ccz $ state the qubit is part of and $D \in \{ A, B, C \}$ distinguishes the 3-qubits within a $\ccz $ state.  The $BU$ subscript highlights that these are input noisy state qubits possibly produced by \BUTOFF.  We also have $3(m+k)$ qubits that we call factory qubits and label $(D, i)_f$ with a subscript $f$ for \textit{f}actory and where $D \in \{ A, B, C \}$ and $i \in [ 1, m+k ]$.  \end{definition}
Notice that the qubit labels assume we have made a spacetime tradeoff, so the factory qubits refer to the $3(m+k)$ qubits prepared in a $\ket{+}$ state.  The $3(n-m-k)$ qubits described earlier as being in the $\ket{0}$ state are omitted as they are surplus to requirement. For the code of interest, (recall \cref{eq:Galpha,eq:Gbeta,eq:Ggamma} there are 9 factory qubits and 24 $BU$ qubits, though the $BU$-qubits do not all need to be prepared at the same time and can be encoded in smaller distance codeblocks.   

A circuit using this labeling appears later in \cref{fig:CNOTcircuitforTDTOFF}.

\subsubsection{Injection with \texorpdfstring{$\cnot$}{CNOT} gates and single qubit measurements}
\label{App:PerformVwithCNOT}

To perform the required sequence of $n$ conjugated-$\ccz $ gates from \cref{eq:GenCCZ}, we may implement \cref{algo:CNOTcircuit}. 
\begin{algorithm}[t] 
\SetAlgoLined
\begin{enumerate}
    \item For each $j \in \{ 1, \ldots , n \}$
        \begin{enumerate}
            \item For each $D \in \{ A, B, C \}$ and each $i$ such that $[G^D]_{i,j}=1$ do a $\cnot$ with control $(D, i )_f$ and target $(D, j)_{BU}$.
            \item For each $D \in \{A, B, C \}$ measure magic state qubit $(D, j)_{BU}$ in the $Z$ basis and record the outcome as $m_j^D \in \{ 0,1 \}$. \label{algo:Clifford1}
            \item For each pair $m_j^D, m_j^{D'}=1$, apply a $Z$ correction to every qubit $(D'',p)_f$ for which $G^{D''}_{p,j}=1$. \label{algo:Clifford2}
            \item For each pair $m_j^D, =1$, apply a $CZ$ correction to every pair of qubits $(D',p)_f$ and $(D'',q)_f$ for which $G^{D'}_{p,j}=G^{D'}_{q,j}1$.
    \end{enumerate}
\end{enumerate} 
\caption{A $\cnot$ circuit realizing $V^\dagger\ccz ^{\otimes n }V$ as defined in \cref{App:SpaceTimeTrade}.  Uses a trio of $G$ matrices with $n$ rows and $k+m$ columns. Qubit label convention given in \cref{def:QubitLabels}.} \label{algo:CNOTcircuit}
\end{algorithm}
In  \cref{algo:Clifford1,algo:Clifford2} of \cref{algo:CNOTcircuit}, the indices $(D, D', D'')$ should be read as distinct triples from the set $ \{ A, B, C \}$. 
For example,  if $D=A$ and $D'=C$ then one infers $D''=B$.  Furthermore, these adaptive Clifford corrections commute with the rest of the circuits and so can all be postponed until later.  We illustrate some of the steps in  \cref{fig:CNOTcircuitforTDTOFF}.

Next, we calculate the action of the circuit described by \cref{algo:CNOTcircuit} for one particular $j$ value. With respect to the factory qubit basis states, we have 
\begin{equation} \label{eq:MatchNotations}
  \ket{\mathbf{a}}\ket{\mathbf{b}}\ket{\mathbf{c}} = \ket{(\mathbf{x},\mathbf{u})}\ket{(\mathbf{y},\mathbf{v})}\ket{(\mathbf{z},\mathbf{w})}  ,
\end{equation}
 where we have broken the state up into 3 blocks corresponding to indices $A$, $B$ and $C$.  For example, qubit $(A,i)_f$ is in state $a_i$.  Furthermore,  $a_i$ equals $x_i$ when $i\leq k$ and $u_i$ when $i > k$. For each $D=\{A, B, C \}$, we implement $\cnot$ gates targeted on the magic state qubit $(D,j)_{BU}$ and controlled on qubits $(D,i)_{BU}$ indicated by $[G^D]_{i,j}=1$.   

Therefore, for $D=A$ the target $(A,j)_{BU}$ qubit is flipped precisely when 
\begin{equation}
    \sum_i [G^A]_{i,j} a_i = [\mathbf{a}G^A ]_j = 1 \pmod{2} ,
\end{equation}
where the summation has been changed to matrix multiplication.  Recall $[\mathbf{a}G^A ]_j$ just means the $j^{\mathrm{th}}$ element of vector $\mathbf{a}G^A $.  Similar expressions hold for $D= B,C$.   The $\ccz$ magic state is given by
\begin{equation} \label{eq:CCZstate}
    \ket{\ccz} = 2^{-3/2} \sum_{y_D \in \mathbb{F}_2} (-1)^{y_A y_B y_C} \ket{y_A}\ket{y_B}\ket{y_C} .
\end{equation}
Ignoring $2^{-3/2}$ for brevity, the $\cnot$s of \cref{algo:CNOTcircuit} act as follows on a $\ket{\ccz}\ket{\mathbf{a}}\ket{\mathbf{b}}\ket{\mathbf{c}}$ state
\begin{align} \nonumber
  & \sum_{y_D \in \mathbb{F}_2} (-1)^{y_A y_B y_C} \ket{y_A}\ket{y_B}\ket{y_C}  \ket{\mathbf{a}}\ket{\mathbf{b}}\ket{\mathbf{c}} \\ \nonumber
  & \rightarrow \sum_{y_D \in \mathbb{F}_2}  (-1)^{y_A y_B y_C} \ket{y'_A}\ket{y'_B}\ket{y'_C}\ket{\mathbf{a}}\ket{\mathbf{b}}\ket{\mathbf{c}} ,
\end{align}
with
\begin{align}
 y'_A  & = y_A + [\mathbf{a}G^A ]_j \\ \nonumber
 y'_B  &  =  y_B + [\mathbf{b}G^B ]_j \\ \nonumber
  y'_C  &  =  y_C + [\mathbf{c}G^C ]_j  .
\end{align}
We follow these $\cnot$s by measurement of the $BU$-qubits in the $Z$ basis, which are afterwards discarded. Assuming measurement outcomes $\ket{m^A_j}\ket{m^B_j}\ket{m^C_j}$ then the only non-vanishing terms have $m^{D}_j=y'_D$, so
\begin{align} \label{yalpha}
 y_A  & = m^{A}_j + [\mathbf{a} G^A ]_j \\ \label{ybeta}
 y_B  &  =  m^B_j + [\mathbf{b} G^B ]_j \\ \label{ygamma}
  y_C  &  =  m^C_j + [\mathbf{c} G^C ]_j .
\end{align}
Discarding the $BU$-qubits, we get
\begin{align}
   \ket{CZZ}\ket{\mathbf{a}}\ket{\mathbf{b}}\ket{\mathbf{c}}  & \rightarrow (-1)^{f(\mathbf{m})}
  \ket{\mathbf{a}}\ket{\mathbf{b}}\ket{\mathbf{c}}  ,
\end{align}
where the phase exponent depends on the measurement outcomes $\mathbf{m}=(m^A_j,m^B_j,m^C_j)$ as follows 
\begin{equation} \label{eq:PhaseExp}
    f(\mathbf{m}) = (m^A_j + [\mathbf{a}G^A ]_j )(m^B_j + [\mathbf{b}G^B ]_j)(m^C_j + [\mathbf{c}G^C ]_j) .
\end{equation}
The value of this phase-exponent was originally $y_A y_B y_C$ but with the substitutions determined by \cref{yalpha,ybeta,ygamma} we get expression \cref{eq:PhaseExp}.

In the case of a $\mathbf{m}=(0,0,0) $ projection, we get the phase
\begin{equation}
    f(0,0,0) =  [\mathbf{a}G^A ]_j  [\mathbf{b}G^B ]_j [\mathbf{c}G^C ]_j,
\end{equation}
so that after switching notation by using \cref{eq:MatchNotations} we get the desired phase in \cref{eq:GenCCZ}.  However, for non-zero measurement outcomes we have
\begin{equation}
    f(\mathbf{m}) =  f(0,0,0)  + g(\mathbf{m}) ,
\end{equation}
where $g(\mathbf{m})$ represents the remaining terms in the expansions of \cref{eq:PhaseExp}.  We can see that these remaining terms will be quadratic in the variables $\{ \mathbf{a} , \mathbf{b} , \mathbf{c} \} $ and so represent Clifford corrections: the quadratic terms  correspond to a circuit of $CZ$ gates,  the linear terms correspond to a circuit of $Z$ gates, and the constant term gives a global phase.  

For example, consider the case when $\mathbf{m}=(1,0,0)$ so 
\begin{align}
   g(\mathbf{m}) & =  [\mathbf{b}G^B ]_j [\mathbf{c}G^C ]_j \\ \nonumber
   & = \sum_{p,q}  G^B_{p,j} G^C_{q,j} b_p c_q  .
\end{align}
This is corrected by a $CZ$ between qubits $(B, p)_f$ and $(C, q)_f$ for every $\{p,q\}$ such that $G^B_{p,j}=G^C_{q,j}=1$.  
This correction precisely matches the rule given in \cref{algo:Clifford1,algo:Clifford2} of \cref{algo:CNOTcircuit}.  
It is straightforward but tedious to verify that the corrections of \cref{algo:CNOTcircuit} always give the desired phase needed to cancel $(-1)^{g(\mathbf{m})}$.

\subsubsection{Injection using lattice surgery}
\label{app:GenCCinject}

\begin{figure*}
    \centering
    \includegraphics{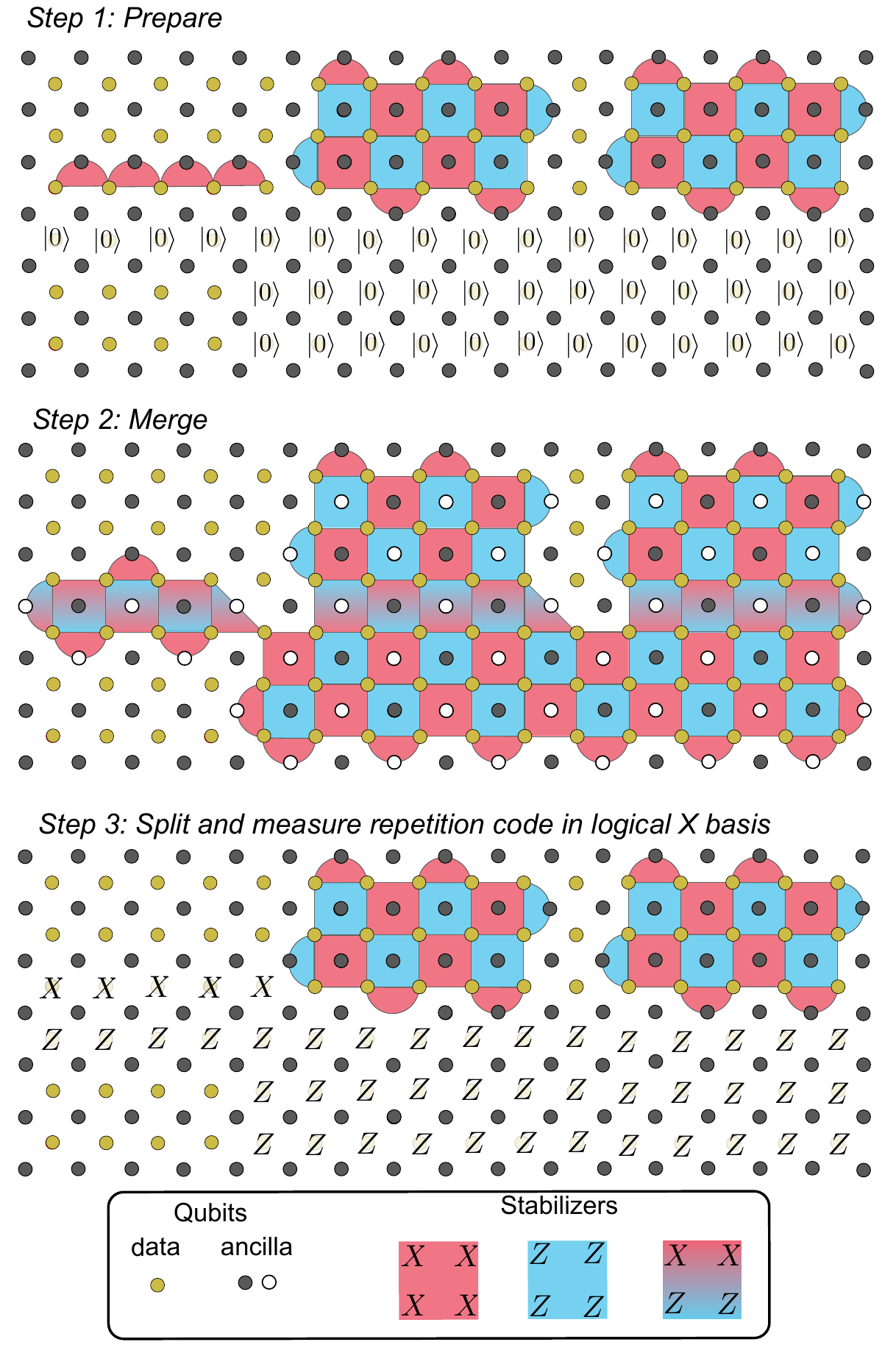}
    \caption{Lattice surgery to measure a multi-patch Pauli measurement between three codes blocks: one repetition code logical qubit and two thin surface code logical qubits. See \cref{fig:LatticeSurgerySimple} for comparison. This provides the principle building block for the execution of \cref{FirstPauliMeasurement}. In \TDTOFF, a multi-patch Pauli measurement is always followed by a single-qubit measurement of the repetition code and here we combine this with the third (split) step of lattice surgery. The gradient coloured squares represent where an $X \otimes X \otimes Z \otimes Z$ stabilizer measurement called a \etc{domain wall}. Multiplying the outcome of the stabilizers labeled with a white dot, gives the outcome of the $Z_L \otimes Z_{L} \otimes Z_{L}$ multi-patch Pauli measurement (this is how we obtain the outcomes labeled $\omega_j^D$ in \cref{FirstPauliMeasurement} of \cref{algo:LatticeSurgery}).  To ensure fault-tolerance of this measurement outcome, we repeat these stabilizer measurements $d_m$ times as discussed in \cref{subsec:CorrErrEdges}. Afterwards all qubits are measured in $Z$ basis, with their product determining the operator $Z_L \otimes \id \otimes \id$ (this is how we obtain the outcomes labeled $m_j^D$ in \cref{SingleQubitPauliMeasurement} of \cref{algo:LatticeSurgery}).  \etc{Since an error during a $Z$ measurement is physically indistinguishable from a bit-flip error prior to the measurement, to fully exploit noise bias we need these failure to be similarly unlikely.  However, in practice, in our noise model we find $Z$ measurement errors are much more likely than bit-flips.  However, since the qubits are idle for a long time after the $Z$ measurement, we can boost their effective fidelity by using an ancilla qubit to perform a non-destructive $Z$ measurement and repeat until the fidelity is suitably high.  As such, here the single qubit $Z$ measurements should be interrupted as repeated non-destructive $Z$ measurements.} We discuss in \cref{App:CliffordNoise} the effect of errors in lattice surgery due to using finite size code blocks. If we wish to instead measure $X_L \otimes Z_{L} \otimes Z_{L}$ then we do not use the \etc{domain wall} on the repetition code block.}
    \label{fig:MultiPatchBasic}
\end{figure*}

Here we provide an alternative formulation of the conjugated $\ccz $ injection from that presented in \cref{App:PerformVwithCNOT}.  Instead of using a $\cnot$ circuit, the injection procedure will be described entirely in terms of multi-qubit Pauli operator measurements, as this is the natural set of operations in lattice surgery implementations. We have already discussed the key ideas of this mapping in \cref{subsec:TimeLikeErrors,Sec:LatticeSurgery}. Here we wish to allow for the option of performing a lattice surgery operation between a repetition code clock (really just a $d_x=1$ surface code) and thin surface codes, and we present an example lattice surgery diagram in \cref{fig:MultiPatchBasic}. 

In general, imagine a circuit that performs the following: (i) do $n$ $\cnot$ gates targeted on qubit 0 and controlled on qubits $1$ to $n$; (ii) measure $Z_0$; (iii) discard qubit zero.  This is equivalent to the following measurement driven procedure: (i') measure multi-qubit Pauli $\prod_{j=0}^n Z_j$; (ii') measure single-qubit Pauli $X_0$ and discard; (iii')  if second step gives $``-1"$ outcome perform a $\prod_{j=1}^n Z_j$ correction.  In the bottom diagram of~\ref{fig:TOFCircuitInject}, we prove equivalence of these approaches through a series of circuit identities (illustrated for the $n=3$ case).  Applying this equivalence to \cref{algo:CNOTcircuit} we obtain \cref{algo:LatticeSurgery}.

\begin{algorithm}[t] 
\SetAlgoLined 
\begin{enumerate}
    \item For each $j \in \{ 1, \ldots , n \}$
        \begin{enumerate}  
            \item \label{FirstPauliMeasurement} For each $D \in \{ A, B, C \}$: measure a multi-qubit $Z$ operator, with support on $(D, j)_{BU}$ and $(D, i)_f$ for every $i$ such that $[G^D]_{i,j}=1$ and record the outcome as $\omega_{j}^D \in \{0,1 \}$.
            \label{SingleQubitPauliMeasurement}
            \item For each $\omega_{j}^D=1$ ,  record a $Z$ correction to qubit $(D, i)_f$ for every $i$ such that $[G^D]_{i,j}=1$.
            \item  For each $D \in \{A, B, C \}$ measure the single-qubit Pauli $(D, j)_{BU}$ in the $X$ basis and record the outcome as $m_j^D \in \{ 0,1 \}$. 
            \item For each pair $m_j^D, m_j^{D'}=1$, record a $Z$ correction to every qubit $(D'',p)_f$ for which $G^{D''}_{p,j}=1$. 
            \item For each pair $m_j^D, =1$, record a $CZ$ correction to every pair of qubits $(D',p)_f$ and $(D'',q)_f$ for which $G^{D'}_{p,j}=G^{D''}_{q,j}=1$.
        \end{enumerate}
\end{enumerate} 
\caption{A Pauli-measurement scheme realizing $V^\dagger \ccz ^{\otimes n }V$ as defined in \cref{App:SpaceTimeTrade}.  Uses a trio of $G^D$ matrices with $n$ rows and $k+m$ columns.  Qubit label convention given in \cref{def:QubitLabels}.} \label{algo:LatticeSurgery}
\end{algorithm}
Note that in a Pauli measurement scheme, we never perform the Clifford corrections.  Rather whenever there is a subsequent Pauli measurement $P$, if the Clifford correction register contains $C$, we instead measure $C P C^\dagger$.  The corrections in \cref{algo:LatticeSurgery} commute with all the measurements here, and so can be postponed until later.

\subsection{Error propagation and detection}
\label{App:ErrorProp}

Here we discuss the effect of noisy $\ket{\ccz}$ states used in the \TDTOFF protocol.  For now, we assume all encoded Clifford gates are ideal, but later we will relax this assumption.

To be precise regarding error propagation we introduce the following language
\begin{definition} \label{def:errors}
Given a $\ket{\ccz}$ magic state, we say it has error pattern $\mathbf{e}=(e^A, e^B, e^C) \in \mathbb{F}_2^3$ error if it is in the state
\begin{align}
    E \ket{\ccz} & = 
    Z[\mathbf{e}] \ket{CZZ} \\
    & = (Z^{e^A} \otimes Z^{e^B} \otimes Z^{e^C}) \ket{\ccz} . 
\end{align}
Given $n$ such states, for each $j\in [1, n]$ we use $\mathbf{e}_{j}=(e^A_j, e^B_j, e^C_j)$ to denote the error for the $j^{\mathrm{th}}$ $\ket{\ccz}$ state, so that
\begin{equation}
    E \ket{\ccz}^{\otimes n} = \bigotimes_{j=1}^n \left( Z[\mathbf{e}_j] \ket{\ccz} \right).
\end{equation}
We say an error has $w$ fault-locations if there are $w$ non-zero $\mathbf{e}_{j}$.  Furthermore, for $D \in \{ A, B, C\}$ we define
\begin{equation}
    \mathbf{e}^D=(e^D_1, e^D_2 , \ldots , e^D_n) .
\end{equation}
\end{definition}

The distinction between our notion of fault-locations and the usual Hamming weight of the concatenated string $(\mathbf{e}_1,\ldots,\mathbf{e}_n)$ is important because many methods of preparing a noisy $\ket{\ccz}$ state will lead to errors such as $Z \otimes Z \otimes Z$ that could have a comparable probability to a single qubit error $Z \otimes \id \otimes \id$.  Indeed, we will typically be interested in knowing how many $\ket{\ccz}$ states are affected by an arbitrary error, though we assume errors are uncorrelated between different $\ket{\ccz}$ states. 
Errors propagate as follows
\begin{claim}[How errors propagate] \label{lem:ErrorProp}
Consider an implementation of \cref{algo:LatticeSurgery} using noisy $\ccz $ states with Pauli $Z$ error described by $\{ \mathbf{e}^A,   \mathbf{e}^B,  \mathbf{e}^C \}$  as in \cref{def:errors}. For each $D \in \{A, B , C \}$, let 
\begin{equation}
     \mathbf{w}^D = \mathbf{e}^D  G^D .
\end{equation}
The output of \cref{algo:LatticeSurgery} differs from the ideal case by an error $Z[\mathbf{w}^A] \otimes Z[\mathbf{w}^B] \otimes Z[\mathbf{w}^C] $ on the factory qubits and where the tensor product represents the three different codeblocks. Identifying the last $m$ qubits of each block as check qubits, we can partition the $\mathbf{w}^D$ into two parts as follows
\begin{align} \label{ErrorComps}
    \mathbf{u}^D & = \mathbf{e}^D  G^D_1, \\
    \mathbf{v}^D & = \mathbf{e}^D  G^D_0.
\end{align}
\end{claim}
 \cref{lem:ErrorProp} tells us that $Z$ errors propagate through  \cref{algo:LatticeSurgery} in a manner that is isomorphic to their propagation through error correction codes represented by the corresponding $G$-matrices. 

We can prove \cref{lem:ErrorProp} by considering how a single $Z$ error on a $BU$-qubit propagates onto a factory qubit under \cref{algo:LatticeSurgery}.  
Since an error on a factory qubit propagates to the end of the circuit, they compose independently.  
Consult the last circuit of \cref{fig:TOFCircuitInject} and consider a $Z$ error on the top qubit.  
It commutes with the multi-qubit Pauli $Z$ measurement but flips the final single qubit $X$ measurement.  
The outcome for this $X$ measurement decides whether to apply $Z$ to the qubits below.  
In other words, a $Z$ on the top qubit propagates to all the qubits below. 
In \cref{algo:LatticeSurgery}, when operating on qubit $(D,j)_{BU}$ we apply the circuit of \cref{fig:TOFCircuitInject} to sets of factory qubits identified by $(D, i)_f$ whenever $G_{i,j}^D=1$. 
Therefore, a $Z$ error on $(D,j)_{BU}$ occurs whenever $\textbf{e}^{D}_j=1$ and will propagate to every $(D, i)_f$ for which $G_{i,j}^D=1$.  
Summing over all $j \in [1,n]$ magic state injections, factory qubit $(D, i)_f$ will have a $Z$ error if 
\begin{equation}
\sum_j \textbf{e}^{D}_j G_{i,j}^D=\left[  \textbf{e}^{D} G^D \right]_i=1 \pmod{2}.
\end{equation}
Since the $i^{\mathrm{th}}$ qubit is $Z$-flipped according to the $i^{\mathrm{th}}$ element of vector $\mathbf{w}^D:=\textbf{e}^{D} G^D $, this vector describes the $Z$-error distribution on factory block $D$. 
Splitting  $G^D$ into its block matrix components $G^D_1$ and $G^D_0$ gives \cref{ErrorComps}.

We have already described the main components of the magic state distillation routine, but for completeness we recap how they fit together in \cref{algo:FullLatticeSurgery}.
\begin{algorithm}[t] 
\SetAlgoLined
\begin{enumerate}
    \item Prepare $3 k$ factory qubits $(D, j)_f$ in the $\ket{+}$ state.
    \item Prepare $n$ noisy $\ccz $ magic states (e.g. using \BUTOFF);
    \item Perform injections using \cref{algo:LatticeSurgery}.
    \item Measure $3m$ check qubits $(D, i)_f$ for all $i=m,\ldots, m+k$ and ACCEPT on $\ket{+}$ for every outcome.
\end{enumerate}
\caption{A complete magic state distillation routine using a space-time tradeoff and multi-qubit Pauli measurements.  It assumes a trio of $G^D$ matrices of size $n \times (m+k) $ representing $[[n,k,d]]$ codes with $\ccz $ transversality as in \cref{lem:examples}.  We gave suitable $G^D$ matrices in \cref{eq:Galpha,eq:Gbeta,eq:Ggamma} for which we have $n=8$, $k=2$, $d=2$ and $m=1$. Qubit label convention given in \cref{def:QubitLabels}} \label{algo:FullLatticeSurgery}
\end{algorithm}
 Combining our previous results, we have that 

\begin{claim}[Distillation] \label{lem:DistillationWorks}
Consider an implementation of \cref{algo:FullLatticeSurgery} using $G^D$ matrices of size $n \times (m+k) $ satisfying \cref{lem:examples} and using $n$ noisy $\ccz $ states with Pauli $Z$ error described by $\{ \mathbf{e}^A,   \mathbf{e}^B,  \mathbf{e}^C \}$  as in \cref{def:errors}.  The protocol will ACCEPT whenever
\begin{align} 
    \mathbf{v}^D & = \mathbf{e}^D  G^D_0 = \textbf{0}
\end{align}
for every $D \in \{ A, B, C \}$.  Furthermore, provided for every $D \in \{ A, B, C \}$ we have 
\begin{align} 
    \mathbf{u}^D & = \mathbf{e}^D  G^D_1 = \textbf{0},
\end{align}
the protocol outputs $\ket{\ccz}^{\otimes k}$. 
Furthermore, if the $j^{\mathrm{th}}$ $\ket{\ccz}$ state has error $Z[\mathbf{e}_j]$ with probability $\mathbb{P}_j(\mathbf{e}_j)$ then the probability of passing the error detection test is
\begin{equation}
    P_{\mathrm{acc}} = \sum_{ \mathbf{e}^D : [ \mathbf{e}^D G^D_0 = \mathbf{0} ] \forall D} \prod_j \mathbb{P}_j(\mathbf{e}_j) ,
\end{equation}
and the output fidelity is
\begin{equation}
F = \frac{1}{P_{\mathrm{acc}}} \left( \sum_{ \mathbf{e}^D : [ \mathbf{e}^D G^D_1 = \mathbf{0} ] \forall D} \prod_j \mathbb{P}_j(\mathbf{e}_j) \right) .
\end{equation}
\end{claim}

First consider when there are no $Z$ errors. From \cref{app:GenCCinject} we see that \cref{algo:LatticeSurgery} will (when there is no $Z$ noise) apply $\ccz ^{\otimes k}$ to the $3k$ qubits labeled $(D, i)_f$ with $i \leq k$.  The check qubits with $i > k$ are unaffected.  Therefore, the check qubits should still be in the $\ket{+}$ state and give ``+1'' in response to an $X$ measurement.  This confirms that the protocol acts correctly in the ideal case.  

When there are one or more $Z$ errors,  \cref{lem:ErrorProp} shows that the check qubits remain unflipped if and only if  $\mathbf{u}^D= \mathbf{e}^D G_0 = \mathbf{0}$ for all $D$.  Furthermore, if $\mathbf{v}^D= \mathbf{e}^D G_1 = \mathbf{0}$ then \cref{lem:ErrorProp} tells us whether there are no $Z$ errors propagated onto the factory qubits forming the output $\ket{\ccz ^{\otimes k}}$ state.  The formulae for $P_{\mathrm{acc}}$ and $F$ follow by simply summing over the probabilities of these events.

For the remainder of this subsection, we consider the special case when $G^D_0=(1,1,\ldots, 1)$ as we have in \cref{eq:Galpha,eq:Gbeta,eq:Ggamma}.  Then, the state will pass the error detection test whenever
\begin{align} 
    \mathbf{v}^D & = \mathbf{e}^D  G^D_0 = \sum_j \mathbf{e}^D_j  = \mathbf{0}.
\end{align}
If there are no fault-locations so $\mathbf{e}_j=0$ for all $j$, then the protocol will ACCEPT. If there is a single fault location, so a single $j$ for which $\mathbf{e}_j=(e_j^A, e_j^B, e_j^C) \neq (0,0,0)$ then the error must be detected as there is no chance for cancellation.  If there are two fault-locations for which $\mathbf{e}_{j} \neq \mathbf{0}$ and $\mathbf{e}_{i} \neq \mathbf{0}$ then the errors will go undetected only if they cancel exactly, so $\mathbf{e}_{j} = \mathbf{e}_{i} $.  Therefore, to leading order 
\begin{equation}
    P_{\mathrm{acc}} = \prod_{j=1}^n \mathbb{P}_j(\mathbf{0}) + \sum_{\{i, j\} \subset [1,n], \mathbf{e} \neq \mathbf{0} } \mathbb{P}_i(\mathbf{e}) \mathbb{P}_j(\mathbf{e}) \prod_{\ell \neq i, j} \mathbb{P}_\ell (\mathbf{0}) + \ldots .
\end{equation}
For instance, let us consider an i.i.d depolarizing noise model such that $\mathbb{P}_j(\mathbf{0})=1-\epsilon$ and  $\mathbb{P}_j(\mathbf{e} \neq \mathbf{0})=\epsilon/7$. There are 7 types of fault $\mathbf{e} \neq \mathbf{0}$ and 28 pairs of possible locations, making 196 different undetected two fault-location errors, so that 
\begin{equation}
    P_{\mathrm{acc}} = (1-\epsilon)^8 + 196 \left( \frac{\epsilon}{7} \right)^2 (1-\epsilon)^6 + \ldots .
\end{equation}
To leading order, the infidelity $1-F$ is upper bounded by the probability of an undetected two fault-location error,
\begin{equation} \label{eq:InfidelityBound}
    1-F \leq  196 \left( \frac{\epsilon}{7} \right)^2 (1-\epsilon)^6 + \ldots .
\end{equation}
However, some undetected two fault-location errors will not lead to an output error (i.e. when $[ \mathbf{e}^D G^D_1 = \mathbf{0} ] \forall D $ ). For the $G^D$ matrices of interest (\cref{eq:Galpha,eq:Gbeta,eq:Ggamma}), by brute force counting we find that 184 of the undetected 196 two fault-location errors will lead to an error.  The 12 harmless faults are listed in \cref{tab:HarmlessFaults} and will return to play an important role in noise tailoring of  \cref{App:Noisetailoring}.
Therefore, we can tighten \cref{eq:InfidelityBound} to
\begin{align}
    1-F & \leq  184 \left( \frac{\epsilon}{7} \right)^2 (1-\epsilon)^6 + \ldots \\ 
    & \sim 3.755 \epsilon^2 + O(\epsilon^3).
\end{align}
Therefore, we have quadratic error suppression with quite a small constant factor for depolarizing noise.  In the main text, we usually quote the error per $\tof$ state and since the protocol outputs two $\tof$ states, we have
$\epsilon_{TD}:=\frac{1}{2}(1-F)$.  For the depolarizing noise model this leads to:
\begin{equation}
    \epsilon_{TD} = \sim 1.878 \epsilon^2 + O(\epsilon^3).
\end{equation}

\subsubsection{Truncation errors}
\label{App:Truncation}

While we give expressions up to second order, these summations can be easily performed to higher order and any truncation error can be controlled.  If we perform calculations up to $t_{\mathrm{max}}$ fault-locations, then the truncation error can be easily upper-bounded by assuming that every error above the cut-off leads to an undetected output error so that we have the rigorous bound
\begin{equation}
   1-F \leq (1-F_{t_{\mathrm{max}}}) + \sum_{t=t_{\mathrm{max}}+1}^8 \binom{8}{t} 7^t \left( \frac{\epsilon}{7} \right)^t (1-\epsilon)^{8-t} ,
\end{equation}
where $ (1-F_{t_{\mathrm{max}}})$ is a estimate counting up to $t_{\mathrm{max}}$ fault-locations and the additional summation is our bound on the truncation error. In all subsequent numerical calculations we have confirmed the possible truncation error is many orders of magnitude smaller than the estimated error.  For instance, using $t_{\mathrm{max}}=3$ then for $\epsilon \leq 10^{-4}$ the truncation error is no more than $3 \cdot 10^{-16}$ and therefore negligible.  

In practice, the error distribution from \BUTOFF is far from depolarizing and this is further skewed when we account for Clifford noise (see \cref{App:CliffordNoise}).  However, truncation error can be estimated of any noise model and controlled in the above manner. Furthermore, one can also tailor the protocol to the noise profile (see \cref{App:Noisetailoring}).  

\subsubsection{Generic noise}
\label{App:genericnoise}

We have show \cref{algo:FullLatticeSurgery} tolerates $Z$ error noise.  Next, we show it also tolerates $X$ noise on the noisy $\ket{\ccz}$ states.  Abstracting away the details of \cref{lem:DistillationWorks}, the protocol maps pure states as follows
\begin{equation}\label{eq:PureMap}
   Z[\mathbf{e}] \ket{\ccz}^{\otimes n} \rightarrow  \mathrm{det}(\mathbf{e})   Z[ \nu(\mathbf{e})] \ket{\ccz}^{\otimes k}  ,
\end{equation}
where $\mathrm{det}(\mathbf{e})=0,1$ depending on whether the error $\mathbf{e}$ is detected or not, and the output error is some function $\nu$ of $\mathbf{e}$.  Formulae for $\mathrm{det}$ and $\nu$ can be extracted from \cref{lem:DistillationWorks}, but here it is useful to ignore these details.  Going to density matrices, we can write
\begin{align}
   \rho & := \kb{\ccz}{\ccz}^{\otimes n} \\ \nonumber
   \sigma & := \kb{\ccz}{\ccz}^{\otimes k} .
\end{align}
Because $Z[\mathbf{e}] \ket{\ccz}^{\otimes n}$ form an orthonormal basis, any input mixed state can be written as
\begin{align}
   \tilde{\rho} & := \sum_{\mathbf{e}, \mathbf{f}} A_{\mathbf{e},\mathbf{f}}  Z[\mathbf{e}] \rho Z[\mathbf{f}] .
\end{align}
If the state suffered stochastic $Z$ noise then it would be diagonal with respect to this basis, so $A_{\mathbf{e},\mathbf{f}}=0$ whenever $\mathbf{e} \neq \mathbf{f}$.  If there are off-diagonal elements $A_{\mathbf{e},\mathbf{f}} \neq 0$ these could be eliminated by applying a random twirl using the Clifford operators that stabilize $\ket{\ccz}$.  However, this would add unnecessary Clifford gates as these off-diagonals are unimportant, as we now show.

\begin{table*}[t]
\begin{tabular}{c c c c} 
    \toprule
    Fault type $\mathbf{e}_i=\mathbf{e}_j=\mathbf{e}=$ & $(1,0,0)$ & $(0,1,0)$ & $(0,0,1)$  \\ 
    \midrule
    Fault locations $\{i,j\}=$ & $\{1,2\} , \{3,6\}, \{4,5 \},\{7,8 \}$ & $\{1,5\}, \{2,4\}, \{3,7 \} , \{6,8 \}$ & 
    $ \{ 1,3 \}, \{ 2,6 \}, \{ 4,8 \}, \{ 5,7 \}$
    \\ 
    \bottomrule 
\end{tabular}
\caption{A list of the errors with two fault-locations that are undetected but do not cause a logical fault when executing \cref{algo:FullLatticeSurgery} with $G^D$-matrices as in \cref{eq:Galpha,eq:Gbeta,eq:Ggamma}.  The errors follow the notation of \cref{def:errors}.  For example, $\mathbf{e}=(1,0,0)$ corresponds to $Z \otimes \id \otimes \id$ and is undetected yet harmless when it acts on $BU$-qubits $(A,1)_{BU}$ and $(A,2)_{BU}$ .  This is a direct consequence of $(1,1,0,0,0,0,0,0)G^A_1=(0,0)$ that can be confirmed by inspection of \cref{eq:Galpha}. Notice that only unit vector $\mathbf{e}$ appears in this list of fault types. }  \label{tab:HarmlessFaults}
    \centering
    \tabcolsep=4pt
    \begin{tabular}{c cccccccc} \toprule
$j$ & 1 & 2 & 3 & 4 & 5 & 6 & 7 & 8   \\ \midrule
 $M_j$ & $\left(
\begin{array}{ccc}
 1 & 0 & 0 \\
 0 & 0 & 1 \\
 0 & 1 & 0 \\
\end{array}
\right)$ & $\left(
\begin{array}{ccc}
 1 & 0 & 0 \\
 0 & 1 & 0 \\
 0 & 0 & 1 \\
\end{array}
\right)$ & $\left(
\begin{array}{ccc}
 0 & 1 & 0 \\
 1 & 0 & 0 \\
 0 & 0 & 1 \\
\end{array}
\right)$ & $\left(
\begin{array}{ccc}
 0 & 0 & 1 \\
 0 & 1 & 0 \\
 1 & 0 & 0 \\
\end{array}
\right)$ & $\left(
\begin{array}{ccc}
 1 & 0 & 1 \\
 0 & 0 & 1 \\
 0 & 1 & 0 \\
\end{array}
\right)$ & $\left(
\begin{array}{ccc}
 1 & 1 & 0 \\
 1 & 0 & 0 \\
 0 & 0 & 1 \\
\end{array}
\right)$ & $\left(
\begin{array}{ccc}
 0 & 1 & 0 \\
 0 & 0 & 1 \\
 1 & 0 & 0 \\
\end{array}
\right)$ & $\left(
\begin{array}{ccc}
 0 & 0 & 1 \\
 0 & 1 & 0 \\
 1 & 0 & 0 \\
\end{array}
\right)$ \\[16pt] 
$(1,0,0)$ & $\left(
\begin{array}{ccc}
 1 & 0 & 0 \\
\end{array}
\right)$ & $\left(
\begin{array}{ccc}
 1 & 0 & 0 \\
\end{array}
\right)$ & $\left(
\begin{array}{ccc}
 0 & 1 & 0 \\
\end{array}
\right)$ & $\left(
\begin{array}{ccc}
 0 & 0 & 1 \\
\end{array}
\right)$ & $\left(
\begin{array}{ccc}
 1 & 0 & 1 \\
\end{array}
\right)$ & $\left(
\begin{array}{ccc}
 1 & 1 & 0 \\
\end{array}
\right)$ & $\left(
\begin{array}{ccc}
 0 & 1 & 0 \\
\end{array}
\right)$ & $\left(
\begin{array}{ccc}
 0 & 0 & 1 \\
\end{array}
\right)$ \\ \bottomrule
\end{tabular}
    \caption{A set of transformation matrices $M_j$ that represent Clifford symmetries as defined in \cref{eq:CliffSym}.  For every distinct pair of indices $\{i,j \}$ they satisfy either condition $1^\star$ or condition $2^\star$ as stated in the proof of \cref{claim:Taylored}.  In particular, the only pairs for which condition $1^\star$ does not hold are $\{ 1,2\}$,$\{ 3,7\}$ and $\{4,8 \}$.  However, for these three cases the fault pattern is one of the harmless cases listed in \cref{tab:HarmlessFaults}.   }
    \label{tab:CliffordSymmetries}
\end{table*}
\tabcolsep=6pt

By \cref{eq:PureMap} and linearity, we have
\begin{align}
   \tilde{\rho} \rightarrow    \tilde{\sigma} & = \sum_{\mathbf{e}, \mathbf{f}} \mathrm{det}(\mathbf{e})\mathrm{det}(\mathbf{f}) A_{\mathbf{e},\mathbf{f}}   Z[\nu(\mathbf{e})] \sigma Z[ \nu( \mathbf{f} )] .
\end{align}
Because any physical process does not increase the trace of any terms, there is no way for off-diagonal elements (with $\mathbf{e} \neq \mathbf{f}$) to be mapped to on-diagonal elements (with $\nu(\mathbf{e}) \neq \nu(\mathbf{f})$).  Since the success probability and fidelity only depend on the output diagonal elements, we conclude that our figures of merit only depend on the diagonal $A_{\mathbf{e},\mathbf{e}}$ elements.  In other words, the success probability and output fidelity are unchanged whether or not we twirl the initial state.  In all numerics presented, whenever the input magic states suffer a mix of $Z$ and $X$ noise, we have calculated the exact $\tilde{\rho}$ matrix, extracted the diagonal elements and used them to build an equivalent stochastic $Z$ noise model.  Consequently, any error on a single $\ket{\ccz}$ state appears as a stochastic mixture of $Z$ errors at 1 fault location.

After the protocol is complete, we can twirl the output states to ensure that the infidelity matches the trace norm error of the output states.  Though again, this twirl is never actually performed but included into the Clifford record to modify Pauli-measurements used to inject the magic state into the algorithm.

\subsection{Noise tailoring through Clifford symmetries}
\label{App:Noisetailoring}

There is some freedom in how injections are scheduled and whether to include certain Clifford gates in \TDTOFF protocol. A $\ket{\ccz}$ gate is invariant under permutation of qubits $A$, $B$ and $C$.  More generally, there are Clifford symmetries $C$ such that $C\ket{\ccz}=\ket{\ccz}$.  A permutation is a sort of Clifford symmetry, but one that can be realized at no further gate count.

As such, we can add Cliffords or freely permute some of the indices in \cref{algo:LatticeSurgery}.  In the ideal case, with no errors, these symmetry operations have no effect.  However, they can change the noise model.  For qubits with depolarizing noise, the noisy state is invariant under all these symmetries.  However, for \BUTOFF the output noise model is very asymmetric and highly skewed towards a $Z$ error on qubit $A$ and so applying symmetry operations can  change the protocol's performance.  Here we explain the idea of noise tailoring through symmetries and find that the change can be dramatic.  Indeed, while the protocol usually quadratically suppresses errors, we can tailor the noise for cubic suppression of 1 error type. To make a clean statement we consider a toy noise model.
\begin{claim} \label{claim:Taylored}
Consider a noise model on $\ket{\ccz}$ states such that for every $j$ it experiences error $Z[\mathbf{e}_j]$ (recall \cref{def:errors}) with probability
\begin{equation}
    \mathbb{P}_{j}(\mathbf{e}_j)  := \begin{cases}
    1-\epsilon_1 -\epsilon_2 & \mbox{ if } \mathbf{e}_j=(0,0,0) \\
   \epsilon_1  & \mbox{ if } \mathbf{e}_j=(1,0,0)  \\
   (\epsilon_2 / 6) &  \mbox{ otherwise }
    \end{cases} ,
\end{equation}
where $\epsilon_2 \ll \epsilon_1$.  Directly applying \cref{lem:DistillationWorks} leads to an output infidelity of $O( \epsilon_1^2) + O(\epsilon_2^2)+O( \epsilon_1 \epsilon_2)$. However, there exists a set of Clifford symmetries $\{ C_j \}$ such that $C_j \ket{\ccz}=\ket{\ccz}$ and if applied at the start of the protocol lead to an output infidelity of $O( \epsilon_1^3) + O( \epsilon_1 \epsilon_2)+ O(\epsilon_2^2)$.
\end{claim}

Consider a set of Clifford symmetries such that 
\begin{equation} \label{eq:CliffSym}
C_j Z[\mathbf{e}_j] C_j^\dagger= \pm Z[\mathbf{e}_j M_j] ,
\end{equation}
where $M_j$ is an invertible $3 \times 3$ binary matrix and $\mathbf{e}_j M_j$ represents matrix multiplication.  The $\pm$ phase will depend on $Z[\mathbf{e}_j]$ but is irrelevant to our analysis. For example, if $C_j$ permutes qubits in Hilbert space then $M_j$ represents the permutation of the indices.  Then applying $C_j$ to the input magic states generates a new probability distribution for $Z$ errors
\begin{equation}
    \mathbb{P}'_{j}(\mathbf{e}_j M_j)  :=  \mathbb{P}_{j}(\mathbf{e}_j ).
\end{equation}
Using that $M$ must be invertible, we equivalently have
\begin{equation}
    \mathbb{P}'_{j}(\mathbf{e}_j )  :=  \mathbb{P}_{j}(\mathbf{e}_jM_j^{-1} ).
\end{equation}
Only errors with two fault-locations contribution second order contributions to the output infidelity.  Recall from \cref{App:ErrorProp} that for such an error to go undetected, we must have that $\mathbf{e}_i=\mathbf{e}_j=: \mathbf{e} \neq 0$ for some distinct pair $\{i,j \}$. We have introduce the shorthand $\mathbf{e}$ for whatever nonzero error type is under consideration. This occurs with probability 
\begin{equation}
    \mathbb{P}'_{i}(\mathbf{e} M_i ) \mathbb{P}'_{j}(\mathbf{e} M_j )  \mathbb{P}'_{j}(\mathbf{0})^6 =  \mathbb{P}_{i}(\mathbf{e}M_i^{-1} ) \mathbb{P}_{j}(\mathbf{e}M_j^{-1}  )  \mathbb{P}_{j}(\mathbf{0})^6  .
\end{equation}
This probability is of size $O(\epsilon_1^2)$ if
\begin{equation} \label{eq:MijCond}
    \mathbf{e} M_i^{-1}=\mathbf{e} M_j^{-1}=(1,0,0) ,
\end{equation}
and otherwise the probability is smaller: either $O(\epsilon_1 \epsilon_2)$, $O(\epsilon_2^2)$ or zero. Inverting again, \cref{eq:MijCond} can be converted into 
\begin{equation}
    \mathbf{e}=(1,0,0)M_i=(1,0,0)M_j.
\end{equation}
It follows that to achieve $O(\epsilon_1^3)$ scaling of output infidelity, we require that for every $\{i,j \}$ pair either 
\begin{enumerate}
    \item[($1^\star$)] $(1,0,0)M_i \neq (1,0,0)M_j$ ;
    \item[($2^\star$)] \textit{or} if $(1,0,0)M_i = (1,0,0)M_j$ then fault $\mathbf{e}=(1,0,0)M_j$ corresponds to one of the harmless errors listed in \cref{tab:HarmlessFaults}.
\end{enumerate}
There are 7 different possible values of $(1,0,0)M_j$ but 8 different $j$ indices, so it is (narrowly) not possible to use condition $1^\star$ alone.  However, it is possible to find a set of Clifford symmetries such that some pairs $\{i,j \}$  are covered by condition $1^\star$ and some pairs $\{i,j \}$  are covered by condition $2^\star$.  

We provide such an $\{ M_j \}$ set in \cref{tab:CliffordSymmetries}, which suffices to prove \cref{claim:Taylored}.  Furthermore, this set can be implemented especially easily. Consulting \cref{tab:CliffordSymmetries} we find that the Clifford symmetries for indices $\{1,2,3,4,7,8\}$ all correspond to permutations of indices and so can all be performed in software.  The only exceptions are indices $\{5,6\}$ that correspond to a $W=\cnot_{B,A}X_A$ gate followed by an index permutation. In other words, $W$ bit-flips qubit $A$ if qubit $B$ is in the $\ket{0}$ state.  Consequently, the state $\ket{1,1,1}$ is invariant under $W$ and other computational basis states are permuted. Therefore, $W$ is a Clifford symmetry of the $\ket{\ccz}$ state since all terms except $\ket{1,1,1}$ carry the same amplitude and phase.  Since $\ket{\tof}=H_C \ket{\ccz}$ and $[W,H_C]=0$, we know $W$ also stabilizes $\ket{\tof}$. Furthermore, when conjugating a $Z$ error we have that \cref{eq:CliffSym} takes the form
\begin{equation}
    W Z[\mathbf{e}_j]W^\dagger = W (-1)^{e_j^A}Z[\mathbf{e}_j M_{W} ] ,
\end{equation}
where
\begin{equation}
    M_W = \left( \begin{array}{ccc}
        1 & 1 & 0  \\
        0 & 1 & 0 \\
        0 & 0 & 1 \\
    \end{array}\right) .
\end{equation}
Permuting qubit indices after $W$ corresponds to swapping columns of $M_W$. We get $M_5$ of \cref{claim:Taylored} by swapping columns 2 and 3 of $M_W$.   We get $M_6$ of \cref{tab:CliffordSymmetries} by swapping columns 1 and 2 of $M_W$. 

The actual noise distribution output from \BUTOFF is not exactly the toy noise model of \cref{tab:CliffordSymmetries} but it shares the feature that $(Z \otimes \id \otimes \id)$ errors dominate. In all numerics presented, we use the Clifford symmetry operations of \cref{claim:Taylored} and \cref{tab:CliffordSymmetries} but analyzed using the correct \BUTOFF noise model. 

Implementing $W$ on repetition encoded qubits $A$ and $B$ is straightforward because $W$ is transversal and the codeblocks are adjacent to each other in the proposed layout.

\begin{figure*}
    \centering
    \includegraphics[width=1.7\columnwidth]{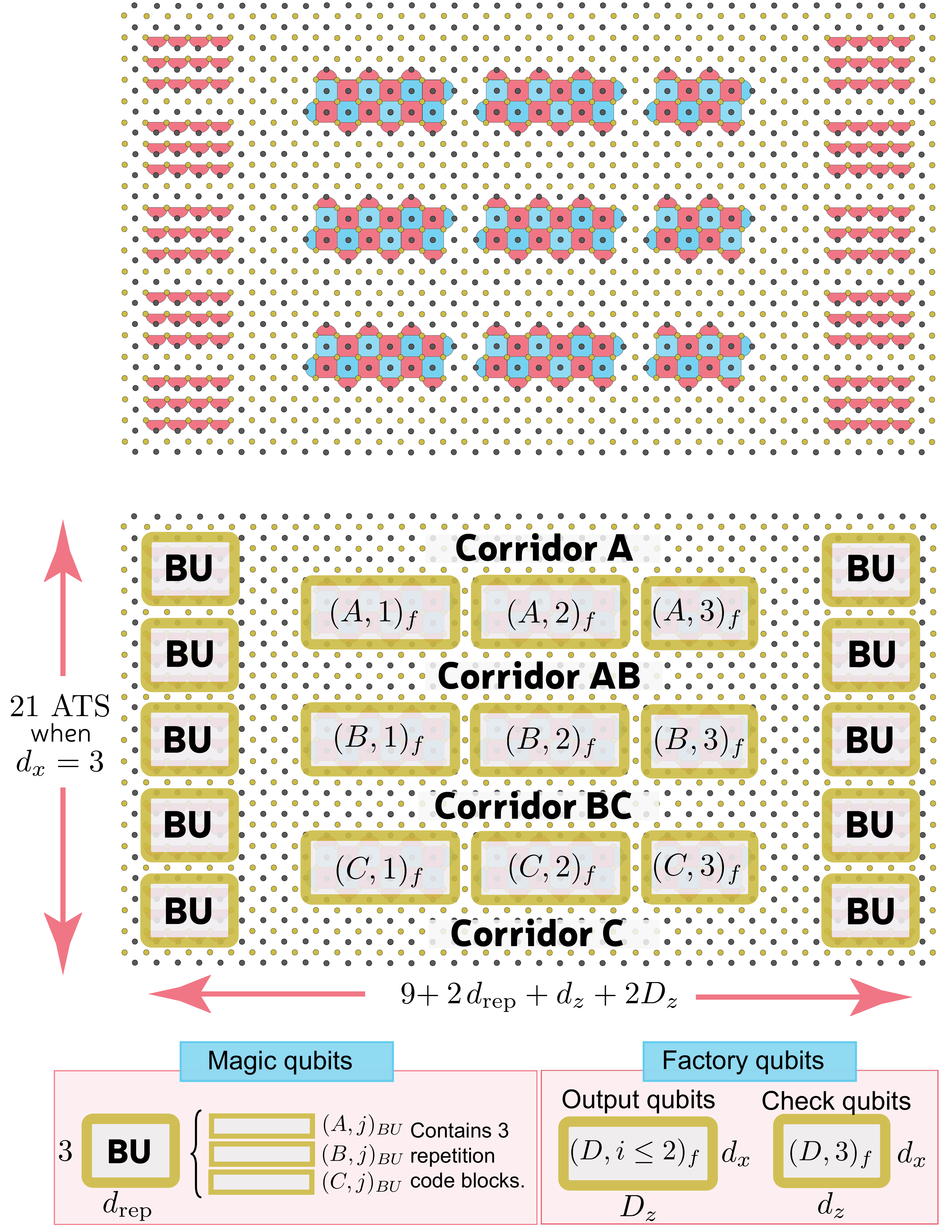}
    \caption{A 2D layout for realising $8 \tof \rightarrow 2 \tof$ distillation via lattice surgery using a mixture of repetition codes and thin surface code.  Example dimensions shown here with: encoding distances $d_x=3$, $d_{\mathrm{rep}}=5$, $d_z=5$ and $D_z=7$; and $M=10$ \BUTOFF modules.  Additional space between codeblocks is provided for lattice surgery and routing between code blocks (see \cref{subsec:TimeLikeErrors} and \cref{fig:MultiPatchBasic}).  We give explicit locations for the 9 factory qubits with labels $(D, i)_f$ following \cref{def:QubitLabels}.  The modules labeled BU consists of 3 repetition codes and provide space to attempt a noisy $\ket{\tof}$ preparation using the \BUTOFF protocol.  Note that \BUTOFF is executed with a distance $d_{BU}$ repetition code (typically we set $d_{BU}=5,7$) and then immediately grow to distance $d_z > d_{BU}$.}
    \label{fig:BigLayoutLabelled}
\end{figure*}

\begin{table*}[t]
\begin{tabular}{c cx{4cm} cx{4cm} cx{4cm} cx{4cm}} 
\toprule
 \textbf{Factory clock} & \textbf{Corridor A} & \textbf{Corridor AB} & \textbf{Corridor BC} & \textbf{Corridor C} \\ \midrule
 1 & 
$\begin{array}{c}
 \text{X(}\text{C},4)_{BU} \\
 \text{Z(}\text{A},3)_f \\
 \text{then} \\
 \text{Z(}\text{C},4)_{BU} \\
\end{array}$
 & 
$\begin{array}{c}
 \text{X(}\text{C},3)_{BU} \\
 \text{Z(}\text{B},1)_f \\
 \text{Z(}\text{B},2)_f \\
 \text{Z(}\text{B},3)_f \\
 \text{then} \\
 \text{Z(}\text{C},3)_{BU} \\
\end{array}$
 & 
$\begin{array}{c}
 \text{X(}\text{C},2)_{BU} \\
 \text{Z(}\text{C},2)_f \\
 \text{Z(}\text{C},3)_f \\
 \text{then} \\
 \text{Z(}\text{C},2)_{BU} \\
\end{array}$
 & 
$\begin{array}{c}
 \text{X(}\text{C},1)_{BU} \\
 \text{Z(}\text{C},1)_f \\
 \text{Z(}\text{C},2)_f \\
 \text{Z(}\text{C},3)_f \\
 \text{then} \\
 \text{Z(}\text{C},1)_{BU} \\
\end{array}$
 \\ \midrule
 2 & 
$\begin{array}{c}
 \text{Z(}\text{B},1)_{BU} \\
 \text{Z(}\text{A},0)_f \\
 \text{Z(}\text{A},1)_f \\
 \text{Z(}\text{A},2)_f \\
 \text{then} \\
 \text{X(}\text{B},1)_{BU} \\
\end{array}$
 & 
$\begin{array}{c}
 \text{Z(}\text{B},2)_{BU} \\
 \text{Z(}\text{B},0)_f \\
 \text{Z(}\text{B},2)_f \\
 \text{then} \\
 \text{X(}\text{B},2)_{BU} \\
\end{array}$
 & 
$\begin{array}{c}
 \text{Z(}\text{B},4)_{BU} \\
 \text{Z(}\text{B},0)_f \\
 \text{Z(}\text{B},2)_f \\
 \text{then} \\
 \text{X(}\text{B},4)_{BU} \\
\end{array}$
 & 
$\begin{array}{c}
 \text{Z(}\text{B},3)_{BU} \\
 \text{Z(}\text{C},1)_f \\
 \text{Z(}\text{C},2)_f \\
 \text{then} \\
 \text{X(}\text{B},3)_{BU} \\
\end{array}$
 \\ \midrule
 3 & 
$\begin{array}{c}
 \text{Z(}\text{A},3)_{BU} \\
 \text{Z(}\text{A},0)_f \\
 \text{Z(}\text{A},2)_f \\
 \text{then} \\
 \text{X(}\text{A},3)_{BU} \\
\end{array}$
 & 
$\begin{array}{c}
 \text{Z(}\text{A},2)_{BU} \\
 \text{Z(}\text{A},0)_f \\
 \text{Z(}\text{A},1)_f \\
 \text{Z(}\text{A},2)_f \\
 \text{then} \\
 \text{X(}\text{A},2)_{BU} \\
\end{array}$
 & 
$\begin{array}{c}
 \text{Z(}\text{A},1)_{BU} \\
 \text{Z(}\text{B},0)_f \\
 \text{Z(}\text{B},1)_f \\
 \text{Z(}\text{B},2)_f \\
 \text{then} \\
 \text{X(}\text{A},1)_{BU} \\
\end{array}$
 & 
$\begin{array}{c}
 \text{Z(}\text{A},4)_{BU} \\
 \text{Z(}\text{C},1)_f \\
 \text{Z(}\text{C},2)_f \\
 \text{then} \\
 \text{X(}\text{A},4)_{BU} \\
\end{array}$
 \\ \hline
 4 & 
$\begin{array}{c}
 \text{X(}\text{C},8)_{BU} \\
 \text{Z(}\text{A},3)_f \\
 \text{then} \\
 \text{Z(}\text{C},8)_{BU} \\
\end{array}$
 & 
$\begin{array}{c}
 \text{X(}\text{C},6)_{BU} \\
 \text{Z(}\text{B},2)_f \\
 \text{Z(}\text{B},3)_f \\
 \text{then} \\
 \text{Z(}\text{C},6)_{BU} \\
\end{array}$
 & 
$\begin{array}{c}
 \text{X(}\text{C},7)_{BU} \\
 \text{Z(}\text{B},1)_f \\
 \text{Z(}\text{B},3)_f \\
 \text{then} \\
 \text{Z(}\text{C},7)_{BU} \\
\end{array}$
 & 
$\begin{array}{c}
 \text{X(}\text{C},5)_{BU} \\
 \text{Z(}\text{C},1)_f \\
 \text{Z(}\text{C},3)_f \\
 \text{then} \\
 \text{Z(}\text{C},5)_{BU} \\
\end{array}$
 \\ \midrule
 5 & 
$\begin{array}{c}
 \text{Z(}\text{B},5)_{BU} \\
 \text{Z(}\text{A},0)_f \\
 \text{Z(}\text{A},1)_f \\
 \text{Z(}\text{A},2)_f \\
 \text{then} \\
 \text{X(}\text{B},5)_{BU} \\
\end{array}$
 & 
$\begin{array}{c}
 \text{Z(}\text{B},7)_{BU} \\
 \text{Z(}\text{A},1)_f \\
 \text{Z(}\text{A},2)_f \\
 \text{then} \\
 \text{X(}\text{B},7)_{BU} \\
\end{array}$
 & 
$\begin{array}{c}
 \text{Z(}\text{B},8)_{BU} \\
 \text{Z(}\text{B},2)_f \\
 \text{then} \\
 \text{X(}\text{B},8)_{BU} \\
\end{array}$
 & 
$\begin{array}{c}
 \text{Z(}\text{B},6)_{BU} \\
 \text{Z(}\text{C},2)_f \\
 \text{then} \\
 \text{X(}\text{B},6)_{BU} \\
\end{array}$
 \\ \midrule
 6 & 
$\begin{array}{c}
 \text{Z(}\text{A},6)_{BU} \\
 \text{Z(}\text{A},0)_f \\
 \text{Z(}\text{A},2)_f \\
 \text{then} \\
 \text{X(}\text{A},6)_{BU} \\
\end{array}$
 & 
$\begin{array}{c}
 \text{Z(}\text{A},5)_{BU} \\
 \text{Z(}\text{B},1)_f \\
 \text{Z(}\text{B},2)_f \\
 \text{then} \\
 \text{X(}\text{A},5)_{BU} \\
\end{array}$
 & 
$\begin{array}{c}
 \text{Z(}\text{A},7)_{BU} \\
 \text{Z(}\text{C},2)_f \\
 \text{then} \\
 \text{X(}\text{A},7)_{BU} \\
\end{array}$
 & 
$\begin{array}{c}
 \text{Z(}\text{A},8)_{BU} \\
 \text{Z(}\text{C},2)_f \\
 \text{then} \\
 \text{X(}\text{A},8)_{BU} \\
\end{array}$
 \\ \midrule
 7 & 
$\begin{array}{c}
 \text{Setup} \\
 \text{exit} \\
\end{array}$
 & 
$\begin{array}{c}
 \text{X(A,3})_f \\
 \text{Clifford} \\
 \text{corrected} \\
\end{array}$
 & 
$\begin{array}{c}
 \text{X(B,3})_f \\
 \text{Clifford} \\
 \text{corrected} \\
\end{array}$
 & 
$\begin{array}{c}
 \text{X(C,3})_f \\
 \text{Clifford} \\
 \text{corrected} \\
\end{array}$
 \\ \midrule
 8 & exit & exit & exit & exit \\ \bottomrule
\end{tabular}
   \caption{The final form of our \TDTOFF protocol for one full cycle of the factory.  It executes a variant of \cref{algo:FullLatticeSurgery} that has been modified according to the qubit permutations required for noise-tailoring (see \cref{App:Noisetailoring}) and embedded within the 2D layout of \cref{fig:BigLayoutLabelled}.  Each cell for factory clocks 1-6 has the form $A$ then $B$.  Instruction $A$ specifies a multi-qubit Pauli operator using the qubit notation of \cref{def:QubitLabels}.  For example, $X(C, 4)_{BU}, Z(A, 3)_f$ means measure the operators $X \otimes Z$ where the $X$ acts on magic input labeled $(C, 4)_{BU}$ and the $Z$ acts on factory qubit $(A, 3)_f$. Instruction $B$ specifies a single-qubit measurement of a magic input qubit. The $B$ instructions can be realized with physical single-qubit measurements that takes a single surface code cycle. As such, $B$ instructions require negligible time compared to the $A$ instructions, so we present both $a$ and $B$ within a single Factory clock step that has duration $d_m +1$. In factory clock steps 1 and 4, the role of $Z$ and $X$ are swapped on the magic state qubits to account for the Hadamard difference between $\ket{\ccz}$ and $\ket{\tof}$. The column headers ``Corridor" indicate which Corridor from \cref{fig:BigLayoutLabelled} is used to realize the multi-qubit Pauli measurement since lattice surgery requires some workspace to operate.  Note that Corridor AB can only be used to access factory qubit with labels of the form $(A, i)_f$  or $(B, i)_f$.  The column headers also list which factory blocks $\{A, B, C \}$ the Corridor can be used to access and this constraint it respected in this schedule. Notice that multi-qubit measurements of the form $X(C, 4)_{BU}, Z(A, 3)_f$ involve different capital letter indices on the factory and magic qubits.  In contrast,  \cref{FirstPauliMeasurement} of \cref{algo:LatticeSurgery} describes multi-qubit Pauli measurements with matching capital letter indices.  This is due to the permutation operations required for noise tailoring (see \cref{App:Noisetailoring}). In particular, when performing measurements with the $j=4$ index, the matrix $M_4$ of \cref{tab:CliffordSymmetries} instructs us to swap the $A$ and $C$ indices for the magic state qubit.  In the cases of $M_6$ and $M_7$, these are decomposed into a single Clifford gate $W$ and a permutation.  The above table only accounts for the permutation, with the Clifford performed on the input magic state qubits prior to injection into \TDTOFF.  Factory clock times 1-3 correspond to batch 1, so that measurements involve only magic state qubits of the form $(D, i)_{BU}$ with $i \in [1,4]$.  Factory clock times 4-6 correspond to batch 2, so that measurements involve only magic state qubits of the form $(D, i)_{BU}$ with $i \in [5,8]$.  The importance of batching and the related issue of \BUTOFF scheduling is discussed in \cref{App:FactoryLayout}. ``Exit" refers to factory qubits moving out of the factory.}
    \label{tab:FactoryClock}
\end{table*}

\subsection{Factory layout and scheduling}
\label{App:FactoryLayout}

We see from \cref{fig:MultiPatchBasic} that lattice surgery require some additional workspace to connect the various codeblocks.   \cref{fig:BigLayoutLabelled} presents a 2D layout to realize \TDTOFF using lattice surgery, including all necessary workspace.  If bit-flips are sufficiently small, then the factory can be realized completely with repetition codes.  In the regime where bit-flips are rare but not completely negligible, we use a mix of repetition codeblocks (for the $BU$-qubits) and $d_x=3$ thin surface codes (for the factory qubits) to tolerate a single physical bit-flip anywhere in the factory. Additional bit-flip protection could be achieved by increasing the $X$ distance of all code blocks and/or performing two rounds of \TDTOFF. Here we only describe a single round and primarily focus on the version using $d_x=3$ surface code blocks.

For now, we assume a supply of $\tof$ states generated from \BUTOFF. Then we can schedule the main \TDTOFF steps as listed in \cref{tab:FactoryClock}. The required 8 input $\ket{\tof}$ are divided into 2 batches of 4.  How quickly can a batch of 4 input $\ket{\tof}$ magic states be injected?  Each $\ket{\tof}$ state comprises 3 qubits, so there are a total of $12=3\times 4$ multi-patch Pauli measurements needed per batch. These can be partly parallelised. \cref{fig:BigLayoutLabelled} shows 4 horizontal empty regions that we will call access corridors labeled $\{A, AB, BC, C\}$.  This allows us to perform 4 multi-patch Pauli measurement in parallel.  There are some constraints on which multi-patch Pauli measurements are performed (further discussion in the caption of ~\cref{fig:BigLayoutLabelled}).  The first batch is injected in factory clock steps 1-3. The second batch is injected in factory clock steps 4-6. Factory clock step 7 performs the measurement of the check qubits, and starts the process of exiting some factory qubits out of the factory.  Factory clock step 8 completes the process of exiting the factory qubits.  Each factory clock step takes a time $(d_m+1)T_{\mathrm{surf}}$ where $T_{\mathrm{surf}}$ is the duration of 1 surface code cycle and $d_m$ is the number of surface code cycle used per multi-qubit Pauli measurement.  The $``+1"$ in $(d_m+1)$ provides time to perform high fidelity single qubit measurements and reset between rounds of multi-qubit Pauli measurement. Roughly, a single execution of \TDTOFF takes time $8(d_m+1)T_{\mathrm{surf}}$, though small extra additive timecosts may be incurred to execute \BUTOFF, which we discuss next.

\begin{table*}[t]
    \centering 
        \begin{tabular}{p{8cm}cx{2cm}cx{2cm}cx{2cm}} \toprule
        \textbf{Fault source and remarks} & \textbf{Propagated} & \textbf{Risk} & \textbf{Suppressing}   \\  
        & & & \textbf{parameter}  \\ \midrule
        $Z$-logical errors on repetition codes during storage & Backwards & not critical & $d_{\mathrm{rep}}$ \\ \midrule
           $Z$-logical errors on factory qubits during storage & Forwards & critical & $D_z$  \\ \midrule
            $Z$-logical errors on check qubits during storage & Forwards & not critical & $d_z$  \\ \midrule
             $X$ logical on repetition codes during storage & Backwards & not critical & $|\alpha|^2$  \\ \midrule
           $X$ logical on surface codes factory qubits & Stuck  & critical & $d_x$, $|\alpha|^2$  \\ \midrule
           Timelike error during lattice surgery multi-patch measurement. \textit{Remarks:} This flips multi-qubit measurement outcome (denoted $\omega^D_j$ in \cref{algo:LatticeSurgery}) but is equivalent to Pauli error on input magic state. See \cref{subsec:TimeLikeErrors} for details.  & Backwards & not critical & $d_m$    \\ \midrule
           Measurement failure when reseting after lattice surgery. \textit{Remarks:} This flips some single Pauli measurement outcome $m^D_j$ in \cref{algo:LatticeSurgery}. Equivalent to Pauli error on input magic state. & Backwards & not critical & $|\alpha|^2$   \\ \bottomrule
        \end{tabular}
        \caption{Fault sources due to imperfect Cliffords.  Each error is either propagated forwards or backwards, or it is stuck. We sum the probability of all stuck errors and add to the overall infidelity of \TDTOFF. Backwards propagated errors modify the noise distribution on the input magic states. Forwards propagated are handled by modifying the formulae (see \cref{eq:ModifiedAccept,eq:ModFidelity}) for the infidelity and acceptance probability. An error is a critical risk it occurs with probability $p$ and contributes to the overall infidelity with probability $O(p)$ rather than $O(p^2)$.  Every error source can be exponentially suppressed some parameter, where $\{d_{\mathrm{rep}},d_z, D_Z , d_x  \}$ are code distance illustrated in \cref{fig:BigLayoutLabelled}; $d_m$ is the measurement distance denoting the number of surface code cycles used during lattice surgery (see \cref{subsec:TimeLikeErrors}); and $|\alpha|^2$ is the mean phonon number in the cat code qubit.  For critical risk errors, the associated parameter is typically set higher than the parameters set for non-critical errors. In particular, the parameters $\{ d_z, d_m , d_{\mathrm{rep}} \}$ can be safely set at about half the value of $D_z$ though our actual choice is determined by numerical search.}
        \label{tab:CliffordFaults}
    \end{table*}

The \BUTOFF protocol can have a fairly high failure probability, labeled here by $F_{BU}$.  This failure probability depends on the repetition code distance $d_{BU}$ used in \BUTOFF. To boost the probability of having ample supply of states from \BUTOFF, we add redundancy in both time and space. Our illustrations show $M=20$ modules for \BUTOFF, but we only need 8 input $\ket{\tof}$ or $\ket{\ccz}$ states for the protocol.  Not all 8 input $\ket{\tof}$ or $\ket{\ccz}$ states need to exist at the same time as they are split into two batches.  Rather we aim to prepare 4 $\ket{\tof}$ at the start of factory clock steps 1 and 4. Therefore, during the factory clock steps 4-8 (a total time of $5(d_m+1)T_{\mathrm{surf}}$), we need to prepare 4 $\ket{\tof}$ for the first batch of the next round of \TDTOFF. During the factory clock steps 1-3 (a total time of $3(d_m+1)T_{\mathrm{surf}}$), we need to prepare 4 $\ket{\tof}$ for the second batch in the current round of \TDTOFF. Let us focus our discussion on preparation during steps 1-3 as this is the bottleneck point.      Furthermore, our schedule requires that, of these 4 $\ket{\tof}$ states, 2 are located on the left and 2 are located on the right.  Considering just one side, we have $M/2$ \BUTOFF modules. Of these $M/2$ modules, 2 are busy storing $\ket{\tof}$ states and performing the required lattice surgery operations.  This leaves $(M-1)/2$ modules responsible for preparing 2 $\ket{\tof}$ states.  Each attempt at \BUTOFF takes a time 
\begin{equation}
    T_{BU} =  2 d_{BU} T_{\mathrm{rep}} + \frac{d_{BU}+1}{2}(2+d_{BU}+1 )T_{\mathrm{cnot}} ,
\end{equation}
where $T_{\cnot}$ is the optimal time for a $\cnot$ gate and $T_{\mathrm{rep}}$ is the time for a repetition code cycle.  Therefore, steps 1-3 provide enough time to fit in $R := \lfloor 3(d_m+1)T_{\mathrm{surf}}/ T_{BU} \rfloor $ repeated attempts at \BUTOFF.  Given $R$ temporally multiplexed attempts, each \BUTOFF module has its failure probability reduced from $F_{BU}$ to $\tilde{F}_{BU}:=F_{BU}^R$.  Each side fails if there are zero or one module successes of the $(M-1)/2$ modules, which occurs with probability
\begin{align} 
\nonumber
    F_{\mathrm{side}} & = \tilde{F}_{BU}^{(M-1)/2} + \frac{M-1}{2} \tilde{F}_{BU}^{(M-1)/2}(1-\tilde{F}_{BU}) \\  \label{Fside}
    & = F_{BU}^{R(M-1)/2} + \frac{M-1}{2} F_{BU}^{R(M-1)/2}(1-F_{BU}^R) .
\end{align}
For instance, executing \BUTOFF at distance 5 and using $d_m=15$  surface code cycles per lattice surgery operation we have $R=3$ attempts at \BUTOFF (assuming $\kappa_1/\kappa_2 = 10^{-5}$ and $|\alpha|^2=8$).  If $F_{BU}=0.447$ then the temporal redundancy reduces this to $\tilde{F}_{BU}=F_{BU}^3=0.089$.  Providing $M=10$ modules in total, there is $(M-1)/2=3$ available spatial redundancy on each side, which further suppresses the failure probability to $ F_{\mathrm{side}}=0.023$.  This is already quite low.  We can further reduce the failure probability by either: increasing space cost $M$; or inserting a small number $Q$ additional rounds of $\BUTOFF$ between steps 3 and 4. In the latter case, the runtime of \TDTOFF is extended to
\begin{equation}
  T_{TD} =   Q T_{BU} +  8 (d_m + 1 ) T_{\mathrm{surf}},
\end{equation}
where we have assumed $Q T_{BU} \leq 2 (d_m + 1 )$.  This further reduces $F_{\mathrm{side}}$.  A coarse bound is obtained by replacing $R \rightarrow R+Q$ in \cref{Fside}, though actually the suppression is slightly better as there are now $M/2$ modules available for the $Q$ attempts.   We do not wish to set $Q$ too high, as the additional delay leads to logical error accumulation due to finite distance choices.

Whenever \BUTOFF fails to proceed the required $\ket{\tof}$ states, we count this as a failure of the whole \TDTOFF protocol. However, we use sufficient redundancy that such occurrences are very rare.   Typically, we set $Q=1$ or $Q=2$, and we use $M=10$ when $d_{BU}=5$ and $M=20$ when $d_{BU}=7$. 

An additional consideration is that a lattice dislocation is used when performing a multi-qubit Pauli measurement including a $Z_L$ on a repetition encoded logical qubit (see \cref{fig:MultiPatchBasic}).  This dislocation uses a small amount of additional space.  However, when using $\ket{\tof}$ input states (instead of $\ket{\ccz}$) the third qubit differs by a Hadamard and so the protocol is adjusted to measure $X_L$ and a dislocation is not required. For this reason, we inject the qubits in reverse order: $(C,j)_{BU}$, $(B,j)_{BU}$ then $(A,j)_{BU}$. After $(C,j)_{BU}$ is injected (without needing a dislocation) some space is freed-up for dislocations to be used, enabling $(B,j)_{BU}$ and $(A,j)_{BU}$ to be injected.

\subsection{Clifford noise}
\label{App:CliffordNoise}

\begin{table*} \centering
\begin{tabular}{cx{1.5cm}cx{1.5cm}cx{1.5cm}cx{1.5cm}cx{1cm}ccccc} \toprule 
 $\epsilon _{\text{TD}}$ & \# ATS  & $P_{\text{ACC}}$ (\%) & Time/Tof ($\mu  s$)  & $d_{\text{BU}}$ & $d_{\text{rep}}$ & $d_z$ & $D_z$ & $d_x$ & $d_m$ \\ 
 \midrule
 $7.6 * 10^{-6}$ & 1176 & 93 & 1602 & 5 & 5 & 7 & 15 & 3 & 11 \\
 $7.3 * 10^{-6}$ & 1218 & 73 & 1871 & 5 & 5 & 9 & 15 & 3 & 10 \\
 $4.8 * 10^{-6}$ & 1218 & 94 & 1723 & 5 & 5 & 9 & 15 & 3 & 12 \\
 $3.7 * 10^{-6}$ & 1260 & 93 & 1734 & 5 & 5 & 7 & 17 & 3 & 12 \\
 $1.5 * 10^{-6}$ & 1302 & 94 & 1724 & 5 & 5 & 9 & 17 & 3 & 12 \\
 $9.9 * 10^{-7}$ & 1386 & 98 & 1766 & 5 & 5 & 9 & 19 & 3 & 13 \\
 $8.6 * 10^{-7}$ & 1386 & 94 & 1721 & 5 & 7 & 9 & 17 & 3 & 12 \\
 $8.2 * 10^{-7}$ & 1428 & 94 & 1723 & 5 & 5 & 11 & 19 & 3 & 12 \\
 $6.1 * 10^{-7}$ & 1428 & 98 & 1889 & 5 & 5 & 11 & 19 & 3 & 14 \\
 $2.6 * 10^{-7}$ & 1470 & 98 & 1762 & 5 & 7 & 9 & 19 & 3 & 13 \\
 $1.3 * 10^{-7}$ & 1512 & 99 & 1886 & 5 & 7 & 11 & 19 & 3 & 14 \\
 $5.6 * 10^{-8}$ & 1596 & 99 & 1886 & 5 & 7 & 11 & 21 & 3 & 14 \\
 $4.1 * 10^{-8}$ & 1596 & 99 & 2137 & 5 & 7 & 11 & 21 & 3 & 16 \\
 $3.3 * 10^{-8}$ & 1680 & 99 & 2011 & 5 & 7 & 11 & 23 & 3 & 15 \\
 $2.8 * 10^{-8}$ & 1680 & 99 & 2262 & 5 & 7 & 11 & 23 & 3 & 17 \\
 $1.7 * 10^{-8}$ & 1722 & 99 & 2136 & 5 & 7 & 13 & 23 & 3 & 16 \\
 $1.5 * 10^{-8}$ & 1722 & 99 & 2388 & 5 & 7 & 13 & 23 & 3 & 18 \\
 $1.3 * 10^{-8}$ & 1806 & 99 & 2388 & 5 & 7 & 13 & 25 & 3 & 18 \\
 $1.3 * 10^{-8}$ & 1890 & 99 & 2513 & 5 & 7 & 13 & 27 & 3 & 19 \\
 $1.1 * 10^{-8}$ & 1890 & 99 & 2388 & 5 & 9 & 13 & 25 & 3 & 18 \\
 $1.1 * 10^{-8}$ & 1932 & 99 & 2388 & 5 & 7 & 15 & 27 & 3 & 18 \\
 $1.1 * 10^{-8}$ & 1932 & 99 & 2639 & 5 & 7 & 15 & 27 & 3 & 20 \\
 $1.1 * 10^{-8}$ & 1974 & 99 & 2513 & 5 & 9 & 13 & 27 & 3 & 19 \\
 $1.1 * 10^{-8}$ & 2016 & 99 & 2639 & 5 & 7 & 15 & 29 & 3 & 20 \\
 $9.0 * 10^{-9}$ & 2016 & 99 & 2388 & 5 & 9 & 15 & 27 & 3 & 18 \\
 $8.9 * 10^{-9}$ & 2016 & 99 & 2639 & 5 & 9 & 15 & 27 & 3 & 20 \\
 $8.8 * 10^{-9}$ & 2100 & 99 & 2639 & 5 & 9 & 15 & 29 & 3 & 20 \\
 $8.5 * 10^{-9}$ & 2226 & 99 & 2639 & 5 & 9 & 17 & 31 & 3 & 20 \\
 $8.5 * 10^{-9}$ & 2436 & 99 & 2639 & 5 & 9 & 19 & 35 & 3 & 20 \\
 $8.5 * 10^{-9}$ & 2604 & 99 & 2764 & 5 & 9 & 19 & 39 & 3 & 21 \\
 \bottomrule 
\end{tabular}
\caption{Assuming $\kappa_1 / \kappa_2 = 10^{-5}$, $\kappa_\phi=0$ and $| \alpha |^2 =8$. Performance of optimized \TDTOFF factory using \BUTOFF with $d_{BU}=5$ and $M_{BU}=10$. Zero-dephasing noise.}
\label{tab:FactoryCost1}
\end{table*}

Perhaps one of the most importance aspects of magic state factory design is the choice of distance for various code blocks.  It is possible to use much smaller code distances within the factory than used inside the main algorithm. Using finite code distances leads to noisy Clifford gates, noisy lattice surgery operations and non-negligible memory noise.  This needs to be accounted for in addition to the error estimated by \cref{lem:DistillationWorks} under the assumption of ideal Cliffords.  Indeed, typically Clifford noise is the dominate source of errors and the error of \cref{lem:DistillationWorks} should instead be regarded as the minimum achievable error (with 1 round of \TDTOFF) in the limit of infinite code distances. 

Some of the relevant spatial code distance parameters are shown in \cref{fig:BigLayoutLabelled}.  An important additional quantity is the ``measurement distance" $d_m$ that is increased to suppress the effect of timelike errors during lattice surgery (see \cref{subsec:TimeLikeErrors} for further details).  A common choice in the literature is to set $d_m=\mathrm{max}[d_z,d_x]$, but this is by no means necessary or optimal. 

Rather than a Monte Carlo simulation of Clifford noise, we perform a computer-assisted analytical analysis. It is helpful to distinguish critical and non-critical faults. We say a Clifford fault is a critical risk if (assuming no other errors occur) it leads to an undetected fault on the output magic states. Conversely, a fault is a non-critical risk if it will be detected (assuming no other errors).  All sources of Clifford noise can be grouped into one of four classes
\begin{enumerate}
    \item Backwards propagating and not critical:  these are errors that can be commuted towards the start of the circuit, so that they act on a single noisy input $\vert \tof \rangle$ state.  If $\rho$ is the density matrix with only noise from \BUTOFF, the backwards propagating noise is applied so $\rho \rightarrow \rho'$. Then the effective $Z$ logical error distribution is determined from $\rho'$ using the procedure of \cref{App:genericnoise}.
    \item Forwards propagating and not critical: these errors can be commuted to the end of the circuit, so that they act on the check qubits in the factory just before they are measured.  
    \item Forwards propagating and critical: these errors can be commuted to the end of the circuit, so that they act on the output magic state qubits.
    \item Stuck errors and potentially critical: these are errors that are difficult to commute forwards or backwards through the circuit.  We sum the probability of these events and add it to the error rate on the output magic states.
\end{enumerate}
Our treatment of stuck errors means that we obtain an upper bound on the performance. One might be concerned that this bound is loose, but in practice the stuck errors are very rare and not, therefore, of major importance.  Indeed, if we instead attempted a Monte Carlo simulation, the statistical variance in the error estimate would exceed that of the total stuck error probability.  Therefore, our computer-assisted analytical analysis leads to more accurate results than Monte Carlo methods.  We further remark that while a mild amount of truncation of higher order processes is employed, we use the procedure of \cref{App:Truncation} to monitor this truncation error and verify that it is negligible.

We list all the source of imperfections in \cref{tab:CliffordFaults} and describe the propagation type and risk level.  Let us assume that backwards propagation has been performed and we have accounted for the effect of noise tailoring (recall \cref{App:Noisetailoring}) on the error distribution.  Following earlier notation of \cref{def:errors} and \cref{App:Noisetailoring}, we say $j^{\mathrm{th}}$ noisy $\tof$ state suffers fault $Z[\mathbf{e}_j]$ with probability $\mathbb{P}_j(\mathbf{e}_j)$ that we precompute.  Then, without any other noise sources, \cref{lem:DistillationWorks} would describe the acceptance probability and output infidelity. However, the factory qubits may be affected by some forwarded propagated error $Z[(\tilde{\mathbf{w}}^A,\tilde{\mathbf{w}}^B, \tilde{\mathbf{w}}^C )]$ where the labels $\{A,B,C \}$ refer to the 3 different blocks of factory qubits.  We used similar notation, without the tilde, in \cref{lem:ErrorProp} to describe how errors due to input magic states impact the protocol.  To combine with the forwarded propagated errors we simply replace $\mathbf{w} \rightarrow \mathbf{w} + \tilde{\mathbf{w}}$ to add the effect of the forwarded propagated errors and follow this modification through the analysis of \cref{lem:DistillationWorks}.  As we did earlier, it will be useful to split $\mathbf{w}=(\mathbf{v},\mathbf{u} )$ to distinguish errors on check qubits and output qubits. If the forwarded propagated error $\tilde{\mathbf{w}}$ on each block occurs with some probability $\mathbb{F}(\tilde{\mathbf{w}})$ then the results of \cref{lem:DistillationWorks} modify to
\begin{equation} \label{eq:ModifiedAccept}
    P_{\mathrm{acc}} = \sum_{
    \substack{ \tilde{\mathbf{w}}^D,\mathbf{e}^D  \\ [ \mathbf{e}^D G^D_0 = \tilde{\mathbf{v}}^D ]  \forall D}}
   \prod_{\substack{1 \leq j \leq 8 \\ D \in \{ A,B,C\} }} \mathbb{P}_j(\mathbf{e}_j)  \mathbb{F}(\tilde{\mathbf{w}}^D)  ,
\end{equation}
and 
\begin{equation} \label{eq:ModFidelity}
F = \frac{1}{P_{\mathrm{acc}}} \sum_{\substack{\tilde{\mathbf{w}}^D, \mathbf{e}^D \\ [ \tilde{\mathbf{e}}^D G^D_1 = \tilde{\mathbf{u}}^D ] \forall D} }   \prod_{\substack{1 \leq j \leq 8 \\ D \in \{ A,B,C\} }} \mathbb{P}_j(\mathbf{e}_j)  \mathbb{F}(\tilde{\mathbf{w}}^D)  .
\end{equation}
There are three important changes here.  First, in both equations we have summed over forwards propagated errors and weighted by the appropriate probability. In the acceptance probability the summation constraint $[ \mathbf{e}^D G^D_0 = \mathbf{0} ]  \forall D$ has been replaced by $[ \mathbf{e}^D G^D_0 = \tilde{\mathbf{v}}^D ]  \forall D$ since to pass the check measurement any forwards propagated error $\tilde{\mathbf{v}}^D $ must cancel (therefore equal) some other error to go undetected.  Similarly, in the fidelity expression we have replaced  $[ \mathbf{e}^D G^D_1 = \mathbf{0} ] \forall D$ with $[\mathbf{e}^D G^D_1 = \tilde{\mathbf{u}}^D ] \forall D$ because to contribute to the fidelity any forwarded propagated error $\tilde{\mathbf{u}}^D $ must cancel (therefore equal) some other error.

Calculating the expressions for $\mathbb{P}_j$, $\mathbb{F}$, performing the summation and adding the stuck error events is too involved to perform by hand. But it is relatively straightforward for a symbolic mathematics package such as Mathematica. Optimizing over various error suppressing parameters, we find the factory designs that achieve a certain target error per Toffoli at the minimum qubit and ATS cost (without making significant sacrifices to acceptance probabilities) and present results in \cref{tab:FactoryCost1}.

\subsection{The fidelity bottleneck in \TDTOFF}
\label{App:LimitingFactors}

 There are many contributing sources of error to the results presented in \cref{tab:FactorySummary,tab:FactoryCost1}.  In the main text summary of \cref{tab:FactorySummary}, the lowest reported Toffoli error probability was $2.3*10^{-9}$.  Here we discuss which error sources are the bottleneck factor that limit us from reaching higher fidelities with  \TDTOFF protocol.  We conclude with a discussion how the presented \TDTOFF protocol can be adapted to pass this bottleneck. 
 
Let us consider the process of bit-flip errors on noisy input $\tof$ states encoded in repetition codes.  For hardware parameters $\kappa_1/ \kappa_2 =10^{-5}$, $\kappa_\phi=0$ and $|\alpha^2|=8$, the lowest total error rate is $\delta:=2.7*10^{-8}$ per repetition code cycle (recall \cref{fig:TotalRepetitionCodeFail}). If a single $\tof$ state is stored for $r$ repetition code cycles, then roughly the accumulated error on qubits $B$ and $C$ is $ \delta r$ and this gives an additional contribution of $2 \delta r$ to the non-dominate noise contributions $\epsilon_2$.  Returning to our benchmark example, the output error is (roughly) lower bounded by 
\begin{align}
 C \epsilon_1  \epsilon_2' & \sim C \epsilon_1 \left( \epsilon_2 + 2 \delta r \right) \\ \nonumber
 & = C(2 * 10^{-5}) \left( 7.5*10^{-9} + 2*2.7*10^{-8} r \right).
\end{align}
where $C$ is some constant that depends on the exact details of the noise profile and we have discussed examples where $C \sim 2 $ and $C \sim 8$. Furthermore, for the larger factory examples in \cref{tab:FactorySummary} repetition codes could be in storage for as long as $r \sim 200$ repetition code cycles.  Together this approximate accounting indicates (with $r=200$ and $C=5$) that we should not expect output infidelities lower than $\sim 1.1*10^{-9}$. Of course, this is a rough estimation of one error source, just to provide the reader with some intuition.  Rather, for a precise accounting of all error sources the lowest observed infidelity was $2.5*10^{-9}$. However, there are several straightforward routes to reaching even lower infidelities. By converting immediately after \BUTOFF from repetition code to thin surface code we would reduce the time exposed to bit-flip errors. For our benchmark example, encoding directly into surface codes should enable us to get much closer to $1.2*10^{-12}$ infidelity (the ideal Clifford limit for noise tailored \TDTOFF).  Ultimately, arbitrarily high fidelities can be reached by concatenating \TDTOFF, though resource costs jump substantially with each level of concatenation. Alternatively, better fidelities could be reached it hardware parameters could be improved by either further suppressing bit-flips by increasing $|\alpha|^2$ or decreasing $\kappa_1 / \kappa_2$.  

\bibliographystyle{apsrev4-2}
\bibliography{refs}

\end{document}